%% file: thesisproposal.tex
\documentclass[11pt, a4paper, oneside]{Thesis} 

\graphicspath{{./Pictures/}} 
\usepackage{ tipa }
\usepackage[square, numbers, comma, sort&compress]{natbib} 
\hypersetup{urlcolor=black, colorlinks=true} 
\title{\ttitle} 

\usepackage{tikz}
\usepackage{setspace}
\usepackage{listings}
\usepackage{adjustbox}
\usepackage{pbox}
\usepackage{multirow}
\usepackage{braket}
\usepackage{amsmath}
\usepackage{slashed}

\definecolor{OrangeGSSI}{RGB}{237,113,45}
\definecolor{blue}{RGB}{25,25,112}

\thesistitle{Feasibility of a directional solar neutrino measurement with the CYGNO/INITIUM experiment} 
\supervisor{Prof. ssa Elisabetta \textsc{Baracchini}} 
\examiner{Dr. Name  \textsc{Surname}} 
\degree{Doctor of Philosophy} 
\authors{Samuele \textsc{Torelli}} 

\usepackage{pifont}
\usepackage{xcolor,colortbl}
\usepackage{flushend}
\usepackage[normalem]{ulem} 
\usepackage{multicol}
\usepackage{xcolor}
\usepackage{lineno}

\begin{document}
\frontmatter 

\setstretch{1.3} 

\fancyhead{} 
\rhead{\thepage} 
\lhead{} 

\newcommand{\HRule}{\rule{\linewidth}{0.5mm}} 

\hypersetup{pdftitle={\ttitle}}
\hypersetup{pdfsubject=\subjectname}
\hypersetup{pdfauthor=\authornames}
\hypersetup{pdfkeywords=\keywordnames}


\begin{titlepage}
\begin{center}

\includegraphics[width=0.4\textwidth]{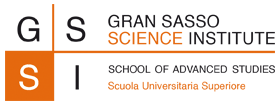}~\\[1cm]
\textsc{\Large Doctoral Thesis Research Proposal}\\[0.5cm] 

\HRule \\[0.1cm] 
{\huge \bfseries Feasibility of a directional solar neutrino\\[0.2cm]measurement with the \\[0.8cm] CYGNO/INITIUM experiment  }\\[0.3cm] 
\HRule \\[0.9cm] 

{\Large \textsc{PhD Program in Particle and Astroparticle Physics: XXXV cycle}}\\[2cm]

\begin{minipage}[t]{0.4\textwidth}
\begin{flushleft} \large
\emph{Author:}\\
\bigskip \authornames \\
\href{mailto:name.surname@gssi.infn.it}{samuele.torelli@gssi.it}
\end{flushleft}
\end{minipage}
\begin{minipage}[t]{0.5\textwidth}
\begin{flushright} \large
\emph{Thesis Advisor:} \\
\bigskip \supname \\
\href{mailto:name.surname@gssi.infn.it}{elisabetta.baracchini@gssi.it} \\
\bigskip \bigskip
\end{flushright}
\end{minipage}\\[2.2cm]
 

{\large \today}\\[2.2cm] 

\univname \\
\addressnames

\vfill
\end{center}

\end{titlepage}

\addtotoc{Abstract} 

\abstract 
Over the past five decades, solar neutrino research has been pivotal in driving significant scientific advancements, enriching our comprehension of both neutrino characteristics and solar processes. Despite numerous experiments dedicated to solar neutrino detection, a segment of the lower pp spectrum remains unexplored, while the precision of measurements from the CNO cycle remains insufficient to resolve the solar abundance problem determined by the discrepancy between the data gathered from helioseismology and the forecasts generated by stellar interior models for the Sun. The CYGNO/INITIUM experiment aims to deploy a large 30 m$^3$ directional detector for rare event searches focusing on Dark Matter. Recently, in the CYGNUS collaboration, there has been consideration for employing these time projection chamber technology in solar neutrino directional detection trough neutrino-electron elastic scattering. This is due to their potential to conduct low-threshold, high-precision measurements with spectroscopic neutrino energy reconstruction on an event-by-event basis driven by the kinematic. However, so far, no experiments have been investigated on the feasibility of this measurement using actual detector performances and background levels. Such a detector already with a volume of $\mathcal{O}(10)$ m$^3$ could perform an observation of solar neutrino from the pp chain with an unprecedented low threshold, while with larger volumes it could measure the CNO cycle eventually solving the solar metallicity problem.\\
The detector will consist of a gaseous time projection chamber filled with a Helium-CF$_4$ gas mixture, featuring an amplification stage of the primary ionization signal comprised of a triple stack of Gas Electron Multipliers with optical readout consisting of scientific complementary metal-oxide semiconductor (sCMOS) cameras and photomultiplier tubes. This configuration enables the imaging of recoil tracks, allowing the measurement of energy, direction, and track topology on an event-by-event basis. 
In synergy with CYGNO, the INITIUM project aims at developing negative ions operation in the context of the CYGNO time projection chamber with optical. Employing negative ions as charge carriers leads to a reduced diffusion, which implies better tracking performance and improved directionality capabilities.\\
This thesis delves into the feasibility of a directional solar neutrino measurement interacting trough elastic scattering with electrons in the medium by employing the CYGNO Time Projection Chamber approach.\\
To study the feasibility of this measurement, extensive research has been conducted on the detector's response to low-energy electron recoils. The study has been done with the latest and largest CYGNO's prototype known as LIME, featuring 50 cm drift length, a primary charge amplification area consisting of a triple stack of 33$\times$33 cm$^2$ Gas Electron Multipliers, and the light is readout by a sCMOS camera and 4 PMTs. To characterize the response to low-energy electron recoil, the detector has been exposed to X-Rays at different energies, and an assessment of the detector's light response and energy resolutions was carried out. 
The $_{s}\mathcal{P}\textrtaill{lot}$ tool has been used to unfold a set of 9 topological track shape variables for the signal for further comparison with the simulation. A $\sim$13\% energy resolution above 6 keV has been obtained.\\
Reducing electron cloud diffusion during drift towards the amplification stage is crucial for accurately measuring the topological information of recoil tracks. To implement and study Negative Ion Drift operation at atmospheric pressure with optical readout, within the INITIUM project, several measurements with negative ions as charge carriers have been performed with the MANGO prototype. Introducing highly electronegative gases causes electrons produced by ionizing particles to be captured by the electronegative molecules within a range of few micrometers. 
With the addition of a small amount of electronegative gaseous dopant, negative ions operations have been successfully performed, with the achievement of a transverse diffusion coefficient as low as 45 $\mu$m/$\sqrt{\text{cm}}$ with a drift field of 600 V/cm, one of the smallest ever measured in a gas detector.\\
Additionally, to better understand the detector response, and to study the angular resolution performances for low-energy electron recoil, a simulation aiming at replicating sCMOS-like images of tracks as they would appear in the detector has been developed. The simulation has been extensively tested on low-energy electron recoil in this thesis, and 
through a comparison between simulated electron recoils and real data, it has been demonstrated the capability to correctly reproduce the data in light response and energy resolution, but also in the reproduction of 9 topological track shape variables. \\
Subsequently, an algorithm to reconstruct the electron recoil initial direction has been developed for this thesis and tested on simulated tracks simulating. 
The results revealed an angular resolution on electron recoil tracks, defined as the standard deviation of the difference distribution between the measured production angle and the true production angle, of approximately 28° at 20 keV, which improves to approximately 13° at 70 keV. While for tracks with diffusion simulated under negative ions operations, it is obtained an improved angular resolution, ranging from approximately 20° at 20 keV to around 10° at 60 keV.\\
Finally, a sensitivity analysis of solar neutrino observation has been conducted on a CYGNO 30 m$^3$ experiment with the obtained resolutions. 
The background level for such a detector has been assessed through a GEANT4 simulation supposing it to be realized with an optimal geometry and with the most radiopure materials developed to these days. The analysis has been conducted employing a purely Bayesian statistical approach. The study reveals that a CYGNO 30 m$^3$ detector could achieve directional observation of solar neutrinos originating from the pp chain in $\sim$5.5 years with a confidence level above 3$\sigma$ if the background level can be constrained below 1760 ev/y. This corresponds to the ability of identifying 165 neutrino signal over, 9680 background events.

\clearpage 

\chapter*{Acknowledgements}
This project has received fundings under the European Union’s Horizon 2020 research and innovation programme from the European Research Council (ERC) grant agreement No 818744 and from the Italian Ministry of Education, University and Research through the project PRIN: Progetti di Ricerca di Rilevante Interesse Nazionale “Zero Radioactivity in Future experiment” (Prot. 2017T54J9J)

\clearpage 


%
%


\pagestyle{fancy} 

\lhead{\emph{Contents}} 
\tableofcontents 

\setstretch{1.3} 
%
%
%


\mainmatter 

\pagestyle{fancy} 


\input{chapters/introduction}

\input{chapters/directionaldetection}

\input{chapters/TheCYGNOINITUMExperiment}

\input{chapters/LIMEPrototype}

\input{chapters/MANGOProtRD}

\input{chapters/simulation}

\input{chapters/directionality}

\input{chapters/neutrinodetection}

\input{chapters/Conclusions}


\addtocontents{toc}{\vspace{2em}} 
\appendix 


\input{./appendices/appendixA}

\input{./appendices/appendixB}

\input{./appendices/appendixC}
\input{./appendices/appendixD}
\input{./appendices/appendixE}

%

\clearpage


\label{Bibliography}

\lhead{\emph{Bibliography}} 

\bibliographystyle{unsrtnat} 

\bibliography{Bibliography} 

\end{document}

%% file: chapters/introduction.tex
\chapter*{Introduction}
\addcontentsline{toc}{chapter}{Introduction}
\label{chap:intro}

Over the past 50 years, the study of solar neutrinos has been instrumental in driving significant scientific breakthroughs, contributing to our understanding of both neutrino properties and the Sun itself. Notably, solar neutrinos provided compelling evidence for the nuclear fusion processes powering the Sun. Additionally, solar neutrino experiments played a pivotal role in the discovery of neutrino flavor oscillations, revealing that neutrinos possess mass, contrary to the expectations of the Standard Model of Particle Physics. The Sun generates energy by fusing four protons into a helium nucleus through a series of nuclear reactions occurring in its core. Approximately 99\% of the Sun's energy originates from the fusion of protons in the pp-chain, while the remaining 1\% arises from nuclear reactions driven by the presence of carbon, nitrogen, and oxygen in the CNO cycle. Due to their small cross-section, neutrinos produced in these processes can traverse the Sun's interior and serve as direct messengers of the nuclear reactions occurring in its core, reaching the Earth.\\
Despite numerous experiments that have performed measurement of solar neutrino fluxes, a portion of the pp spectrum at lower energy remains unexplored, and the precision of measurements of neutrino flux from the CNO cycle is still insufficient to address the solar metallicity problem. To overcome this challenge, the proposal of employing a large-volume gaseous Time Projection Chamber as a directional detector has recently emerged. High-granularity tracking Time Projection Chambers could in fact achieve precision measurements of solar neutrinos interacting via neutrino-electron elastic scattering by measuring both the energy and direction of the scattered electron. Directionality offers the advantage of tolerating a large amount of background while maintaining performance comparable to non-directional detectors. Furthermore, the ability to measure the energy and direction of each scattered electron enables event-by-event reconstruction of the neutrino energy, allowing for a direct measurement of the neutrino spectrum.\\
The CYGNO/INITIUM project aims to develop an advanced optical readout gaseous Time Projection Chamber for directional Dark Matter searches and solar neutrino scattering on nuclei. The project is based at the Laboratori Nazionali del Gran Sasso. Notable features of CYGNO include the use of an optical readout system composed of sCMOS cameras and PMTs, combined with an amplification stage comprising multiple Gas Electron Multipliers. By combining the information of the high-granularity camera (x-y) with the fast sampling of the PMT (z), it is possible to perform 3D track reconstruction. The detector operates with a He:CF$_4$ gas mixture at local atmospheric pressure (900 mbar) and room temperature. 
Minimizing electron cloud diffusion as electrons drift towards the amplification stage is essential for precisely capturing the topological details of recoil tracks. In synergy with the CYGNO experiment, the ERC Consolidator Grant INITIUM project aims at implementing Negative Ion Drift operation within the CYGNO Time Projection Chamber 3D optical readout approach. By introducing an electronegative gas like $SF_6$, the primary ionization charge is captured by these molecules, leading to the drift of ions instead of electrons. This approach significantly minimizes diffusion, leading to better tracking performances, and thus improved directional performances.\\ 
The motivation behind developing such a detector is driven by its implications for Dark Matter physics. CYGNO/INITIUM is a project aiming to construct a 30 m$^3$ Time Projection Chamber primarily for directional Dark Matter searches. However, a recent proposal within the CYGNUS protocolcollaboration suggests utilizing such a detector for directional measurements of solar neutrinos trough neutrino-electron elastic scattering, and it is foreseen that already a $\mathcal{O}(10)$ m$^3$ detector could provide a first detection of neutrino originating from the pp chain.\\
This thesis work investigates the potential of conducting a directional measurement of solar neutrinos originating from the pp chain within the framework of the CYGNO/INITIUM project, utilizing a future 30 m$^3$ experiment. 
Thanks to its ability to tolerate a significant background levels given the directional capabilities, a 30 m$^3$ CYGNO detector, despite its relatively small exposure compared to other neutrino detectors at the same volume, could perform an observation of solar neutrino from the pp chain with lower thresholds than ever achieved before. Furthermore, currently, there are no operational detectors capable of performing this measurement. Such a measurement, still important by itself, is particularly significant as it could highlight the advantages of the directional detection technique, paving the way for the development of larger Time Projection Chamber detectors capable of possibly improving the precision on the measurement of the pp flux, and accurately measuring the flux of neutrinos from the CNO cycle. Moreover, for neutrinos produced in the Sun, knowing the direction of the incident particle, and measuring the energy and angle of the scattered electron closes the interaction kinematics. Consequently, it becomes feasible to reconstruct the neutrino energy on a per-event basis, allowing for a direct measurement of the neutrino flux. This method, for a $\mathcal{O}(100-1000)$ m$^3$ detector, can allow to resolve the $^7$Be and $pep$ neutrino fluxes as lines on top of the CNO continuum spectrum, possibly allowing for a precise measurement of the CNO neutrinos capable of resolving the solar abundance problem.
The sensitivity on the detection of these solar neutrinos with a Time Projection Chamber depends on the detector response, thus the performance of the detector in terms of energy response, energy resolution, and angular resolution of the electron recoils of interest, as well as the level of electron recoils background expected. Thus, in this thesis, these aspects have been extensively studied and applied to the case of physics mentioned above.
Throughout this work, the performance of the CYGNO approach has been investigated in terms of reconstructing electron recoil events. For the energy response, this analysis was conducted using real data collected from a detector resembling the modules to be used in the construction of the larger detector. However, due to the inability to perform directional calibration using real data, the study of directional capabilities was carried out using Monte Carlo simulations. To validate the assumption that Monte Carlo simulations can accurately represent the detector response, a thorough data/Montecarlo comparison was performed on the reconstructed tracks, confirming its adequacy. Subsequently, an algorithm for electron recoil directional reconstruction was developed, and the expected background for CYGNO-30 was estimated using GEANT4 simulations, leading to the determination of the sensitivity to this measurement.

In the initial Chapter of this thesis (Chap. \ref{chap:sota}) a full overview on solar neutrinos will be presented, covering an outline of solar neutrinos within the Standard Model, followed by an examination of their production mechanisms within the Sun, leading up to the present state of solar neutrino detection. Subsequently, the advantages and features of directional detection in the field of solar neutrinos will be discussed. The thesis will continue with an in-depth description of the CYGNO/INITIUM experiment (Chap. \ref{chap:CYGNO}). In this Chapter, a thorough explanation of the CYGNO technique will be provided, accompanied by a timeline of the project and a description of how the ERC Consolidator Grant INITIUM project fits in this context in synergy with CYGNO with the goal of improving its tracking performances. Furthermore, the Chapter will conclude with an overview of directional Dark Matter searches, which is the primary objective for which CYGNO/INITIUM has been proposed. The subsequent Chapter (Chapter \ref{chap:LIMEDetector}) will offer a comprehensive insight into the meticulous analysis developed for this thesis of the detector's response to low-energy electron recoil tracks, which mirror the same tracks expected from neutrino interaction. Following that, as part of the author thesis project, Chapter \ref{sec:NID} will provide a detailed description of the detector setup, the analysis methodology and the results obtained concerning Negative Ions Drift operations performed with the smallest prototype of INITIUM, named MANGO.
Continuing, the thesis will detail the track simulation algorithm developed for the CYGNO experiment in Chapter \ref{chap:Simulation}. The simulation, starting from a simulated electron or nuclear recoil, is able to accurately produce a sCMOS-like image of the track resembling a picture from the data. The Chapter will provide a comprehensive explanation of the simulation algorithm, including a description of the parameters and their optimized values. Furthermore, in the Chapter a comparison between data and simulated tracks of low-energy electron recoil in terms of light response, energy resolution, and a set of nine track topological variables will be presented. The data-Monte Carlo comparison has been developed to validate the simulation in the context of this thesis work, and it's of fundamental importance since the angular resolution performances will be studied on simulated tracks. In Chapter \ref{chap:directionality} the focus shifts towards the directionality measurement of low energy electron recoil. In the Chapter, the challenges in performing this measurement will be explained, followed by a description of the algorithm developed within this thesis for electron recoil initial direction determination. In this context, the algorithm optimization will be illustrated, together with the performances of angular resolution on low-energy electron recoil and the algorithm efficiency. Additionally, the directionality performances for tracks simulated with NID diffusion will be presented to study the effect of a lower diffusion on the directionality performances. Finally, the sensitivity for a directional measurement of solar neutrino from the pp chain, which is the core result of this thesis work, will be illustrated utilizing a fully Bayesian statistical approach (Chap. \ref{chap:solarnu}). The Chapter will start from the illustration of the signal model, followed by the calculation of the expected signal event rate. It will proceed towards the assumption done for the analysis, with the description of the simulation implemented to estimate the expected background for a 30 $m^3$ prototype. The Chapter will finally end with the sensitivity analysis done employing a Bayesian approach by means of toy Montecarlo considering both NID and ED directional performances.

%% file: chapters/directionaldetection.tex
\chapter{Solar neutrino directional detection}
\label{chap:sota}
The study of solar neutrinos has been a key motivation for advancements in neutrino and solar physics over the past half-century. This field serves as an intersection between particle physics, nuclear physics, and astrophysics, necessitating collaborative efforts from these three communities. The Super-Kamiokande \cite{Super-Kamiokande:2002weg} and the Borexino \cite{BorexinoScheme} experiments have extensively measured solar neutrino in different energy ranges and exploited different detection techniques.
However, the spectrum of solar neutrinos has not been fully explored in the low-energy part, and the measurement of the CNO cycle has not been performed with enough precision to address the solar metallicity problem.
Both these measurements could be improved with a detector featuring low-threshold and high capabilities of background discrimination. A Time Projection Chamber (TPC) developed for rare events searches, as a directional detector, would be suitable for performing high-precision measurements of solar neutrinos. This new approach, with a very low detection threshold, would be capable of measuring the neutrino produced in the Sun down to 50 keV on the neutrino energy covering the lower part of the pp chain thanks to the directional capabilities. Moreover, once the directionality approach is explored in depth, a larger detector could provide a precision measurement of neutrino from the CNO cycle. In this thesis, a high-precision innovative approach for solar neutrino measurement by means of a Time Projection Chamber, with a very low threshold and directional capability is proposed.
In this chapter, a description of neutrinos in the standard model framework will be provided (Sec. $\ref{sec:stdmodelNu}$), followed by an illustration of neutrino production mechanisms in the Sun (Sec. \ref{sec:solarnuprod}).  The chapter will proceed with an examination of the interaction mechanisms (Sec. \ref{sec:solarnuInt}) of neutrinos in the energy range typical of those produced in the Sun, moving on to illustrate the experimental challenges in solar neutrino detection (Sec. \ref{sec:solarnuchallenges}) and the techniques employed up to now (Sec. \ref{sec:solarnudetectiontec}). Finally, the current status of solar neutrino measurement will be outlined (Sec. \ref{sec:NuCurrentSttus}), along with the illustration of the directional TPC approach and the features of the directionality approach for this type of measurement (Sec. \ref{sec:nuTPC}).

\section{Neutrinos in the Standard Model}
\label{sec:stdmodelNu}
Unlike most particles discovered in the early years of particle physics, neutrinos were postulated theoretically before their initial experimental observation. The observation of the continuous spectrum in $\beta$ decays by Hahn and Meitner at the beginning of the second decade of the 20$^{th}$ century, later measured also by Chadwick in the same years, seemed to imply that, differently to $\alpha$ and $\gamma$ decays, in the $\beta$ decay process energy was not conserved \cite{RevModPhys.20.278, Franklin2009}. To preserve the conservation of energy, Wolfgang Pauli proposed a “remedy” to solve this problem in 1930. In his renowned letter addressed to participants of a meeting in Tübingen, he suggested the existence of a new, extremely light, and challenging to detect neutral particle having spin 1/2 (for angular momentum conservation rules). This particle would be emitted in the $\beta$ decay process, together with the electron, thereby accounting for the apparent missing energy.
An advancement on this topic was brought by Enrico Fermi, with the introduction of the four-fermion $\beta$ decay Hamiltonian in 1934 \cite{Fermi1934TentativoDU}, which transformed Pauli's initial conjecture into a predictive theory. This theory modeled gained increasing acceptance within the scientific community and reached its peak with the first observation of antineutrinos at the Savannah River reactor by Cowan and Reines in 1953 \cite{Cowan:1956rrn}. For the experiment, a nuclear reactor was used as a source of $\bar{\nu_e}$ with a flux of $\sim 10^{17}$ $\nu$/s/cm$^2$. An anti-neutrino, interacting with a proton in two water tanks via inverse beta decay $\bar{\nu}_e + \text{p} \rightarrow \text{n} +e^+$, produces a positron and a neutron.\\ 
Substantial advancements in particle physics have been achieved since then, propelled by a combination of groundbreaking experiments and theoretical contributions from luminaries such as Glashow, Salam, and Weinberg \cite{BILENKY198273}, Feynman \cite{feynman2010quantum}, Gell-Mann \cite{Gell-Mann:1964ewy}, Higgs \cite{Higgs:1964pj}, and numerous others. This progress culminated in the establishment of a comprehensive theoretical framework, known as the Standard Model (SM) of Particle Physics \cite{Gaillard_1999}. The SM successfully integrates the current understanding of the strong, weak, and electromagnetic interactions, emerging as a highly effective theory capable of explaining nearly all experimental results to date while also predicting diverse phenomena.\\
Within the framework of the Standard Model, there exist twelve elementary particles with spin $-\frac{1}{2}$ \cite{Butterworth:2016mrp}. These comprise six quarks, which interact not only through electromagnetic and weak forces but also through the strong force, and six leptons, which experience weak interaction, with the charged ones also interacting electromagnetically. Quarks and leptons are organized into three generations or families. Each family in the leptonic sector consists of a charged lepton ($e$, $\mu$, $\tau$) and a neutrino associated with the corresponding flavor ($\nu_e$, $\nu_{\mu}$, $\nu_{\tau}$).
Neutrinos, since they are electrically neutral particles, exclusively interact via weak interaction. As early as 1956, Wu et al. demonstrated the violation of parity in weak processes involving charge exchange \cite{MadWu:1957my}. This violation implied that only particles (or antiparticles) in a left-handed (or right-handed) helicity state will interact. Goldhaber et al. subsequently confirmed this in 1958 by showing that neutrinos are predominantly produced left-handed \cite{GoldhaberNuHeli}. This observation led to the conclusion that right-handed neutrinos, even existing, can not interact.\\
In the SM model, weak and electromagnetic interactions are combined in a $SU(2)_L \times U(1)_Y$ symmetry group and described within the electroweak theory \cite{SALAM1964168,Quigg:2002td}. The $SU(2)_L$ group represents the symmetry under weak isospin (I) transformations, with the index L meaning that the group operates on the left-handed chiral part of the fermion field. Conversely, the symmetry under the hypercharge Y is represented by the group $U(1)_Y$. This allows the expression of the electric charge generator Q to be defined as $Q = I_3 + Y /2$, with $Y$ proportional to the identity and $I_3$ being the third component of the generator of the weak isospin group.
\begin{table}
    \centering
    \includegraphics[width=0.6\linewidth]{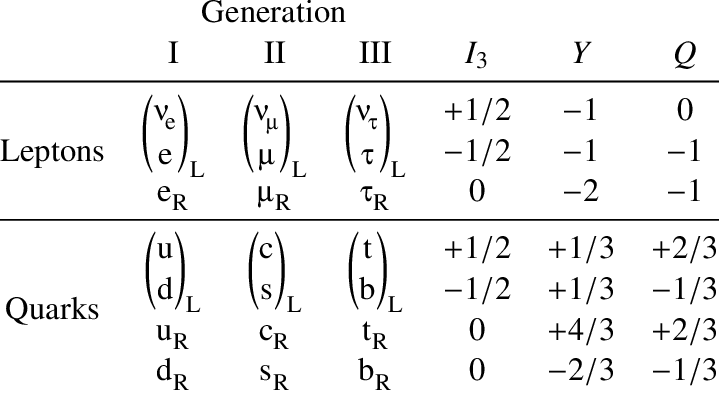}
    \caption{In table, the fermions in the Standard Model are presented, with information on isospin ($I_3$), hypercharge (Y), and electrical charge (Q).}
    \label{tab:fermionsSM}
\end{table}
The left-handed leptons consist of isospin doublets, represented as ($\nu_l$, $l$), with identical hypercharges, whereas the right-handed components exist as singlets under $SU(2)$. \\ The invariance of the theory under local gauge transformations brings to the introduction of four massless bosonic fields: a vector boson from the $U(1)$ group, acting as an isospin singlet, and an isospin triplet of vector bosons associated with the $SU(2)$ generators. With the spontaneous symmetry breaking, the generator of the $U(1)$ group, $B_{\mu}$, and the ones of $SU(2)$, $W_{\mu}^i={W_{\mu}^1,W_{\mu}^2,W_{\mu}^3}$ can be arranged to preserve the invariance of the Lagrangian under group transformation giving rise to the photon $\gamma$ and the carriers of the weak force $W^{\pm}$ and $Z_0$. The charged vector boson W can be expressed as 
\begin{equation}
    W^{\pm}_{\mu}=1/\sqrt{2}(W_\mu^1\mp i W_\mu^2)
\end{equation} 
with mass $m_W = gv/2$ where $g$ is the weak isospin coupling of the $SU(2)_L$ and $v$ is the Higgs field vacuum expectation value. The photon and the electrically neutral $Z_0$ boson can be expressed respectively as
\begin{equation}
    A_{\mu} = \frac{1}{\sqrt{g^2+g'^2}}(g'W_{\mu}^3+gB_{\mu})
\end{equation}
with $g'$ being the weak hypercharge coupling associated to $U(1)_Y$, with mass $m_A=0$, and 
\begin{equation}
    Z_{\mu}=\frac{1}{\sqrt{g^2+g'^2}}(gW_\mu^3-g'B_{\mu})
\end{equation}
with mass $m_Z=v/2\sqrt{g^2+g'^2}$ \cite{BETTINELLI_2009}. It can be observed that by defining $\cos(\theta_{W}) = \frac{g}{\sqrt{g^2+g'^2}}$ and $\sin(\theta_{W}) = \frac{g'}{\sqrt{g^2+g'^2}}$, the fields $A_\mu$ and $Z_{\mu}$ are expressed as a planar rotation of $B_{\mu}$ and $W_{\mu}^3$ by the angle $\theta_{W}$ called Weinberg angle. The masses of the weak interaction bosons are respectively $m_{W}$ = 80.4 GeV, $m_{Z}$ = 91.2 GeV (from \cite{PhysRevD.98.030001}), while it is 0 for the photon. These high masses of the weak interaction bosons result in a remarkably short interaction range, historically leading to the name $weak$ for this force. Additionally to the $W$ and $Z$ bosons, the Higgs field provides mass to all the Standard Model fermions. This occurs through the Yukawa coupling of the right-handed fermion with its left-handed doublet and the Higgs field. As right-handed neutrinos are not foreseen in the model, neutrinos remain massless in the Standard Model.\\
Left-handed neutrinos $\nu_L$ interact with other particles of the Standard Model via $weak$ interaction mediated by a $W_{\pm}$ or a $Z_{0}$ boson. Interactions mediated by $W_{\pm}$ are named charged-current (CC) interactions, coupling neutrinos to their isospin counterparts. This process is described by the interaction of the charged leptonic weak current with the $W$ field in the Lagrangian \cite{horejsi2022fundamentals} :
\begin{equation}
    \mathcal{L}^{CC} = -\frac{g}{2\sqrt{2}}\left( 2\sum_{\alpha}^{e,\mu,\tau}{\bar{\nu}_{\alpha L}\gamma^{\mu}\ell_{\alpha L}} \right)W_{\mu} + h.c
    \label{eq:cclagr}
\end{equation}
where g is the weak coupling constant, $\nu_{\alpha L}$ is the left-handed neutrino field associated to the leptonic flavor $\alpha$, $\bar{\nu}_{\alpha L}=\nu_{\alpha L}^{\dag}\gamma^0$ , $\gamma_{\mu}$ with $\mu=0,1,2,3$ are the Dirac matrices, $\ell_{\alpha L}$ is the left-handed leptonic field associated to the leptonic flavor $\alpha$, $W_{\mu}$ is the W field, and $h.c.$ represent the hermitian conjugate term. Charged-current interactions, which mediate the $\beta$ decay, were directly observed in the context of neutrino discovery in 1956.  \\
The neutral-current (NC) process is mediated by the $Z_0$ boson and is described by the neutral current Lagrangian \cite{horejsi2022fundamentals}: 
\begin{equation}
    \mathcal{L}^{NC} = -\frac{g}{2\cos(\theta_W)} \left( \sum_{\alpha}^{e,\mu,\tau}{ g_L^{\nu}\bar{\nu}_{\alpha L}\gamma^{\mu}\nu_{\alpha L} +g_L^l\bar{\ell}_{\alpha L} \gamma^{\mu}\ell_{\alpha L} + g_R^\ell \bar{\ell}_{\alpha R} \gamma^{\mu}\ell_{\alpha R} } \right) Z_{\mu} + h.c.
\end{equation}
where $\ell_{\alpha L/R}$ are the lepton fields, and $Z_{\mu}$ is the bosonic field. 
Since the $Z_0$ boson is generated by the superposition of two different fields $U(1)$ and $SU(2)$ is not purely left-handed, and involves also the right-handed fermions with different couplings which are represented by the term $g_{L/R}^{\nu/\ell}$. The first observation of neutral current interactions was made by the Gargamelle experiment at CERN in 1973 \cite{HASERT1973138}. The Gargamelle experiment consisted of a 12 m$^3$ heavy liquid Freon bubble chamber, and the detector was exposed to a high-energy muon neutrino beam.  Aside from the charge current interaction caused by the minimal contamination of electron neutrinos in the beam, resulting in only a negligible background, the characteristic observed signature for the neutral current interaction was a high energy electron recoil produced inside the detector compatible with a scattering from a neutrino from the beam. Since muon neutrino couldn't interact through charge current with electrons of the medium, the interaction was attributed to a neutral current. 


\subsection{Neutrino oscillations}
\label{sec:nuoscillation}
At the beginning of 1950 expecting that the Sun operates through nuclear fusion reactions that should produce neutrinos, experiments were constructed to detect these neutrinos.
The initial radiochemical experiments measuring neutrinos originating from nuclear reactions in the Sun produced unexpected results, observing a deficit of neutrinos significantly below expectations. This discrepancy is known as the solar neutrino problem \cite{Nakahata:2022xvq}. The first attempt to explain this deficit was proposed in 1957 by Bruno Pontecorvo \cite{Pontecorvo:1957qd}, considering an analogy with the theoretical explanation of the oscillation observed in $K_0-\bar{K}_0$ meson system \cite{BESHTOEV2011276}. Specifically, it was hypothesized that contrary to the predictions of the Standard Model, neutrinos could possess a small mass, and the eigenstates of weak interaction could not be the same as the mass eigenstates involved in their propagation. This mismatch in eigenstates implies that a neutrino produced in a specific flavor can be observed as a neutrino of a different flavor.\\
Within this framework, it is possible to represent the flavor eigenstates $\ket{\nu_{\alpha}}$ ($\alpha=e,\mu,\tau$) as a linear combination of Hamiltonian eigenstates $\ket{\nu_k}$ ($k=1,2,3$). Each Hamiltonian eigenstate is associated with a specific mass $m_k$, such that
\begin{equation}
    \mathcal{H}\ket{\nu_k} = E_k\ket{\nu_k}
\end{equation}
where $E_{k} = \sqrt{\textbf{p}^2 +m_k^2}$ is the energy of the particle and \textbf{p} is the tri-momentum.
A state with flavor $\alpha$ can hence be described as the superposition of the mass eigenstates as:
\begin{equation}
    \ket{\nu_{\alpha}} = \sum_{i=0}^{3}U^*_{\alpha k}\ket{\nu_k}
    \label{eq:Combination}
\end{equation}
where $U^*$ is the mixing matrix, which is a 3$\times$3 unitary matrix named afterward Pontecorvo, Maki, Nakagawa, and Sakata matrix (PMNS matrix), and the $^*$ represents the complex conjugate.
From these equations the temporal evolution of the state $\ket{\nu_{\alpha}(t)}$ can be described as:
\begin{equation}
    \ket{\nu_{\alpha}(t)}=\sum_{k=1}^{3}U^*_{\alpha k}e^{-iE_kt}\ket{\nu_k}
    \label{eq:TempEvo}
\end{equation}
Since U is a unitary matrix the Eq. \ref{eq:Combination} can be easily inverted and the Hamiltonian eigenstates can be expressed as a combination of the flavor ones:
\begin{equation}
    \ket{\nu_k} = \sum_{\alpha}^{e,\mu,\tau} U_{k\alpha}\ket{\nu_{\alpha}}
    \label{eq:CombinationInv}
\end{equation}
By substituting the expression of $\nu_{k}$ in Eq. \ref{eq:CombinationInv} into the equation \ref{eq:TempEvo} the evolution of the flavor eigenstate can be obtained
\begin{equation}
    \ket{\nu_{\alpha}(t)} = \sum_{\beta}^{e,\mu,\tau}\left( 
    \sum_{k=1}^{3} U^*_{\alpha k}e^{-iE_kt} U_{k\beta}  \right)\ket{\nu_{\beta}}
\end{equation}
The equation demonstrates that if U is not diagonal, the eigenstate $\nu_{\alpha}$ becomes a combination of all flavor eigenstates, allowing neutrinos to oscillate between different flavors. Therefore, the probability of observing a neutrino with flavor $\beta$ given its production with flavor $\alpha$ is given by
\begin{equation}
    P(\nu_{\alpha}\rightarrow \nu_{\beta}) (t) = | \braket{\nu_\beta}{\nu_\alpha(t)} |^2 = \sum_{k,j}U^*_{\alpha k} U_{\beta k}U_{\alpha j} U^*_{\beta j} e^{-i(E_k-E_j)t}
\end{equation}
In the case of ultra-relativistic neutrinos (which can always be considered true given their very small masses), the relation $E_{k}\simeq E+m^2_k/2E$ holds, with $E=|\textbf{p}|$. By rewriting the time t as the traveled distance L and factorizing the energy, the equation becomes
\begin{equation}
    P(\nu_{\alpha}\rightarrow \nu_{\beta}) (E,L) = \sum_{k,j}U^*_{\alpha k} U_{\beta k}U_{\alpha j} U^*_{\beta j} e^{-i\frac{\Delta m^2_{kj}L}{2E}}
    \label{eq:osc}
\end{equation}
with $\Delta m^2_{kj}$ is the difference of the mass square $m^2_k-m^2_j$.\\ The observation of neutrino oscillations was achieved through concurrent measurements of solar and atmospheric neutrinos conducted by the SNO and Kamiokande experiments \cite{Ahmad_2001}. Since then, this phenomenon has been investigated under various conditions utilizing diverse neutrino sources. This extensive research has yielded precise measurements of both the mixing parameters and the mass-squared differences of the Hamiltonian eigenstates. However, determining the absolute values of the neutrino masses remains one of the primary unresolved questions in neutrino physics.\\
The squared-mass differences accountable for the oscillation of solar neutrinos are $\Delta m^2_{12} = 7.53 \pm 0.18 \times 10^{-5} \text{eV}^2$ \cite{PDGneutrinoPar}, while those responsible for the oscillation of atmospheric neutrinos are $\Delta m^2_{31} \simeq \Delta m^2_{32} = 2.44 \pm 0.06 \times 10^{-3} eV^2$ \cite{PDGneutrinoPar}.\\ 
In numerous scenarios, such as solar neutrino, given the small angle $\theta_{13} \sim 8.7$° \cite{PDGneutrinoPar} compared to the others, and the closeness in mass of two of the mass states compared to the third, it is feasible to simplify the three-neutrino model and characterize the oscillation using a more straightforward effective approach that involves only two neutrinos. Within this framework, the mixing matrix U can be represented as a 2D rotation matrix that depends upon a single mixing angle
\begin{equation}
    U=\begin{pmatrix}
    \cos(\theta_{mix}) & \sin(\theta_{mix}) \\ 
    -\sin(\theta_{mix}) & \cos(\theta_{mix}) 
    \end{pmatrix}
\end{equation}
By rewriting Eq. \ref{eq:osc} within this framework, and considering the two neutrino flavors as $\ket{\nu_e}$ and $\ket{\nu_\mu}$ with the two mass eigenstate $\ket{\nu_1}$ and $\ket{\nu_2}$ with square mass difference $\Delta m_{12}^2$, it can be obtained \cite{Vissani}:
\begin{equation}
    P(\nu_{\alpha}\rightarrow \nu_{\beta})(E,L) = \delta_{\alpha\beta} -(2\delta_{\alpha\beta}-1)\sin^2(2\theta_{mix})\sin^2\left(\frac{\Delta m^2_{12}L}{4E}\right)
    \label{eq:osc2nu}
\end{equation}
where $\alpha,\beta = \{e,\mu\}$, and the $\delta$ is the Kronecker $\delta$ with $\delta_{ee} = \delta_{\mu\mu} =1$ and 0 otherwise. 
The Eq. \ref{eq:osc2nu} expresses the oscillation probability in the two flavors case for a neutrino with energy E, after traveling a distance L.
The oscillation probability depends on two main parameters: $\sin^2(\Delta m^2L/4E)$, which describes the frequency of the oscillations for a distance L and the neutrino kinetic energy E (from here the term "oscillations"), and $\sin^2(2\theta)$ which defines the amplitude of these oscillations.

\section{Mechanisms of neutrino production in the Sun}
\label{sec:solarnuprod}
The formation of a star starts with the gravitational collapse of a molecular cloud, primarily consisting of molecular Hydrogen, with trace amounts of other elements. As gravitational forces compress the cloud, it fragments into denser regions known as protostellar cores. Within these cores, gravitational contraction heats and compresses the gas, triggering nuclear fusion in the core once temperatures and pressures reach sufficient levels. This fusion process generates energy, balancing the inward gravitational force and leading to the establishment of hydrostatic equilibrium, thus giving rise to a stable, luminous star. As the star evolves, it exhausts its Hydrogen fuel in the core and begins fusing predominately Helium into heavier elements like Carbon, Oxygen, and eventually up to Iron. However, when a star reaches the fusion of Iron, a critical threshold is reached. Fusion of elements heavier than Iron requires more energy than they release, resulting in an endothermic reaction. Therefore, Iron fusion does not provide the energy needed to support the star against gravitational collapse. Subsequent stages of the star future depend on the star mass \cite{carroll2017introduction}. \\
The concept of the Sun being powered by nuclear reactions traces back to the beginning of 1920 \cite{eddington_1988}. However, it wasn't until Gamow theorized the tunnel effect in 1928 \cite{1928ZPhy5204G} that this idea was refined to offer a precise explanation of energy production in stars. In subsequent years, various authors explained solar energy production through fusion mechanisms which starting from four protons lead to the formation of a Helium nucleus, releasing 26.7 MeV of energy \cite{PhysRev53595,chandrasekhar1957introduction}. Two mechanisms for this conversion, both proposed in the late 1930s, emerged. The first, introduced by H. Bethe in \cite{PhysRev.54.248}, is the fusion chain involving two protons (pp chain), which starts with the combination of those into a Deuteron, is followed by a series of nuclear processes (Sec. \ref{sec:ppchain}). The second series of fusion process, introduced by Weitzsäacker \cite{von1937elementumwandlungen} and Bethe \cite{PhysRev.55.434}, is the CNO cycle, where the proton fusion reactions are catalyzed by the presence of $^{12}$C. Nowadays, these processes are well known and will be described in the following sections.\\ 
From \cite{bahcall1989neutrino} the reaction rate per unit of volume inside the star, given two nuclei A and B can be computed as:
\begin{equation}
    R_{A,B} = \frac{n(A)n(B)}{1+\delta_{AB}} \cdot \Bar{\sigma v}_{AB}
\end{equation}
where n is the density of nuclei of a given species, $\bar{\sigma v}$ is the cross-section times the velocity of the nuclei averaged on their thermal distribution, and the term $\delta_{AB}$ is the Kronecker delta and accounts for identical nuclei to renormalize the density. 
For nuclei A and B to interact, they have to surpass the repulsive Coulomb barrier. However, the thermal energy in a stellar environment is typically insufficient for the nuclei to overcome this barrier in an efficient way to balance the gravitational pressure. These interactions, although, occur through the quantum tunneling effect. This effect is a phenomenon where a particle, traverses through a potential energy barrier. This enables the particles to traverse the barrier of potential with a certain probability, even if they don't have sufficient energy to do it.\\ 
These phenomena are collectively described by the standard solar model, which provides the most accurate description of the Sun, taking into account the various physical processes and boundary conditions that must be satisfied to reproduce the Sun's current state \cite{bahcall1989neutrino}. Since the model is calibrated to match solar conditions, it heavily relies on the measured properties of the Sun, particularly the abundance of heavy elements. Some of these measurements rely on modeling the temperature and density profiles by studying the absorption line intensities with element abundances \cite{2005ASPC..336...25A}. Another method involves using the study of the composition of meteorites as a reference for refractory heavy metals present in the solar system \cite{Asplund_2009}. The standard solar model is constructed based on two different sets of metal abundances: the GS98 \cite{1998SSRv...85..161G} catalog for older photospheric analysis methods and the AGSS09 \cite{doi:10.1146/annurev.astro.46.060407.145222,Vinyoles_2017} for meteorite values for refractory elements. These two methods, both based on well reliable techniques, provide different results. The standard solar model constructed with GS98 abundances is termed the High-metallicity Standard Solar Model (HZ SSM), while the one with AGSS09 is called the Low-metallicity Standard Solar Model (LZ SSM) due to its lower metal content. These two models predict different abundances of heavy metal levels, and this discrepancy gives rise to what is called the solar metallicity puzzle \cite{Bahcall_2005}. Differences in metal abundance affect the Sun's seismic behavior but also influence temperature profiles, which, in turn, affect solar neutrino production rates. Predicted solar neutrino fluxes differ between the HZ and LZ SSMs, as it will be shown later in tab \ref{tab:ppfluxes} and \ref{tab:fluxCNO}. 
The cross-sections of nuclear processes occurring in the Sun are crucial components in determining the neutrino flux to the Earth. These indeed, despite having an uncertainty of $\mathcal{O}(1\%)$, represent the largest source of error in the neutrino flux prediction \cite{BAHCALL19981}. The measurement of these cross-sections is challenging due to the low reaction rates at these energies. 
To address this, measurements are conducted in underground laboratories where cosmic ray background is significantly reduced \cite{Best_2016}. An example of this is the LUNA-MV \cite{Prati_2020} facility underground at Gran Sasso Laboratories, specifically designed to study ion collision and to determine with high precision the cross-sections of nuclei interactions. The previous version of LUNA-MV, LUNA in its first phase has measured the cross-section of the process $^3\text{He}(^3\text{He},2p)^4\text{He}$ \cite{ARPESELLA1996452} in the pp chain, and the cross-section of one of the fundamental processes of the CNO cycle $^{14}\text{N}(p,\gamma)^{15}\text{O}$ \cite{BEMMERER2006297}.
Other cross-section of mechanisms investigated by Luna are the one of $^3\text{He}(^4\text{He},\gamma)^7\text{Be}$ \cite{FedericaConfortola_2006}, $^{17}\text{O}(p,\gamma)^{18}\text{F}$  \cite{PhysRevLett.109.202501}, and $^{13}\text{C}(\alpha,n)^{16}\text{O}$ \cite{LUNA13C}.
In cases where direct probing of the energy range of interest is not feasible, measured cross-sections are extrapolated, or theoretical calculations are performed when possible. In the following sections, a detailed description of the pp and the CNO cycle will be provided.

\subsection{The proton-proton chain}
\label{sec:ppchain}
\begin{figure}
    \centering
    \includegraphics[width=0.75\linewidth]{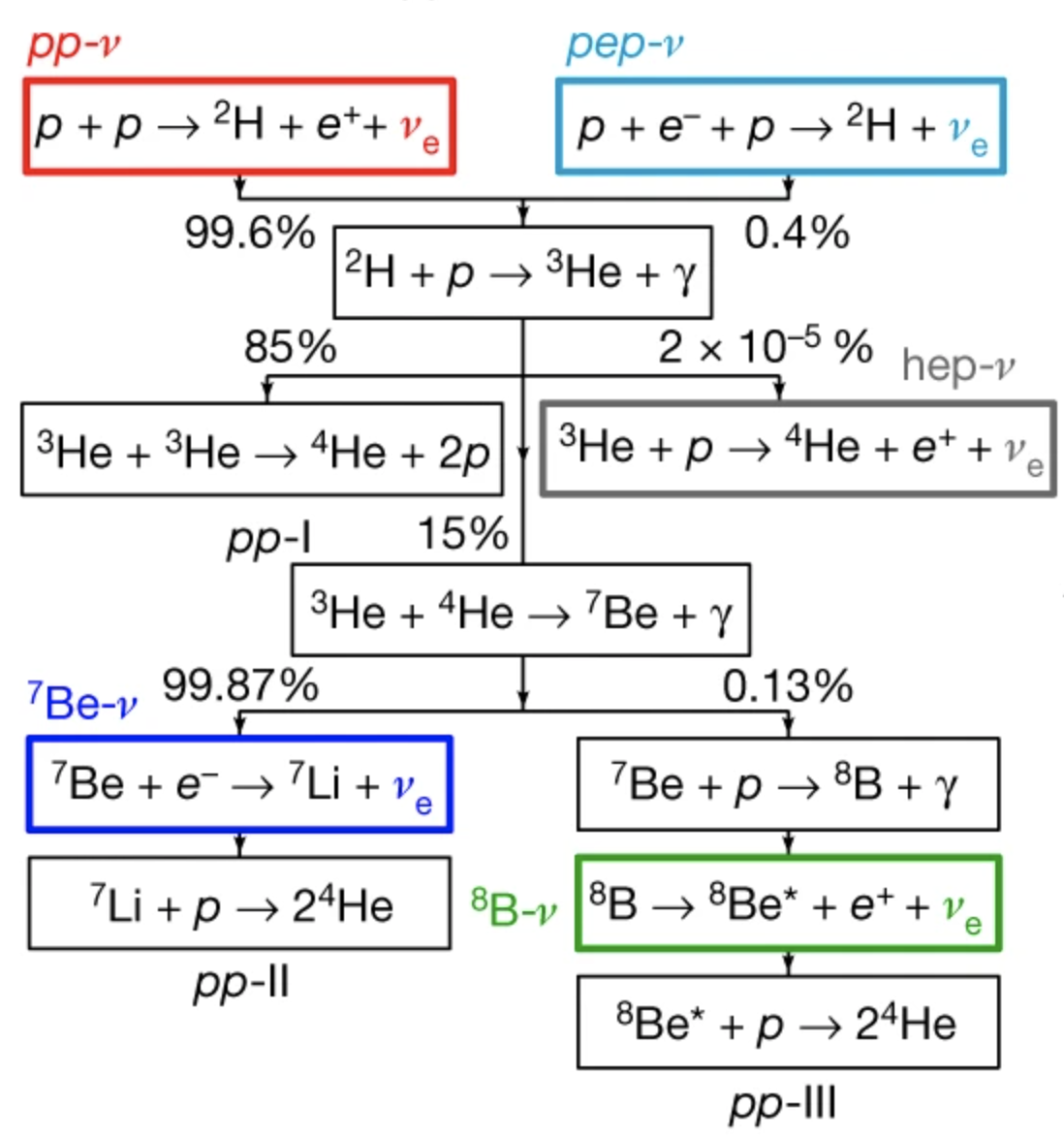}
    \caption{Scheme of the sequence of the proton-proton chain. The fractions on the arrow are the branching ratio of the process. The colored boxes around the processes highlight the ones in which a neutrino is emitted, together with the name associated to the emission. Picture from \cite{borexino2018comprehensive}.}
    \label{fig:ppchainnu}
\end{figure}

A detailed scheme of the proton-proton ($pp$) chain is presented in Fig. \ref{fig:ppchainnu}.\\
The proton-proton chain is the primary nuclear fusion process that powers young Sun-like stars, including about 99\% of its total energy output.
In the initial stage of the chain, two protons undergo the $pp$ fusion reaction to form as a final result a $^{2}$H, a positron, and an electron neutrino, with total energy released under the form of kinetic energy (Q value) of the reaction of 0.42 MeV.
\begin{equation}
    \textbf{pp} \ \ \ \ \ \text{p}+\text{p} \rightarrow\  ^2\text{H}+e^++\nu_e \ \ \ \ \ Q=0.42\  \text{MeV} 
\end{equation}
The energy released is distributed among the reaction products, leading to a continuum neutrino spectrum with endpoint at the Q value.
This is the main reaction of the whole pp chain, ruling the rate of all subsequent processes in the pp chain, with a branching ratio of 99.6\%. The cross-section of this process can be precisely calculated using the theory of weak interactions \cite{Adelberger_2011}.\\ Alternatively, a Deuteron can be generated in the $pep$ reaction, which engages two protons and an electron. Also, this reaction includes the production of an electron neutrino and has a total Q value of 1.44 MeV. 
\begin{equation}
    \textbf{pep} \ \ \ \ \ \text{p}+\text{e}^-+\text{p}\rightarrow\  ^2\text{H} + \nu_e \ \ \ \ \ Q=1.44\  \text{MeV}
\end{equation}
In this process, by kinematic the entire energy is transferred to the neutrino, resulting in the production of monochromatic neutrinos with an energy of 1.44 MeV. However, this process only contributes to a small fraction of Deuteron production, as accurate calculations based on weak interaction theory reveal its branching ratio to be of 0.4\% \cite{Adelberger_2011}.\\
The Deuteron produced by the previous processes undergoes then proton absorption, and it is transformed into a $^3$He, with an associated production of a photon, releasing an energy of 1.59 MeV 
\begin{equation}
   ^{\textbf{3}}\textbf{He} \ \ \ \ \ ^2\text{H} + \text{p} \rightarrow\  ^3\text{He} +\gamma \ \ \ \ \ Q = 1.59\ \text{MeV}
\end{equation} 
From this step, the $pp$ chain can advance in three distinct directions, resulting in three separate branches of the chain, determined by the interactions involving $^3$He. These branches are denoted as $pp$I, $^7$Be, and $hep$.\\
In the $pp$I branch, with a branching ratio of 85\%, an $^3$He undergoes fusion with another $^3$He, giving rise to a $^4$He and generating two free protons with a total Q value of 12.86 MeV
\begin{equation}
    \textbf{ppI} \ \ \ \ \ ^3\text{He}+\ ^3\text{He} \rightarrow\  ^4\text{He} + \text{p} + \text{p} \ \ \ \ \ Q=12.86 \text{MeV}
\end{equation}
Alternatively, the $^3$He can undergo fusion with a proton producing a $^4$He, a positron, and an electron neutrino
\begin{equation}
    \textbf{hep} \ \ \ \ \ ^3\text{He}+p\rightarrow \ ^4\text{He} +\text{e}^+ + \nu_e \ \ \ \ \ Q= 19.79 \text{MeV}
\end{equation}
This is the process involving the highest energies in the chain with a Q value of 19.79 MeV, despite it is very rare and happens only with a branching ratio of $10^{-5}$ \%. 
As a third alternative, the chain can proceed towards the production of a $^7_4$Be with the fusion of the $^3$He with an $^4$He. In this process along with the production of the $^7$Be a photon is emitted with a total Q value of 1.59 MeV and a branching ratio of 15\%
\begin{equation}
 ^{\textbf{7}}\textbf{Be} \ \ \ \ \   ^3\text{He} + ^4\text{He} \rightarrow\  ^7\text{Be}+ \gamma \ \ \ \ \  Q= 1.59 \text{MeV}
\end{equation}
After the $^7$Be production, the chain further split towards the branches $pp$II and $pp$III. The $^7$Be produced with a branching ratio of 99.87\% undergoes electron capture, with electrons present in the solar plasma entering the so-called $pp$II chain. As a result of this process, a $^7$Li is produced together with the emission of an electron neutrino, for a total Q value of 0.862 MeV
\begin{equation}
    \textbf{ppII} \ \ \ \ \ ^7\text{Be} +e^- \rightarrow \ ^7\text{Li} + \nu_e \ \ \ \ \ Q=0.862 \ \text{MeV}
\end{equation}
In this reaction, the $\nu_e$ by kinematic is emitted with a monochromatic energy. Nonetheless, from the reaction, with a 10\% probability the $^7$Li can be produced in an excited state with an energy of 487 keV. Thus, the neutrino can be emitted with an energy of 862 keV if the $^7$Li is not produced in the exited state or of 384 keV otherwise. This effect gives rise to the presence of two lines in the neutrino spectrum associated with this process. The $^7$Li subsequently can interact with a proton disintegrating into $^4$He, with a total Q value of 17.34 MeV
\begin{equation}
    \textbf{ppII} \ \ \ \ \ ^7\text{Li}+p \rightarrow \ ^4\text{He} + \ ^4\text{He} \ \ \ \ \ Q=17.34 MeV 
\end{equation}

Alternatively to the electron capture, with 0.13\% branching ratio, the $^7$Be can interact with a proton opening the branch of the $pp$III. As a result of this interaction, an $^8$B is produced together with a $\gamma$ with a total Q value of 0.14 MeV
\begin{equation}
   \textbf{ppIII} \ \ \ \ \ ^7\text{Be} +\text{p} \rightarrow \   ^8\text{B} +\gamma \ \ \ \ \ Q=0.14\  \text{MeV}
\end{equation}
The $^8$B produced is unstable and undergoes beta decay into $^8$Be with the emission of a positron and a neutrino with a total Q value of 17.98 MeV
\begin{equation}
   \textbf{ppIII} \ \ \ \ \ ^8\text{B} \rightarrow \ ^8\text{Be} +\text{e}^+ +\nu_e \ \ \ \ \ Q=17.98\  \text{MeV}
\end{equation}
These high-energy neutrinos with a beta decay spectrum and an endpoint at a very high Q value are the ones observed in the first solar neutrino observation from the Homestake experiment in 1960 \cite{PhysRevLett.20.1205}. Finally, the $^8$Be further decays into two $^4$He, ending the $pp$III chain 
\begin{equation}
   \textbf{ppIII} \ \ \ \ \ ^8\text{Be} \rightarrow\   ^4\text{He} + ^4\text{He} \ \ \ \ \ Q=0.14\  \text{MeV}
\end{equation}
All the $^4$He produced at the termination of the chain will then take again part in the chain.\\
The neutrino fluxes predicted are reported in table \ref{tab:ppfluxes}, for the two different solar models. 
\begin{table}[]
    \centering
\begin{tabular}{lcc}
\hline
$\nu$ flux & GS98 & AGSS09 \\\hline 
pp & $5.98(1 \pm 0.006)$ & $6.03(1 \pm 0.006)$ \\
pep & $1.44(1 \pm 0.012)$ & $1.47(1 \pm 0.012)$ \\
hep & $8.04(1 \pm 0.30)$ & $8.31(1 \pm 0.30)$ \\
$^7\mathrm{Be}$ & $5.00(1 \pm 0.07)$ & $4.56(1 \pm 0.07)$ \\
$^8\mathrm{B}$ & $5.58(1 \pm 0.14)$ & $4.59(1 \pm 0.14)$\\ \hline
\end{tabular}
    \caption{Neutrino fluxes from different reactions in units of $10^{10}$ (pp), $10^9$ ($^7$Be), $10^8$ (pep), $10^8$ ($^8$B), and $10^3$ (hep) cm\textsuperscript{-2} s\textsuperscript{-1}, predicted by the two different solar models. Values from \cite{Solarnufluxes}.}
    \label{tab:ppfluxes}
\end{table}

\subsection{The CNO cycle}
\begin{figure}
    \centering
    \includegraphics[width=0.7\linewidth]{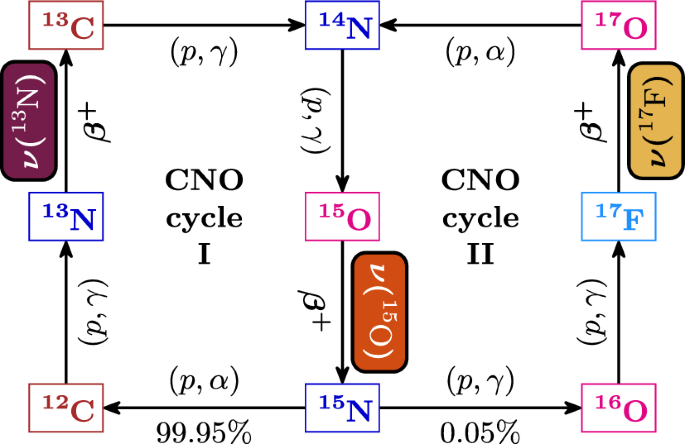}
    \caption{Sequence of the CNO nuclear reactions in the Sun, with the contributions in which the neutrino are produced highlighted. Picture from \cite{Agostini_2020Bi}.}
    \label{fig:CNOCycle}
\end{figure}

The energy generation in stars, arising from the conversion of four protons into a $^4$He nucleus, can also occur through a series of nuclear reactions in which Carbon, Nitrogen, and Oxygen (CNO) act as catalysts. 
Involving elements heavier than those considered in the pp chain, the Coulomb barrier influencing fusion reactions is higher, leading to a pronounced dependence on the local temperature of the star, due to the energy needed to overcome the barrier. 
The energy production rate of the CNO cycle scales $\propto T^{16.7}$, while that of the pp chain scales roughly as $\propto T^4$. Another crucial factor influencing the energy production from the CNO cycle is the abundance of Carbon, Nitrogen, and Oxygen (CNO elements) in the star core \cite{NPofS}. Being the Sun a young star, the pp chain assumes dominance as the primary source of energy production (99\%), but with the advancing of age, as the star ages and accumulates heavier elements, the CNO cycle emerges as the predominant mechanism for energy production at high temperatures. At the current status, the CNO contributes only for 1\% of the energy production in the Sun \cite{BorexinoCNO}.\\
While four distinct cycles of nuclear reactions involving various isotopes are theoretically possible in stellar environments, the primary source of energy in the Sun resulting from the fusion catalyzed by the CNO is the CNO-I cycle, while the contribution of the CNO-II cycle is minor (see Fig. \ref{fig:CNOCycle}). The impact of the other cycles is negligible in the case of the Sun.\\ 
Within the series of reactions comprising the Carbon-Nitrogen-Oxygen cycle, the collective transformation of four protons into a Helium nucleus, two positrons, and two electron neutrinos is catalyzed through the involvement of $^{12}$C, the predominant heavy isotope under typical stellar conditions. The overall reaction of the cycle, with a total Q value equal to the one of the pp chain of 26.7 MeV, is
\begin{equation}
    \textbf{CNO} \ \ \ \ \ ^{12}\text{C} +4\text{p} \rightarrow ^{12}\text{C} + ^4\text{He} +2\text{e}^+ + 2\nu_e
\end{equation}
The CNO cycle consists of two sub-cycles: the CN cycle (CNO-I) and the subordinate NO (CNO-II) cycle. Both of the cycles have a common branch (middle column of Fig. \ref{fig:CNOCycle}). As can be seen from Fig. \ref{fig:CNOCycle}, the breaking point towards CN or NO is when $^{15}$N is produced. $^{15}$N indeed, with a branching ratio of 0.05\% can absorb a proton and undergo deexcitation into $^{16}O$. Alternatively, with a 99.95\% BR, it can absorb a proton and decay with the emission of an alpha particle into $^{12}C$. Anyway, both these two branches will re-converge into the production of $^{14}$N again towards $^{15}$N.\\
Starting from $^{12}$C (bottom left of Fig.\ref{fig:CNOCycle}) the reactions in the CNO-I cycle are:
\begin{equation}
\begin{alignedat}{2}
     ^{12}\text{C} + \text{p} &\rightarrow \ ^{13}\text{N} +\gamma \ \ \ \ \ \ \ \ &&Q = 1.94\  \text{MeV} \\
     ^{13}\text{N} &\rightarrow\  ^{13}\text{C}+\text{e}^++\nu_e \ \ \ \ \  \ \ \ &&Q = 1.19\ \text{MeV} \\
     ^{13}\text{C} + \text{p} &\rightarrow\  ^{14}\text{N} +\gamma \ \ \ \ \ \ \ \ \ \ &&Q = 7.75\  \text{MeV}\\
     ^{14}\text{N} + \text{p} &\rightarrow\  ^{15}\text{O} +\gamma \ \ \ \ \ \ \ \ \ &&Q = 7.39\  \text{MeV}\\
     ^{15}\text{O} &\rightarrow\  ^{15}\text{N} +\text{e}^+ + \nu_e \ \ \ \ \ \ \ \ \ &&Q = 1.73\  \text{MeV}\\
     ^{15}\text{N} + \text{p} &\rightarrow\  ^{13}\text{N} +\gamma \ \ \ \ \ \ \ \ \ \ &&Q =  1.94\  \text{MeV}\\
     ^{15}\text{N} + \text{p}&\rightarrow\  ^{16}\text{O}^* \rightarrow\  ^{12}\text{C} + ^{4}\text{He} \ \ \ \ \ \ \  \ \ &&Q = 4.97\  \text{MeV}
\end{alignedat}
\end{equation}
In the CNO-I cycle, neutrinos are released through the $\beta$ decay of $^{13}$N and $^{15}$O, exhibiting a continuous spectrum with endpoints of 1.19 MeV and 1.73 MeV, respectively.
In the last passage, the $^{16}$O$^*$ undergoes fragmentation into a $^{4}\text{He}$ and a $^{12}\text{C}$. However, with a 0.05\% BR, it can undergo deexcitation with the emission of a gamma, opening the road to the secondary CNO-II cycle. The reactions in the CNO-II cycle are:
\begin{equation}
    \begin{alignedat}{2}
        ^{15}\text{N} + \text{p} &\rightarrow\  ^{16}\text{O}^* \rightarrow \ ^{16}\text{O} +\gamma \ \ \ \ \ \ \ \ &&Q = 12.13\  \text{MeV} \\
        ^{16}\text{O} + \text{p} &\rightarrow \ ^{17}\text{F} +\gamma \ \ \ \ \ \ \ \ &&Q = 0.06\  \text{MeV} \\
        ^{17}\text{F} &\rightarrow \ ^{17}\text{O} +\text{e}^+ +\nu_e \ \ \ \ \ \ \ \ &&Q = 1.74\  \text{MeV} \\
        ^{17}\text{O} + \text{p} &\rightarrow \ ^{14}\text{N} + ^{4}He \ \ \ \ \ \ \ \ &&Q = 1.20\  \text{MeV} \\
    \end{alignedat}
\end{equation} 
Due to the low probability of the initial deexcitation reaction with respect to fragmentation, the CNO-II cycle contributes to only a small fraction of the overall energy generated within the CNO mechanism. In this cycle, depicted in Fig. \ref{fig:CNOCycle}, right part, neutrinos are generated through the $\beta$ decay of $^{17}$F, with an endpoint of 1.74 MeV. Neutrinos from the CNO cycle have been recently observed by the Borexino experiment \cite{BorexinoCNO}. The fluxes are shown in table \ref{tab:fluxCNO}.\\
\begin{table}[]
    \centering
\begin{tabular}{lcc}
\hline
$\nu$ flux & GS98 & AGSS09 \\
\hline${ }^{13} \mathrm{N}$ & $2.96(1 \pm 0.14)$ & $2.17(1 \pm 0.14)$ \\
${ }^{15} \mathrm{O}$ & $2.23(1 \pm 0.15)$ & $1.56(1 \pm 0.15)$ \\
${ }^{17} \mathrm{F}$ & $5.52(1 \pm 0.17)$ & $3.40(1 \pm 0.16)$\\ \hline
\end{tabular}
    \caption{Neutrino fluxes for the different reaction of the CNO cycle presented in units of $10^8$ ($^{13}$N, $^{15}$O), and $10^8$ ($^{17}$F) cm\textsuperscript{-2} s\textsuperscript{-1} predicted by the two different solar models. Values from \cite{Solarnufluxes}.}
    \label{tab:fluxCNO}
\end{table}

\begin{figure}
    \centering
    \includegraphics[width=0.75\linewidth]{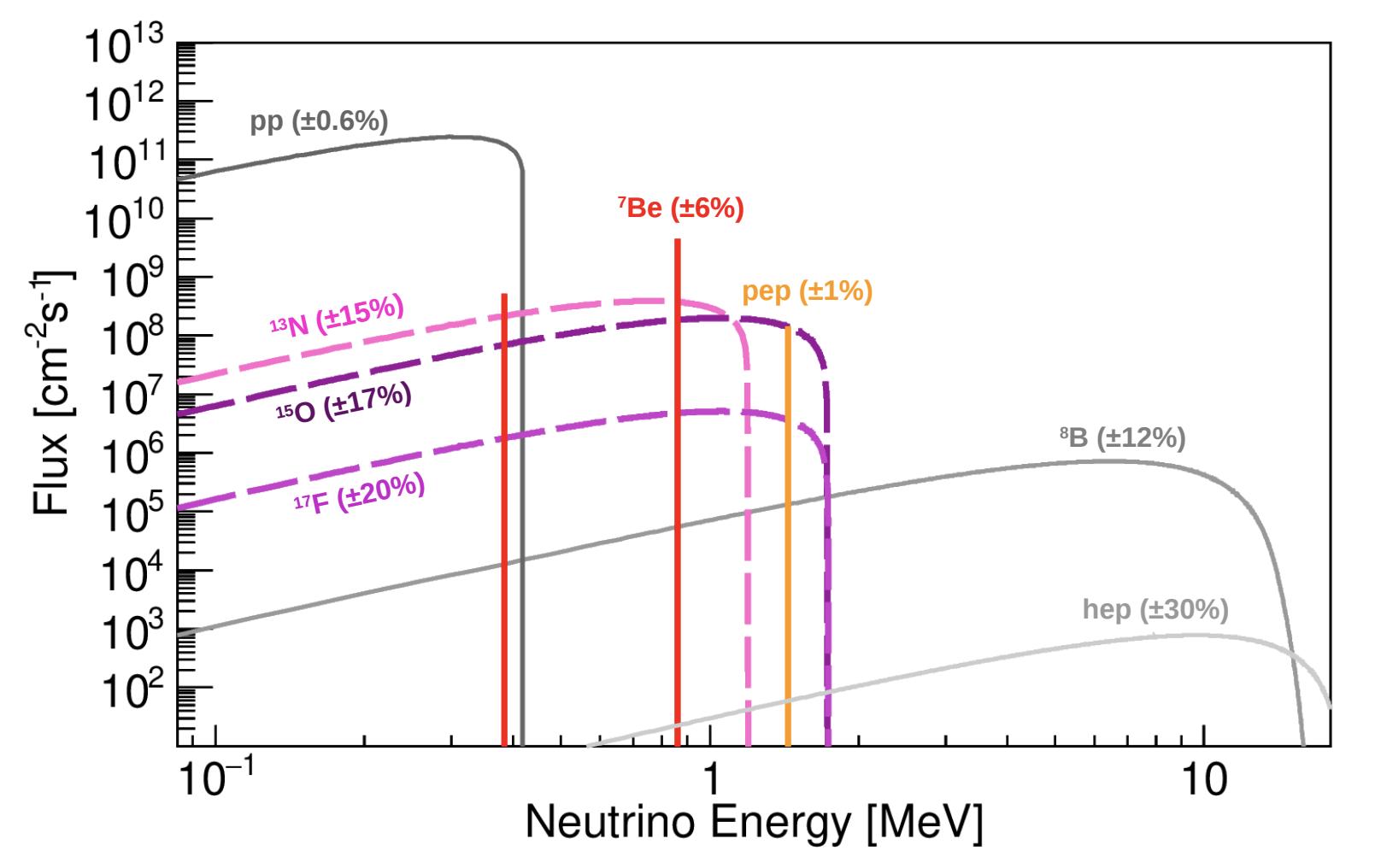}
    \caption{Theoretical spectral distributions of solar neutrinos originating from the pp chain (depicted by solid lines) and CNO cycle (represented by dashed lines). The values enclosed in parentheses indicate the relative uncertainties associated with the predictions. Plot from \cite{borexino2018comprehensive}.}
    \label{fig:SolarNuFull}
\end{figure}

The full spectrum of solar neutrinos from the Sun is shown in Fig. \ref{fig:SolarNuFull}.

\section{Scattering interactions of Solar Neutrinos}
\label{sec:solarnuInt}
Given the discussion in Sec. \ref{sec:stdmodelNu} and the energies involved ($<$20 MeV), neutrinos produced in the Sun, once they reach Earth, interact trough scattering interaction in 2 main processes: neutrino-electron elastics scattering ($\nu e$ES), coherent elastic neutrino-nucleus scattering (CE$\nu$NS). In the first process, neutrinos can interact with electrons, transferring part of their energy to them.
The second interaction occurs through a neutral current process wherein the neutrino scatters coherently with the entire nucleus $A$ in the process $\nu + A \rightarrow \nu + A$.
These processes will be described in more detail in the following paragraphs.

\subsection{Neutrino elastic scattering on electrons ($\nu e$ES)}
\label{sec:NueES}
\begin{figure}
    \centering
    \includegraphics[scale=.3]{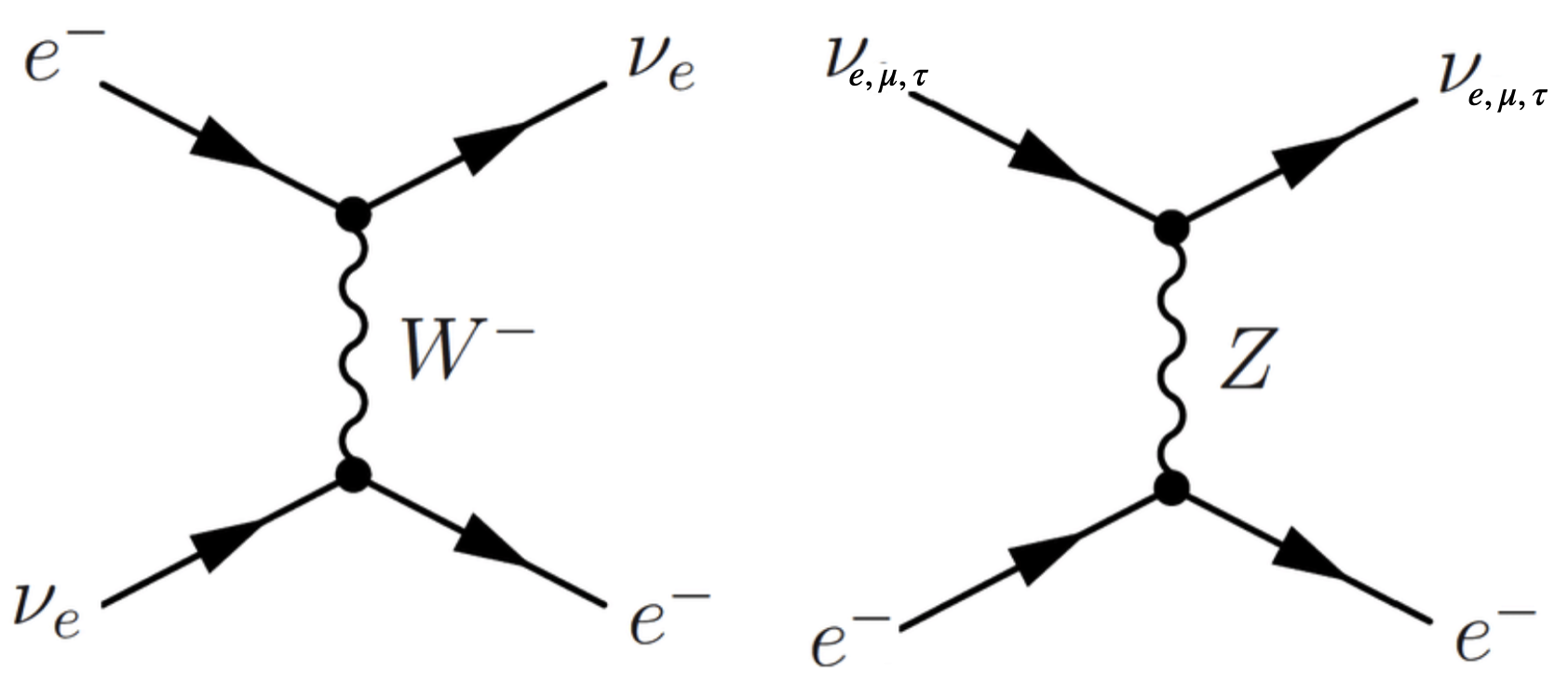}
    \caption{Left: Feynman diagram of electron neutrino interacting on an electron through a charge current interaction. Right: neutrino interacting on an electron through a neutral current interaction.}
    \label{fig:nuIntOne}
\end{figure}

Neutrino-electron elastic scattering ($\nu$eES) constitutes a purely leptonic process in which a neutrino scatters off an electron through the exchange of a virtual vector boson.
The elastic neutrino-electron scattering kinematic can be completely characterized by a single kinematic variable, for example, by $\theta$, representing the angle of the outgoing particle with respect to the initial neutrino direction. Given a neutrino with energy $E_\nu$, the kinetic energy $T'_e(\theta)$ transferred to a free electron at rest in the laboratory reference frame, which recoils with an angle $\theta$ with respect to the initial neutrino direction is given by the equation:
\begin{equation}
        T'_e(\theta)= \frac{2 E_\nu^2 m_e cos^2(\theta)}{(E_\nu+m_e)^2-E_\nu^2 cos^2(\theta)}
        \label{eq:kinematic}
\end{equation}
where $m_e$ is the mass of the electron. The maximum value for $T'_e$ is obtained for $\theta=0$, corresponding to the value at which the electron is emitted forward, and the neutrino is back-scattered. The corresponding value is:
\begin{equation}
        T_{e,Max}=\frac{2E_\nu^2}{m_e+2E_\nu}
    \label{TMax}
\end{equation}
To conserve energy and momentum the electron will be primarily scattered in a semisphere cut by the plane perpendicular to the neutrino incoming direction, on the side pointed by the neutrino direction.\\
In calculating the cross-section, it is crucial to consider that the interaction can take place through either a charged current process, mediated by a $W^-$ boson, or a neutral current process, mediated by a $Z^0$ boson. A depiction of these two processes under the form of Feynman diagrams is presented in Fig. \ref{fig:nuIntOne}.
The first interaction can exclusively occur with electron neutrinos $\nu_e$, whereas the second interaction is possible with any neutrino flavor. Due to neutrino oscillation, electron neutrinos produced in the Sun can interact in the detector as neutrinos with a flavor different from the electronic one $\nu_{\mu,\tau}$. This interaction involves a distinct cross-section due to the diverse nuclear matrix elements entering the calculation.\\
Specifically, the differential cross-section $\frac{d\sigma_{\nu_e-e}(E_\nu,T'e)}{dT'{e}}$ for elastic scattering of $\nu_e$ on electrons, considering both charge current and neutral current interactions, can be expressed as a function of the neutrino energy $E_{\nu}$ and the kinetic energy of the outgoing electron $T'_{e}$ in the following way \cite{toft}:  
\begin{align}
\begin{split}
        \frac{d\sigma_{\nu_e-e}(E_\nu,T'_e)}{dT'_e}=\frac{G_F^2m_e}{2\pi}&\left\{(2+g_V+g_A)^2 +(g_V-g_A)^2\left(1-\frac{T'_e}{E_\nu}\right)^2\right.-\\
        &-\left.(g_V-g_A)(g_V+g_A+2)\frac{m_eT'_e}{E_\nu^2}\right\}
    \label{eq:DiffCS}
\end{split}
\end{align}
where $G_F$ is the Fermi coupling constant, $g_V$ and $g_{A}$ are the axial and vectorial coupling, which can be expressed in terms of the Weinberg angle $\theta_{W}$ as $g_V = g_L+g_r$ and $g_A = g_L-g_R$, with $g_L = -1/2 + \sin^2(\theta_W)$ and $g_R = \sin^2(\theta_W)$. The detailed calculation of the cross-section is reported in Appendix \ref{app:appendixC}.
By considering only the neutral current interaction, the differential cross-section $\frac{d\sigma_{\nu_{\mu,\tau}-e}(E_\nu,T'_e)}{dT'_e}$ for non-electron neutrino can be written as:
\begin{equation}
\small
        \frac{d\sigma_{\nu_{\mu,\tau}-e}(E_\nu,T'_e)}{dT'_e}=\frac{G_F^2m_e}{2\pi}\left\{(g_V+g_A)^2+(g_V-g_A)^2\left(1-\frac{T'_e}{E_\nu}\right)^2-(g_V^2-g_A^2)\frac{m_eT'_e}{E_\nu^2}\right\}
    \label{muCS}
\end{equation}

By considering a detection threshold on the electron kinetic energy of $T'_{e,Thr}$, the total cross-section can be obtained by integrating the Eq. \ref{eq:DiffCS} and Eq. \ref{muCS} from this minimum value of the electron kinetic energy (the threshold), up to the maximum value reported in Eq. \ref{TMax}. The total cross-sections as a function of the neutrino energy for both processes can be obtained as:  
\begin{equation}
\sigma_{\nu_{(\mu,\tau/e)-e}}(E_\nu)=\int_{T'_{e,Thr}}^{T'_{e,Max}}{\frac{d\sigma_{\nu_{(\mu,\tau/e)-e}}(E_\nu,T'_e)}{dT'_e} dT'_e}    
\end{equation}
The two integrated total cross-sections describing the interaction of a neutrino of energy E with a threshold on the scatter $T'_{e,Thr}$ are: 
\begin{multline}
        \sigma_{\nu_e}(E_\nu)=\frac{G_F^2m_e}{2\pi}\left\{(g_V+g_A+2)^2\left[\frac{2E_\nu^2}{(m_e+2E_\nu)}-T'_{e,Thr}\right]+\right.\\ 
        -(g_V-g_A)^2\frac{E_\nu}{3}\left[   \left(1-\frac{2E_\nu}{m_e+2E_\nu}\right)^3 -  \left(1-\frac{T'_{e,Thr}}{E_\nu}\right)^3 \right]+\\ -(g_V-g_A)\left.(g_V+g_A+2)\frac{m_e}{2}\left[ \frac{4E_\nu^2}{(m_e+2E_\nu)^2}-\frac{{T'_{e,Thr}}^2 }{E_\nu^2} \right] \right\}
        \label{CsIntegrNue}
\end{multline}
for an electron neutrino, and
\begin{multline}
        \sigma_{\nu_{\mu,\tau}}(E_\nu)=\frac{G_F^2m_e}{2\pi}\left\{(g_V+g_A)^2\left[\frac{2E_\nu^2}{(m_e+2E_\nu)}-T'_{e,Thr}\right]+\right.\\ 
        -(g_V-g_A)^2\frac{E_\nu}{3}\left[   \left(1-\frac{2E_\nu}{m_e+2E_\nu}\right)^3 -  \left(1-\frac{T'_{e,Thr}}{E_\nu}\right)^3 \right]+\\ -(g_V^2-g_A^2)\frac{m_e}{2}\left[ \frac{4E_\nu^2}{(m_e+2E_\nu)^2}-\left.\frac{{T'_{e,Thr}}^2 }{E_\nu^2} \right] \right\}
        \label{CsIntegrNumu}
\end{multline}
for a neutrino with a flavor different from the electronic one.\\ It can be noticed that in case of no threshold and $E_\nu>>m_e$, the Eq. \ref{CsIntegrNue} reduces to 
\begin{equation}
    \sigma(E_\nu)\simeq \frac{G_F^2 m_e}{2\pi}\left\{(g_V+g_A+2)^2+\frac{(g_V-g_A)^2}{3}\right\}E_\nu
\end{equation}
and inserting numerical values and multiplying by $(\hbar c)^2=3.8\cdot 10^{-28}$ GeV$^2$ cm$^2$ to convert GeV$^{-2}$ in cm$^2$, the expression becomes
\begin{equation}
    \sigma(E_\nu)\simeq 9.7\cdot 10^{-42} E_\nu \frac{cm^2}{GeV} = 0.97 \cdot 10^{-44} E_\nu \frac{cm^2}{MeV}
    \label{sec:SigmaNumbers}
\end{equation}
yielding the same result as in \cite{Vissani}. The detailed calculation on the cross-section carried out by the author can be found in Appendix \ref{app:appendixC}.

\subsection{Coherent elastic neutrino-nucleus scattering}
\label{sec:NuCoerent}
\begin{figure}
    \centering
    \includegraphics[width=0.3\linewidth]{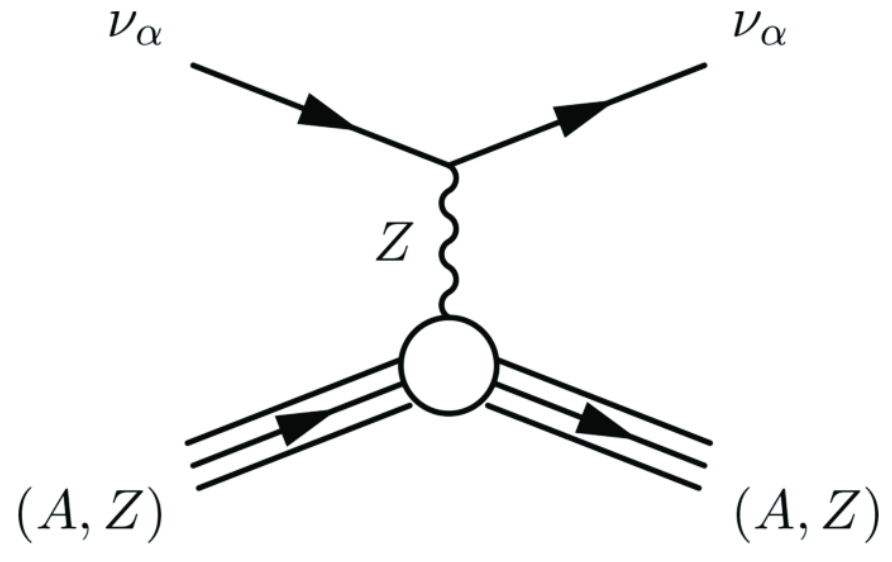}
    \caption{Feynman diagram of CE$\nu$NS process.}
    \label{fig:FenynmanDiacCEvNSIBD}
\end{figure}

Coherent elastic neutrino-nucleus scattering is a neutral-current phenomenon occurring when the momentum exchange between neutrinos and nuclei is smaller than the inverse of the nuclear size (nucleus De Broglie wavelength).
The process involves the scattering of a neutrino on a nucleus, treating this last one as a single particle. This is illustrated in terms of Feynman diagram in the left image of Fig. \ref{fig:FenynmanDiacCEvNSIBD}. 
In the Standard Model, CE$\nu$NS is primarily described by the neutral current interaction between neutrinos and nuclei. Due to the characteristics of SM couplings, the process cross-section scales proportionally to the square of the neutron number \cite{CEVNSN2}. Furthermore, the cross-section increases with the square of the neutrino energy, and the same does the maximum energy of the recoil. However, increasing the energy of the neutrino above 50 MeV, would not increase the recoil energy since it would lead to the loss of coherence in the scattering process \cite{Bednyakov_2018}.
The order zero cross-section of this process can be written as 
\begin{equation}
    \frac{d\sigma}{dT} = \frac{G_F^2M}{4\pi}\left(1-\frac{MT}{2E_{\nu}^2} - \frac{T}{E_\nu}\right) Q_W^2 [F_W(q^2)]^2
\end{equation}
where $M$ denotes the mass of the nucleus, $T$ represents its recoil energy, $F_W(q^2)$ is the weak form factor dependent on the nuclear density distribution of protons and neutrons, and $Q_W$ is the weak charge of the nucleus defined as $Q_W = Z(1-4\sin^2\theta_W)-N$, where $Z$ and $N$ are the respective numbers of protons and neutrons within the nucleus. Since the term in parentheses in the $Q_W$ equation is small due to the accidental suppression of the proton weak charge (approximately $10^{-1}$), the cross-section with nuclei is predominantly influenced by the $N^2$ term. Therefore, the cross-section is enhanced with nuclei featuring a higher number of neutrons. A more detailed calculation of the cross-section with all the corrections can be found in \cite{Abdullah:2022zue}.\\
The kinematics of this process mirrors that of electron scattering on a neutrino, with the distinction that the mass featured in the Eq. \ref{eq:kinematic} is not that of the electron but rather the nucleus one ($\mathcal{O}(10^4-10^5)$ times larger). 
Thus, despite having a relatively large total cross-section within neutrino interactions, usually greater than $10^{-41}$ cm$^2$ for $E_{\nu}>$5 MeV ($10^3$ times higher than $\nu_eeES$), thanks to the $N^2$ dependence of the cross-section, detecting CE$\nu$NS has been historically challenging due to the low energy transferred to the nucleus, typically on the order of few keV.
Despite its high cross-section and the large availability of neutrino sources, the first CE$\nu$NS observations have been performed only in 2017, from the COHERENT collaboration given the difficulties of the detection of such low energy nuclear recoils \cite{science.aao0990}. The experiment has bees carried out with a beam of proton impinging on a Hg target, producing a 1.7 $\times$ 10$^{11}$ $\nu_\mu$/cm$^2$/s neutrino beam from stopped pions produced in proton-mercury interactions. The $\pi^+$ stopped in matter produces a $\sim 30$ MeV muon neutrinos from their decays. The muon from $\pi^+$ decay travels approximately one-tenth of a millimeter before decaying at rest, resulting in the additional production of a $\bar{\nu_\mu}$ and $\nu_e$, each with a well-defined energy spectrum. The detector was constructed using a 14.57-kg CsI(Na) crystal, and the measurement involved comparing data collected during beam-on and beam-off periods.
From the comparison of the data background subtracted for the beam off and beam on samples, a rejection of the background-only hypothesis was obtained with a significance of 6.7$\sigma$ which leads to the claim of the observation \cite{Scholberg:2018vwg}. This first observation was followed by the employment of a liquid argon detector \cite{PhysRevLett.126.012002} confirming the $N^2$ behavior of the cross-section. 
This process has yet to be observed with reactor and solar neutrinos because in both cases, the neutrino energy is less than 20 MeV (less than 10 MeV for reactors), leading to the production of even lower-energy electron recoils. Furthermore, given that the flux of solar neutrinos from $^{8}B$, the most energetic source with a considerable flux, is only of the order of $10^6$/cm$^2$/s (which is $10^5$ times less the one employed from COHERENT), a very large exposure would be necessary to detect such neutrinos thought CE$\nu$NS process. The second and third generations of dark matter (DM) experiments will indeed encounter neutrinos interacting through Coherent Elastic Neutrino-Nucleus Scattering (CE$\nu$NS) as a significant source of background within the experiments.

\begin{figure}
    \centering
    \includegraphics[width=0.75\linewidth]{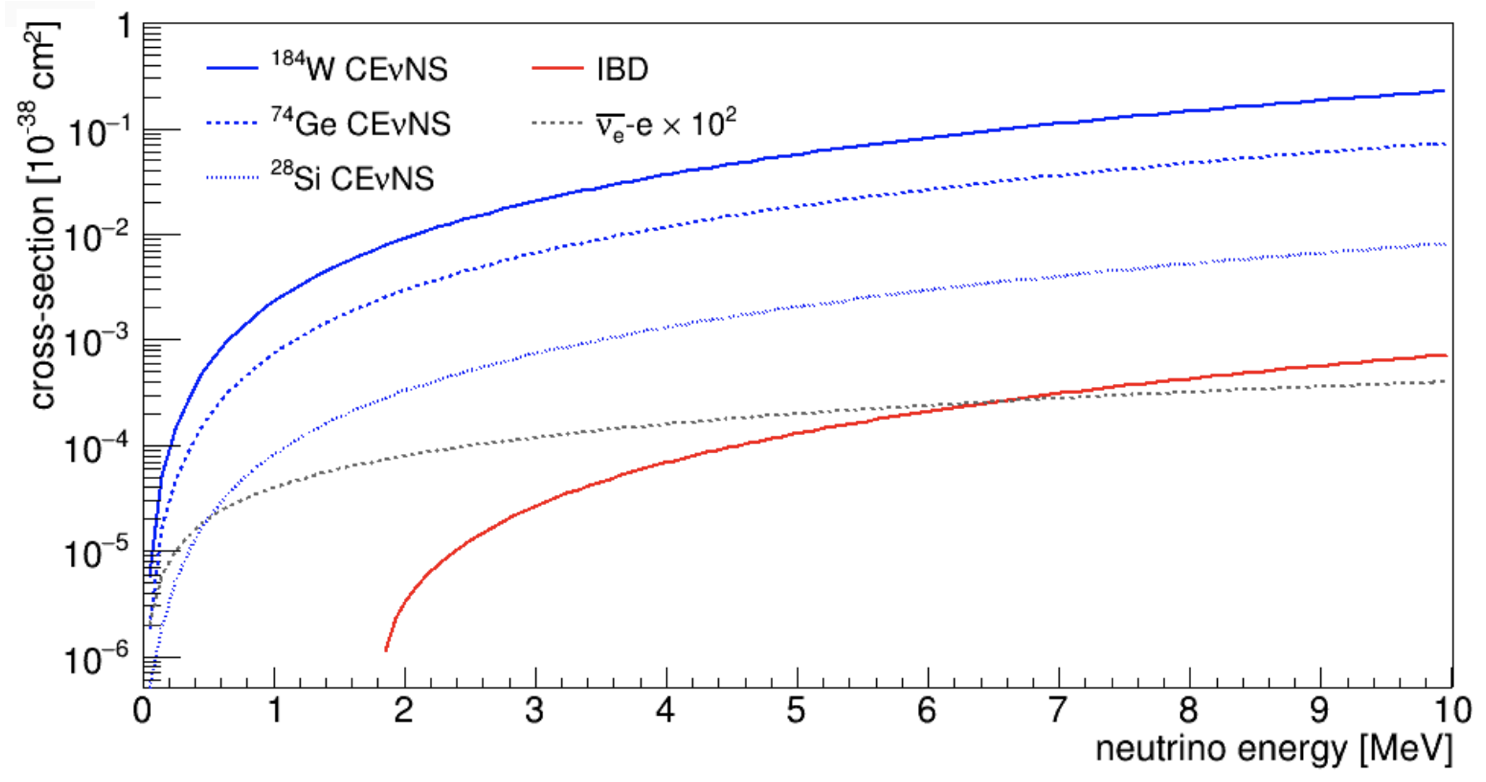}
    \caption{In figure, the cross-sections for $\nu e$ES, and CE$\nu$NS on different targets as a function of the energy are shown. In the plot, also the interaction via inverse beta decay is shown. Plot from \cite{nucrossSec}.}
    \label{fig:DifferentCs}
\end{figure}

A comparison of the cross-section for different processes is shown in Fig. \ref{fig:DifferentCs}.

\section{Solar neutrino measurement experimental challenges}
\label{sec:solarnuchallenges}
\begin{figure}
    \centering
    \includegraphics[width=1\linewidth]{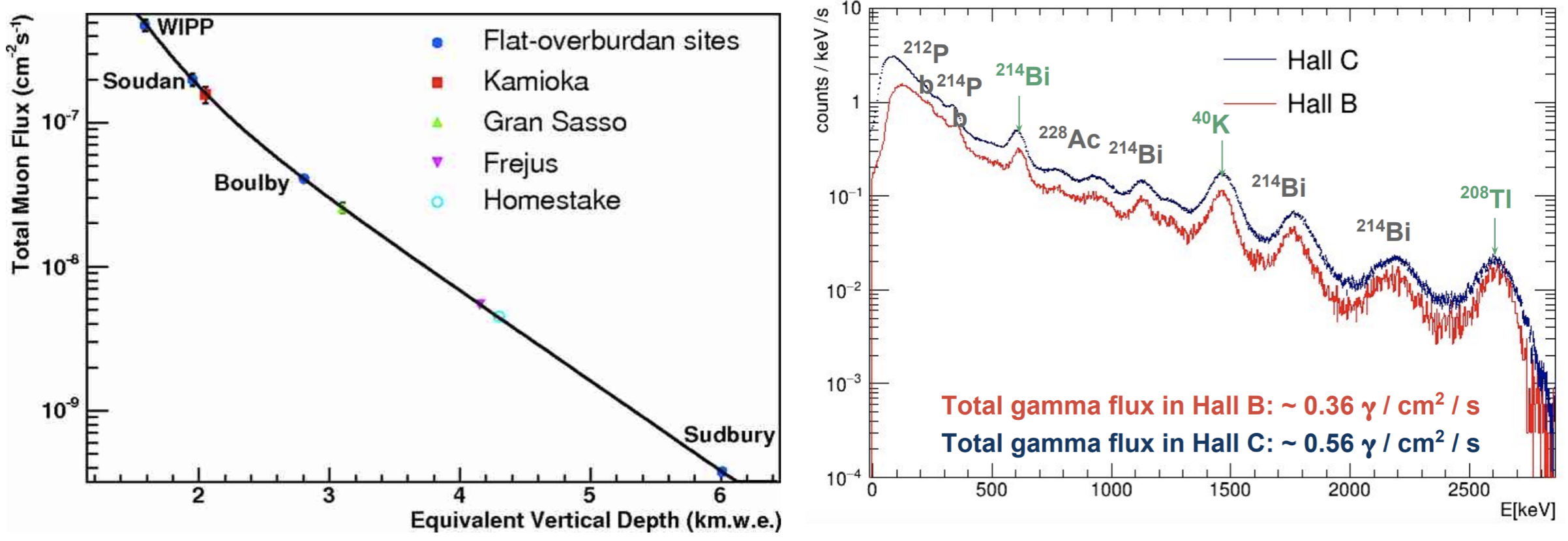}
    \caption{Left: cosmic rays flux as a function of the depth of the laboratory in km of water equivalent. Plot from \cite{Cosmics_fluxUnder}. Right: Gamma spectrum from environmental natural radioactivity. Plot from \cite{SabregammaSpec}.}
    \label{fig:CRGammaBkg}
\end{figure}

The particle interaction rate $R$ of a particle beam with a flux $\phi$ and a cross-section $\sigma$, impinging on a detector with number of target $N_t$ (electrons for $\nu e$ES, and nuclei for CE$\nu$NS) is given by the product of these quantities $R=\phi\cdot\sigma \cdot N_t$.  
Neutrino interacts only through weak interaction, with cross-sections for elastic scattering lower than $10^{-42}$ cm$^2$ as shown in Sec. \ref{sec:solarnuInt}. Despite the very high neutrino fluxes produced in the Sun, reaching on Earth levels of $10^9- 10^{11}\  cm^{-2}s^{-1}$ (Sec. \ref{sec:solarnuprod}), the expected number of interactions can be of $\mathcal{O}(1)$ event/day/ton, making the direct detection a search for very rare events. These events need to be discerned from cosmic rays, as well as natural environmental and material radioactivity, which generate electromagnetic radiation at rates typically $10^6$-$10^8$ times higher than an anticipated neutrino signal. Hence, experiments require adequate shielding and active measures to identify and suppress background signals during analysis.\\ 
To mitigate the impact of cosmic rays background, experiments are typically situated in underground facilities, of which notable examples are the Laboratori Nazionali del Gran Sasso (LNGS) and the Kamioka mine in Japan, among others. The muon flux reduction as a function of the depth of the laboratory is shown in Fig. \ref{fig:CRGammaBkg}.
Although these underground facilities shields from cosmic rays, they still exhibit an intrinsic environmental radioactivity originating from the surrounding rocks. The radioactive chains of Thorium and Uranium are present in nearly all rocks. These chains encompass a diverse array of elements, and due to secular equilibrium, even isotopes with very short half-lives play significant roles. Notably, key contributors include $^{238}$U, $^{234}$Th, $^{232}$Th, $^{226}$Rn, $^{214}$Pb, $^{214}$Bi, $^{227}$Ac, $^{220}$Rn, $^{212}$Pb, and $^{212}$Bi. Additionally, the presence of the radioactive element $^{40}$K is common. These isotopes undergo various decay processes, emitting alpha particles, electrons, positrons, gamma rays, and neutron. Among the particles emitted, alphas, positrons, and electrons can be more readily suppressed due to their continuous energy loss, typically remaining confined within the rock. Conversely, gamma rays, predominantly emitted during the de-excitation of a nucleus following radioactive decay, can traverse the rock, and penetrate the active volume of experiments. Within the active volume, these gamma rays can generate electron recoils (ERs) interacting through Compton scattering or photoelectric effect that exactly mimics a neutrino-electron recoil. Another source of background is represented by neutrons. The source of emitted neutrons can be categorized into two types: radiogenic, stemming from radioactivity, and cosmogenic, triggered by muons. Radiogenic neutrons may arise from natural fission in very high atomic number (A) elements, or via an ($\alpha$,n) reaction, where an $\alpha$ particle interacts with a material of high atomic number (Z), resulting in neutron emission \cite{osti_15215}. On the other hand, cosmogenic neutrons are generated through spallation processes that occur after collisions with muons \cite{Kudryavtsev_2008}.\\
To mitigate the radiation reaching the sensitive components of the detectors, underground experiments incorporate shielding materials. These shields may be passive, offering stopping power properties, or active, outfitted with instrumentation to detect signals arising from energy deposition in the shield material. This enables the characterization of external background and active suppression through vetoes. Commonly, Copper and Lead serve as passive shielding with high Z to halt gamma rays and beta particles. Materials rich in Hydrogen, such as water or polyethylene, are employed to decelerate and absorb neutrons due to their high kinematically favored interaction with Hydrogen. Additionally, scintillators and water Cherenkov detectors equipped with photon detectors represent potential solutions for active shielding configurations \cite{Alimonti_2009}. Using the first method, it is indeed feasible to identify particles originating from radioactive decays and discard events resulting from secondary decays in radioactive chains. An example of this is the $^{40}K$ decay which after undergoing a first electron capture emits a cascade of low energy gammas, followed by a 1460 keV gamma from the excitation of $^{40}Ar^*$ which has a half-life of 1.12 ps \cite{FIORENTINI_2007}. By tagging the second gamma with an active veto it is possible to reject the first inside the detector and vice versa \cite{DAngelo:2022kse}. A Cherenkov detector can be instead employed to tag muons to suppress muon-induced backgrounds. When the muon crosses the water volume, it produces a cone of Cherenkov light through all the water volume, which is a clear signal of the muon passage. After the muon has been identified the following events if compatible with muons-induced events can be rejected as background \cite{Aprile_2014_Muon}.\\
Electromagnetic backgrounds may also arise from the materials used in the construction of both the shielding and the detector itself. These backgrounds are commonly referred to as internal backgrounds. Given their proximity to the sensitive volume, blocking them is challenging, making it crucial to minimize their impact by enhancing the radiopurity of the employed materials \cite{EFCopper, Aprile_2017}. These involve mostly isotopes like $^{40}$K and the decay chains of $^{238}$U, $^{235}$U, and $^{237}$Th, which are commonly present as contaminants, other typical sources of contamination include isotopes such as $^{60}$Co, $^{57}$Co, $^{58}$Co, and $^{54}$Mn. These last isotopes result from the activation of Copper and Iron used in the shielding and structural components due to interactions with muon-induced neutrons. To handle this background source, it is essential to conduct a thorough characterization of the radioactivity of the detector components by precisely measuring their radioactivity and employing ultrapure materials for the realization of the detector and shielding. Since this background cannot be reduced, the solution is to model it in a very detailed way and, any signal must be identified as an excess of events over the expected background, as outlined in \cite{Redchuk:2020hjv}.\\

\section{Solar neutrino detection techniques}
\label{sec:solarnudetectiontec}
Neutrino detection encompasses a diverse range of techniques, driven by technological development, each offering unique advantages and challenges. Radiochemical experiments, such as the pioneering Homestake Solar Neutrino Detector, utilize specific materials and chemical processes to detect solar neutrinos through inverse $\beta$ decay. Water Cherenkov detectors, such as Kamiokande and Super-Kamiokande, detect ultra-relativistic charged leptons generated in neutrino interactions by leveraging Cherenkov radiation, which allows for directional tracking. Liquid scintillator experiments, including Borexino and KamLAND, focus on measuring low-energy solar neutrinos through neutrino electron elastic scattering with a precise electron energy determination, well-controlled background, and relatively low threshold. The main solar neutrino detection techniques are illustrated in the following:
\begin{itemize}
    \item \textbf{Radiochemical experiments:} Radiochemical detection of neutrinos is performed through the inverse $\beta$ decay process ($\nu_e + \text{n} \rightarrow e^-+\text{p}$) within a target chemical element characterized by atomic numbers (A, Z). Here, the target element undergoes transmutation into another chemical element, denoted as (A, Z+1) with the resulting nuclide radioactive, with a half-life of $T_{1/2}$. Detecting neutrinos using this method necessitates sensitive radiochemical techniques to separate and eliminate the few atoms of the product element (Z+1) from the initial target (Z). Operational runs typically follow a batch mode, with exposure durations on the order of 2$T_{1/2}$. During these runs, approximately 10 product atoms (depending on the exposure) are isolated and collected from a pool of around $10^{30}$ target atoms, employing chemical techniques, achieving a separation efficiency of $\sim$90\% \cite{ANSELMANN1993117}. The purified product is then separated from contaminants, particularly radioactive ones like U, Th, and Rn. In experiments involving Chlorine and Gallium, the product atoms are converted into suitable gaseous chemical forms, such as $^{37}$Ar and $^{71}$GeH$_4$, respectively to facilitate separation. These are subsequently placed within high-efficiency, low-background gas-filled proportional counters. The radioactive nature of the product nuclides is unambiguously identified through their energy spectra and half-lives \cite{Richard_L_Hahn_2008}. Since the final products of the reactions are extracted with a fixed cadence of 2$T_{1/2}$ and analyzed only at that point this type of experiment provides an integrated measurement of the solar neutrino flux above a particular energy threshold of the IBD. However, these techniques do not allow for the separation of the contributions from different components of the solar neutrino spectrum since no energy and real-time information is provided. The first solar neutrino experiment, Homestake Solar Neutrino Detector \cite{Lande:1990ct}, situated in the Homestake Gold Mine in South Dakota, employed this detection technique, specifically detecting solar neutrinos through the reaction $^{37}$Cl($\nu_e$, $e^-$)$^{37}$Ar, which has an energy threshold of 0.814 MeV. The Homestake Chlorine experiment was followed by two experiments employing the inverse $\beta$ decay of $^{71}$Ga, through the process $^{71}$Ga($\nu_e$,$e^-$)$^{71}$Ge, GALLEX\cite{1999127} and GNO\cite{Altmann_2005}. This process having an energy threshold of 0.233 MeV allows for sensitivity to all solar neutrinos fluxes, encompassing the abundant and low-energy pp neutrinos emitted by the Sun. 
    
    \item \textbf{Water Cherenkov detectors:} Water Cherenkov detectors enable the measurement of neutrinos by tracking the ultra-relativistic charged leptons produced through neutrino interactions in the detector volume.
    A charged particle traveling through a medium with a velocity greater than the speed of light in that medium emits a luminous radiation called Cherenkov radiation. The reason for this emission can be attributed to the polarization and depolarization effects of the medium associated with the passage of the charge. These charge motions generate a series of spherical waves, which travel in a non-dispersive medium with a group velocity $v_g=\frac{c}{n}$, where $n>1$ is the refractive index on the medium whose envelope constitutes a conical wavefront. By simple geometric construction, the angle of Cherenkov light emission is given by the relation $\cos{\theta_c} =\frac{1}{\beta n}$ which immediately reveals that the Cherenkov effect has a threshold given by $\beta_{th} = \frac{1}{n}$ and that for relativistic particles ($\beta \approx 1$), the emission angle is independent of momentum and equals $\cos{\theta_c} = \frac{1}{n}$. Since the light cone is emitted symmetrically around the particle trajectory, by reconstructing the light profile it is possible to infer the particle direction.\\ 
    This capability distinguishes tracks with directions pointing back to the Sun from the isotropic background. Typically utilizing water as the active material, these detectors have the advantage of a large active volume (several kilotons) at a relatively low cost. However, a notable drawback is the relatively high threshold ($>$ MeV), limiting the measurement to $^8$B neutrinos.  It is important to note that, unlike radiochemical experiments that rely on purely charged-current interactions, the elastic scattering of neutrinos on electrons involves all neutrino flavors, with the cross-section of $\nu_e$ being approximately six times larger than that of $\nu_{\mu,\tau}$. Within the category of water Cherenkov detectors, the KamioKande experiment (1983–1995) \cite{Kamiokande-II:1991pyu} and its successor, Super-Kamiokande \cite{PhysRevD.73.112001} (1996–ongoing), play a pivotal role. These detectors are situated at the Kamioka mine in Japan. Another notable water Cherenkov neutrino detector is the Sudbury Neutrino Observatory (SNO) experiment \cite{BOGER2000172}, located in a mine near Sudbury, Canada.
    
    \item \textbf{Liquid scintillator experiments:} Following the confirmation of neutrino oscillation through precise measurements of $^8$B neutrinos by SNO and Super-Kamiokande, extensive efforts were made to utilize large liquid scintillator experiments. The primary objective was to measure the low-energy parts of the solar neutrino spectrum. These detectors have the capability to measure the recoil energy of electrons generated in neutrino electron elastic scattering with an exceptionally low energy threshold. However, a major challenge in employing this detection technique arises from the fact that the background resulting from radioactive contaminants cannot be separated from the signal component, and under normal conditions, would far surpass the solar neutrino signal by orders of magnitude. One of the experiments utilizing liquid scintillators to measure solar neutrinos is Borexino \cite{Alimonti_2009}, which has been operational since 2007 at the INFN Gran Sasso National Laboratories in Italy. Another neutrino experiment employing liquid scintillator is KamLAND \cite{suekane2004overview}, situated in the same experimental cave that previously hosted Kamiokande in the Kamioka mine in Japan.
\end{itemize}

\section{Current status of solar neutrino measurement}
\label{sec:NuCurrentSttus}
Significant advancements in solar neutrino measurements have been led by the Kamiokande experiment at the Kamioka mine in Japan and the Borexino experiment installed underground at Laboratori Nazionali del Gran Sasso. 
Kamiokande relies on the detection of Cherenkov radiation emitted by recoil electrons resulting from neutrino–electron scattering. Given the characteristic emission direction of Cherenkov radiation, Kamiokande exhibits sensitivity to the direction of solar neutrino interactions in the very high part of the spectrum. On the other hand, the Borexino experiment employs several tons of liquid scintillators and utilizes scintillation light produced by charged particle interaction to precisely measure the energy of the recoil electrons down to low energy.

\subsection{Super-Kamiokande - Water Cherenkov detector}
\begin{figure}
    \centering
    \includegraphics[width=0.7\linewidth]{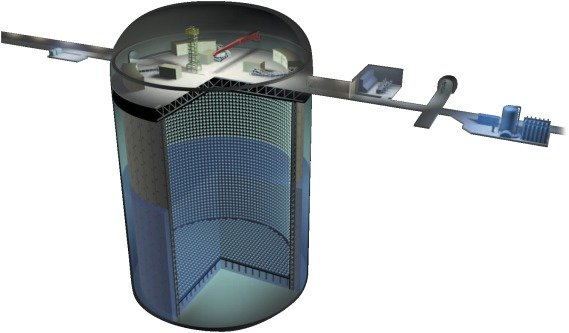}
    \caption{Diagram illustrating the Super-Kamiokande detector. The top section of the image shows the control services of the detector, while the lower section depicts the water tank surrounded by PMTs. Plot from \cite{SKamokandescheme}.}
    \label{fig:SKScheme}
\end{figure}

\begin{figure}
    \centering
    \includegraphics[width=0.75\linewidth]{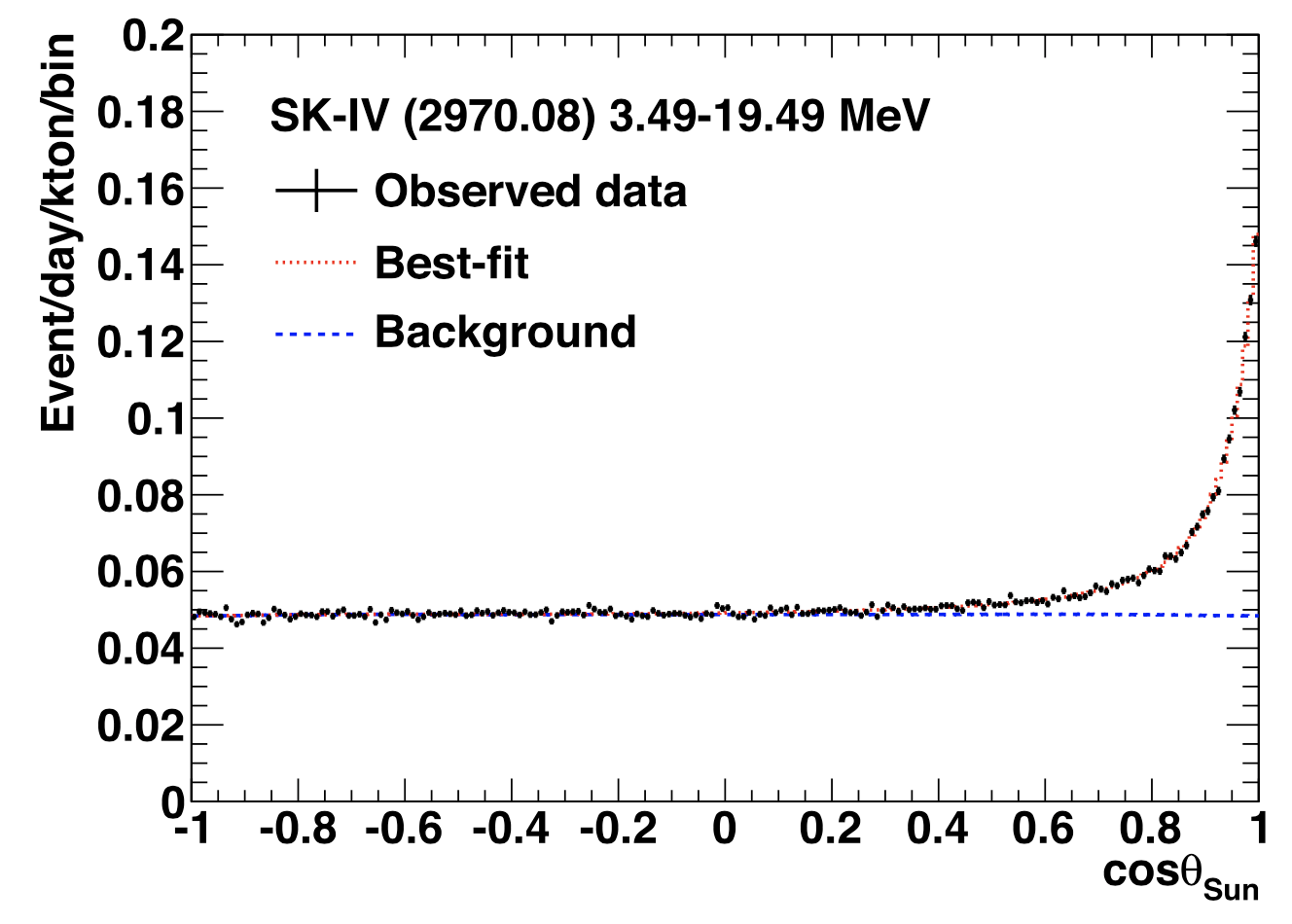}
    \caption{Electron recoil angular distribution obtained in SK-IV in the full energy range 3.49-19.49 MeV. The black crosses are the observed data, the yellow dashed line is the best fit for S+B and the blue line represents the background. Plot from \cite{Abe_2016}.}
    \label{fig:SKResults}
\end{figure}

The Super-Kamiokande (SK) detector, situated 1000 m underground in the Kamioka mine in Japan, is a water Cherenkov detector with a total mass of 50,000 tons. The detector consists of a tank of 39.3 m in diameter and 41.4 m in height filled with ultrapure water. The detector comprises an inner active volume of 32,000 tons, observed by 11,146 photomultipliers (PMTs) with a diameter of 20 inches each. For solar neutrino measurements, to remove events originating from the material surrounding the water, the fiducial volume is set at 22,500 tons, defined as the region more than 2 m away from the surface of the PMTs \cite{Super-Kamiokande:2002weg}. A scheme of the Super-Kamiokande detector is shown in Fig. \ref{fig:SKScheme}.
Cherenkov radiation is produced when charged particles traverse a medium at a velocity $v$ higher than the speed of light in that medium.  For example, water has a refractive index $n$ of 1.33, this results in a light speed in the medium of only 0.75$c$, where $c$ is the speed of light in vacuum. The emitted light falls within the UV range and is emitted at an angle $\theta_{c}$ with respect to the particle trajectory, with $\cos(\theta_c)=1/(\beta n)$, where $\beta = v/c$.  Consequently, the intrinsic threshold for Cherenkov light production of an electron in water is established at 175 keV for an electron. However, due to factors such as the large size of the detector, the minimum length that an electron must travel to provide a detectable signal, light loss in water, PMT noise, and considerations regarding detection efficiency, SK operates with an electron energy threshold of $E_{th}>3.5 MeV$ \cite{Abe_2016}. This threshold makes Kamiokande sensitive exclusively to neutrinos originating from the $^8B$ process. Given that the energy of the solar neutrino from $^8$B is significantly greater than the mass of an electron, the trajectory of a scattered electron results in being aligned with the direction of the incoming neutrino. Thanks to the identification of the Cherenkov ring pattern, SK is sensitive to the direction of the incoming neutrino. \\
The cumulative solar neutrino data obtained so far comprises 93,555 events recorded over a span of 5,480 days, from May 1996 to December 2017 (SK-I for 1,496 days, SK-II for 791 days, SK-II for 548 days, and SK-IV for 2,860 days) \cite{SKResults}.
The recoil electrons $\cos{\theta_{Sun}}$ distribution, where theta is calculated with respect to the Sun's direction, is shown in Fig. \ref{fig:SKResults}. The solar neutrino events can be clearly seen as a peak at $\cos(\theta_{Sun}) =1 $ over a clear flat distribution represented by the background. 
SK has provided the most accurate measurement of the neutrino flux originating from $^8B$ emission, as reported in \cite{Nakahata:2022xvq}
\begin{equation}
    \phi^{ES}( ^8B) = 2.35\pm 0.01\pm 0.04 \cdot 10^6 \ \text{cm}^{-2}\text{s}^{-1}
\end{equation}
where ES stands for elastic scattering. The uncertainties reported are statistical first and then systematic.
Furthermore, in the domain of solar neutrinos, Super-Kamiokande has contributed to significant findings on neutrino oscillation. The joint analysis of SK's measurement of $^8\text{B}$ solar neutrinos, along with data from the Sudbury Neutrino Observatory, has yielded the most stringent constraint on the mixing angle between the mass eigenstate 1 and 2, $\sin^2(\theta_{12})$. When combined with results from the KamLAND experiment \cite{2004physics4071S}, this provides the best measurement of the oscillation parameters $\Delta m^2_{21}$ and $\sin^2(\theta_{12})$ \cite{SajjadAthar:2021prg}. 

\subsection{Borexino - Liquid scintillator detector}
\begin{figure}
    \centering
    \includegraphics[width=1\linewidth]{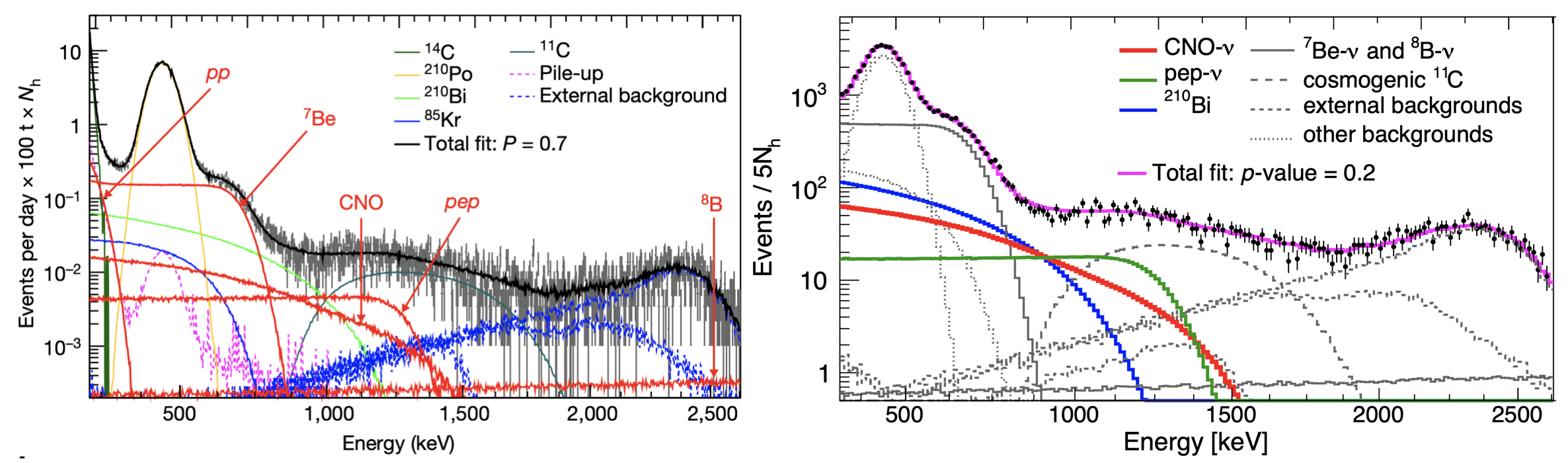}
    \caption{Left: Event distribution in Borexino Phase-II utilized for a thorough measurement of solar neutrinos originating from the pp chain. The graph illustrates the data, best-fit representation, and all signal sources. Right: Spectral fit of Borexino Phase-III data, which leads to the CNO cycle flux measurement. Plots from \cite{borexino2018comprehensive} and \cite{BorexinoCNO} respectively.}
    \label{fig:BorexinoSpectra}
\end{figure}

The Borexino experiment is a liquid scintillator detector that in its activity has provided some of the most precise and important results in solar neutrino measurement, thanks to its exceptional radiopurity and control and knowledge of the detector response. The experiment has operated since 2007 underground at Laboratori Nazionali del Gran Sasso, and it was primarily designed to measure monochromatic neutrinos from $^7Be$. However, thanks to the exceptional radiopurity level which has also improved in time, it was able to perform many groundbreaking measurements of neutrinos from the Sun.
\begin{figure}
    \centering
    \includegraphics[width=0.7\linewidth]{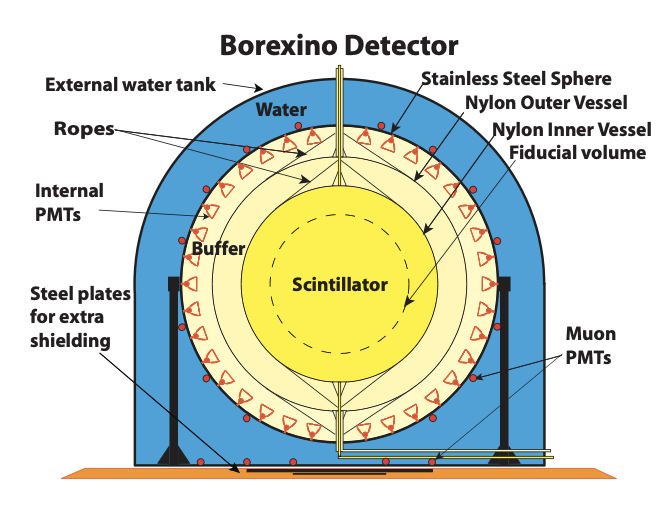}
    \caption{Schematic of the Borexino detector highlighting its components. The diagram illustrates the scintillator volume enclosed by the nylon inner vessel, surrounded by the buffer liquid within the stainless steel sphere. Internal PMTs mounted on the stainless steel sphere facing the liquid scintillator are depicted, along with the external water tank housing the muon PMT for cosmic ray rejection. Plot from \cite{BorexinoScheme}. }
    \label{fig:BorexinoScheme}
\end{figure}
The central part of Borexino consists of an outer undivided stainless steel sphere, functioning as both the housing for the scintillator and the structural support for the photomultipliers. Within this sphere, two nylon vessels partition the scintillator volume into three concentric shells with radii of 4.25 m, 5.50 m, and 6.85 m, the latter corresponding to the stainless steel sphere radius itself. The liquid scintillator solution, specifically PC (pseudocumene, 1,2,4-trimethylbenzene), with a small amount of PPO
(2,5-diphenyloxazole), is contained within the inner nylon vessel. The second and third shells are also filled with PC but diluted with a quencher to reduce the scintillation light, leading to a decrease in the trigger rate. The first layer is the detector sensitive part, while the second and third layer acts as active shielding for the inner part of the detector. The scintillation light is captured by 2212 photomultipliers affixed to the inner surface of the stainless steel sphere. The stainless steel sphere is housed within a tank filled with ultra-pure water, to further shield it from gammas and neutrons, and external muons \cite{Alimonti_2009}. A scheme of the Borexino detector is shown in Fig. \ref{fig:BorexinoScheme}.
In the initial phase of Borexino, the most precise measurement of neutrinos from $^8B$ was performed, yielding a final unoscillated flux result of
\begin{equation}
    \phi^{ES}( ^8B) = 2.57^{+0.17+0.07}_{-0.18-0.07} \cdot 10^6 cm^{-2} s^{-1}
\end{equation}
where ES stands for elastic scattering. The first error source is statistical, and the second is systematic. 
This reported value contains also statistics from the second phase \cite{Agostini_20208B}.
Thanks to the unprecedented, and enhanced radiopurity achieved in the second phase, a comprehensive measurement of solar neutrinos from the pp-chain was performed with a threshold of 165 keV on the electron energy. Through a simultaneous fit of the data encompassing all neutrino components, the following fluxes have been determined \cite{Redchuk:2020hjv}:
\begin{equation}
    \begin{split}
    &\phi^{ES}(pp) = (6.1\pm0.5^{+0.3}_{-0.5})\cdot 10^{10} \ \text{cm}^{-2}\text{s}^{-1}\\
    &\phi^{ES}(^7\text{Be}) = (4.99\pm0.11^{+0.06}_{-0.08})\cdot 10^9 \ \text{cm}^{-2}\text{s}^{-1}\\
    &\phi^{ES}(pep_{HZ})=(1.27\pm0.19^{+0.08}_{-0.12})\cdot 10^8 \ \text{cm}^{-2}\text{s}^{-1}\\
    &\phi^{ES}(pep_{LZ})=(1.39\pm0.19^{+0.08}_{-0.13})\cdot 10^8 \ \text{cm}^{-2}\text{s}^{-1}
    \end{split}
\end{equation}
The two values presented for the pep emission were computed by constraining the CNO flux at the theoretical value under the two distinct hypotheses of low metallicity (LZ) and high metallicity (HZ). Only an upper limit was set to neutrino fluxes from the CNO cycle and the $hep$. The fitted spectrum is depicted in the left section of Fig. \ref{fig:BorexinoSpectra}. 
The measurement of the CNO flux was carried out during the third phase of Borexino. During this phase, the detector was stabilized in temperature, with the resulting suppression of the convective motions of the scintillator liquid. These motions were carrying over $^{210}$Po originating from the vessel surface, driving out of equilibrium the inner radioactive contaminants. In thermal stability conditions, with the detector free from additional out-of-equilibrium elements, it was possible to perform a precise measurement of $^{210}$Bi contamination through the alpha decay rate of $^{210}$Po. The constraint on the amount of $^{210}$Bi plays a crucial role in the overall measurement, since it's strongly correlated with the CNO electron recoil spectrum \cite{Agostini_2020Bi}. The measured neutrino flux is 
\begin{equation}
    \phi^{ES}(CNO) = (6.6^{+2.0}_{-0.9}\cdot 10^8)\text{cm}^{-2}\text{s}^{-1}
\end{equation}
The complete spectrum, along with the various components of neutrino signals and background, is shown in Fig. \ref{fig:BorexinoSpectra}, right plot. As can be seen from the plot, the constraints on the $^{210}$Bi contamination are of paramount importance for the detection of the CNO neutrinos in the energy range 0.8-1.0 MeV.\\
The Borexino detector can rely on a very high target mass, which contributes to ample statistics from neutrino interactions. However, due to the elevated presence of $^{14}C$ in the organic scintillator which undergoes beta decay with endpoint at 156 keV, and the dominance of pile-up events (events occurring so closely in time that they cannot be distinguished) in the low energy region of the spectrum, Borexino has operated with a threshold of 165 keV \cite{Borexino:2015axw}. This threshold corresponds to an effective threshold of approximately 300 keV on the neutrino energy, leaving the lower part of the pp spectrum yet unexplored.

Currently, the Borexino collaboration, recognizing the advantages of the directionality in signal/background discrimination, is actively investing efforts in electron recoil direction reconstruction to enhance their sensitivity in solar neutrino detection \cite{PhysRevLett.128.091803,PhysRevD.105.052002}. The accomplishment was made possible within the range of 0.54-0.74 MeV thanks to the capability of discriminating Cherenkov light from scintillation light, and served as a proof of principle for future hybrid detectors \cite{PhysRevLett.128.091803}. With this achievement, the possibility of combining the directional advantages of a Cherenkov detector with the energy resolution and low energy threshold of a liquid scintillator detector has been demonstrated in a small range of energy.

\section{Neutrino directional detection with gaseous TPC}
\label{sec:nuTPC}
Although the Borexino observation of neutrino from the pp chain has been performed, the lower part of the energy spectrum remain unmeasured, and the flux uncertainty remain around 10\%. Additionally, the observation of the CNO flux has been a groundbreaking result in the field of neutrino physics, but it lacks the sensitivity to address the solar metallicity problem. Therefore, within a context where more in-depth measurement of solar neutrinos could offer insights into new physics results, attention should shift toward precision detection with experiments featuring low thresholds and directional capabilities.\\
A gaseous Time Projection Chamber can be an ideal detector for this purpose. Despite the trade-off of the need for large volumes to achieve significant exposures, a gaseous TPC developed for rare events searches offers the capability of 3D track reconstruction, enabling the measurement of electron and nuclear recoil directions down to a few tens of keV. The ability to reconstruct the recoil direction enhances signal sensitivity by exploiting the angular distribution of the signal which differs from the background one, which translates into a higher background tolerance.
Furthermore, unlike existing detectors designed specifically for neutrino detection or Dark Matter detection, a TPC would exhibit much more versatility, being sensitive to both very low-energy nuclear recoils from CE$\nu$NS and DM, and energy electron recoils from the whole solar neutrino spectrum simultaneously. This would make a high-precision TPC a suitable candidate for next-generation experiments.

\begin{figure}
    \centering
    \includegraphics[width=0.5\linewidth]{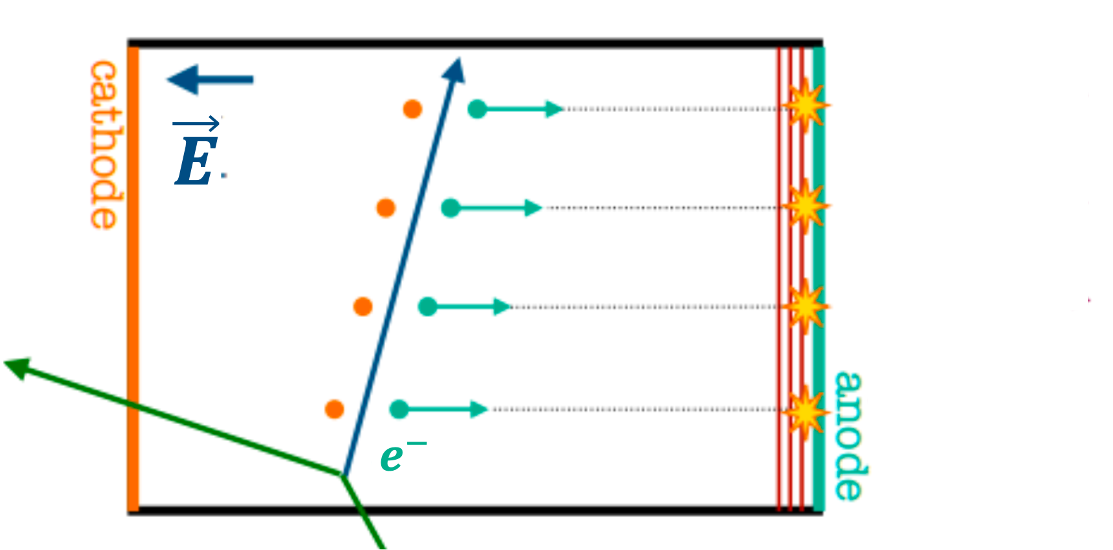}
    \caption{Scheme of a Time Projection Chamber}
    \label{fig:TPCScheme}
\end{figure}

A gaseous Time Projection Chamber consists of a volume filled with gas, serving as a sensitive target for the detector. Within the gas, an electric field is generated by the voltage difference between a cathode positioned at one extremity of the gas volume, and an anode place on the opposite side of the cathode with respect to the gas volume. When a charged particle with enough energy crosses the gas volume, it ionizes the gas, producing free electron-ion pairs along its path. These electrons are then guided to the anode by the electric field, preserving their relative positions during the drift. Subsequently, depending on the instrumentation at the anode, the electrons undergo amplification in different way and are readout using various methods. As a result, it becomes possible to measure the energy released by the particle in the detector, along with the three-dimensional profile of the track. The scheme of a TPC is shown in Fig. \ref{fig:TPCScheme}.
The complete three-dimensional trajectory of a recoil can be reconstructed by merging the two-dimensional measurement of the ionization charge distribution at the readout plane with the third dimension obtained through sampling the collected signal over time. Detecting the entire track profile allows for the reconstruction of the initial direction of the recoil, which, in turn, can be correlated with the direction of the incoming particle. A TPC with high-granularity readout is additionally the most mature technology for track direction reconstruction of low-energy recoils \cite{NRDirect}. Moreover, such a feature is possible only in gaseous targets, since in solid or liquid material, an electron with an energy of $\sim$100 keV would travel a distance of less than 100 $\mu$m suffering additional high multiple scattering. The track length would be even shorter for a nuclear recoil. On the contrary, in gas, such an electron recoil would travel tens of cm. Furthermore, in the context of the TPC for rare events searches, given the small energies needed to ionize the gas, together with the high gain provided by the amplification system and the low noise of the readout, presents a very low threshold achievable of $\mathcal{O}$(1) keV \cite{Baracchini_2020234}, making them suitable for low energy events detection.

\subsection{A first proposal}
\label{sec:firstproposal}
\begin{figure}
    \centering
    \includegraphics[width=0.85\linewidth]{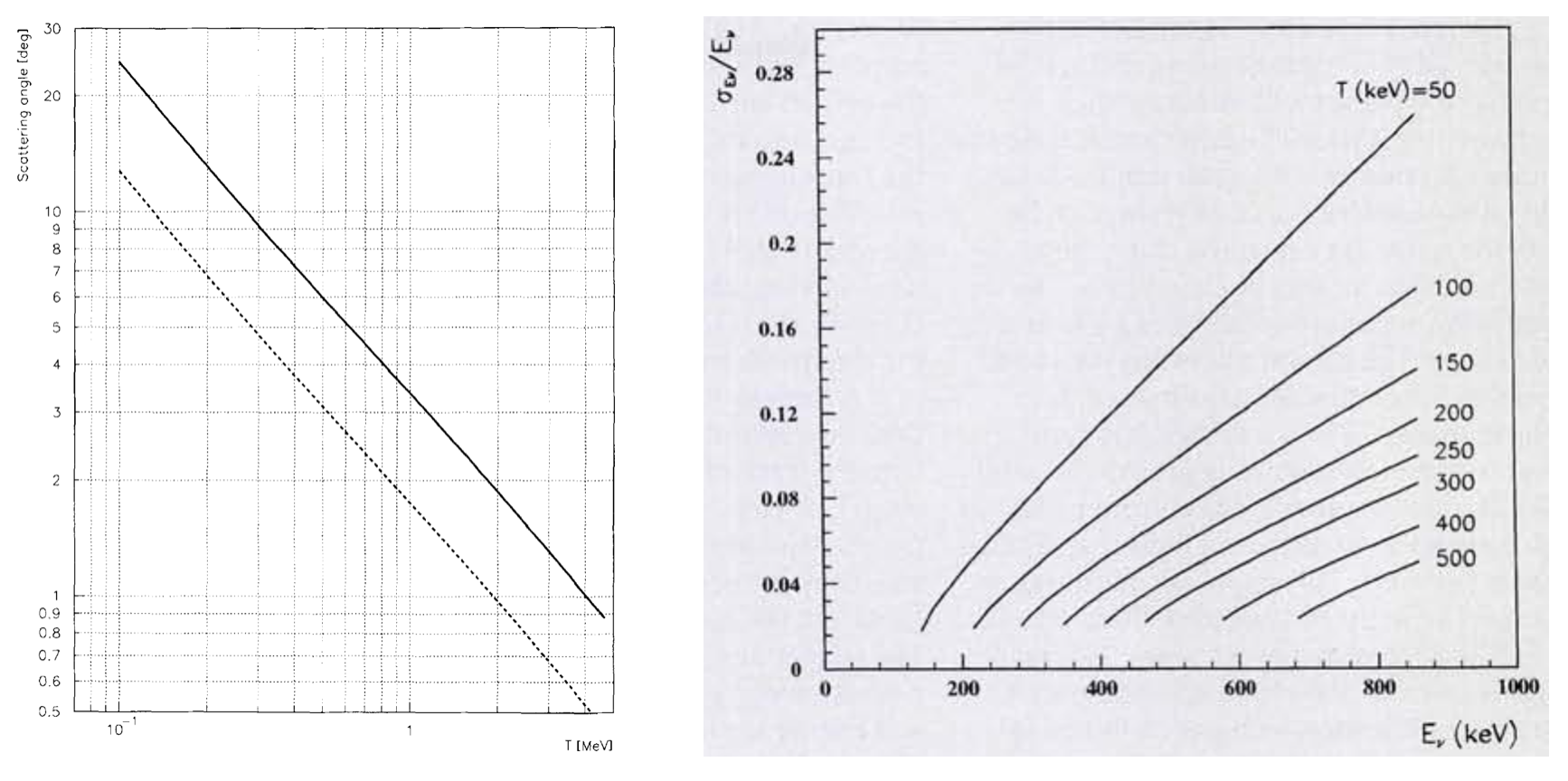}
    \caption{Left: RMS of the track scattering angle after traveling 2 cm in pure CF$_4$ configuration (full line) and in He:CH$_4$ (dotted line). Right: Resolution on the reconstructed neutrino energy $\sigma_{E_{\nu}}/E_{\nu}$ as a function of the neutrino energy $E_{\nu}$ in pure CF$_4$. The simulated behavior is shown for different electron recoil energy $T$. Plots from \cite{Arpesella:1996uc} and \cite{HELLAZNuDet} respectively.}
    \label{fig:HELLAZPerf}
\end{figure}

\begin{figure}
    \centering
    \includegraphics[width=0.85\linewidth]{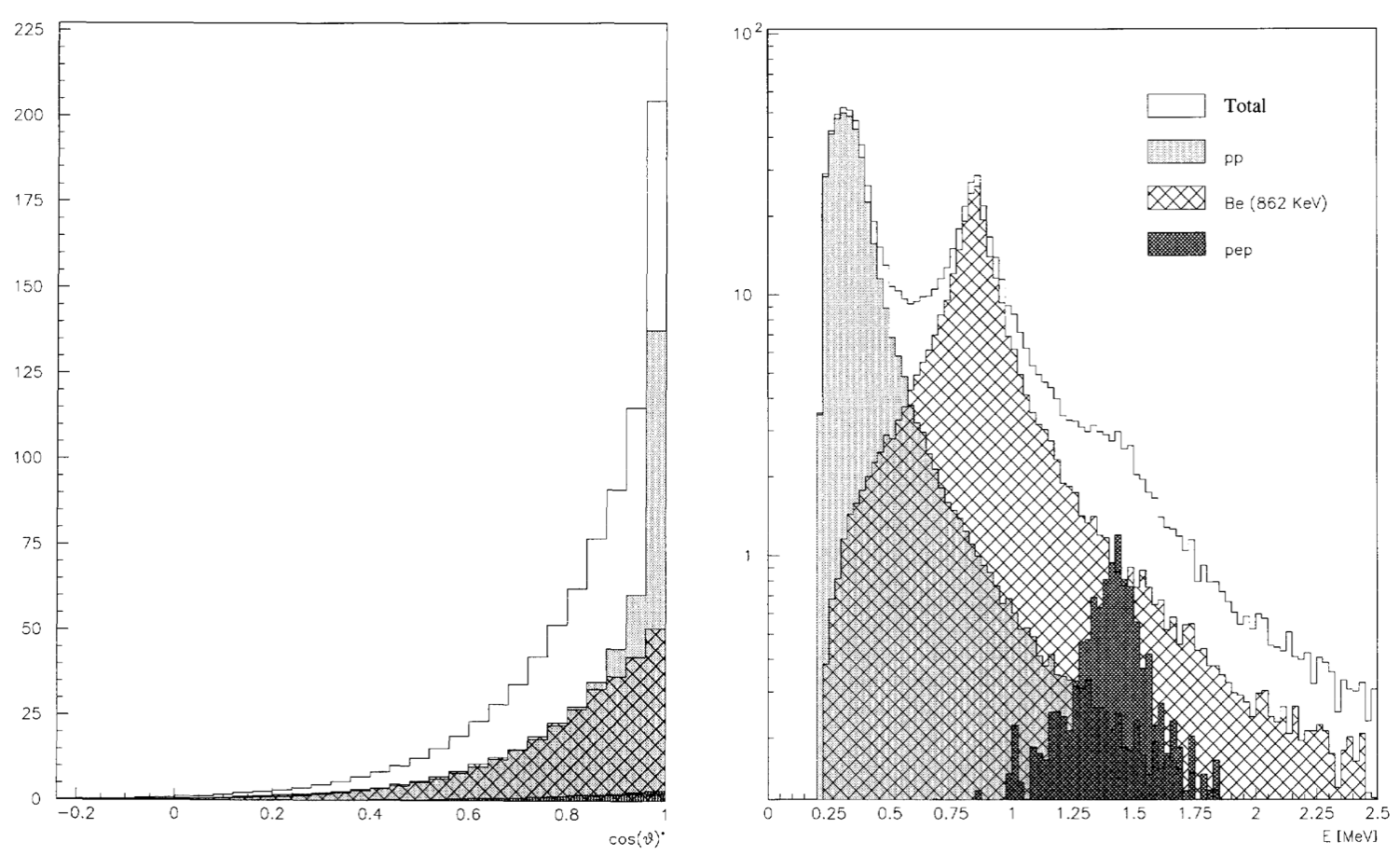}
    \caption{Left: simulated $\cos(\theta)^*$ distribution of ER direction, where the $*$ denotes that the angle is calculated with respect to the Sun direction in pure CF$_4$. The different contributions are reported in the right plot. Right: distribution of reconstructed neutrino energy starting from the simulated $\theta$ and T in pure CF$_4$. Plots from \cite{Arpesella:1996uc}. }
    \label{fig:NuRecoAngleSpec}
\end{figure}

\begin{figure}
    \centering
    \includegraphics[width=0.75\linewidth]{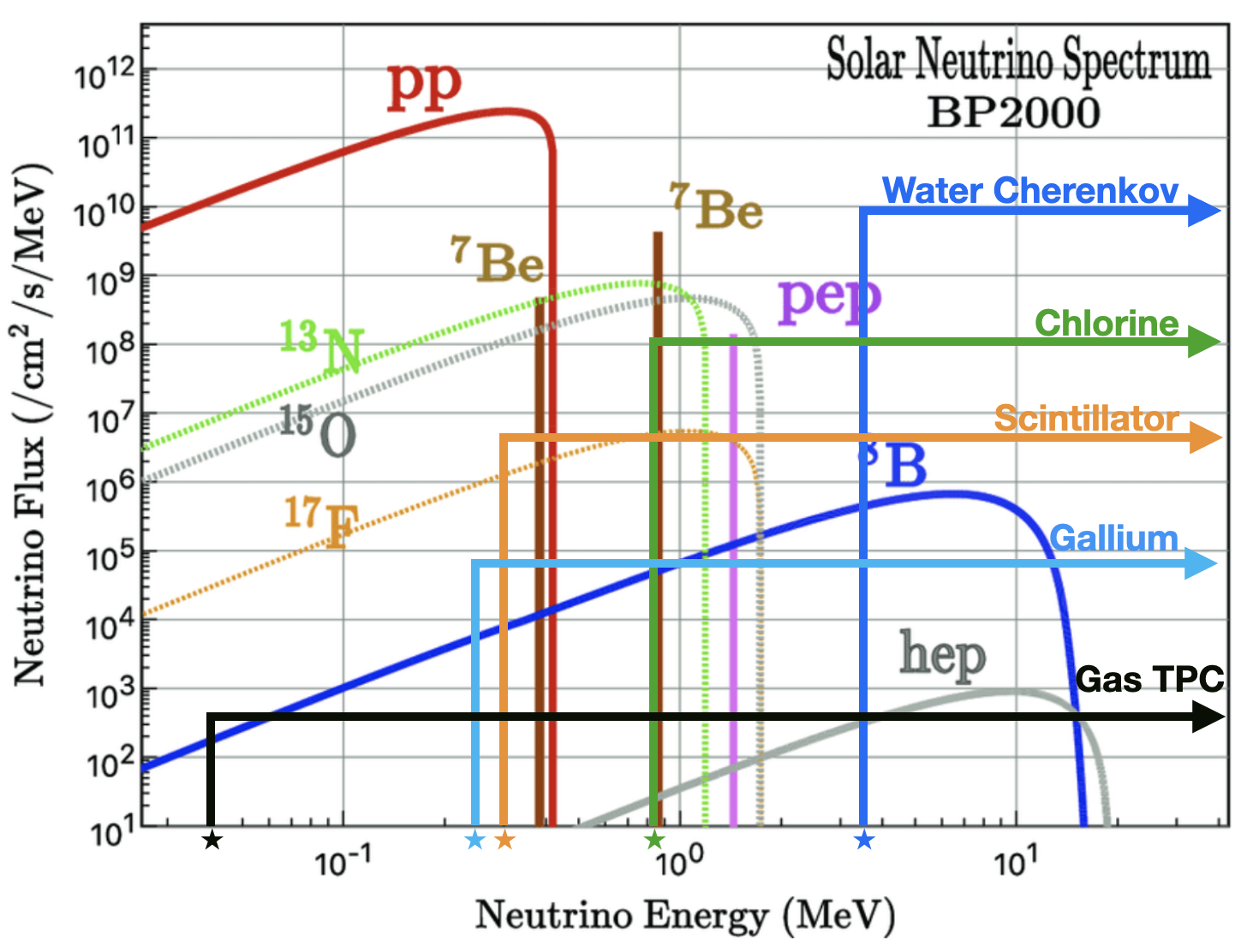}
    \caption{Threshold on the neutrino energy for different technologies employed in neutrino detection. Plot adapted from \cite{PlotSolarNU}.}
    \label{fig:solarnuthr}
\end{figure}

The proposal of performing a solar neutrino measurement trough neutrino-electron elastic scattering ($\nu e$ES) with a large gaseous Time Projection Chamber was presented for the first time in 1990 \cite{HighRateNu} after the initial $^8{B}$ solar neutrino measurements with the radiochemical experiments (Sec. \ref{sec:solarnudetectiontec}). At that time, the observed fluxes with radiochemical experiment represented only 30-50\% of model predictions. Data analyses suggested that the $^{7}$Be solar neutrinos flux was either very low or zero, while the $^8$B neutrino flux was approximately 50\% of expectations. However, this posed a puzzle as the production of $^8$B proceeds through $^7$Be. Additionally, the GALLEX and SAGE experiments (Sec. \ref{sec:solarnudetectiontec}), which measured the neutrino flux down to the pp energies, detected inconsistencies in the neutrino fluxes with respect to the solar model prediction. These measurements suggested that the model may be incorrect, or an alternative explanation, necessitates the consideration of new physics. The proposal that neutrinos can oscillate between different flavors, was already taken into account but not demonstrated, and was based on the hypothesis that an inverse $\beta$ decay detector would only detect $\nu_e$.
In this context, the first real-time experiment Kamiokande was proposed for the observation of solar neutrino from the $^8B$, but at the time, only 80 $\nu_e$ elastic scatterings were foreseen due to the very high threshold of 7 MeV because of the efficiency of Cherenkov light detection. At the same time, the second generation of radiochemical experiments utilizing inverse $\beta$ decay on Gallium held the promise of a considerably lower detection threshold, but still with lack of energy measurement and real-time detection capabilities. Moreover, these detectors were sensitive only to the electronic neutrino flavor. Additionally, the third generation water Cherenkov experiment approved, SNO and Super-Kamiokande together with the liquid argon detector ICARUS were all relying on a threshold $>$5 MeV. Finally, the radioactivity content of the proposed Borex and Borexino detector, with trimethylborate (TMB) scintillating target, was thought to certainly prevent the pp observation, eventually allowing the detection of the $^{7}Be$ neutrino depending on the development in TMB purification methods \cite{HighRateNu}.\\ Thus, a detector with a low threshold, real-time interaction information, and the ability to measure both the direction and the energy of the scattered electron was proposed. In an initial iteration of the detector, Seguinot, Ypsilantis, and Zichichi \cite{Arzarello:1994jv} suggested employing a 2000 m$^3$ Time Projection Chamber filled with He at 5 bar pressure and a temperature of 77 K to increase the density and reduce Radon contamination in the gas. The detector geometry was proposed as two back-to-back cylindrical TPCs with a common cathode, a 10 m drift length (distance between cathode and anode), and a 5.5 m radius readout area. The total mass of the target was foreseen to be 6 tons. Subsequently, the use of a CF$_4$ gas mixture was suggested due to the advantages of a higher electron density, allowing operation at atmospheric pressure while still maintaining a total mass of 7.4 tons \cite{Arpesella:1996uc}. The detector was designed to be read out by a multiwire chamber (MWC) due to its ability to achieve high gain and its high sensitivity to single electrons. These capabilities were thought to provide the ample sampling required to determine track direction and to identify both the head and tail of a track \cite{HELLAZNuDet}. With this detector, the expected track RMS scattering angle as a function of the energy is reported together with the expected energy resolution on the neutrino energy as a function of the neutrino energy for different electron scattered energies. The plots are shown in Fig. \ref{fig:HELLAZPerf}.
With the capacity to reconstruct the neutrino energy on an event-by-event basis, given the kinematic, there was the potential to reconstruct the $^{7}Be\ (862\ keV)$ and $pep$ spectral lines (not observed at that time). The simulated angular distribution of the electron recoil from the different contributions is shown in Fig. \ref{fig:NuRecoAngleSpec} together with the simulated reconstructed neutrino spectrum. 
As can be seen from the plot, with a threshold as low as 100 keV, the observation of the pp chain could be effectively carried out.
A graphic comparison of the region of the solar spectrum that different detection methods can probe is shown in Fig. \ref{fig:solarnuthr}.\\ However, at that time, the TPC technology was not mature enough. For example, the suggestion to use multiwire proportional chambers as readouts with large granularity, combined with the effect of diffusion and readout, would not have led to the reported angular resolution expected. Moreover, the performances claimed have been studied only on simulation with limited computation power, and never been experimentally demonstrated. As a result, this initial proposal was not pursued further, in favor of the development of more established technologies at that time such as scintillating and water Cherenkov detectors.


\subsection{Directionality advantages in neutrino detection}
\label{sec:diradvantages}
The initial direction track reconstruction lead to very significant advantages like \cite{ARPESELLA1996333}:
\begin{itemize}
    \item \textbf{Unambiguous signal identification}: an excess of electron recoils in the opposite direction of the Sun can be unambiguously attributed to particles coming from the Sun.
    \item \textbf{Online measurement of the background}: for the same reason as before, part of the angular spectrum which contains a significantly small amount of signal events can be used to accurately estimate the flat background.
    \item \textbf{Solar neutrino spectroscopy}: Knowing the direction of the incoming particle, and measuring the energy and the angle of the electron with respect to the direction of the incoming neutrino closes the kinematic. This allows for an unambiguous determination of the neutrino energy event by event, leading to the reconstruction of the solar spectrum. 
\end{itemize}

\label{sec:bkgRejection}
As a matter of fact, the directional capability in a solar neutrino measurement offers the advantage of correlating the direction of the scattered electron with the source direction. Given the kinematics, the electron is expected to be scattered in the opposite hemisphere relative to the source.  As a first approximation, it can be expected the angular distribution of internal and external backgrounds to be isotropic. Electron recoils from eventual localized sources of gamma tend to be flattened as the angular recoil distribution is transformed into Solar coordinates due to Earth's rotation around its axis. Furthermore, the inclination of Earth's rotation axis relative to the Solar system plane contributes to further blurring local anisotropies in the background angular distribution. Hence, similarly to Kamiokande (Fig. \ref{fig:SKResults}), the quest for the neutrino signal involves identifying an excess of events concentrated toward the opposite direction of the Sun on top of a uniform background. This is, of course, more effective than relying only on the energy spectrum and searching for a falling exponential distribution on top of another. Moreover, given the peaked angular distribution of the signal, specific regions in the angular space could exhibit a notable lack of signal events, providing a region, useful for evaluating the intrinsic flat background component. The existence of these regions facilitates background modeling and contributes to an improved signal-to-noise ratio, as the signal can be searched out from this region. Furthermore, directionality enables event discrimination, allowing for greater tolerance to the background while maintaining the same precision in flux measurements.

\begin{figure}
    \centering
    \includegraphics[width=0.5\linewidth]{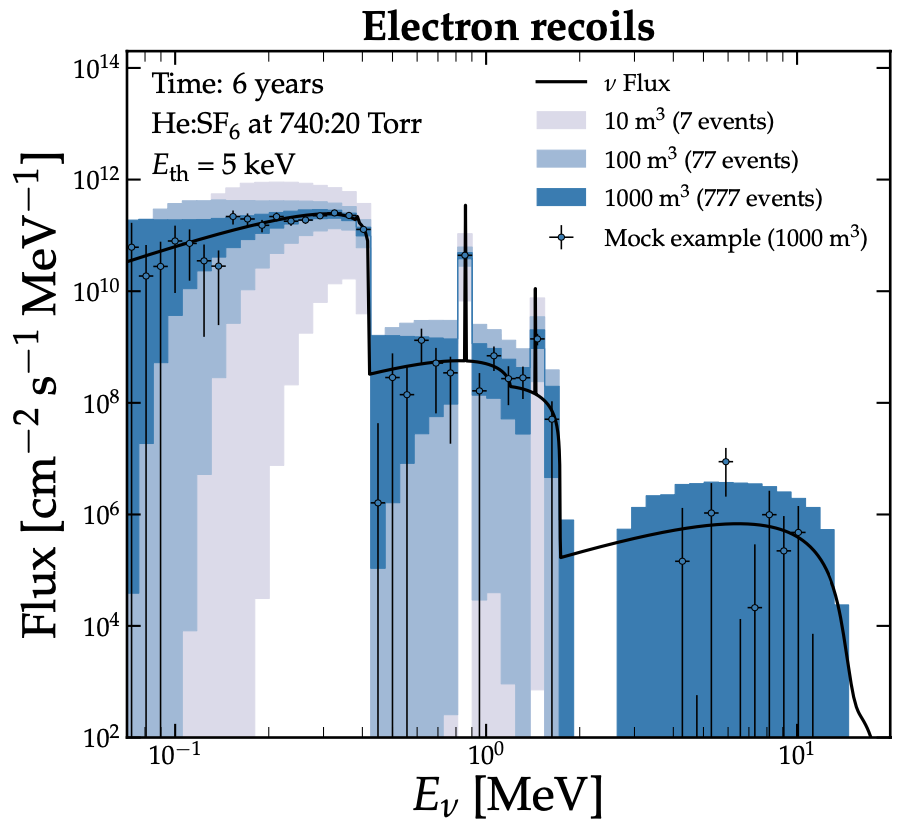}
    \caption{Solar neutrino spectra reconstructed from recoil energies and direction information from $\nu$eES events taken as an example. In the plot, a threshold of 5 keV is assumed for a detector with 740:20 Torr He:SF$_6$ at 1 atmosphere. The shading, ranging from light to dark, indicates the 95\% contours for increasing TPC volume. Preliminary plot from \cite{SolrNuDir}.}
    \label{fig:somefig}
\end{figure}

\label{sec:spectroscopy}
The kinematic parameters characterizing $\nu eES$ include the energy of the incident neutrino, the energy of the outgoing electron, and the scattering angle ($\theta_e$) of the electron relative to the initial neutrino trajectory, assuming knowledge of the incoming neutrino direction. 
As described in section \ref{sec:NueES} the kinematic process of $\nu eES$ is determined by a single free parameter, $\theta_e$, once this is set, there exists a unique combination of values for ($E_{\nu}$,$E_{e}$) that can satisfy the energy and momentum conservation principles.\\ Generally, the direction of the primary particle is not known, but in the case of solar neutrino searches (known source) it can be hypothesized, considering it as the line connecting the Sun and the Earth. Under this assumption, measuring the angle of the electron recoil relative to this line and its energy completes the kinematic information. Consequently, there is a unique value of ($E_{\nu}$) that can give rise to such a recoil. By inverting Eq. \ref{eq:kinematic}, the solution for $E_{\nu}$ can be derived.
\begin{equation}
    E_{\nu,Reco}=\frac{-m_eT_e-\sqrt{T_e^2m_e^2\cos(\theta)^2+2T_em_e^3\cos(\theta)^2}}{(T_e-T_e\cos(\theta)^2-2m_e\cos(\theta)^2)}
    \label{eq:enureco}
\end{equation}
Hence, with the ability to measure all the parameters of the kinematics, the reconstruction of the initial neutrino energy becomes feasible on an event-by-event basis. This real-time detection method stands in contrast to the approach employed by Borexino, which statistically calculates the pp spectrum. 
Additionally, as shown in the right plot of Fig. \ref{fig:BorexinoSpectra}, the $pep$ and $^{7}$Be neutrinos represent significant sources of background for CNO measurements. These neutrino interactions generate a continuous spectrum that overlaps with the CNO spectrum, and in Borexino, the maximum sensitivity is only in a small region around 1 MeV. Utilizing an event-by-event spectroscopy technique would enable the reconstruction of these events as lines, facilitating the identification of the CNO spectrum and flux measurement. In Fig. \ref{fig:somefig}, an illustration of neutrino spectrum reconstructed from $\nu_e$ elastic scattering events using recoil energies and direction information is presented as an example for a detector with 740:20 Torr He:SF$_6$ gas mixture at 1 atmosphere, a threshold of 5 keV and 6 y exposure. The ration in Torr refers to partial pressures.
At present indeed, the Borexino measurement of the combined CNO flux falls below the level required for a definitive resolution of the solar metallicity problem. Therefore, additional data would undoubtedly be beneficial for its resolution, further strengthening the motivation for the development of ton-scale gas TPCs as well as the exploration of DM parameter space beyond the neutrino floor \cite{SolrNuDir}.

\subsection{A CYGNUS proposal}
\begin{figure}
    \centering
    \includegraphics[width=0.7\linewidth]{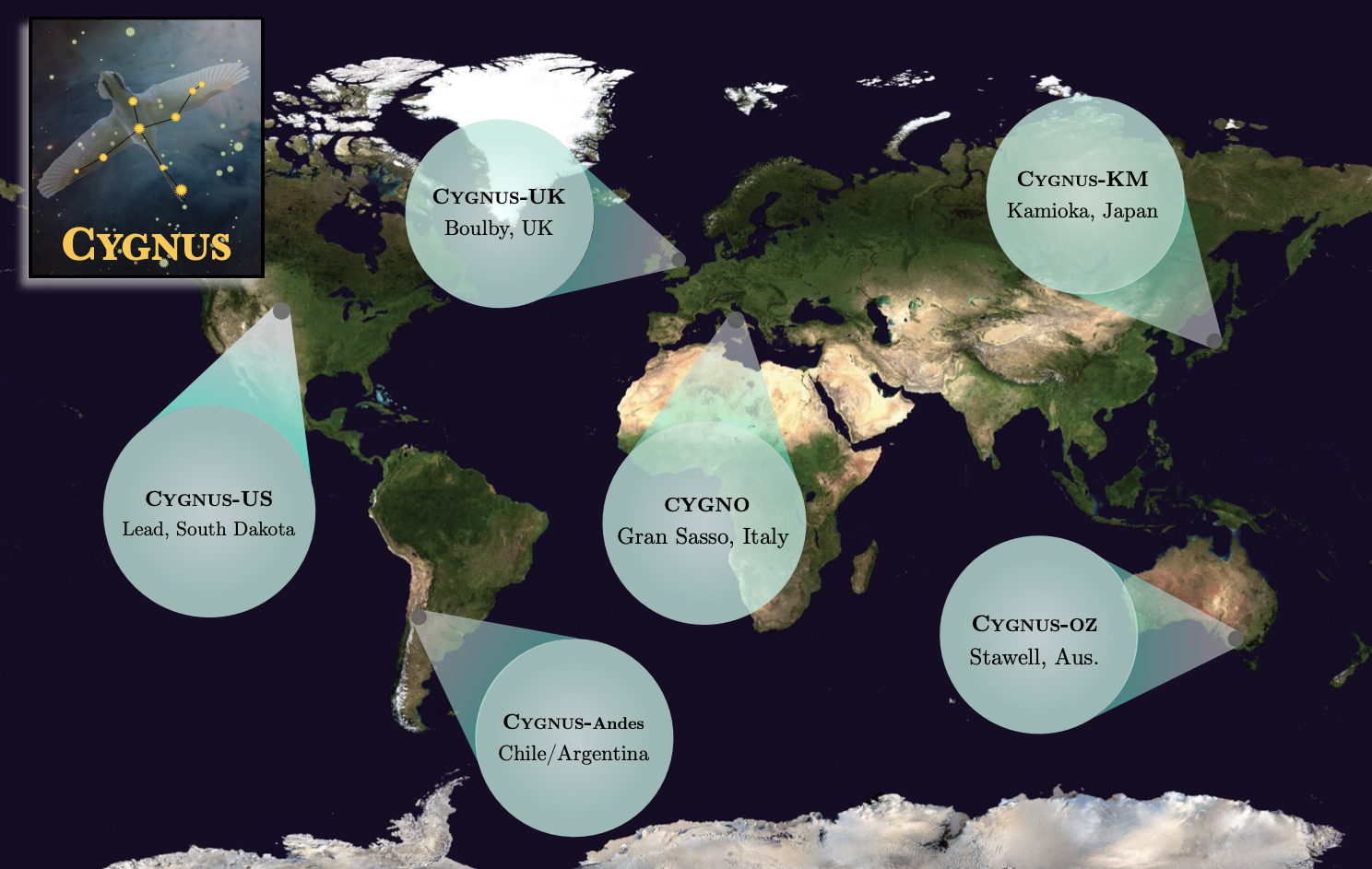}
    \caption{Different research groups of the CYGNUS collaboration. Plot from \cite{vahsen2020cygnus}}
    \label{fig:CYGNUSMulti}
\end{figure}

\begin{figure}
    \centering
    \includegraphics[width=0.7\linewidth]{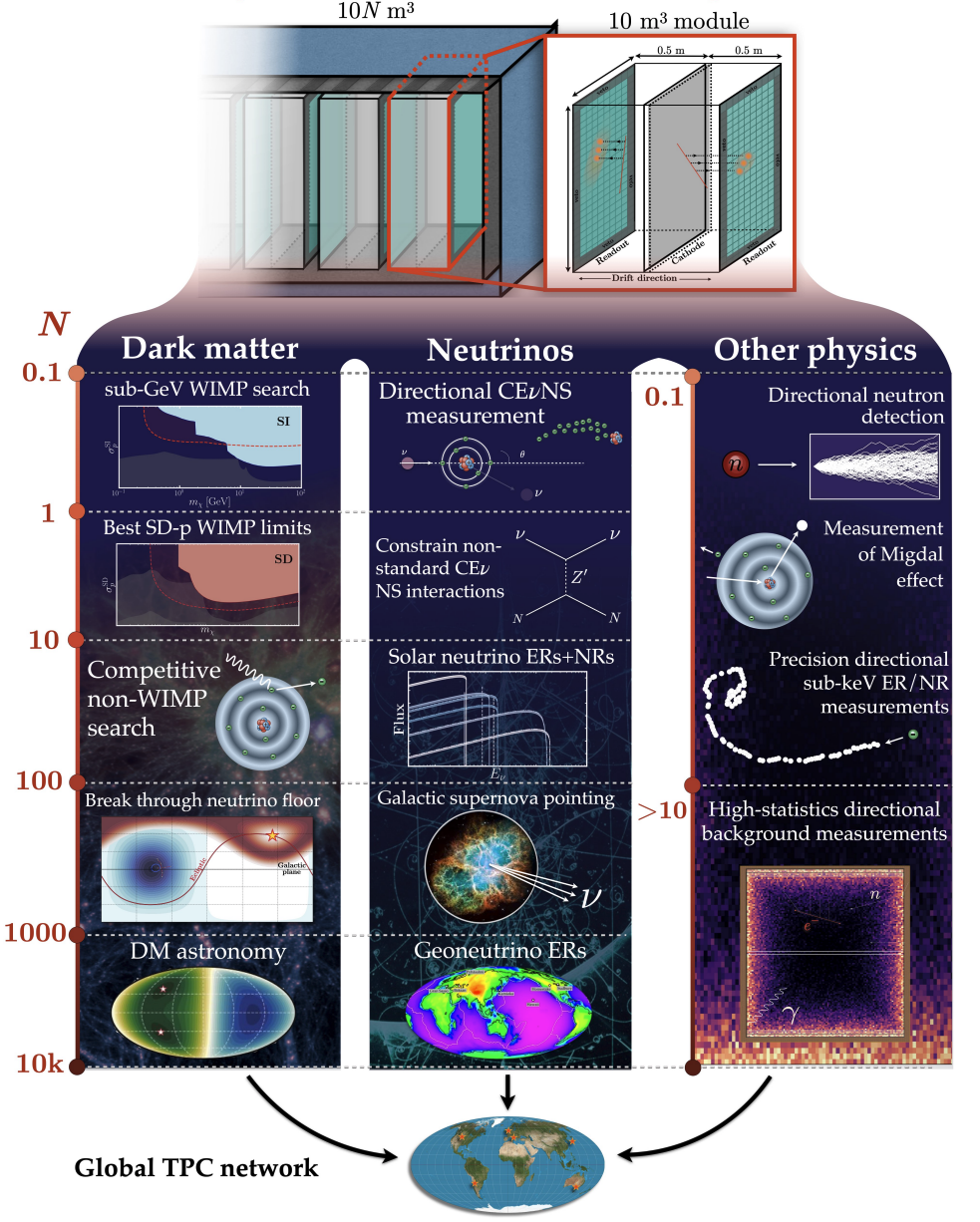}
    \caption{Overview of the physics cases for a directional gas TPC, divided into Dark Matter studies, neutrino physics, and various fundamental and applied physics scenarios. The representation is in function of the required total size of a gas TPC experiment, denoted by N, representing the number of 10 $m^3$ TPC modules operating near atmospheric pressure. The volumes are presented with an order of magnitude precision. A distinct scale for N is employed in the 'other physics' column, reflecting the potential achievement of these goals with significantly smaller-scale experiments. Plot from \cite{Vahsen_2021CYGNUS}.}
    \label{fig:WorldNuObservatory}
\end{figure}

Subsequent advancements in TPC development for high-energy physics \cite{ALME2010316} led to a rapid improvement in this technology. Additionally, the latest development towards precision TPCs for rare event searches, involving studies on different gas mixtures, amplification systems, and readouts, has further pushed this technology to a point where it can effectively be employed in rare event studies \cite{Battat_2021,Shimada:2023vky,VAHSEN201595,TAO2022165412,Leyton_2016}.
The development of a gaseous TPC for rare events searches is strongly driven by Dark Matter in the context of directional Dark Matter searches (illustrated in detail in Sec. \ref{subsec:wimp}). DM, under the hypothesis of this consisting of Weakly Interacting Massive Particles with masses of $\mathcal{O}$(10-100 GeV) should interact inside the detector producing $\mathcal{O}$(1-10 keV) nuclear recoils. With the next generations of experiments reaching mass scales of $\mathcal{O}(100)$ Ton, the experiments will start to become sensitive to solar neutrino interaction through coherent elastic scattering on nuclei producing signals that resemble DM ones, generating the so-called "neutrino fog" below which the realization and construction of larger detectors aimed at enhancing sensitivity to Dark Matter by increasing exposure face significant challenges and are not particularly convenient \cite{O_Hare_2021Nufog}. To circumvent this problem in an efficient way, recoil-based experiments would require some form of information to discriminate between the Dark Matter and neutrino events beyond just recoil energy. Previous work has shown that the directional recoil information is enough to efficiently push through the neutrino fog \cite{O_Hare_2015,PhysRevD.102.063024}.\\
For this kind of physics, moving toward operation at atmospheric pressure, contained drift lengths, and a modular configuration, a gas experiment eventually approaching the ton-scale in target mass, and the possibility of performing a solar neutrino measurement appears feasible. Such a program is the primary aim of the CYGNUS project \cite{vahsen2020cygnus}. CYGNUS aims to establish a network of TPC for a global recoil observatory with directional sensitivity to both Dark Matter and neutrinos. Indeed, given the requirement needed, this last measurement can be performed with the same technology developed for directional DM searches. An illustration of the international collaboration and the experiments carrying on this development is shown in Fig. \ref{fig:CYGNUSMulti}
As the detector volume and consequently exposure increase, it gains sensitivity to various physics scenarios. These include directional CE$\nu$NS measurements using neutrino beams at lower volumes, up to Supernovae pointing, and geoneutrinos with higher detection volumes \cite{Vahsen_2021CYGNUS}. Leveraging the modularity of the detection approach, different modules can be deployed in underground laboratories worldwide, operating as a network and eventually employing different technologies. Fig. \ref{fig:WorldNuObservatory} illustrates the diverse physics cases for this experiment as a function of the exposed target mass.
\\
Within a CYGNUS detector, the highest energy fluxes of $^8$B and $hep$ neutrinos will generate nuclear recoils through CE$\nu$NS interaction, establishing them as significant backgrounds for forthcoming Dark Matter searches. However, owing to the directional capabilities of CYGNUS, these CE$\nu$NS recoils can be distinguished from those generated by DM, making them a new case of physics instead of a background.
For a fluorine-based target, most CE$\nu$NS events resulting from solar neutrinos interaction will occur below 10 keV \cite{vahsen2020cygnus}. By considering a worst-case scenario, with a threshold of 8 keV, as reported in \cite{vahsen2020cygnus}, a CYGNUS 1000 m$^3$ could detect O(10) CE$\nu$NS events over a time span of a few years. With a more optimistic but still achievable threshold of 1 keV on nuclear recoil, the number of nuclear recoils detected from neutrino interaction could increase to 30–50, depending on the percentage of the fluorine-based gas with respect to helium \cite{vahsen2020cygnus}.\\
Nevertheless, even prior to the availability of very large-scale and low-threshold gas TPC experiments, there are opportunities for engaging in innovative neutrino physics studying solar neutrinos trough $\nu e$ES with intermediary $\mathcal{O}$(10) m$^3$ experiments developed within the CYGNUS collaboration. The measurement could be performed by detecting $\mathcal{O}(10)$ ev/y and discriminating those from the background leveraging the directionality feature and allowing for the pp solar neutrino spectroscopy on an event-by-event basis \cite{vahsen2020cygnus}.
The more advantageous kinematics of elastic neutrino-electron scattering with respect to the CE$\nu$NS one result in recoil energies ranging from tens to thousands of keV. Consequently, an experiment with thresholds of approximately 10 keV would be capable of observing the most intense - but lowest energy - flux of solar neutrinos originating from pp reactions. Additionally, given the kinematic, the directions of electron recoils would also be stronger peaked towards the Sun direction (Fig. \ref{fig:NuRecoAngleSpec}) compared to CE$\nu$NS providing background rejection. Although most of the events would originate from pp neutrinos, whose flux is relatively well-understood but still not measured with a precision better than 10\%, with a $\mathcal{O}$(1000) m$^3$ detector a measurement of the CNO flux could be performed. A more precise determination of the CNO flux could help address a longstanding puzzle related to the Sun's heavy element composition, often referred to as the solar metallicity problem \cite{2020arXiv200406365V}. A directional measurement, as illustrated earlier, would not only facilitate the independent and less degenerate measurement of multiple fluxes but also at significantly lower energies compared to, for example, Borexino, whose threshold is approximately 160 keV. At the moment, Borexino's measurement of the combined CNO flux falls short of the level required to conclusively resolve the solar metallicity problem. Therefore, additional data would undoubtedly be beneficial, making a compelling argument for ton-scale gas TPCs. 
\begin{figure}
    \centering
    \includegraphics[width=1.\linewidth]{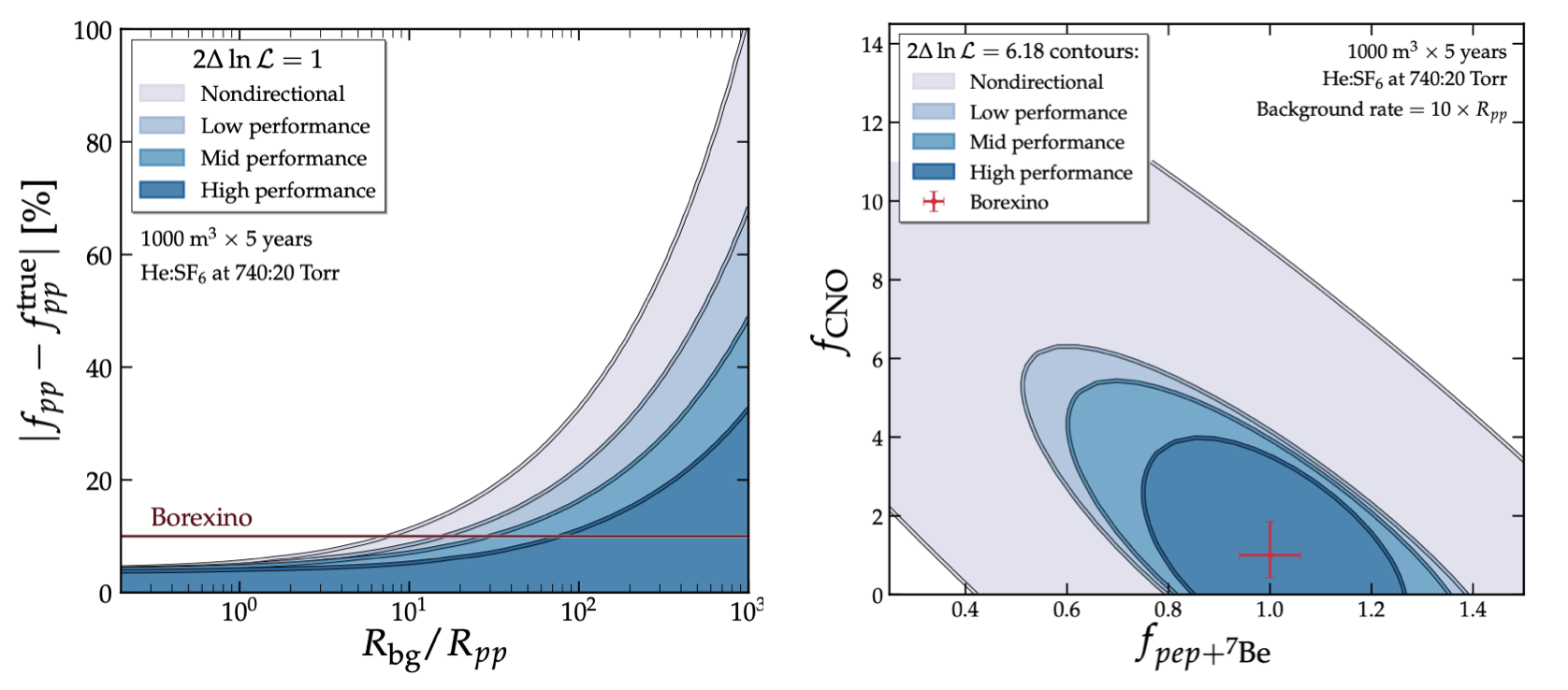}
    \caption{Left: precision relative to a pp neutrino flux measurement against the total non-neutrino electron background rate. The background is assumed to be uniformly distributed in energy and directionally uniform. Right: the expected 2$\sigma$ sensitivity for a combined estimate of the CNO and pep+$^7$Be fluxes. The background rate is set at 10 times the pp electron recoil rate. The different shadowed contours represent varying degrees of directional sensitivity, spanning from an optimistic projection for electron recoil energy/angular resolution based on gas simulations (darkest blue) to a scenario devoid of directional sensitivity altogether (lightest blue). Plots from \cite{vahsen2020cygnus}.}
    \label{fig:ppCNONeutrino}
\end{figure}
Fig. \ref{fig:ppCNONeutrino} in the left plot illustrates the expected relative precision to a pp neutrino flux as a function of the total non-neutrino electron background rate, while in the right plot the expected 2$\sigma$ sensitivity for a combined measurement of the CNO and pep+$^7$Be fluxes, with the background rate set at 10 times the pp electron recoil rate in shown. The estimations are presented for a 1000 m$^3$ experiment employing a gas mixture of 740:20 Torr of He:SF$_6$ and operating for five years. The four scenarios in the plot represent four general performance levels for track orientation recognition, angular resolution, and energy resolution. The lightest blue represents a lack of directional sensitivity, while the darkest blue reflects a high-performance benchmark constrained by factors like multiple scattering, diffusion, and readout resolution. These factors would cap the angular resolution at approximately 15° around 50 keV. While these performance benchmarks serve as a proof of concept, they clearly illustrate that enhancing directional performances can substantially enhance the capability to measure these fluxes, even without increasing in the event rates. Remarkably, as shown in \ref{fig:ppCNONeutrino}, in the right plot high-performance detectors can achieve Borexino's precision with a background-to-signal ratio of $R_{bg}/R_{pp}=80$ For context, Borexino observed 134 signal counts/day/(100 t) from pp interactions, with a background rate in the low energy region (0.19–2.93 MeV) of only 317 counts/day/(100 t) with a ratio $R_{bg}/R_{pp}\leq 2.3$.\\
Simultaneously, a CYGNUS $O(1000) $m$^3$ detector can explore the Spin-Independent sub-10 GeV region of Dark Matter masses below the neutrino floor, discriminating an eventual interaction involving DM from a neutrino one.\\
These capabilities constitute a compelling motivation for the development of a ton-scale TPC \cite{vahsen2020cygnus,SolrNuDir}.\\ 
Finally, as concluded in \cite{vahsen2020cygnus}, a TPC detector designed for rare events with a volume approaching $\mathcal{O}$(10) m$^3$ can detect neutrinos from the pp chain. The effectiveness of directional TPCs in this context relies heavily on detector performance and background levels. Hence, a focused study was undertaken on a specific detector to gain a thorough understanding.

\subsubsection{The CYGNO approach}
The CYGNO experiment (a CYGNus module with Optical readout), described in large detail in Chap. \ref{chap:CYGNO}, is being developed by a distinct collaboration that is working in the context of the broader CYGNUS collaboration \cite{instruments6010006}. The objective of CYGNO is to construct a high-resolution 3D gas TPC operating at atmospheric pressure, employing a helium-carbon tetrafluoride (He:CF$_4$) gas mixture for electron drift. The experiment will be situated underground at LNGS and will be employed in the directional detection of low-mass Dark Matter and solar neutrino spectroscopy. To realize these goals, CYGNO plans to utilize Active Pixel Sensors based on scientific CMOS technology (sCMOS) and photomultiplier tubes coupled to a triple-thin Gas Electron Multiplier (GEM) amplification stage. The high charge amplification attained in this setup, together with the relatively large light output attributed to the presence of CF$_4$, combined with the high granularity and high sensitivity of the sCMOS approach and the fast sampling in z allows for high precision 3D-tracking down to very low energy. Concurrently, the collaboration, in synergy with CYGNO, and in the context of the ERC Consolidator Grant INITIUM project, is also advancing the development of Negative Ion Drift (NID) operation at atmospheric pressure utilizing the CYGNO 3D optical readout method (more details in Chap. \ref{sec:NID}). 
The goal of the CYGNO experiment is to build a 30 m$^3$ experiment for directional DM searches. With a volume of 30 m$^3$, CYGNO can offer significant competitiveness in Dark Matter searches, particularly in detecting Spin-Dependent interactions. The high fluorine content would allow for the exploration of parameter space in the cross-section-mass parameter space that remains unexplored to date (see Sec. \ref{sec:DMSensitivity}).\\ 
Given that, the target electron density in the standard CYGNO gas mixture of He:CF$_4$ 60:40 (volume ration) is approximately six times higher than the one utilized for the plots shown in Fig. \ref{fig:ppCNONeutrino}, CYGNO-30 could potentially achieve the first directional observation of solar neutrinos from the pp chain. As discussed in the preceding section, given the target mass, a detector of this volume would only be sensitive to pp neutrinos. Therefore, in this thesis, a comprehensive feasibility study on CYGNO-like 30 m$^3$ detector capabilities to conduct a directional observation of neutrinos from the pp chain has been developed. This has been done by evaluating the resolution capabilities in energy and direction of such a detector, and the background level needed for carrying out this measurement, using the optical TPC approach. Moreover, the impact on the measurement of the better tracking performances carried out by NID operation has been evaluated.
This measurement, crucial for validating the detection technique, would pave the way for the advancement of larger TPC detectors employing varied amplification and readout technologies. These detectors could enable precise measurements of the CNO cycle, geoneutrinos, and core-collapse supernova neutrinos detection, potentially enabling supernova pointing as well \cite{vahsen2020cygnus}.

%% file: chapters/TheCYGNOINITUMExperiment.tex
\chapter{The CYGNO/INITIUM project}
\label{chap:CYGNO}
The CYGNO experiment is developing an innovative approach for directional Dark Matter (DM) and rare events searches at low energy. CYGNO employs a high-resolution Time Projection Chamber (TPC) with optical readout, operating with a 
Helium- Tetrafluoromethane gas mixture at atmospheric pressure and room temperature. The setup uses Gas Electron Multipliers (GEM) \cite{SAULI20162,SAULI1997531} coupled to optical sensors for amplification of the primary ionization charge and readout, respectively. In particular, to image a large detection area with high granularity using a reduced number of readout channels scientific CMOS (sCMOS) cameras are used. The combination of sCMOS camera images with the signals provided by photomultiplier tubes (PMTs) allows for particle 3D reconstruction with directional capabilities and sense recognition, at particle energies down to $\mathcal{O}(1)$ keV. The peculiar goal of the ERC Consolidator Grant INITIUM is to develop Negative Ion Drift operation (see Sec. \ref{sec:NID}) at atmospheric pressure within the CYGNO experimental approach. By introducing a small amount of highly electronegative gas, the primary electrons generated are absorbed by the gas molecules, thereby generating anions that will act as the charge carriers instead of electrons. This will lead to reduced diffusion, which translates into better tracking performances.
The two projects results therefore extremely synergic and the development and characterization of their prototypes and detectors proceeded together.\\ 
A complete overview of the CYGNO experiment can be found in \cite{instruments6010006}. Since the beginning of the project, many prototypes have been developed with increasing dimensions. This approach aims to solidify the experimental technique, refine the performances, and study the scalability of the optical gas TPC approach. The objective of CYGNO is to construct a detector with a volume of $\mathcal{O}(10-100)\ m^3$ in underground facilities, to be competitive in Dark Matter searches and solar neutrino spectroscopy with the unique feature of the directional sensitivity. Additionally, CYGNO is part of the CYGNUS project, which aims to establish a multi-site Galactic Recoil Observatory, that can probe Dark Matter hypotheses beyond the "neutrino fog" (once "neutrino floor") \cite{O_Hare_2021}, and measure the nuclear and electron elastic scattering of neutrinos from the Sun and supernovae with a directional approach. Currently, the CYGNO collaboration comprises approximately 50 researchers from institutions across four different countries: Italy, Portugal, United Kingdom, and Brazil. 
In this chapter, the CYGNO experimental approach is presented (Sec. \ref{sec:CYGNOTec}) together with the choice of the gas mixture (Sec. \ref{sec:gasmixture}), the amplification stage (Sec. \ref{sec:GEM}), and the readout strategy (Sec. \ref{sec:camera} and Sec. \ref{sec:PMT}). The staged CYGNO approach and its timeline are discussed in (Sec. \ref{sec:Timeline}), and its synergy with the INITIUM project is illustrated in Sec. \ref{sec:INITIUM}. Finally, a brief illustration of the primary science case of the CYGNO/INITIUM project for directional Dark Matter searches and its motivation is presented in Sec. \ref{sec:PhysCase}.

\section{The CYGNO experimental technique}
\label{sec:CYGNOTec}
\begin{figure}
    \centering  \includegraphics[scale=.5]{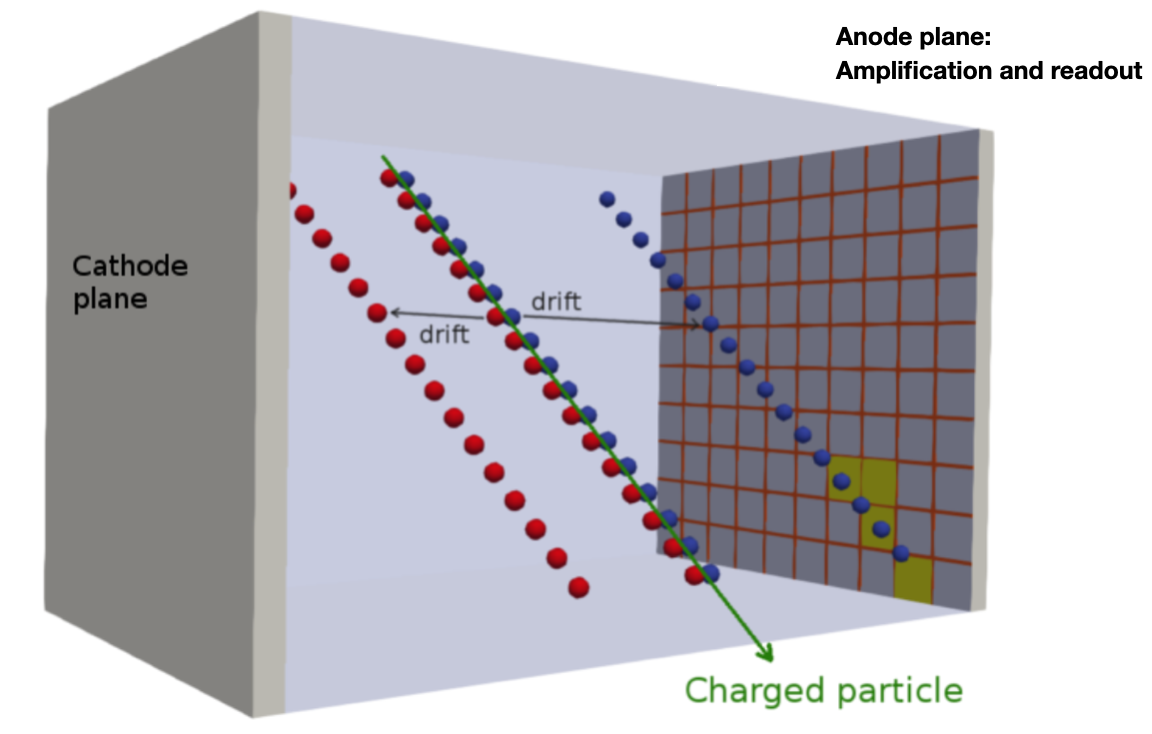}
    \caption{Illustration of a Time projection chamber highlighting its key features.}
    \label{fig:tpcscheme}
\end{figure}

TPCs are detectors consisting of a sensitive volume enclosed between two electrodes: a cathode and an anode. These elements generate an electric field inside the sensitive volume, that is maintained homogeneous by a field cage composed of metallic ring kept at a progressively lower tension along the drift direction \cite{10.1063/1.2994775, Nygren:1974nfi, ATWOOD1991446}. Typically, the anode of the TPC is instrumented with a readout system (but also the cathode can be). In Fig. \ref{fig:tpcscheme} a schematic view of a TPC is shown, with its key components. 
The passage of an ionizing particle inside the sensitive volume creates a trail of electrons-ions pair along its trajectory. The electric field in the active gas volume makes the electrons drift toward the anode. Since the charge generated during ionization is often too small for direct detection, a stage of amplification is commonly placed at the anode to amplify the primary electron charge. The amplified signal can be recorded using various technologies. In CYGNO, light is generated during the primary electrons amplification process through the interaction of them with CF$_4$ molecules. This light produced in this way is then collected by a scientific CMOS cameras and PMTs.
TPCs are by construction 3D detectors, with low energy threshold, due to the small ionizing power needed to create an electron-ion pair ($\mathcal{O}(10)$ eV), and thus suitable for low energy electron and nuclear recoil tracking. The detector concept of CYGNO integrates 4 pioneering elements, illustrated in the following points:
\begin{itemize}
    \item The target gas comprises a mixture of He and CF$_4$. This multi-target gas mixture enables simultaneous sensitivity to both Spin-Independent (SI) and Spin-Dependent (SD) coupling to DM (see Sec. \ref{sec:WIMPInteraction}). Additionally, CF$_4$ has scintillation properties which are optimal for an optical readout given its high light yield and the emission wavelength in the visible range. The addition of a 60\% Helium content in volume to the CF$_4$ ensures a low gas density, roughly comparable to that of air, and ensures stability in long-term detector operations. This allows electron and nuclear recoils to travel relatively longer distances, facilitating their direction reconstruction.
    \item The amplification stage consists of three 50 $\mu m$ thick Gas Electron Multipliers. These devices are a type of Micro-Pattern Gas Detectors (MPGD), which provide significant charge amplification gain, enabling low energy thresholds, and offer high granularity, aligning well with the sCMOS camera sensor and optical system used in the CYGNO setup \cite{knoll2010radiation, SAULI20162}. Moreover, GEMs are one of the few MPGDs that can be manufactured in large dimensions, a crucial requirement for the upcoming stages of the experiment (Sec. \ref{sec:CYGNOFut}).
    \item The scintillation light produced by the CF$_4$ is optically read out using both PMTs and sCMOS cameras. Optical sensors, in general, offer a significantly higher signal-to-noise ratio compared to charge detectors, enhancing detection efficiency. sCMOS cameras provide a highly granular 2D readout at a reasonable cost, with a high quantum efficiency in the optical band, making them suitable for imaging large areas when coupled with appropriate lenses (Sec. \ref{sec:camera}). 
    With their long acquisition times, sCMOS cameras effectively yield a 2D projection of all the recoil tracks, on the GEM plane (x-y plane), allowing for the extraction of tracks topology and directional information. 
    The PMTs sample the light produced by the track while primary electrons gradually pass through the amplification plane. The PMT (Sec. \ref{sec:PMT}), thus, is able to measure the relative extension of the track along the drift direction (i.e. z coordinate).
    The combination of the sCMOS camera with PMT enables the accurate measure of the track energy using two distinct photosensors. 
    While the solid angle coverage is limited, affecting the number of photons reaching the optical sensors, this is effectively balanced by the considerable gain achievable through the GEM amplification structure (Sec. \ref{sec:GEM}).
    \item The ability to operate at atmospheric pressure. It ensures a favorable volume-to-target-mass ratio, simultaneously reducing engineering demands for the realization of a vessel able to mechanically tolerate pressure differences, and minimize internal backgrounds. Among all the directional detectors, CYGNO stands out as the only directional DM search detector to operate at atmospheric pressure. This enhances the experiment's exposure without compromising the recoil track lengths, given the low gas density.
\end{itemize}

\subsection{Choice of the gas mixture}
\label{sec:gasmixture}
\begin{figure}
    \centering
    \includegraphics[scale=.40]{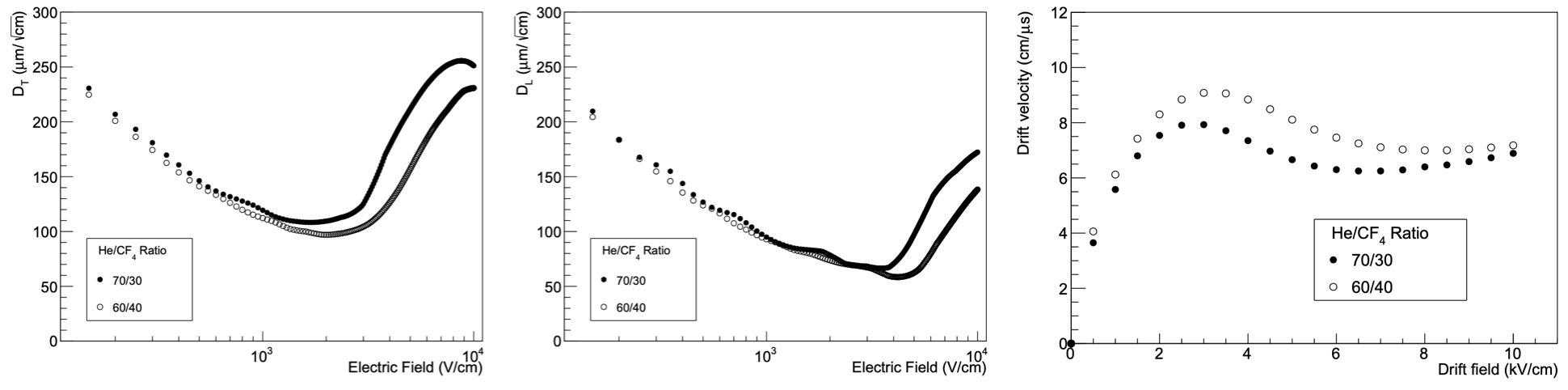}
    \caption{Left/Center: Transversal/Longitudinal diffusion for electrons in a He:CF$_4$ mixture 60:40 and 70:30 simulated with Garfield++. Right: plot of the electron drift velocity as a function of the drift field intensity from Garfield++ simulation. Plots from \cite{Baracchini_2020_Stab}.}
    \label{fig:diffvel}
\end{figure}

\begin{figure}
    \centering
    \includegraphics[scale=.3]{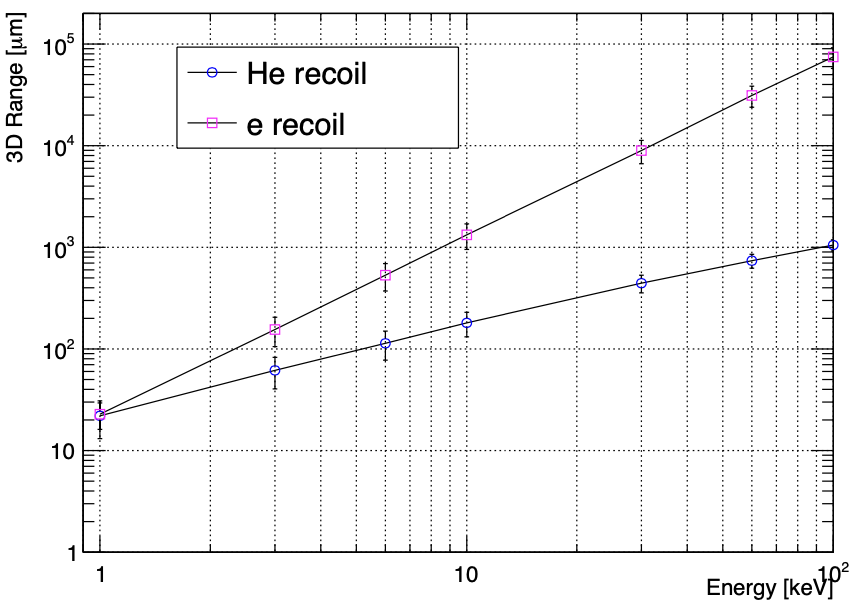}
    \caption{Average 3D range for He recoils and electrons recoils as a function of their kinetic energy.}
    \label{fig:ranges}
\end{figure}

The gas mixture selection takes a central role as it strongly influences the experimental properties of the detector and delineates the target for the physics searches. Tetrafluoromethane, CF$_4$, is a scintillating gas known for its high light yield within the visible spectrum, even when combined with noble elements \cite{LMMargato_2013,FRAGA200388,AMorozov_2012,KMPZMT_CF4Trasnp}. The emitted light spectrum of the He:CF$_4$ gas mixture, shown together with the sCMOS camera and PMT sensitivities in Fig. \ref{fig:QE}, exhibits two distinct continua, the first peaking around 290 nm (UV region) and the second at 620 nm (visible region). Emission within the visible orange region stems from the de-excitation process of a Rydberg state of the neutral CF$_3^{*-}$, resulting from CF$_4$ fragmentation, with an associated energy threshold of 12.5 eV. Conversely, the dissociative ionization threshold is 15.9 eV, and numerous resulting ionic fragments from CF$_4$ fragmentation, emit within the UV band. Gas mixtures incorporating CF$_4$ display remarkably low transversal and longitudinal diffusion during the drift phase \cite{SAULI20162}, attributable to the substantial characteristic low scattering cross-sections of electrons with CF$_4$.\\
A low diffusion coefficient of the electrons in the gas is a factor that permits to maintain high-quality 3D tracking of a Time Projection Chamber. Having a lower diffusion allows the realization of a detector with a longer drift field (thus a larger volume) while maintaining the same tracking performance. In the CYGNO experiment, Helium (He) is combined with CF$_4$ in the gas mixture. Helium, being a noble gas, is chemically stable and inert, making it practical for use in particle detectors. Additionally, when combined with other molecular gases, Helium can provide large electron amplification, as evidenced in prior studies \cite{SAULI20162}. The strategic combination of the very low density of Helium with CF$_4$ scintillation properties allows CYGNO to operate at atmospheric pressure while maintaining an overall target density only approximately twice as large as that of the NEWAGE experiment, which operates with pure CF$_4$ at 100 Torr of pressure \cite{10.1093/ptep/ptab053}. \\
A study of optimization was carried out to determine the optimal He:CF$_4$ ratio that would enhance the gain and light production in the experimental approach of CYGNO, while simultaneously minimizing diffusion and detector instabilities. A ratio of 60:40 was identified as the most suitable, as reported in \cite{Baracchini_2020_Stab,Campagnola:2313231}. The transversal and longitudinal diffusion coefficient per unit of length traveled by the electrons, and the drift velocity as a function of the electric field, calculated using the Garfield++ software of simulation \cite{Veenhof:1993hz,VEENHOF1998726} are reported in Fig. \ref{fig:diffvel}.
From the simulation \cite{Baracchini_2020_Stab}, a diffusion coefficient for both longitudinal and transversal diffusion below 140 $\mu$m/$\sqrt{\text{cm}}$ in a range of electric field of $[0.5, 1]$ kV/cm is expected. This can be compared with the diffusion measured in Ar:CF$_4$ of around 600 $\mu m/\sqrt{\text{cm}}$\cite{ParticleDataGroup:2020ssz}. The drift velocity is estimated to be $\sim3-6\  \text{cm}/\mu$s in the same range. The energy required in the gas mixture to produce a single electron-ion pair (namely the W-value) results in average $\sim 46.3$ eV \cite{Sharma:1998xw}. To further reduce diffusion and increasing the spatial resolution of the track, Negative Ion Drift studies are in progress in the context of the INITIUM project. The characteristics of this peculiar modification in the TPC operation are described in Sec. \ref{sec:NID}, including the first successful demonstration of its feasibility at atmospheric pressure within the CYGNO readout strategy. \\
Finally, the electron and He ion recoil ranges have been simulated using the GEANT4 toolkit \cite{AGOSTINELLI2003250} and the SRIM software \cite{Ziegler1985}, respectively in the CYGNO gas mixture.
The mean 3D ranges, namely the distance in a straight line between production and absorption points, as a function of the kinetic energy of He nuclei and electrons are shown in Fig. \ref{fig:ranges}. It can be observed from the plot that the range of Helium recoils is below a few millimeters up to 100 keV, resulting in the creation of a dense and bright spot slightly asymmetric in the recoil direction that preserves the directional information. Contrarily, electrons can travel several cm inside the gas volume before losing their entire energy.

\subsection{Triple GEM amplification system}
\label{sec:GEM}
Among the devices capable of providing high gains in gas, GEMs are highly performing. GEMs represent a specific implementation of Micro Pattern Gas Detectors, initially designed to enhance spatial granularity and efficiency with respect to wire chamber detectors, especially in the presence of high particle fluxes \cite{knoll2010radiation,SAULI20162}. This device was invented by Sauli in 1997, and an extensive discussion about its characteristics and features can be found in \cite{SAULI20162}. A GEM is composed of a thin, strongly insulating polymer layer coated with conductive metal on the two sides, with a high density of equispaced holes. When a potential difference is applied across the two electrodes, high electric fields of approximately $\mathcal{O}$(10) kV/cm are produced within the holes. These fields are strong enough to trigger avalanches when a single electron crosses the holes. The ions generated during the avalanche are primarily captured by the upper metallic surface, mitigating ion back-flow in the drift region and ensuring stable operations, even in high-rate environments \cite{Tripathy:2021puz}. The materials used to fabricate GEMs can vary, but the most commonly used materials include fiberglass or Kapton$^{TM}$ for the insulating layer and copper for the coating. However, R\&Ds are currently been carried on to find alternative materials for the insulating layers of GEMs to reduce their radioactivity content \cite{pr11041215}.  GEMs can be manufactured with various thicknesses, from 50 $\mu$m up to 1 mm. For 50 $\mu$m thick GEMs, the hole diameter is typically 70 $\mu$m wide, as smaller holes do not significantly enhance the effective gain due to an exceedingly high charge density in the hole \cite{SAULI20162}. The pitch between holes is typically 140 $\mu$m. A hole of this size is realized through a technique called etching, consisting of precise chemical excavation of tiny holes in materials. The etching procedure applied to GEMs yields to a double-conical geometry for the hole, with the slimmer diameter at its center. The precise localization of avalanches within these GEM holes enables exceptional granularity, especially pronounced in the 50 $\mu$m thin ones, largely preserving the x-y topological characteristics of the primary electron cloud. The secondary electrons generated by a GEM during the amplification process can either be collected by an anode or directed to another GEM for additional amplification. Indeed, the use of multiple GEMs stack has been explored and researched for various applications, achieving gains of $\mathcal{O}(10^4-10^6)$ \cite{SAULI20162,GEMOptRead}.\\ As will be illustrated in Sec. \ref{sec:optics}, the optical system coupled to the sCMOS camera to image a large area, results in a notable reduction in signal intensity due to its limited solid angle coverage. For this reason, the amplification stage of CYGNO consists of a stack of three 50 $\mu$m GEMs in order to achieve a very large gain, which implies more light, to compensate for its loss. Ongoing R\&D efforts focusing on further optimizing and customizing the amplification structure for the CYGNO experiment
are in progress.

\subsection{Optical readout}
\label{sec:optics}
Luminescence in gases is a thoroughly investigated and identified phenomenon. When particles interact with the gas, they can ionize atoms and molecules and also excite them. As these atoms and molecules undergo de-excitation, they emit photons. The quantity and spectrum of the produced light are significantly influenced by the gas mixture content, its density, and the presence and intensity of the electric field applied. In typical gas mixtures with CF$_4$ content, the emission can range between $10^{-2}$ and $10^{-1}$ photons per electron produced in the avalanche process during the amplification \cite{knoll2010radiation,LMMargato_2013,FRAGA200388,MONTEIRO201218}. \\
The optical readout approach consists in detecting the light produced along with the electronic avalanches rather than the charge, a concept pioneered in 90s studies involving parallel plates detector \cite{optReadFirst}. Charge readout poses challenges for detecting extremely low energy deposits, since it is often accompanied by electronic noise. The mitigation of this noise requires foresight in the design of the detector and the electric circuits. On the other hand, optical sensors can rely on significantly higher signal-to-noise ratios due to the extremely lower noise level. Moreover, the placement of the readout sensors outside the sensitive volume minimizes radioactivity and gas mixture contamination. Nevertheless, this positioning advantage comes with the trade-off of the reduction of the total light acquired, due to limited solid angle coverage which is caused by the lower flux going further from the light source. \\ 
As a directional detector, CYGNO's objective is to measure both the direction and energy of recoils. To achieve optimal assessment of the recoil direction, a comprehensive 3D analysis of the track is crucial. This goal in CYGNO is achieved through the simultaneous employment of two optical sensors: a scientific CMOS camera that measures the light produced along the x-y projection of the tracks, and a set of PMTs that measure the integrated light and samples the track development along the drift direction (z-axis).

\subsubsection{The sCMOS camera}
\label{sec:camera}
\begin{figure}
    \centering
    \includegraphics[scale=.23]{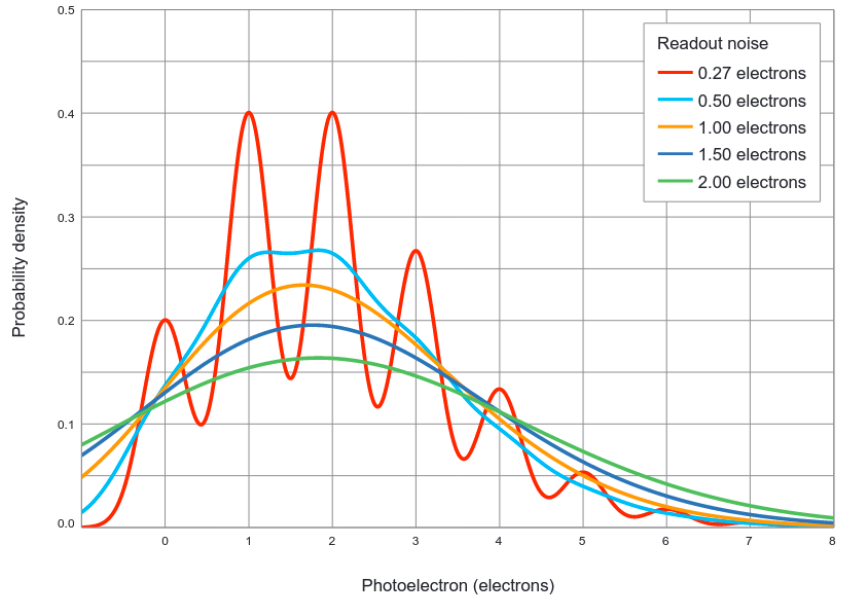}
    \caption{In figure, the probability density function of the number of photoelectrons for different readout noise (different colors) is shown. Plot from \cite{QuestCamera}.}
    \label{fig:ReadNoise}
\end{figure}

\begin{figure}
    \centering
    \includegraphics[scale=.4]{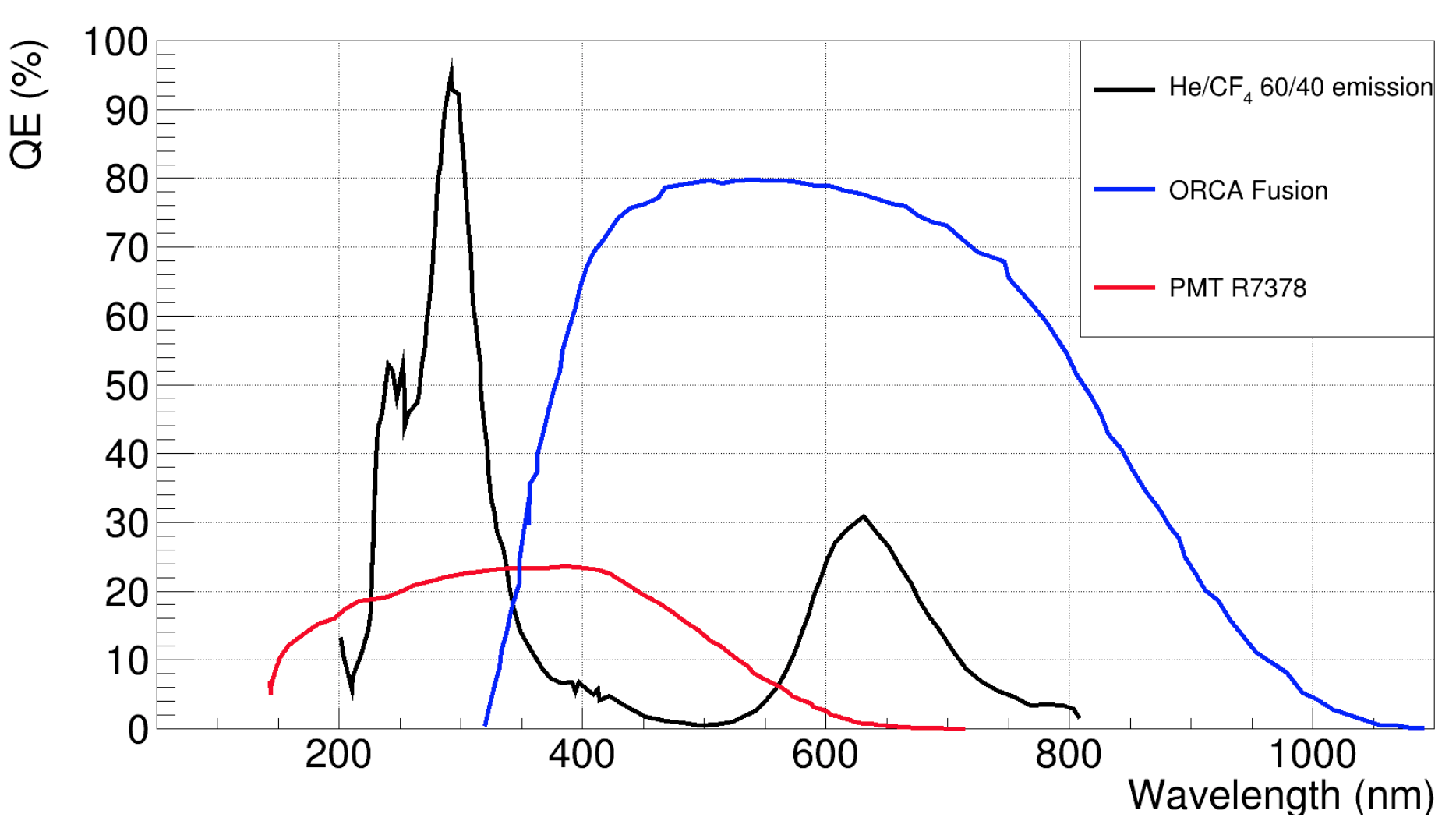}
    \caption{The quantum efficiency as a function of the wavelength is shown for the Orca Fusion camera (blue) and the PMT R7378 (red) from Hamamatsu, together with the emission spectrum of the He:CF$_4$ 60:40 gas mixture is presented (black).}
    \label{fig:QE}
\end{figure}

Leveraging on the benefits of low noise, high granularity, and significant potential for advancements, scientific CMOS (sCMOS) has been established for the readout system in the CYGNO experiment.
CYGNO utilizes sCMOS cameras manufactured by the Hamamatsu Corporation \cite{HamamatsuPhot}. Over time, the collaboration tested various models, with increasing performances, and their main features are shown in Table \ref{tab:camersa}. The different acquisition modes, Fast, Slow, and Ultra Quiet (U.Q), are related to the time that the camera employs to expose the sensor to the light. The evolution of technology is evident in the increase in pixel number, reduction in pixel size, and most importantly reduction of noise. Specifically, the noise RMS in the Orca QUEST\textsuperscript{TM} model is impressively low at 0.27 electrons, enabling the precise counting of photons converted in a pixel.
\begin{table}
    \centering
    \begin{tabular}{|c|c|c|c|}
    \hline
        Camera Model & Orca Flash\textsuperscript{TM} & Orca Fusion\textsuperscript{TM} & Orca Quest\textsuperscript{TM} \\ \hline
        Number of pixels & 2048$\times$2048 & 2304$\times$2304 & 4096$\times$2304 \\ \hline
        Pixel Size ($\mu$m$^2$) & 6.5$\times$6.5 & 6.5$\times$6.5 & 4.6$\times$4.6 \\ \hline
        RMS Noise (Fast mode) & 1.6 e$^{-}$ & 1.4 e$^{-}$ & x \\ \hline
        RMS Noise (Slow mode) & 1.4 e$^{-}$ & 1.0 e$^{-}$ & 0.43 e$^{-}$ \\ \hline
        RMS Noise (U.Q. mode) & x & 0.7 e$^{-}$ & 0.27 e$^{-}$ \\ \hline
        Max Frame Rate (Fast) & 100 fps & 89.1 fps & x \\ \hline
        Max Frame Rate (Slow) & 30 fps & 23.2 fps & 120 fps \\ \hline
        Max Frame Rate (U.Q.) & x & 5.4 fps & 5 fps \\ \hline
    \end{tabular}
    \caption{Characteristics of the different sCMOS cameras from Hamamatsu tested by the CYGNO collaboration.}
    \label{tab:camersa}
\end{table}
Fig. \ref{fig:ReadNoise} shows the simulated p.d.f. of the camera response obtained for a Poissonian distribution of photoelectrons produced per pixel (with an average number of photoelectrons $\lambda=2$) convoluted with the noise of the camera for different noise level. 
As can be seen from the picture, the readout noise level of 0.27 e$^-$ RMS (indicated by the red line) is the only one enabling differentiation of signals arising from the different numbers of photons, allowing for direct photon counting. Hence, this new class of sCMOS is often referred to as quantitative CMOS (qCMOS) cameras.\\
sCMOS cameras exhibit high sensitivity within the optical spectrum, while they are insensitive to UV and IR.
The Quantum Efficiency (QE) can peak at around 80\% at approximately 600 nm. This matches well with the emission peak in the visible range for the He:CF$_4$ gas mixture used in CYGNO. In Fig. \ref{fig:QE}, the Quantum Efficiency of the ORCA Fusion is shown, together with the QE of the PMT R7378, a bialkali borosilicate PMT \cite{PMTHamam} used in LIME (Sec. \ref{sec:LIME}). Additionally, the emission spectrum of the CYGNO gas mixture, as measured in \cite{FRAGA200388}, is presented. It's worth noting that the sCMOS sensor exhibits a significantly higher QE compared to conventional PMTs.

\subsubsection{Optical system}
Contemporary sCMOS cameras equipped with 2D arrays of millions of pixels enable the imaging of extended areas when coupled to suitable optical systems. Specifically, areas as large as $\mathcal{O}$(1) m$^2$ can be imaged with a single camera while maintaining excellent granularity (area seen by each pixel) of the same order magnitude of the segmentation of the GEM amplification stage.
\begin{figure}
    \centering
    \includegraphics[width=0.75\linewidth]{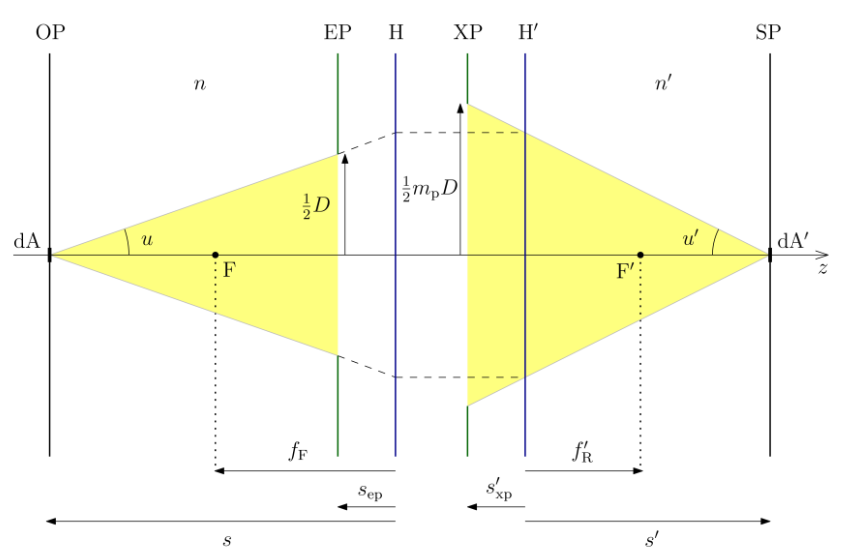}
    \caption{Diagram illustrating a lens in the Gaussian approximation, adapted from \cite{10.1088/978-0-7503-2558-5}. The yellow cones represent the photons emitted from the object plane (OP) and focused onto the sensor plane (SP). Plot from \cite{DhoTesi}.}
    \label{fig:OptSystem}
\end{figure}
Fig. \ref{fig:OptSystem}, shows a schematic representation of an optical system with a thick lens in Gaussian approximation projected onto the plane formed by the optical axis (connecting object and sensor and passing through the center of the lens) and the perpendicular axis \cite{10.1088/978-0-7503-2558-5}.
In the picture, OP designates the object plane, housing the focused source, while SP denotes the sensor plane, i.e. the location. of the photosensor. EP and XP represent the entrance and exit planes, respectively, through which photons enter and exit the lens. H and H' designate the hyperfocal planes, representing the position of a thin lens possessing the optical characteristics of a thick lens. D represents the aperture of the lens, determining the surface area designated to receive photons. The infinitesimal element of area emitting photons is denoted as dA, with u representing the tangent of the opening angle. The variables s and s' indicate the distances between the object and the hyperfocal plane H, and between the sensor and the hyperfocal plane H', respectively. Additionally, $f_F$ represents the focal distance, and $F$ represents the focal point. The illustration depicts the operation of a typical thick lens. Photons within the yellow region on the left of EP are captured by the lens and focused onto the sensor. When OP, EP, XP, and SP are parallel, the flux of photons ($\phi$) reaching the lens, emitted from the area $dA$, can be calculated as:
\begin{equation}
    \phi = \pi L u^2 \ dA
    \label{eq:flux}
\end{equation}
where L is the absolute luminosity of the source \cite{10.1088/978-0-7503-2558-5}. The lens's coverage of the solid angle fraction, denoted as $\Omega_f$, can be found by using eq. \ref{eq:flux} and dividing it by the total solid angle of photon emission, which is $4\pi$.
\begin{equation}
    \Omega_f = \frac{\phi}{L\ dA} \frac{1}{4\pi} = \pi u^2 \frac{1}{4\pi} = \frac{(D/2)^2}{4s^2}
    \label{eq:solidang}
\end{equation}
The relation between distances $s$ and $s'$, and the magnification I (the process of enlarging the apparent size) can be treated as analogous to the linear optics under the condition $s$ $>>$ $s_{ep}$, where $s_{ep}$ represents the distance between the entrance plane and the hyperfocal plane. In the context of CYGNO optical system, this condition holds, and therefore:
\begin{equation}
    1/f = 1/s +1/s' \ \ \ \ \ \ \ \ I = y'/y = s'/s
    \label{eq:magnific}
\end{equation}
with y and y' the size of the object and the size of the image respectively \cite{1987opt2.bookH}. By combining eq. \ref{eq:solidang} and eq. \ref{eq:magnific} the following expression for the solid angle can be derived:
\begin{equation}
    \Omega_f = \frac{1}{( 4N(\frac{1}{I}+1) )^2}
    \label{eq:solidangle}
\end{equation}
where N is the focal number defined as $N=D/f$.
The lens employed for all CYGNO detectors is a Schneider Xenon lens with a focal length $f$ of 25.6 mm, an aperture ratio $N$ (f-number) of 0.95, and an optical transparency of 0.85 in the visible range. To capture images of 34.9$\times$34.9 $cm^2$ areas using this optical system, the sensor must be placed 623 mm away from the surface of the closest GEM. 
The standard solid angle seen by the Schneider lens and, for instance, the ORCA Fusion camera (as shown in the second row of Table \ref{tab:camersa}) during the imaging of a 34.9$\times$34.9 $cm^2$ area is on the order of $10^{-4}$. This factor leads to a strong reduction in the number of photons reaching the sensor. Consequently, a very high gain is necessary to reach O(keV) energy threshold, and this is the reason for the choice of the triple GEMs (Sec. \ref{sec:GEM}) stack configuration chosen within the CYGNO project.

\subsubsection{The Photomultiplier Tube}
\label{sec:PMT}
\begin{figure}
    \centering
    \includegraphics[scale=1.2]{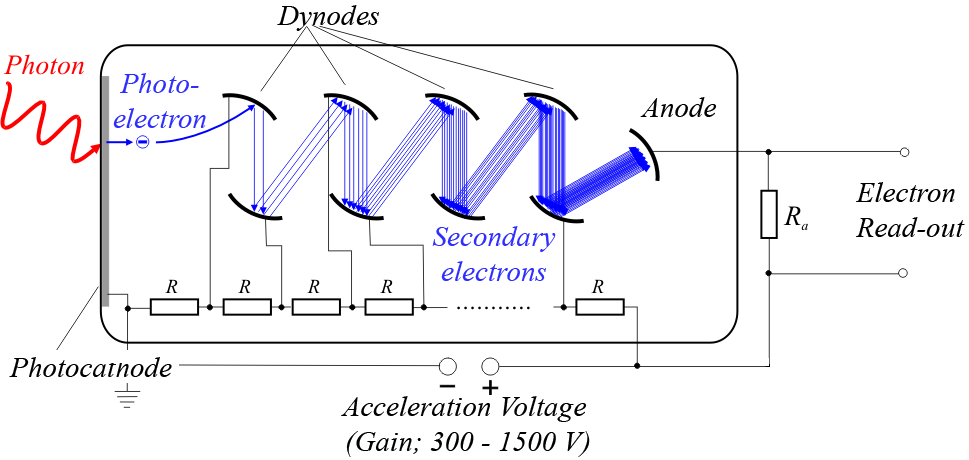}
    \caption{Working scheme of a PMT with all its key components highlighted. Plot from \cite{PMTScheme}.}
    \label{fig:PMTScheme}
\end{figure}

\begin{figure}
    \centering
    \includegraphics[scale=.2]{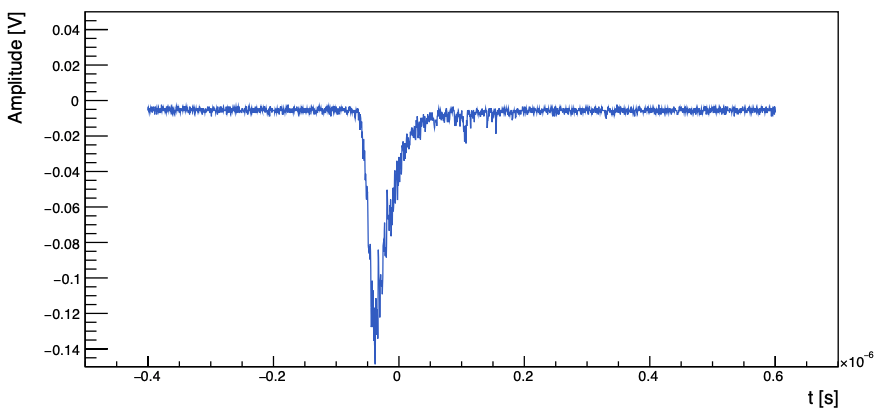}
    \includegraphics[scale=.2]{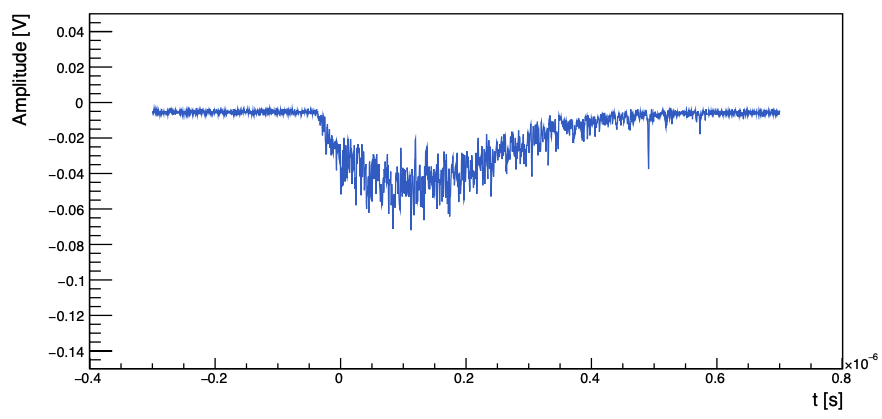}
    \caption{Left: PMT waveform of an alpha track parallel to the GEM plane. Right: signal of a track tilted with respect to the GEM plane. In the first case, since the track is parallel to the GEM plane, the electrons drifting will arrive at the amplification stage in a very short time window, thus the whole amount of light will be released in that small time window. In the second case, being tilted with respect to the GEM plane the electrons will reach the GEM plane for amplification and light production in a long time window, thus the light produced will be distributed in time.}
    \label{fig:PMTSign}
\end{figure}

The Photomultiplier Tube is a photon-detecting device with origins dating back to the '30s \cite{PMTHamam}. Its main feature is the single photon sensitivity together with the capability of maintaining a wide dynamic range. The PMT consists of a glass tube under vacuum, comprising an input window covering a photocathode on one side and a readout anode on the other. The photocathode is made of a highly photosensitive material, and it is responsible for converting the photons impinging on it into electrons through the photoelectric effect. The primary electrons produced by the photocathode are driven by an electric field to the amplification stage. Within the tube, a series of electron multipliers known as dynodes are present. These dynodes are kept at progressively increasing voltage levels and an electric field is created between these. The electrons driven by the electric fields are guided towards the series of dynodes. In each interaction of the electron with the dynodes, three to four additional electrons are emitted by the dynode for every incident electron. This avalanche process results in a significant amplification of the number of primary electrons. At the end of this chain, an anode collects the amplified electron signal and it is readout. Considering the PMT pulse rise time (of $\mathcal{O}$(ns)) and the very high photoelectron numbers, the time resolution of a PMT can arrive at values below $\mathcal{O}$(1) ns \cite{Liu:2008uda}. A scheme of the PMT is shown in Fig. \ref{fig:PMTScheme}.
The PMT gain, which corresponds to the power of amplify the primary signal, can be parameterized as $G=(\Delta V)^{\alpha N}$, where $\Delta V$ is the operative voltage applied, $\alpha$ is a coefficient that depends on the dynode material (ranging from 0.6 to 0.8) and $N$ is the number of dynodes. The dynamic range and gain achieved as high as $10^6$ depends on the number of dynodes, their composition, and the voltage divider. The quantum efficiency for different wavelengths of the PMT depends on the materials used for the window and photocathode. Typically, windows are crafted from borosilicate glass, allowing photons above 300 nm to pass through, and can be produced with relatively low radioactive materials. The most common materials used for the production of photocathode are bialkali, typically Sb-K-Cs, named by the presence of more than one alkali element. The QE of these photocathodes reaches its peak at 400 nm, remaining sensitive up to 650 nm. Alternative materials such as GaAsP(Cs) exhibit higher QE within the visible range. In any case, the PMT quantum efficiency typically does not exceed $\mathcal{O}$(30\%) at the peak.\\
The overall charge readout at the anode is directly proportional to the number of photons converted, providing thus a measurement of the energy released. Given the typically high gain of PMTs, the energy resolution depends mainly on the fluctuation of the primaries and not on the amplification.
In the CYGNO gas mixture, with the typical drift field, where the electron drift velocity is $\sim$5 cm/$\mu$s, a sampling rate of approximately O(1) GS/s for the PMT waveform provides a z-direction granularity of about 50 $\mu$m.
Given the very high time resolution of the PMT signal, this allows the study of the track's progression along the drift direction. This capability can enable not only the determination of the recoil's inclination with respect to the GEM plane but can also provide essential insights into the track topology, crucial for recognizing direction and orientation. As an example, Fig. \ref{fig:PMTSign} shows on the left (right) a typical PMT waveform for a track parallel to (inclined with respect to) the GEMs plane.

\section{The CYGNO timeline}
\label{sec:Timeline}
The CYGNO project's final aim is to implement a large-scale detector for the directional detection of DM and Solar neutrino spectroscopy. To strengthen the experimental technique, enhance performances, and explore the scalability of the optical gas TPC approach, multiple prototypes have been developed.
\begin{figure}
    \centering
    \includegraphics[scale=.3]{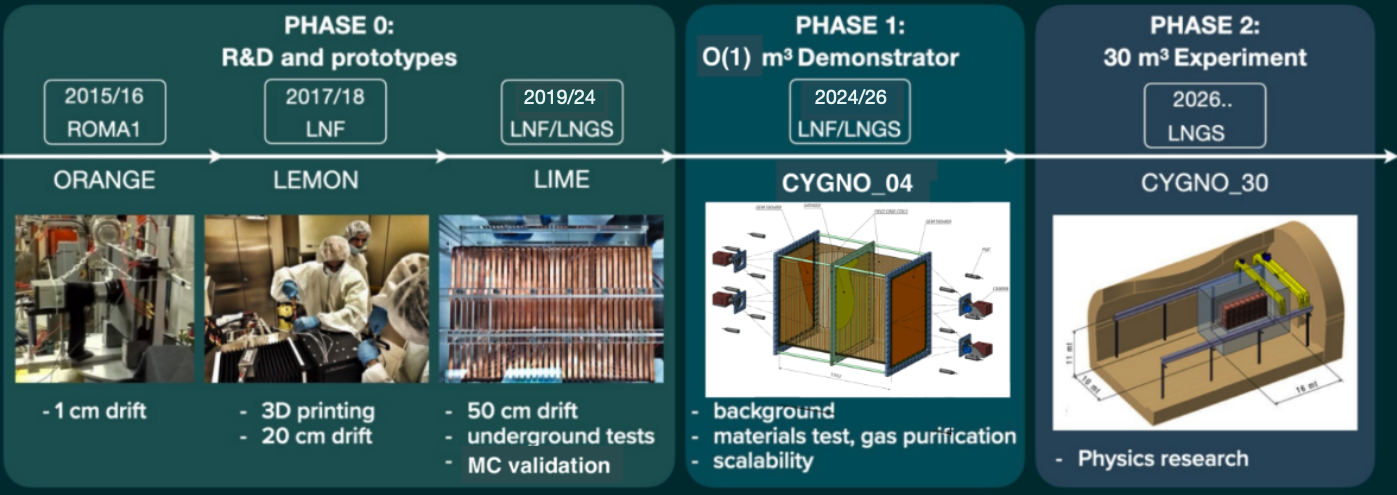}
    \caption{Timeline of the CYGNO experiment.}
    \label{fig:CYGNOTimeline}
\end{figure}
At the moment the experiment is in the PHASE\_0, focused on research and development, during which various prototypes were constructed and utilized to evaluate and enhance different aspects of the CYGNO detector concept. The subsequent phase, PHASE\_1, aims to validate the scalability of the detection technique by employing a detector with a volume of $\mathcal{O}$(1) cubic meter, featuring a modular readout system. This detector will also serve as a test for the radiopurity of the materials used. The ultimate objective is to progress to PHASE\_2, with the realization of a large-scale detector with a sensitive volume of 30 $m^3$. This phase will be the formal beginning of the CYGNO experiment which will give a significant contribution to Dark Matter searches and neutrino spectroscopy.
In Fig. \ref{fig:CYGNOTimeline}, the project's road map is shown.\\
The initial prototype, ORANGE, consisted of a 10×10 cm$^2$ area with a 1 cm drift gap, demonstrating the detection feasibility of O(1) keV energy deposits in the gas by coupling an sCMOS camera to a GEM amplification stage. Additionally, it evaluated the potential for 3D reconstruction by incorporating a PMT. Following ORANGE, the subsequent prototype LEMOn featured a 20 cm drift gap and a more expansive readout area of 20×24 cm$^2$. LEMOn was employed to investigate the detector’s performance concerning a larger drift distance and readout area.
In parallel, a prototype named MANGO, with a 10$\times$10 cm$^2$ readout area and an adjustable drift region, was built in 2018. MANGO's purpose was to conduct R\&D tests on different gas mixtures and amplification structures. In 2019, the construction of the LIME prototype concluded. LIME having a 50 cm drift and a 33$\times$33 cm$^2$ readout area detector represents the foundation for the larger-scale experiment. Following an overground commissioning phase, LIME was relocated underground at Laboratori Nazionali del Gran Sasso (LNGS) in June 2022 to undergo a comprehensive characterization of the detector in a future realistic environment. All these prototypes collectively constitute CYGNO PHASE\_0, aiming to affirm the optical readout's full capabilities using APS commercial cameras and PMTs, as well as validate the Monte Carlo simulation of anticipated backgrounds. 
Subsequent sections will delve into the latest prototypes and the primary results achieved with them.

\subsection{Early PAHSE\_0: first prototypes}
Three primary prototypes have been designed to explore and refine the CYGNO detection technique. This section discusses the first two prototypes, ORANGE and LEMOn. Both of them were readout with an ORCA-Flash 4.0\textsuperscript{TM} CMOS camera, with a light response of 0.9 counts/photons \cite{Marafini_2015}. The camera mounted a Schneider Xenon lens.

\subsubsection{The ORANGE prototype} The initial prototype developed by the CYGNO collaboration, known as ORANGE (Optically ReAdout GEM), was designed to showcase the feasibility of three-dimensional tracking via optical readout of GEMs. It featured a triple stack of 10$\times$10 cm$^2$ GEMs spaced 2 mm apart to amplify charge generated by events within a 1 cm drift gap, resulting in a total sensitive volume of 100 cm$^3$. The light produced was collected by the Hamamatsu ORCA Flash 4.0\textsuperscript{TM} positioned 20 cm from the final GEM, with a solid angle of 3.46$\times$10$^{-2}$. A scheme of the ORANGE experiment is shown in the left plot of Fig. \ref{fig:ORLEMON}. ORANGE served as an initial evaluation platform for reconstructing track images \cite{Campagnola:2313231}, offering an initial assessment of light yield performance \cite{Marafini_2015} and showcasing the first instance of three-dimensional track measurement using PMTs within the CYGNO experiment \cite{Antochi_2018}.\\

\subsubsection{The LEMOn prototype} 
\begin{figure}
    \centering
    \includegraphics[width=1.\linewidth]{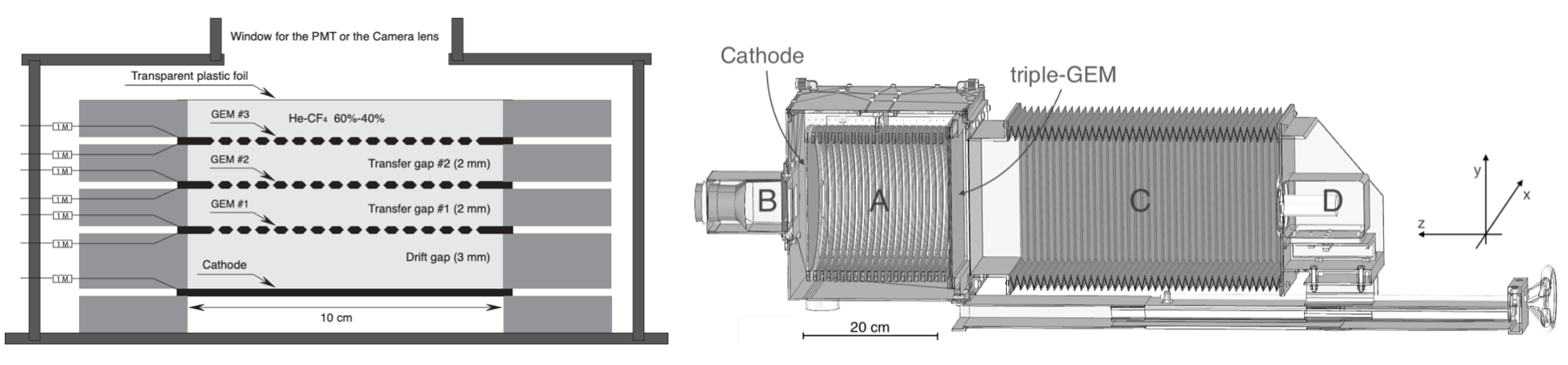}
    \caption{Left: sketch of the Orange detector. The triple GEM stack, the window for the camera, and the cathode are shown. Right: scheme of the LEMON experiment. The sensitive volume with the field cafe (A), the photomultiplier at the cathode (B), the adaptable bellow (C), and the sCMOS camera (D) are shown.}
    \label{fig:ORLEMON}
\end{figure}

The LEMOn (Long Elliptical MOdule) detector, consists of a 7-liter active volume detector featuring a 20 cm drift length and a 24$\times$20 cm$^2$ readout area, all housed within a gas-tight acrylic vessel. Within the chamber, an ellipsoidal field cage, comprised of silver wires supported by 3D-printed plastic structures with a 1 cm pitch, ensures uniformity of the drift field across the whole 20 cm drift length. The cathode is constructed using an ATLAS Micromegas mesh \cite{Kuger_2016}, featuring a wire diameter of 30 $\mu$m and a pitch of 50 $\mu$m. The amplification stage incorporates three 24$\times$20 cm$^2$ GEMs placed at a distance of 2 mm from each other. The detector is optically readout from a Hamamatsu ORCA Flash 4.0\textsuperscript{TM} coupled through a TEDLAR transparent window and an adjustable bellow which isolates from external light and at the same time allows the camera to be moved to keep the focus. Positioned approximately 50 cm from the final GEM, the camera images an area measuring 25.6$\times$25.6 cm$^2$, corresponding to an effective area seen by each pixel of 125$\times$125 $\mu$m$^2$. Behind the transparent cathode of LEMOn, an HZC Photonics XP3392 PMT is positioned. A scheme of the LEMOn experiment is shown in the left plot of Fig. \ref{fig:ORLEMON}. The LEMOn prototype served as a validator of the detector technique on a medium-size scale, evaluating the energy resolution \cite{Baracchini_2020_Stab}, studying the detector stability \cite{Baracchini_2020_Stab}, providing a preliminary estimation of the absolute z of Minimum Ionizing Particles (MIP) tracks \cite{Costa_2019}, and conduct a first estimate of electron/nuclear recoil identification capabilities \cite{Costa_2019,Baracchini:2020nut}. Moreover, a threshold of 2 keV while maintaining a rate of fake cluster of $<10$/y have been obtained \cite{Costa_2019}. With LEMOn, additionally, it was demonstrated that on a medium-sized CYGNO detector, the efficiency in detecting 5.9 keV electron recoils is 100\% if the drift field is kept above 300 V/cm, for drift distances up to 20 cm \cite{Baracchini_2020_Stab}. More details on the LEMOn experiment can be found in \cite{Costa_2019,ANTOCHI2021165209,8123941}.
\\

\begin{figure}
    \centering
    \includegraphics[scale=.4]{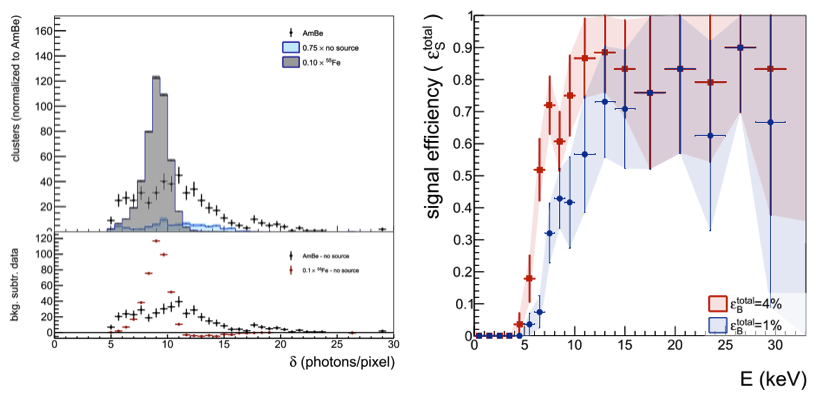}
    \caption{Left: distribution of the delta variable for background run, $^{241}$AmBe run and $^{55}$Fe run. The scale factors on the distributions are for live-time normalization purposes. Right: plot of the signal efficiency as a function of the energy for the two different cut $\delta>10 / 11$ with efficiency on background identification of $4-1\%$ respectively. Plots from \cite{Baracchini:2020nut}.}
    \label{fig:ERNRDiscr}
\end{figure}

\textbf{ER/NR discrimination} The capability to distinguish nuclear recoil and electron recoil is of fundamental importance for rare events searches. Indeed, low energy electron recoil in CYGNO can mimic a neutron/WIMP interaction at very low energies. Thanks to the high granularity TPC approach, it is possible to use a large amount of information on the track to efficiently discriminate between nuclear recoil (NR) and electron recoil (ER).
To assess the capability of ER/NR discrimination, LEMOn has been exposed to $^{55}$Fe source and $^{241}$AmBe source producing neutrons. The detector was shielded with lead in order to minimize the contribution from external gamma.
The measurement has been performed overground at Laboratori Nazionali Frascati. This implied a high occupancy of cosmic rays in the tracks among which the interesting events must be found. To discriminate the interesting events of nuclear recoil from $^{55}$Fe spots, cosmics, and high-energy electrons, the shape variables of the tracks have been used (Detailed definition in Sec. \ref{sec:splot}). Of particular relevance are the track length and the track slimness (ratio width/length) which can be used to separate short and round tracks of NR and $^{55}$Fe ER, from extended tracks of cosmics and ER. The former tracks indeed appear as a round, thick, and dense light spot, while the latter, with a much lower energy ionization profile, appear as long and slim tracks.
What mainly discriminates ER and NR is the energy ionization profile $dE/ dx$. This second category of tracks has a much higher $dE/ dx$. 
A variable called light density $\delta$, proportional to the $dE/ dx$, and defined as the total number of photons in the cluster divided by the total number of pixels, can thus be employed to discriminate ER and NR. 
As can be seen from the left plot of Fig. \ref{fig:ERNRDiscr} the light density variable can be obtained to separate the NR from the $^{241}$AmBe run and the ER from the $^{55}Fe$ run. In particular, with two different selections on $\delta$ considered, $\delta>10$ and $\delta>11$ an efficiency in $^{55}$Fe tracks rejection respectively of 96\% and 99\% can be obtained while keeping an efficiency on the NR signal of 50\% and 40\% respectively \cite{Baracchini:2020nut}.
The same efficiency has been computed for the two $\delta$ selections as a function of visible energy, and the results are reported in the right plot of Fig. \ref{fig:ERNRDiscr}. This measurement highlights the discrimination capability of the TPC optical approach. Despite the selection cut used being very loose and basic, a rejection power of $\sim 10^2$ can be reached below 10 keV. A multivariate analysis based on machine learning techniques for ER/NR discrimination is in progress and could exploit the full topological information to improve CYGNO discrimination capabilities.

\subsection{End of PHASE\_0: The LIME prototype}
\label{sec:LIME}
\begin{figure}
    \centering
    \includegraphics[scale=.4]{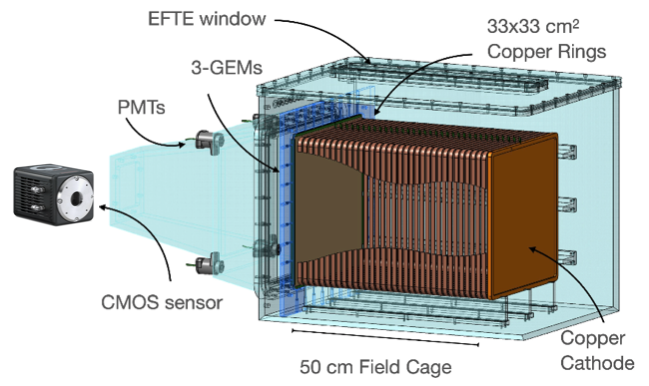}
    \includegraphics[scale=.2]{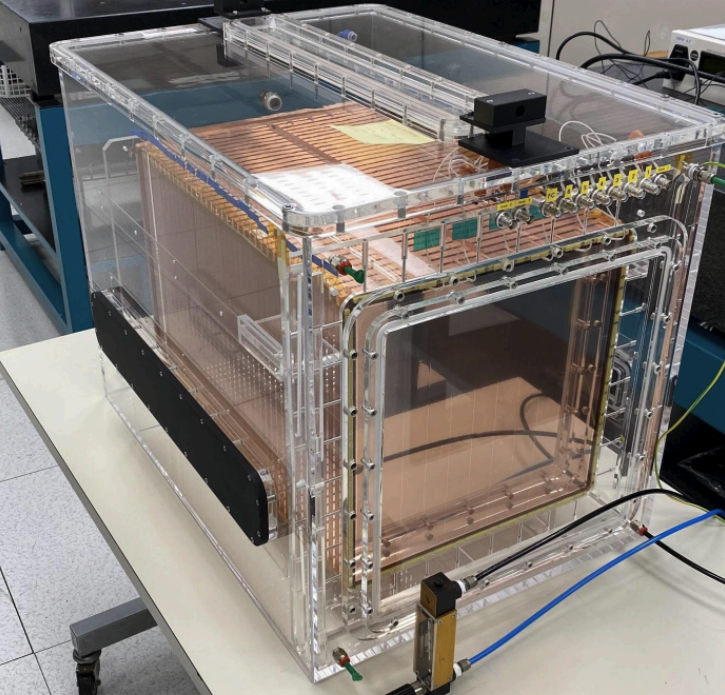}
    \caption{Left: a scheme of the LIME prototype. The arrangement in the acrylic vessel, the copper rings, and the optical readout are depicted. On top also the EFTE window is shown. Right: a picture of the LIME detector inside the acrylic vessel.}
    \label{fig:LimeScheme}
\end{figure}

The Long Imaging Module (LIME) is the largest prototype developed under the CYGNO experimental approach, representing a key milestone in the R\&D phase of the CYGNO project as the peak of the PHASE\_0. A scheme of LIME is shown in Fig. \ref{fig:LimeScheme}. 
The LIME TPC sensitive volume is contained between a copper cathode and an amplification stage composed of a triple-thin GEM stack. The light produced during the amplification process is detected by an sCMOS camera and four PMTs positioned around the camera, at the corners of a square facing the GEMs. The field cage has a square cross-section with each side measuring 330 mm and a drift length of 488 mm. It consists of 34 rounded square coils, each 10 mm wide, positioned 4 mm apart from each other, with a pitch of 14 mm. These coils are electrically interconnected via 100 M$\Omega$ resistors to ensure a uniform and controlled electric field within the gas vessel. The total sensitive volume amounts to approximately 50 liters. At one end of the field cage, a 0.5 mm thin copper cathode, with a frame matching the size of the field cage coils, establishes the base potential for the drift field. At the opposite side with respect to the cathode, at the anode, a stack of three standard thin GEMs (each 50 $\mu$m thick), covering an area of 33$\times$33 cm$^2$ is present. The holes have a diameter of 50 $\mu$m, and they are distributed with a pitch of 140 $\mu$m. The GEMs are spaced 2 mm one from the other, and the first GEM is 7 mm from the first ring of the field cage.
Being LIME the end of the PHASE\_0, these specific dimensions have been selected to replicate and test the drift length of the 0.4 m$^3$ CYGNO-04 detector in an overground and underground environment (Sec. \ref{sec:CYGNOFut}). The detector is enclosed in a vessel consisting of a 100 liters box made by 10 mm thick PMMA (Polymethyl methacrylate), ensuring durability and structural integrity. Additionally, an external Faraday cage, made of aluminum, shields the detector from any potential electromagnetic interference. Positioned at the top side of the vessel is a narrow opening of 5 cm in width and 50 cm in length that extends along the drift distance, sealed with a thin 150 $\mu$m ETFE window. This configuration allows for the passage of low-energy photons, reaching down to the keV energy range, into the gas volume, which allows for the use of external artificial radioactive sources for calibration purposes. 
On the same side of the vessel containing the GEM stack, a black PMMA conical structure is affixed to accommodate the optical sensors. This setup comprises 4 Hamamatsu R7378 PMTs, each with a diameter of 22 mm, in addition to an ORCA-Fusion\textsuperscript{TM} sCMOS camera. The camera, featuring 2304$\times$2304 pixels, with each pixel having an active area of 6.5$\times$6.5 $\mu$m$^2$, mounts a Schneider lens featuring a focal length of 25.6 mm and an aperture of 0.95, the camera is placed at a distance of 623 mm from the GEMs. The sCMOS camera, with a QE of around 80\% in the 450-630 nm range of wavelengths, aligns well with the emission spectrum of the He:CF$_4$ gas mixture (Fig \ref{fig:QE}). In this setup, the sensor covers a surface area of 34.9$\times$34.9 cm$^2$, resulting in each pixel covering an area of 152$\times$152 $\mu$m$^2$. The geometrical acceptance (solid angle) of the lens in this setup is calculated to be $\Omega$ = 1.2 $\times$ 10$^{-4}$.\\
After a first phase of overground commissioning \cite{refId0}, LIME was transported and installed in the underground facility of the Laboratori Nazionali del Gran Sasso (LNGS), where it is currently operational in a low background environment. The primary objectives of LIME are validating the capabilities of the CYGNO approach within an underground setting, confirming the accuracy of Monte Carlo (MC) simulations to model the detector background, and testing both the construction and operational protocols for the realization of PHASE\_1 (Sec. \ref{sec:CYGNOFut}). Additionally, it will deliver a precise, spectral, and directional measurement of the environmental neutron flux at LNGS.

\subsubsection{LIME overground commissioning}
\label{sec:overgroundstudies}
Before its underground installation at the LNGS, LIME underwent a commissioning phase at the overground LNF, to properly characterize the detector response. This step aimed to evaluate and document the studies on threshold, determination of absolute position along the drift direction, light output, and energy response. Numerous tests and measurements were conducted, comprehensively characterizing the detector's performance and establishing standardized data collection procedures for all subsequent campaigns. The subsequent section offers an overview of the results acquired during this commissioning phase. Additionally, a comprehensive analysis of the response to multi-energy X-Rays, being a core argument of this thesis, will be provided in greater detail in Chapter \ref{chap:LIMEDetector}. A comprehensive description of the result to which the author contributed can be found in \cite{refId0}.

\textbf{Optical effects}
Due to the lens effect, the light gathered from the image's outer regions radially diminishes while moving away from the center of the images. This effect is called vignetting effect. It has been observed that in LIME, this effect causes a light reduction up to approximately 20\% compared to the center of the image \cite{Goossens_2019}. This phenomenon was investigated by capturing images of a uniformly lit white surface and rotating the camera around the optical axis to eliminate any bias in the light's direction onto the sensor. Subsequently, a vignetting map was constructed from these images and is employed in all data analyses to correct for this radial inconsistency in light intensity. This effect, and how the correction is applied, is discussed more in detail in section \ref{sec:vignetting}.

\textbf{Light output}
The process of energy calibration, which translates the camera sensor counts into energy measurements, relies on a $^{55}$Fe source serving as a reference standard. Upon interaction with the gas, the 5.9 keV X-Rays emitted by the source, interacting via photoelectric effect generate spot-like tracks in the image, typically spanning around 20 mm$^2$ after factoring in diffusion effects and sensor granularity. The integral counts of the pixels within the reconstructed spots are fitted with a Gaussian function. The mean of this Gaussian is interpreted as the average light integral corresponding to a 5.9 keV deposit, serving as the calibration factor for light yield (LY). The standard deviation (sigma) of the Gaussian is utilized to estimate the energy resolution. Although the LY may fluctuate over time due to factors such as pressure and temperature, it typically remains around 10$^3$ counts/keV.

\textbf{Detection threshold and energy resolution at 6 keV}
The track reconstruction software sometimes groups pixels originating from noise over-fluctuations in the sensor. From time to time, these over-fluctuated groups of pixels can overcome the threshold and bypass the zero suppression process in the reconstruction. Pedestal runs, empty of real track signals, reveal that the contribution of these spurious clusters can be neglected beyond 400 camera counts belonging to the same cluster. Setting a threshold on clusters with a number of counts above this value is equivalent to setting a threshold in energy roughly at 0.5 keV. Analysis of the $^{55}$Fe spots also demonstrated an average energy resolution of approximately 14\% for clusters produced across the entire LIME volume. Implementation of a Multi-Variate Analysis (MVA) based on machine learning, correcting for light yield variation in the (x, y) plane mainly due to nonuniformity in the GEM response, results in an improvement of up to 10\% in energy resolution.

\begin{figure}
    \centering
    \includegraphics[width=1\linewidth]{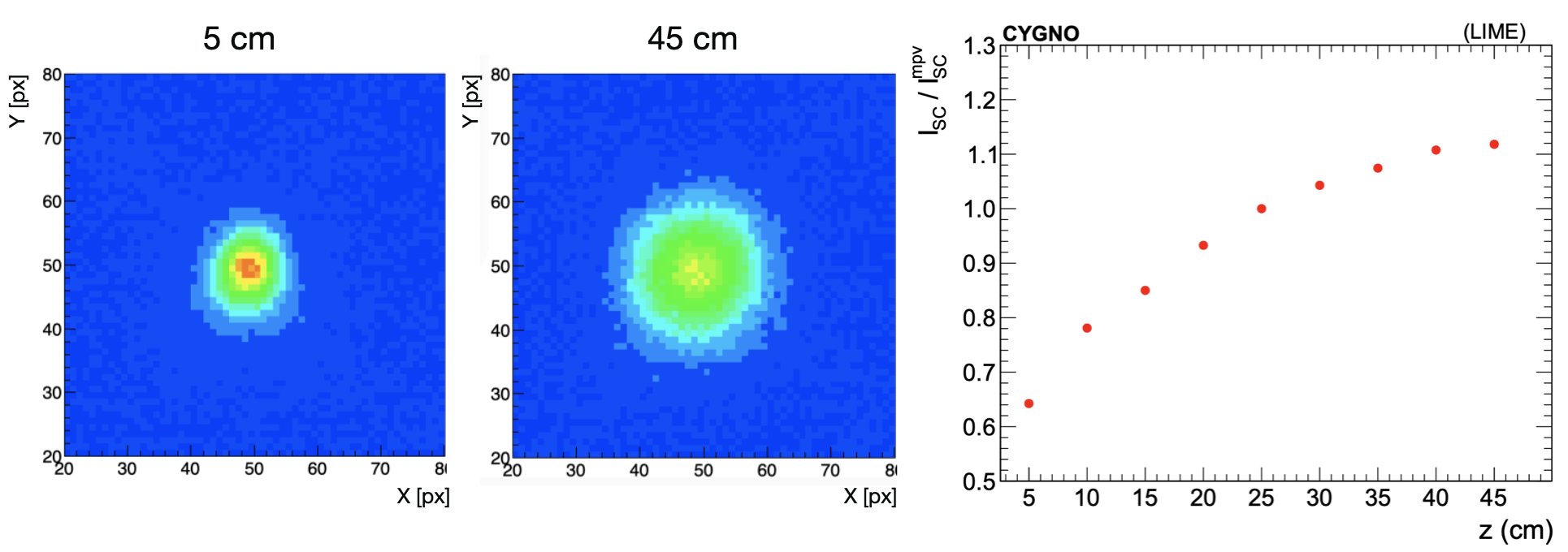}
    \caption{Left: Two iron spots acquired with the source at 5 cm and 45 cm distance from the GEM. The x and y axes are both in pixels. Right: Plot of the iron light integral $I_{SC}$ normalized to the value at 25 cm $I^{mpv}_{SC}$ as a function of the source position. Plot from \cite{refId0}.}
    \label{fig:saturationPlot}
\end{figure}

\begin{figure}
    \centering
    \includegraphics[scale=.4]{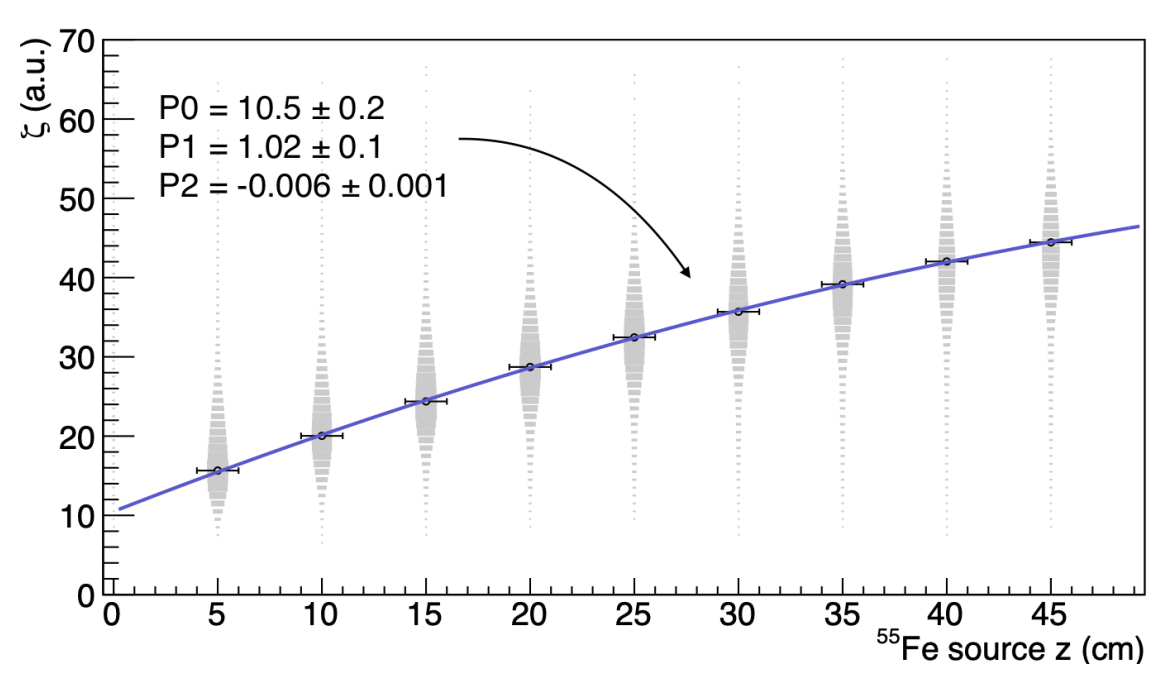}
    \caption{In figure the $\zeta$ distributions as a function of the z position of the $^{55}$Fe source are shown (gray bands). The mean values of the distributions (blue dot) fitted with a quadratic equation (black line) are also present together with the result of the fit. Plot from \cite{refId0}.}
    \label{fig:XiVsZ}
\end{figure}

\textbf{GEM gain saturation} The high quantity of charge generated in the electron avalanches can significantly diminish the electric field within the GEM holes, particularly on the last GEM, thereby reducing the effective gain provided by the device. This saturation phenomenon hinges on the total density of charge entering the GEM, which depends on the total charge produced by the interaction and the surface area over which ionization electrons are dispersed (thus the dE/dx). The exposition of LIME to the $^{55}$Fe source placed at various distances from the GEMs allowed to measure and quantify this effect as a variation in the response of iron spots at different distances. Tracks produced in proximity to the GEMs intensify the saturation effect, leading to a lower light yield. This occurs because, with the track diffusion in the drift region, the primary charge is spread both transversely and longitudinally over a larger area, resulting in a less dense electron cloud reaching the GEMs. Thus, closer tracks, diffusing less will be more subject to saturation, while on the contrary tracks produced farther from the GEM plane will be less subject to this. The dependence of the saturation as a function of the source position is shown in Fig. \ref{fig:saturationPlot}, together with two iron spots at different drift distances as an example. Counterintuitively, due to the lower energy loss per unit path length, saturation is expected to be less pronounced for higher-energy electron recoils, resulting in a more sparse distribution of ionization charge compared to the denser spot-like tracks. These tracks, composed of a long tail and a head in correspondence with the Bragg peak (more detail in Sec. \ref{sec:electronspattern}) will be indeed dominated by a non-saturated tail and only a small saturated region in correspondence with the end of the track.
\textbf{Absolute z position} Due to diffusion within the gas medium, the configuration of the primary ionization electron cloud reaching the Gas Electron Multipliers varies in a related way with the distance covered during drift, denoted by the absolute z position of the event. Previous studies, performed with the LEMOn prototype \cite{ANTOCHI2021165209}, demonstrated this relationship for ultra-relativistic electrons. Analyzing the track's width, expressed as the standard deviation $\sigma_T$ of its Gaussian transverse profile, with the z distance from the GEM plane, a clear correlation has been found between these variables. With LIME, studying spot-like events generated by the $^{55}$Fe X-Ray source, various track shape parameters have been examined to identify those most sensitive to the absolute z position. It was found that the product between the track transverse profile, and the RMS of the counts per pixel within the spots, denoted as $\zeta$, exhibits the most effective and clear dependence from the z position of the event, as depicted in Fig. \ref{fig:XiVsZ}. Despite the clear correlation, the $\zeta$ distribution feature extended tails in all instances, and the accuracy of the absolute z estimation deteriorates with increasing distance from the GEMs, with a resolution between 4 cm for smaller drift distances up to 8 cm for the larger ones. This investigation represents a preliminary analysis of the absolute z coordinate determination and the analysis remains in progress, also with the employment of alternative methodologies.

\subsubsection{LIME Underground}
\label{sec:LIMEUnderground}
\begin{figure}
    \centering
    \includegraphics[scale=.4]{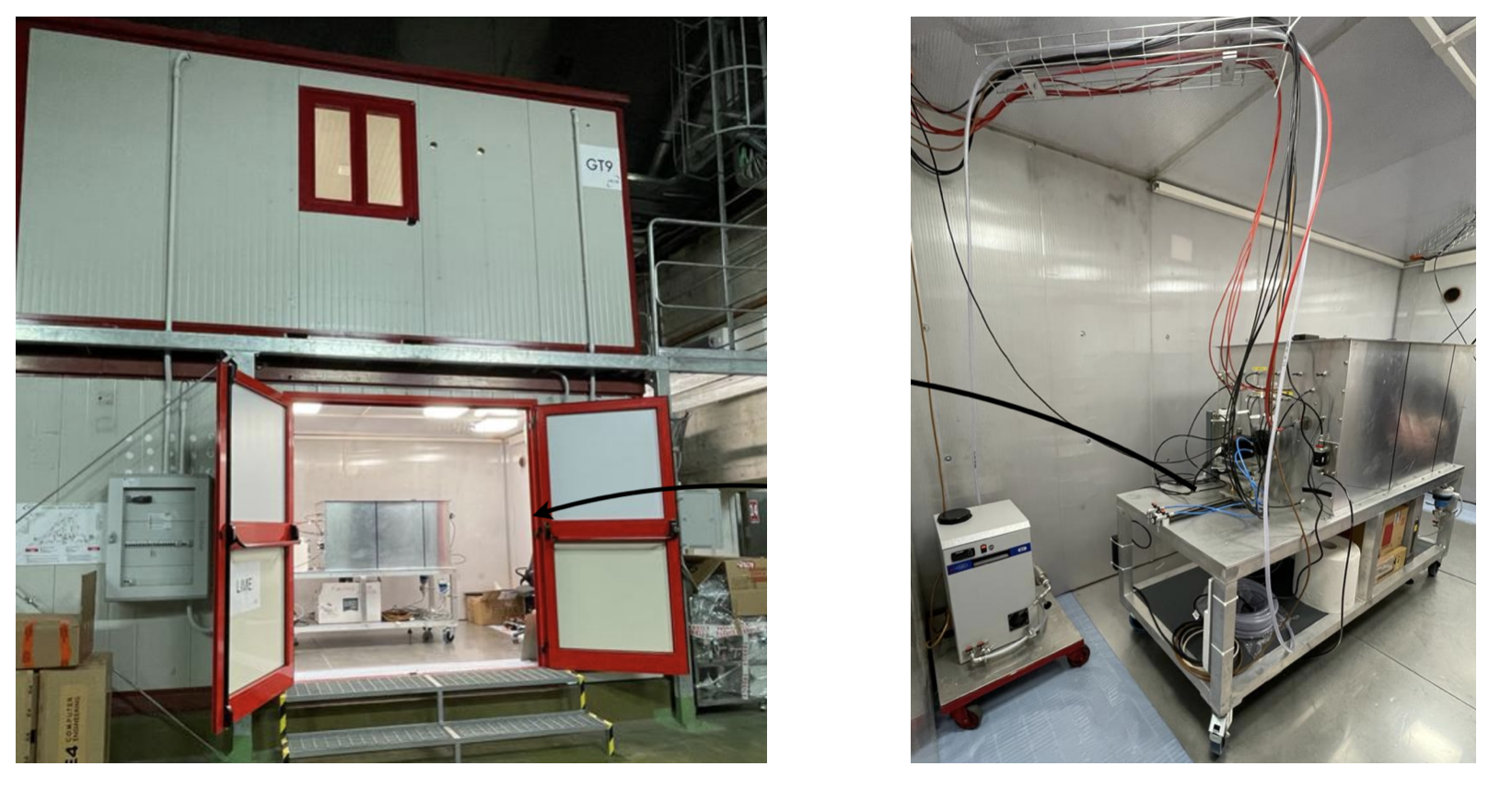}
    \caption{Left: Picture of the box where lime is installed underground at LNGS. Right: Picture of LIME inside the box at LNGS.}
    \label{fig:LimeUndergr}
\end{figure}

One of the primary objectives of the LIME prototype is to evaluate its performance in the low-background environment of the Laboratori Nazionali del Gran Sasso (LNGS), where it has been in operation since February 2022. The experimental setup is situated in a container positioned within the corridor between Hall B and Hall A, directly in front of the DAMA/LIBRA experiment. The container spans two levels: the lower floor houses the detector and any future shielding, while the upper floor accommodates the control room, which also houses the data acquisition (DAQ) servers, modules, and high-voltage supplies. In Fig. \ref{fig:LimeUndergr}, two images of the underground LIME site are shown. The gas system is located just outside the experimental area and is provided by an Air Liquide GAS system. This system manages the mixture's flow to the detector, purifies it from impurities, recirculates it, and facilitates the recovery for the disposal of greenhouse gases.
Understanding and characterizing the radioactive background is of paramount importance, particularly when searching for rare events. To address this, a comprehensive Monte Carlo simulation of the expected background in LIME was conducted using the GEANT4 software\cite{AGOSTINELLI2003250}. This simulation incorporated the CAD engineering design of the prototype, including various shielding configurations. The external radioactive background, originating from sources outside the detector, was simulated using neutron and gamma flux spectra measured at the Gran Sasso National Laboratory (LNGS) by other experiments \cite{DEBICKI2009429,Belli1989DeepUN}. The radioactivity of the detector components used to build LIME, which includes cathode, copper shielding, field rings, resistors, GEM, acrylic vessel, camera lens, and body, has been measured from the LNGS low radioactivity laboratory with a high purity Germanium detector, and the results have been used for the internal radioactivity simulation. \\
Notably, the field rings, resistors, GEMs, and the acrylic vessel are among the primary contributors to the radioactivity budget, with a smaller impact also observed from the camera lens and body. Addressing which are the most radioactive materials is crucial for future detector development.\\
The underground installation and operation of LIME has several purposes, including background studies, the validation of MC simulations, and neutron flux measurements. As each of these goals necessitates different configurations and background levels, the underground LIME program has been implemented in a phased manner, with additional shielding layers progressively added around the detector. The planned phases include:
\begin{outline}
    \1 \textbf{Run0} No shielding:
        the initial phase aims to commission LIME following its underground installation. The detector is installed inside an aluminum Faraday cage utilized during the overground tests at LNF. This marks the initial assessment of the detector's performance in an underground setting, characterized by the low-rate conditions characteristic of the conduction of a DM experiment, in contraposition with the overground operation where image occupancy is predominated by cosmic rays.
    \1 \textbf{Run1} No shielding:
        following the commissioning of the detector, the initial data collection phase focuses on analyzing and thoroughly characterizing the external backgrounds. In this regard, during Run 1, no additional shielding is introduced to LIME, which remains housed within the aluminum Faraday cage. In this condition, the external background is still the dominant background contribution. The predominant background arises from the natural gamma flux within the laboratory environment.
    \1 \textbf{Run2} 4 cm Cu shielding:
        the copper shielding effectively attenuates external gamma radiation by approximately a factor of 50. This leads to a notable decrease in the event rate within the detector and reduces the occupancy of tracks in the images. In this phase, the external background is still dominating the internal one.
    \1 \textbf{Run3} 10 cm Cu shielding:
        with this shielding configuration, external background sources are suppressed to a level comparable to that of internal radioactivity from the detector material, and it allows for its characterization. In this same configuration, an assessment of the environmental neutron flux is scheduled after Run 4, aiming to provide a more precise characterization of the nuclear recoil background induced by neutrons for upcoming experiments, including CYGNO. This measurement aims at renewing old measurements performed on this topic \cite{DEBICKI2009429,Belli1989DeepUN,Bruno_2019}. This measurement will be done in the context of the PRIN-funded project “Zero radioactivity for future experiments”. During this run, data have been also acquired with a $^{152}$Eu source at different distances to study the effect of saturation on MIP-like particles.
    \1 \textbf{Run4} 10 cm Cu + 40 cm water:
        water tanks are positioned around LIME, offering a 40 cm layer of shielding rich in Hydrogen to moderate environmental neutrons and reduce the background of nuclear recoils induced by external sources.
\end{outline}

\subsection{PAHSE\_1: CYGNO-04}
\label{sec:CYGNO04}
\begin{figure}
    \centering
    \includegraphics[scale = .4]{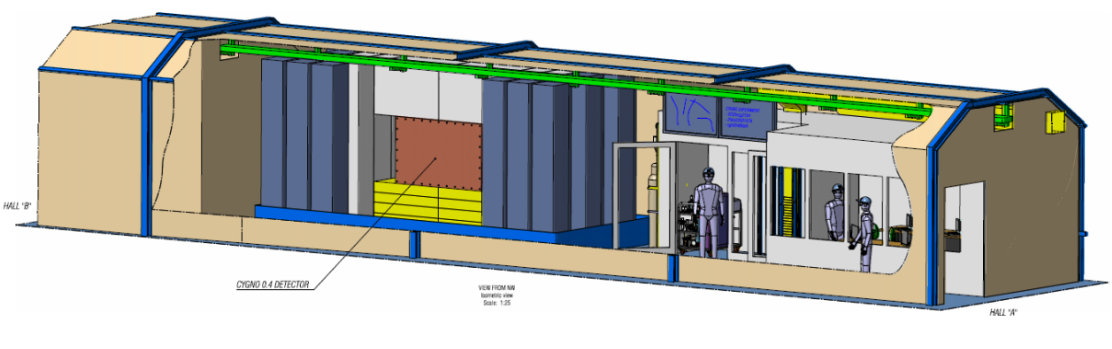}
    \caption{3D Scheme of the detector and the services adapted to the Hall F experimental area at LNGS}
    \label{fig:CYGNO04HallF}
\end{figure}

\begin{figure}
    \centering
    \includegraphics[scale=.4]{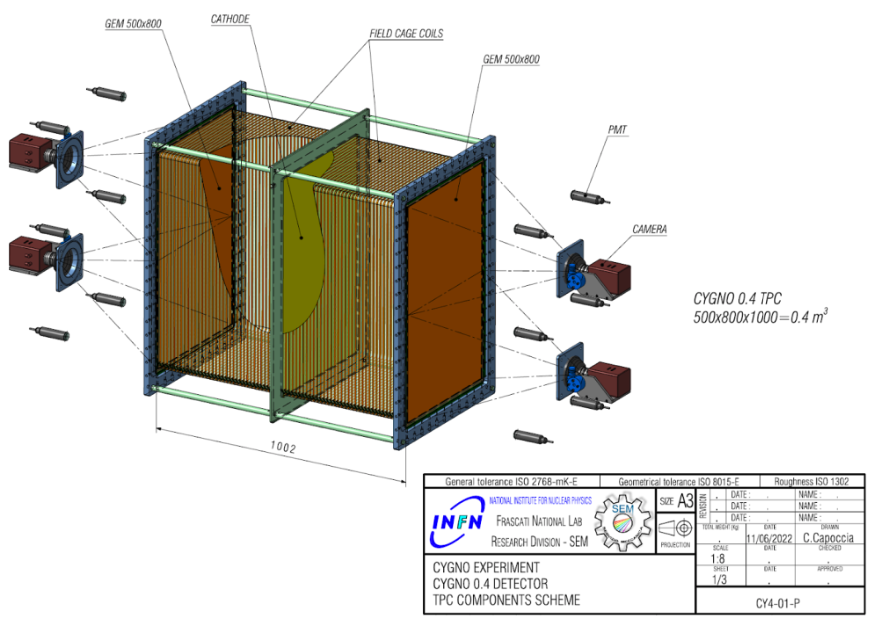}
    \caption{In figure the schematic representation of the CYGNO-04 detector illustrating all its components is shown.}
    \label{fig:cygno04scheme}
\end{figure}

Drawing on the knowledge and insights gained from the construction, underground deployment, and operational phase of the LIME prototype, the CYGNO collaboration is advancing towards the development of a 0.4 $m^3$ experimental demonstrator for PHASE\_1, namely, CYGNO-04. The core objectives of PHASE\_1 encompass realization of a detector employing materials with low radioactivity in the framework of a realistic experimental setup and scale, all while demonstrating the scalability and the strength of the detection technique towards a larger PHASE\_2 detector. The CYGNO-04 technical design report \cite{giovanni_mazzitelli_2023_76967} was submitted to the LNGS administration and the pertinent funding agencies in July 2022 and has received approval from LNGS management and the funding agency. The CYGNO-04 detector will be installed in Hall F, between Hall A and Hall B in 2024 at LNGS. A scheme of the experiment setup, comprising its services and control room, is shown in Fig. \ref{fig:CYGNO04HallF}.\\
The detector will consist of two back-to-back TPCs with a common cathode, each of 50 cm drift length. Leveraging the experience acquired from the DRIFT collaboration, the cathode will be made of a 0.9 $\mu m$ aluminized Mylar foil \cite{Daw:2011wq,Battat_2015}. The use of aluminum coating is advantageous as it does not introduce radioactive materials. Moreover, the film's thinness also maximizes the possibility for alpha particles coming from eventual radioactive contaminants to escape and deposit their energy in the fiducial volume, so that they can be tagged and excluded from the analysis of the events. In order to further mitigate internal background interference, the field cage will be fabricated from a single 35 $\mu$m Kapton foil, with thin 50 $\mu$m copper layers printed on it that operate as rings. The copper layer will have a thickness in z of 5 cm, and they will be spaced 5 cm apart. This method, previously used by the DRIFT collaboration, eliminates the need for ceramic resistors typically found in conventional field cages which are highly radioactive and allows for their substitution with smaller SMD resistors. Simultaneously, it minimizes the copper content and thus the presence of radioactive materials. The amplification stage will comprise a triple thin 50 um GEM stack of 50$\times$80 cm$^2$ area. To further mitigate the radioactive contamination of this component, a cleaning technique based on deionized water baths, developed by the TREX collaboration \cite{Castel_2019}, will be employed to remove radioactive contaminants from the GEM surface resulting from the GEM chemical fabrication process \cite{Castel_2019}.
The detector will be enclosed in a PMMA acrylic vessel for the purpose of reducing material radioactivity, preventing gas contamination, and ensuring electrical insulation from the internal electrodes.
The optical readout configuration includes six PMTs and two sCMOS cameras on each side to cover the entire amplification area. The selected PMT model is the Hamamatsu R7378A, which offers a 15\% quantum efficiency, particularly at long wavelengths, such as 550 nm \cite{Antochi_2018}. Given the usual low radioactivity of PMTs and their positioning at a significant distance, their contribution to the background levels can be considered negligible. As sCMOS cameras, the Hamamatsu ORCA QUEST will be used. This innovative sensor exhibits enhanced noise performance in comparison to the ORCA Fusion (Sec. \ref{sec:camera}), with approximately 0.27 electrons RMS. This allows each pixel to perform single-photon counting while maintaining an effective granularity of approximately 200$\times$200 $\mu$m per pixel in the CYGNO-04 configuration. Due to the significant $^{40}$K contamination observed in the camera lens, the CYGNO-04 project is currently collaborating with external companies to develop customized lenses made from ultra-pure silica (SUPRASIL). This initiative is expected to lead to the production of lenses with radioactive contamination from these components by approximately a factor of 10$^4$ less. 
A detailed scheme of the detector with all the components is shown in Fig. \ref{fig:cygno04scheme}.\\
A preliminary GEANT4 simulation was conducted to assess the expected radioactive background in CYGNO-04, considering both external and internal material radioactivity contributions. The purpose of this simulation was to perform a preliminary selection of the materials to use as shielding. The initial choice for the shield, which also considers the constraints of available space in Hall F, consists of a 10 cm layer of copper enclosed in 1.1 meters of water. This configuration is foreseen to reduce the contribution to the total background from external sources to approximately 1/20 of the internal one.

\subsection{PAHSE\_2: CYGNO-30}
\label{sec:CYGNOFut}
\begin{figure}
    \centering
    \includegraphics[scale=.35]{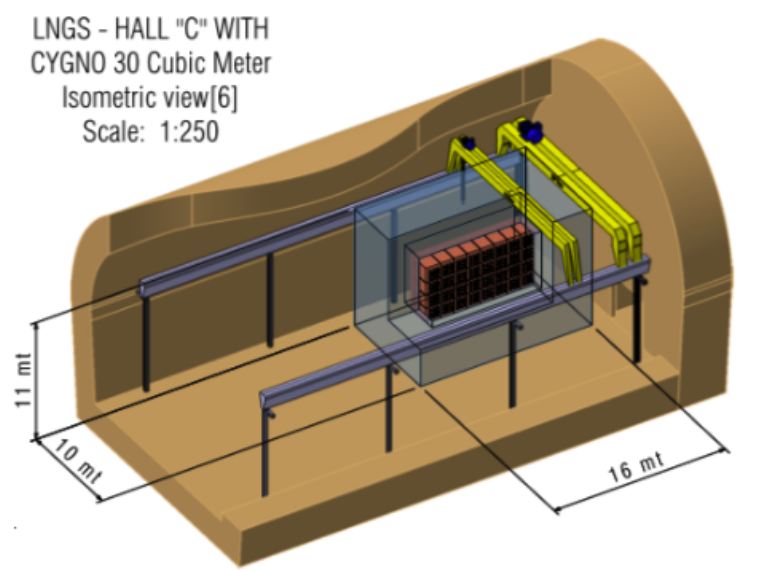}
    \caption{Sketch of a 30 $m^3$ CYGNO PAHSE\_2 detector in Hall C at LNGS.}
    \label{fig:CYGNO30}
\end{figure}

Given the modularity of the CYGNO approach, a CYGNO-30 detector could be, in a first concept, composed of 75 CYGNO-04 modules. These modules will be arranged in three vertically stacked rows, and each row will be composed of 25 modules. A conceptual design of the CYGNO-30 detector will be presented in Chap. \ref{chap:solarnu}, in the context of the background studies for the sensitivity to the directional solar neutrino measurement. This detector setup could be accommodated in Hall C of LNGS, with additional passive shielding implemented to mitigate external background interference.
The CYGNO experiment has been developed for directional Dark Matter searches (see Sec. \ref{subsec:wimp} for details). Dark matter is a hypothetical type of matter that seems to have no interaction with light or the electromagnetic field. Its existence is suggested by gravitational effects that cannot be fully explained by general relativity without the presence of additional matter beyond what is observable. Under the hypothesis of DM to consist of WIMP (Sec. \ref{sec:WIMPS}) it is expected to be distributed in halo around galaxies (Sec. \ref{sec:DMEvidences}). The solar system, to which the Earth belong, during its revolution motion around the center of the galaxy is expected to experience an apparent DM wind coming from the direction of the Cygnus constellation (Sec. \ref{sec:WIMPDirectional}). This WIMP coming all from the same apparent direction would produce in the detector nuclear recoil that by kinematic will feature an angular distribution peaked in the opposite side of the Cygnus constellation over a flat background. In this context, measuring the direction of nuclear recoil would lead to great advantages (Sec. \ref{sec:diradvantages}). First of all detecting an excess of event from the Cygnus constellation would result in a positive identification of Dark Matter since no background whatsoever can mimic this angular distribution feature. Secondly, this would allow for a high background rejection with the directionality power, including neutrino from the Sun that interacting through CE$\nu$NS can mimic a DM interaction giving rise to nuclear recoil which can still be separated from a DM signal since these recoils will backpoint towards the Sun.
In this context, a CYGNO-30 experiment, with its directional performances, has the potential to make a substantial contribution to the Dark Matter searches in the range of masses below 10 GeV, encompassing both Spin-Independent and Spin-Dependent coupling interactions (Sec. \ref{sec:WIMPInteraction}). Also, in case of other experiments would claim a DM observation, CYGNO's directional capabilities would play a decisive role in definitively confirming the Galactic origin of an eventual DM signal. 
Directionality doesn't exhaust its potential in DM research but can instead open doors to multiple cases of physics where the signal has a directionally different distribution from the background.
Apart from its contribution to Dark Matter research, such an experiment can also have relevance in the field of neutrino physics.
It is on this basis that the feasibility study of this thesis has been developed, to demonstrate that an experiment like CYGNO-30 could provide significant contributions not only to DM research but also to the study of solar neutrinos.
Indeed, as extensively illustrated in Chap.\ref{chap:sota}, CYGNO-30 could open the way for a directional measurement of solar neutrinos from the pp chain via elastic scattering on electrons, potentially expanding the energy range of the Borexino findings \cite{borexino2018comprehensive} down to $\mathcal{O}(10)$ keV on the electron recoil energy. Moreover, traditional Dark Matter experiments, which lack directional sensitivity in their event detection, are unable to differentiate between electron recoils induced by solar neutrinos and those arising from natural radioactivity \cite{instruments6010006}. In contrast, a directional detector such as CYGNO can exploit the directional information to discriminate electron recoils resulting from solar neutrino interactions. These interactions exhibit a highly peaked asymmetric angular distribution, over the flat angular distribution of electron recoils coming from background sources.\\
A sketch of a possible configuration of a 30 $m^3$ CYGNO PHASE\_2 detector assembled in Hall C of the underground laboratories at LNGS is shown in Fig. \ref{fig:CYGNO30}.

\subsubsection{R\&D lines for CYGNO PHASE\_2}
\label{sec:redphase2}
The CYGNO Collaboration is currently engaged in numerous R\&D projects aimed at optimizing the capabilities of the selected 3D optical readout TPC approach. Applying an additional electric field of approximately $\mathcal{O}$(10) kV/cm following the amplification stage can lead to an augmentation in the overall light yield \cite{Baracchini_2020EL}. This adjustment enables a reduction in the gain of the GEM while maintaining consistent light output, thereby mitigating the effects of charge gain saturation. Alternatively, maintaining high charge gain while boosting light output facilitates a reduction in the energy threshold. This enhancement is realized by introducing an additional electrode, such as a metallic mesh or ITO glass, post the GEM amplification stage to allow light transparency. By applying a high voltage between this final stage and the last GEM across a small gap, a high electric field of 10 kV/cm or higher is generated, further accelerating the amplified electrons. The resulting electron energies may induce CF$_4$ fragmentation into neutral CF$_3^*$, accompanied by the emission of additional light during its de-excitation. While additional ionization is also produced, albeit to a lesser extent compared to the extra light emission, studies have shown \cite{DhoTesi} that the energy resolution remains largely unaffected, while a doubling in light output is observed.\\ Moreover, INITIUM (Innovative Negative Ion Time Projection Chamber for Underground Dark Matter Searches) is an ERC Consolidator Grant initiative aimed at investigating the functionality of a TPC akin to CYGNO, employing negative ion drift operation \cite{MARTOFF20003551NID,Ohnuki_2001} instead of the conventional electron drift for ionization charge transport. A detailed description of the INITIUM project will be given in Sec. \ref{sec:INITIUM} and a full overview of Negative Ions Drift (NID) studies and results will be given in Chap. \ref{sec:NID}.\\
Still, on the R\&D side, the addition of a Hydrogen-rich gas has been evaluated in CYGNO. The inclusion of Hydrogen-rich gases in the CYGNO mixture can enhance sensitivity to lower WIMP masses due to the improved kinematic matching of nuclei and WIMPs in the $\mathcal{O}$(1) GeV mass range, as detailed in Section 2.1.2. Various hydrocarbons, such as Methane and Isobutane, underwent investigation within the CYGNO Collaboration using a specialized setup at the University of Coimbra (Portugal). A compact prototype named APPLE, featuring a 6 mm drift gap and a thin GEM for amplification, was employed to assess the light yield from a $^{55}$Fe X-Ray source. Secondary scintillation light was detected using a LAPD, while charge collection occurred at the base of the GEM, followed by amplification and analysis via a multichannel analyzer. Promising outcomes were achieved, particularly with the addition of minor amounts of isobutane (i-C4H10) to the standard He:CF$_4$ 60:40 CYGNO gas mixture, ranging from 1\% to 5\% \cite{MONTEIRO201218}. The number of avalanche electrons demonstrated an increase with higher percentages of isobutane while maintaining a consistent energy resolution of around 14\%. Conversely, the light output declined with rising isobutane concentrations; however, even with 5\% isobutane, only a threefold decrease in light yield was observed compared to the standard CYGNO mixture. Future plans include optimizing the CF$_4$ content to offset the decrease in light yield and exploring alternative hydrocarbons that may not compromise light output. These promising findings suggest that incorporating hydrocarbons holds the potential for enhancing CYGNO's capability to explore lower WIMP masses.

\section{The INITIUM Project}
\label{sec:INITIUM}
In synergy with CYGNO, INITIUM (Innovative Negative Ion Time projection chamber for Underground dark Matter searches) is an ERC Consolidator Grant designed to develop Negative Ion Drift operation at atmospheric pressure within the CYGNO optical approach. Described extensively in Chap. \ref{sec:NID}, NID is a modification of conventional TPC operation involving the introduction of a small amount of a highly electronegative dopant \cite{Martoff:2000wi,Ohnuki_2001} in the gas mixture. This dopant captures primary electrons within a few hundred micrometers, creating anions that act as charge carriers instead of electrons. These anions, which can be created of different species, drift towards the anode, where their additional electron is stripped, allowing this to initiate a standard electron avalanche. The anions, featuring a higher mass than the electron one, thermalize with the molecule of the gas during the drift process, leading to a much reduced diffusion. As it will be shown in Chap. \ref{sec:NID}, with a small addition of SF$_6$ in the CYGNO gas mixture, the diffusion improves from 140 $\mu$m/$\sqrt{\text{cm}}$ to 45 $\mu$m/$\sqrt{\text{cm}}$ at 600 V/cm. As the mobility of anions depends on mass, the difference in the time of arrival among anions of different species effectively serves as a means to measure the position of the event along the drift direction \cite{Snowden-Ifft:2013vua}.
With these two features, the readout planes of a Negative Ions TPC (NI-TPC) can image a larger volume compared to conventional TPC approaches while maintaining high tracking performances, resulting in lower backgrounds and costs per unit mass. 
The objective of INITIUM is to study and employ a scintillating gas mixture comprising Helium (He), Carbon Tetrafluoride (CF$_4$), and Sulfur Hexafluoride (SF$_6$) at atmospheric pressure, with a small amount of SF$_6$ content, for Negative Ion Drift with optical readout. The potential achievement of NID within the optical approach would highly enhance the detector tracking capabilities. Beyond the clear improvement in tracking on the x and y coordinates, attributed to the minimized diffusion, these enhancements would extend to the reconstruction of tracks along the drift direction. Given that the drift velocity of NID is approximately $10^3$ times smaller than that of electrons, sampling the track along the z-direction with the charge moving slower can improve the resolution in the z. Currently, cameras with a rate high enough to sample the light produced by the drifting charge exhibit low resolution and high noise, making them unsuitable for low-energy rare event searches. With a reduction of the drift velocity, a lower rate would be needed to perform this task. Considering the rapid advancement of sCMOS technology, there is the possibility of making progress in a short timeframe, potentially enabling the realization of a high-framerate camera that can sample the track in the z-direction through multiple image acquisition.

\subsection{The MANGO prototype}
\label{sec:MANGO}
\begin{figure}
    \centering
    \includegraphics[scale = .3]{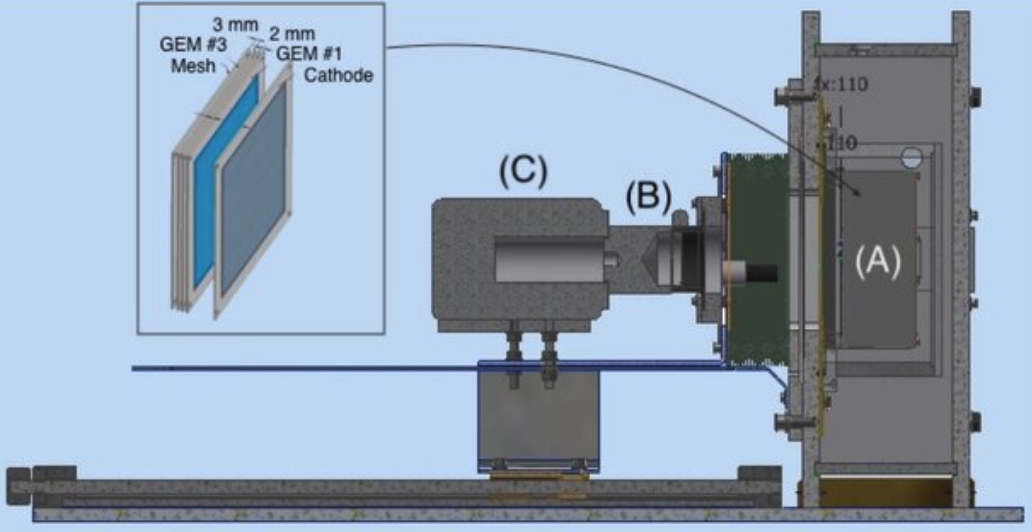}
    \caption{Scheme of the mango detector. The sensitive volume (A), the adaptable bellow (B), and the CMOS camera (C) are shown. In the square, a scheme of the TPC with the cathode and the three GEMs is shown. Plot from \cite{Baracchini_2020EL}.}
    \label{fig:MANScheme}
\end{figure}

The Multipurpose Apparatus for Negative Ion Studies with GEM Optical Readout (MANGO) is the detector employed in all the studies foreseen in the INITIUM (Sec. \ref{sec:INITIUM}) project and in other R\&D project for CYGNO. The detector is a compact TPC featuring a 10$\times$10 cm$^2$ amplification area and an adjustable drift region \cite{Baracchini_2020EL}. MANGO fits in the CYGNO context as an experimental platform for different purposes, including testing different amplification systems and gas mixtures, testing the electroluminescence for enhanced light production in a CF$_4$ based gas mixture, and investigating the feasibility of negative ions drift operation with optical readout.\\
The amplification stage is constructed using a series of GEMs, each measuring 10$\times$10 cm$^2$ and spaced 2 mm apart, with a field of 2.5 kV/cm applied between the GEMs, named transfer field. MANGO is typically equipped with triple 50 $\mu$m thin GEMs, although various thicknesses and stacking configurations have been subjected to testing (Sec. \ref{sec:gemgasmix}). The GEMs are labeled from 1 to 3, with GEM1 positioned closest to the drift region. These GEMs are supplied power by the HVGEM \cite{CORRADI200796}, a customized high-voltage source developed by the Laboratori Nazionali di Frascati (LNF). The HVGEM serves as an active high-voltage divider and is designed to provide power to seven separate and independent channels, the 3 GEMs, and 1 cathode when operated in a 1 cm drift configuration (lower voltages required). In longer drift configurations, the cathode is powered with a separate power supply. Additionally, for the electroluminescence studies positioned at a distance of $\Delta$z = 3 mm from the main setup, there is a metallic mesh adopted from an ATLAS MicroMegas. This mesh, comprising 30 $\mu$m diameter metallic wires at a 50 $\mu$m pitch, yields a transparency of approximately 0.55 and is strategically placed to induce an electric field beneath the electrode of the last GEM amplification stage called induction field. In the last studies, the metallic mesh was substituted with an ITO glass, which features a higher transparency of 0.9.\\
MANGO is commonly operated in continuous gas flow mode and is housed within a light-tight enclosure made of 3D-printed black plastic. This enclosure contains an airtight acrylic vessel that holds the gas. A slim, highly transparent (transparency $>$ 0.9) Mylar window isolates the gas detector from the optical readout system. The light is readout by a Hamamatsu Orca Fusion\textsuperscript{TM} sCMOS camera \ref{sec:camera} and a PMT Hamamatsu H3164-10 \cite{PMTMAN}. The optical readout components are positioned at a distance of (20.5 $\pm$ 0.3) cm and are focused on the closest GEM plane. The camera is coupled to the detector and isolated from the light through an adaptable bellow. In this configuration, the camera images an area of approximately 11.3$\times$11.3 cm$^2$, leading to an effective pixel size of around 49$\times$49 $\mu$m$^2$. A scheme of the MANGO detector is shown in Fig. $\ref{fig:MANScheme}$
The drift region of MANGO can be extended up to 5 cm in this configuration, through the use of a field cage. The 5 cm field cage consists of a cylindrical structure comprised of 1 mm silver wires enclosed within 0.5 cm thick polycarbonate field rings, with an internal diameter of 7.4 cm and spaced at 1 cm intervals. The silver wires are connected through 1 G$\Omega$ resistors to segment the high voltage (HV) provided by means of a CAEN N15701 power supply, ensuring a consistent drift field distribution.\\ For the negative ions drift studies, a longer field cage, featuring 15 cm in length and built exactly as the previous one, has been used. In this setup, the TPC structure is housed within a 150-liter stainless steel vacuum chamber, positioned so that the amplification plane is adjacent to a quartz window with 90\% transparency, mounted on the keg structure. The same readout in this configuration has been used.

\subsubsection{Test on different GEMs configurations and gas mixture}
\label{sec:gemgasmix}
\begin{figure}
    \centering
    \includegraphics[scale=.35]{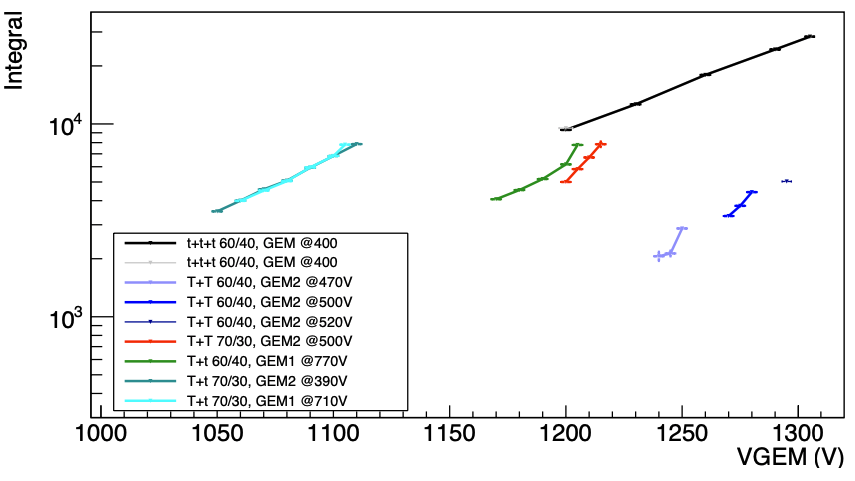}
    \caption{In figure, the light integral as a function of the total voltage applied to the GEM for different GEMs stacking configurations and gas mixture is shown. Plot from \cite{DhoTesi}.}
    \label{fig:LightVsV}
\end{figure}

\begin{figure}
    \centering
    \includegraphics[scale=.45]{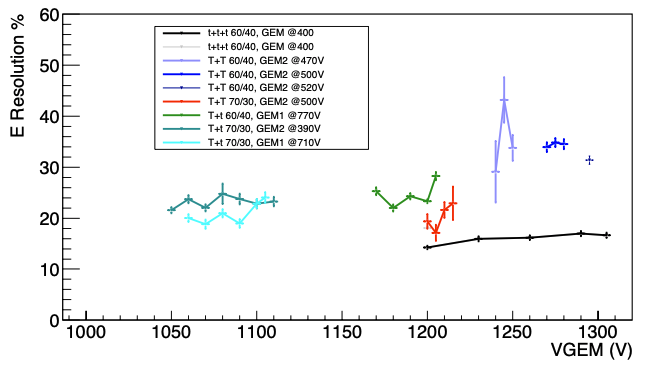}
    \caption{The energy resolution at 5.9 keV as a function of the total voltage applied to the GEMs for different configurations is shown. Plot from \cite{DhoTesi}.}
    \label{fig:EresoIron}
\end{figure}

\begin{figure}
    \centering
    \includegraphics[scale=.35]{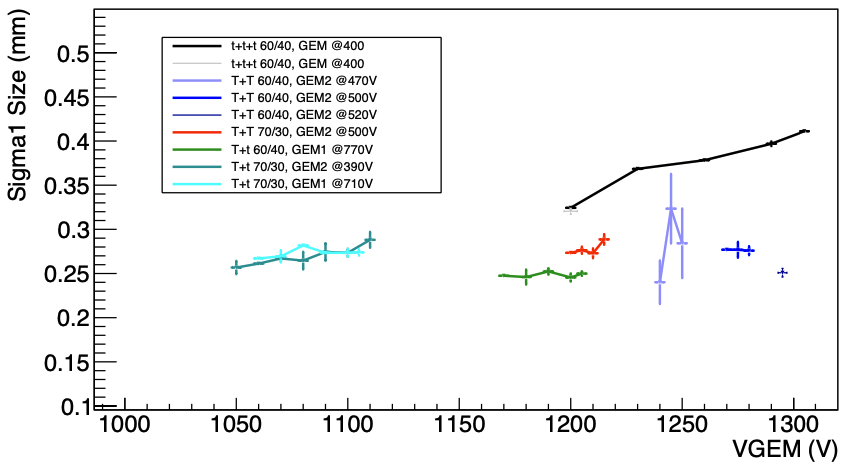}
    \caption{In figure, the average sigma of the iron spots as a function of the different GEM voltages for the different configurations is shown. Plot from \cite{DhoTesi}.}
    \label{fig:diffsigmaFe}
\end{figure}

A comprehensive study of the detector performance involving various He:CF$_4$ gas mixture ratios (e.g., 60/40, 70/30) and different GEM thicknesses and stacking configurations is conducted with the detector in the 1 cm drift length configuration. Two categories of GEMs have been utilized: one featuring thin 50 $\mu$m GEMs with 70 $\mu$m radius holes and 140 $\mu$m pitch (referred to as "t"), and the other consisting of thicker 125 $\mu$m GEMs with 175 $\mu$m radius holes and 350 $\mu$m pitch (referred to as "T"). All measurements are conducted under atmospheric pressure conditions at the Laboratori Nazionali del Gran Sasso (LNGS), situated at an elevation of approximately 1000 meters above sea level (a.s.l.), which corresponds to a pressure of (900 $\pm$ 7) mbar. In these studies, a $^{55}$Fe X-Ray source with an activity of approximately 480 kBq is utilized to produce 5.9 keV electrons trough photoelectric effect within the active gas volume of the MANGO detector. The relatively low source activity enables the reconstruction of each individual $^{55}$Fe track in the sCMOS images. 
The light integral distribution of the reconstructed events is fitted with a Gaussian function, and the fitted mean value is considered as the light integral. The light integral as a function of the total voltage applied to the GEM is shown in Fig. \ref{fig:LightVsV}.
Each set of GEM stacking configurations (i.e., ttt, Tt, and TT) exhibits a consistent gain slope. In contrast, the Helium to CF$_4$ ratio affects the voltage required on the GEMs to achieve a similar gain, with a higher Helium content allowing for lower voltages. The highest gain and light output are attained with the three thin GEM configurations (ttt), with integral values averaging about three times higher than Tt and up to approximately 10 times higher than TT. From the $\sigma$ of the Gaussian used to fit the $^{55}Fe$ peak, the energy resolution at 5.9 keV has been studied in the different configurations. The energy resolution, defined as the ratio $\sigma/\mu$, as a function of the total GEM voltage applied, is shown in Fig. \ref{fig:EresoIron}.
An analysis of the $^{55}$Fe spots in the gas also allows for a measurement of the contributions from the amplification stage to overall track diffusion, given the very small drift length of 1 cm. This has been measured as a function of the total GEM voltages and considering all the possible combinations of GEMs and gas mixtures. The diffusion has been measured by fitting with a Gaussian the projection on x and y of the light contained in the iron spot. The final value of the sigma has been obtained by averaging these quantities. The results are shown in Fig. \ref{fig:diffsigmaFe}.
From the plot, it can be observed that as expected, the configurations with two amplification stages perform better in diffusion with respect to the one with 3 GEMs. This confirms the assumption that each stage contributes independently to the total diffusion. Another enforcement of this argument is that the diffusion with two GEMs is around 2/3 of the one with 3 gems.
Moreover, the Tt configuration performs better than the TT due to the closer hole pitch of t GEMs. Indeed, the t GEM has a hole pitch of 140 $\mu$m while the T one has a pitch of 350 $\mu$m. The camera, with a granularity of 49$\times$49 $\mu$m$^2$ seen by each pixel, is indeed sensible to this feature.

\section{CYGNO for directional Dark Matter searches}
\label{sec:PhysCase}
This thesis extensively explores the use of CYGNO for the directional detection of solar neutrinos. Nonetheless, originally the CYGNO/INTIUM project has been developed and designed for directional Dark Matter searches, in particular in the form of Weakly Interactive Massive Particles (WIMP). This Section, for completeness, will illustrate the experimental observations that lead to the DM paradigm (Sec. \ref{sec:DMEvidences}), the expected characteristics of WIMPs particles  that could explain this hypothesis (Sec. \ref{sec:WIMPS}) and their possible interaction with standard matter (Sec. \ref{sec:WIMPInteraction}). It will further discuss the kinematic of the interaction (Sec. \ref{sec:WIMPKinematic}) and directional signature of a DM signal (Sec. \ref{sec:WIMPDirectional}), focusing on the advantages of the use of directionality (Sec. \ref{sec:diradvantagesDM}) and the expected CYGNO-30 sensitivity to this science case (Sec \ref{sec:DMSensitivity}).

\subsection{Dark matter}
\label{subsec:wimp}
In recent decades, a well-established paradigm in modern physics has emerged, driven by many astrophysical and cosmological observations indicating a discrepancy in the current understanding of the Universe. These measurements suggest the existence of an elusive component beyond the known mediators and particles, acting like a mass and commonly referred to as Dark Matter. Despite years of experimental endeavors providing insights into its phenomenology, the true nature of Dark Matter remains unsolved. One of the leading hypotheses is that Dark Matter is made by particles and contributes to approximately 84\% of the total mass of the universe \cite{Planck:2018vyg}, playing a fundamental role in shaping the organized structures of the Universe \cite{Frenk_2012}. 

\subsubsection{Dark Matter evidences}
\label{sec:DMEvidences}
\begin{figure}
    \centering
    \includegraphics[width=1\linewidth]{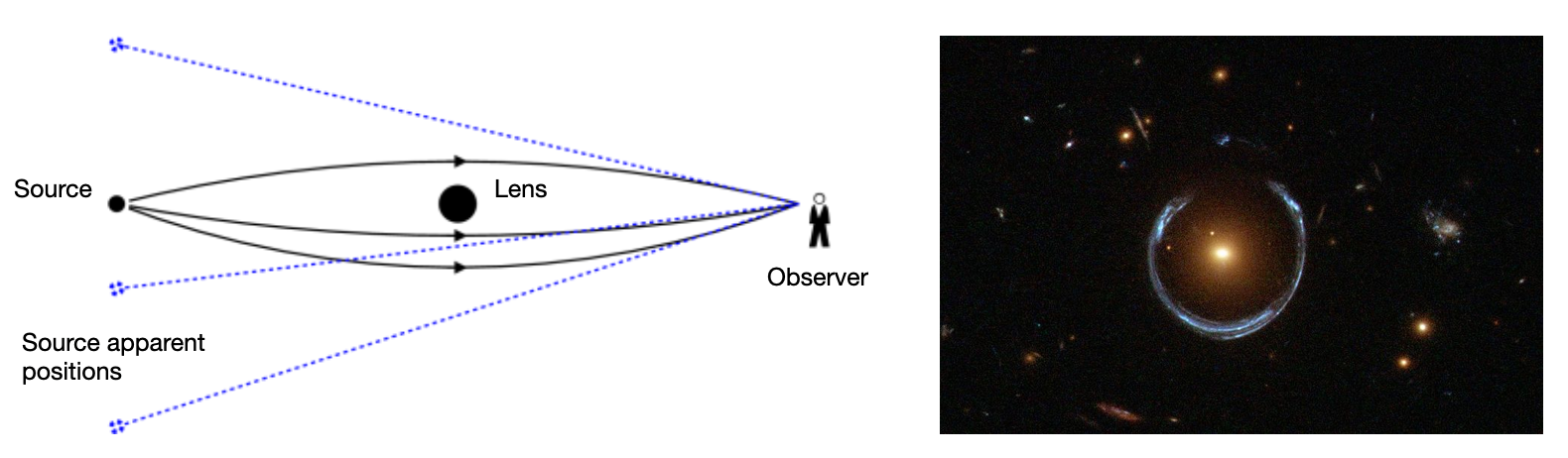}
    \caption{Left: scheme of the gravitational lensing effect. Right: Einstein ring produced by the distortion of the light emitted by a galaxy due to the presence of a high mass object. This causes the formation of multiple images of the galaxy along a ring around the object. Plots from \cite{Imm1} and \cite{Imm2} respectively. }
    \label{fig:GravLensing}
\end{figure}

\begin{figure}
    \centering
    \includegraphics[width=0.75\linewidth]{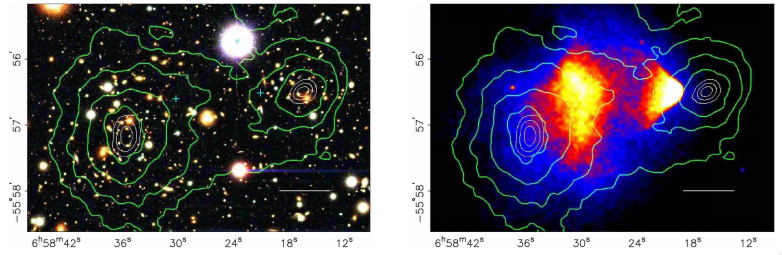}
    \caption{Optical image captured by Magellan (left) and X-Ray image captured by Chandra (right) of the galaxy cluster 1E 0657-56, also known as the "Bullet cluster. The contours in green illustrate the mass distribution as determined through weak lensing measurements. Plots from \cite{Bertoni2015DarkMP}.}
    \label{fig:BulletCluster}
\end{figure}

Indications of a predominant unknown component in the Universe denoted as Dark Matter, have been observed across various astronomical phenomena, in scales ranging from individual galaxies to the greatest cosmic structures. \\
The most convincing and well-established evidence for Dark Matter comes from studying the rotation curves of spiral galaxies, including our own Milky Way \cite{10.1093/mnras/249.3.523,Strigari_2013}. In these galaxies, a significant portion of stars resides in the central region known as the bulge. Surrounding the bulge, flat spiral arms, comprising stars and gas clouds, orbit in approximately circular paths. The rotation curve represents the profile of circular velocities for the stars and gas clouds across the radial distance from the galaxy's center. In this context, in which a rotating galaxy is viewed as a closed system, an object with mass $m$ at a distance $R$ from the center experiences a centripetal force, which is equivalent to the gravitational force exerted by the total mass within the same radius, denoted as $M(R)$:
\begin{equation}
    m\frac{v^2}{R} = \frac{GM(R)m}{R^2}
\end{equation}
where $v$ is the tangential velocity of the object having a mass $m$ and G is the universal gravitational constant. Thus, the rotational velocity of a galaxy's components can be formulated in terms of both $R$ and the total mass $M$ as:
\begin{equation}
    v = \sqrt{\frac{GM(R)}{R}}
\end{equation}
It is reasonable to make the assumption that the most significant portion of a galaxy's mass is situated within a characteristic radius $R_c$, as estimated by measurements ranging from 5 to 10 $kpc$ \cite{Strigari_2013}. Thus, for $R<R_C$ assuming uniform mass density, $M(R)$ increases linearly with volume, causing the velocity to rise linearly with $R$. Conversely, for $R>R_c$, where the mass
M is constant, the velocity decreases with $\sqrt{R}$. The rotation curve can be determined by observing the Doppler shift of the 21 cm emission line of H I. A comprehensive set of rotation curves, extending well beyond the optical limit and reaching 200 kpc, has been measured, revealing consistent findings. 
\begin{figure}
    \centering
    \includegraphics[width=0.5\linewidth]{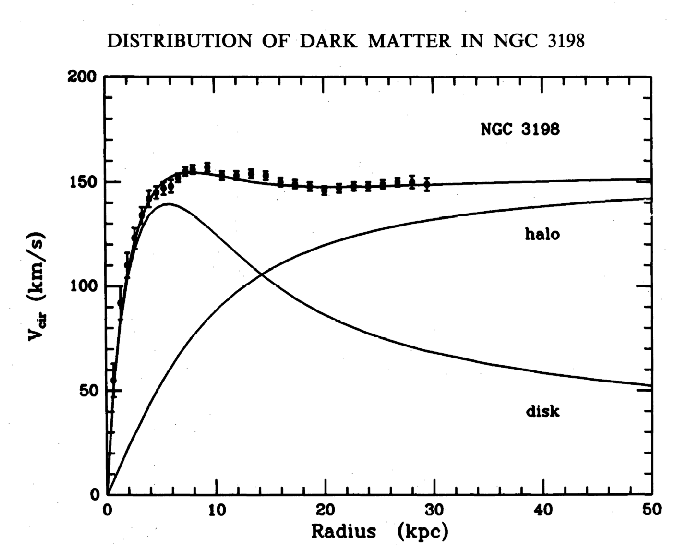}
    \caption{The figure shows the rotation curve measurement of the galaxy NGC6503, sourced from \cite{10.1093/mnras/249.3.523}. The solid line corresponds to the overall fit of the data, while the dashed line represents the gas component, the dotted line corresponds to the visible component, and the dash-dotted line illustrates the Dark Matter component.}
    \label{fig:galortcurve}
\end{figure}
The velocity rotation curve for the galaxy NGC6503 is shown in Fig. \ref{fig:galortcurve}. The observed orbital velocity initially follows a linear increase at small radii but unexpectedly remains constant at larger radii, contradicting predictions. This flatness hints at the presence of an unseen mass in the form of a dark galactic halo surrounding spiral galaxies, exhibiting no electromagnetic interaction. This matter halo seems to conform, at a basic level, to a spherical distribution with a density proportional to $R^{-2}$.\\
Further confirmation of the existence of Dark Matter is derived from observations of gravitational lenses \cite{Bartelmann:2010fz}. This phenomenon, predicted by General Relativity, arises due to spacetime distortions caused by the distribution of energy and mass within it. When a significant mass is concentrated in a small region, the spacetime around it distorts, causing the light rays to be deflected in the direction of the mass. Hence, objects positioned beyond galaxies or galaxy clusters in our line of sight may appear distorted or elongated when viewed from Earth. Analyzing these distorted images enables the estimation of the mass of the object situated in between. A scheme illustrating the effect is presented in the left plot of Fig. \ref{fig:GravLensing}, while the right plot displays an image of light deflection resulting in multiple images of the original object at the source when a massive object is positioned between an observer and the photon source. 
In this scenario, the distribution of mass between the observer and the source of photons is called lens. The angle of deflection with respect to the straight path of the light $\theta$, experienced by a light ray passing at a distance R from a lens with mass M can be expressed in the Schwarzschild approximation \cite{wald2010general} as: 
\begin{equation}
    \theta = \sqrt{\frac{4GM}{Rc^2}}
\end{equation}
where G is the universal gravitational constant and c is the speed of light. From this law, the measurement of the deflection angle allows for inferring the object mass M. 
The effects of gravitational lensing vary depending on the mass and distribution of the lens. Strong lensing occurs when the lens has a large mass and minimal change in position relative to Earth's velocity, allowing telescopes to resolve the effects in single image \cite{Clowe_2006}. Weak lensing, on the other hand, is caused by smaller masses and results in a distortion of galaxy images due to the bending of light \cite{Huterer_2010}. Microlensing occurs when the mass of the lens is so small that it doesn't visibly deform images but affects the intensity of light from luminous sources\cite{Wegg_2016}.\\
Various experimental and analytical techniques enable the assessment of mass distributions between the observer and luminous sources based on the light emitted by the object. A compelling evidence for the existence of missing mass in galaxy clusters can be derived from the study of merging clusters. By comparing the X-Ray emission from hot gases with the mass distribution inferred from gravitational lensing, it becomes evident that while stars, due to their small size relative to the cluster's dimensions, are not expected to collide, the intergalactic hot gas predominantly interacts electromagnetically. As a result of the interaction, the gas decelerates in the direction of the galaxies motion, leading to an increase in temperature and a strengthening in X-Ray emission. In Fig. \ref{fig:BulletCluster}, the example of the Bullet cluster is presented \cite{Bertoni2015DarkMP}. 
The red halo indicates the regions of strong X-Ray emission due to the gas interaction, while the green contour lines show the gravitational interacting content estimated through lensing. The noticeable displacement between these two measurements strongly reinforces the hypothesis that additional mass, beyond what is visibly apparent, must be present. Furthermore, it suggests that Dark Matter is non-collisional, remaining unaffected by the interaction of the two clusters. Other proofs of DM existence are the Cosmic Microwave Background anisotropies \cite{Hu_2002} due to gravitational potential, the formation of large scale structures \cite{Kunz_2016} and the measurements of cosmological-related quantities \cite{OZER1987776}.

\subsubsection{Weakly Interacting Massive Particles}
\label{sec:WIMPS}
The Standard Model (SM) of particle physics \cite{Gaillard:1998ui} is the most accurate model to describe particles and their interactions. The last triumph in the Standard Model is the identification of the Higgs Boson at LHC in 2012 \cite{201230}, reaffirming the model's validity and coherence. Nevertheless, the SM is acknowledged to have various limitations, including issues such as Higgs mass corrections, neutrino masses, strong CP violation, matter-antimatter asymmetry, and the absence of gravitational interaction \cite{ellis2002limits,Lee_2021}. Thus, an extension of the SM is needed to account for these limitations.
In the hypothesis that Dark Matter is composed of particles, these must have the following characteristics: it must be a non-baryonic particle, and it must have an abundance in the universe $\sim$6 times the one of the baryonic matter. The particle must be electrically neutral, and non-interacting under strong interaction. Furthermore, it must interact through gravitational interaction, or at most, through weak interaction. Lastly, the Dark Matter candidate must exhibit stability or have an exceptionally long lifetime, greater than the age of the universe \cite{doi:10.1073/pnas.1516944112}. Due to these attributes, the modeling of a Dark Matter particle inevitably requires an extension of the Standard Model, since none of the particles within the Standard Model fits these constraints. \\
Weakly Interacting Massive Particles stand among the most testable and well-supported candidates. WIMPs are not predicted in the Standard Model of particle physics, but they can be consistent candidates in beyond the Standard Model scenarios, such as Supersymmetric theories. Moreover, they are compatible candidates in the standard cosmological theory as relic matter from the early universe, since they reproduce the correct abundance predicted by the model under the assumption of an interaction cross-section on the scale of weak interactions \cite{Scherrer:1985zt,Bernal_2017}. The fundamental premise of WIMPs is that they are neutral particles with interactions with ordinary matter occurring at or below the scale of the weak force, having a mass of the order of the weak scale 1 GeV - 10 TeV \cite{Roszkowski_2018}. Given this range of mass, WIMPs are searched through their interactions with nuclei.

\subsubsection{WIMPS interaction with ordinary matter}
\label{sec:WIMPInteraction}
From observations of the rotation curve of our galaxy, it has been observed that it is also dominated by Dark Matter. Therefore, as the Earth moves with the solar system within this halo of DM, an interaction of DM with ordinary matter is expected, hoping for some other interaction beyond the gravitational one. Thus, if this is possible, the interaction rate of DM scattering with nuclei is determined by the product of the number of nucleons in the target, the DM flux, and the interaction cross-section, all integrated over the kinematics \cite{PhysRevD.66.103513}:
\begin{equation}
    \frac{d R}{d q^2 d \Omega}=\frac{N_0}{A_{\text {mol }}} \frac{\rho_0}{m_\chi} \int_{v_{\min }(q)} \frac{d \sigma}{d q^2 d \Omega} v f(\vec{v}) d^3 v
\end{equation}
where $N_0$ represents Avogadro's number, $\rho_0$ the DM density at Earth, $A_{\text{mol}}$ the molar mass of the target, $\sigma$ the cross-section between the WIMP and the nucleus, $m_{\chi}$ the mass of the WIMP, and $f(\vec{v})$ the distribution of velocities of the WIMPs in the Galactic Rest Frame.
Hence, this rate is composed of key components, namely: the velocity distribution, the cross-section, and the kinematics.
In the assumption of DM to be constituted by WIMPs, the Milky Way is situated in a halo consisting of these cold, non-relativistic Dark Matter particles. The DM halo of the Milky Way is commonly described by the Standard Halo Model (SHM) \cite{2008gady.book}. The SHM consider an isotropic, isothermal sphere of DM particles with a density profile given by $\rho(r) \propto 1/r^2$. Assuming collisionless particles and an isothermal system, the velocity density distribution can be described using a Maxwell-Boltzmann distribution: 
\begin{equation}
    f(\vec{v})= \begin{cases}\alpha e^{-\frac{v^2}{v_p^2}} & \text { if }|\vec{v}|<v_{e s c} \\ 0 & \text { if }|\vec{v}|>v_{e s c}\end{cases}
\end{equation}
where $v_p$ is the most probable value of the velocity distribution of WIMPs near the Earth, $v_{esc}$ is the local escape velocity and $\alpha$ is a normalization factor. Under the assumption of an ideally flat rotation curve for the Galaxy, the value of $v_p$ can be determined at different distances, and it is observed to be approximately 230 km/s at the Earth's radius \cite{Baxter_2021,Eilers_2019}. The velocity $v_{esc}$ is determined using data from the Gaia telescope and RAVE survey, where it is found to be 544 km/s \cite{Piffl_2014,Monari_2018}. In this halo, the density of Dark Matter at the Solar distance from the center of the Galaxy, denoted as $\rho_0$, is estimated to be around 0.3 GeV $cm^{-3}$ \cite{Baxter_2021}.\\
The nature and specifics of the interaction between WIMPs and Standard Model matter remain unknown. As a practical approximation, Fermi's Golden Rule can be employed. Consequently, the factor $d\sigma$ can be approximated by separating the energy dependence of the differential cross-section into an independent term $\sigma_{WA}$ and another term, $S(q)$, which encompasses the entire dependence on the transfer momentum q and the form factor. The differential cross-section can thus be written as:
\begin{equation}
    \frac{d\sigma}{dq^2}= \frac{1}{\pi v^2} |\mathcal{M}|^2 = \frac{\sigma_{WA}S(q)}{4\mu_A^2v^2}
\end{equation}
where v is the WIMP velocity (flux factor), $\mathcal{M}$ is the matrix therm of the interaction and $\mu_A$ is the reduced mass WIMP-Nucleus:
\begin{equation}
    \mu_A=\frac{m_Xm_A}{m_X+m_A}
\end{equation}
The nucleons that constitute ordinary matter are non-relativistic. Simultaneously, in the assumption of the SHM, WIMPs are required to be non-relativistic with respect to the Earth reference frame, with a local velocity peak of $\sim$230 km/s. The De Broglie length associated with 10 GeV particle, moving at that velocity can be estimated as the ratio between the Plank constant and the particle momentum $\lambda_{DM}=h/p \sim 160\  fm$. Conversely, the typical nuclear dimension is on the order of a few femtometers, at least one order of magnitude smaller than $\lambda_{DM}$. Consequently, it is reasonable to assume that the WIMP perceives the nucleus as a whole. Thus, it is expected DM to interact with nuclei coherently, and due to the low energy transfer, the scattering is considered elastic.
The actual nature of DM coupling is not known, but it is possible to make a simplified standard assumption that it can possess Spin-Independent and Spin-Dependent interactions. These components account for the coupling to the number of nucleons and spin of the nucleus, respectively \cite{Kurylov_2004}.
A generic expression for the cross-section can be formulated as follows:
\begin{equation}
    \sigma_{WA} = \sigma_{WA,SI} + \sigma_{WA,SD}
\end{equation}
The SI term can be parameterized as a function of the proton and neutron content of the target nucleus:
\begin{equation}
    \sigma_{WA,SI} = \frac{4\mu_A^2}{\pi}(Zf_p +(A-Z)f_n)^2
\end{equation}
In the equation Z and A represent respectively the atomic number and the mass number of the target, and $f_p$ and $f_n$ represent the effective Spin-Independent coupling of Dark Matter to neutrons and protons. A reasonable assumption, supported by various models, is that WIMPs have the same coupling to proton and neutrons, resulting in the two terms $f_n$ and $f_p$ to be equal. Under this assumption, the previous equation reduces to:
\begin{equation}
    \sigma_{WA,SI}=\frac{4\mu_A^2}{\pi} A^2f_n^2 = \sigma_{n,SI}\frac{\mu_A^2}{\mu_n^2}A^2
\end{equation}
Where $\mu_n$ is the reduced mass of DM with the single nucleon constituting the target. Thus, the cross-section with a generic nucleus can be represented as a function of the cross-section with the single nucleon of the nucleus, denoted as $\sigma_{n,SI}$. Since the cross-section scales $\propto A^2$, consistently with the expectation of a coherent scattering where the contribution from all the nucleons are summed up together, it became evident that the interaction with heavy nuclei is favored. \\
The cross-section for the SD interaction can be written as:
\begin{equation} 
    \sigma_{WA,SD} = \frac{32G_F^2\mu_A^2}{\pi} \frac{J+1}{J}(a_p\langle S_p\rangle+a_n\langle S_n\rangle)
\end{equation}
where J is the nuclear spin, $\langle S_p\rangle$ and $\langle S_n\rangle$ are the expectation value of the proton and the neutron spin in the nucleus, and $a_p$ and $a_n$ are the effective SD coupling of WIMPs to proton and neutron respectively. The cross-section is highly dependent by the spin state of the nucleus. A disparity in the number of protons or neutrons inside the nucleus is necessary to achieve a significant expectation value of spin and, consequently, sensitivity to Spin-Dependent interactions, given that the contributions of various nucleons typically balance each other \cite{Tovey_2000}.
\begin{table}
    \centering
\begin{tabular}{|c|c|c|c|c|c|c|c|}
\hline Nucleus & $Z$ & Odd Nuc. & $J$ & $\left\langle S_p\right\rangle$ & $\left\langle S_n\right\rangle$ & $\frac{4\left\langle S_p\right\rangle(J+1)}{3 J}$ & $\frac{4\left\langle S_n\right\rangle(J+1)}{3 J}$ \\
\hline${ }^1 \mathrm{H}$ & 1 & $\mathrm{p}$ & $1 / 2$ & 0.500 & 0.0 & 1.0 & 0 \\
${ }^{19} \mathrm{~F}$ & 9 & $\mathrm{p}$ & $1 / 2$ & 0.447 & -0.004 & $9.1 \times 10^{-1}$ & $6.4 \times 10^{-5}$ \\
${ }^{73} \mathrm{Ge}$ & 32 & $\mathrm{n}$ & $9 / 2$ & 0.030 & 0.378 & $1.5 \times 10^{-3}$ & $2.3 \times 10^{-1}$ \\
${ }^{129} \mathrm{Xe}$ & 54 & $\mathrm{n}$ & $1 / 2$ & 0.028 & 0.359 & $3.1 \times 10^{-3}$ & $5.2 \times 10^{-1}$ \\
\hline
\end{tabular}
\caption{Table presenting a summary of nuclear spin characteristics for a selection of pertinent nuclei with respect to their sensitivity to Spin-Dependent interactions. From \cite{Tovey_2000}.}
\label{tab:SDcoreff}
\end{table}
The values of the coefficients which have a role in SD coupling for different elements, typically used as targets in DM detectors, are reported in table \ref{tab:SDcoreff}. As described in \cite{Schnee:2011ooa} the proton and neutron Spin-Dependent cross-section can be rewritten as:
\begin{equation}
    \sigma_{p,SD}=\frac{24 G_F^2\mu_p^2a_p^2}{\pi} \ \ \ \ \ \ \sigma_{n,SD}=\frac{24 G_F^2\mu_n^2a_n^2}{\pi}
\end{equation}
with $\mu_n$ and $\mu_p$ being the reduced mass between DM, and neutron and proton, respectively.
Combining the previous equations it is possible to obtain the full cross-section of DM with a nucleus with an odd number of protons:
\begin{equation}
    \sigma_{W A}=\sigma_{n, S I} \frac{\mu_A^2}{\mu_p^2} A^2+\sigma_{p, S D} \frac{\mu_A^2}{\mu_p^2} \frac{4\left\langle S_p\right\rangle(J+1)}{3 J}
\end{equation}
It can be observed that the SD term, $[4\left\langle S_p\right\rangle(J+1)]/3 J$, commonly smaller than 1, is considerably less than the SI term. Consequently, if $\sigma_{n,SD}$ is of the same order as $\sigma_{p,SD}$, the SI interaction tends to dominate for most targets, making the SD component negligible. As a result, experiments establishing constraints on the WIMP-to-nucleon cross-section distinguish between the SI and SD scenarios. Considering the mass range of WIMPs (around 1-100 GeV) and their velocity, the interaction of a Dark Matter particle with an atomic nucleus would produce a nuclear recoil of approximately 1-100 keV.

\subsubsection{Kinematic of WIMP interaction}
\label{sec:WIMPKinematic}
The kinematic of elastic scattering between two particles can be analyzed broadly as a relativistic collision. According to the Standard Halo Model (SHM), WIMPs are anticipated to possess an average velocity on the order of $10^{-3}c$, where $c$ is the speed of light. The thermal energy of a nucleus of approximately $\mathcal{O}$(10) GeV mass at room temperature is approximately of $10^{-5}c$. Therefore, it is reasonable to envisage a Dark Matter particle $\chi$ with momentum $p_0$ and velocity $v$ colliding with a stationary nucleus A. The nucleus subsequently recoils with momentum $q$ at an angle $\theta$ with respect to the direction of the incoming DM particle, due to the elastic collision. According to the conservation laws the following equation is valid:
\begin{equation}
    P_0^{\mu} + Q_0^{\mu} = P'^{\mu} + Q^{\mu}
\end{equation}
being these quantities in order the 4-momenta of the incoming Dark Matter particle, the nucleus at rest, the outgoing Dark Matter particle, and the scattered nucleus. The 3-momentum
$q$ of the recoiling nucleus can be expressed as:
\begin{equation}
    q=\frac{2 m_A\left(\sqrt{p_0^2+m_\chi^2}+m_A\right) p_0 \cos \theta}{\left(\sqrt{p_0^2+m_\chi^2}+m_A\right)^2-p_0^2 \cos ^2 \theta}
    \label{eq:kinem}
\end{equation}
Since the particles involved in WIMPs to nuclei scattering are non-relativistic, the following approximation holds: $m_{\chi},m_A>>p_0$, where $m_{\chi}$ and $m_A$ are respectively the masses of the Dark Matter particle and the recoil nucleus. Under this assumption, Eq. \ref{eq:kinem} can be simplified to:
\begin{equation}
    q=2v\mu_A\cos(\theta)
\end{equation}
with $\mu_A$ being the reduced mass of the system $\chi-A$.
As the possible WIMP mass varies significantly across orders of magnitude, the efficiency of momentum transfer to the nucleus relies on the masses of both particles. The maximum energy achievable by the recoil is:
\begin{equation}
    E_{\text {max }}=\frac{q_{\max }^2}{2 m_A}=\frac{2 \mu_A^2 v^2}{m_A} \equiv r \frac{1}{2} m_\chi v^2
\end{equation}
where $r$ is a factor representing the efficiency in the transferred momentum, defined as $r=4 \frac{m_\chi m_A}{\left(m_\chi+m_A\right)^2}$. At $m_A = m_{\chi}$, the efficiency is at its maximum, $r = 1$. This illustrates the highest sensitivity of a particular target to WIMPs with masses equivalent to the atoms comprising the target. For instance, to enhance sensitivity to WIMP masses near $1\  \text{GeV}$, light elements such as Helium or Hydrogen are the most suitable options. Another important relation is the minimum velocity required for a WIMP to produce a recoil energy $E$. Indeed, for a given $E$ and scattering angle $\theta$, different velocities $v$ of a DM particle can induce the same $E$. The minimum velocity needed to produce a recoil energy $E$ is: $v_{min} = \sqrt{\frac{m_A E}{2 \cdot \mu_A}}$.

\subsubsection{Directional signature}
\label{sec:WIMPDirectional}
\begin{figure}
    \centering
    \includegraphics[width=0.65\linewidth]{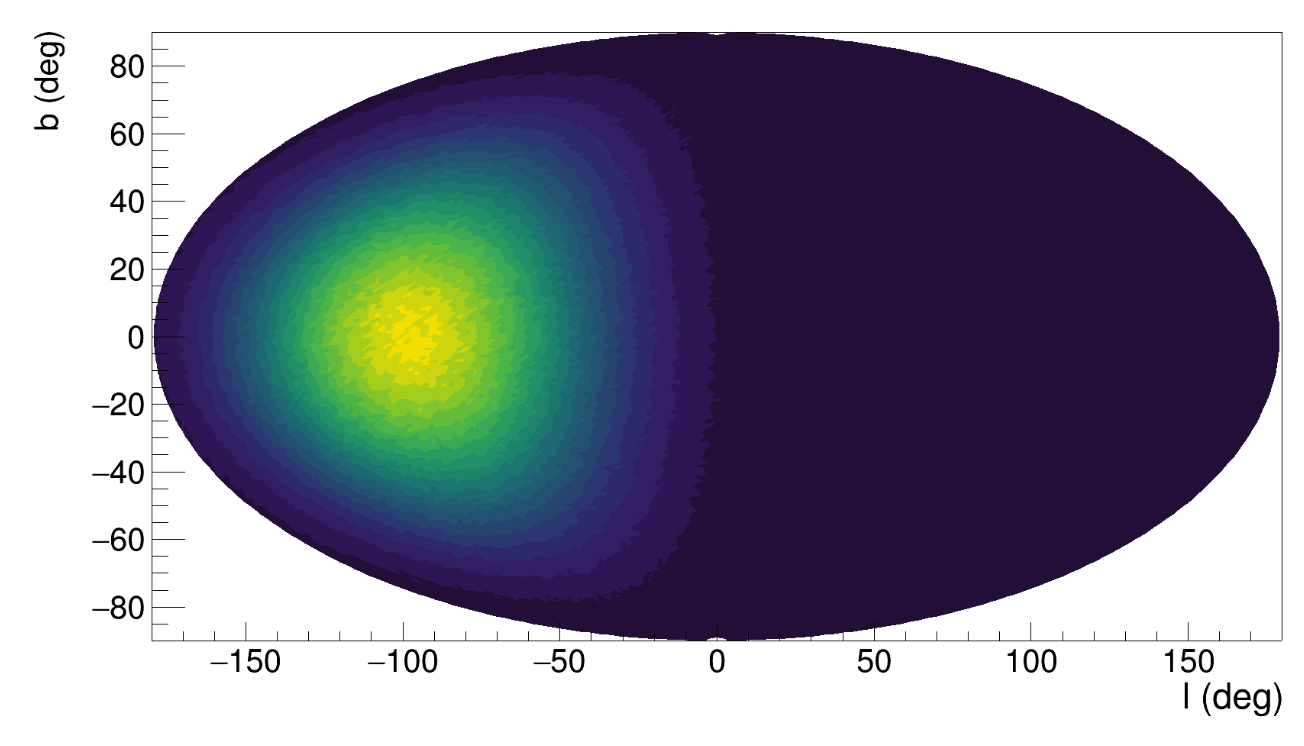}
    \caption{Angular distribution of recoils resulting from WIMP interaction, based on Standard Halo Model (SHM) assumptions. The distribution of F recoil induced by 10 GeV WIMPs, showcasing the characteristic dipole structure, is shown. Plot from \cite{Baracchini_2020_Gio}.}
    \label{fig:recoilangdep}
\end{figure}

The Earth orbits the center of the galaxy at a speed of about 230 km/s towards the Cygnus constellation, experiencing an apparent Dark Matter wind due to its motion in the Dark Matter halo. With the provided flux, local Dark Matter density, and a cross-section on the order of weak interaction, the expected interaction rate of Dark Matter is below approximately $\mathcal{O}(1)$ event/kg/year. However, these events need to be discerned from a background of nuclear and electron recoils, which are approximately $10^6$-$10^9$ times higher. For this reason, DM detection faces similar challenges reported for neutrino detection in Sec. \ref{sec:solarnuchallenges}. The determination of the recoil directions caused by WIMPs has the potential to significantly enhance the quest for Dark Matter. This improvement ranges from offering a potential means for positively identifying WIMPs to facilitating the discrimination of the signal from background events. \\
The Earth's co-motion with the Sun around the galaxy center introduces a directional dependence in the interaction rate, peaking in the average tangential direction of the Sun with respect to its motion around the center of the galaxy (approximately toward the Cygnus constellation). The angular dependence is more clear moving to galactic coordinates. The Galactic coordinate system is a right-handed celestial coordinate system in spherical coordinates, centered on the Sun. The origin is oriented toward the center of the Milky Way, and the x-axis aligns with the Galactic plane. The Galactic coordinates utilized consist of the Galactic longitude $l$ (x) and the Galactic latitude $b$ (y), exhibiting a behavior analogous to terrestrial longitude and latitude. In this coordinate system, the Cygnus constellation is positioned approximately at (l, b) $\simeq$ (-81, 0.5). The differential interaction rate per unit of solid angle is expressed in the equation:
\begin{equation}
    \frac{dR}{d\cos{\gamma}} \propto \int_{E_{thr}}^{E_{max}}{ e^{-\frac{(v_{lab}\cos{\gamma}-v_{min})^2}{vp^2}} dE }  
\end{equation}
where, $E_{thr}$ is the energy threshold of the detector and $E_{\text {max }}=\frac{1}{2} m_\chi r\left(v_{l a b}(t) \cos \gamma+v_{\text {esc }}\right)^2$ is the maximum energy of the recoil, $v_{lab}$ is the velocity of the lab averaged in galactic coordinates, $v_{min}$ is the minimum velocity required for the DM particle to produce a recoil, and $\gamma$ is the angle between the recoil direction and the lab direction. This recoil rate is shown in galactic coordinates in Fig. \ref{fig:recoilangdep} for a WIMP mass of 10 GeV on Fluorine is shown.
The plots distinctly reveal the anisotropic characteristics of the angular distribution of recoils caused by WIMP interactions, showcasing an excess of events at negative longitudes, opposite to the Sun's motion. This signature exhibits a directional correlation with an astrophysical source, a unique feature that cannot be replicated by any background (helped by the rotation of Earth around its axis). Consequently, this provides the potential for a positive claim of a Dark Matter signal. Plot from \cite{Baracchini_2020_Gio}.

\subsubsection{Directionality advantages}
\label{sec:diradvantagesDM}
The additional information gained from measuring the recoil direction, alongside energy, can be crucial for the research, identification, and analysis of Dark Matter. \\ A directional detector inherently offers more detailed information about each recoil compared to a non-directional detector. The extra angular information enhances the capability to distinguish between signal and background when the background has an angular distribution different from that of the WIMP recoil signal. Hence, directional detectors can establish more robust exclusion limits compared to non-directional detectors, given the same exposure \cite{vahsen2020cygnus}. Neutrinos originating from the Sun, diffused supernovae, and the atmosphere, interacting trough CE$\nu$NS, as described in Sec. \ref{sec:NuCoerent}, can produce nuclear recoil which can closely mimic a Dark Matter signal when only the nuclear recoil event energy distribution is utilized for discrimination. Enhancing the understanding of neutrino fluxes, employing multiple target elements, and detecting a large number of events can help break through the neutrino floor by providing more details on the recoil spectrum generated by the coherent elastic neutrino-nucleus scattering (CE$\nu$NS) interaction \cite{Ruppin_2014}. Nevertheless, it has been extensively demonstrated that measuring the direction of the recoil is the most effective approach to overcome this issue, especially for solar neutrinos \cite{vahsen2020cygnus,O_Hare_2015}. 
\begin{figure}
    \centering
    \includegraphics[width=0.75\linewidth]{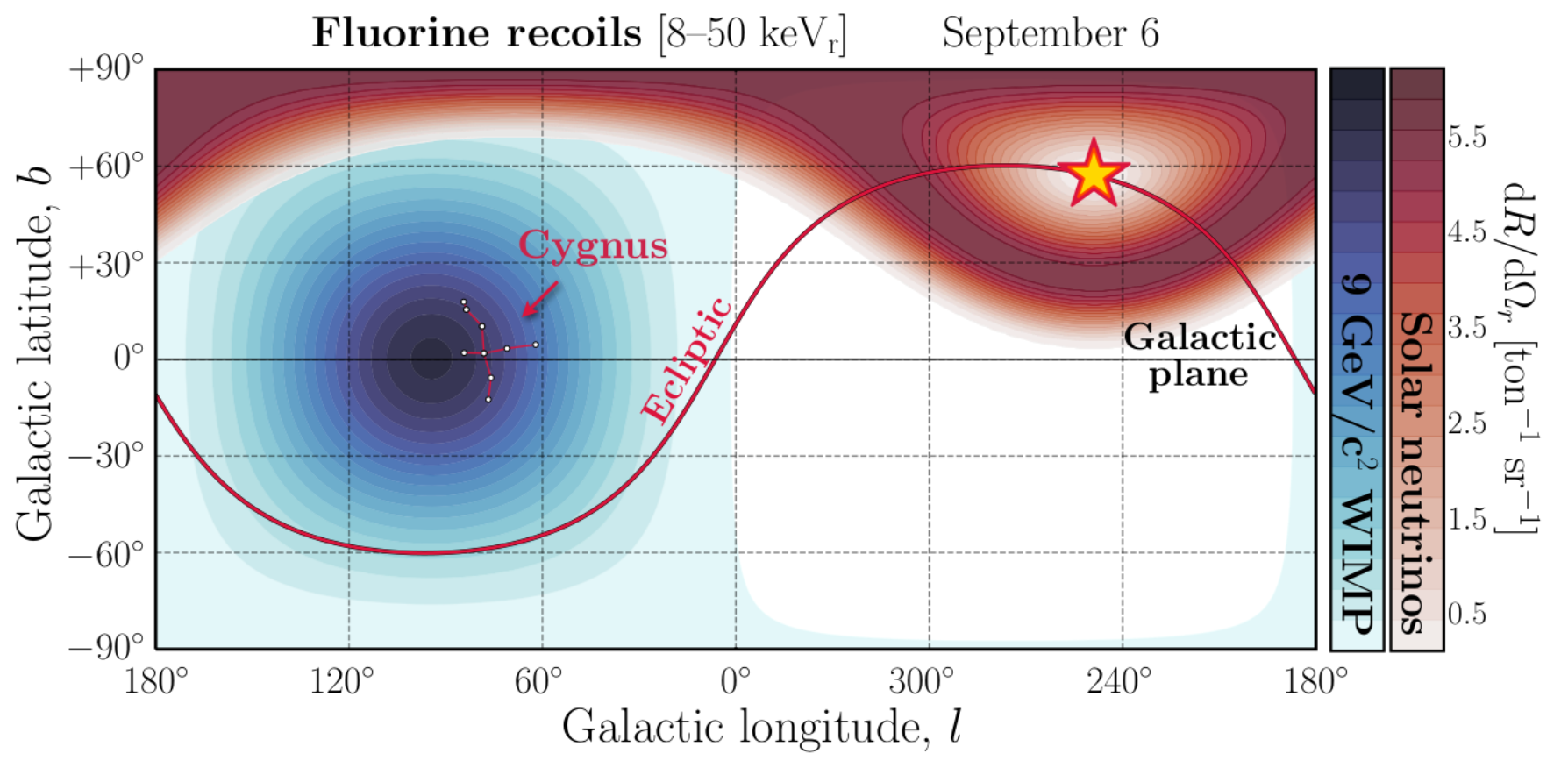}
    \caption{Angular distribution of nuclear recoils induced by WIMPs and CE$\nu$NS from solar neutrino in galactic coordinates. The red line represents the ecliptic path, highlighting minimal overlap between the two distributions throughout the year. Plot from \cite{vahsen2020cygnus}.}
    \label{fig:nusunDM}
\end{figure}
Fig. \ref{fig:nusunDM} illustrates the angular distribution of DM-induced recoils, peaked at the Cygnus constellation position, and those produced by solar neutrinos on September 6th (the day of maximal separation). As outlined in \cite{OHare:2015utx}, to effectively mitigate the neutrino background, directional detectors must meet certain technical criteria, including a time resolution of a few hours or less, energy resolution lower than 20\%, a nuclear recoil orientation recognition for events above 75\%, and a resolution in angle lower than 30 degrees. 

\subsubsection{CYGNO sensitivity to DM searches}
\label{sec:DMSensitivity}
\begin{figure}
    \centering
    \includegraphics[width=1.\linewidth]{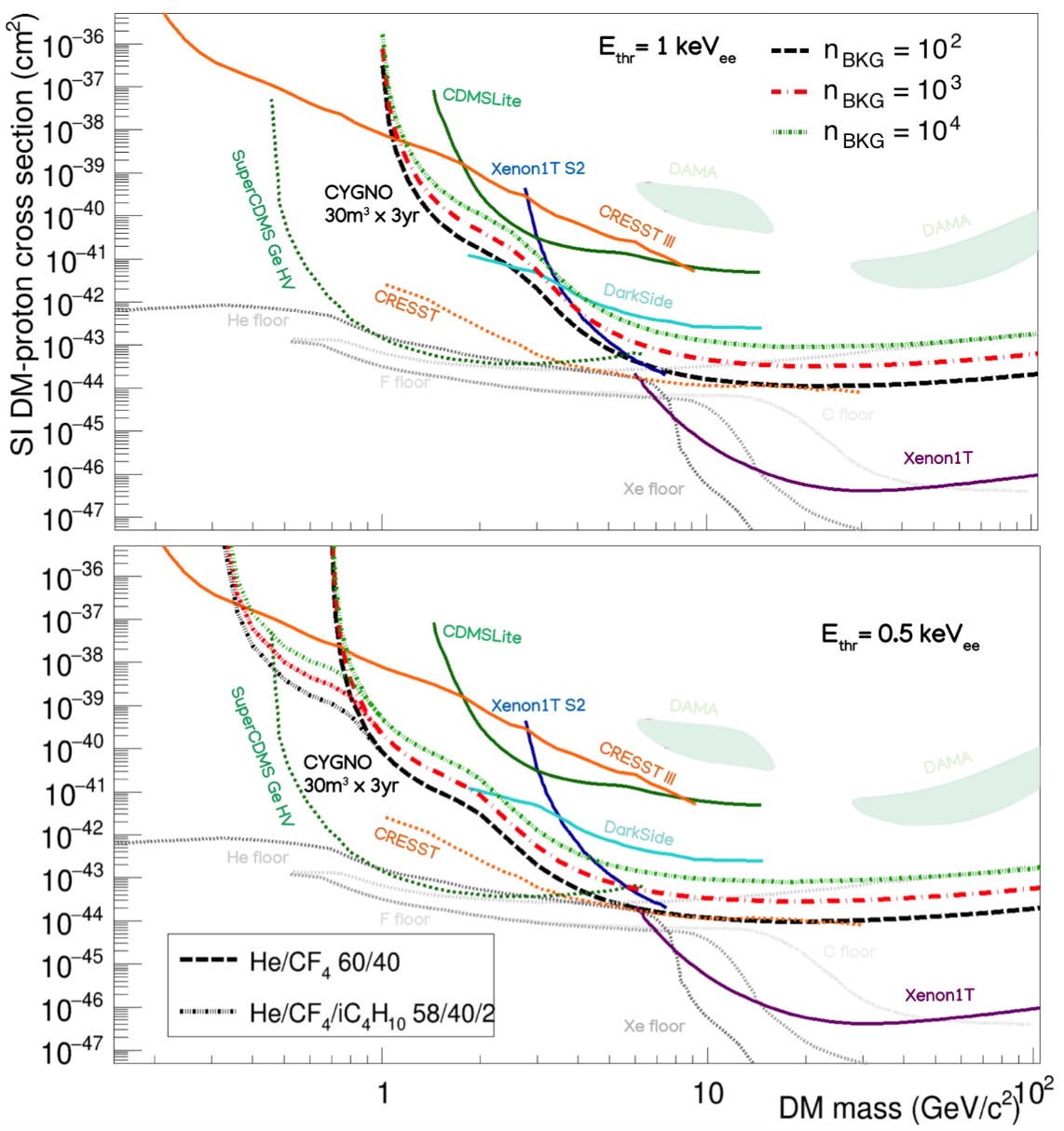}
    \caption{Sensitivity analysis for the Spin-Independent WIMP-nucleon cross-section in a 30 $m^3$ CYGNO detector over a 3-year exposure, considering various background levels and an operational threshold of 1 keV (upper plot) and 0.5 keV (lower plot). Dashed curves represent a He:CF$_4$ 60:40 detector with background levels of 100 (black), 1000 (red), and 10,000 (dark green). The dotted curves display sensitivity for a He:CF$_4$:isobutane 58:40:2 mixture. Plot from \cite{DhoTesi}.}
    \label{fig:SIDMSensitivity}
\end{figure}

\begin{figure}
    \centering
    \includegraphics[width=1.\linewidth]{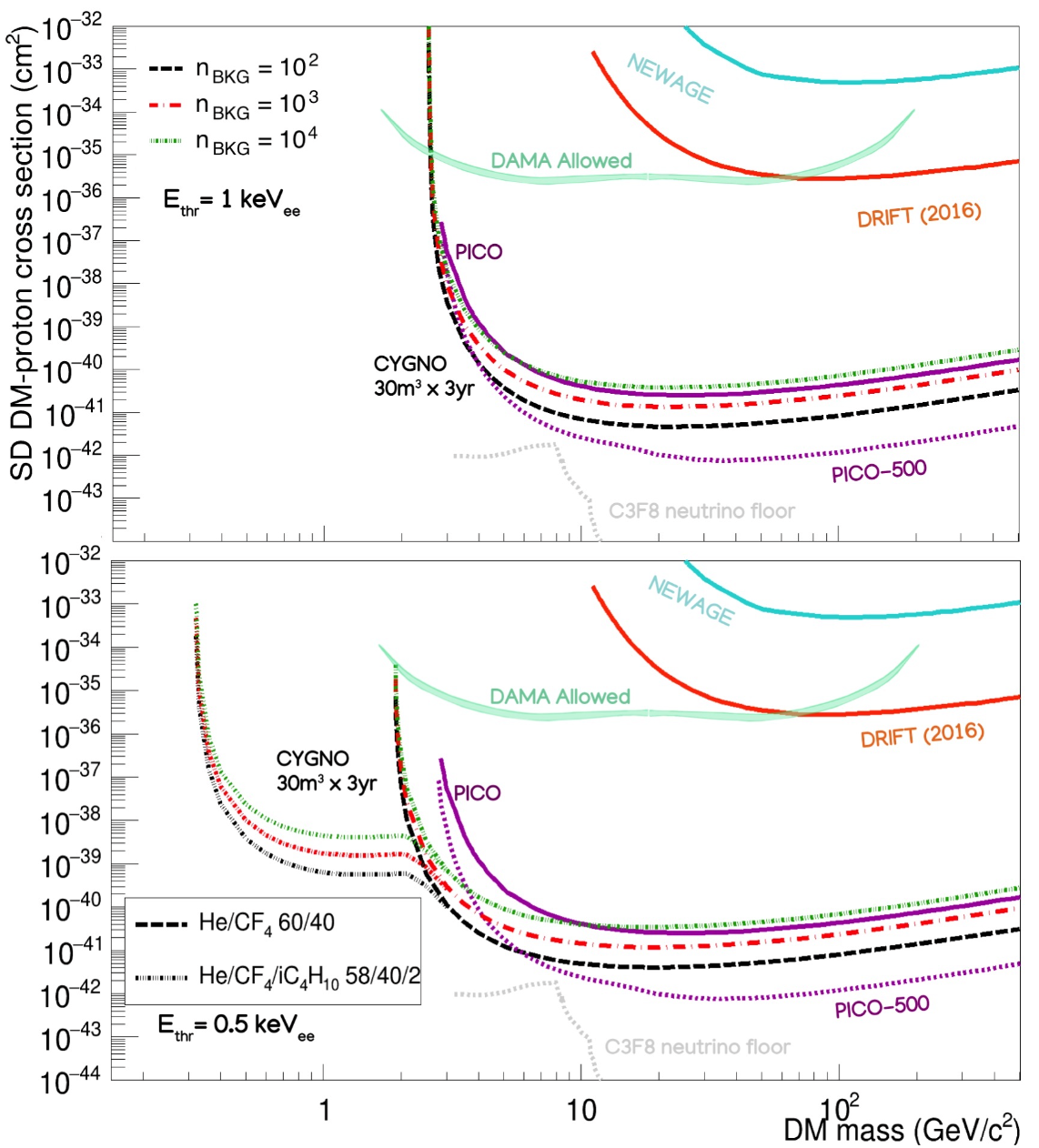}
    \caption{Sensitivity to Spin-Dependent WIMP–proton cross-sections for the 30 $m^3$ CYGNO detector with a 3-year exposure, considering various background level scenarios and operating thresholds of 1 keV (top plot) and 0.5 keV (bottom plot). The dashed curves represent scenarios with $N_{bkg}$ = 100 (black), 1000 (red), and 10,000 (dark green) events. The dotted curves depict sensitivity for a He:CF$_4$:isobutane 58/40/2 mixture. Plot from \cite{DhoTesi}.}
    \label{fig:SDDMSensitivity}
\end{figure}

A statistical analysis utilizing the Bayesian approach was conducted to assess the sensitivity of the CYGNO PHASE\_2 30 $m^3$ experiment for WIMP searches in the presence of background. In this sensitivity analysis, two energy thresholds were considered: a conservative 1 keV$_{ee}$, supported by published results \cite{Costa_2019}, and an improved value of 0.5 keV$_{ee}$, extrapolated from the enhanced performances observed with the PHASE\_0 LIME prototype (Sec. \ref{sec:overgroundstudies}). To convert these thresholds into nuclear recoil energy, a SRIM simulation \cite{ZIEGLER20101818} was developed to assess the quenching factors (QF) for the elements present in our gas mixture, including Hydrogen (see Sec. \ref{sec:redphase2}). Nuclear recoils typically result in a reduced response in target materials compared to electron recoils of equivalent energy. This discrepancy primarily arises from the relatively small portion of the recoil energy transferred to electrons. An energy-dependent quenching factor can be established as the ratio of the ionization induced from a nuclear recoil to that from an electron recoil at the same energy \cite{PhysRevC.88.035806}. The QFs for H, He, C, and F in He/CF$_4$ 60/40 at 1 atm were evaluated as a function of the nuclear recoil energy E [keV$_{nr}$]. 
Consequently, effective energy thresholds of 1.4 (0.8) keV$_{r}$ for H, 2.1 (1.2) keV$_{r}$ for He, 3.1 (1.8) keV$_{r}$ for C, and 3.8 (2.2) keV$_{r}$ for F were obtained for a 1 (0.5) keV$_{r}$ energy deposit. 
The angular distributions of the signal were then computed 
in galactic coordinates, based on the following studies \cite{LEWIN199687,PhysRevD.66.103513, Baxter_2021} and disregarding Earth's motion, as it has been demonstrated to have secondary significance in the angular distribution. While CYGNO's directional capabilities are still under evaluation, an angular resolution of 30 degrees across the entire detectable range was assumed for this study, based on literature \cite{KNakamura_2012} and the CYGNUS simulation \cite{vahsen2020cygnus}. It is also assumed a full head-tail recognition down to the 1 keV$_{ee}$ energy threshold. Predicting the number of expected background events for CYGNO PHASE\_2 is challenging at this project stage. This quantity will depend on the results of PHASE\_0 and PHASE\_1. Hence, various background scenarios are simulated, encompassing 100, 1000, and 10,000 events per year.\\
In the upper section of Fig. \ref{fig:SIDMSensitivity}, the expected SI limits for a 30 $m^3$ CYGNO PHASE\_2 experiment over a 3-year exposure are presented, considering various background scenarios and a 1 keV energy threshold. The lower part of displays the potential sensitivity with an operational threshold of 0.5 keV, along with the outcomes achievable with a Hydrogen-rich gas mixture containing 2\% isobutane content. While future experiments like SuperCDMS \cite{Agnese_2018}, CRESST \cite{Abdelhameed_2019}, Darkside low-mass \cite{manthos2023darkside20k}, and NEWS-G may potentially reach these regions, their operation modes often entail significant compromises in background tolerance and discrimination capabilities, sometimes even relinquishing them altogether. Achieving this goal necessitates stringent criteria for the radio purity of materials and accurate estimation of expected backgrounds. In fact, any signal observed in this range by these experiments would be challenging to interpret definitively as a DM signal. In contrast, CYGNO's utilization of the directional properties of recoils offers a pivotal and conclusive means for positively identifying DM. In this simplified scenario, CYGNO-30 could emerge as a competitive contender, matching the current most stringent limit for WIMP masses ranging from 1-5 GeV under optimal background conditions.\\
The plot for the SD sensitivity is shown in Fig. \ref{fig:SDDMSensitivity}. The improvements that can be obtained with an operational threshold of 0.5 keV are shown on the bottom side of Fig. \ref{fig:SDDMSensitivity}.
Fluorine-based gas in CYGNO enables efficient sensitivity to Spin-Dependent coupling (Sec. \ref{sec:WIMPInteraction}). With its robust coupling and abundant fluorine, coupled with the lightweight nature of Hydrogen (assuming the addition of isobutane to the gas mixture), CYGNO-30 can explore vast regions not yet excluded by the PICO experiment under low-background conditions. The PICO experiment, which currently holds the highest sensitivity among all existing and planned experiments investigating Spin-Dependent coupling \cite{PICO:2023uff}, employs an energy threshold approach. This approach means that signal observation does not provide information about the energy of the detected nuclear recoil, preventing the translation of observations into constraints in the mass versus coupling parameter space. Moreover, experiments employing this methodology are unable to measure the energy of an interaction and instead rely solely on counting statistics for the search for WIMPs.
Consequently, PICO lacks confirmation of the Galactic origin of the detected signal, unlike CYGNO.

%% file: chapters/LIMEPrototype.tex
\chapter{Analysis of LIME response to low energy electron recoils}
\label{chap:LIMEDetector}
This chapter presents the initial part of the thesis author's work, focusing on the characterization of the detector response to low-energy electrons with an original analysis developed by the author.
As extensively described in Chap.\ref{chap:sota} solar neutrinos interacting through elastic scattering $\nu e$ES in the detector, produces an electron recoil. Thus, to characterize the signal expected from neutrino interaction, extensive studies on low-energy electron recoil have been performed on the LIME detector.
LIME (Sec. \ref{sec:LIME}) is the largest and latest prototype developed by the CYGNO collaboration. The TPC has a rectangular-shaped sensitive volume, exhibiting a large readout area measuring 33×33 cm$^2$ and a drift length of 50 cm, with a total active volume of 50 liters. The cathode and the squared field rings that constitute the field cage are constructed from Copper. The rings are positioned with a pitch of 16 mm and interconnected via ceramic resistors to create the voltage partitioning, for maintaining constant and uniform the drift field. The amplification stage relies on a triple stack of GEMs, each with a thickness of 50 $\mu$m, with an area of 33$\times$33 cm$^2$. The GEM area is optically readout, employing a Hamamatsu Orca-Fusion $\textsuperscript{TM}$ sCMOS camera and 4 Hamamatsu R7378A8 PMTs placed at the 4 corners of the detector. The LIME detector moreover will be the base for the realization of the CYGNO-04 demonstrator, which will be then the basic module for the realization of a CYGNO 30 $m^3$ experiment. Since CYGNO-04 will show a similar granularity and the same drift length as LIME, comparable performances are expected for CYGNO-04. For this reason, the response of a CYGNO detector to low-energy electron recoil has been studied on LIME, and employed in the feasibility study for solar neutrino detection (Chap. \ref{chap:solarnu}). This analysis on low-energy electron recoil will additionally be used to validate the simulation (Chap. \ref{chap:Simulation}) on which then the directionality algorithm will be implemented, and the directionality performances will be studied (Chap. \ref{chap:directionality}).
Moreover, in the context of the CYGNO-04 detector, the analysis of the LIME performances is essential to validate the efforts made and affirm the scientific and technological decisions leading toward the development of the CYGNO-04 detector.\\
This chapter outlines the algorithm employed for track identification and reconstruction from the sCMOS image (Sec. \ref{chap:reco}). Following that, a comprehensive description of the dedicated analysis developed in the context of this thesis to study the detector light response and energy resolution to low-energy electron recoils at different energies is presented (Sec. \ref{sec:XRayAn}). 
\section{The track reconstruction algorithm}
\label{chap:reco}
\begin{figure}
    \centering
    \includegraphics[width=0.5\linewidth]{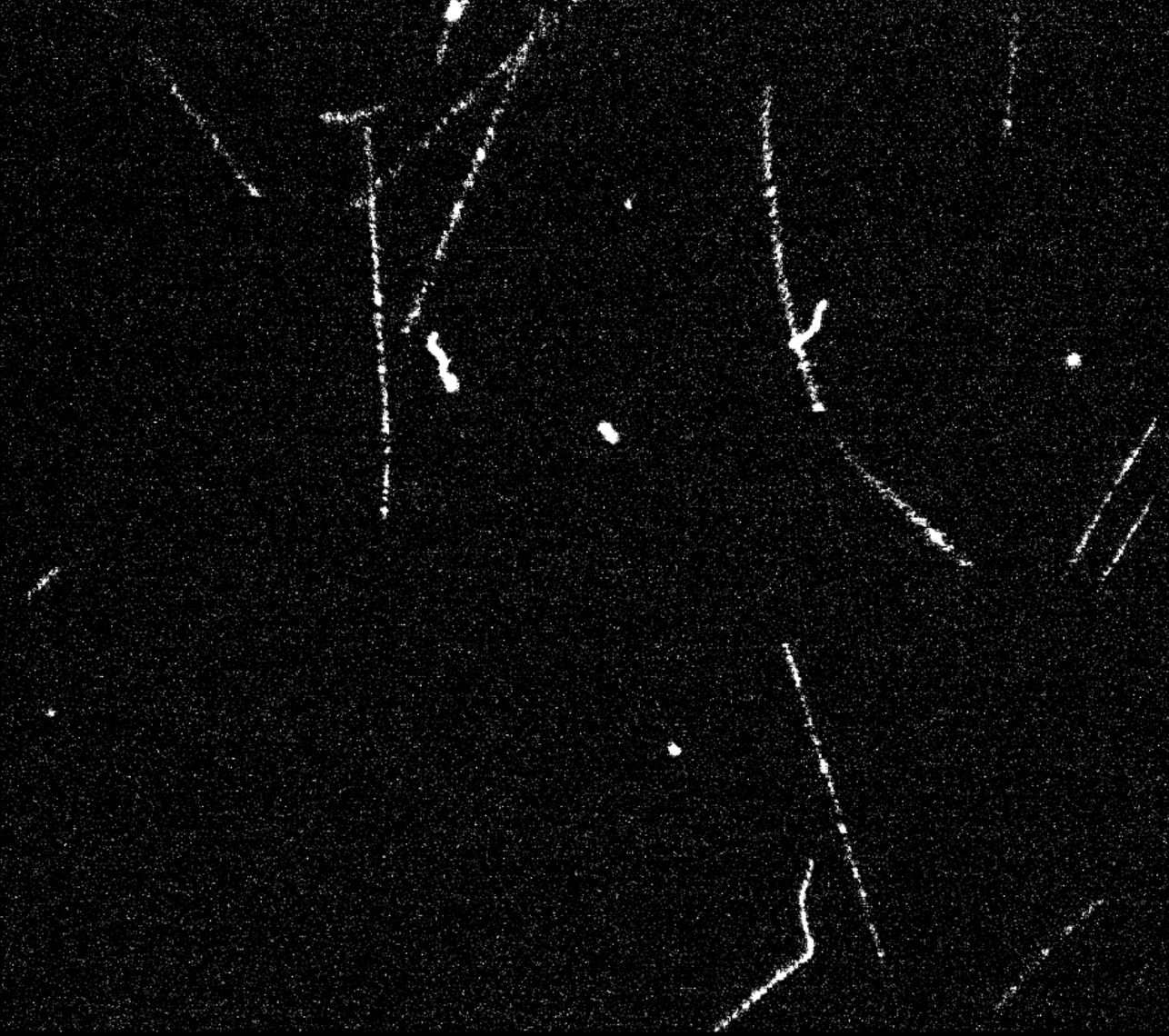}
    \caption{In figure, an event acquired with LIME overground is shown. The image is 2304$\times$2304 pixels.}
    \label{fig:LIMEEvent}
\end{figure}
As can be seen in Fig. \ref{fig:LIMEEvent} particle interactions in the detector result in light clusters on a sCMOS picture. To perform an analysis on the tracks, these must be identified and extracted from the picture. To reconstruct these tracks, a dedicated dynamic algorithm has been developed by the CYGNO collaboration \cite{Amaro_2023}. The algorithm is designed for efficient reconstruction of the tracks released by different kinds of particles, each showing distinct energy release patterns and geometric characteristics. It is indeed able to recognize very low-energy nuclear recoils consisting of tiny dense light spots, long and straight tracks induced by cosmic rays crossing the detector volume, or curly with non-constant energy ionization profiles tracks generated by electrons crossing the sensitive volume. The reconstruction algorithm needs therefore to efficiently be able to reconstruct all these different patterns. 
For these reasons, the analysis of the sCMOS images proceeds through four steps, by firstly dealing with the intrinsic noise sensor (Sec. \ref{sec:prelim}) and the optical distortion of the system (Sec. \ref{sec:vignetting}), and subsequently finding basic clusters from the single small deposit (sec \ref{sec:dbscan}), which are eventually merged into the full tracks that represent the output of the reconstruction (Sec. \ref{sec:finaldbscan}).
The output generated by the reconstruction code comprises a set of x-y coordinates of the pixel constituting each track along with their respective intensities. The sum of the pixel intensity provides a quantity proportional to the energy $E$ of the track. Moreover, from the information on the single pixels of the tracks, important topological quantities can be derived. These include, among others, the specific ionization along the track $\Delta E/\Delta l$, where $l$ is the track length, or the total light density $\delta=E/n_{pixels}$, where $n_{pixels}$ is the track total number of pixels. This is a distinctive aspect of this kind of detector, which allows for the track direction and orientation recognition, and particle discrimination by leveraging these topological track variables.

\subsection{Sensor noise reduction}
\label{sec:prelim}

\begin{figure}
    \centering
    \includegraphics[scale=.5]{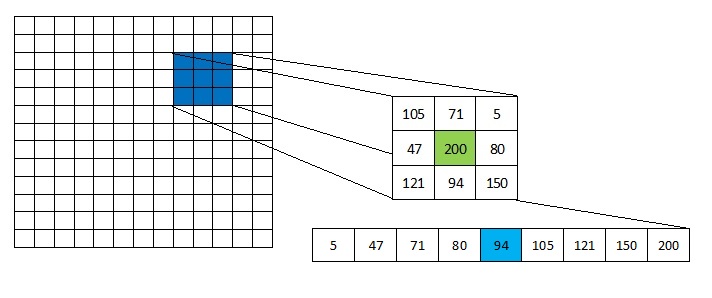}
    \caption{In the picture, the working mechanism of the median filter is shown. The central pixel of value 200 is replaced by the median value of all the pixels in a 3$\times$3 square around the central one, in this case, 94. Plot from \cite{MedianFilterScheme}.}
    \label{fig:medianfilter}
\end{figure}

A sCMOS picture acquired by the Fusion ORCA\textsuperscript{TM} camera consists of an array of 2304$\times$2304 pixels with intensity expressed in ADC counts. The amplification process of photoelectron conversion to an electric signal, the heat-generated ionization in the silicon sensor, and the electrical connections are all sources of noise \cite{Lundberg2002NoiseSI}. 
To handle this noise, the set of images taken with GEMs at 200 V, commonly referred to as “pedestal”, is used. In the absence of any source of amplification, the images acquired with the camera contain only counts related to the pixel noise. Consequently, from those images, it is possible to perform a statistical analysis to characterize the mean value and the RMS of the noise present in each pixel in such a way to handle it.
From these pedestal images, a pedestal map has been built. This map consists of two separate 2304$\times$2304 matrices containing the information of the mean value $\bar{N}_{i,j}$ and the RMS $\sigma_{N i,j}$ of each pixel $i,j$.\\
In order to reduce the noise, the pedestal map of the mean is subtracted pixel by pixel to each sCMOS image. Subsequently, to further reduce the noise, a zero suppression operation has been performed on the pedestal subtracted image. By defining $C_{ij}$ the content of each pixel after the pedestal subtraction, the condition for the pixel to survive the zero suppression is 
\begin{equation}
    C_{i,j}> n_\sigma \cdot \sigma_{Nij}
\end{equation}
where $n_\sigma$ is a parameter of the algorithm that is optimized before the reconstruction process depending on the noise condition of the dataset. For the overground LIME data, the parameter $n_\sigma$ was just set above the noise level, specifically to $n_\sigma$=1.2. This choice was made to ensure good detection efficiency while preventing overloading of the reconstruction algorithm with pixels dominated by noise.
Images are subsequently rebinned (4$\times$4) for CPU reasons into images of 576$\times$576 macro-pixel. The counts of each macro-pixel are calculated as the average of the counts from the 4$\times$4 pixels composing it. 
The rebinned image is then processed using a median filter with a kernel 3$\times$3, a local image filtering technique commonly used for noise reduction purposes. For each pixel of the image, a square of 3$\times$3 pixels (the kernel) around the designed one is considered. The value of the designed pixel is then substituted in a new image with the median value of all the pixels in the 3$\times$3 square \cite{Welk_2016}.
The working scheme of the median filter is illustrated in Fig. \ref{fig:medianfilter}.

\subsection{Vignetting correction}
\label{sec:vignetting}
\begin{figure}
    \centering
    \includegraphics[width=1\linewidth]{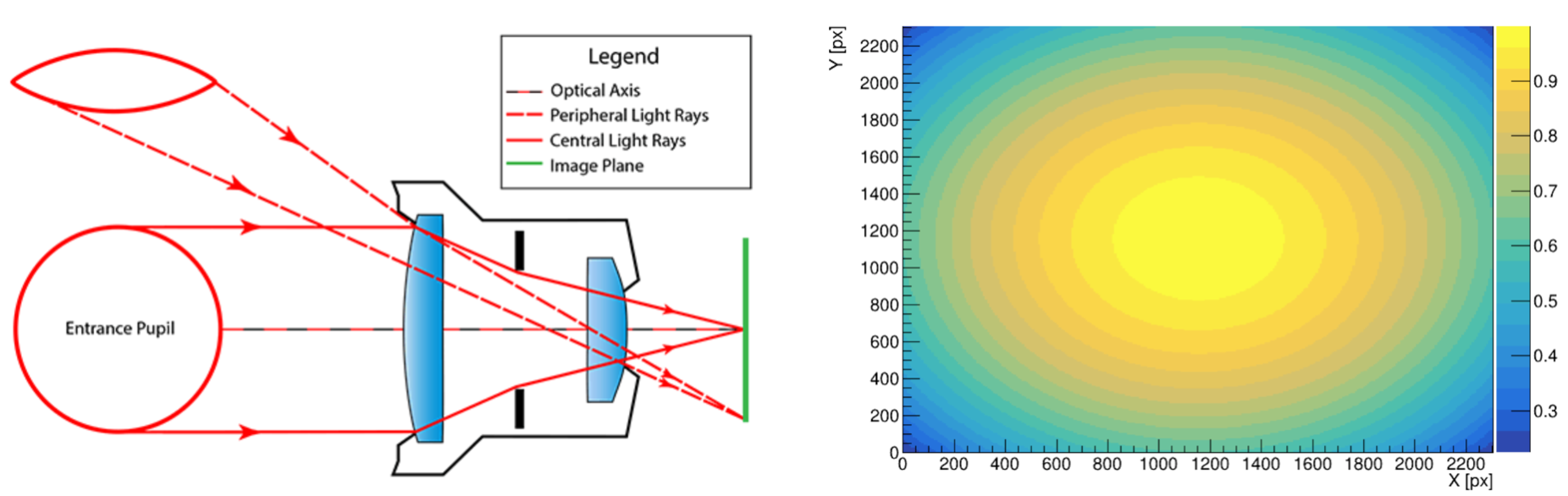}
    \caption{The picture describes the vignetting effect. The light coming from the points of an object off-axis is reduced due to the solid angle. Picture from \cite{SchemeVign}. Right: map of the vignetting effect obtained exposing the camera to a ”uniformly illuminated” setup.}
    \label{fig:vignettingphen}
\end{figure}

The resulting image is then corrected for the vignetting effect caused by the camera lenses.
In the presence of extended sources, part of the source will be off-axis with respect to the axis of the camera.
In an optical system like the one used in CYGNO, the illuminance on the sensor plane for these off-axis regions is diminished compared to the axial value (more details in Sec. \ref{sec:optics}), determined by the angle $\phi$ between the optical axis and the direction connecting the source with the lens entrance's center. The photon flux $\Phi$ from these off-axis sources reaching the lens is expressed as a function of the angle $\phi$ that the off-axis point forms with respect to the optical axis as $\Phi = \pi L dA u^2  \cos^4(\phi)$ where the variables are defined in Sec. \ref{sec:optics}. Thus, the farther the source is from the optical axis, the lower the photon flux. This phenomenon, known as natural vignetting effect, significantly influences CYGNO image analysis, leading to an x-y nonuniformity in light yield, with the image corners exhibiting lower intensity despite the source's equal luminosity. A scheme of this effect is shown in the right plot of Fig. \ref{fig:vignettingphen}.
More in detail, the intensity $I$ of the light due to this effect is maximal when the object is on-axis $\phi$=0 and decreases proportionally to $I\propto cos(\phi)^4$. The reduction can arrive at up to 20\% of the original light intensity in the corner of the image in the LIME camera-lens configuration \cite{10.1088/978-0-7503-2558-5}.

In order to properly correct for this effect on the acquired images, the vignetting has been characterized by exposing the camera to a white wall illuminated with light to simulate a “uniformly illuminated” setup. Different pictures of the wall have been acquired with the camera in a horizontal position and rotated with steps of 90° around the axis passing by the center of the camera and perpendicular to the lens. Subsequently, a vignetting map, shown in the right plot of Fig. \ref{fig:vignettingphen}, has been constructed by averaging the light intensity in the pictures and normalizing it to the maximum value of the ADC counts. 
The vignetting correction is then applied to the sCMOS data images by multiplying each pixel $i,j$ by the value $1/V_{i,j}$, where $V(i,j)$ is the value of the vignetting map in the pixel $i,j$.

\subsection{The iDBSCAN algorithm}
\label{sec:dbscan}
\begin{figure}
    \centering
    \includegraphics[scale=.6]{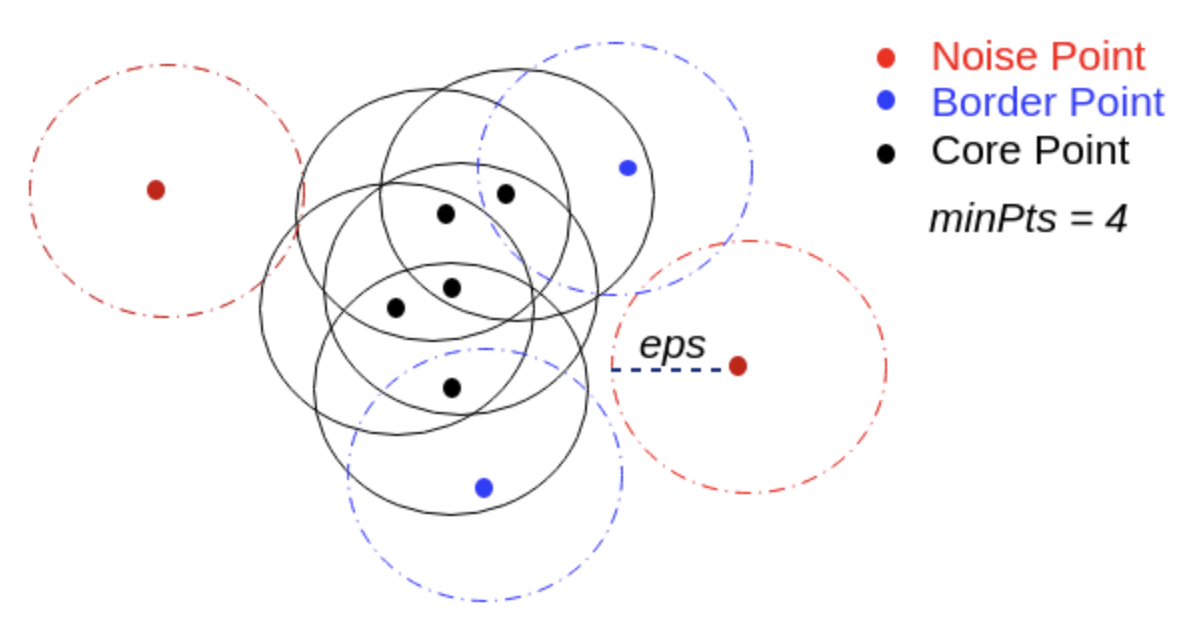}
    \caption{In figure, the DBSCAN working principle is shown. The black points having at least 4 points within a radius $\epsilon$ are marked as core points. The blue points are found close to a core point but do not satisfy the requirement of the minimum number of points, so they are marked as border points. The red points are noise. Plot from \cite{su11174718}.}
    \label{fig:DBSCAN}
\end{figure}

Once the preliminary operations are concluded, the official CYGNO reconstruction \cite{Baracchini:2020nut} code based on the Density-Based Spatial Clustering of Applications with Noise (DBSCAN) \cite{scikit-learn} algorithm is used to identify and reconstruct the light clusters. 
DBSCAN is capable of clustering high-density areas separated by low-density areas within a picture. The clustering operation relies on both the distance between pixels and their intensity as rules for cluster formation. 
These operations are performed driven by two parameters: $\epsilon$ and $N_{min}$. Whenever the number of points within a radius $\epsilon$ (in pixel) centered in a selected point randomly chosen (seed) is greater than $N_{min}$, the clustering process starts. The seed point is marked as part of the cluster, and the sampling procedure proceeds by iterating through every point found within $\epsilon$ in the previous iteration. A point is labeled as a core point if it has $N_{min}$ neighbor within a radius $\epsilon$, if this condition is not satisfied, a point is marked as a border-point. Any points not satisfy the condition of core-point or border-point are marked as noise. The cluster is finally formed by the union of core points and border points \cite{scikit-learn}. A scheme of the DBSCAN working principle is shown in Fig. \ref{fig:DBSCAN}.
In CYGNO, alongside the information on the coordinate of a point, the intensity information is also used. Therefore, an intensity-based version of the DBSCAN (iDBSCAN) algorithm is used.
In iDBSCAN the condition of the minimum number of points $N_{min}$ to form a cluster is integrated with the requirement of a minimum intensity among neighbor points. As demonstrated in \cite{Baracchini_2020}, the iDBSCAN algorithm, on CYGNO images, shows an improved background rejection efficiency compared to both the standard DBSCAN and a Nearest Neighbor Clustering method.

\subsection{Cluster identification}
\label{sec:finaldbscan}
\begin{figure}
    \centering
    \includegraphics[scale=.6]{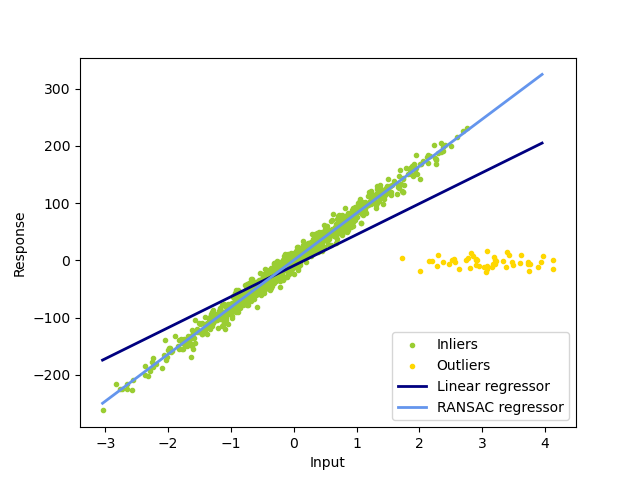}
    \caption{Comparison of a linear and a RANSAC regression. The outlier points influence the goodness of the linear regression, while RANSAC shows robust fitting on the inliers. Plot from \cite{RanSacScheme}.}
    \label{fig:RANSAC}
\end{figure}

To handle the high pileup and to reconstruct long tracks generated by cosmics and high energy electrons, a dedicated algorithm based on iDBSCAN is used to reconstruct and remove this source of background. These types of tracks can be approximated by linear and high-degree polynomial curves. As a result, the directional information of the cluster is employed to enhance the algorithm ability in the identification of this background source.
A key component of this directional iDBSCAN (iDDBSCAN) is the RANdom SAmple Consensus (RANSAC) algorithm. RANSAC is an iterative method used to determine the optimal parameters of a model in a set of data containing a large proportion of outliers. This is realized by (1) selecting a random subset of data of size $m$ from the overall dataset of size $M$, (2) fitting the selected data with the model, and (3) evaluating the number $i$ of inliers points which are within a distance $\omega$ from the fitted model. The sequence of steps 1 to 3 can be iterated $N$ times. The model with the highest number of $i$ is the most suitable one for accurately representing the data, omitting the effect of the outliers \cite{RANSAC}. The difference between a linear fit and a RANSAC regression is shown in Fig. \ref{fig:RANSAC}. 
\\
It can be estimated that given $w$ as the actual fraction of inliers, $p$ the desired probability of finding a subset of data without outliers, the average number of iterations $k$ required for this condition is:  
\begin{equation}
    k= \frac{log(1-p)}{log(1-w^n)}
\end{equation}
with $n<w \cdot M$. To provide an idea, given a model with $w=0.8$, by choosing $n=10$ and $p=0.9$ the resulting number of iterations is $k=20$.

\begin{figure}
    \centering
    \includegraphics[scale=.19]{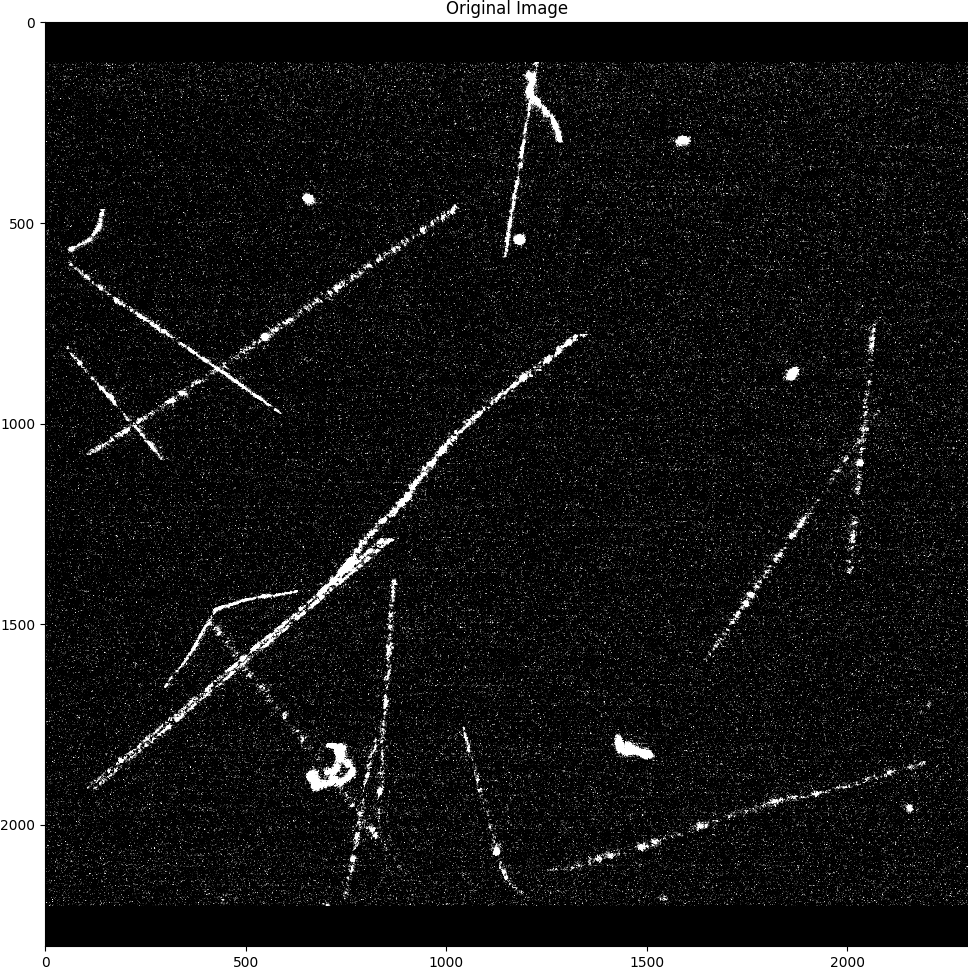}
    \includegraphics[scale=.19]{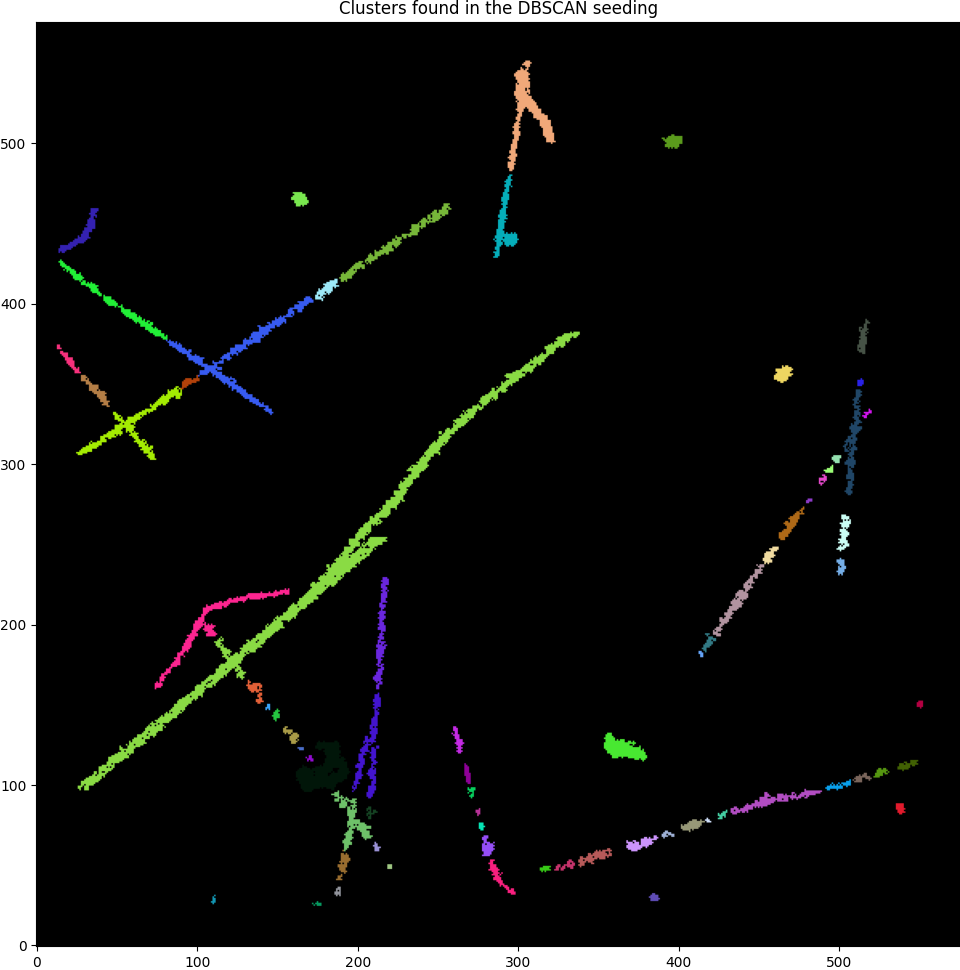}
    \includegraphics[scale=.19]{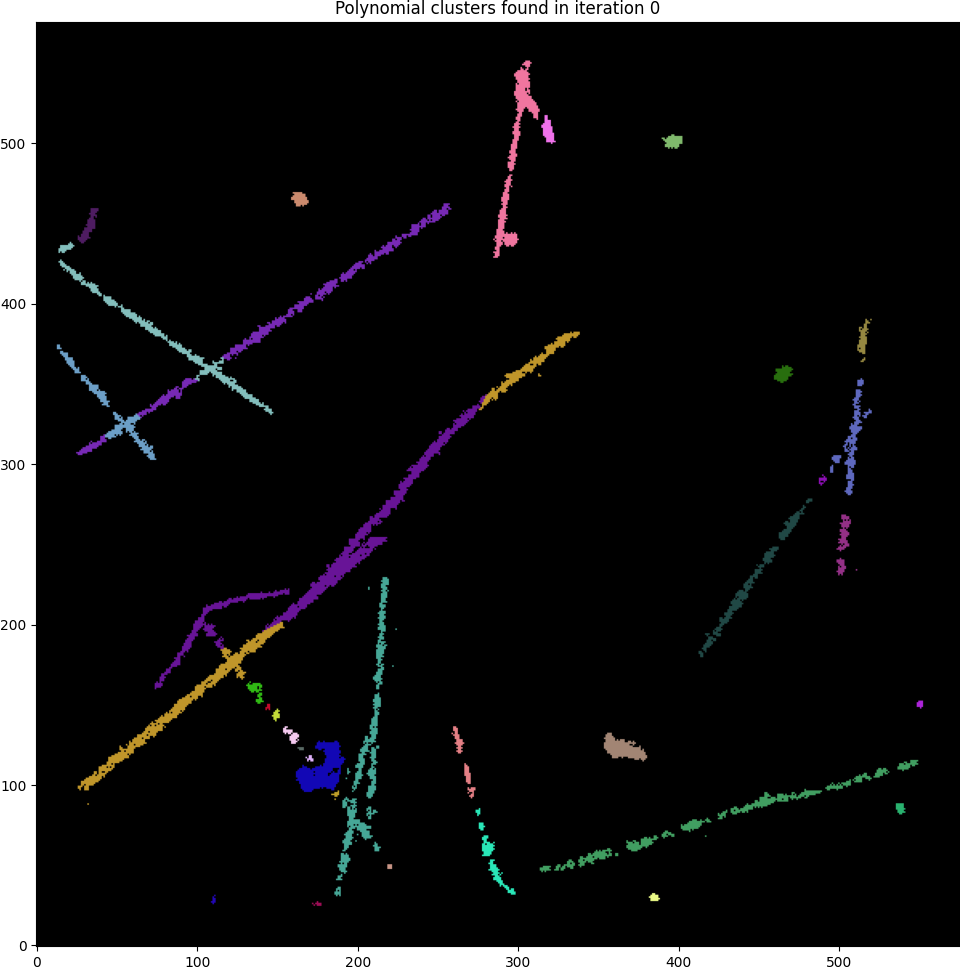}
    \caption{Left: Original picture from sCMOS camera, center: iDBSCAN seeding, right: iDDBSCAN clustering. The x and y axes are expressed in pixels.}
    \label{fig:picReco}
\end{figure}

In light of this, the algorithm performs a first seeding, applying an iDBSCAN iteration to the whole image. After this procedure, each identified cluster with coordinates ($x_i,y_i$) is tested with a linear-RANSAC ($y(x)$). The parameters used are $m=1/2\cdot M$ (50\% pixels), and $\omega$=1 MAD (Median Absolute Deviation), where 1 MAD is defined as 
\begin{equation}
    1 MAD = median(|y_i-median(y_i)|)
\end{equation} 
The procedure is terminated, and the cluster is marked as directional if the inline ratio reaches $i/M>0.8$. If no directional cluster is found, the algorithm execution is concluded, and the result is given by the iDBSCAN clustering. 
If a directional match is found, the algorithm proceeds with the directional clustering. For a cluster marked as directional, the iDBSCAN procedure is applied. Subsequently, a third-order polynomial RANSAC is executed. iDBSCAN is then executed in a region of size $\omega$ centered around the optimal third-order polynomial model found for the cluster. For the directional searches, a radius $\epsilon_{dir}$ different from $\epsilon$ is used. The procedure continues until no more pixels can be added to the cluster. Once the procedure is ended, it is repeated for any other directional cluster and continues until all directional clusters are processed. 
After this procedure, an additional step for noise reduction is performed. Any pixels with a certain distance $\rho_{isolation}$ from a directional cluster reconstructed are marked as noise, and are excluded in the final step. This is to prevent additional cluster formation close to cosmics and high-energy electrons. Since these tracks are very long, the probability of finding noise around that can become part of the cluster is not negligible. 
Finally, iDBSCAN is applied to all the pixels not included in a directional cluster and not marked as noise. After an exhaustive parameter optimization and validation, described in detail in \cite{Amaro_2023}, the X-Ray runs reconstruction has been performed. 
In Fig. \ref{fig:picReco}, from left to right, are shown: the original image, the tracks reconstructed with iDBSCAN, and the tracks reconstructed with iDDBSCAN are shown for an event taken as example. Pixels with the same color belong to a single track.
The pixels x-y positions, and the number of counts of each pixel associated with each track, are then used to analyze the data.

\section{Analysis of multi-energy electron recoil}
\label{sec:XRayAn}
In order to evaluate the CYGNO experimental approach capability to detect low-energy electron recoils, LIME was exposed to X-Rays with energies from 3.5 keV to 45 keV, and its response to them was thoroughly analyzed. The analysis of the data presented in this thesis has been done to study in detail the detector response to ER at the energy of interest for the neutrino physics case. In particular, the aim of the analysis is to study the light response and the energy resolution for ER at different energies.
In this section, the experimental setup and dataset acquired (Sec. \ref{sec:setupXray}), and the energy response linearity and resolution evaluated from this data (Sec. \ref{sec:analysis}) will be presented.

\subsection{Experimental Setup}
\label{sec:setupXray}
\begin{figure}
    \centering
    \includegraphics[scale=.4]{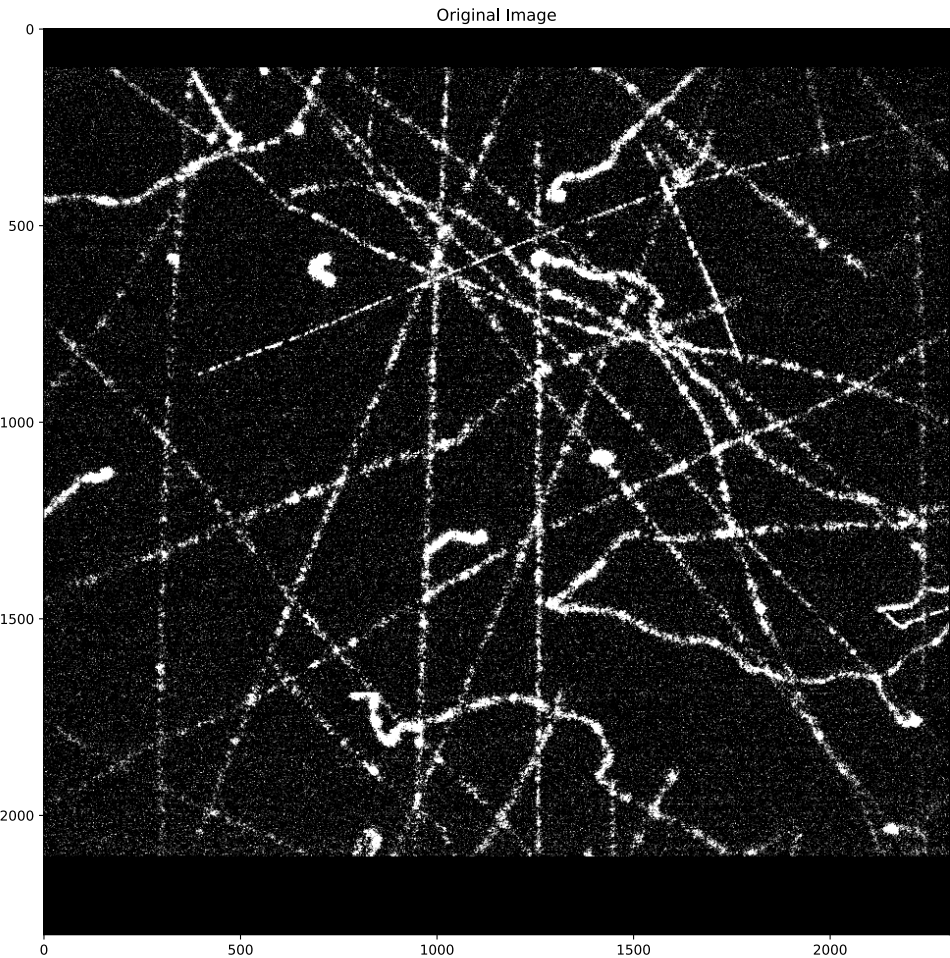}
    \includegraphics[scale=.28]{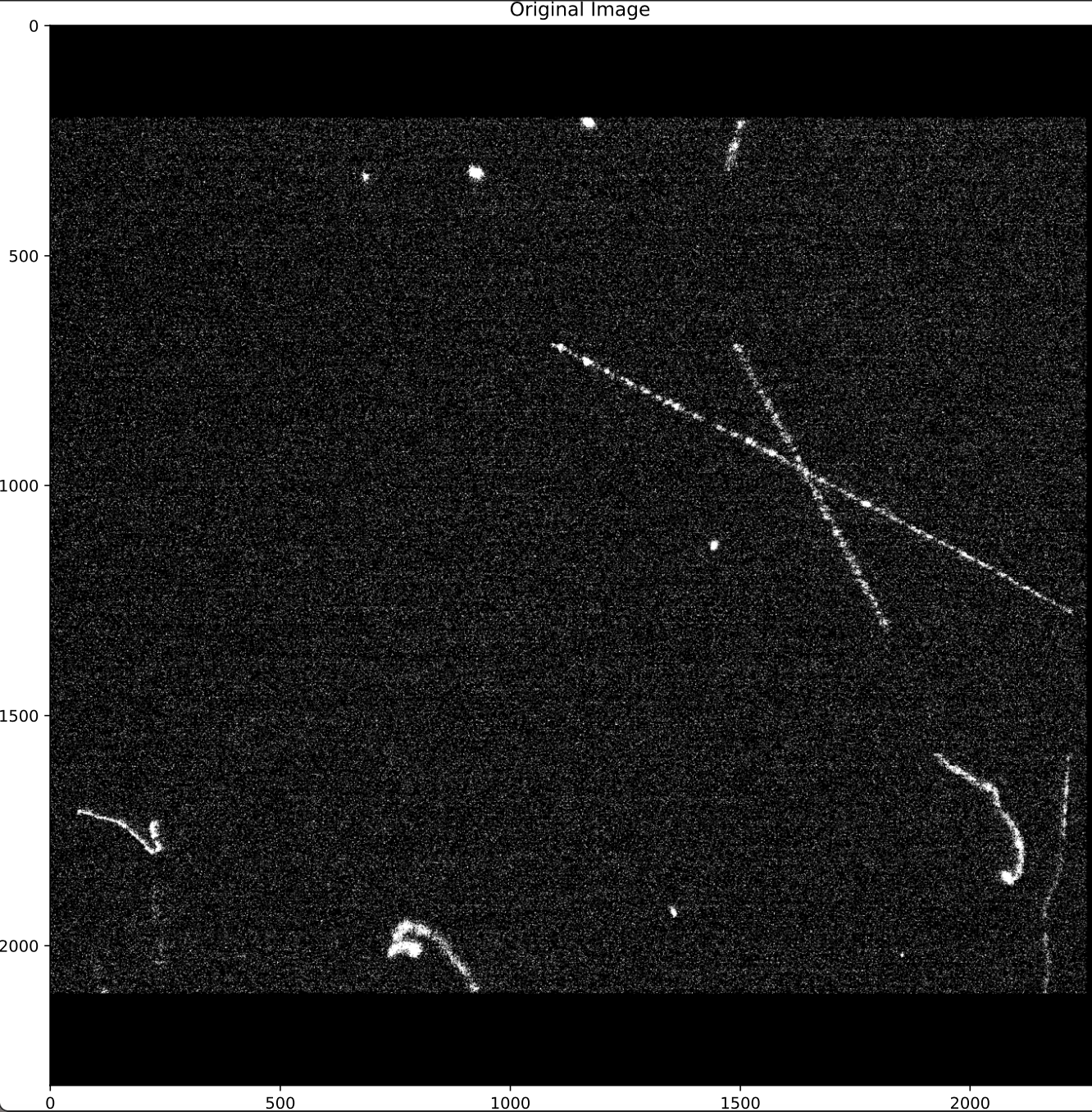}
    \caption{Pictures of natural radioactivity acquired with the sCMOS with 300/50ms exposure left/right }
    \label{fig:expdiff}
\end{figure}

\begin{figure}
    \centering
    \includegraphics[scale=.45]{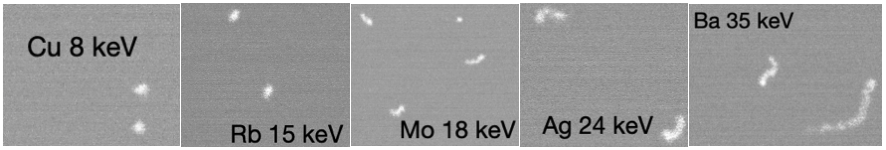}
    \caption{In figure, tracks of electron produced by gammas at different energy are shown.}
    \label{fig:multiEletracks}
\end{figure}

The data taking was carried out using the collaboration largest prototype, LIME, which is extensively described in Sec. \ref{sec:LIME}, located above ground at LNF. It was filled with a gas mixture of He:CF$_4$  (60:40), and it was operated with a constant flux of 9 L/h, for the whole data taking. This is done to ensure the purity of the gas mixture and the flux and the right percentage composition are monitored by a gas mixer. The detector was configured to its standard operational settings: the voltage difference on the three GEMs was set $V_{G1}=V_{G2}=V_{G3}=440$V, the transfer field between the GEMs have been set to $V_{tr1}$ = $V_{tr2}$ =2.5 kV/cm and the drift field was set to $V_{drift}=940$ V/cm. For the cooling of the camera, the water cooling system was employed, using water at 25°. This ensures a constant temperature on the sensor to avoid noise variation. During the operation, the camera was operated in continuous acquisition mode and the images were acquired with the Hamamatsu Hokawo software. The data taking was performed overground and without a shield from external radiation. This implies a high occupancy of tracks in the sCMOS images. This effect of the occupancy can be mitigated by adjusting the exposure time on the camera. The difference between a sCMOS image acquired with 300ms and 50ms exposure is shown in Fig. \ref{fig:expdiff}. For this reason, an acquisition time of 50 ms was chosen to handle the high occupancy of cosmics and electrons from natural radioactivity in the images. 
Prior to data acquisition, the voltage differences GEMs were lowered to 200 V, and a set of 400 pictures at the same exposure was acquired in order to measure the camera's noise. The pedestal will be later subtracted at the reconstruction level from the sCMOS picture acquired, as explained in detail in Chap. \ref{chap:reco}. 
The signal in the detector was induced in the detector by means of X-Rays at different energies, produced by a source at 25 cm from the GEMs, in the central part of the detector. Different radioactive sources were employed for X-Ray production: an Amersham $^{241}$Am-based source, which generates monochromatic X-Rays by emitting alpha particles onto different materials (see Appendix \ref{app:appendixB}), and the $^{55}$Fe source. Additionally, the $5.9$ keV X-Rays from the $^{55}$Fe source were used to irradiate a sample of Ca, producing X-Rays at lower energies. The energy in detail are shown in Tab \ref{tab:kalphaenergies}, and a set of pictures of the tracks are shown in Fig. \ref{fig:multiEletracks}.
\begin{table}[!ht]
    \centering
    \begin{tabular}{|l|l|l|l|l|l|l|l|l|l|}
    \hline
        Target & Ca & Ti & Fe & Cu & Rb & Mo & Ag & Ba & Tb \\ \hline
        k$_{\alpha}$ [keV] & 3.69 & 4.51 & 5.9 & 8.04 & 13.37 & 17.44 & 22.01 & 32.06 & 44.48 \\ \hline
    \end{tabular}
    \caption{Table summarizing the $k_{\alpha}$ emission line for different materials. Details in Appendix \ref{app:appendixB}.}
    \label{tab:kalphaenergies}
\end{table}
As can be seen from the figure, at low energies the tracks appear round, dominated by diffusion, while with increasing energy they start to exhibit a tail.
For each energy level, a set of 4000 pictures was acquired. Detailed information about the sources and X-Ray production methods is provided in Appendix \ref{app:appendixB}.

\subsection{Data analysis}
\label{sec:X-RayAnalisys}
When the detector is exposed to a monochromatic source of X-Ray, these interacting in the detector volume through a photoelectric effect produce an electron of the same energy inside the detector. The light produced at the amplification process and collected by the sCMOS is proportional to the energy of the electron recoil.
The light distribution of these electron recoils in principle should be a peaked delta, but with the energy resolution effect, the energy distribution is smeared, giving rise to a Gaussian signal. This signal will stand on top of a background of ER which is indistinguishable from the signal. Thus, the way of evaluating the signal response is to use a Maximum Likelihood Fit (MLF) composed of a signal and a background component. Moreover, this approach allows for the use of the $_{s}\mathcal{P}\textrtaill{lot}$ (Sec. \ref{sec:splot}) to extract the track shape variables for the signal component that will be used in data Montecarlo comparison (Sec. \ref{sec:datamccomparison}). This last method indeed poses his basis on the unbinned MLF technique.
\subsubsection{Data selection and fit models}
For each dataset, a very loose data selection has been made as follows:
\begin{itemize}
\setlength\itemsep{0em} 
    \item  light integral $>$ 1000 ADC to suppress the noise
    \item  track length $<$ 400 pixels to remove very long tracks (mainly cosmics)
    \item track barycenter within 800 pixels from the center of the picture to limit the vignetting effect
\end{itemize}
where the light integral is the sum of the counts in the pixels associated with the tracks, the track length is defined as the dimensions of the major axis of the cluster determined through principal component analysis, and the barycenter is calculated as the average position of all the pixels in x and y.
Considering the very low attenuation length of gammas from Ca, the majority of the interactions in the detector occurred very close to the top border of the detector. Therefore, a different geometrical selection has been applied to Ca data: $900<x<1400$ pixels and $2000<y<2200$ pixels. The x and y axes are defined as in a standard Cartesian reference frame.\\
The distribution of light integral from the monochromatic X-Ray data has been fitted with two exponential to model the background, and a Gaussian to model the signal.
A part of the background, as shown in \cite{Costa_2019}, is composed of fake clusters, with falling exponential energy distribution. The second exponential function is used instead to model the real physics background \cite{EISLER1983421}, originating from electrons and cosmics interacting in the detector.
To model the signal, depending on the case, 1 or 2 Gaussian are used. X-Rays with energies higher than 8 keV are capable of extracting additional electrons from the Copper comprising the field cage bars. For this reason, the signal coming from these sources is expected to be composed of two components, one centered at the energy of the primary X-Ray and a second one at 8 keV, which is the X-Ray emission line from Cu. 
For the $^{55}$Fe, given the high intensity of the source, there was a non-negligible probability of having two very close Iron spots in space. In such an instance, the reconstruction algorithm, merging the close spots, gives rise to a second peak at an energy double the Iron one. 
For Ti, even if the $^{55}$Fe source was shielded, a large amount of residual $^{55}$Fe was present in the pictures. In the analysis phase, the Ti Gaussian peak was observed to be completely covered by the Iron one, leading to a non-trustable result. The data from Ti will not be considered in this thesis.

\subsubsection{Unbinned maximum likelihood fit}
\label{sec:likelihood}
\label{sec:linereso}
The technique used to perform the fit of the energy spectrum is the unbinned maximum likelihood fit. This different technique, which has been developed for this thesis work, with respect to the one published in \cite{refId0}, allows for the statistical subtraction of the background contribution via the $_{s}\mathcal{P}\textrtaill{lot}$ formalism as described in section \ref{sec:splot} and in more detail in appendix \ref{app:appendixA}. Moreover, a Gaussian function to model the signal has been used instead of a Cruijff function \cite{PhysRevLett.111.151801} since the long tails of this last function would have led to the introduction of a large background component in the signal distribution obtained with the $_{s}\mathcal{P}\textrtaill{lot}$.\\
Considering a random variable x characterized by a probability density function (p.d.f.) $f(x|\Vec{\theta}\ )$, with $f$ known but any or part of the parameters of the function $\Vec{\theta}=(\theta_1,\theta_2,...,\theta_n)$ unknown, the maximum likelihood method can be used to obtain the best estimation of $\Vec{\theta}$ $\rightarrow$ $\Vec{\theta}_{best}$ from the data. The resulting p.d.f. $f(x|\Vec{\theta}_{best}\ )$ is the one which effectively describe the data distribution \cite{lista2017statistical}. 
For a set of N events, the joint probability of each $x_i$ to be within $[x_i,x_i+dx_i]$ will be equal to
\begin{equation}
    p(x_i \in [x_i,x_i+dx_i] \ \forall i) = \prod_{i=0}^{N}{f(x_i,\Vec{\theta}\ )dx_i}
\end{equation}

If the hypothesis on $f$ and the values of $\Vec{\theta}$ are correct in describing the data, it is expected a higher probability for the data measured. In the opposite case, the probability will be lower. On these bases, it can be defined a likelihood function, which represents the likelihood of a given set of data to be described by the p.d.f. with given parameters:
\begin{equation}
    \mathcal{L}(\Vec{\theta}\ ) = \prod_{i=0}^{N}{f(x_i,\Vec{\theta\ })}
\end{equation}
Once the experiment is done, the set of $\Vec{x}$, being the data acquired, are fixed numbers, thus $\mathcal{L}$ is only a function of the parameters $\Vec{\theta}$. Maximizing the likelihood on the data with respect to $\Vec{\theta}$ gives the best estimate for the parameters
\cite{Cowan:1998ji}. From an operative point of view, the negative logarithm of the likelihood function, denoted as $-\log(\mathcal{L})$, is typically utilized. This choice is made due to the wide range of values that $\mathcal{L}$ can encompass, and it is easier to work with sums than with products ($log(ab)=log(a)+log(b)$). Thus, it will be more efficient and more stable for a calculator to minimize this quantity.
In case of a single Gaussian signal ($s$) with mean $\mu$ and standard deviation $\sigma$, and 2 exponential backgrounds ($b_1$,$b_2$) with parameters $\alpha_1$ and $\alpha_2$, the respective p.d.f. for $s$ and $b$, namely $f_s$ and $f_b$ can be written as
\begin{align}
 &f_s(x_i| \mu,\sigma)=\frac{1}{\sqrt{2\pi\sigma^2}}e^{-\frac{1}{2}\left(\frac{x_i-\mu}{\sigma}\right)^2} \\  
 &f_{b}(x_i| \alpha_1,\alpha_2,c) = c\cdot \frac{1}{\alpha_1}e^{-\frac{x_i}{\alpha_1}}+(1-c)\cdot \frac{1}{\alpha_2}e^{-\frac{x_i}{\alpha_2}} \ \ with \ \  0<c<1
\end{align}

Here, the parameter $c$ takes into account the probability for an event to belong to the first type of background or the second. Using the expression $f_{b}=c \cdot f_{b1}(x_i,\alpha_1) + (1-c) \cdot f_{b2}(x_i,\alpha_2)$ guarantees the normalization of $f_b$ and allows to use an overall $N_b$ parameter for the total number of background event.
Defining $N_s$ and $N_b$ respectively the number of events for $s$ and $b$, and $\Vec{\theta}=(N_b,N_s,\mu,\sigma,\alpha_1,\alpha_2,c)$, the p.d.f. which describes the total distribution, can be written as
\begin{equation}
    f(x_i,\Vec{\theta})=P_s(N_s,N_b)f_s(x_i,\mu,\sigma) + P_b(N_s,N_b) \cdot f_{b}(x_i,\alpha_1,\alpha_2,c)
    \label{fgen}
\end{equation}
where $P_s(N_s,N_b)$ and $P_b(N_s,N_b)$ are respectively the probability for an event of being signal and background. These quantities, which will be equal to the fraction of events of one kind, normalized to the total number of events, can be expressed as 
\begin{equation}
    P_s(N_s,N_b) = \frac{N_s}{N_s+N_b} \ \ \ P_b(N_s,N_b) = \frac{N_b}{N_s+N_b}
\end{equation}
Finally, the likelihood function can be written as 
\begin{equation}
    {\mathcal{L}(\Vec{\theta} \ ) = \prod_{i=0}^{N}{\left(\frac{N_s}{N_s+N_b} \cdot \frac{1}{\sqrt{2\pi\sigma^2}}e^{-\frac{1}{2}\left(\frac{x_i-\mu}{\sigma}\right)^2} + \frac{N_b}{N_s+N_b} \cdot\left( c\cdot \frac{1}{\alpha_1}e^{-\frac{x_i}{\alpha_1}}+(1-c)\cdot \frac{1}{\alpha_2}e^{-\frac{x_i}{\alpha_2}}\right)\right)}}
\label{eq:lik}
\end{equation}
By minimizing \ref{eq:lik} on the data the values of the model parameters which better represent the data are found.
The case of two Gaussian signals can be obtained by modifying equation \ref{fgen} into 
\begin{equation}
    f(x_i,\Vec{\theta})=\sum_{j=1}^{2}P_{sj}(N_{s1},N_{s2},N_b)f_{sj}(x_i,\mu_j,\sigma_j) + P_b(N_s,N_b) \cdot f_{b}(x_i,\alpha_1,\alpha_2,c)
\end{equation}
with $\Vec{\theta}=(N_b,N_{s1},N_{s2},\mu_1,\sigma_1,\mu_2,\sigma_2,\alpha_1,\alpha_2,c)$, and $P_{s1/s2/b}(N_{s1},N_{s2},N_b)=\frac{N_{s1/s2/b}}{N_{s1}+N_{s2}+Nb}$.

\begin{figure}
    \centering
    \includegraphics[width=1\linewidth]{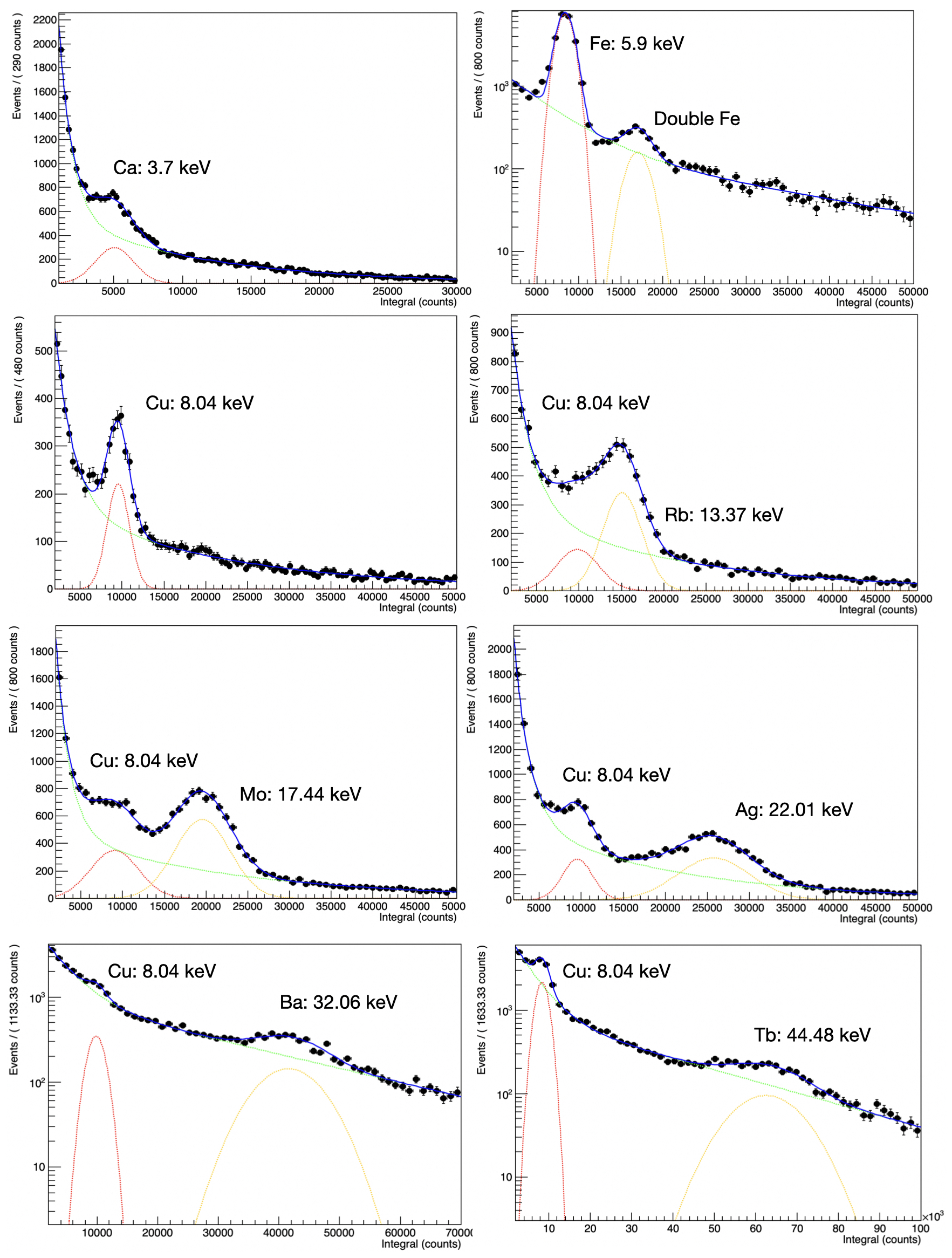}
    \caption{The results of the fit are shown from top to bottom and from left to right in order of energy: Ca, Fe, Cu, Rb, Mo, Ag, Ba, and Tb. In the plots, the black dot represents the data in a binned representation, the green curve is the background function, the red and orange dotted curves are the signal, while the blue line is the sum of $s$ and $b$.}
    \label{fig:fits}
\end{figure}

\label{sec:analysis}
For the analysis, the modeling, and the fit, the $RooFit$ toolkit has been employed \cite{Verkerke:2003ir}. The minimizer used to find the minimum of $-log(\mathcal{L})$ is MIGRAD from the MINUIT package of RooFit \cite{James:1994vla}.
The results of the fit are shown in Fig. \ref{fig:fits}, and the parameters are reported in Tab. \ref{tab:fitRes}.

\begin{table}
    \centering
    \adjustbox{max width=\textwidth}{
    \begin{tabular}{|l|l|l|l|l|l|l|l|}
    \hline
        Source  & $\alpha_1$ & $\alpha_2$ & $c$ & $\mu_1$ & $\sigma_1$ & $\mu_2$ & $\sigma_2$ \\ \hline
        Ca & -9.5(4)$\times$10$^{-4}$ & -9.7(2)$\times$10$^{-5}$ & 0.247(8) & 5.06(8)$\times$$10^{3}$ & 1.39(7)$\times$$10^{3}$ & x & x \\ \hline
        Fe & -1.9(1)$\times$10$^{-4}$ & -3.8(4)$\times$10$^{-5}$ & 0.56(3) & 8.336(7)$\times$$10^3$ & 9.53(7)$\times$10$^2$ & 1.70(1)$\times$10$^4$ & 1.4(1)$\times$10$^3$ \\ \hline
        Cu & -4.5(4)$\times$10$^{-4}$ & -5.1(2)$\times$10$^{-5}$ & 0.21(2) & 9.54(7)$\times$$10^3$ & 1.34(8)$\times$10$^2$ & x & x \\ \hline
        Rb & -4.2(5)$\times$10$^{-4}$ & -4.9(3)$\times$10$^{-5}$ & 0.22(2) & 9.8(8)$\times$$10^3$ & 2.49(4)$\times$$10^3$ & 1.51(3)$\times$$10^4$ & 2.3(2)$\times$$10^3$ \\ \hline
        Mo & -6.0(7)$\times$$10^{-4}$ & -4.7(3)$\times$$10^{-5}$ & 0.20(3) & 9.0(2)$\times$$10^3$ & 2.8(3)$\times$$10^3$ & 1.97(1)$\times$$10^4$ & 2.9(3)$\times$$10^3$ \\ \hline
        Ag & -5.3(4)$\times$10$^{-4}$ & -5.4(2)$\times$10$^{-5}$ & 0.20(2) & 9.5(1)$\times$10$^3$ & 1.8(1)$\times$10$^3$ & 2.57(1)$\times$10$^4$ & 4.3(2)$\times$10$^3$ \\ \hline
        Ba & -2.4(1)$\times$10$^{-4}$ & -3.82(9)$\times$10$^{-5}$ & 0.36(2) & 9.8(2)$\times$10$^3$ & 1.4(2)$\times$10$^3$ & 4.15(3)$\times$10$^4$ & 5.28(4)$\times$10$^3$ \\ \hline
        Tb & -2.06(6)$\times$10$^{-4}$ & -3.16(8)$\times$10$^{-5}$ & 0.47(1) & 8.38(4)$\times$$10^3$ & 1.54(5)$\times$10$^3$ & 6.25(5)$\times$10$^4$ & 8.3(6)$\times$10$^3$ \\ \hline
    \end{tabular}}
    \caption{In the table the parameters of the fits are reported. In order the values are the two coefficiente of the exponentials used to model the background, the relative weight of the two background components, and the mean and the sigma od the gaussians fitting the signals. The numbers between parenthesis represent the error on the last digit.}
    \label{tab:fitRes}
\end{table}

\begin{figure}
    \centering
    \includegraphics[width=1\linewidth]{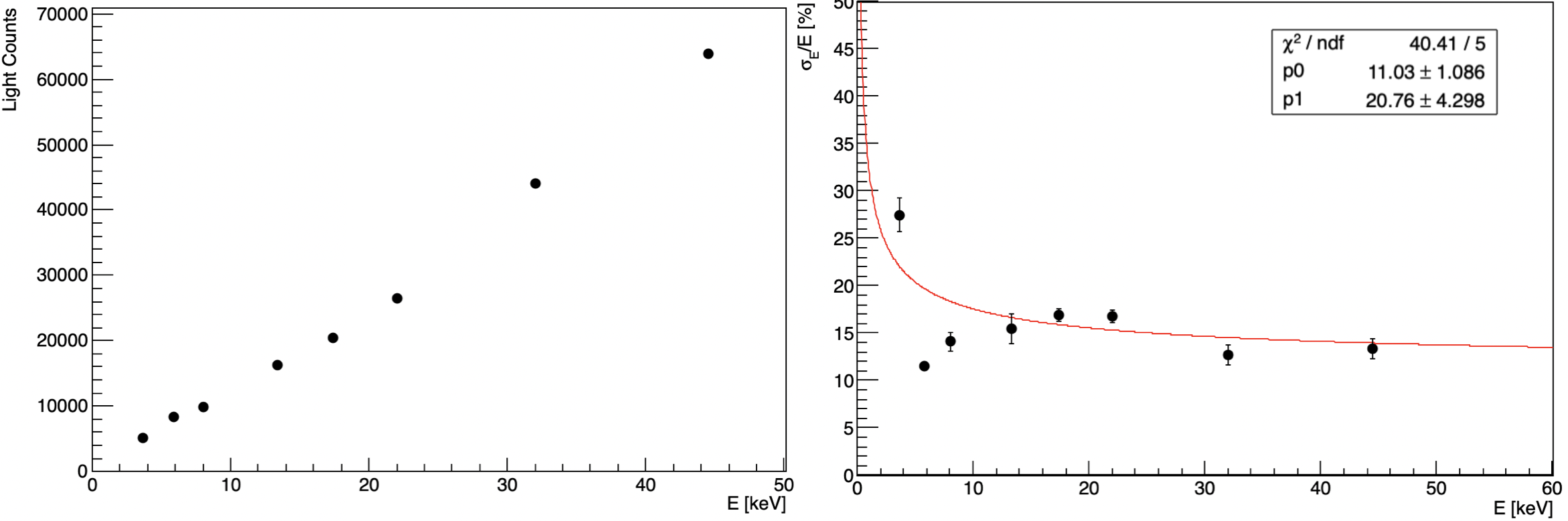}
    \caption{Left: Plot of the mean value of the light response as a function of the energy. Right: detector energy resolution, reported in percentage ($\sigma_E/E$), as a function of the energy.}
    \label{fig:linereso}
\end{figure}

Starting with a mixed sample of signal (composed of X-Rays with known energy) and background, the MLF technique has provided an estimate of the parameters characterizing LIME response to the signal in terms of mean count response ($\mu$) and count dispersion ($\sigma$). By relating the estimate of the light integral to the true energy of the X-Rays, it is possible to characterize the response and the energy resolution.  
From the fit results, the response in the energy of the detector has been characterized by plotting the mean count response as a function of the energy. Moreover, the percentage energy resolution of the detector has been reported as the ratio of the Gaussian standard deviation divided by its mean value. The plots of the detector response and energy resolution as a function of the energy are reported in Fig. \ref{fig:linereso}.

As described in \ref{sec:overgroundstudies}, the saturation affect most tracks of electron recoil at lower energy.
To model the resolution and interpolate its value at higher energies, the data has been fitted with $\sigma_{E}/E = p_0 + p_1/\sqrt{E}$. The point at 6 keV has been excluded from the fit since the parameter of the reconstruction code has been optimized for that energy. This could potentially bias the fit result.

%% file: chapters/MANGOProtRD.tex

\chapter{Negative Ions Drift studies}
\label{sec:NID}
The doctoral research program of the author of this thesis was funded by the INITIUM project under the European Union’s Horizon 2020 research and innovation program from the European Research Council (ERC) grant agreement No. 818744. Therefore, as part of his thesis, the author contributed to the implementation of Negative Ions Drift operation with the MANGO prototype at LNGS working on the hardware side, and to the data analysis for characterizing the Negative Ions Drift operations.\\
One significant limitation of large gaseous TPCs is that the electron cloud, drifting to the amplification plane, often degrades the intrinsic characteristics of the initial ionization track created by the particle interacting with the gas, mainly due to diffusion. There are two main contributions to the diffusion: one occurring during the particle's drift within the sensitive TPC volume, and the other during the subsequent amplification processes. A method used to reduce transverse diffusion (along the GEM plane) is to apply a uniform magnetic along the drift direction \cite{blum2008particle}. However, within the framework of rare events searches, the scale of the experiments would necessitate magnets with excessively large dimensions, strongly increasing the complexity of the experiment. Additionally, the longitudinal diffusion would be worsened, posing further challenges.
Instead, the potentiality to decrease the diffusion by introducing a small percentage of electronegative gas to switch to Negative Ions Drift (NID) operation has been extensively studied, especially in these last years with the advancement of TPC for DM detection. This peculiar TPC operating mode enables the creation of detectors featuring longer drift distances, while at the same time showing improved tracking capabilities compared to electron drift (ED) operations. Moreover, it offers advantages in terms of volume scalability and reduced background contamination. Typically, negative ions have been employed at low pressure ($<$100 tor), with few exceptions close to atmospheric pressure \cite{Baracchini_2018NID,LIGTENBERG2021165706}, but never at atmospheric pressure with an optical readout.\\
The results of NID operation at atmospheric pressure with a TPC optically readout obtained are reported in this thesis, together with in-deep transverse diffusion studies slightly below atmospheric pressure.
In this chapter, the operation with Negative Ions Drift and the advantages carried by operating a TPC in this way will be illustrated (Sec. \ref{sec:NIDoperations}), followed by an overview of the state of the art of NID operations (Sec. \ref{sec:Stateofart}). Subsequently, the results of NID operations at atmospheric pressure with optical readout obtained by the CYGNO collaboration will be illustrated (Sec. \ref{sec:NIDAtmosf}), followed by detailed studies on ions transverse diffusion performed on sCMOS pictures slightly below atmospheric pressure (Sec. \ref{sec:imAnalysis}).

\section{Negative ion drift operations}
\label{sec:NIDoperations}
The addition of an electronegative species to the gas mixture has the potential to modify the operation of a gas detector with a drift field used to transport the primary charge to the amplification plane, thereby modifying the traditional operation of a Time Projection Chamber \cite{Martoff:2000wi,Ohnuki_2001}. Specifically, free electrons can be promptly captured by these electronegative dopants, typically in the range of $\mathcal{O}$(10 $\mu m$), resulting in the formation of negative ions. Guided by the applied electric field, these anions drift towards the anode, assuming the role of effective charge carriers in place of electrons. Upon reaching the anode, the intense electric fields stimulate the release of the additional electron, initiating a standard electron avalanche. This modified TPC operation offers two key advantages: a reduction in transversal diffusion, leading to better tracking performances, and the ability to fiducialize the detector.\\

\textbf{Diffusion reduction} A key benefit of Negative Ion Drift operation is the minimized diffusion during the drift. This is because the diffusion coefficient of electrons and ions in gas $\xi_T$ is proportional to $\xi_T\propto\sqrt{\bar{E}}$, where $\bar{E}$ is the average energy of the drifting particle \cite{blum2008particle,Ohnuki_2001}. Electrons are significantly lighter than gas molecules by orders of magnitude $>10^4$, leading to an inefficient momentum transfer during collisions. Consequently, when electrons experience collisions with particles in the medium, their outgoing direction is randomized, and they preserve most of the kinetic energy acquired from the electric field in the time between two consecutive collisions. Conversely, the anion masses are comparable to the one of the gas molecules, enabling them to thermalize and disperse the energy acquired from the drift field in an efficient way. By employing this approach, diffusion can be minimized to the thermal limit \cite{blum2008particle}, resulting in diffusion orders of magnitude smaller \cite{DRIFT:2014bny}. The approximate formula in the low field limit describing the transversal diffusion $\xi_T$ of a charged cloud traveling a distance $L$ under the influence of an electric field $E$ is \cite{blum2008particle}:
\begin{equation}
    \xi_T^2 = \frac{2k_BTL}{eE}
    \label{eq:diffequation}
\end{equation}
where $k_B$ is the Boltzmann constant, $T$ is the temperature of the gas in Kelvin, and $e$ is the electric charge. Minimizing diffusion significantly enhances directional capabilities and background rejection, as the track topology is less affected by charge dispersion. The maximum drift length in a Time Projection Chamber, essential to increase the detector mass without increasing the detector price and radioactivity contribution, is directly linked to diffusion properties. At greater distances, the smearing effect becomes pronounced, causing all track types to appear as blurred, large spots. Negative ions can therefore allow for an increase in the field cage length of the detector without compromising performance compared to regular Electron Drift operations.\\

\textbf{Fiducialization} The fiducialization refers to the capability to identify the 3D position of an event in a detector. In a TPC featuring an optical readout, the most challenging position to measure is along the drift direction $z$. Through Negative Ion Drift operation, it has been proven that multiple species of negative ion carriers, each with different masses, can be generated in the primary electrons capture process \cite{Snowden-Ifft:2013vua}. Among these various species, one primarily contributes to the charge transport, making the others referred to as “minority carriers”. Given that anion mobility is influenced by mass, and all these distinct species originate in proximity to the initial position of the primary charges, the discrepancy in the time of arrival of various anions effectively can be used as a means to measure the absolute position of the event along the z-axis. The third coordinate can be determined using the formula: 
\begin{equation}
    z = \frac{v_m v_M}{v_m+v_M}\Delta t
\end{equation}
where $v_m$ represents the velocity of the minority carrier, $v_M$ is the velocity of the majority one, and $\Delta t$ is the arrival time difference between the two carriers.\\


\section{State of the art}
\label{sec:Stateofart}
The exploration of the potential utilization of Negative Ion Drift operation in Time Projection Chambers to mitigate diffusion was initiated by Martoff, Snowden, Ohnuki, and Spooner in 2000 \cite{Martoff:2000wi}. This involved the introduction of CS$_2$ electronegative gas to Xenon or Argon within a Multi Wire Proportional Chamber (MWPC) at very low pressure ($\sim$ 40 Torr) \cite{MARTOFF20003551NID}. The reported diffusion coefficients in this paper were remarkably low, signifying a notable reduction in diffusion compared to ED operations. Subsequent investigations into the negative ion properties of CS$_2$ in conjunction with various gases, such as He, CF$_4$, and Ar at pressures of 40 Torr with MWPC, confirmed the ionic nature of the charge carriers with reduced diffusion \cite{MARTOFF2005551bar, Ohnuki_2001,ALNER2004644}. The NID operation has also been demonstrated at close to atmospheric pressure in MWPCs, in a gas mixture with Helium \cite{Dion:2009bti}, and recently at 1030 mbar with a gas mixture of Ar:iC$_4$H$_{10}$:CS$_2$ with a GridPix charge readout \cite{LIGTENBERG2021165706}. Studies have been conducted using Nitromethane (CH$_3$NO$_2$) in combination with CO2 as a NID carrier at low pressure ( $\sim$ 70 Torr) \cite{Dion:2009bti}.\\
The breakthrough in Negative Ion Drift operation within Time Projection Chambers took place with the successful operations employing SF$_6$. This breakthrough was realized in 2016, wherein pure SF$_6$ at pressures between 20 and 40 Torr was utilized in a TPC equipped with a 400 $\mu$m thick Gas Electron Multiplier and charge readout \cite{Phan_2017}. This setup not only achieved negative ion drift operation but also demonstrated the feasibility of fiducialization with minority carrier transport.
This accomplishment is highly significant, given that SF$_6$ is inert and non-toxic, in contrast to the highly toxic nature of CS$_2$ and Nitromethane. Moreover, SF$_6$ is easily manageable, widely used in industry, and commonly employed in gaseous detectors due to its ability to capture electrons, thereby preventing large discharges, as observed in Resistive Plate Chambers (RPCs).
Further studies with pure SF$_6$ at 20 Torr were conducted by the NEWAGE collaboration, utilizing Gas Electron Multipliers and charge readout. These investigations demonstrated fiducialization with minority carriers, achieving a 130 $\mu$m resolution in z for alpha particles \cite{Ikeda_2020}. Additionally, NID operation with SF$_6$ was validated using triple thin GEMs and TimePix2 charge readout \cite{Baracchini_2018NID}. In these studies, pure SF$_6$ was operated at pressures of 75, 100, and 150 Torr, with reduced mobilities measured independently of the pressure. SF$_6$ was also employed in mixtures, including Ar:CO$_2$:SF$_6$ (192:85:93) Torr and He:CF$_4$:SF$_6$ (360:240:10) Torr. Lastly, NID operations have been successfully demonstrated possible with optical readout in a gas mixture of CF$_4$:CS$_2$ at a pressure range of 50-150 Torr, utilizing a glass GEM for amplification and a sCMOS for light detection \cite{NIDOpticalLowP}.\\
In the subsequent sections, we present the results of NID operations conducted at nearly atmospheric pressures using a gas mixture of He:CF$_4$:SF$_6$ (59.2:39.2:1.6), employing a triple thin GEM amplification system and optical readout. It is noteworthy that this is the first time that negative ion operations are successfully conducted at nearly atmospheric pressure with optical readout. A recap of the status of NID operations highlighting the different pressures and readout used is shown in table \ref{tab:NIDOPerations}.

\begin{table}[h]
    \centering
    \scalebox{0.75}{
    \begin{tabular}{|l|l|c|}
    \hline
         Pressure & Charge readout & Optical readout \\ \hline
         &  \pbox{30cm}{$\bullet$\ Concept demonstrated in 2000\\ at 40 Torr CS$_2$ with MWPC\cite{MARTOFF20003551NID}\\} &  \\ 
        Low & \pbox{20cm}{$\bullet$\ Pioneered in a actual experiment by DRIFT\\ with CS$_2$:CF$_4$:0$_2$ at 40 Torr with MWPC\cite{snowdenifft2013discovery}\\} & \pbox{20cm}{$\bullet$\ 50-150 Tor CF$_4$:CS$_2$ with\\ glass GEM and CMOS \cite{NIDOpticalLowP}} \\ 
        ~ & \pbox{20cm}{$\bullet$\ 20-40 Tor pure SF$_6$ in 2017 with THGEM \cite{Phan_2017}\\} & ~ \\ 
        ~ & \pbox{20cm}{$\bullet$\ 20 Tor pure SF$_6$ with THGEM-multiwire\\ and mUPIC in 2020\cite{Ikeda_2020}}  & ~ \\ 
        ~ & ~ & ~\\ \hline
        & \pbox{30cm}{$\bullet$\ Demonstrated in 2010 in He:CS$_2$ and\\ CO$_2$:Ne:CHNO$_2$ with GEMs and MWPC\cite{Dion:2009bti}\\} & ~ \\ 
        Atmospheric & \pbox{30cm}{$\bullet$\ In 2017 at 610 Tor of He:CF$_4$:SF$_6$\\ with GEMs and TimePix2 \cite{Baracchini_2018NID}\\} & THIS WORK \\ 
        ~ &  \pbox{30cm}{$\bullet$\ In 2021 in Ar:iC$_4$H$_{10}$:CS$_2$ \\ with GridPix (Ingrid + TimePix3\cite{LIGTENBERG2021165706})} & ~ \\ \hline
    \end{tabular}
    }
    \caption{Table depicting the status of NID operations categorized by pressure of the gas and readout used.}
    \label{tab:NIDOPerations}
\end{table}

\section{NID at atmospheric pressure with optical readout}
\label{sec:NIDAtmosf}
\begin{figure}
    \centering
    \includegraphics[scale=.4]{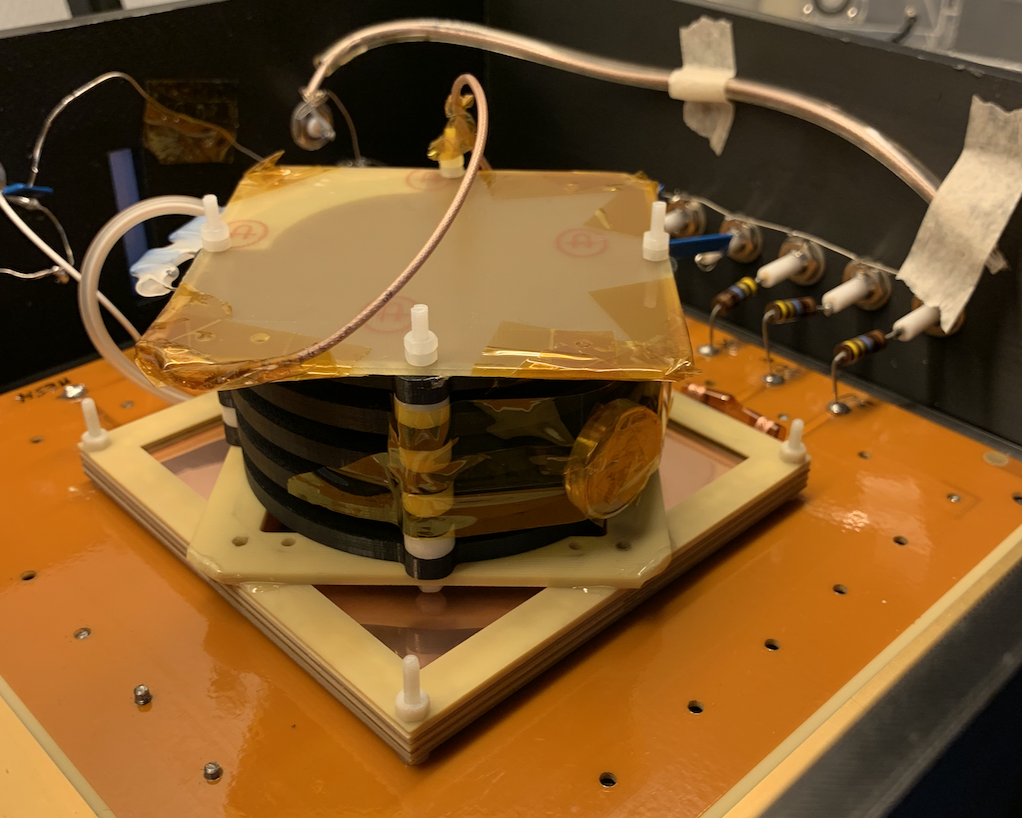}
    \caption{Picture of the MANGO detector with the 5 cm field cage setup. The disk on the side of the field cage is the alpha source employed.}
    \label{fig:MANGOFcSmall}
\end{figure}

The MANGO prototype, detailed in Sec. \ref{sec:MANGO}, was utilized for studying NID operation at atmospheric pressure with optical readout. The detector was set up in the configuration with a 5 cm drift length, corresponding to the maximum feasible within the gas-tight acrylic internal vessel. The operations have been performed at the atmospheric pressure of LNGS measured to be 900 $\pm$ 7 mbar (corresponding to 684 $\pm$ 5 Torr). In the detector two gas mixtures: He:CF$_4$ 60:40 (for ED operations) and He:CF$_4$:SF$_6$ 59.2:39.2:1.6 (for NID operations) have been utilized, the same employed in \cite{Baracchini_2018NID}, in continuous flux mode. A $^{241}$Am source emitting 5.485 MeV alpha particles was positioned between the field cage rings to generate tracks within the detector active volume at a distance from the GEM of 2.5 cm. A picture of the detector in this configuration with the alpha source installed is shown in \ref{fig:MANGOFcSmall}
The signals from the PMT and the output of the charge-sensitive preamplifier (CAEN A422A2 with 300 $\mu$s decay time), connected to the bottom electrode of GEM3 (the last GEM, counting from the cathode side), were readout using a highly performing Teledyne LeCroy oscilloscope.

\subsection{Anion velocity and mobility studies}
\begin{figure}
    \centering
    \includegraphics[width=1\linewidth]{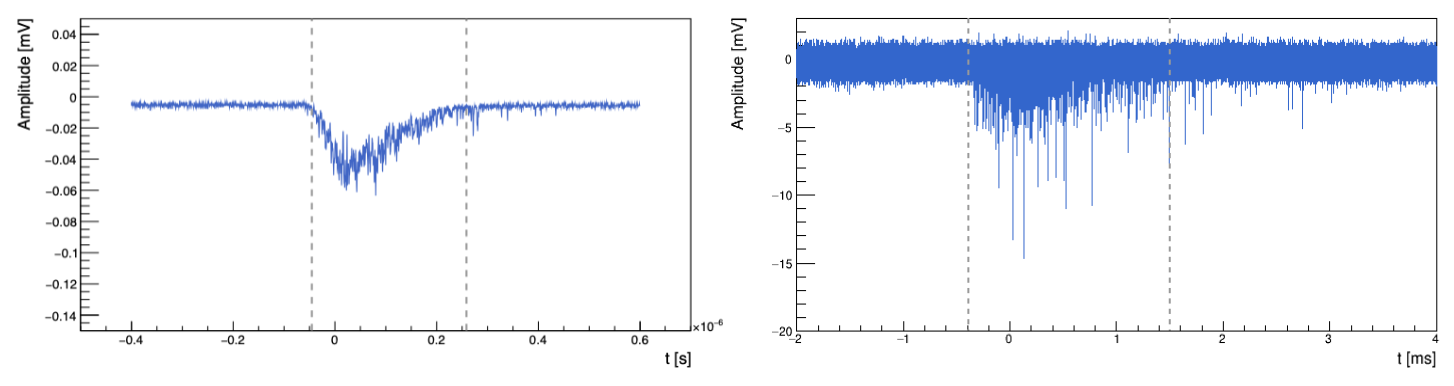}
    \caption{Comparison of PMT waveforms depicting tilted alpha tracks acquired with the ED (left) and NID (right) mixtures with a drift field of 400 V/cm. It is worth noting the very different timescale.}
    \label{fig:AlphaWave}
\end{figure}

\begin{figure}
    \centering
    \includegraphics[scale=.23]{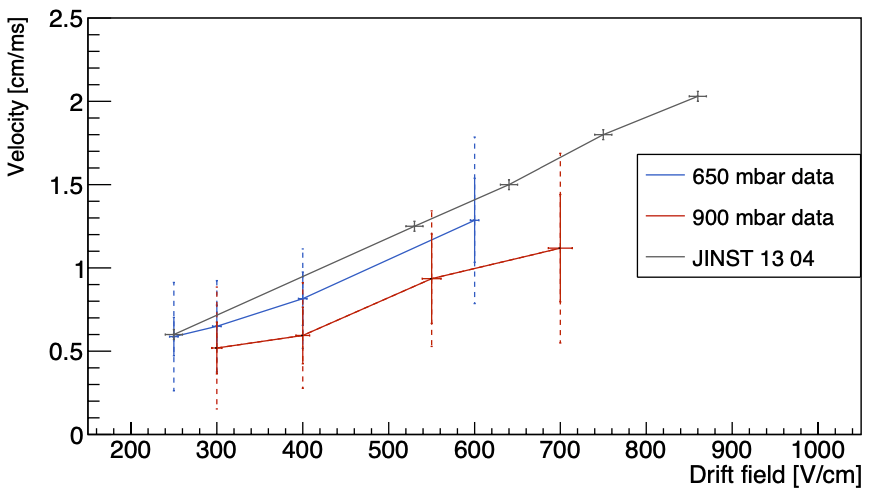}
    \includegraphics[scale=.20]{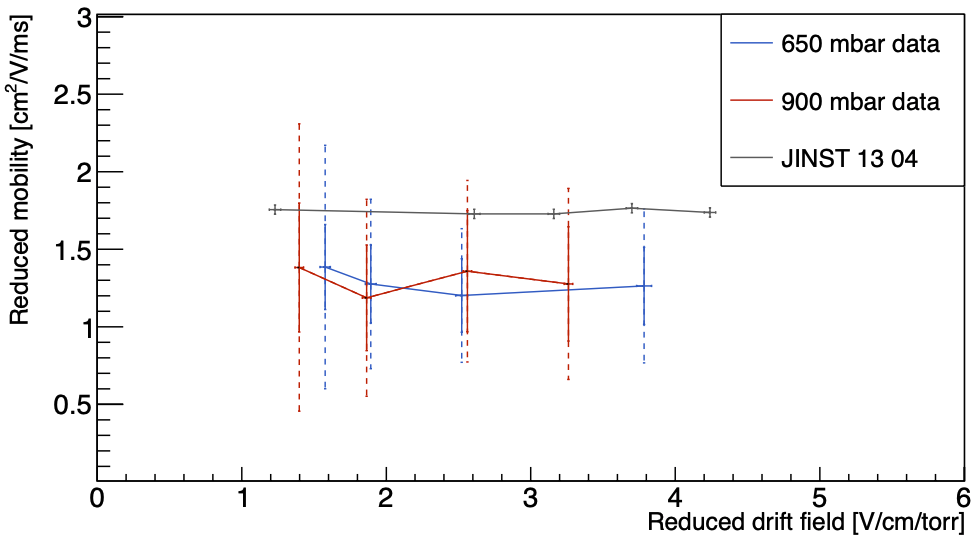}
    \caption{Left: plot of the drift velocity as a function of the drift field applied for two different operating pressures, and the comparison with the work done in \cite{Baracchini_2018NID}. Right: plot of the reduced mobility calculated as a function of the reduced field for two different operating pressures, and the comparison with \cite{Baracchini_2018NID}.}
    \label{fig:NIDVelocity}
\end{figure}

To determine the drift velocity and mobility of the gas mixtures under investigation, the PMT signal induced by alpha particles has been analyzed. Drift fields of 300 V/cm, 400 V/cm, 500 V/cm, and 700 V/cm were tested. In the case of the ED setup, the GEMs were operated at 310 V each, while for the NID setup, GEM1/GEM2/GEM3 were operated at 550/545/540 V, respectively.\\
To measure the drift velocity, given the lack of information on the instant when an alpha particle ionizes the gas in the MANGO configuration, the time length $\Delta T$ of the PMT waveforms produced has been analyzed. Specifically, the time extension of the signal is given by the product of the charge drift velocity and the extension of the alpha track projected in the direction of the drift field, $\Delta z$. With the ED gas mixture, by measuring $\Delta T$, the $\Delta z$ spanned by the track can be calculated since the electron drift velocity is known (Sec. \ref{sec:gasmixture}). By assessing $\Delta T$ with the He:CF$_4$:SF$_6$ 59.2:39.2:1.6 gas mixture, the anion drift velocity can be determined. In this process, the source has been not moved from its original position. An example of the waveform of two tilted alpha tracks for NID and ED is shown in figure \ref{fig:AlphaWave}. As far as the writer is aware, this marks the first observation of a NID signal using a PMT.
The PMT signals for the ED configuration are captured with a sampling rate of 400 ps/pt and a negative edge threshold set at 40 mV on the PMT waveform amplitude. To estimate the duration of the PMT waveform in the ED mode, the start (and end) of the signal are defined where the absolute amplitude exceeds (falls below) 3 times the root mean square (RMS) noise of the waveform, calculated from the first 100 bins lacking any signal. By averaging the time extension of the ED PMT waveform $\bar{\Delta T}$ and considering the simulated drift velocity $v_{drift}^{ED}$ for He:CF$_4$ 60:40 at 900 mbar, the average longitudinal extent $\bar{\Delta z}$ covered by the alpha tracks in the drift direction in this setup is determined to be (0.7 $\pm$ 0.2) cm.\\ The prompt response of the PMT (Sec. \ref{sec:PMT}) connected with the particularly slow drift velocity of NID anions, results in a notably sparse signal, differing significantly from that of the ED (Fig. \ref{fig:AlphaWave}). These NID signals, comprising numerous small amplitude peaks ranging in amplitude from 5 to 30 mV and lasting a few nanoseconds, extend over several milliseconds, depending on the applied drift field. It has been hypothesized that each of these minor peaks corresponds to the arrival of one or a few primary ionization clusters at the amplification plane, identifiable individually due to the slow drift velocity of the anions. Although delving into this subject is beyond this thesis focus, it's worth noting how this characteristic could facilitate cluster counting, an analytical method known for its potential to enhance energy resolution and particle identification capabilities \cite{FISCHLE1991202}, particularly in Helium-based detectors \cite{Cataldi:1996mz}.\\
As each primary cluster signal closely resembles the single photoelectron background signal, a simple threshold-based trigger applied to the PMT proves ineffective in capturing NID waveforms. To address this issue, NID events are recorded by applying a 250 mV positive edge threshold to the GEM preamplifier signal output. This threshold, coupled with the long and continuous alpha track generated by the preamplifier's extended decay time, facilitates easy triggering. Analyzing NID waveforms presents a more complex challenge due to the difficulty in delineating the signal's start and end points. Each peak in the waveform resembles noise, and their spacing is on the order of microseconds. Consequently, to analyze this peculiar waveforms, an original algorithm was developed to manage the irregular and discrete nature of the NID signals. The RMS of the electronic noise is assessed based on the first 500 bins of the waveform, which are expected to be devoid of any signal. From the original waveforms, only peaks exceeding 6 times the RMS, equivalent to approximately 2.5 mV, are retained for subsequent analyses. These selected peaks are utilized to construct a histogram of the rebinned waveform, comprising a total of 150 bins covering a time extension of 10 ms. Within this rebinned histogram, the beginning (end) of the signal is defined when two consecutive bins surpass (fall below) 10 mV. The systematic uncertainties linked to this analytical approach are determined by varying the number of bins in the rebinned histogram from 150 to 250 and modifying the threshold on the two consecutive bins from 5 mV to 15 mV. Using the average measured time extension of NID signals and considering the known $\Delta z$ extent of the alpha setup determined from the ED data, the NID drift velocity is derived as a function of the applied drift field. The left panel of Figure \ref{fig:NIDVelocity} illustrates this relationship for the four applied drift fields. Since the systematic uncertainties, as determined above, may not conform to standard error theory, the complete analysis to determine the drift velocity is repeated under different conditions of binning and threshold, as described for estimating the systematic uncertainties. A confidence interval of 68\% is employed to estimate the systematic effects from the distribution of the obtained drift velocities.
The measured drift velocities of approximately 1 cm/ms clearly indicate that the He:CF$_4$:SF$_6$ 59.2:39.2:1.6 gas mixture at 900 mbar enables NID operations. The findings from a corresponding analysis conducted with data collected at 650 mbar (as part of the diffusion studies on sCMOS images, see Sec. \ref{sec:imAnalysis}) are presented in the same figure and lead to similar conclusions. The reduced mobility, derived from these velocities, is compared with the findings reported in \cite{Baracchini_2018NID} concerning a He:CF$_4$:SF$_6$ 360:240:10 Torr gas mixture, measured with pixel charge readout, as illustrated in the right panel of Figure \ref{fig:NIDVelocity}. The comparison highlights the consistency of the three measurements within their respective uncertainties. The larger systematic uncertainties in the measurements presented here, in contrast to those of \cite{Baracchini_2018NID}, are attributed to the methodological differences. Specifically, the latter study benefits from the use of a beam trigger signal to precisely define the event start, a technique inherently offering greater precision than relying on the extent of alpha tracks in the drift direction.

\section{NID diffusion measurement at 650 mbar}
\label{sec:imAnalysis}
\begin{figure}
    \centering
    \includegraphics[width=0.75\linewidth]{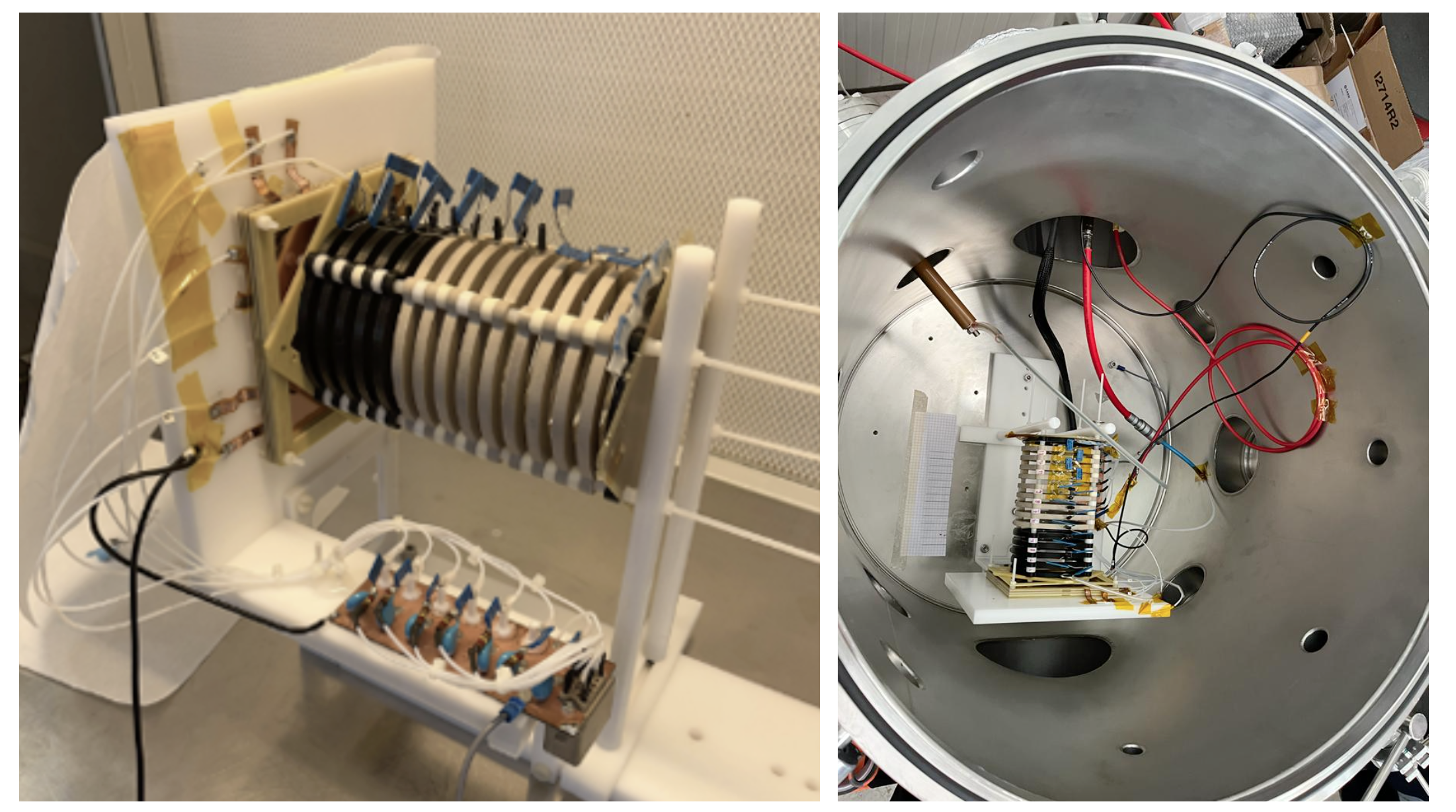}
    \caption{Picture showcasing the MANGO in the keg configuration. On the left, a close-up of the GEM stack, featuring a 15 cm long field cage and structural support. On the right, the identical setup placed within the 150 liter stainless steel vacuum vessel.}
    \label{fig:Mangokeg}
\end{figure}

The limited maximum drift length of 5 cm achievable within the standard MANGO gas-tight acrylic internal vessel is insufficient for conducting diffusion measurements during drift, as diffusion scales with the square root of the distance traveled. To address this limitation, the MANGO amplification stage was outfitted with a 15 cm long field cage, with an identical design with respect to the 5 cm version described in \ref{sec:MANGO}. Additionally, the TPC structure was installed in an existing 150 L stainless steel vacuum vessel, the same one used for the charge pixel readout measurement reported in \cite{Baracchini_2018NID}. A picture of MANGO with a detail on the 15 cm field cage, together with a picture of it inside the keg, is shown in figure \ref{fig:Mangokeg}.
In this new setup, called MANGOk (MANGO in a keg), the photodetectors (i.e. the sCMOS camera and the PMT) are positioned outside the vacuum vessel facing the MANGOk amplification plane through a quartz window with 90\% transparency. The sCMOS camera is situated approximately 26.6 $\pm$ 0.4 cm away from the last GEM. With this arrangement, the camera captures an area of 14.1$\times$14.1 cm$^2$, resulting in an effective pixel size of 61$\times$61 $\mu$m$^2$. Consequently, there is about a threefold reduction in the solid angle covered by the camera sensor with respect to the standard MANGO setup, and thus the measured light yield.\\
To operate in NID mode with a sufficiently high light yield at the sCMOS sensor for measuring diffusion over large distances even with weak electric drift fields, the operational pressure is reduced to (650 $\pm$ 1) mbar (corresponding to (494.0 $\pm$ 0.7 ) Torr) to enhance the gain. Following the vessel sealing operation, the air inside it is pumped out to reach a pressure of less than 0.1 mbar using a dry scroll vacuum pump. After a period of vacuum, the vessel is filled with the selected gas mixture for each measurement. The GEMs are set to operate at 310 V each in the ED case and at 535/530/525 V for GEM1/GEM2/GEM3 in the NID case. These operating GEM voltages are selected to achieve a comparable light yield for both mixtures within this setup. The utilization of the $^{241}$Am source mirrored the procedures outlined in Sec. \ref{sec:NIDAtmosf}; however, this time, the source underwent additional collimation to minimize the dispersion of alpha particles in the longitudinal direction. The acquisition and triggering strategy remained identical to the one detailed in Sec. \ref{sec:NIDAtmosf}. 

\subsection{Diffusion measurement}
\begin{figure}
    \centering
    \includegraphics[width=0.75\linewidth]{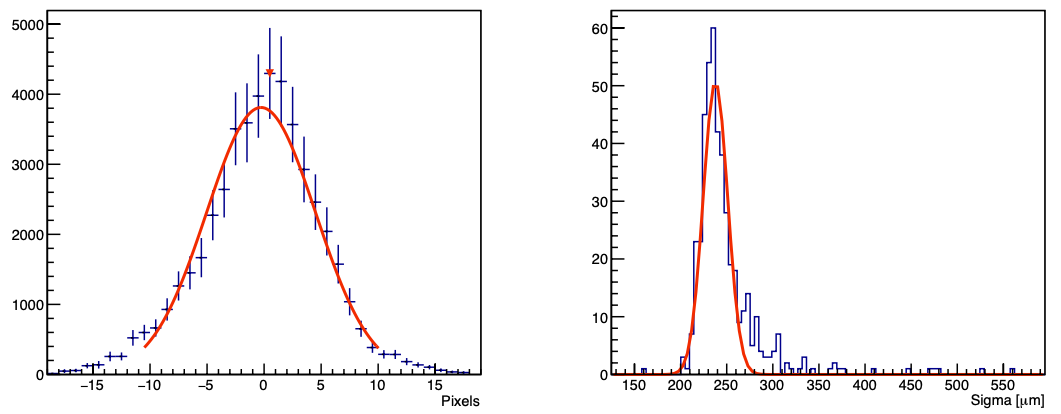}
    \caption{Left: example of transverse profile of an alpha particle that meets the selection criteria outlined in the text, with a Gaussian fit overlaid. Right: distribution of the sigma values obtained from the Gaussian fits.}
    \label{fig:AlphaTranProf}
\end{figure}

To assess the diffusion characteristics of the gas mixtures utilized, data collection is conducted at six distinct drift distances (2.5, 3.5, 4.5, 6.5, 9.5, and 12.5 cm from the GEMs), each under varying drift fields, in the range from 150 V/cm to 600 V/cm, for both the ED and NID gas mixtures at 650 mbar. The collimation of the source ensures the generation of alpha tracks that traverse the plane as perpendicularly as possible to the drift direction. Analysis of the sCMOS images capturing these alpha tracks yields information on diffusion. Alpha particles, owing to their high energy, typically travel along nearly linear paths until they exhaust their energy. Consequently, they leave behind highly intense and straight tracks in an sCMOS image. Given that the intrinsic transverse dimension of the primary electrons produced by an alpha's passage through the gas is on the order of a few tens of microns, the measured transverse profile dimension of the alpha track offers insights into the diffusion during drift and at the amplification stage.\\
The sCMOS tracks have been reconstructed with the standard CYGNO reconstruction (Sec. \ref{chap:reco}). Alpha tracks have been selected by imposing criteria such that the track length exceeds 1.2 cm and the ratio of track width to length is below 0.3, aimed at excluding noise and improperly reconstructed events. The endpoints (i.e., start and end) of the tracks meeting the above criteria are removed from the calculation of the track transverse profile to prevent potential interference from multiple scattering within the gas (end of the track) and irregularities at the detector boundaries due to field effect (start of the track). This is achieved by truncating the tracks within a 0.61 cm radius centered at the intensity barycenter, computed from the complete original track. A Principal Component Analysis (PCA) is utilized on the intensity-weighted array of pixels belonging to this truncated selection to ascertain the principal axes of the track. Subsequently, the pixels of the truncated track are projected onto the major (minor) axis to generate a one-dimensional longitudinal (transverse) profile. A one-dimensional peak detection algorithm utilizing the TSpectrum ROOT class \cite{BRUN199781} with an amplitude threshold of 10\% is utilized on the transverse profile of the truncated tracks. The criterion for selection requires the number of identified peaks to be exactly one, thereby filtering out any overlapping tracks and ensuring the selection of single tracks. The transverse profile of the truncated tracks that meet this criterion is subjected to fitting with a Gaussian distribution. Further elimination of multiple overlapping tracks (resulting from pileup or misreconstructed images) is achieved by imposing a criterion where the $\chi^2$/nDOF of the Gaussian fit is required to be less than 5. Figure \ref{fig:AlphaTranProf}, left plot, illustrates an example of such a transverse profile of an alpha particle, with the Gaussian fit overlaid.
The distribution of $\sigma$ values derived from Gaussian fitting applied to these chosen cropped tracks is then analyzed by fitting them with a Gaussian distribution. The mean of this Gaussian is thus used as a measurement of the track diffusion. An illustration of the $\sigma$ distribution is presented in the right panel of Figure \ref{fig:AlphaTranProf}.

\subsection{Estimation of the transverse diffusion}
\label{sec:TrasnvDiff}
\begin{figure}
    \centering
    \includegraphics[width=0.9\linewidth]{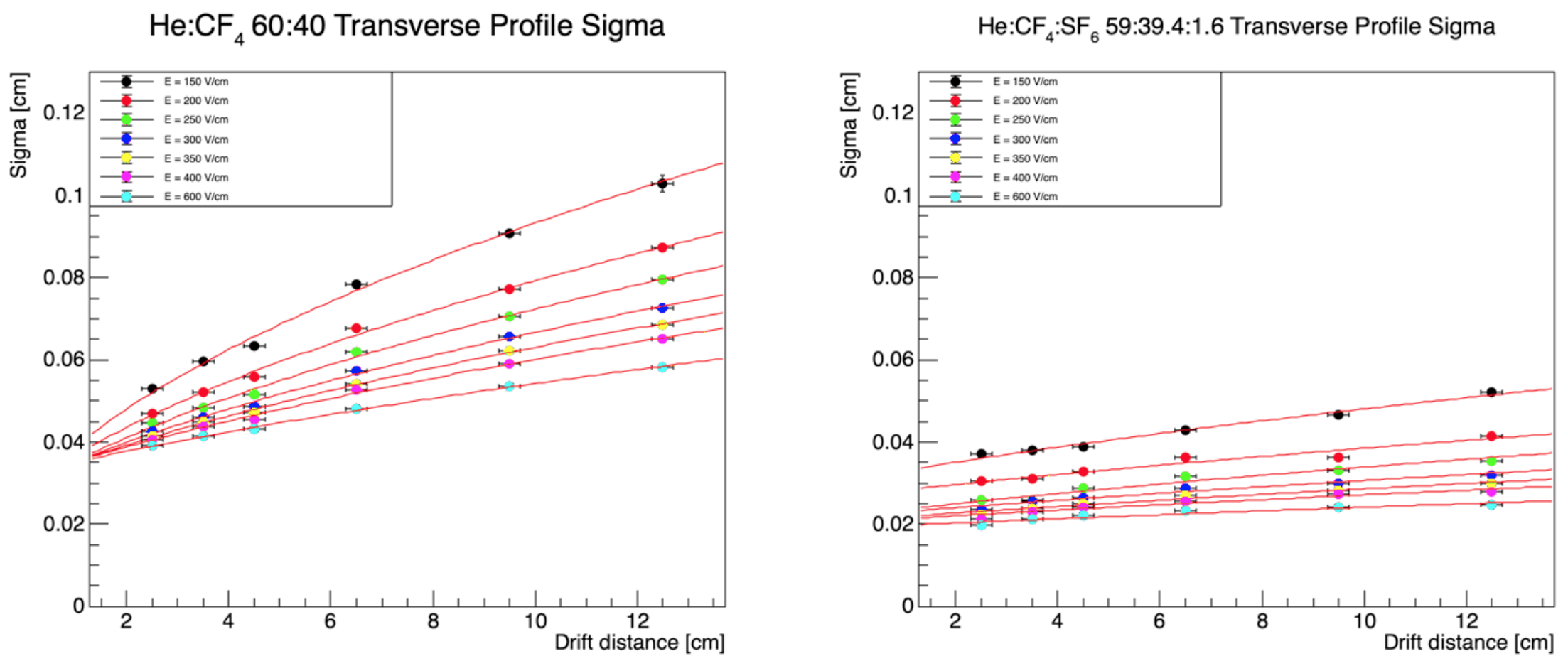}
    \caption{Measurement of diffusion as a function of drift distance under various applied drift fields. The left panel illustrates ED, while the right panel represents NID. Each panel includes a fit made with Eq. \ref{eq:diffequation2}. Different colors in the legend denote the varying drift fields applied in both plots.}
    \label{fig:sigmavsZ}
\end{figure}

The diffusion measured as a function of the drift distances for different values of the electric field are shown in figure \ref{fig:sigmavsZ}.
The observed $\sigma_{meas}$ values are fitted using the equation:
\begin{equation}
    \sigma_{meas}=\sqrt{\sigma_{0T}^2+\xi_T^2L}
    \label{eq:diffequation2}
\end{equation}
where, $\sigma_{0T}$ denotes the constant diffusion experienced within the GEM amplification plane and in the primary interaction in the gas, which is independent of the drift distance. The term $\xi_T$ represents the diffusion coefficient, and L signifies the drift distance from the GEM \cite{blum2008particle}. The outcomes of the Eq. \ref{eq:diffequation2} fitting the data are reported in Tab. \ref{tab:coeffValues}. 

\begin{figure}
    \centering
    \includegraphics[width=0.9\linewidth]{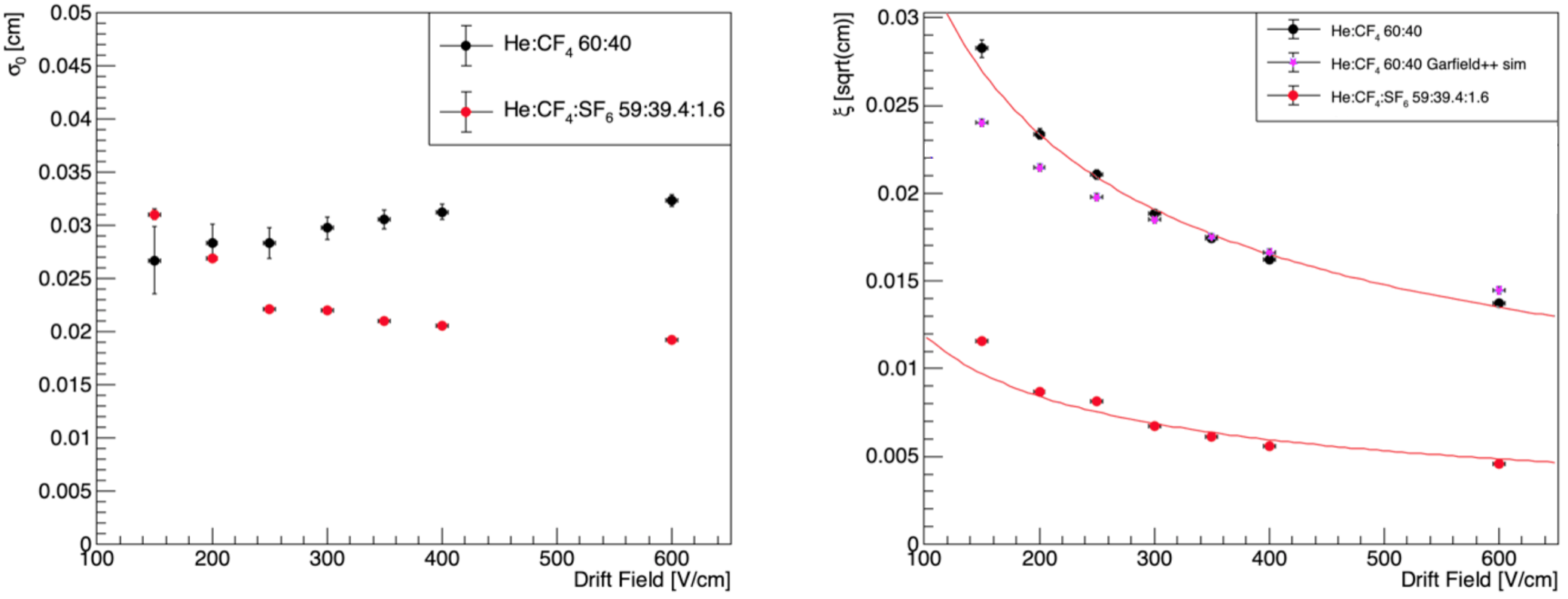}
    \caption{Fitted values of $\sigma_{0T}$ (left) and $\xi_T$ (right) derived from Eq. \ref{eq:diffequation2} plotted against the applied drift field for ED (black) and NID (red). The right panel also includes the Garfield++ simulation results for the ED gas mixture in pink.}
    \label{fig:DiffVsDrift}
\end{figure}

\begin{table}
\centering
\begin{tabular}{|c|c|c|c|c|}
\hline Drift field $[\mathrm{V} / \mathrm{cm}]$ & $\sigma_{0T}^{E D}[\mu \mathrm{m}]$ & $\xi_T^{E D}[\mu \mathrm{m} / \sqrt{\mathrm{cm}}]$ & $\sigma_{0T}^{N I D}[\mu \mathrm{m}]$ & $\xi_T^{N I D}[\mu \mathrm{m} / \sqrt{\mathrm{cm}}]$ \\
\hline \hline 150 & $300 \pm 100$ & $280 \pm 20$ & $320 \pm 30$ & $110 \pm 10$ \\
200 & $290 \pm 60$ & $230 \pm 10$ & $260 \pm 30$ & $90 \pm 20$ \\
250 & $290 \pm 60$ & $210 \pm 10$ & $220 \pm 20$ & $81 \pm 10$ \\
300 & $300 \pm 40$ & $190 \pm 10$ & $220 \pm 20$ & $68 \pm 10$ \\
350 & $300 \pm 40$ & $170 \pm 10$ & $210 \pm 20$ & $60 \pm 10$ \\
400 & $310 \pm 30$ & $160 \pm 10$ & $210 \pm 20$ & $56 \pm 9$ \\
600 & $320 \pm 20$ & $140 \pm 10$ & $200 \pm 20$ & $45 \pm 10$ \\
\hline
\end{tabular}
\caption{In table, the values of $\sigma_{0T}$ and $\xi_T$ measured for electron drift and Negative Ions Drift for the different drift field values are reported.}
\label{tab:coeffValues}
\end{table}
The variation of the derived values for $\sigma_{0T}$ and $\xi_T$ based on the applied drift field is depicted in the left and right panels of Fig. \ref{fig:DiffVsDrift}, with ED data represented in black and NID data in red. Moreover, the right panel of Fig. \ref{fig:DiffVsDrift} illustrates the diffusion expected for the ED mixture simulated via Garfield++ software in pink, which aligns well with the observed diffusion of the ED mixture.
Both the diffusion coefficients $\xi_T$ for ED and NID decrease as the drift field increases, consistent with the well-established 1/ $\sqrt{E}$ dependence \cite{knoll2010radiation,blum2008particle}. While the observed diffusion for ED, as foreseen from theoretical considerations and simulations, remains relatively high, the NID mixture exhibits a remarkably small diffusion coefficient, $\leq$ 50 $\mu$m/$\sqrt{cm}$ at a 600 V/cm drift field. Another noteworthy observation, illustrated in the left panel of Fig. \ref{fig:DiffVsDrift}, is the significant reduction in the constant diffusion term $\sigma_{0T}$ associated with the GEMs amplification plane during NID operation compared to ED. This reduction amounts to approximately one-third, decreasing from an average of 300 $\mu$m to 200 $\mu$m, and remains consistent regardless of the applied drift field. This effect is thought to happen due to the absence of saturation in NID operations. As explained in Sec. \ref{sec:overgroundstudies} the saturation happens when a high charge density crosses the GEM hole. In the case of NID operations, the drift speed is a factor $10^3$ smaller, this would also imply a reduction in the charge density leading to the absence of saturation.
Overall, the NID operations described in this study demonstrate the potential to achieve a reduced transverse diffusion during drift down to 45 $\mu$m/cm. This finding aligns with the smallest transverse diffusion coefficient ever measured in a gas detector, to the best of the author's knowledge, as reported in \cite{MARTOFF20003551NID} using a 40 Torr CS$_2$-based gas mixture.

\subsection{Further crosscheck}
\begin{figure}
    \centering
    \includegraphics[width=0.75\linewidth]{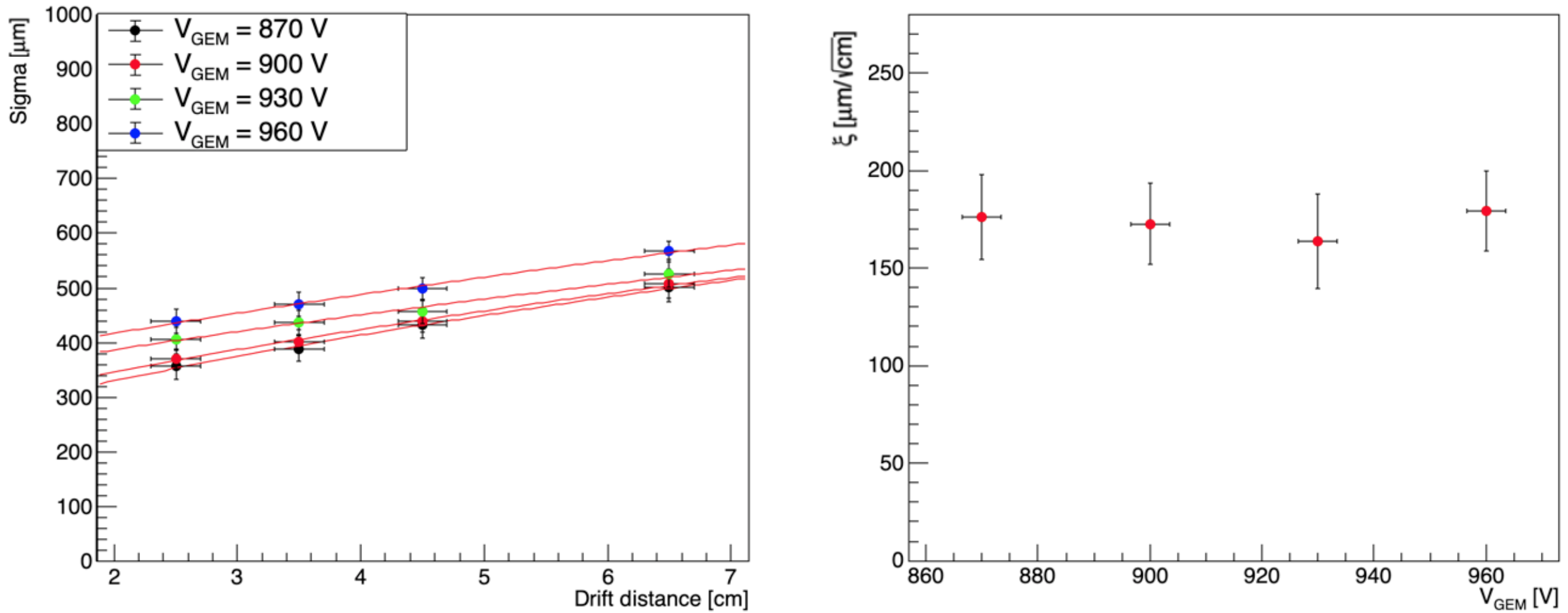}
    \caption{Left: Observed transverse diffusion in the ED mixture plotted vs the drift distance under a constant drift field of 400 V/cm, with various voltages $V_{GEM}$ applied (refer to the legend for details). Right: Fitted $\xi_T$ values from Eq. \ref{eq:diffequation2} depicted as a function of $V_{GEM}$ corresponding to the data presented in the left plot.}
    \label{fig:DiffusionVsVoltage}
\end{figure}

To further verify the analysis methodology, the experimental setup, and the presented results, the diffusion behavior of the ED mixture is investigated for different applied voltage on the GEMs $V_{GEM}$. The objective of this validation is to confirm that the analysis methodology remains robust across varying levels of total light (and charge) generated in the GEM amplification plane, even at higher gains compared to the operational voltage utilized in the preceding measurements. To demonstrate this, a constant drift field of 400 V/cm is maintained, while the voltage applied on the GEMs is adjusted from 290 V to 320 V on each GEM, resulting in a total voltage range of 870 V to 960 V. Fig. \ref{fig:DiffusionVsVoltage} illustrates on the left the observed transverse diffusion for the ED mixture at 400 V/cm across different drift distances, corresponding to four distinct voltage configurations. The right panel of Fig. \ref{fig:DiffusionVsVoltage} illustrates the extracted transverse diffusion coefficient $\xi_T$ plotted against $V_{GEM}$, indicating no correlation with the gas gain (and therefore light generated). This observation reinforces the reliability of these measurements.

Moreover, the light output of both the ED and NID configurations has been examined at the designated $V_{GEM}$ operating voltages used in the MANGOk setup and during data collection, across different applied drift fields and distances ranging from 2.5 cm to 12.5 cm. This was done to validate that the diffusion measurements remain unaffected by potential systematic errors associated with varying drift lengths and fields. 
\begin{figure}
    \centering
    \includegraphics[width=0.75\linewidth]{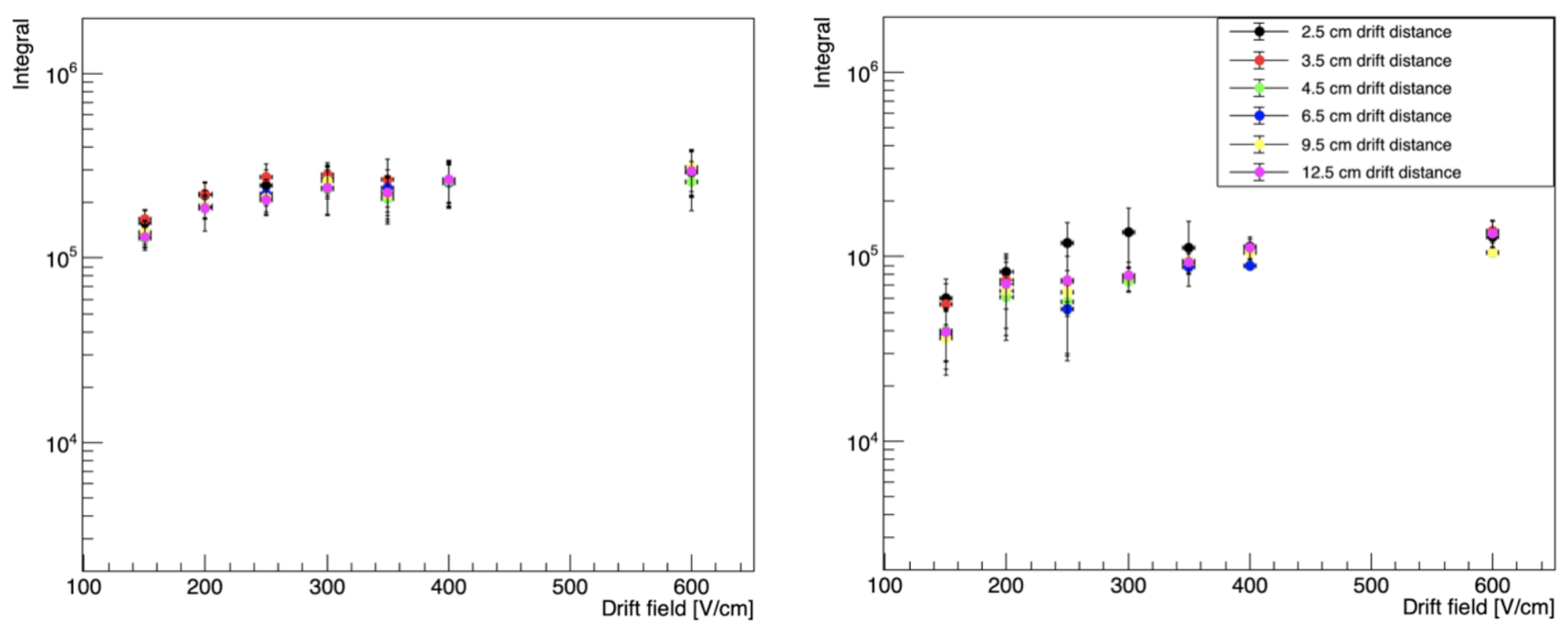}
    \caption{Light integral plotted against the applied drift field, with ED on the left and NID gas mixture on the right. The legend indicates the drift distance for each measurement in both plots.}
    \label{fig:LightVsPar}
\end{figure}
Fig. \ref{fig:LightVsPar} illustrates this comparison, displaying the results for ED on the left and for the NID gas mixture on the right. These results indicate that the light output remains consistent between the ED and NID operating conditions and is largely independent of the variation of the parameters.

\section{Final Discussion}
The findings presented in this chapter mark the initial validation of NID functionality with optical readout under LNGS atmospheric pressure (900 mbar). Additionally, they represent the primary assessment of the transverse diffusion of NID using a He:CF$_4$:SF$_6$ 59.2:39.2:1.6 mixture at 650 mbar. This advancement represents a significant milestone for directional rare events exploration and low-diffusion TPC applications, extending beyond optical readout technologies. 
In the future, there are plans for a systematic investigation into the gain and directional capabilities by varying the gas constituent ratios and employing different amplification structures. This aims to enhance light gain and delve deeper into the diffusion properties of He:CF$_4$:SF$_6$ mixtures. Potential amplification structures for testing include thicker GEMs, such as those combinations outlined in Sec. \ref{sec:gemgasmix}, COBRA structures, modifications of standard GEMs \cite{F_D_Amaro_2010}, and ThickGEM-multiwire hybrids \cite{EZERIBE2021164847}. Additionally, there are plans to conduct tests using charge readout via Timepix3 \cite{LIGTENBERG2021165706} and potentially explore He:SF$_6$ gas mixtures with varying SF$_6$ concentrations.\\
By reducing the drift transverse diffusion by a factor of 3 (from 140 $\mu$m/cm for ED to 45 $\mu$m/cm for NID) and decreasing transverse diffusion within the amplification stage by 100 $\mu$m, a threefold increase in drift length could be achieved within the CYGNO framework, while preserving the same tracking performance. This advancement promises improved tracking capabilities, directionality for both ER and NR, as well as enhanced compactness and scalability.
Currently, the major drawback in NID operations is the low gain which corresponds to a low light production and a low signal-to-noise ratio. However, this study represents only a first investigation of Negative Ions Drift operations with optical readout. Future tests with different amplification structures could allow for obtaining a gain in NID operations comparable to the one obtained with ED operation at the standard operative voltage.


%% file: chapters/simulation.tex
\chapter{Electron recoils sCMOS images simulation}
\label{chap:Simulation}
\begin{figure}
    \centering
    \includegraphics[width=0.75\linewidth]{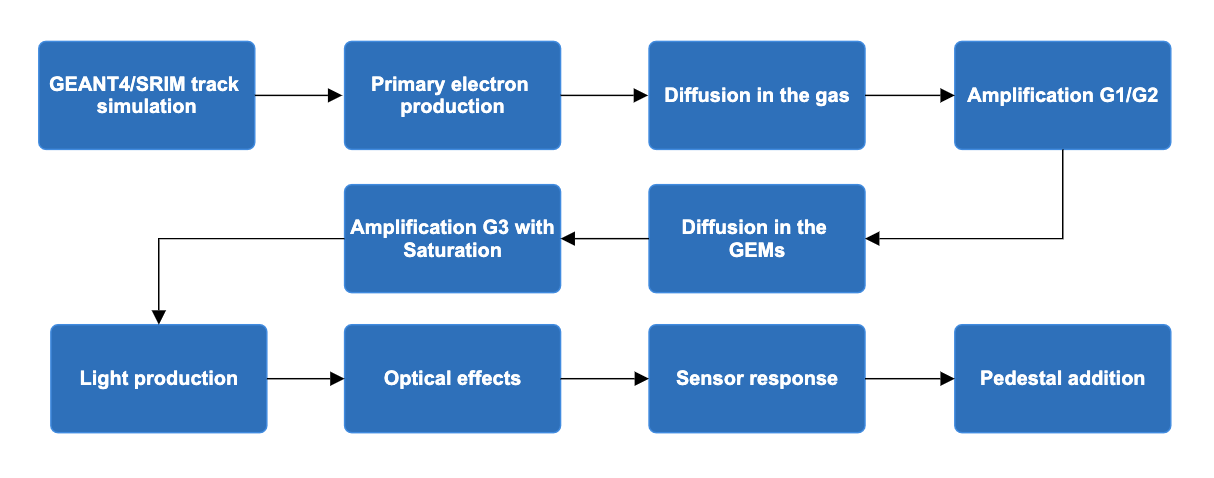}
    \caption{Working scheme of the simulation}
    \label{fig:workingflow}
\end{figure}

While the CMOS image simulation algorithm had already been developed by the collaboration, the author contributed by incorporating the optical vignetting effect into the simulation and conducting the data-Montecarlo comparison analysis.
The simulation of sCMOS images of electron and nuclear recoil is of paramount importance for optimizing the detector performance and fully understanding its response. For this reason, the collaboration has dedicated significant efforts to developing a simulation able to generate sCMOS-like images of electron and nuclear recoil tracks, with the production process accounting for all detector effects and readout response. The simulation process begins with a GEANT4 or SRIM simulated track for electron or nuclear recoil events, respectively, and subsequently the following detector effects are applied. As a consequence of the ionization process in the detector, primary electrons are produced with Poissonian statistics. These electrons then drift to the amplification plane, and in this process, they are subject to diffusion. Subsequently, the electrons undergo amplification in the GEM planes. In this process, the further diffusion due to the GEM contribution, the fluctuation in the GEM gain, and the saturation effect play a decisive role in the amplification process. At the amplification stage light is produced, and it is recorded by the sCMOS after the fluctuation in the light production and the lens optical effects.
Extensive parameter tuning in the simulation has led to an agreement between data acquired with the LIME prototype and the simulated tracks. This achievement was realized through continuous comparison with the data throughout the entire optimization process. 
The detailed comparison between data and simulation, in both detector response and track topological shape variables, studied in this thesis, is fundamental for this work.
Indeed, in the absence of low-energy electron sources with known directionality, simulated tracks have been used to assess the angular resolution performance for low-energy electron recoils. \\ 
In this chapter, the primary track production for electron recoils and nuclear recoils will be illustrated (Sec. \ref{sec:G4Simulation_digi}), followed by a comprehensive description of the simulation algorithm and all the processes included (Sec. \ref{sed:digialgo}). Finally, after an optimization of the simulation parameters, as an original work carried out for this thesis, a comparison analysis between the data collected with LIME (analyzed in Chap. \ref{chap:LIMEDetector}) and the tracks simulated, has been conducted and the results shown (Sec. \ref{sec:datamccomparison}).\\ The working flow of the simulation is shown in Fig. \ref{fig:workingflow}.

\section{Primary track simulation} 
\label{sec:G4Simulation_digi}
\begin{figure}
    \centering
    \includegraphics[scale=.4]{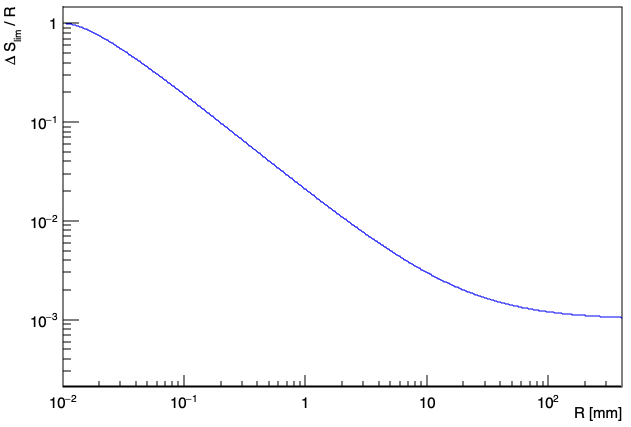}
    \caption{In figure, the function $\Delta S_{lim}/R$ vs R is shown for the actual parameters used in the simulation.}
    \label{fig:MaxStepVsR}
\end{figure}

The interactions of electrons with the CYGNO gas mixture are simulated with the GEANT4 software \cite{AGOSTINELLI2003250,GEANT4Dev,GEANT4Phys}. GEANT4, short for GEometry ANd Tracking, serves as a platform for simulating the passage of particles through matter using Monte Carlo methods. In GEANT4, particles are generated from a single point, and their trajectory within a given material is determined by modeling the applicable physics processes. Each physics process, such as electron ionization in gas, is represented by a C++ class, enabling the computation of the probability of interaction (mean free path) and the generation of the particle's final state. Different models can describe each process, and a single particle may undergo various processes, such as electron gamma emission through bremsstrahlung and delta rays production in gas. Each secondary particle produced follows the same computational approach. Tracking continues until particles are stopped or exit the simulation volume. Throughout the simulation, various physics quantities (energy, position, energy deposit, volume crossed, etc.) are accessible and can be extracted, used, and modified as needed. GEANT4 provides a range of particle types, with characteristics and interactions already defined. Moreover, the toolkit incorporates pre-implemented physics processes that can be used throughout the simulation.\\
In the simulation, tracks production has been performed in a gas mixture consisting of He:CF$_4$ in a 60:40 ratio in volume at atmospheric pressure, with a total gas density calculated by the percentage-weighted density of the single gasses.\\
A crucial point of the simulation is the parameters of the step function for the energy loss. Since GEANT4 operates with the finite element method (FEM), the particle interaction with matter is processed by discrete steps. Processes involving continuous energy loss necessitate to be processed by performing small steps to account for the dependence of the cross-section on the energy. Indeed, a step too large would not accurately approximate the cross-section, assumed constant during the step. Conversely, processing the whole track development with excessively small steps became computationally unfeasible. To address this, a smooth step function controlled by two parameters dynamically regulates the maximum size of the step based on the track range at that energy.
The equation of the step size limitation $\Delta S_{lim}$ is given by:
\begin{equation}
    \Delta S_{lim} = \alpha_R R +\rho_R(1-\alpha_R)(2-\frac{\rho_R}{R})
\end{equation}
where $\alpha_R$ is the $dRoverRange$ parameter, $\rho_R$ is the $finalRange$ parameter, and $R$ is the track range at the energy of the current step. From the equation, it can be observed that $\alpha_R$ represents the limit of $\Delta S_{lim}$ at high energy, while $\rho_R$ represents the value of the range at which the maximum $\Delta S_{lim} \sim R$ \cite{GEANT4Dev}\cite{GEANT4Phys}. This quantity needs adjustment, considering that GEANT has primarily been developed for high-energy physics. Therefore, for low-energy applications, it is reasonable to reduce it to achieve denser and more detailed track information.
In this specific case, to have the whole detail on the track topology, the simulation of the electron recoils necessitates very small steps, to ensure enough dense energy release for the track formation and precision in the interaction process calculation. Consequently, upon noticing that the parameters for the step function of GEANT4 were generating tracks that, after digitization, exhibited discontinuities due to energy deposits being too far apart, the parameters for the step function were set to $\alpha_R=0.01$ and $\rho_R=0.001\ mm$. 
A plot $\Delta S_{lim}/R$ as a function of R using the simulation parameters selected for this study is shown in Fig. \ref{fig:MaxStepVsR}. As can be observed at higher energies when the particle range is higher, to still get details on the track interactions the step limitation is kept to a small fraction of the range. With the track losing energy, and the following decrease in the range, the step limitation gets closer to the particle range.
For each particle produced, the information on the coordinates of the track energy deposits $(x,y,z)$, as well as the energy released at each point, $\Delta E$ are stored.\\
In the simulation of nuclear recoils, the Stopping and Range of Ions in Matter (SRIM) \cite{ZIEGLER20101818} software is employed. SRIM comprises a suite of programs extensively utilized in low-energy physics experiments to compute the stopping and range of ions (up to 2 GeV/amu) in materials, employing a quantum mechanical approach to ion-atom collisions (wherein a moving atom is treated as an “ion”, and all target atoms are considered as “atoms”). At its core lies TRIM (TRansport of Ions in Matter), a Monte Carlo (MC) program that models the interactions of energetic ions with amorphous targets. In CYGNO, a dedicated method based on TRIM has been developed to reproduce the 3D shape of nuclear recoils more accurately than what SRIM alone would achieve. \cite{FlaminiaTesi}.

\section{The track digitization algorithm}
\label{sed:digialgo}
Starting from the energy deposits of ER and NR generated with GEANT4 and SRIM, the track digitization algorithm produces the simulated sCMOS images by adding all the detector effects and readout responses to these.
These effects include the fluctuation in primary ionization, charge carriers diffusion, electron absorption during the drift (Sec. \ref{sec:PPabs}), fluctuation in avalanche multiplication, saturation (Sec. \ref{sec:AmpDiff}), fluctuation in photon production, and optical effects induced by the camera lens, and sensor response (Sec. \ref{sec:lightSensresp}). A comprehensive description of all these steps is provided in the following sections. These steps depend on various parameters that have been either fixed based on CYGNO measurements or optimized for those parameters that we couldn't fix from the measurements, as explained in Section \ref{sec:ParOptDigi}.

\subsection{Primary electron production and absorption}
\label{sec:PPabs}
Each primary energy deposit of the GEANT4/SRIM track is characterized by a position in the space $(x,y,z)$ and the energy released in the point $\Delta E$. Each energy deposition results in the production of primary ionization electrons, which are calculated from the mean number of electrons $\bar{N}_e$ produced by ionization. This number is calculated as $\bar{N}_e=\Delta E/W_i$, where $W_i$ is the w-value of the gas, and represents the mean energy required to create an electron-ion pair in the material. As reported in \cite{knoll2010radiation} and \cite{VAHSEN2014111} respectively, the $W_i$ for He and CF$_4$ are $W_i(\text{He}) = 41$ eV/pair and $W_i(\text{CF}_4) = 54$ eV/pair. To determine this value for the gas mixture the weighted average is calculated 
\begin{equation}
W_i(\text{He:CF}_4\ \text{60:40})=0.6\cdot W_i(\text{He}) + 0.4 \cdot  W_i(\text{CF}_4) = 46.2 \ \text{eV/pair}
\end{equation}
Since the ionization process inside a gas is a Poissonian process, the actual number of ionization electrons $N_{e}^{pr}$ is extracted from a Poissonian distribution with mean value $\bar{N}_e$. 
During the drift process towards the GEMs, there exists a non-negligible probability for the electrons to be captured by impurities present in the gas volume. Some gas contamination due to imperfect gas tightness of the detector, especially $O_2$, could influence this process. Given that $O_2$ is highly electronegative \cite{Snowden-Ifft:2013vua}, even a small amount of it can significantly affect the electron attachment, reducing $\lambda_{att}$ to the order of few meters.     
In the simulation, this effect is addressed by introducing this attachment process with a characteristic $\lambda_{att}$. Thus, the number of electrons arriving at the amplification plane is given by 
\begin{equation}
    N_e = N_{e}^{pr} \cdot e^{-\frac{z}{\lambda_{att}}} 
\end{equation}
where $z$ is the distance of the hit from the amplification plane. In \cite{Hilke_2010}, an electron absorption of 3\%/m is reported for a highly pure gas mixture of Ar:CF$4$ with 1 ppm of $O_2$ contamination. Therefore, for the operational conditions of the gas mixture of CYGNO, it is assumed a $\lambda_{att}$ of $\mathcal{O}(1) \ m$ and subsequently optimized to achieve agreement with the data.

\subsection{Electron amplification and diffusion}
\label{sec:AmpDiff}
\begin{figure}
    \centering  \includegraphics[scale=.4]{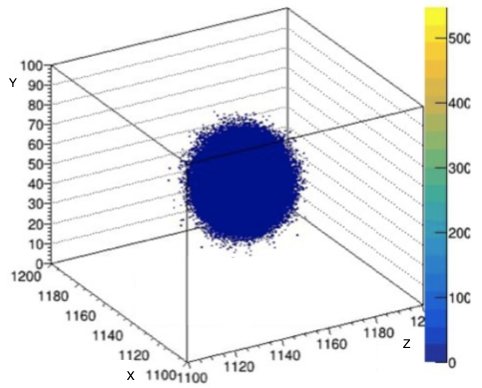}
    \caption{In figure, an example of an electron cloud from a 6 keV electron recoil before entering GEM3 is shown. }
    \label{fig:ecloud}
\end{figure}

The ionization electrons $N_e$ that reach the GEMs undergo amplification. The extremely intense electric fields within the GEM holes, reaching values as high as $E_{GEM}=\Delta V/\Delta x$= $440\ V/50 \ \mu m \sim\ $90 kV/cm, are strong enough to accelerate electrons and provide them an energy sufficient to ionize the gas. This ionization process generates additional free electrons, which, in turn, are accelerated by the electric field, ionizing the gas, via a Townsend avalanche mechanism. Given a number of electrons reaching a GEM hole $n_{in}$ and a number of electrons exiting the GEM hole after amplification $n_{out}$, the gain of the GEM $G$ can be expressed as $G=n_{out}/n_{in}$. The single GEM gain, as measured by our collaboration \cite{Baracchini_2020_Stab} can be parameterized as a function of the GEM voltage as:
\begin{equation}
    G=0.034 \cdot e^{0.021 \cdot \Delta V}
    \label{eq:gemgain}
\end{equation}
Moreover, not all electrons generated during the amplification process may exit the GEM hole, and they could be absorbed by the surface of the GEM. This efficiency, known as extraction efficiency $\epsilon_{extr}$, depends on the GEM voltage and can be parameterized as follows \cite{PinciTesi} :
\begin{equation}
    \epsilon_{extr} = 0.87 \cdot e^{-0.002 \cdot \Delta V}
\end{equation}
Throughout the amplification process, the absorption of electrons during the amplification, together with the imperfection in the GEM manufacturing, can influence the effective field and, consequently, the gain of the GEM, giving rise to fluctuations. As shown in \cite{PinciTesi}, the effective gain of the GEM follows an exponential distribution, with a mean value equivalent to the average gain reported in Eq. \ref{eq:gemgain}.
The gain fluctuations are considered relevant only in the first stage of amplification, where the number of primary electrons is relatively low. Following the first amplification stage, the number of electrons is increased of $\mathcal{O}$(300) (corresponding to a single GEM gain at 440 V). Due to the higher number of electrons, employing the same amplification process with gain fluctuations would have a minimal effect, as the fluctuations would compensate by averaging over all the contributions.
In the second amplification stages thus, these fluctuations are less relevant, computationally expensive to simulate, and hence not taken into account.\\
In the simulation, at each point along the track, containing $N_e$ primary ionization electrons, the number of electrons produced by the amplification of the first GEM, namely G1, $N_{e}^{G1}$ is computed as:
\begin{equation}
    N_e^{G1}= \sum_{k=1}^{N_e}{(G^{G1}_k\cdot \epsilon_{extr}^{G1})}
\end{equation}
where $G^{G1}_k$ is the actual gain of the GEM, extracted from an exponential distribution with mean value $G^{G1}$ (Eq. \ref{eq:gemgain}), while $\epsilon_{extr}^{G1}$ is the extraction efficiency which is related to the probability of an electron to enter the GEM channel.
Since the fluctuations are not considered, the number of electrons at the second amplification stage at each point is given by 
\begin{equation}
    N_e^{G2}=N_e^{G1} (G^{G2}\cdot \epsilon_{extr}^{G2})
\end{equation}
The effect of diffusion is then applied to the electrons exiting the second amplification stage.
As described in Sec. \ref{sec:TrasnvDiff}, during the drift in the gas, primary electrons undergo diffusion. The transversal diffusion $\sigma_T$ can be parameterized in a contribution of diffusion due to the electrons traveling the gas volume $\xi_D$, and a fixed contribution connected to the GEM amplification $\sigma_{T0}$. The effect of the longitudinal diffusion can be modeled in the same way following \cite{blum2008particle}.
\begin{equation}
    \sigma_T= \sqrt{\xi_{T}^2\cdot z +  \sigma_{0T}^2} \ \ \ \ \ \ \ \   \sigma_L = \sqrt{\xi_{L}^2\cdot z +  \sigma_{0L}^2}
\end{equation}
with the diffusion coefficients $\xi_T$ and $\xi_L$ has been determined using a Garfield++ simulation, varying with the applied drift field in \cite{Baracchini_2020_Stab}.
The diffusion effect is included in the simulation by smearing the x, y, and z coordinates of each of the $N_e^{G2}$ electrons at every track interaction point. The smearing is applied using a Gaussian function with $\sigma = \sigma_T$ for the x-y coordinates and $\sigma = \sigma_L$ for the z coordinate. The resulting information is stored in a tridimensional histogram with 3D-pixel (voxel) size $V_x,V_y$ and $V_z$. The voxel sizes on x and y have been chosen to be equal to the area seen by each pixel, while $V_z$ has been chosen to be of the same order of magnitude of the GEM thickness and further optimized as explained in Sec. \ref{sec:ParOptDigi}.  
A picture of the electron cloud diffused before entering the third GEM as simulated for the LIME detector is presented in Fig. \ref{fig:ecloud}.
The third and final stage of amplification, which is subject to saturation, will be treated in detail in the next section.

\subsection{GEM gain saturation}
\label{sec:saturationmodeling}
As discussed in Sec. \ref{sec:overgroundstudies}, the characteristic exponential relationship between gain and GEM voltage is lost when the electron cloud reaches a certain density. Indeed, at the third GEM stage, the charge density can reach a level where the field inside the hole is screened by the electron charge, resulting in a decrease in the electric field.\\ To model this effect, modification of the GEM gain as a function of the charge density has been introduced. This modification is applied only in the third amplification stage, as clarified later, as this effect becomes significant when the electron count reaches the order of $10^5$, occurring specifically during the third amplification step.
Generally speaking, the number of avalanche electrons $dn$ per unit length $dz$ generated by the passage of n electrons in a single GEM hole can be expressed as:
\begin{equation}
    \frac{dn}{dz} = \alpha E n
\label{eq:nprod}
\end{equation}
where $\alpha$ is a normalization constant that depends on the characteristics of the gas, and $E$ is the electric field inside the GEM hole.
As detailed in Section \ref{sec:optics}, CYGNO requires extremely high gains to compensate for the limited solid angle of the optical readout. 
The saturation effect can be modeled by introducing an effective electric field, wherein the shielding effect is modeled with a negative term consisting of the product of a constant $\beta$ and the number of electrons inside the hole.
The effective field can be written as $E_{eff} = E-\beta n$, where $\beta$ is a constant of the model. Introducing the effective field term in \ref{eq:nprod}, the modified equation obtained is:
\begin{equation}
    \frac{dn}{dz} = \alpha(E-\beta n)n
\end{equation}
By integrating the equation the total number of electrons exiting the GEM channel is obtained: 
\begin{equation}
    n_{out} = \frac{n_{in}\cdot e^{\alpha \Delta V}}{1+\beta n_{in}(e^{\alpha \Delta V}-1) }
\label{eq:satgain}
\end{equation}
where $n_{in}$ is the initial number of electrons entering the GEM hole, and $\Delta V$ comes from the integration of the field $E$ along the whole GEM thickness. 
It can be observed from \ref{eq:satgain} that $\beta$ is related to the critical number of electrons. For $n_{in} = 1/\beta$ indeed, the electron number is so high that no amplification is obtained. By redefining $\beta = 1/n_{c}$, where $n_c$ is a critical number of electrons for which there is no gain, and the gain without saturation $g=e^{\alpha \Delta V}$, equation \ref{eq:satgain} can be expressed as:
\begin{equation}
    G(n_{in})=\frac{n_{out}}{n_{in}}=A\cdot \frac{g}{1+\frac{n_{in}}{n_c}(g-1)}
\end{equation}
Where A is an overall normalization parameter of O(1).
Considering a standard capacitance of a GEM hole to be of the order of $C = 5 \cdot 10^{-18}\ F$, the total charge accumulated at the hole surfaces at $\Delta V=440 V$ is given by $Q= C\cdot \Delta V = 2.25 \cdot 10^{-15}$. This charge corresponds to a number of electrons $n = 1.5 \cdot 10^4$. To fully neutralize the charge field inside the GEM hole, the same amount of charge is needed. Therefore, $\beta$ can be expected to be approximately $ 1/n \sim O(10^{-5})$.    

\begin{figure}
    \centering
    \includegraphics[scale=.5]{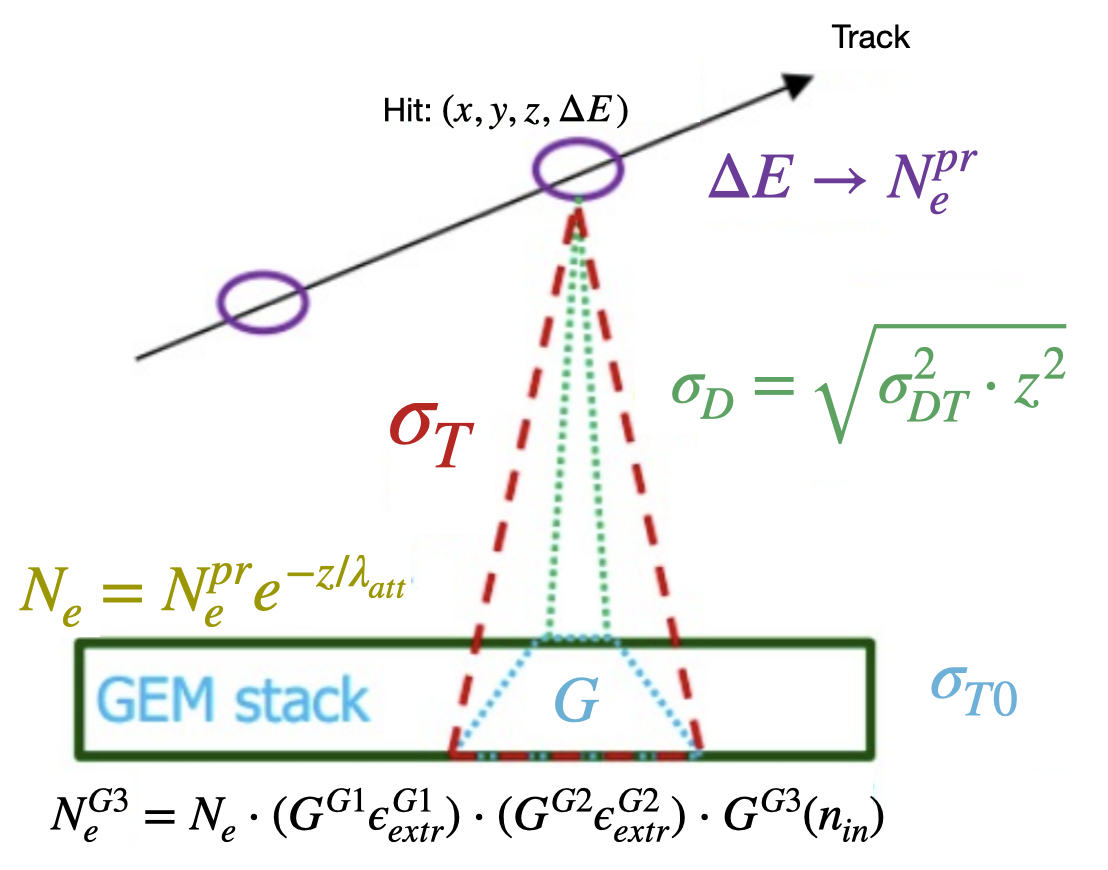}
    \caption{Scheme of the charge production, drift, and amplification process as implemented in the CYGNO simulation. The colors used for the equations are related to the colors of the elements in the scheme.}
    \label{fig:SchemeDigi}
\end{figure}

Thus, in the third amplification stage, the number of electrons in each voxel of the 3D cloud is multiplied by $G^{G3}(n_{in})\cdot \epsilon^{G3}_{extr}$. The total number of electrons out of the third gem $N_{e}^{out}$ at each coordinate x-y is then obtained by projecting the 3D cloud of electrons along the z-axis (on the x-y plane) 
A schematic representation of the steps described in the last two sections is shown in Fig. \ref{fig:SchemeDigi}.

\subsection{Light production and sCMOS sensor response}
\label{sec:lightSensresp}
\begin{figure}
    \centering    
    \includegraphics[scale=.3]{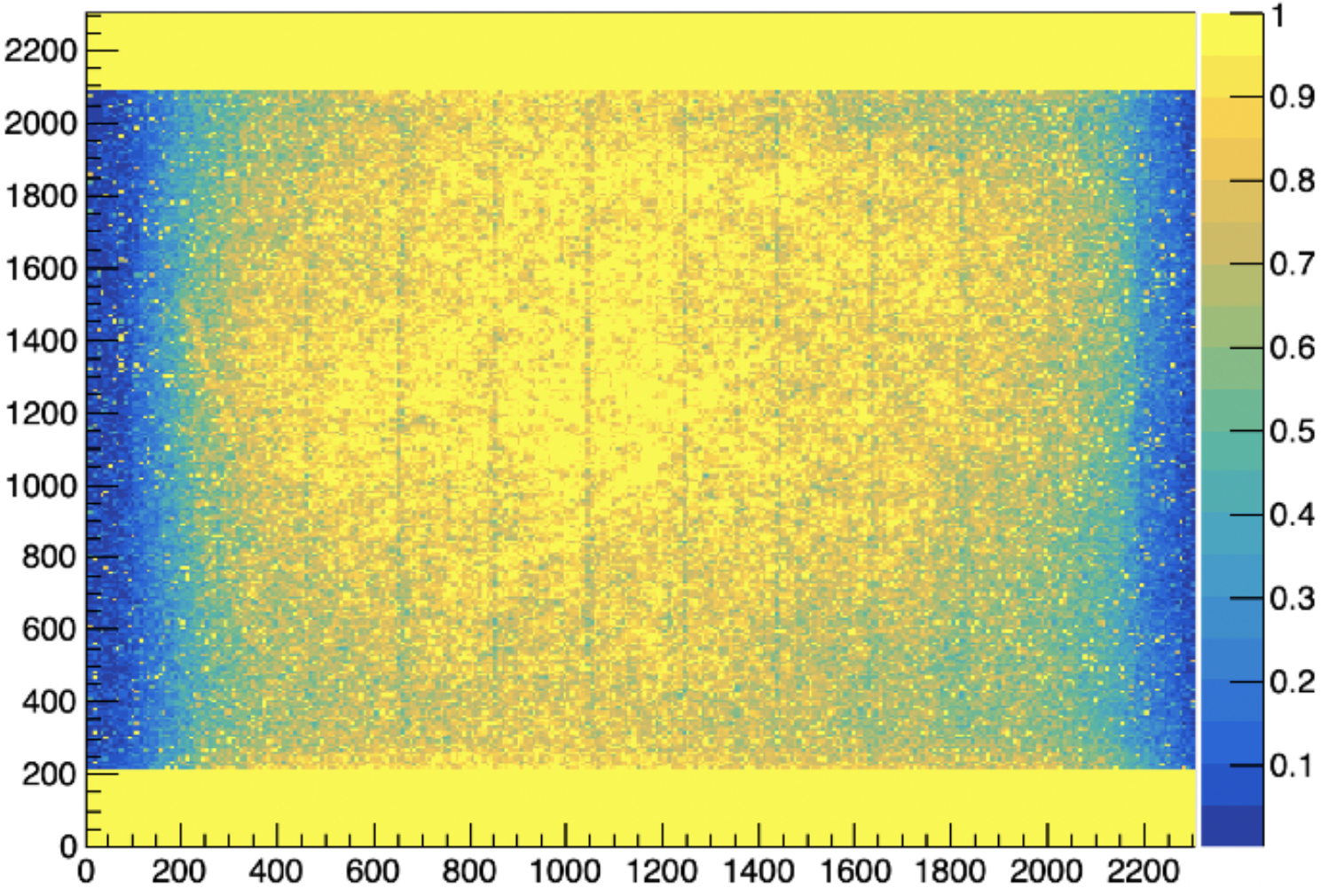}
    \caption{In figure the new vignetting map which includes also the disuniformities of the GEM response along its plane, measured with the LIME detector, is shown.}
    \label{fig:VignLensGEM}
\end{figure}

\begin{figure}
    \centering
    \includegraphics[scale=.25]{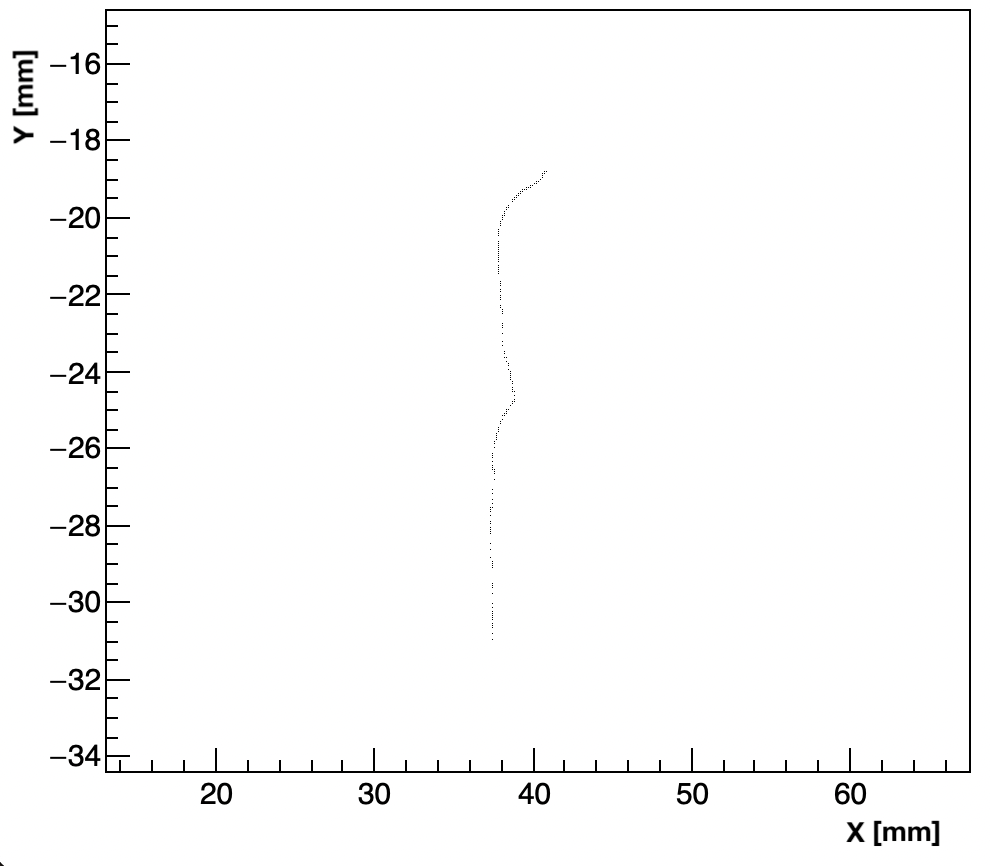}
    \includegraphics[scale=.25]{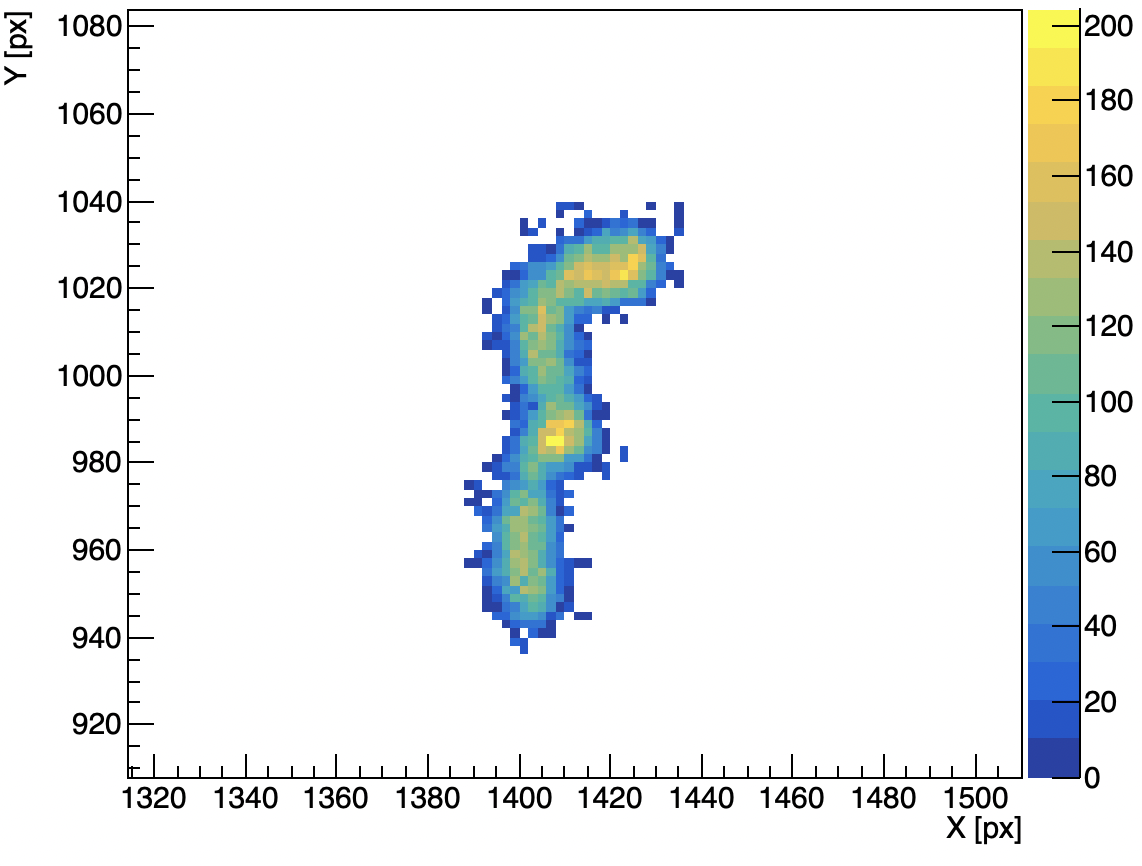}
    \includegraphics[scale=.25]{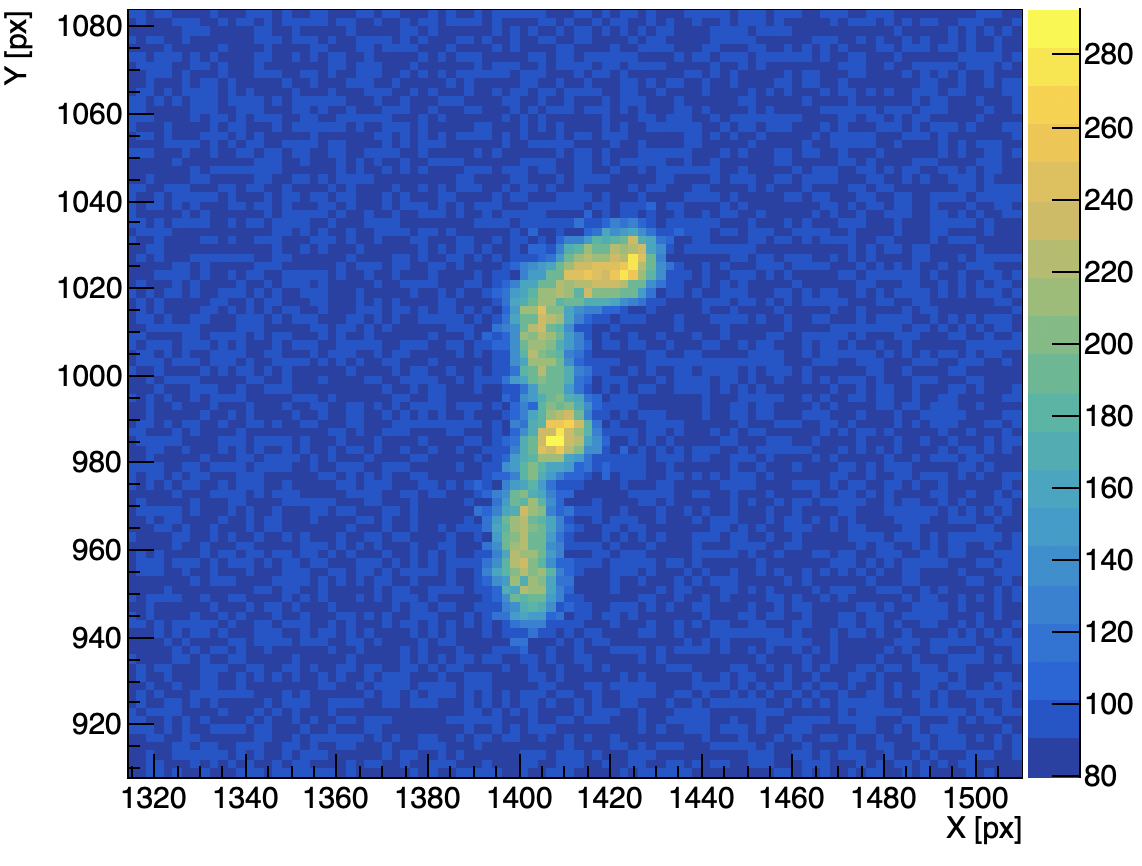}
    \caption{Left: 30 keV ER track simulated with GEANT4. Center: same track digitized without noise addition. Right: track digitized with sCMOS pedestal added.}
    \label{fig:digitprocess}
\end{figure}

Due to the scintillating properties of the He:CF$_4$ gas mixture, as explained in Section \ref{sec:gasmixture}, secondary photons are generated along with avalanche electrons. The number of scintillation photons produced is directly proportional to the number of electrons within the avalanche. Given that the number of electrons in the first two amplification stages is small compared to those coming out from the third GEM, and considering the third GEM cover the first two one, it is reasonable to assume that the photon produced that can reach the optical sensor is proportional only to the number of electrons exiting the third GEM $N_{e}^{out}$.
The overall electron count is then converted into photons, which are subsequently converted into ADC counts of the camera. Since the x-y bin size has the same dimension as the size seen by each pixel of the camera, there is a correspondence 1:1 between bins and pixels. For each bin of the 2D histogram resulting from the projection of the electron cloud along z, the number of photons produced in each bin is calculated. The average number of photon produced in each pixel $\bar{N}_{\gamma}$ is calculated as:   
\begin{equation}
    \bar{N}_{\gamma}=LY \cdot N_{e}^{out} 
\end{equation}
where LY=0.07 $ph/e^-$ is the gas light yield for the He:CF$_4$ gas mixture at atmospheric pressure from \cite{8532631,Marafini_2015,Fraga:2003uu}.
Since the process of photon production involves the breaking of CF$_4$ molecules (Sec. \ref{sec:gasmixture}), it can be treated as a Poissonian process. Consequently, the actual number of photons $N_{\gamma}^{tot}$, is extracted from a Poissonian distribution with mean value $\bar{N}_{\gamma}$.\\ 
The optical system, as detailed in Sec. \ref{sec:optics}, collects the total number of photons produced and directs them to the sensor. However, due to the reduced solid angle, only a fraction of the photons reach the optical system. 
To account for the solid angle $\Omega$, the actual number of photons produced in each bin is then multiplied by the solid angle $N_{\gamma}^{pix} = N_{\gamma}^{tot}\cdot \Omega$ with $\Omega$ being $\Omega = \frac{1}{( 4N(\frac{1}{I}+1) )^2}$, where $N$ is the focal number and $I$ is the magnification (Sec. \ref{sec:optics}).\\
Additionally, as explained in Sec. \ref{sec:vignetting}, the vignetting effect alters the collection of photons, with a reduced acceptance moving towards the borders of the image. Moreover, the nonuniformity of the GEM response along their surface can additionally cause variation in light production. Thus, these effects are subsequently applied to the track pixel intensity by multiplying the number of photons reaching each pixel by the corresponding value of a new vignetting map at that coordinate. To parameterize this effect, a new vignetting map is constructed using natural radioactivity to replicate the assumption of “uniformly illuminated scene”. More in detail, the map is built by summing 1000 images of natural radioactivity, acquired with a long exposure time ($2 s$), after pedestal subtraction. In this way, not only the effect of the natural vignetting but also potential variation of the GEM response along its surface are included in the map. The resulting new vignetting map is shown in Fig. \ref{fig:VignLensGEM}.
The use of the map which includes also the GEMs non-uniformity response is essential for later data-MC comparison. During reconstruction indeed, the correction for the pure lens vignetting has been done with a map realized with the camera facing a white wall, as detailed in Sec \ref{sec:vignetting}. This map does not include the effect of the disuniformity in the GEM response. In this way applying the map which includes the lens vignetting and the disuniformity in the GEM response, and correct with the pure lens vignetting effect ensure the presence of the GEM fluctuation response in both the data and MC for better consistency.\\
In the final step, the sensor response is simulated. Each camera converts the number of photons reaching each pixel into ADC counts using a distinct proportionality factor $C_{conv}$, provided by the camera manufacturer. The track image is subsequently generated by converting the number of photons reaching each pixel into camera ADC counts, defined as $ADC = C_{conv} \cdot N_{\gamma}^{pix}$. Finally, the track created in ADC counts is superimposed on a real sCMOS pedestal to produce the simulated image.
An example of an electron recoil track simulated in GEANT4, digitized, and with pedestal addition after digitization is shown in Fig. \ref{fig:digitprocess}.


\subsection{Simulation parameters for LIME}
\label{sec:ParOptDigi}
The simulation parameters can be categorized into two classes: those that are well-known and measured, and those that have not been measured or are not known a priori and must be assumed and subsequently optimized based on the data. To refine and evaluate the simulation performances, data acquired with $^{55}$Fe source with the LIME prototype at different distances have been utilized to model the effect of saturation and its relationship with diffusion and attenuation length.\\ Among the known parameters there are the diffusion coefficients, specifically the transversal diffusion coefficients $\xi_T$ and $\sigma_{0T}$, which have been experimentally measured. For the longitudinal diffusion, $\xi_L$ and $\sigma_{0L}$, the simulation relies on estimations obtained with Garfield++ (Sec. \ref{sec:gemgasmix}). These parameters have been further optimized to reproduce the iron data. Other measured parameters include the GEMs gain $G$, collection efficiency $\epsilon_{extr}$, light yield of the gas $LY$, and the photon-to-ADC conversion factor $C_{conv}$ which is a known characteristic of the camera, with $C_{conv}$=2 for the Hamamatsu Orca Fusion \cite{FusionCamera} used in LIME.
Among the parameters not known and to be tuned on the data, figures out $\lambda_{att}$, which has been assumed to be $\lambda_{att}=1$ m and then optimized on the data. Other parameters not known a priori are the voxel sizes $V_{x}$, $V_{y}$ and $V_{z}$. These parameters have been set equal to the size of the image seen by a single pixel $V_x=V_y= 0.151$ mm (Sec \ref{sec:LIME}) and $V_z$ to the same order of magnitude of the GEM thickness and further optimized. Due to the absence of precise modeling of saturation, the parameters $A$ and $\beta$ are not known a priori. However, as explained in Sec. \ref{sec:saturationmodeling}, it is reasonable to assume $A$ of $\mathcal{O}$(1) and $\beta \sim 10^{-5}$. \\
After optimizing the parameters to be able to replicate the iron spots at the different drift distances, the optimal values determined are:
\begin{alignat}{4}
    &\sigma_{0T} = 350\  \mu m \ \ \ \ \ \ \ \ \ \ \ \  \  \ && \xi_{T}=115\  \mu m/ \sqrt{cm} \ \ \ \ \ \ \ && \sigma_{0L} =\  260 \mu m \nonumber \\ \ \ &\xi_{L}= 100\  \mu m / \sqrt{cm}  
    && \lambda_{att} = 140\  cm \ \ \ && A=1.5 \nonumber \\ \ \ &\beta=10^{-5} \ \ \ && V_z = 0.1 \ mm \ \ \ && V_x=V_y = 0.151 \ mm \nonumber \\
     &LY = 0.07\ ph/e^- \ \ \ && G(440\ V)= 342 \ \ \ && \epsilon_{extr}(440\  V)= 342.0  
\label{eq:parametes}
\end{alignat}

\section{LIME electron recoils MC simulation validation}
\label{sec:datamccomparison}
\begin{figure}
    \centering
    \includegraphics[width=0.5\linewidth]{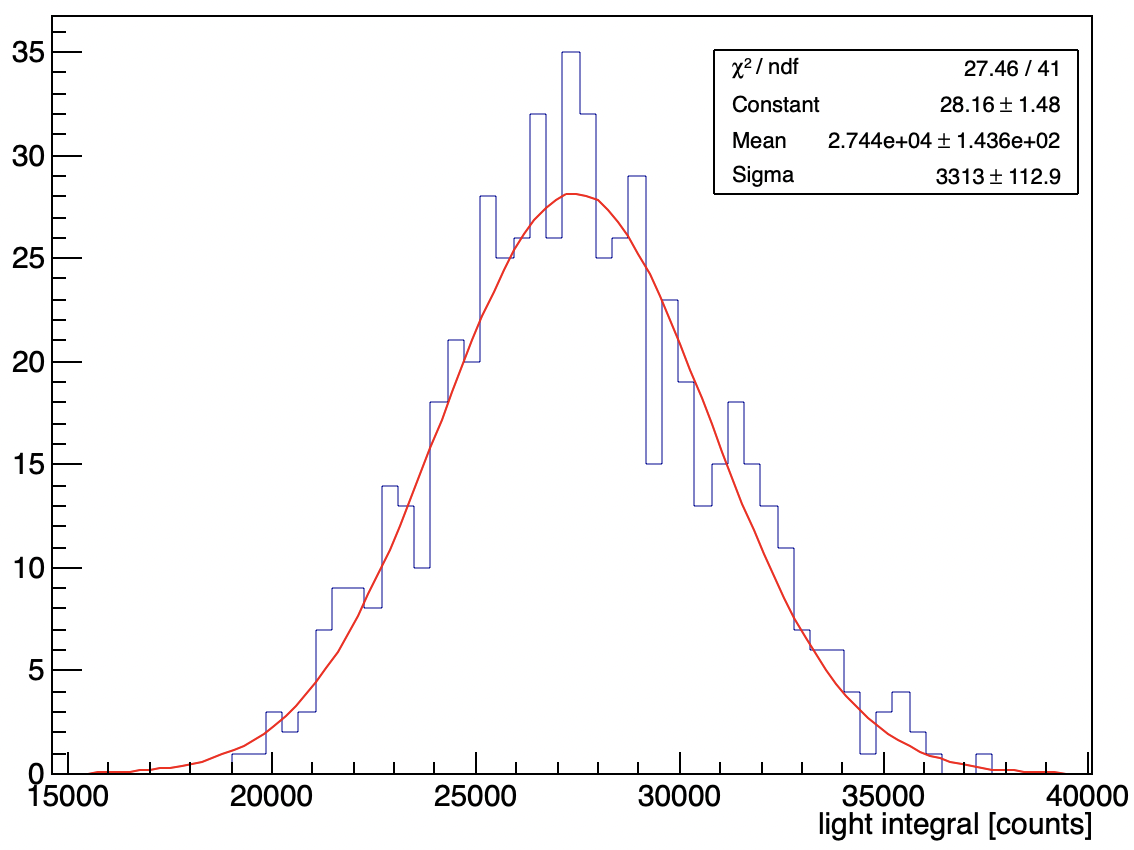}
    \caption{Track light distribution for ER simulated at 22.01 keV. A Gaussian fit is superimposed to the distribution.}
    \label{fig:AgFit}
\end{figure}

In order to validate the MC strategy and the optimized parameters illustrated in Sec. \ref{sec:paroptimization}, MC samples have been generated to reproduce the X-Ray data discussed in Sec. \ref{sec:XRayAn} and extensively compared to these.
For the comparison, a set of 1000 simulated electron tracks was generated at the energy of the $k_\alpha$ shell of the materials. For the simulations, the optimized parameters reported in \ref{eq:parametes} have been used. The tracks have been generated with the following specifications:
\begin{itemize}
    \item Isotropic distribution in direction for both theta, the angle on the GEM plane, and phi, the angle with respect to the GEM plane. 
    \item Uniform distribution within the range of 15-35 cm along the z-axis, to reproduce the distribution in z of the data.
    \item Uniform distribution in space on the x-y plane to reproduce the distribution in x and y of the data.
\end{itemize}
The simulated samples have been reconstructed with the same procedure illustrated in Sec. \ref{chap:reco} and the same parameters used for the X-Ray data.  The tracks found by the reconstruction code have been further selected by applying the cut illustrated in Sec. \ref{sec:X-RayAnalisys}, to maintain consistency with the analysis of real data. The track light distribution for the MC data at different energies is then fitted with a Gaussian. An example of fit to the track light distribution for 22.01 keV simulated electron tracks is shown in \ref{fig:AgFit}.
The mean light response is given by the mean of the Gaussian, while the percentage energy resolution is given by its sigma divided by the mean value.
The data-MC comparison involved assessing both the light response and energy resolution, as well as different topological track shape variables. In particular, the statistical tool $_{s}\mathcal{P}\textrtaill{lot}$ has been utilized to extract the track shape variables of the signal component from the background in the data. These variables have been then compared with those of the simulated tracks.

\subsection{Linearity and energy resolution comparison}
\begin{figure}
    \centering
    \includegraphics[width=1.\linewidth]{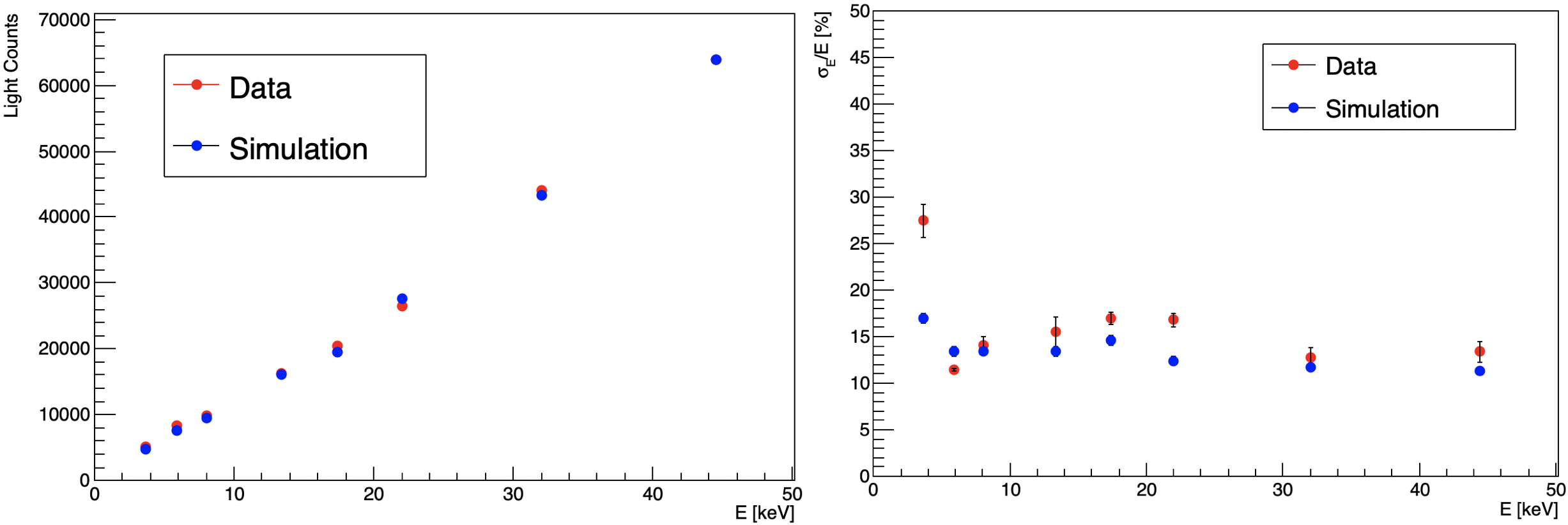}
    \caption{Left: plot of light response for both data and simulation. Right: plot of energy resolution for both data and MC.}
    \label{fig:lineresodatamc}
\end{figure}

The resulting plots showing the light response and energy resolution featuring both data and simulation are shown in Fig. \ref{fig:lineresodatamc}.
The left plot of Fig. \ref{fig:lineresodatamc} shows a very nice consistency between the real and simulated light response for energies between 3.7 and 45 keV. Examining the energy resolutions, an agreement is found for energies above 6 keV, despite some fine-tuning of the parameters is required to reach a perfect agreement in energy resolution. One factor that might create a difference in energy resolution at low energy could be the presence of a high amount of background in the data, since those have been acquired overground. 

\subsection{Track shape variables comparison}
\label{sec:splot}
As mentioned in the introduction, for the development of this thesis, since directional performance can only be estimated through Monte Carlo simulations and not from data, it is crucial that the MC accurately reproduces not only the energy response of electron recoils (ER) in terms of linearity and resolution but also the exact topology of the relevant tracks. For this reason, a series of variables capable of characterizing the topology of the tracks that can be reconstructed from the information provided by the track reconstruction code in the identified images have been identified. 
The chosen variables, employed for the topological comparison between the simulated tracks and the data, include:
\begin{itemize}
    \item \textbf{Track length and width}: The dimensions of both the major and minor axes of each cluster, determined through principal component analysis (PCA) applied to the pixels associated with the track, represent a measurement of the overall shape of the track projected onto the x-y plane.
    
    \item \textbf{Slimness}: Ratio between the track width and the track length, quantifies the roundness of a track.
    
    \item \textbf{Track density}: Ratio between the light contained inside the track and the number of pixels constituting the track, provide a measurement of the light concentration, and it is related to the ionization pattern of the track. 
    
    \item \textbf{Track average specific ionization}: Ratio between the light integral and the track length. Provide the information of how much light is produced in the track length.
    
    \item \textbf{Transverse Gaussian Mean}: The mean position of the Gaussian transverse profile, derived by fitting the projection of the cluster's pixel intensities along the major axis of the track, serves as an indicator of the position of the energy depositions concerning the event's original trajectory in the gas. It primarily reflects the position on which most of the energy is released along the track. 
    
    \item \textbf{Transverse Gaussian Sigma}: The standard deviation of the Gaussian transverse profile, derived by fitting the projection of the cluster's pixel intensities along the major axis of the track, serves as an indicator of the dispersion of the energy depositions concerning the event's original trajectory in the gas. It primarily reflects the diffusion occurring within the gas and the GEMs, as well as the curvature of the track.
    
    \item \textbf{Number of pixels}: The count of pixels of the track exhibiting non-zero intensity following the zero suppression method (Sec. \ref{chap:reco}) is directly related (adjusted by the effective pixel size of 152$\times$152 $\mu$m$^2$) to the spatial spread of the initial energy deposition, affected by diffusion during drift and amplification.
    
    \item \textbf{Track size}: Area of the ellipse in which the track is inscribed. It is related to the total size of the track.
    
\end{itemize}
These quantities can provide information about the track physical dimensions, its extension, as well as light distribution along it.\\
In the Monte Carlo simulations, only pure signal is present, thus an estimation of the tracks shapes variable would be straightforward. In the data instead, the signal is immersed in a background component. Therefore, it is necessary a way to isolate the signal component in order to compare it with the MC. For this purpose, the $_{s}\mathcal{P}\textrtaill{lot}$ technique has been used.
The $_{s}\mathcal{P}\textrtaill{lot}$ technique is used to unfold data distribution. Typically in a dataset, data consists of more than two species (e.g. signal and background), the $_{s}\mathcal{P}\textrtaill{lot}$ technique can be used to statistically separate the contribution of the background from the contribution of the signal for uncorrelated variables. The variables of the data can be divided into two classes. The first class includes the discriminating variables, for which the p.d.f. are known. The second class encloses the control variables, for which the p.d.f. is not known, and the distribution must be determined. The $_{s}\mathcal{P}\textrtaill{lot}$ is a method that estimates in a statistical sense the shape of the distribution of the control variables, relying on the fit of the discriminating variables. For simplicity, it will be considered the case in which there are two species composing the dataset: signal ($s$) and background ($b$).  
For the unfolding, a maximum likelihood fit is performed on the discriminating variable, and for each event, two “sWeights” are assigned: $_{s}\mathcal{P}_s(y)$ and $_{s}\mathcal{P}_b(y)$. These two quantities represent respectively the “sWeight” for $s$ and $b$ calculated on the discriminating variable $y$. Their values are proportional to the probability of the event being attributed to either signal or background.
A detailed calculation of this quantity is reported in appendix \ref{app:appendixA}. The distribution of the control variable for signal and background can be determined by constructing the distribution for each class, weighting each event respectively by its sWeight $_{s}\mathcal{P}_s(y)$ and $_{s}\mathcal{P}_b(y)$ \cite{Pivk_2005}.
Since the sWeigths can be positive or negative, the negative sWeigths contribute to the cancellation of the background components from the pure signal distribution, and vice versa.\\
From an operative point of view, in each X-Ray source dataset, the discriminating variable is the light integral. Contemporary to the unbinned likelihood fit, the \lstinline[basicstyle=\large\ttfamily,language=C++]{RooStats::SPlot}  method has been employed to calculate the sWeights for each event in the dataset, which are then used to build the control variables distributions.

\subsection{Topological track shape variables comparison}
\begin{figure}
    \centering
    \includegraphics[width=1\linewidth]{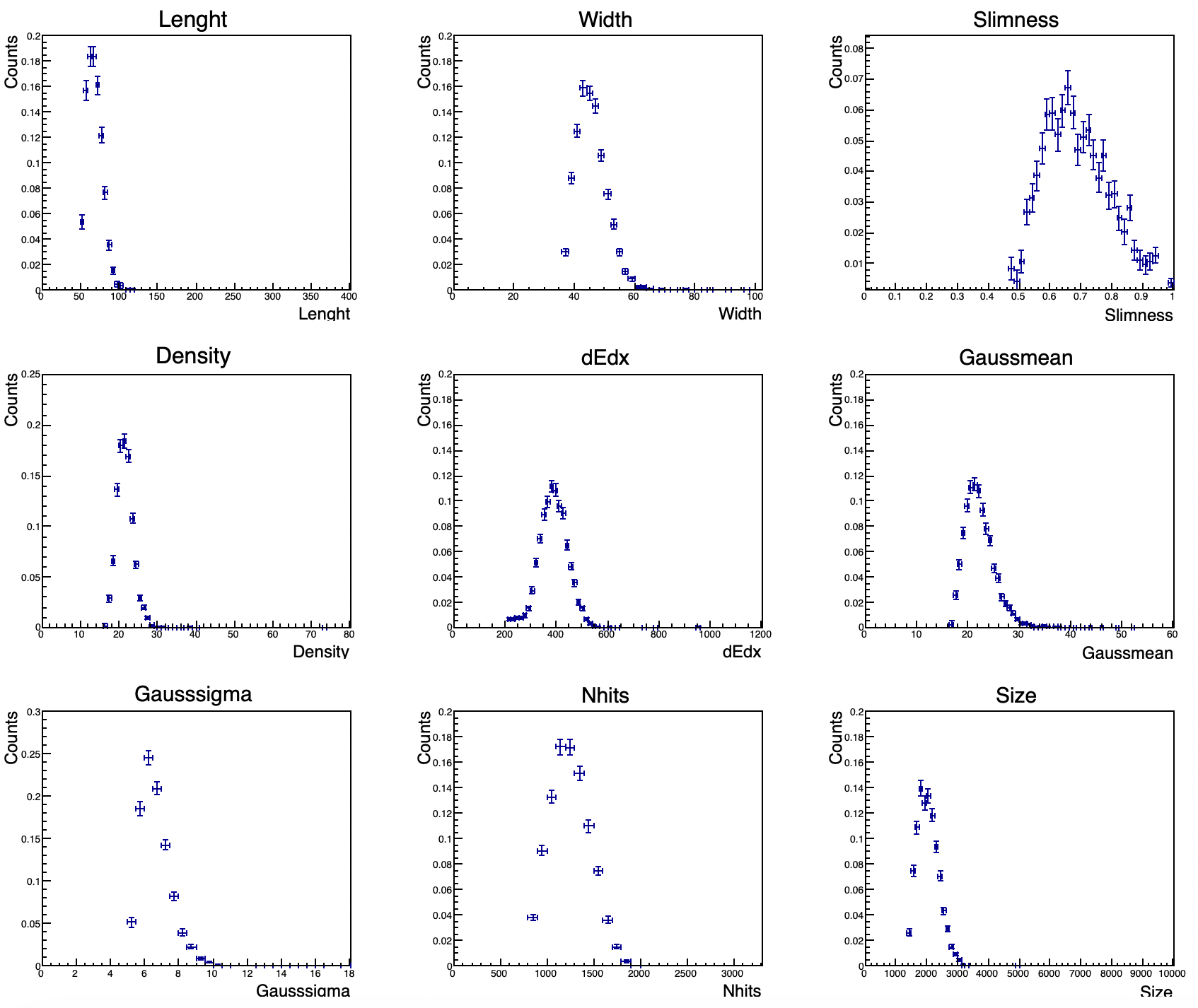}
    \caption{In figure, the unfolded signal distributions for Ag (22.01 keV) run built weighting the quantity of each entry by $_{s}\mathcal{P}_s(y)$ are shown. The distributions are in order from top to bottom and from left to right: track length [px], track width [px], slimness, track density [ADC/px], average specific ionization[ADC/px], Gaussian mean of the transverse profile [px], Gaussian sigma of the transverse profile [px], number of pixels, and track size.}
    \label{fig:SplotAg}
\end{figure}

\begin{figure}
    \centering
    \includegraphics[width=1.0\linewidth]{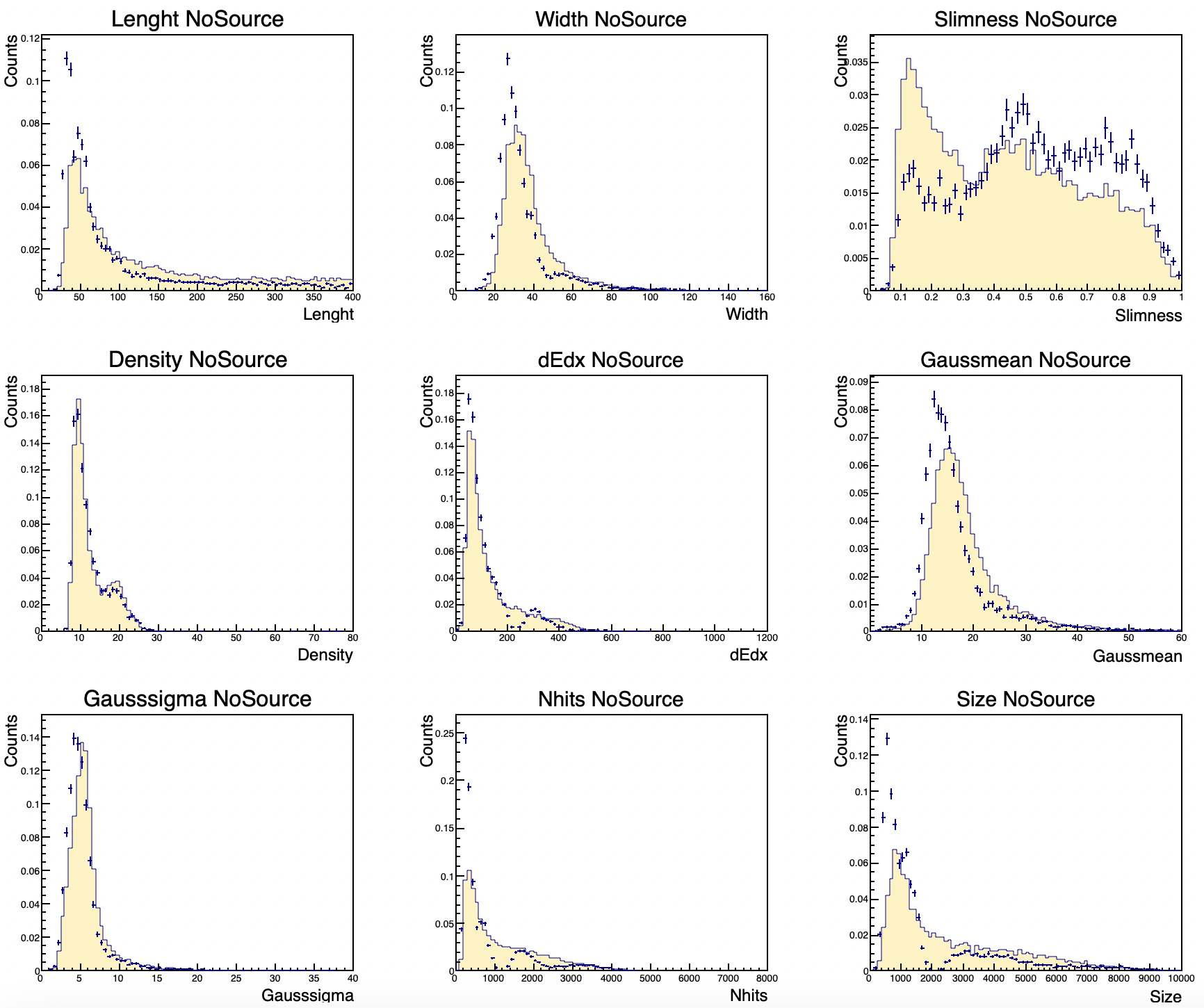}
    \caption{In figure, the unfolded background distributions for the Ag (22.01 keV) run, built weighting the quantity of each entry by $_{s}\mathcal{P}_b(y)$ are shown, and compared with no-source data. The blue cross represents the unfolded background distribution, while the yellow histograms are the no-source data distributions. The distributions are in order from top to bottom and from left to right: track length, track width, slimness, track density, average specific ionization, Gaussian mean of the transverse profile, Gaussian sigma of the transverse profile, number of pixels, and track size.}
    \label{fig:SplotAgBkg}
\end{figure}

\begin{figure}
    \centering    
        \includegraphics[width=1.0\linewidth]{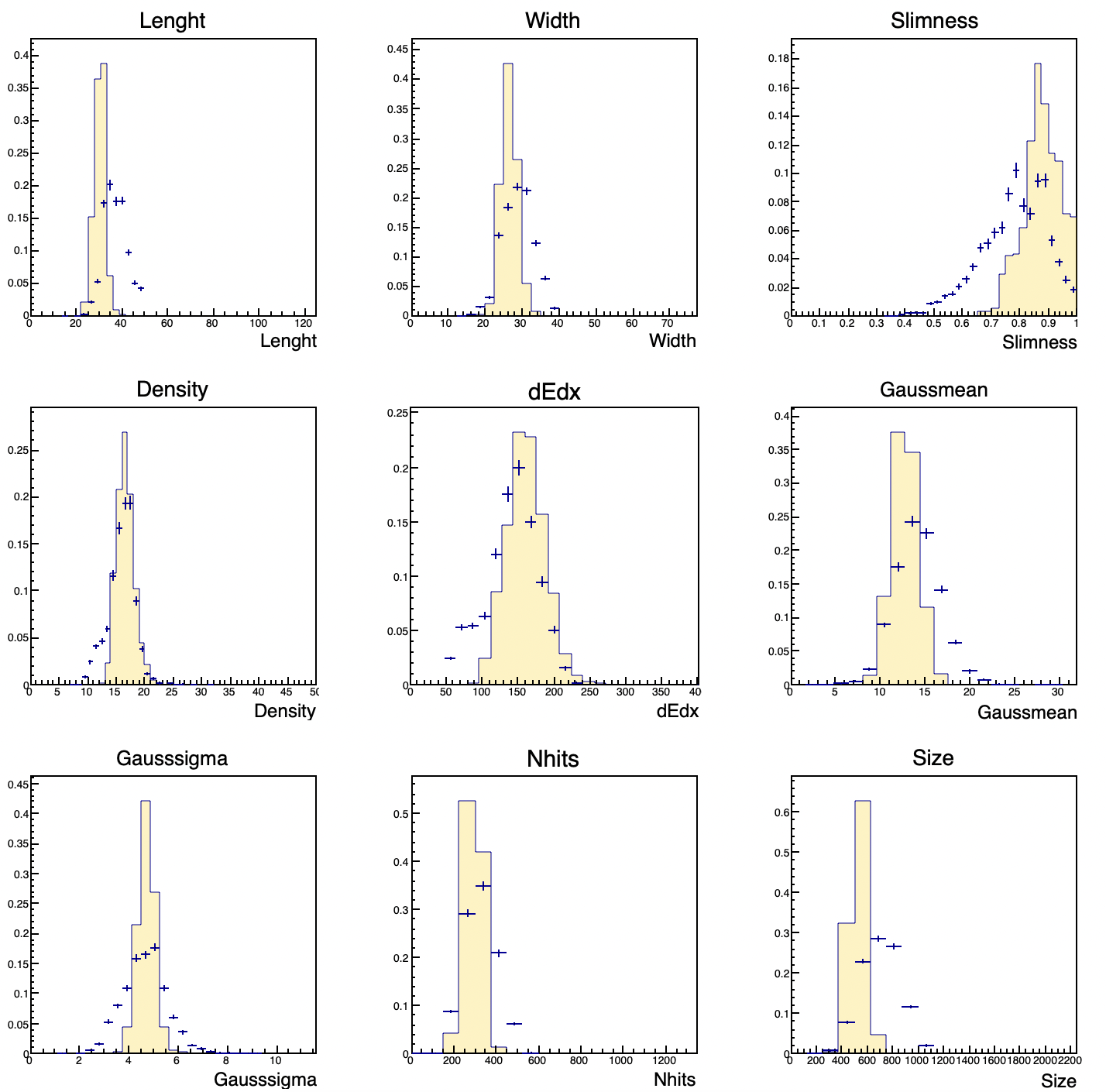}
    \caption{Track shape variables distributions comparison for Ca at 3.7 keV. The histograms are data unfolded signal distributions built weighting the quantity of each entry by $_{s}\mathcal{P}_s(y)$ (blue crosses) and simulated tracks (yellow histograms). The distributions are in order from top to bottom and from left to right: track length, track width, slimness, track density, average specific ionization, Gaussian mean of the transverse profile, Gaussian sigma of the transverse profile, number of pixels, and track size.}
    \label{fig:Cacomp}
\end{figure}

\begin{figure}
    \centering    
        \includegraphics[width=1.0\linewidth]{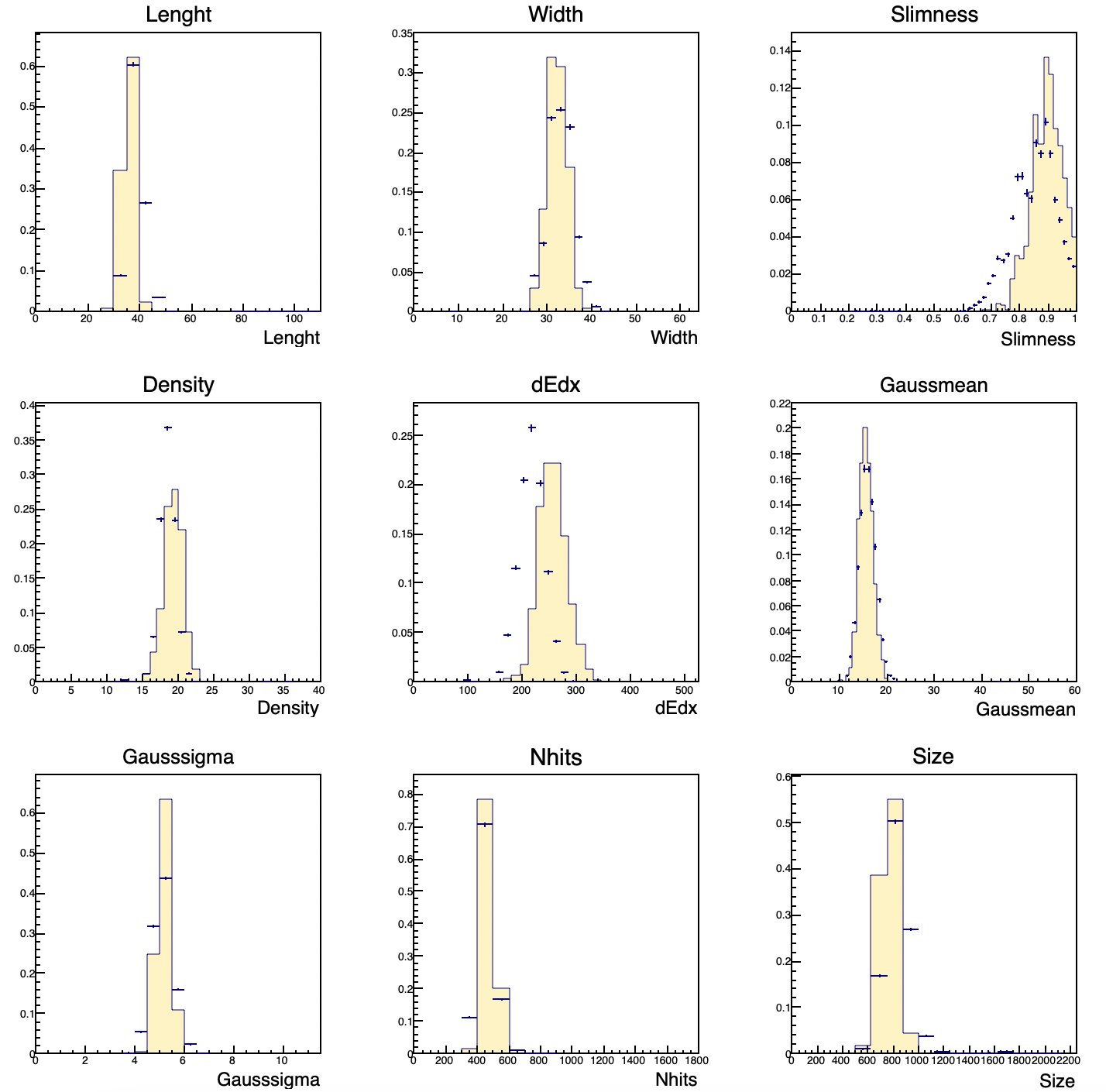}
    \caption{Track shape variables distributions comparison for Fe at 5.9 keV. The histograms are data unfolded signal distributions built weighting the quantity of each entry by $_{s}\mathcal{P}_s(y)$ (blue crosses) and simulated tracks (yellow histograms). The distributions are in order from top to bottom and from left to right: track length, track width, slimness, track density, average specific ionization, Gaussian mean of the transverse profile, Gaussian sigma of the transverse profile, number of pixels, and track size.}
    \label{fig:Fecomp}
\end{figure}

\begin{figure}
    \centering    
        \includegraphics[width=1.0\linewidth]{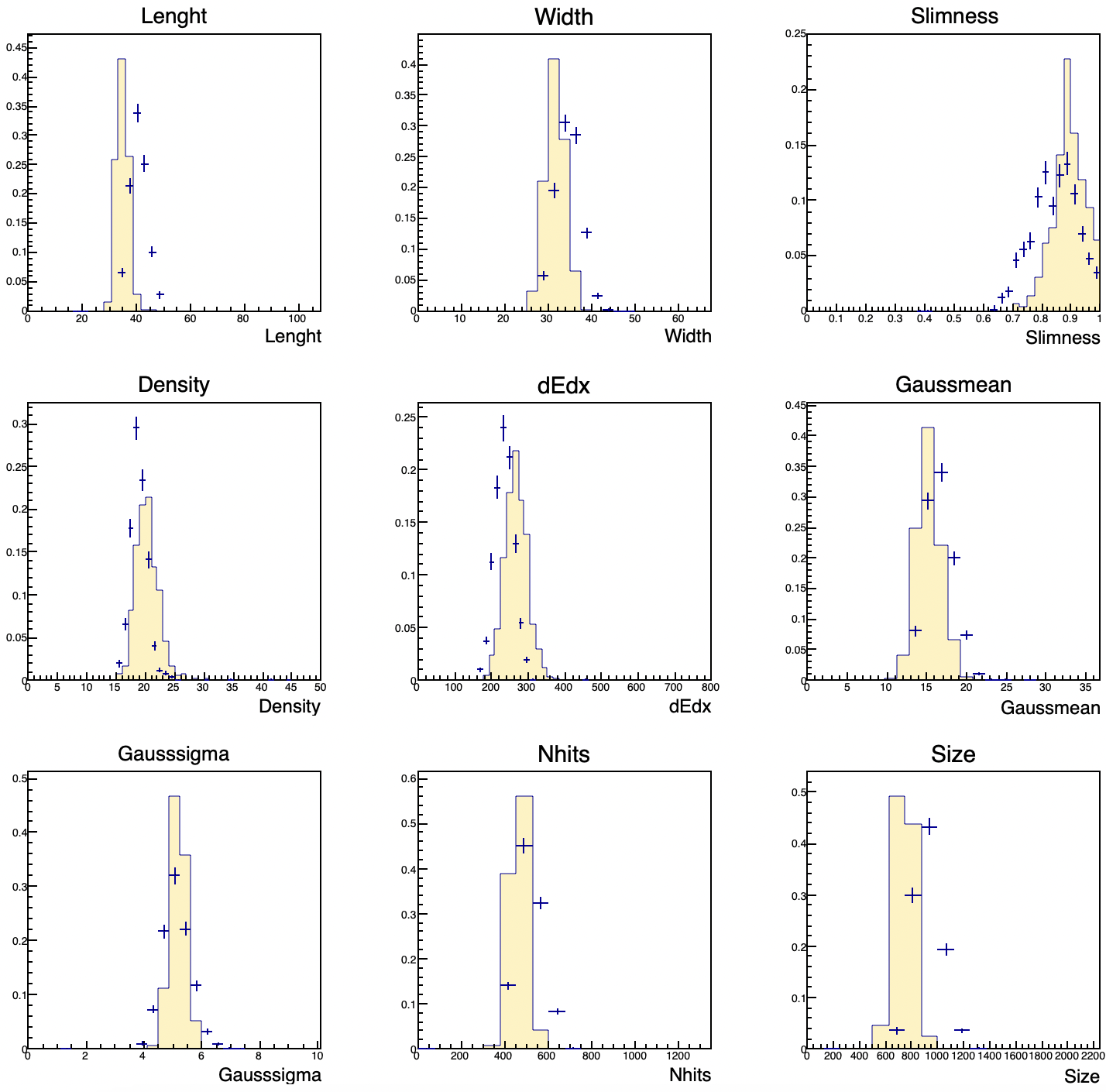}
    \caption{Track shape variables distributions comparison for Cu at 8.04 keV. The histograms are data unfolded signal distributions built weighting the quantity of each entry by $_{s}\mathcal{P}_s(y)$ (blue crosses) and simulated tracks (yellow histograms). The distributions are in order from top to bottom and from left to right: track length, track width, slimness, track density, average specific ionization, Gaussian mean of the transverse profile, Gaussian sigma of the transverse profile, number of pixels, and track size.}
    \label{fig:Cucomp}
\end{figure}

\begin{figure}
    \centering    
        \includegraphics[width=1.0\linewidth]{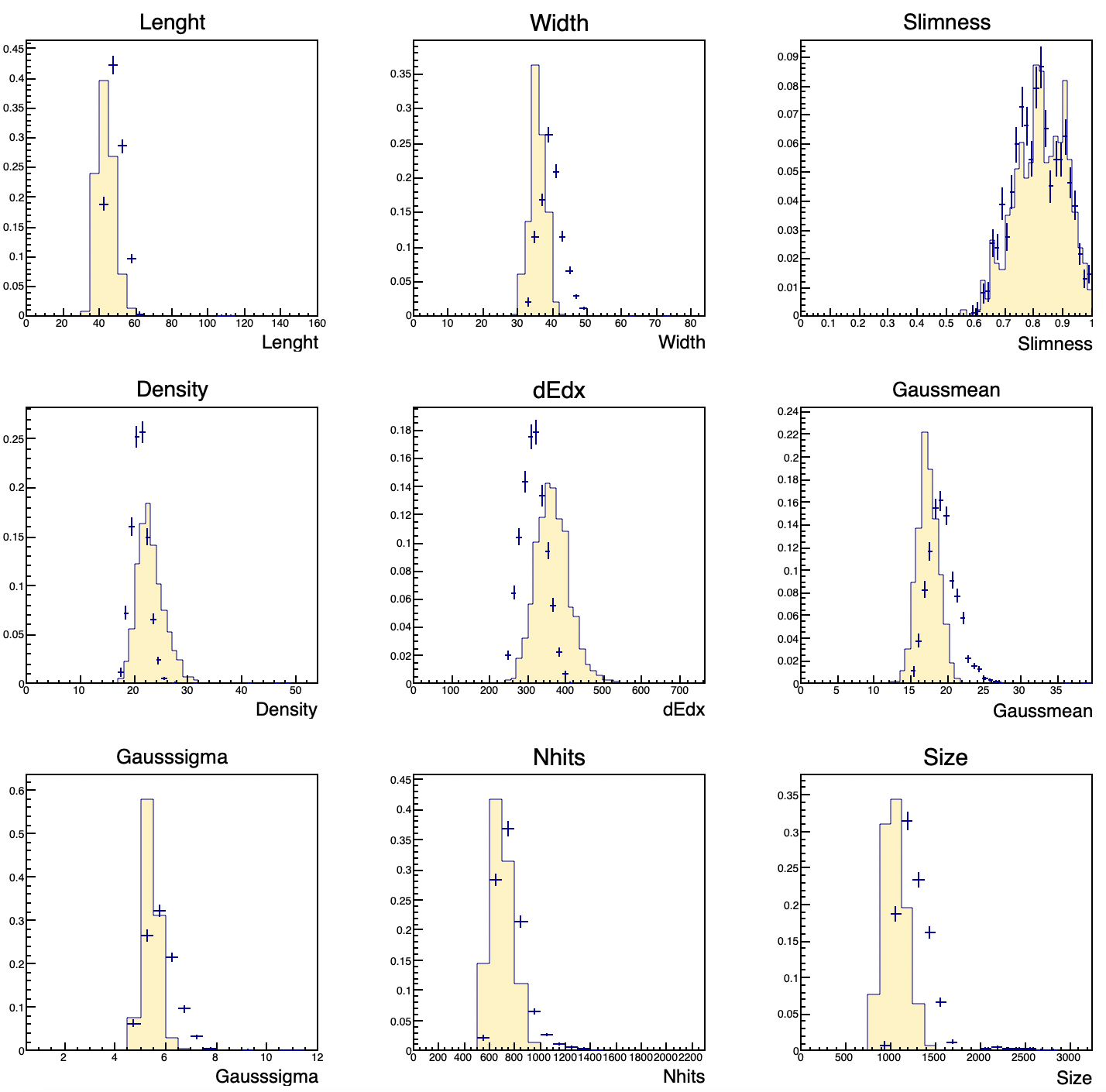}
    \caption{Track shape variables distributions comparison for Rb at 13.37 keV. The histograms are data unfolded signal distributions built weighting the quantity of each entry by $_{s}\mathcal{P}_s(y)$ (blue crosses) and simulated tracks (yellow histograms). The distributions are in order from top to bottom and from left to right: track length, track width, slimness, track density, average specific ionization, Gaussian mean of the transverse profile, Gaussian sigma of the transverse profile, number of pixels, and track size.}
    \label{fig:Rbcomp}
\end{figure}

\begin{figure}
    \centering    
        \includegraphics[width=1.0\linewidth]{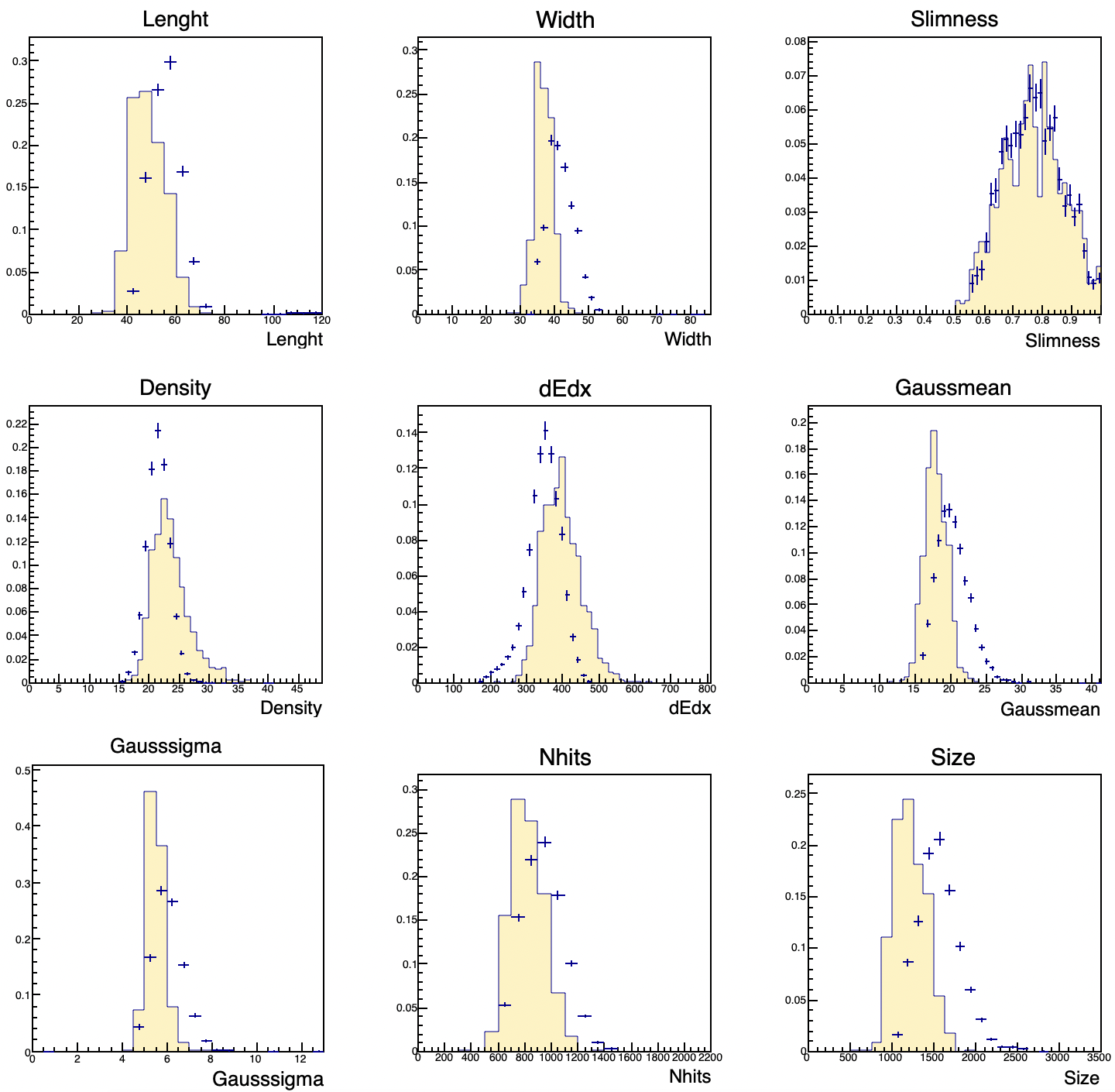}
    \caption{Track shape variables distributions comparison for Mo at 17.44 keV. The histograms are data unfolded signal distributions built weighting the quantity of each entry by $_{s}\mathcal{P}_s(y)$ (blue crosses) and simulated tracks (yellow histograms). The distributions are in order from top to bottom and from left to right: track length, track width, slimness, track density, average specific ionization, Gaussian mean of the transverse profile, Gaussian sigma of the transverse profile, number of pixels, and track size.}
    \label{fig:Mocomp}
\end{figure}

\begin{figure}
    \centering    
        \includegraphics[width=1.0\linewidth]{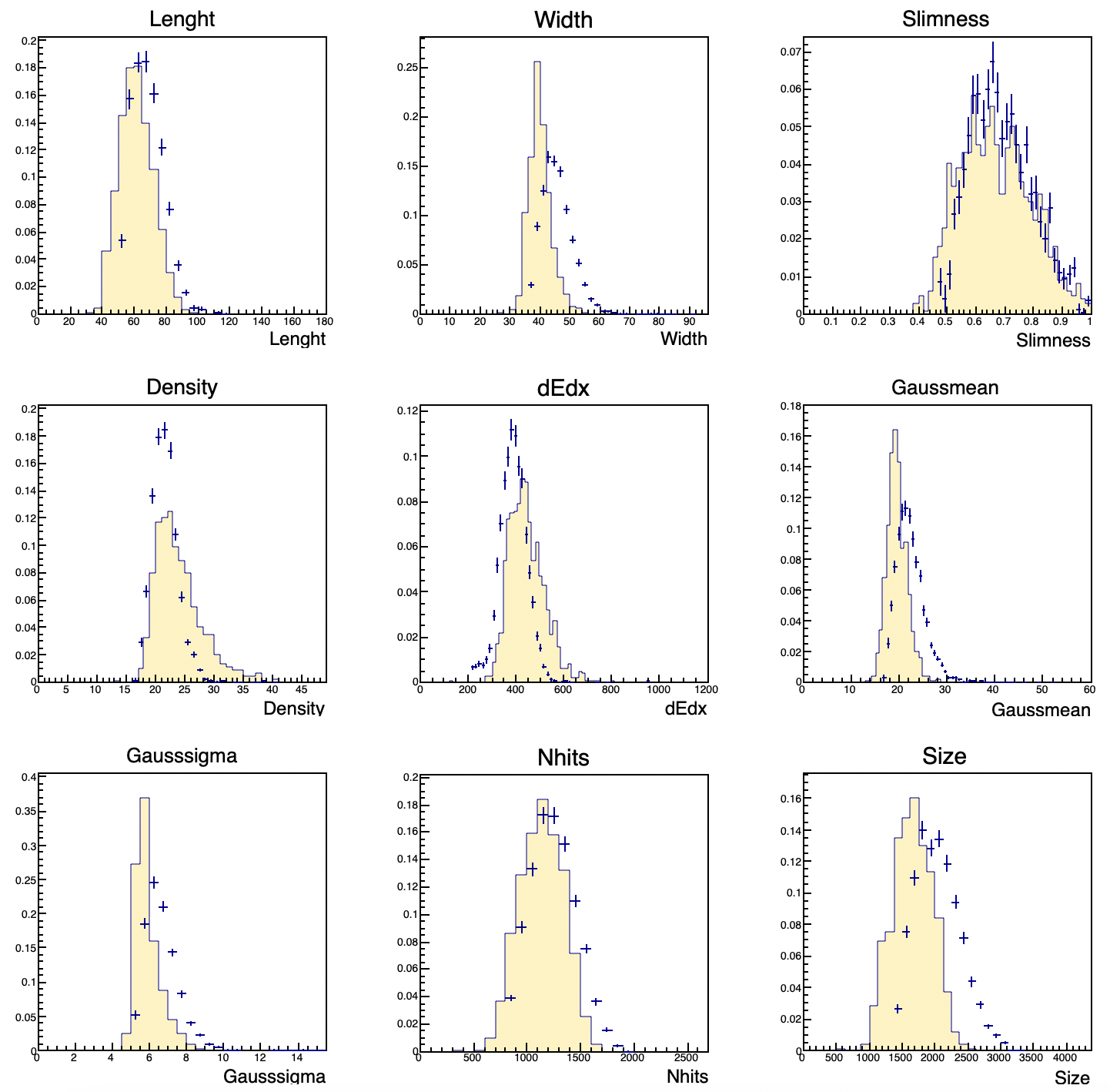}
    \caption{Track shape variables distributions comparison for Ag at 22.01 keV. The histograms are data unfolded signal distributions built weighting the quantity of each entry by $_{s}\mathcal{P}_s(y)$ (blue crosses) and simulated tracks (yellow histograms). The distributions are in order from top to bottom and from left to right: track length, track width, slimness, track density, average specific ionization, Gaussian mean of the transverse profile, Gaussian sigma of the transverse profile, number of pixels, and track size.}
    \label{fig:Agcomp}
\end{figure}

\begin{figure}
    \centering    
        \includegraphics[width=1.0\linewidth]{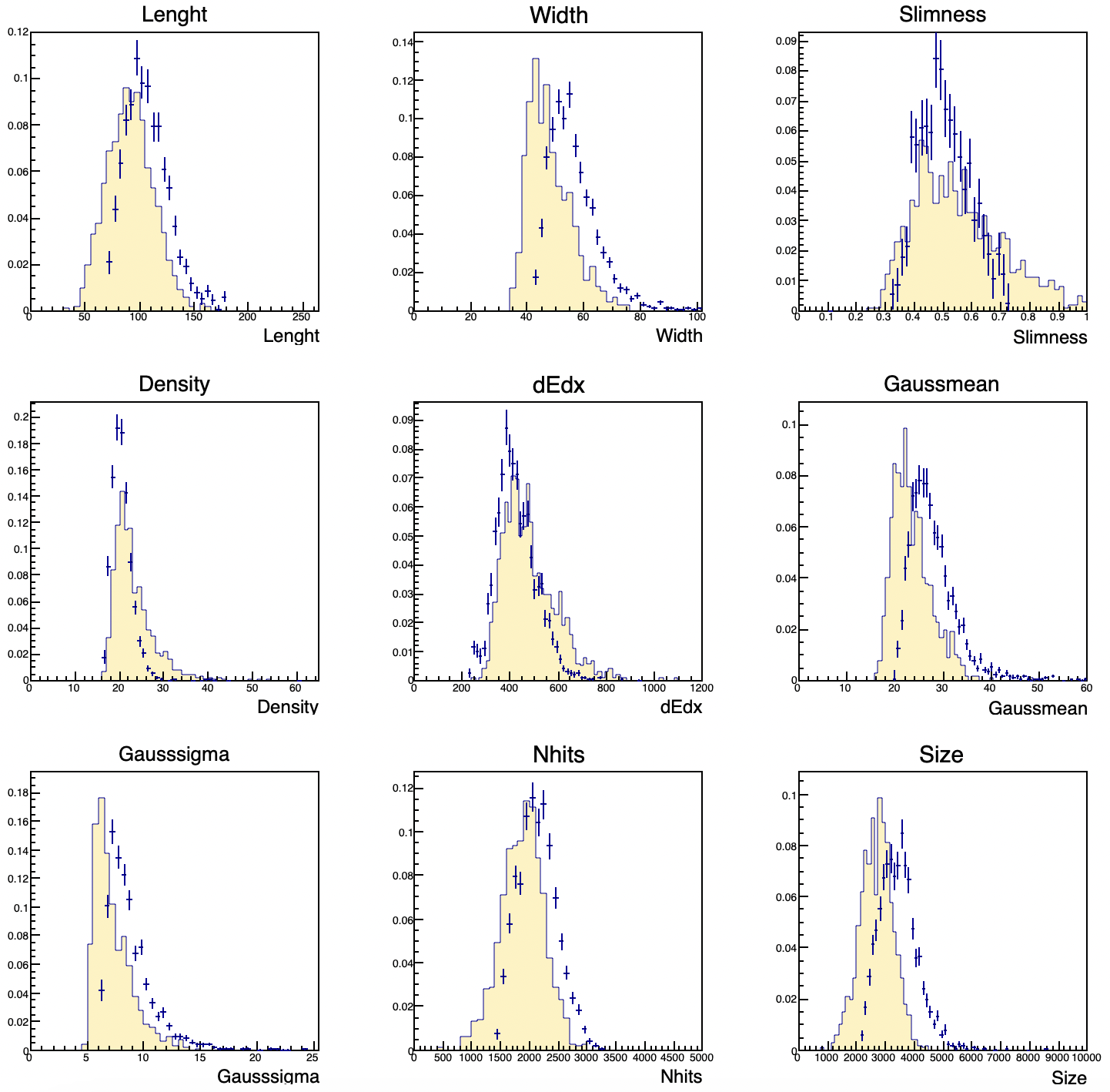}
    \caption{Track shape variables distributions comparison for Ba 32.06 keV. The histograms are data unfolded signal distributions built weighting the quantity of each entry by $_{s}\mathcal{P}_s(y)$ (blue crosses) and simulated tracks (yellow histograms). The distributions are in order from top to bottom and from left to right: track length, track width, slimness, track density, average specific ionization, Gaussian mean of the transverse profile, Gaussian sigma of the transverse profile, number of pixels, and track size.}
    \label{fig:Bacomp}
\end{figure}

\begin{figure}
    \centering    
        \includegraphics[width=1.0\linewidth]{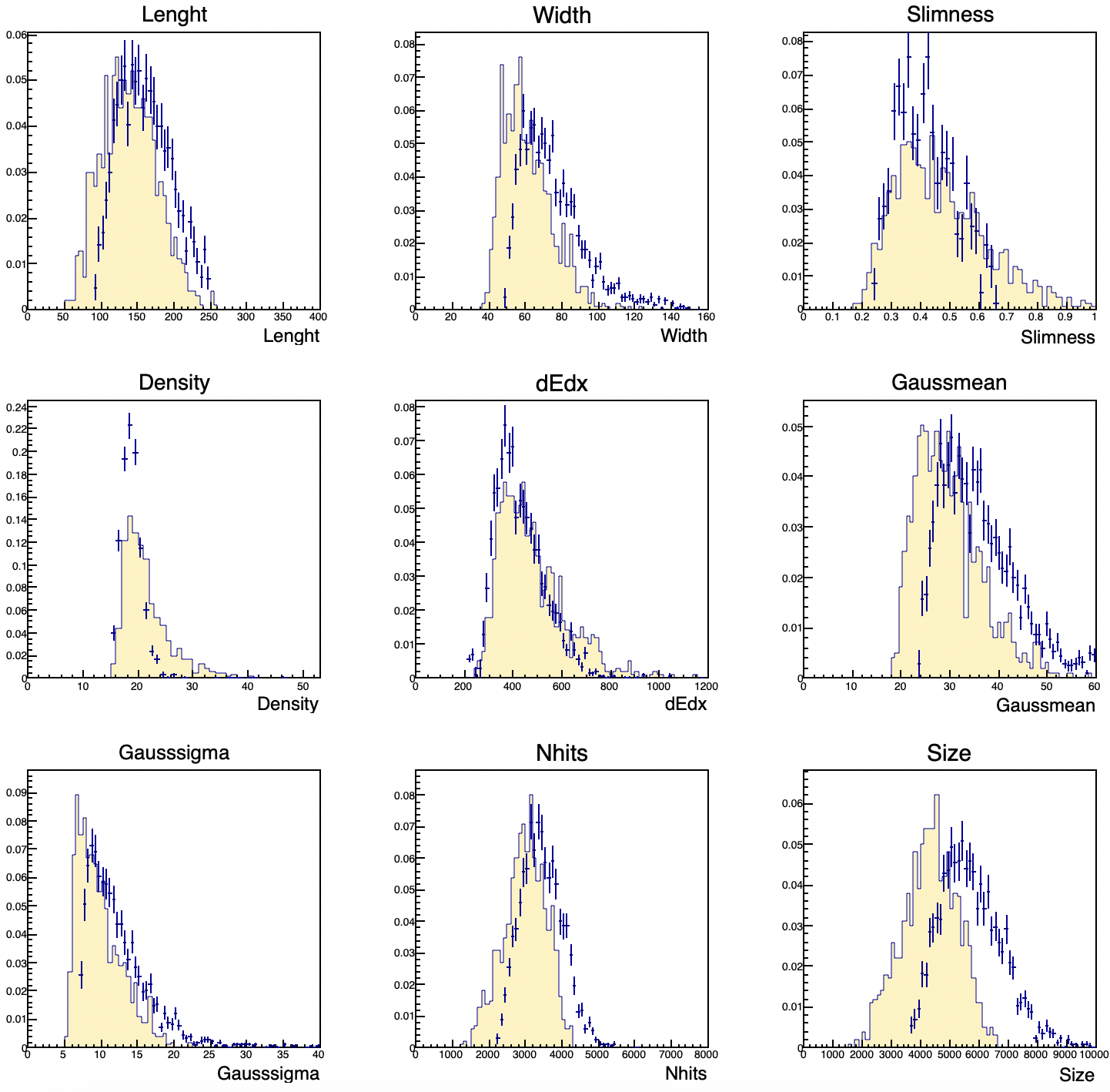}
    \caption{Track shape variables distributions comparison for Tb at 44.48 keV. The histograms are data unfolded signal distributions built weighting the quantity of each entry by $_{s}\mathcal{P}_s(y)$ (blue crosses) and simulated tracks (yellow histograms). The distributions are in order from top to bottom and from left to right: track length, track width, slimness, track density, average specific ionization, Gaussian mean of the transverse profile, Gaussian sigma of the transverse profile, number of pixels, and track size.}
    \label{fig:Tbcomp}
\end{figure}

The topological variables distribution from the Ag (22.01 keV) dataset for the signal component, built by weighting the variable of each event by $_{s}\mathcal{P}_s(y)$ are shown in Fig. \ref{fig:SplotAg} as an example.
Moreover, as a crosscheck, the distribution of pure background unfolded from the Ag run has been compared with the no-source data and a good agreement is found. The plots are shown in Fig. \ref{fig:SplotAgBkg}.
The data-MC comparison has been done for the pure signal component unfolded from the data, and the variables calculated on the simulated tracks. The distributions, normalized to 1, are shown for electron recoils from the different elements from Fig. \ref{fig:Cacomp} to Fig. \ref{fig:Tbcomp}. In all the figures the distributions are in order: track length, track width, slimness, track density, average specific ionization, Gaussian mean of the transverse profile, Gaussian sigma of the transverse profile, number of pixels, and track size.

\begin{figure}
    \centering
    \includegraphics[width=1.\linewidth]{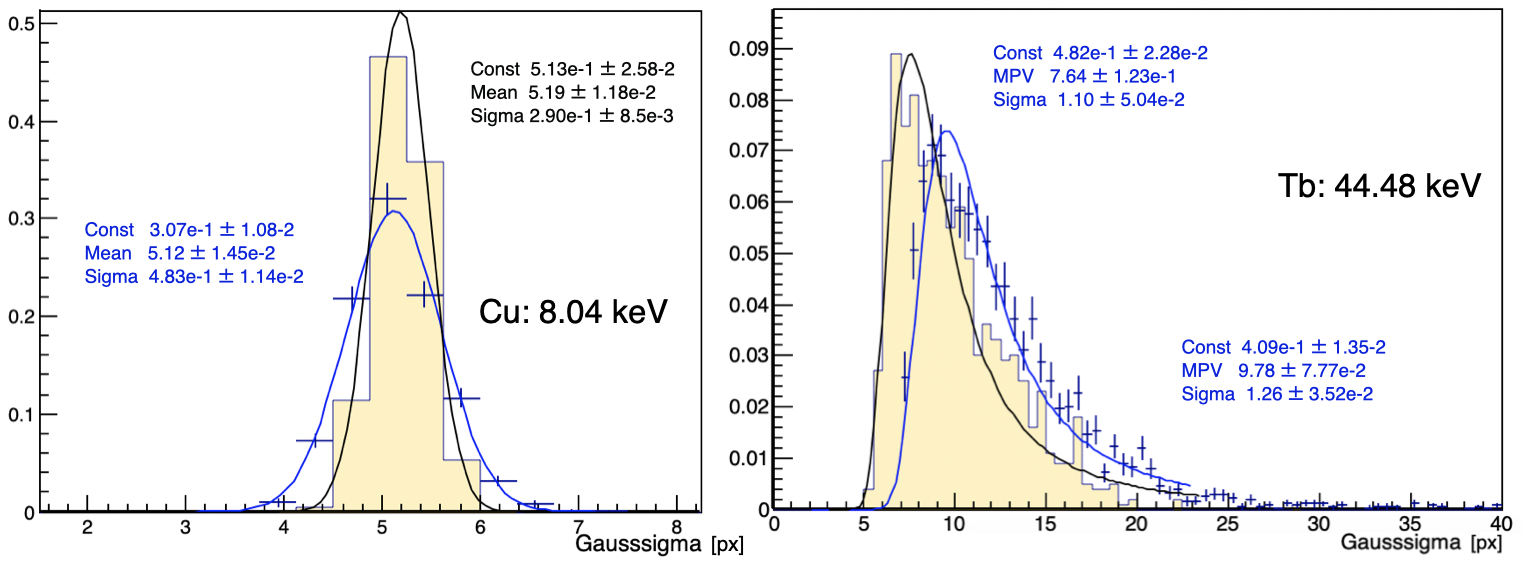}
    \caption{Left: Distributions of the Transverse Gaussian Sigma variable for 8.04 keV electrons. The histograms are data unfolded signal distributions built weighting the quantity of each entry by $_{s}\mathcal{P}_s(y)$ (blue crosses) and simulated tracks (yellow histograms). Both the distribution have been fit with a Gaussian. Left plot: distributions of the same variable for 44.48 keV. The distributions have been fit with a Landau function.}
    \label{fig:gauslandau}
\end{figure}

\begin{figure}
    \centering
    \includegraphics[width=1\linewidth]{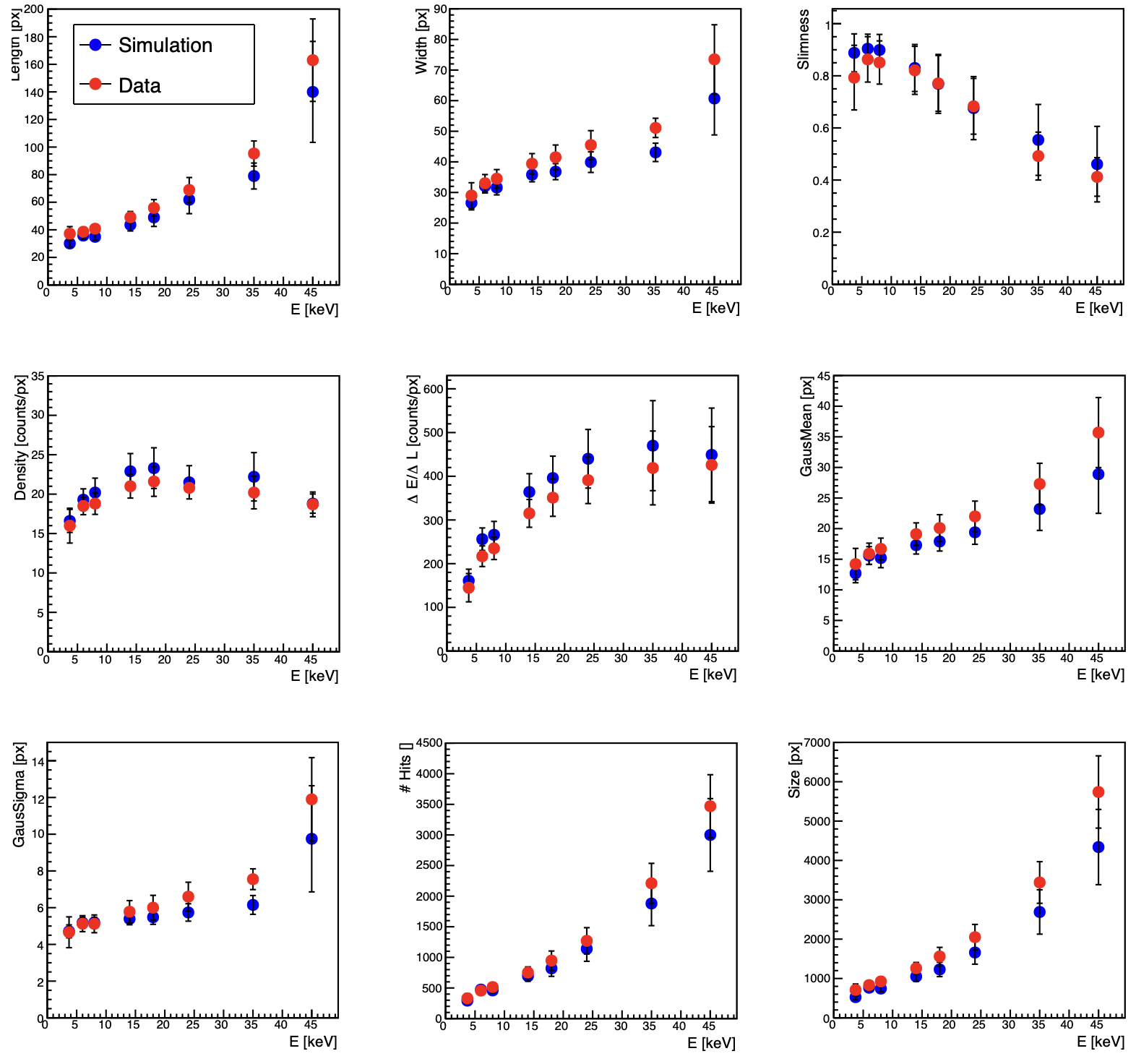}
    \caption{Plot showing the mean values of track shape variables utilized for the data-MC comparison as a function of the energies for the different shape variables considered in the analysis. Blue points depict simulation data, while red points represent actual data.}
    \label{fig:datamccomparisonAll}
\end{figure}

The data and simulation distributions were fitted with either a Gaussian or a Landau function, depending on the case. For the different variables, the mean values with errors, taken as the sigma of the distribution, were then plotted against energy for both data and simulation. Fig. \ref{fig:gauslandau} presents examples of fits at 8.04 keV for the Transverse Gaussian Sigma variable using a Gaussian distribution and at 44.48 keV using a Landau distribution.
The plots in Fig. \ref{fig:datamccomparisonAll} displays the mean values of the distribution along with their associated error represented by the sigma of the fit function as a function of the energy for the different track shape variables.
Despite observing some systematic differences in the distributions between the data and the Monte Carlo simulations, it's meaningful to note the complexity of the simulation, which aims to reproduce the entire electron recoil sCMOS image. Nonetheless, a satisfactory agreement is evident between data and simulation in terms of track shape variables, along with the light response and energy resolution. 
In the future by including other effects in the simulation, such as a modeling of the distortion effect provided by the lenses, a better treatment of the saturation, a better agreement between data and MC can be found.



%% file: chapters/directionality.tex
\chapter{Directionality of low energy electron recoils}
\label{chap:directionality}
This chapter discloses another segment of the author's thesis work, focusing on investigations into the directionality of low-energy electron recoils, as well as examining the angular resolution performance and features of the algorithm.
The reconstruction of the initial direction of a recoil can offer several advantages in physics cases where signal events are expected to have a preferential direction in space as described in detail in Sec. \ref{sec:diradvantages} for neutrinos and in Sec. \ref{sec:diradvantagesDM} for Dark Matter. 
Directional information can in fact be used as an additional handle for background rejection by considering events originating from the particles interaction with a direction compatible with the expected signal. Most importantly, the observation of an excess of events from the expected direction would be a very clear signature for signal identification (Sec.s \ref{sec:diradvantages} and \ref{sec:diradvantagesDM}). 
Ultimately, in the solar neutrino physics case, it enables the reconstruction of the incident particle's initial energy through kinematics, using the direction of the initial particle, as illustrated in Sec. \ref{sec:spectroscopy}.\\
As illustrated in Sec. \ref{sec:nuTPC}, high-precision 3D Time Projection Chamber with high-performance tracking capabilities has been proposed as neutrino detectors, specifically for the solar neutrinos physics case. This is achieved by leveraging on the elastic scattering of solar neutrinos on the gas electrons and measuring the resulting electron recoils (ER) from this interaction. In this context, the ability to measure the direction of the ER and correlate it with the neutrino origin is crucial for discriminating the signal from the isotropic background of electron recoils coming from natural radioactivity.
Measuring the initial direction of low-energy electron recoils in a TPC poses a significant challenge due to the concurrent effect of multiple scattering in the gas and diffusion. The initial direction of the track is rapidly lost due to the electron scattering with the particles in the medium, and the goodness in the determination of the particle interaction point information is affected by the primary charge diffusion.\\
To assess the feasibility of a directional measurement of solar neutrinos with the CYGNO experimental approach, a dedicated algorithm has been developed and characterized, as part of this thesis work, to determine the initial direction of low-energy electron recoils and study the angular resolution obtainable with the detector. Given the difficulties in finding a source of low-energy electron recoil with a known direction, the analysis has been performed on simulated tracks produced with the simulation described in Chap. \ref{chap:Simulation}, which shows a good agreement with LIME data (Sec. \ref{sec:datamccomparison}) in light response, and energy resolution, but also in the track shape variables. As described in Sec. \ref{sec:CYGNOFut}, the construction of a CYGNO-30 experiment will be done in a modular way, employing detectors with identical drift lengths and amplification systems, and comparable granularity to LIME (Section \ref{sec:LIME}). Hence, it is logical to assess the directional capabilities of a CYGNO-30 experiment using simulated tracks generated by a simulation capable of accurately replicating all significant characteristics of ER events in LIME.
A detailed simulation of the PMT response in the CYGNO experimental approach is still under development by the collaboration to these days, thus this study is carried on the sCMOS images only.\\
In this chapter, an examination of the behavior of low-energy electrons interacting in gas will be presented (Sec. \ref{sec:electronspattern}). Following this, an in-depth description of the algorithm developed by the author for reconstructing the initial direction of low-energy electron recoils will be outlined (Sec. \ref{sec:dirAlgo}). Additionally, the process of optimizing algorithm parameters that was performed for this thesis will be detailed (Sec. \ref{sec:paroptimization}), along with the results of angular resolution on low-energy electron recoils (Sec. \ref{sec:angResResults}). Furthermore, an assessment of the algorithm's efficiency has been conducted, along with an exploration of the factors contributing to its inefficiency, and proposals for enhancements (Sec. \ref{sec:algoeff}). Subsequently, as detailed in Chapter \ref{sec:NID}, the collaboration successfully operated the small prototype under Negative Ions Drift operation with optical readout, measuring also the diffusion achievable using Negative Ions as charge carriers. In the subsequent part of the chapter, the results regarding angular resolution for electron recoils simulated with NID diffusion will be outlined (Sec. \ref{sec:NIDAngReso}). The chapter will conclude with a final discussion on the directionality performances (Sec. \ref{sec:Discussion}). To the best of the author's knowledge, this is the first time that the capability of reconstructing the initial directions of low-energy electron recoils within this energy range has been evaluated.

\section{Electron recoils energy loss and straggling}
\label{sec:electronspattern}
\begin{figure}
    \centering
    \includegraphics[width=0.7\linewidth]{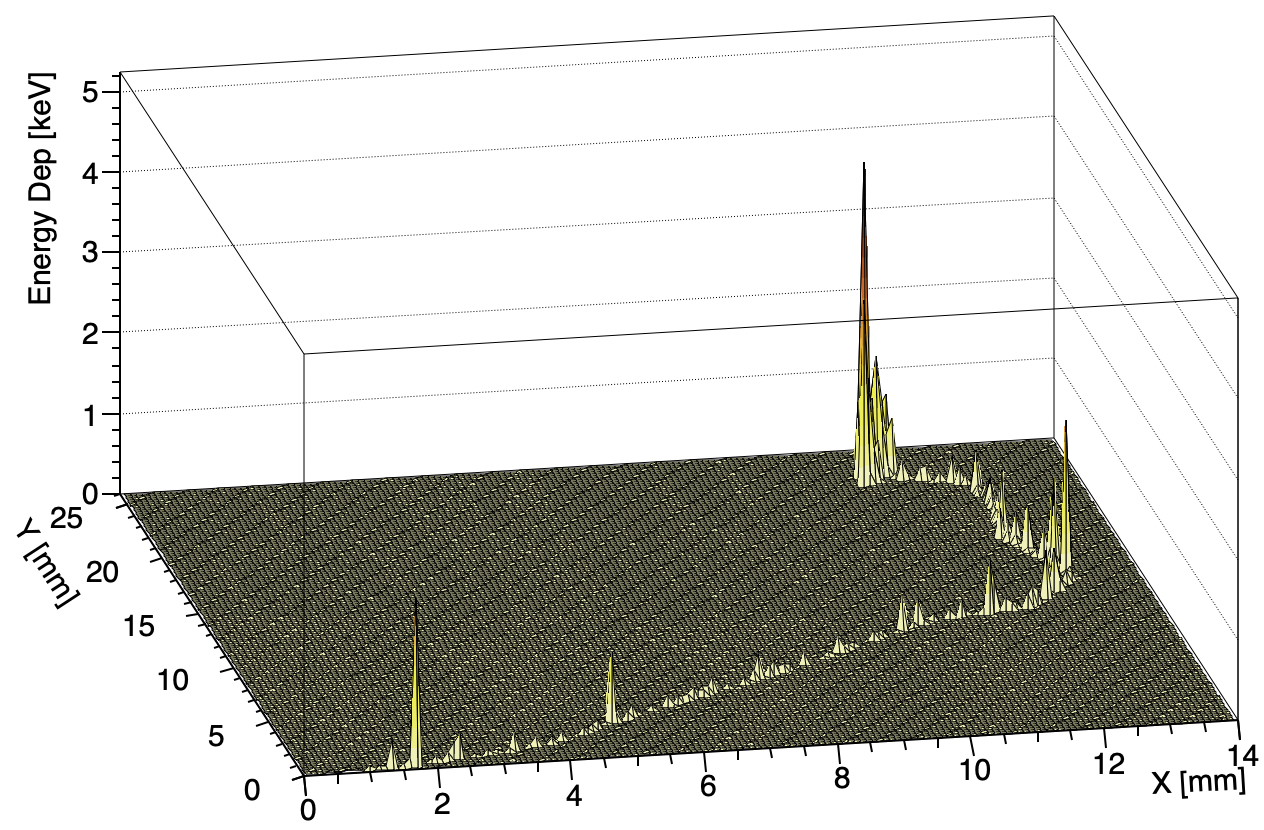}
    \caption{2D projection of a GEANT4 simulated 50 keV electron track produced with initial direction along the positive side of the x-axis. The amplitude of the peaks, extending along the z direction, is proportional to the energy deposited in each point.}
    \label{fig:ER50keV}
\end{figure}

At low energies, electrons primarily dissipate their energy through collisions with the atomic electrons of the medium. Other mechanisms such as radiation losses or Bremsstrahlung become significant only for electron energies well above 1 MeV. When considering electrons, two factors must be considered to adapt the Bethe-Bloch formula (4.1) for accurate energy loss predictions. Firstly, the assumption that the particle's mass is much greater than that of the orbital electrons is invalid, leading to significant changes in the particle's trajectory after each collision. Secondly, the collisions involve interactions between identical, indistinguishable particles. Specifically, the maximum transferable energy in each collision is given by $W_{max}=\frac{E_{kin}}{2}$, where represents the kinetic energy of the incoming electron. The Bethe-Bloch formula for electrons is then expressed as \cite{leo1994techniques}:
\begin{equation}
    -\frac{d E}{d x}=k \frac{1}{\beta^2}\left[\ln \frac{\gamma^2 m_e^2 c^2 v^2(\gamma-1)}{2 I^2}-\frac{2 \gamma-1}{\gamma^2} \ln 2+\frac{8+(\gamma-1)^2}{8 \gamma^2}-\delta-2 \frac{C}{Z}\right]
    \label{eq:PDGEnergyLossInMatter}
\end{equation}
where $k=2\cdot \pi N_A r_e^2m_ec^2\rho Z/A$, $r_e$ is the classical electron radius, $m_e$ is the mass of the electron, $N_A$ is the Avogadro number, $I$ is the ionization potential of the medium, $Z$ is the atomic number of the material, $A$ is its atomic mass, $\rho$ is the mass density of the target material, $z$ is the charge of the incident particle in unit of electron charge, $\beta=v/c$ is the ration between the particle velocity and the speed of light, $\gamma =1/\sqrt{1-\beta^2}$ is the Lorentz factor, $\delta$ is the density correction, and $C$ is the shell correction.
The formula in Eq. \ref{eq:PDGEnergyLossInMatter} should include corrections of elastic collisions with atomic electrons, albeit with significantly lower probability.
As evident from Eq. \ref{eq:PDGEnergyLossInMatter}, the predominant component at low energies is the $1/\beta^2$ term. Consequently, as the electron dissipates energy, its velocity decreases, leading to further higher energy loss. This behavior results in a peak of energy released from the electron at the end of the track called Bragg-Peak. As a consequence of the energy loss, a charged particle moving through a medium experiences numerous small-angle scatters, primarily caused by Coulomb scattering with gas molecules. In the case of many small-angle scatters, both the net scattering and displacement distributions follow a Gaussian pattern. However, occasional “hard” scatters, although less frequent, result in non-Gaussian tails. The Coulomb scattering distributions in these scenarios are described by Molière's theory \cite{PhysRev.89.1256}.
The combination of these effects gives rise to low-energy electron recoil tracks, as depicted in Fig. \ref{fig:ER50keV}. The track is simulated using GEANT4, which accounts for all the effects of electrons scattering in gas. The figure depicts the 2D projection of the electron recoil generated within the CYGNO gas mixture, with the height of the peaks proportional to energy deposited along the track direction. The recoil has been produced with an initial direction along the positive direction of the x-axis.
Observations from the track highlight three crucial points. Firstly, merely tracing a straight line from start to end is not sufficient to establish the recoil direction due to the multiple scattering effect. Secondly, track direction inference is reliable only in the initial part of the track, and the information is rapidly lost thereafter. Thirdly, the variable ionization pattern along the track, culminating in a Bragg-peak-like energy release at the end, offers a method to discern the head and tail of the track.\\
Consequently, a dedicated algorithm leveraging these features has been developed to establish the initial direction of the electron recoil and is illustrated in Sec. \ref{sec:dirAlgo}.  

\section{The directionality algorithm}
\label{sec:dirAlgo}
The algorithm developed in this thesis to infer the initial direction of ER of interest for the directional detection of solar neutrinos has been adapted and optimized starting from a strategy originally developed for X-Ray polarimetry \cite{AstroXRayPol}. X-Ray polarimetry is a field of physics that aims at measuring the polarization of X-Rays coming from astrophysical sources, since this information can provide significant insight into magnetic fields of astrophysical objects. The primary interaction mechanism for photons with energies in the range of a few keV is the photoelectric effect. Specifically, in the case of K-shell absorption, the emission direction of the photoelectron is completely modulated around the polarization axis of the photon, following a cosine squared function of the azimuth angle \cite{2017SPIE10397E..0FS}. Measuring the direction of the electron emitted is thus equivalent to measuring the polarization of the X-Ray if the incoming direction is known \cite{Costa_2022}. Thus, in this field, algorithms have long been developed to accomplish this task.
As shown in \cite{BALDINI2021102628} the algorithm was implemented on tracks detected by the Gas Pixel Detector (GPD) on board of the  Imaging X-Ray Polarimetry Explorer (IXPE). IXPE is a cosmic X-Ray observatory, developed by National Aeronautics and Space Administration (NASA) and Agenzia Spaziale Italiana (ASI), equipped with three identical telescopes aimed at measuring X-Ray polarization. These telescopes consist of three co-aligned mirrors mounted on a 4-meter extender arm, aligned with the spacecraft's pointing axis, directing incoming X-Rays onto three gas pixel detectors \cite{Christensen_2022}. The GPD features a dimension of 15$\times$15 mm$^2$ and a drift length of 10 mm. The detector, filled with He:DME 60:40 (dimethyl ether) at 800 mbar, uses a single GEM as amplification, and it is readout by a custom developed pixelated Application-Specific Integrated Circuit (ASIC) with $\sim$300 pixels with a hexagonal pattern and a 50 $\mu$m pitch covering the side of the GEM \cite{AstroXRayPol}. Due to the single amplification stage and the very small drift length, the Gas Pixel Detector is capable of detecting tracks with directional sensitivity starting from 2 keV. However, due to the constrained size of the sensitive volume, its detection range extends up to 8 keV. Above that energy tracks are not contained anymore. \\ In a detector with the size of LIME, tracks would exhibit a broader energy range (10s-100s keV), resulting in a more various track topology. Therefore, an adaption of the algorithm has been done to correctly handle this wider range of typologies.\\
To determine the initial track direction, the first step is identifying the beginning of the track, i.e. the interaction point (Sec. \ref{sec:IPSearch}). This involves locating the track's extremities, and, as explained in Sec. \ref{sec:electronspattern}, exploiting the asymmetry in the energy release to determine the beginning of the energy release. Once the interaction point is determined, the track's direction is calculated following the light distribution profile from this point, focusing on a small portion of the track before the information is lost due to multiple scattering (Sec. \ref{sec:recoildirection}). \\ 
As illustrated in the introduction, the directionality algorithm has been developed, optimized and tested on a sample of MC simulated ER (Chap. \ref{chap:Simulation}) in a LIME-like detector reconstructed with the procedure discussed in Chap. \ref{chap:reco}. To this aim, the full information of the light intensity measured by each pixel associated with the reconstructed track, together with its x-y position in the image, has been used.

\subsection{Track major axis determination}

\label{sec:MAxis}
\begin{figure}
    \centering
    \includegraphics[scale=.3]{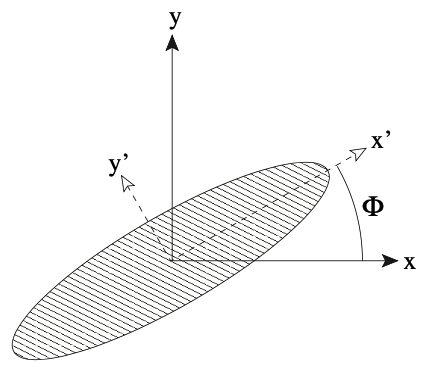}
    \caption{Reference frame used for the calculation of the second moment of the track light distribution. The track is schematized as an elliptical distribution, and the reference frame with the prime index is rotated of an angle $\Phi$ with respect to the main frame around the track barycenter. }
    \label{fig:refFramel}
\end{figure}

To evaluate the track interaction point (IP), the initial step involves the calculation of the track light barycenter, formulated as follows: 
\begin{equation}
    x_c = \frac{\sum_{i=1}^{N_{pix}}{Q_i\cdot x_i}}{\sum_{i=1}^{N_{pix}}{Q_i}} \ \ \ \ \  y_c = \frac{\sum_{i=1}^{N_{pix}}{Q_i\cdot y_i}}{\sum_{i=1}^{N_{pixels}}{Q_i}}
    \label{eq:barycenter}
\end{equation}
where the $Q_i$ represents the light counts inside pixel $i$, $x_i$ and $y_i$ are the coordinates of pixel $i$, and $i$ ranges from 1 to the total number of pixels associated to the track by the reconstruction algorithm, $N_{pix}$.
The track major axis is the determined by finding the line $l$, passing through the barycenter, for which the second moment $M_2$ of the distribution obtained by projecting the track pixels intensity on $l$, is maximum.
In order to do this, a new reference frame is defined by rotating the x-axis of angle $\Phi$ around the track barycenter, as illustrated in Fig. \ref{fig:refFramel}. In this reference frame, the second moment of the distribution can be written as:
\begin{equation}
    M_2(\Phi) = \frac{ \sum_{i=1}^{N_{pix}}{Q_i\cdot (x'_i)^{2}} }{ \sum_{i=1}^{N_{pix}}{Q_i} }
\end{equation}
By rewriting the rotated coordinate ($x'$, $y'$) as a function of the original reference frame ($x$,$y$) and the rotation angle $\Phi$, it is obtained for $x'$:
\begin{equation}
x'= (x-x_c)cos(\Phi) + (y-y_c)sin(\Phi) 
\end{equation}
M2 can be expressed as:
\begin{equation}
    M_2(\Phi) = \frac{ \sum_{i=1}^{N_{pix}}{Q_i\cdot[ (x_i-x_c)cos(\Phi)+(y_i-y_c)sin(\Phi) ]^2 } }{ \sum_{i=1}^{N_{pix}}{Q_i} }
\end{equation}
The major axis can hence be found by determining the angle $\Phi$ that maximize the M2($\Phi$) distribution. This can be inferred by taking the derivative of M2($\Phi$) with respect to $\Phi$ and evaluate the solution for $\Phi_0$ when the derivative is equal to zero.
\begin{equation}
    \frac{d M_2(\Phi)}{d\phi}=0\ \ \  \Rightarrow \ \ \ \phi_0=-\frac{1}{2}\arctan\left(  \frac{2\sum_i{Q_i (x_i-x_c)(y_i-y_c)}}{ \sum_i{ Q_i[(y_i-y_c)^2 -(x_i-x_c)^2 ] } }   \right)
\end{equation}
The angle $\Phi_0$ ranges from -$\pi$/4 to $\pi$/4 and represents either the maximum or minimum value of the second moment within this angular span. The value of M2($\Phi_0$) is then compared to M2($\Phi_0$-$\pi$/2) in the perpendicular direction to determine whether $\Phi_0$ corresponds to the maximum or minimum of M2($\Phi$).
\begin{align}
    M_2(\Phi_0) &= \frac{  \sum_i{  Q_i[ (x_i-x_c)\cos(\Phi_0) + (y_i-y_c)\sin(\Phi_0) ]^2  }  }{  \sum_i{Q_i}  } \\
    M_2(\Phi_0-\pi/2) &= \frac{  \sum_i{  Q_i[ -(x_i-x_c)\sin(\Phi_0) + (y_i-y_c)\cos(\Phi_0) ]^2  }  }{  \sum_i{Q_i}  }
\end{align}
It can be further noticed that if $\Phi_0$ corresponds to the minimum of $M_2(\Phi)$, then by construction $\Phi_0-\pi/2$ will correspond to the maximum and vice versa. Fig. \ref{fig:trackBarMa} shows a simulated 30 keV ER with the calculated barycenter and main axis. From now on, the $\phi_0$ corresponding to the solution for the maximum of $M_2(\Phi)$ will be referred to as $\Phi_{max}$.\\

\subsection{Interaction point determination}
\label{sec:IPSearch}
\begin{figure}
    \centering
    \includegraphics[scale=.4]{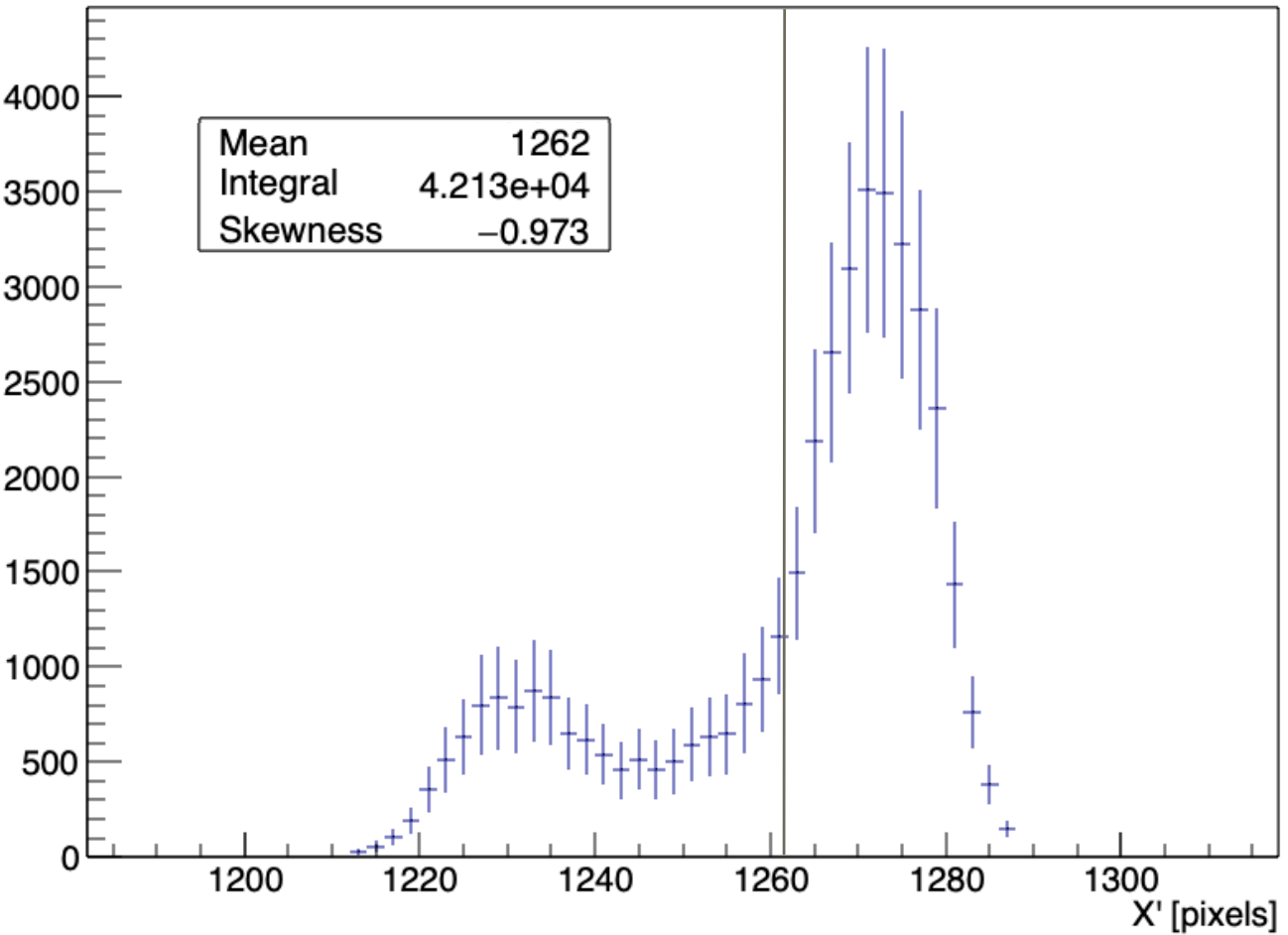}
    \caption{Light distribution of the electron projected along the main axis. The gray line represents the barycenter position along the axis.}
    \label{fig:lightdistrib}
\end{figure}

Once the main track axis has been determined with the procedure described in Sec. \ref{sec:MAxis}, the initial interaction point along this is inferred by exploiting the way in which ERs release energy along their path.
\begin{figure}
    \centering
    \includegraphics[scale=.3]{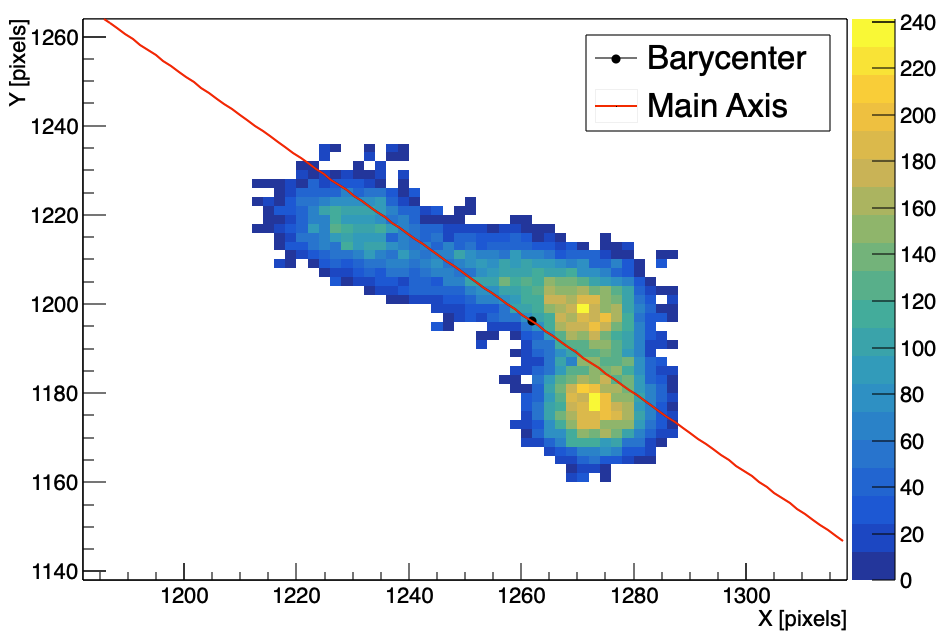}
    \caption{In figure, a simulated 30 keV electron recoil with its barycenter and main axis calculated is shown.}
    \label{fig:trackBarMa}
\end{figure}
Electron recoil at low energy (below the MIP energy, with $\beta \gamma \lesssim 3$), according to Eq. \ref{eq:PDGEnergyLossInMatter} releases their energy in the gas according to the low 
\begin{equation}
    -\frac{dE}{dx} \propto \frac{1}{\beta^2} \propto \frac{1}{E_e}
\end{equation}
This implies that as the energy of the electron recoil decreases, its energy loss becomes more pronounced. Consequently, a significant amount of energy is released over a very short distance at the end of the track. As a result of this, the electron recoil tracks will exhibit a tail with a low light density, and a head with a higher light density. This can be appreciated in Fig. \ref{fig:trackBarMa}, where it can be distinctly observed that one end of the track displays a much brighter light intensity with respect to the other. This localized release of energy is typically referred to as the Bragg peak.
This energy release asymmetry can be quantified by evaluating the third moment of the distribution along the main axis of the track with respect to the barycenter, also known as skewness $M_3$.
The third moment $M_3$ of the light distribution projected on the main axis with respect to the barycenter along $l$, can be expressed as:
\begin{equation}
    M_3(\Phi) = \frac{\sum_{i}{Q_i (x'_{i})^{3} } }{\sum_{i}{Q_i}} = \frac{ \sum_{i}{Q_i\cdot [ (x_i-x_c)\cos(\Phi)+(y_i-y_c)\sin(\Phi) ]^3 } }{ \sum_{i}{Q_i} }
\end{equation}
As an example, the light distribution of the ER in Fig. \ref{fig:trackBarMa} projected along the main axis (shown as the red line in the same figure) is displayed in Fig. \ref{fig:lightdistrib}, clearly showing a negative skewness.
The measurement of skewness allows to determine which points lie on the opposite side of the Bragg peak with respect to an axis passing through the barycenter, and perpendicular to the major axis of the previously defined track.
Once the value of $M_3$ is calculated, each point $i$ on the opposite side of the Bragg peak with respect to the barycenter can be identified by the condition:
\begin{equation}
    \frac{x'_i}{M_3}<0\ \ \ \  \Rightarrow \ \ \ \frac{  (x_i-x_c)\cos(\Phi_{max}) + (y_i-y_c)\sin(\phi_{max})  }{  M_3  }<0 
\end{equation}
This condition is specifically met by every point on the opposite side of the Bragg peak region, with respect to a line perpendicular to the main axis passing through the barycenter.\\
Once identified with this method, the points on the opposite side of the Bragg peak as illustrated earlier, the interaction point needs to be identified from this set, as the points lying at the extremity of this region with respect to the barycenter.
Thus, to identify the points at the extremity of the track on the other side with respect to the Bragg-Peak, a circumference of a radius $r$ centered in the barycenter is considered, and the points meeting the following two conditions are selected:
\begin{equation}
    \left\{\frac{x'_i}{M_3}<0 \right\} \ \ \ \ \  \cup \ \ \ \ \ \left\{ d_{cm}=\sqrt{(x_i-x_c)^2 + (y_i-y_c)^2}\ >r \right\}
\label{eq:conditionSel}
\end{equation}
The first condition of \ref{eq:conditionSel} requires that the point is on the other side of the Bragg peak, while the second condition, different from the original algorithm, requires the point to be at a distance $d_{cm}>r$ from the barycenter. 
The procedure is iterated starting from a small value of $r$, and incrementing its value at each step, until a predetermined number of pixels $N_{pt}$ is left in the selection. The quantity $N_{pt}$ is a model parameter, not known a priori, and its optimization will be treated in Sec. \ref{sec:paroptimization}. The barycenter of the region of the pixels passing the selection, calculated with the same method used to calculate the barycenter of the track (Eq. \ref{eq:barycenter}), is designated as the track interaction point (IP).
\begin{figure}
    \centering
    \includegraphics[scale=.45]{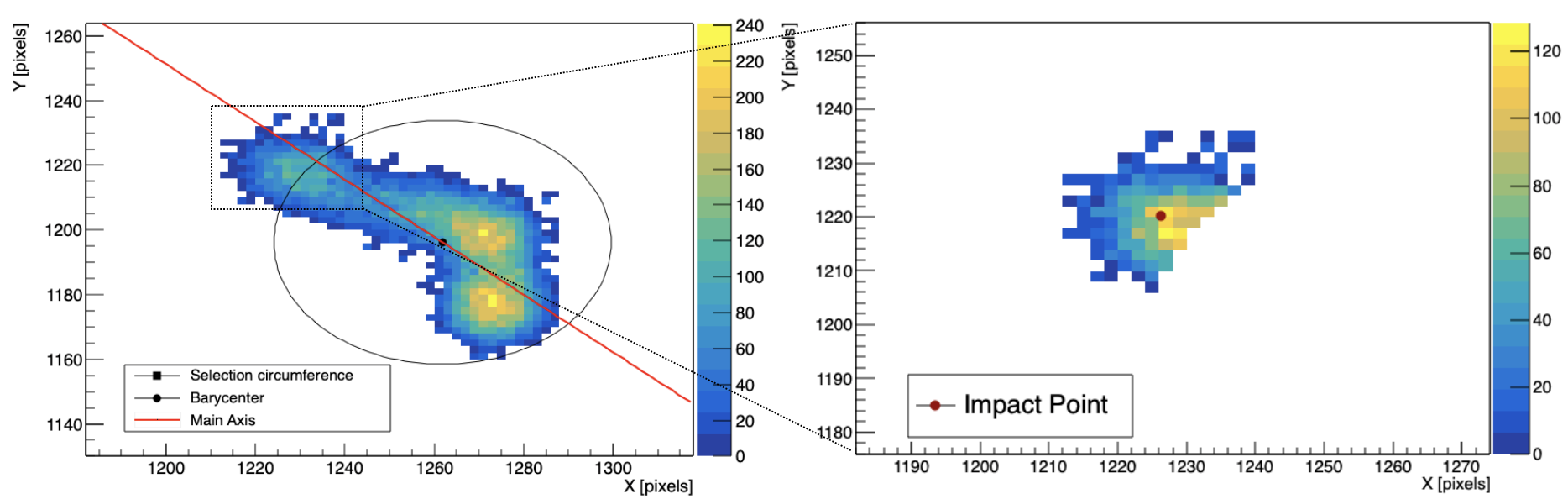}
    \caption{Left: 30 keV electron with main axis, barycenter, and selection circumference shown. Right: zoom of the region delimited by the dashed square in the left plot showing the point passing the selection in Eq. \ref{eq:conditionSel}. The impact point calculated as the weighted average of this interaction region is also shown.}
    \label{fig:IPCal}
\end{figure}
This selection procedure is shown in Fig. \ref{fig:IPCal} on the track of the 30 keV electron. In the left plot, the circumference has the exact radius $r$ for which the number of points satisfying the condition \ref{eq:conditionSel} are precisely $N_{pt}$.
The reason for which the impact point is calculated with a weighted average, rather than relying on the extremity of the track, comes from the transverse diffusion of electrons originating from the real interaction point. As electrons diffuse in all directions from the original interaction point, the electrons distribute in a cloud around it. Therefore, determining the interaction point as the weighted average of the initial part of the track has been demonstrated to be an effective strategy for its identification \cite{AstroXRayPol}.\\

\subsection{Electron recoil direction evaluation}
\label{sec:recoildirection}
\begin{figure}
    \centering
    \includegraphics[scale=.21]{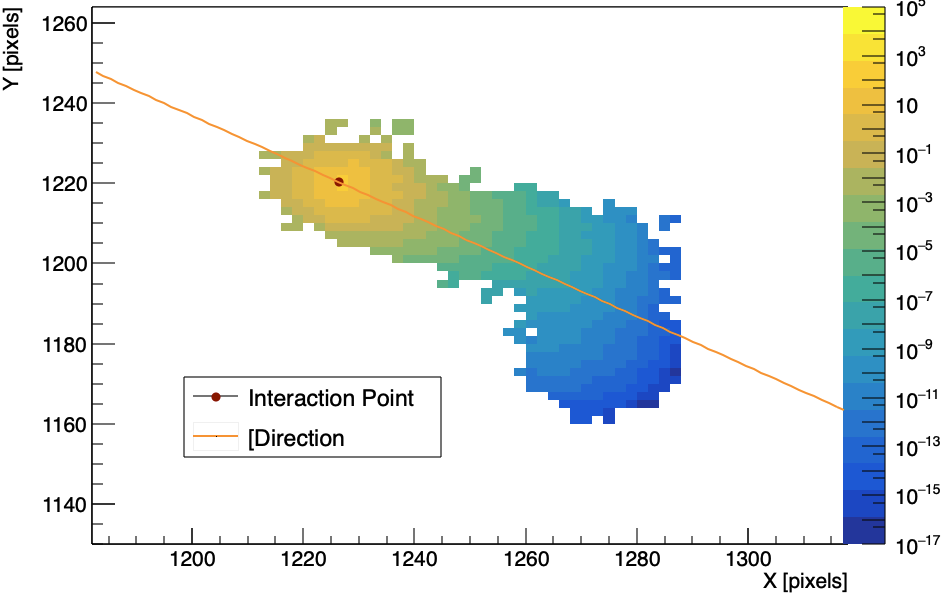}
    \includegraphics[scale=.25]{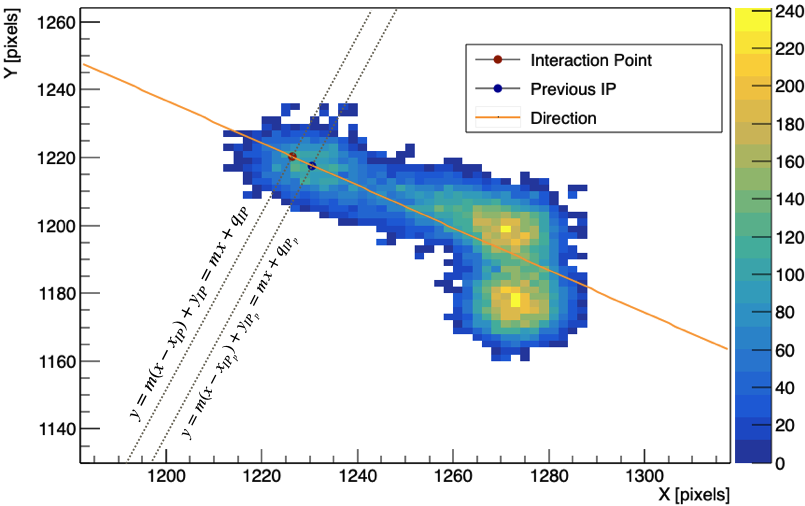}
    \caption{Left: plot of the rescaled track, the interaction point, and the line of the direction of the track is shown. Right: plot describing the procedure to determine the orientation on the track. The $IP$, the $IP_{pr}$, the direction, and the two lines perpendicular to the direction, passing by $IP$ and $IP_{pr}$, used to give an orientation to the track are also shown.}
    \label{fig:RescaledOri}
\end{figure}

\begin{figure}
    \centering
    \includegraphics[scale=.4]{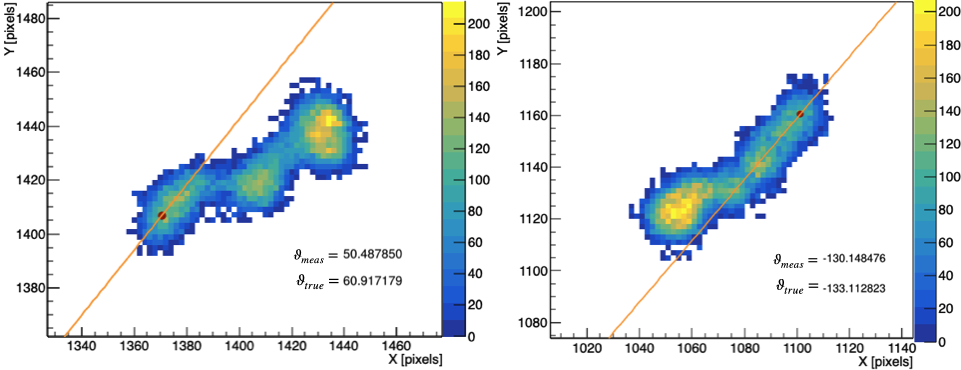}
    \caption{Two 30 keV electrons with direction reconstructed by the algorithm and true direction from the MC are shown in the legend.}
    \label{fig:twotrackexample}
\end{figure}

Once the impact point of the track is identified, the direction is calculated following the light intensity along the track profile.
However, the information on the track direction is contained at the very beginning of the track, and it's progressively lost proceeding along its length. Hence, the intensity of each pixel along the track is scaled by the quantity:
\begin{equation}
    W(d_{ip}) = e^{-d_{ip}/w}
\end{equation}
where $d_{ip}$ represents the distance of the pixel from the interaction point, and $w$ is a normalization factor, which is another free parameter of the algorithm, whose optimization will be discussed in Sec. \ref{sec:paroptimization}. This approach results in an exponential decrease in the weighting of the pixels as the increasing distance from the impact point. In this way, a major importance is given to pixels close to the IP. Subsequently, the track direction is calculated as the main axis of the track passing by the impact point, repeating the same procedure of Sec. \ref{sec:MAxis}. In this case, the calculation involves re-weighted pixels, and the main axis is constrained to pass by the interaction point. An example of 30 keV electron track rescaled with this procedure, with the impact point and the direction calculated, is shown in Fig. \ref{fig:RescaledOri} (left plot).\\
The procedure outlined so far allows to determine the direction of the track, but not its orientation. The orientation is established following the light path of the track with a procedure implemented by the thesis author. To give an orientation to the track, during the process of pixel selection for the impact point calculation, involving the progressively increasing of $r$, the barycenter of the selected pixels is calculated in the iteration in which the double of the number of points with respect to $N_{pt}$ is selected. The barycenter of this region will be called $IP_{pr}$. Selecting a higher number of pixels causes the barycenter of the selected points to trace the light pattern in the track. The information of the barycenter of the selected points in a previous iteration can thus allow for assigning an orientation to the track.
Considering the line that represents the track direction to be expressed as $y=ax+b$, with $a=\tan(\Phi)$, the subsequent step is to calculate the two lines perpendicular to the direction line, passing respectively by $IP$, with coordinates ($x_{IP}$, $y_{IP}$), and $IP_{pr}$, with coordinates ($x_{IP_{Pr}}$, $y_{IP_{Pr}}$). These two lines will have equations:
\begin{align}
IP \rightarrow & y=m(x-x_{IP})+y_{IP} = mx+q_{IP} \\   IP_{Pr} \rightarrow & y=m(x-x_{IP_{Pr}})+y_{IP_{Pr}} = mx+q_{IP_{Pr}}
\end{align}
with $m= -1/a$. Depending on the sign of $a$ and the relative value of $q_{IP}$ with respect to $q_{IP_{pr}}$, the direction with orientation of the track can be taken simply as the $\arctan(a)$ or be corrected by $\pm 180°$ to take into account the orientation on the [-180°,-90°] and [+90°,180°] quadrants. Since these quadrants are not part of the image of the $\arctan()$ function, a correction must be subsequently applied. The correction will be applied on the basis of the criteria outlined in tab. \ref{tab:tabcorrangle}.
\begin{table}
    \centering
    \begin{tabular}{|l|c|c|}
    \hline
        \diagbox[width=\dimexpr \textwidth/8+2\tabcolsep\relax, height=1cm]{ And }{If} & $a>0$ & $a<0$ \\ \hline 
        $q_{IP}>q_{IP_{pr}}$ & $\arctan(a)$-180° & $\arctan(a)$ \\ [1ex] \hline
        $q_{IP}<q_{IP_{pr}}$ & $\arctan(a)$ & $\arctan(a)$+180° \\ [1ex] \hline
    \end{tabular}
    \caption{In the table the calculation of the track direction with orientation is shown for the different sign of $a$ and relative value of $q_{IP}$ and $q_{IP_{pr}}$.}
    \label{tab:tabcorrangle}
\end{table}
A scheme of the procedure applied is shown in Fig. \ref{fig:RescaledOri} (right plot). For the track displayed in the plot, the coefficient of the line direction $a$ is negative, and $q_{IP}>q_{IP_{pr}}$. Thus, the direction with the orientation of the track is straightforwardly determined by $\vartheta=\arctan(a)$. An example of two tracks, with their reconstructed directions and the corresponding Monte Carlo true direction (reported in the legend), is shown in Fig. \ref{fig:twotrackexample}.

\section{Parameters optimization}
\label{sec:paroptimization}
\begin{figure}
    \centering
    \includegraphics[width=0.5\linewidth]{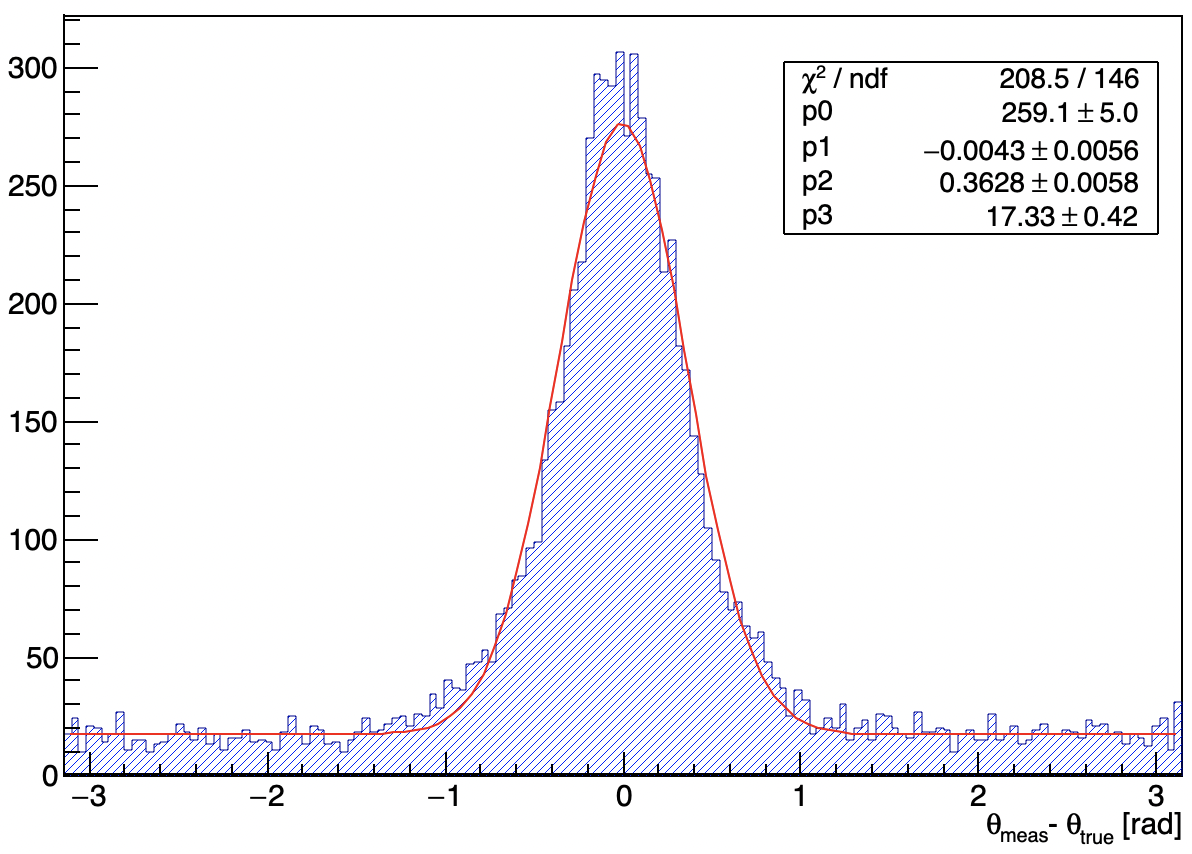}
    \caption{Distribution of $\vartheta_{meas}-\vartheta_{true}$ for 30 keV ER with fit consisting of a Gaussian plus a constant term superimposed. The parameters in the legend are respectively the amplitude, the mean, and the sigma of the Gaussian, and the value of the constant term.}
    \label{fig:gaussiandistrib}
\end{figure}

\begin{figure}
    \centering
    \includegraphics[width=1\linewidth]{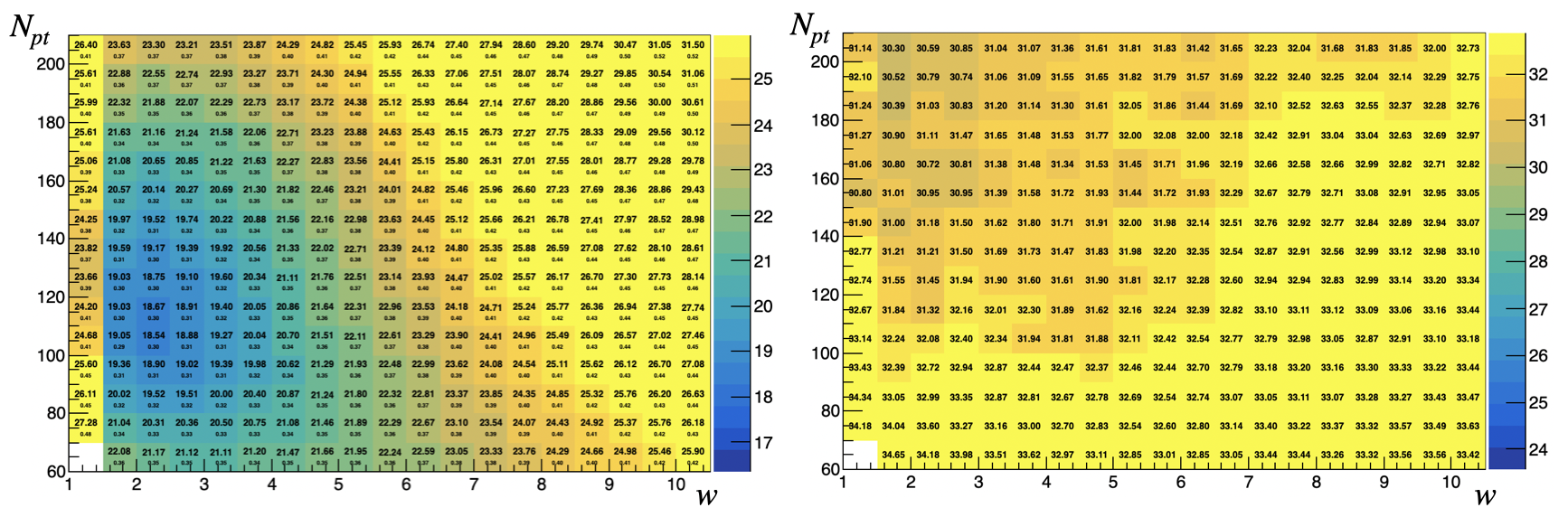}
    \caption{Left: Angular resolution obtained for the 40 keV dataset as a function of the two parameters of the algorithm $N_{pt}$ and $w$. In each bin, the numerical value of angular resolution is reported as the $\sigma$ of the $\theta-\theta_{true}$ distribution, together with the error on the $\sigma$ (smaller number). The colors of the bins follow the scale on the right, where colors towards the blue represent a better angular resolution while towards the yellow a worse one. Right: Plot of the value of the constant term as a function of $N_{pt}$ and $w$.}
    \label{fig:AngResVsPar}
\end{figure}

\begin{figure}
    \centering
    \includegraphics[width=0.6\linewidth]{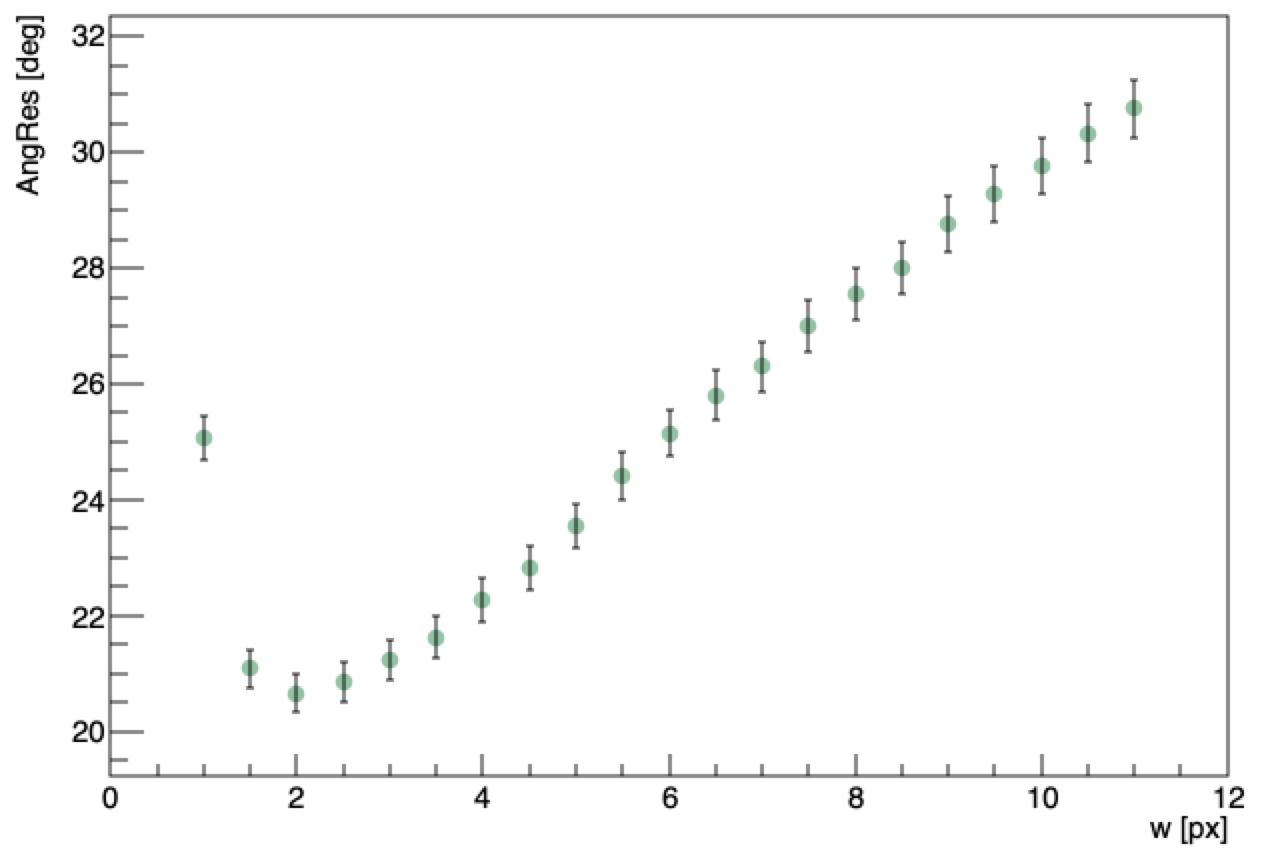}
    \caption{Behavior of angular resolution as a function of the parameter $w$ for 40 keV ER. The values are computed for the optimal value of $N_{pt}$.}
    \label{fig:AngResVsw}
\end{figure}

The algorithm described in Sec. \ref{sec:dirAlgo} depends on two free parameters, $N_{pt}$ and $w$.
Since the optimal values for the parameters $N_{pt}$ and $w$ of the algorithm, for which the best angular resolutions is obtained are not known a priori, these have been optimized on simulated tracks produced in a LIME-like detector. 
A sample of 10,000 ER sCMOS images per energy, at the energies of 16, 18, 20, 22, 24, 28, 32, 36, 40, 50, 60, and 70 keV, with isotropic electron direction, with uniform distribution in drift length within 5 and 45 cm, and uniformly distributed in x and y has been produced. The tracks have been reconstructed using the code illustrated in Sec. \ref{chap:reco}, employing the same parameters as those used for the data.\\
The angular resolution $\sigma_{\vartheta}$ is assessed by calculating the standard deviation $\sigma$ of Gaussian obtained fitting the distribution $\vartheta_{meas}-\vartheta_{true}$, where $\vartheta_{true}$ is the information from the MC-truth.
Given the absence of the third dimension, and, as of now, the algorithm fails in reconstructing the direction of tracks with the tail perpendicular to the GEM plane, or those having an overlap in correspondence with the interaction point. Moreover, because of its intrinsic way of working, the algorithm might fail in handling the topology of the track, failing to identify the impact point.
In such instances, the reconstruction of the track direction may fail, generating in output a random direction. This is addressed by introducing a constant term to the fit. The distribution of $\vartheta_{meas}-\vartheta_{true}$ is shown in figure \ref{fig:gaussiandistrib} for 40 keV electron recoil with fit superimposed.
The implementation of a PMT analysis, enabling tridimensional track reconstruction, is expected to overcome scenarios where the algorithm struggles to reconstruct the track direction. The second case is expected to be solved by the implementation of an improved algorithm that is able to follow the track pathlength for IP determination. The contributions to this flat component will be discussed in detail in Sec. \ref{sec:angResResults}. \\ For optimization purposes, only the sigma of the Gaussian has been used since the flat term amplitude results in being constant as a function of the parameters varying $N_{pt}$ and $w$, in the considered ranges.
To optimize the algorithm's free parameters, a systematic scan of the angular resolution and amplitude of the flat term was conducted by varying $N_{pt}$ and $w$. 
The parameters tested encompass all the combinations of $N_{pt}$ ranging from 60 to 200 in steps of 10, and $w$ ranging from 1 to 10 in steps of 0.5. These values were selected to ensure a sufficiently dense scan.The values of $\sigma_{\vartheta}$ as a function of these two parameters are shown along with their corresponding errors in Fig. \ref{fig:AngResVsPar}, left plot, for 40 keV electrons as an example. The values of the constant term amplitude are also shown in the right plot of figure \ref{fig:AngResVsPar}.
For each value of $N_{pt}$, $\sigma_{\vartheta}$ undergoes rapid improvement with the increasing of $w$, reaches a minimum, and subsequently raises again. This behavior is due to the meaning of the $w$ parameter. If $w$ is excessively small, a portion of the track excessively short, dominated by diffusion, is used to evaluate the direction, resulting in an absence of directional information. Increasing the value of $w$, the length of the track used to evaluate the direction expands, enhancing the sensitivity to directional information up to an optimal value. With the further increase of the value of $w$, portions of the track subject to multiple scattering start to enter into the track direction calculation washing out the directionality information. 
Hence, the optimal value of $w$ for which the best angular resolution is achieved arises from a balance between these effects. The behavior of the angular resolution as a function of $w$ for the optimal $N_{pt}$ value found for ER at 40 keV is shown in Fig. \ref{fig:AngResVsw}.
The best values found for each energy are reported in Tab. \ref{tab:parametersvsE}.
\begin{table}
    \centering
    \begin{tabular}{|l|cccccccccccc|}
    \hline
        E [keV] & 16 & 18 & 20  & 22 & 24 & 28 & 32 & 36 & 40 & 50 & 60 & 70 \\ \hline
        $N_{pt}$ & 100 & 100 & 100 & 100 & 110 & 110 & 110 & 120 & 120 & 120 & 120 & 120 \\ 
        $w$ & 1.5 & 1.5 & 1.5 & 1.5 & 1.5 & 1.5 & 2 & 2 & 2 & 2.5 & 2.5 & 2.5 \\ \hline
    \end{tabular}
    \caption{In the table, the optimal couples of parameters $N_{pt}$ and $w$ which provide the best angular resolution for each energy dataset simulated are shown.}
    \label{tab:parametersvsE}
\end{table}
It can be observed from the table that the value of $N_{pt}$ stays approximately constant for all the energies, showing only a slight increase with rising energy. Oppositely, the optimal value of $w$ exhibits an increase as energy levels rise. This pattern arises because, with the increasing of the energy, the multiple scattering effect became less significant, making it more effective to include a larger track portion of the track in the direction calculation.

\section{Directionality performances on a Lime-like detector}
\label{sec:angResResults}
\begin{figure}
    \centering
    \includegraphics[width=0.75\linewidth]{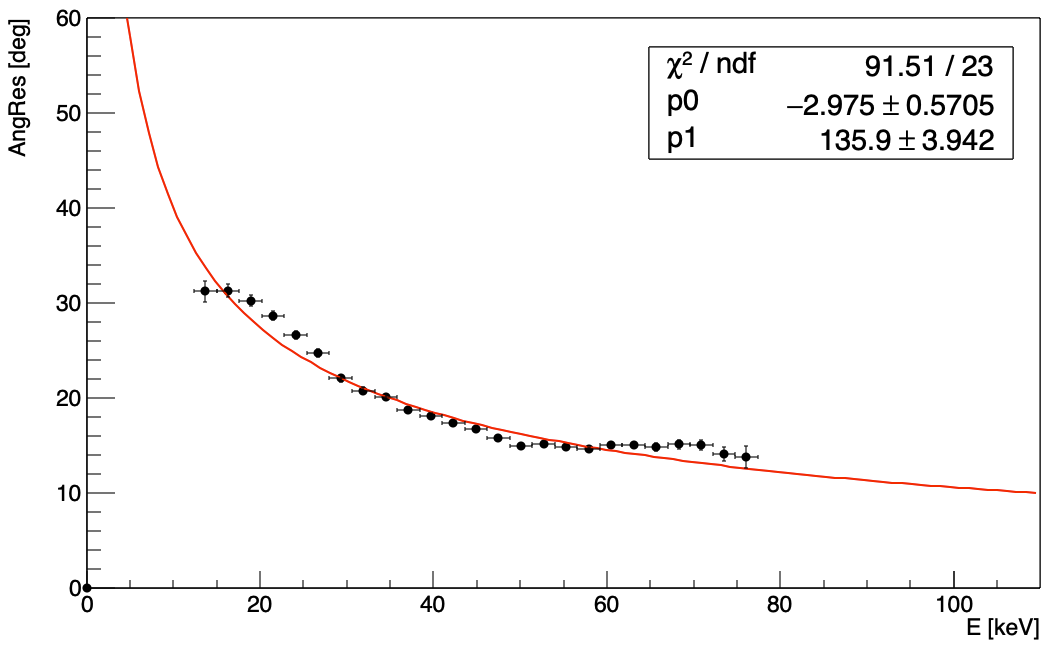}
    \caption{Plot of the angular resolution as a function of the track energy with the fit function superimposed.}
    \label{fig:AngResIPResVsE}
\end{figure}

To evaluate the angular resolution, a data sample of electron tracks uniformly distributed in energy from 16 to 70 keV has been produced in GEANT4, digitized with the simulation code, and reconstructed. The calibration of the track energy has been performed considering only tracks with E$>$16 keV using the curve shown in Fig. \ref{sec:overgroundstudies}. 
\begin{figure}
    \centering
    \includegraphics[width=0.5\linewidth]{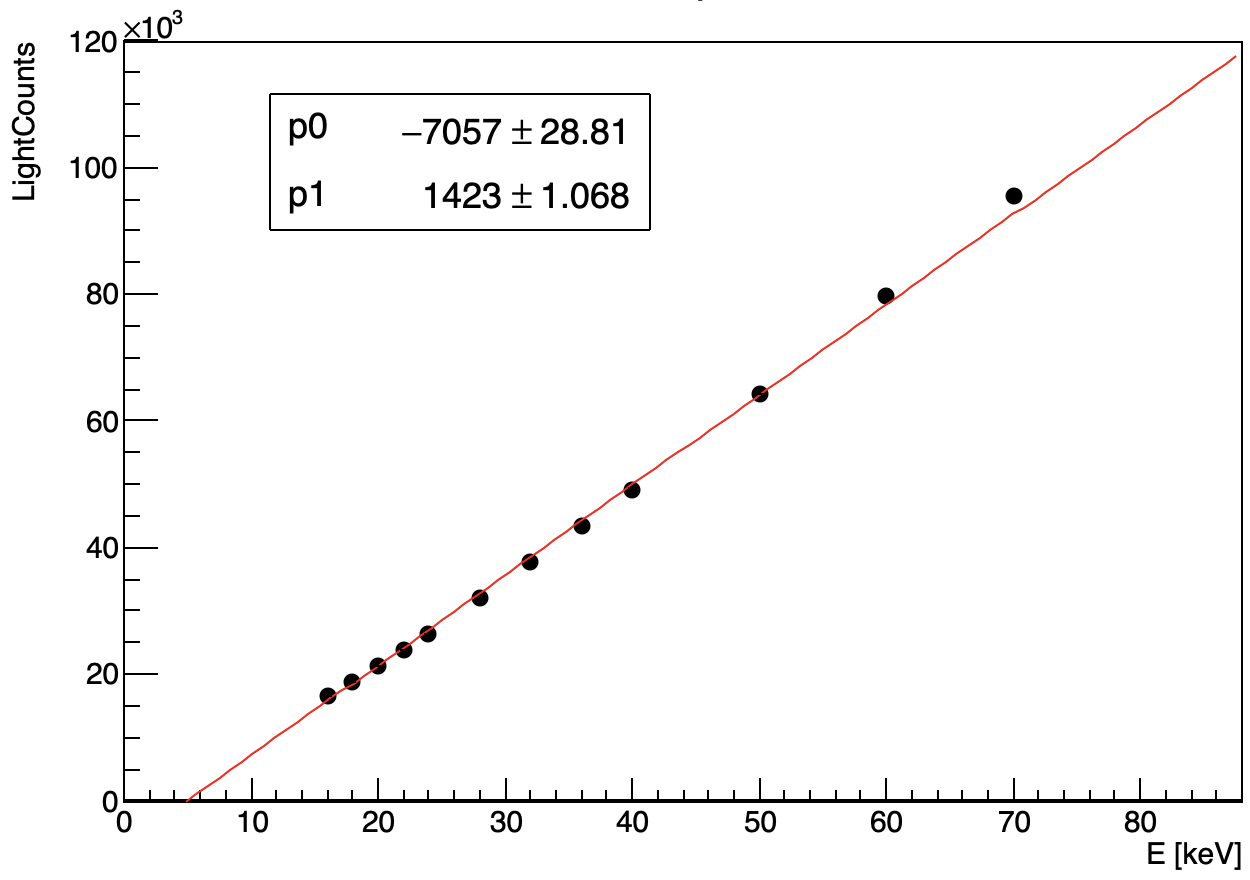}
    \caption{Calibration curve of the light integral response as a function of the track energy produced with the simulated data.}
    \label{fig:calib}
\end{figure}
This can be justified by the fact that with increasing energy the tracks start to exhibit a low energy density tail which do not saturate. This determines a more linear response of the detector with respect to the energy range below 16 keV (see Sec. \ref{sec:LIME}). Moreover, this energy range is the one expected for the solar neutrino measurement and thus, for this specific case, can be treated separately from the low-energy one. Thus, for each electron, the track energy in keV is determined from the light integral using the calibration curve, and the $N_{pt}$ and $w$ values are set as the ones from the closest energy reported in \ref{tab:parametersvsE}. After that, the angular resolution has been evaluated as a function of the energy in bins. Additionally, the impact point resolution is evaluated as the sigma of a Gaussian fit to the distribution of the difference between the measured X and Y coordinates and the ones from the MC truth.
The plot of the behavior of the angular resolution is shown in Fig. \ref{fig:AngResIPResVsE}. 

\begin{figure}
    \centering
    \includegraphics[width=1\linewidth]{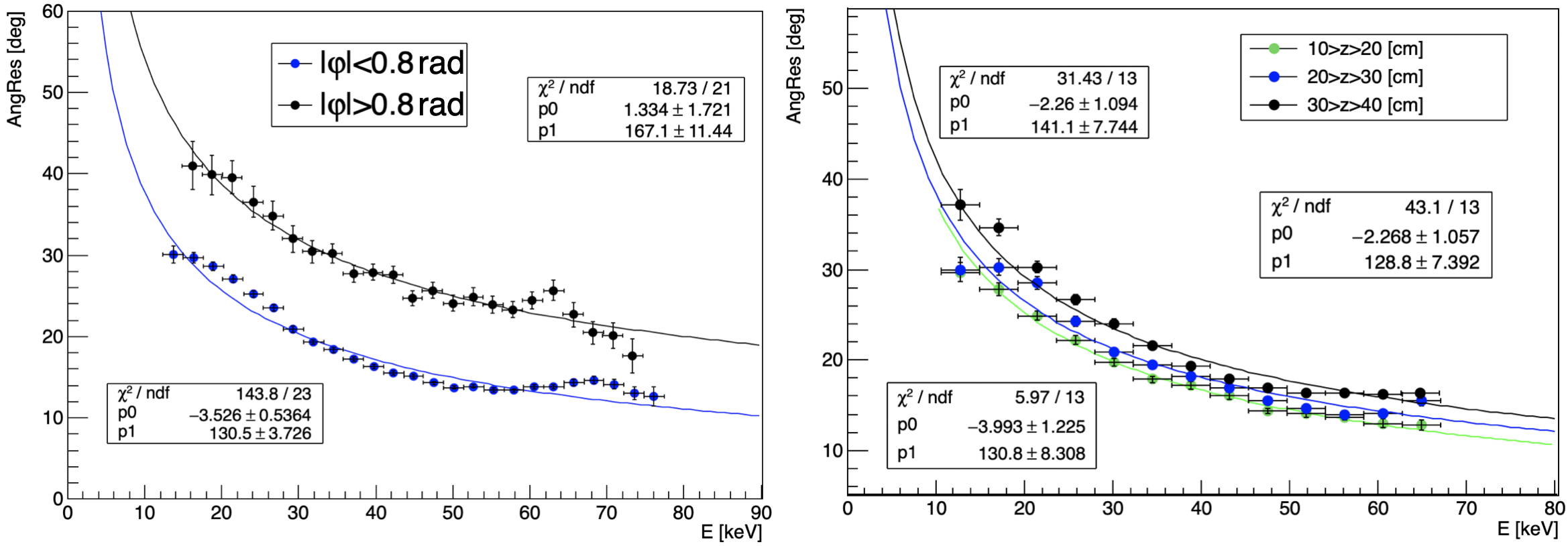}
    \caption{Left: Plot of angular resolution vs energy for tracks selected in different angular ranges with respect to the GEM plane. Right: Plot of angular resolution vs energy for tracks selected in different diffusion ranges.}
    \label{fig:AngResFurther}
\end{figure}

The behavior of angular resolution as a function of the energy has been fit with the functional form $\sigma_{\vartheta}=p_0+p_1/\sqrt{E}$. In the fit, the parameter $p_0$ represents the asymptotic value that the angular resolution can reach, where the multiple scattering becomes totally irrelevant. Despite this parameter being negative, it is expected the angular resolution reach asymptotically a value different from zero. In the high energy regime indeed, the uncertainty on the initial direction should be dominated by the diffusion effect, which affects the track impact point determination, and consequently the angular resolution. This parameterization, together with the assumption that the angular resolution will converge to a value greater than zero at high energy, will be used to infer the angular resolution at higher energies in the feasibility study (Chap \ref{chap:solarnu}). The impact point resolution is approximately constant in energy at $\sim 520\  \mu$m.\\
The angular resolution was additionally studied as a function of the drift distance and the track inclination with respect to the amplification plane to further characterize and understand the directionality performances of this approach. In particular, the right plot of Fig. \ref{fig:AngResFurther} shows the angular resolution behavior in the drift ranges [10-20] cm, [20-30] cm, and [30-40] cm, and it can be evinced that the resolution exhibits only minimal sensitivity to diffusion in this configuration. This is primarily because the track diffusion at the operational drift field of LIME (1 kV/cm) is strongly dominated by the contribution from the amplification plane that is independent of the drift distance (see Sec. \ref{sec:TrasnvDiff}). 
A parameter that affects the angular resolution is, as expected, the inclination of the initial part of the track with respect to the GEM plane. As previously explained, since the analysis is on the bi-dimensional projection of the track, the algorithm is less effective in reconstructing the direction of an electron produced with initial direction towards the perpendicular to the GEM plane. As shown in the left plot of Fig. \ref{fig:AngResFurther}, by defining $\varphi$, as the track production angle with respect to the GEM plane, it can be observed that the angular resolution for tracks with $\varphi >$0.8 rad results significantly degraded with respect to $\varphi <$0.8 rad. This worsening in angular resolution is however expected to be reduced once the PMT information is added to the analysis. 

\section{Algorithm efficiency}
\label{sec:algoeff}
\begin{figure}
    \centering
    \includegraphics[width=.9\linewidth]{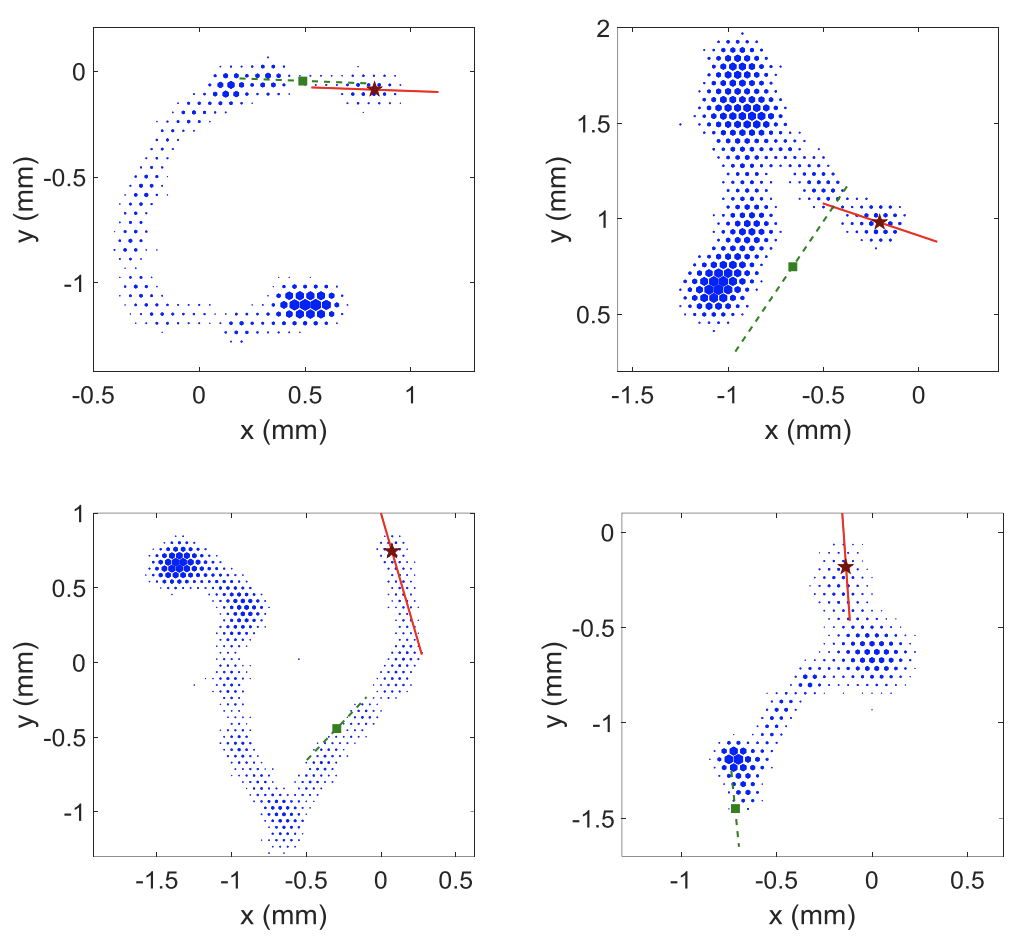}
    \caption{Cases of tracks with topologies where the algorithm fails to correctly identify the interaction point. The red star with the line represents the true track direction and the true IP, while the green square and line represent the one measured by the algorithm. Plots from \cite{LI201762}.}
    \label{fig:algoineff}
\end{figure}

\begin{figure}
    \centering
    \includegraphics[width=1\linewidth]{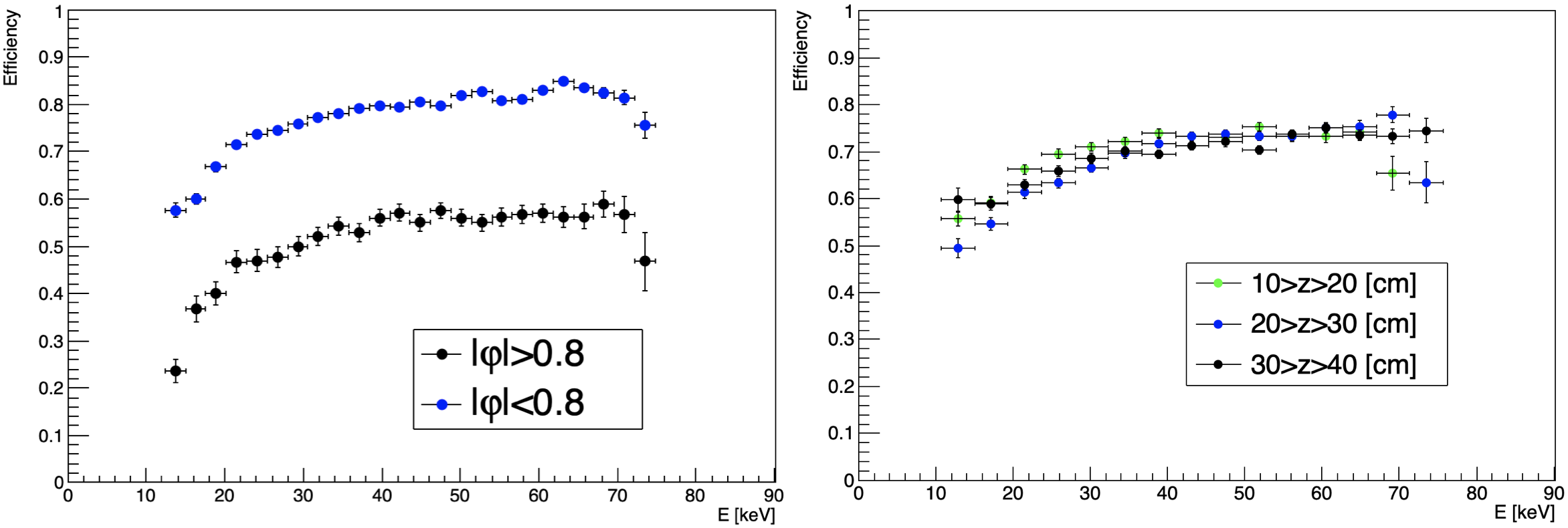}
    \caption{Left: Directionality efficiencies for different production angles with respect to the GEM plane. Left: same for different drift distances.}
    \label{fig:EffFurther}
\end{figure}

Due to its intrinsic functioning and the absence in the analysis of the PMT information, the algorithm exhibits an efficiency in initial direction reconstruction. This inefficiency results in a misidentifying of the impact point, which leads to the reconstruction of a random track direction. This translates into a flat component in the $\theta_{meas}-\theta_{true}$ distribution, as shown in Fig. \ref{fig:gaussiandistrib}. Thus, the algorithm direction reconstruction efficiency is quantified as the ratio between the number of events belonging to the Gaussian part of the fit component and the total number of events. In particular, it is calculated as the integral of the Gaussian of the fit divided by the total area subtend by the fit function.\\ 
The cases in which the original polarimetry algorithm \cite{AstroXRayPol} fails to correctly identify the impact point due to track topology have been analyzed in \cite{LI201762}, and are reported in Fig. \ref{fig:algoineff}. These scenarios encompass instances where the track has a high energy deposit towards the middle due to hard scattering, potentially altering the skewness value. Additionally, there are cases where the track's tail forms a curved arc, leading the algorithm to fail in the selection of the track extremity.
Drawing inspiration from the polarimetry algorithm, the cases in which the algorithm developed for this thesis fail can be considered of the same category.\\ 
Additionally to these cases, there are scenarios where the track initial direction is perpendicular to the GEM plane, which leads to the loss of the impact point in the 2D track projection, making its identification unfeasible without PMT information.\\
\begin{figure}
    \centering
    \includegraphics[width=.5\linewidth]{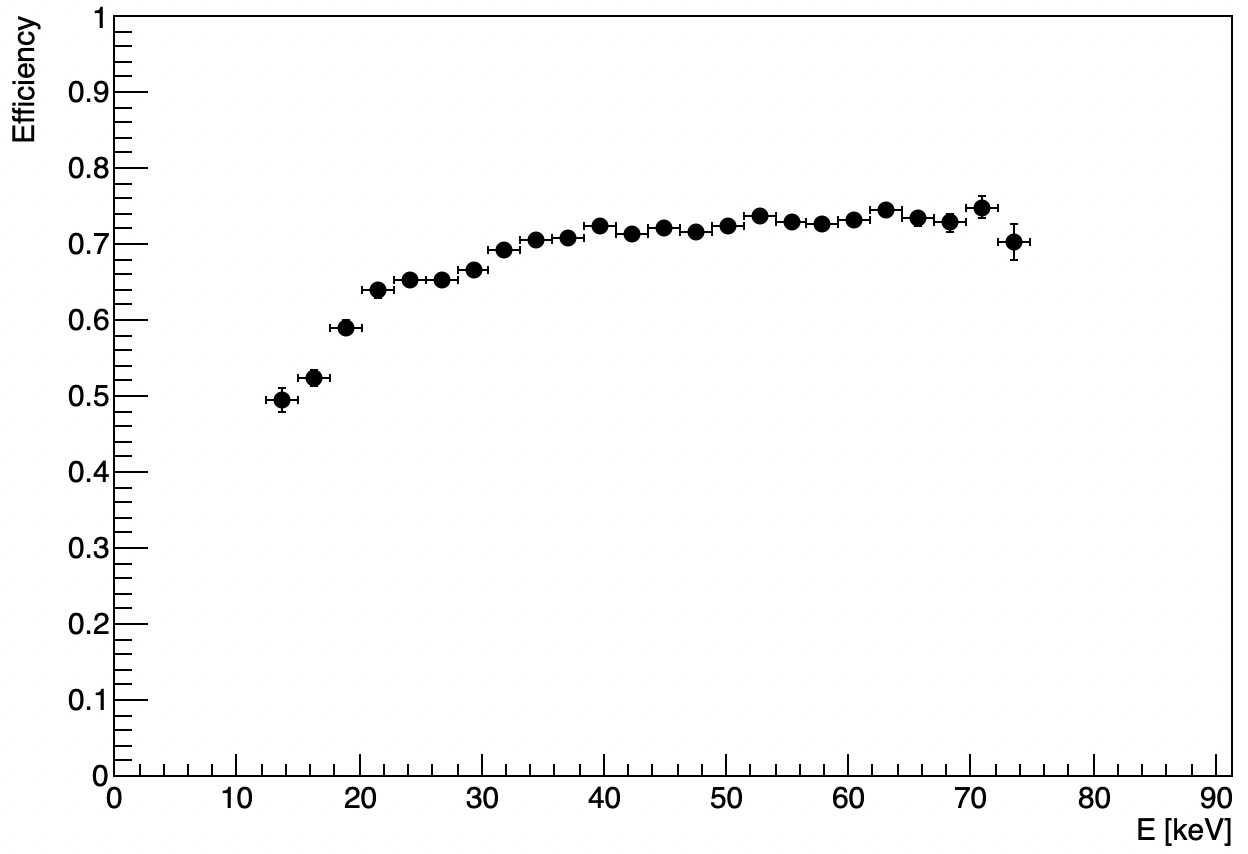}
    \caption{Directional efficiency, defined as the number of events falling within the Gaussian distribution of the $\theta_{meas}-\theta_{true}$ distribution divided by the total number of events, as a function of the electron energy.}
    \label{fig:efficiency}
\end{figure}
The efficiency in directionality reconstruction as a function of the energy is shown in Fig. \ref{fig:efficiency}. As expected, the efficiency increases with energy due to longer tracks, making the head and tail surely more distinguishable. Additionally, shorter tracks, especially those produced towards the GEM plane, pose challenges in retrieving impact point information. These shorter tracks are also the ones for which the directionality calculation is more affected by multiple scattering and diffusion. With the increase in energy, some of the effects impacting shorter tracks are reduced, and the inefficiency component is predominantly dominated by the cases in which the algorithm fails to identify the IP (Fig. \ref{fig:algoineff}), and the instances in which the IP is not identified due to the lack of PMT information in the analysis. These effects, above a certain energy, are energy-independent, resulting in the efficiency converging to a constant value as energy increases.\\
Furthermore, the efficiency has also been studied across different drift distances and emission angles relative to the GEM plane. The plots are shown in Fig. \ref{fig:EffFurther}.
As shown in the left plot, the fact that the analysis is carried out on sCMOS images significantly impacts the efficiency of determining the initial direction of the track, whereas no significant effect from diffusion is observed as shown in the right plot.\\
Efforts from the thesis author are underway to enhance the efficiency of directional reconstruction and solve the cases depicted in Fig. \ref{fig:algoineff} through the development of an algorithm based on path length tracking to identify the correct IP. This method is based on the thinning method implemented in the OpenCV library \cite{opencv_library}. Once the thinned track is obtained, the two endpoints of the tracks are identified, and the light integral is computed in a radius around the endpoint. Thus, the extremity with a lower charge is selected to proceed with the IP computation. Middle endpoints are artifacts of the thinning and anyway, they are not included in the process. 
\begin{figure}
    \centering
    \includegraphics[width=1.\linewidth]{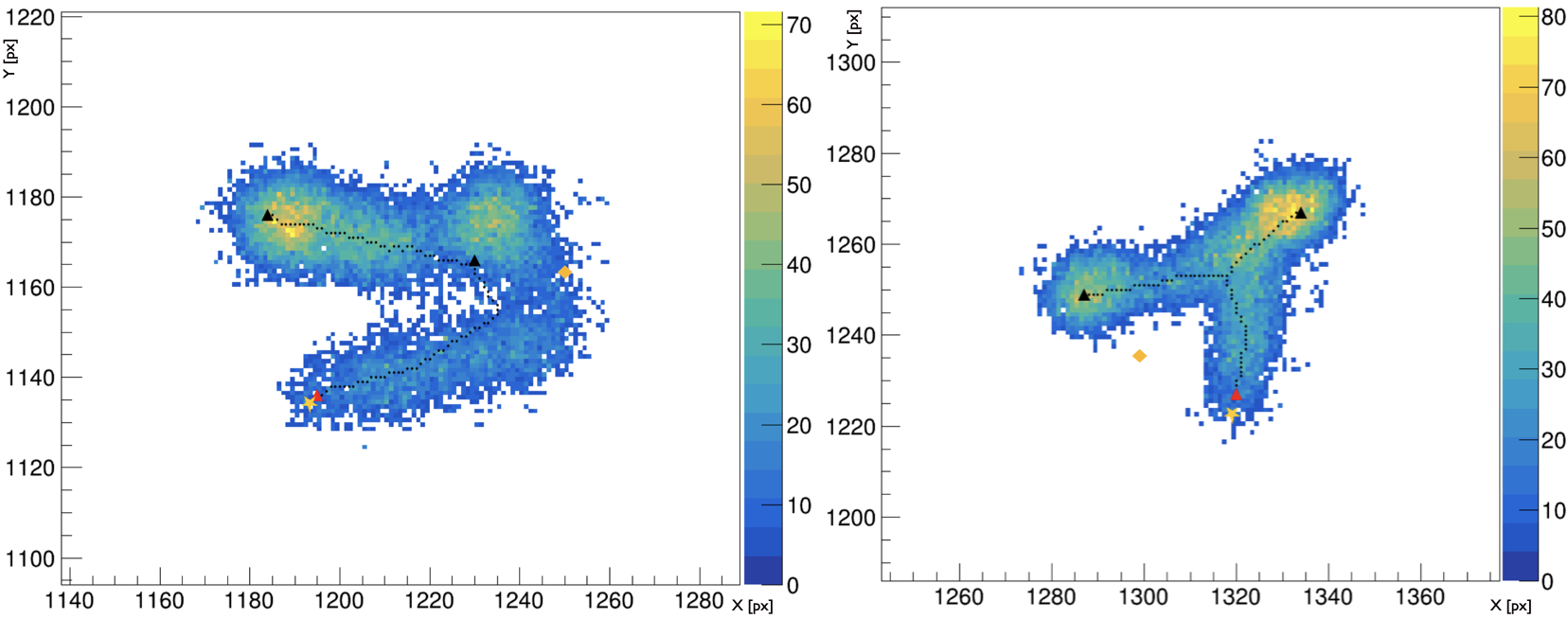}
    \caption{In figure, two examples of tracks in which the improved version of the algorithm is able to correctly identify the IP region. The small black dots represent the thinned track, the triangles represent the endpoints and in particular, the red one represents the IP region. The yellow star is the true IP. While the orange rhombus represents the IP found with the original algorithm.}
    \label{fig:SkCorrectIP}
\end{figure}
Fig. \ref{fig:SkCorrectIP} illustrates two examples where the improved algorithm correctly identifies the interaction point region. Next, the procedure involves selecting a region of the track along the skeleton, beginning from the endpoint within the interaction region, and repeating the same procedure for the IP identification, replacing the track main axis with an axis obtained fitting the first points of the skeleton. Finally, the same procedure for the directionality calculation will be used. This refined approach not only will lead to enhanced efficiency but also has the potential to improve angular resolution performances. However, this new algorithm is still under development.\\
The cases in which the track is produced along the direction perpendicular to the GEM plane can be solved once the PMT information is added. In these cases, the tail of the track, projected on the 2D GEM plane would appear as a round spot, and the direction cannot be measured working only on the 2D information. Once the information on the third dimension is added, the initial track direction with orientation can be measured in 3 dimensions.

\section{Angular resolution for Negative Ion Drift operation}
\label{sec:NIDAngReso}
\begin{figure}
    \centering
    \includegraphics[width=1.\linewidth]{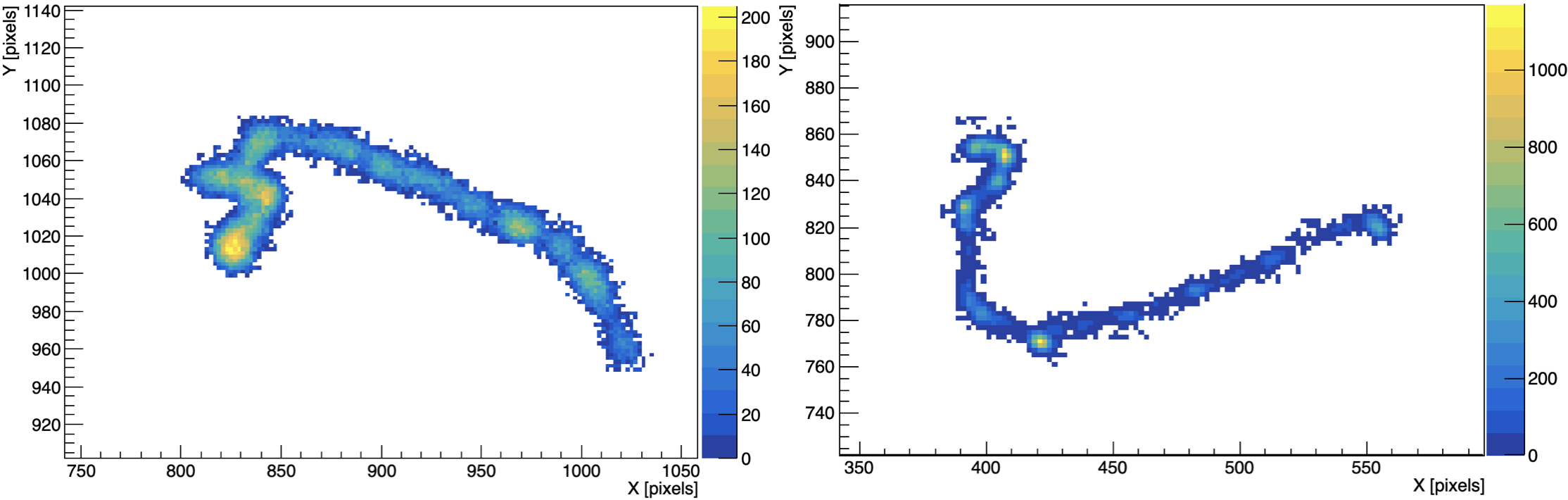}
    \caption{Left: 60 keV ER simulated with ED diffusion. Right: 60 keV ER simulated with NID diffusion.}
    \label{fig:EDNIDComparion}
\end{figure}

\begin{figure}
    \centering
    \includegraphics[width=0.6\linewidth]{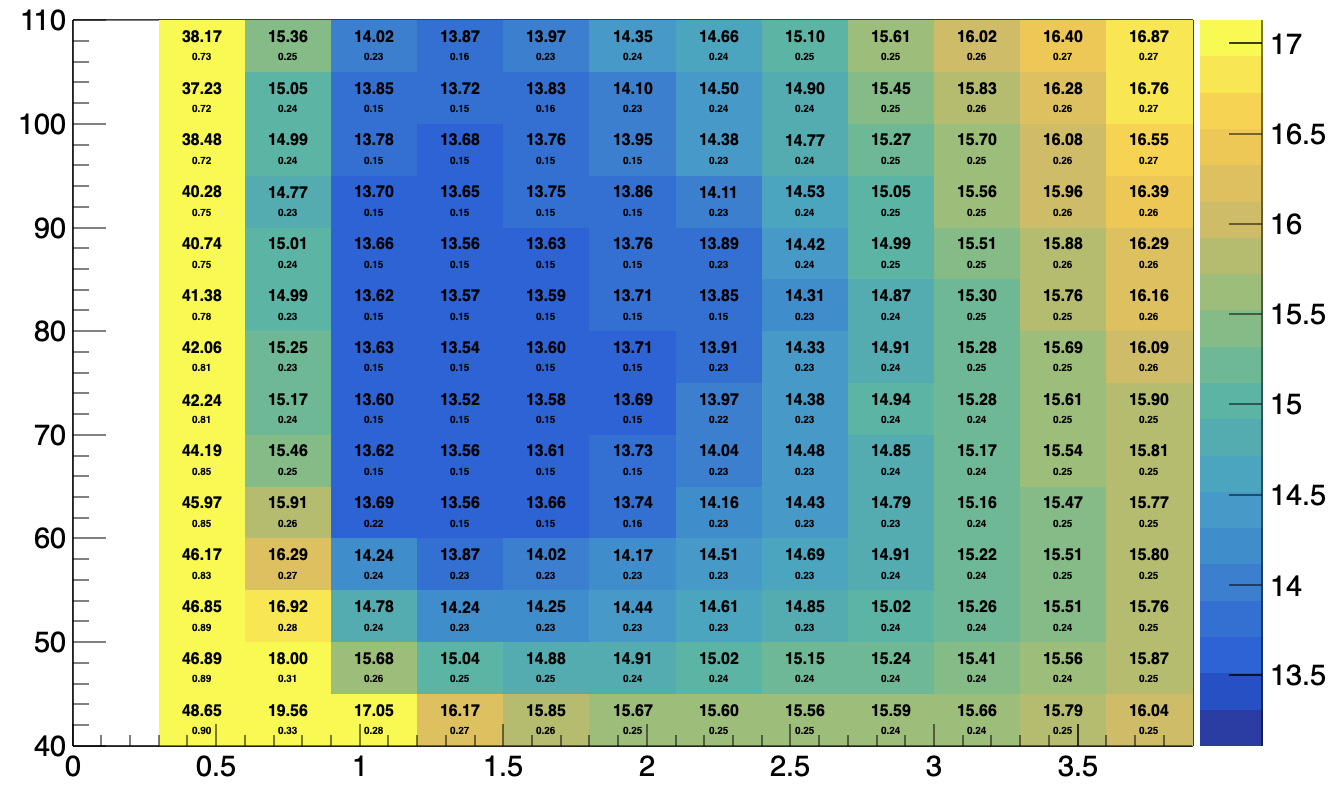}
    \caption{Angular resolution obtained for the 40 keV NID dataset as a function of the two parameters of the algorithm $N_{pt}$ and $w$. In each bin, the numerical value of angular resolution is reported as the $\sigma$ of the $\theta-\theta_{true}$ distribution, together with the error on the $\sigma$ (smaller number). The colors of the bins follow the scale on the right.}
    \label{fig:AngResVsParNID}
\end{figure}

As illustrated in detail in Chap. \ref{sec:NID}, the CYGNO collaboration demonstrated the possibility of achieving negative ion operation at atmospheric pressure with an optical readout by adding to the CYGNO gas mixture a small doping of SF$_6$ and obtained impressive results in the minimization of the track diffusion. For this reason, the study of the achievable angular resolution with the algorithm illustrated in Sec. \ref{sec:dirAlgo} has been repeated on images simulated with the measured transverse diffusion reported for NID operation. In particular, for the simulation, the transversal diffusion value used has been obtained by interpolating the values of $\sigma_{0T}$ and $\xi_{T}$ as a function of the drift field at the operative value of 1 kV/cm. The values used for the track simulation are:
\begin{equation}
    \sigma_{0T} = 173\  \mu m \ \ \  \xi_{T}=37\  \mu m/\sqrt{cm}
\label{eq:parametesNID}
\end{equation}
Given the significantly lower drift speed of negative ions compared to electron drift (by a factor of $\mathcal{O}(10^3)$), the charge flux within the GEM hole is likewise approximately $\mathcal{O}(10^3)$ times smaller. Consequently, it is assumed that no saturation occurs for Negative Ion Detectors (NID) operations. The primary electron longitudinal diffusion, in the simulation, plays a role in longitudinally diffusing the primary charge in the 3D histogram to then apply the saturation. Since the saturation in this case is absent, the longitudinal diffusion doesn't play any role in the track production. For this reason, the longitudinal diffusion term is irrelevant for simulation purposes.
As detailed in Chap. \ref{sec:NID}, the gains obtained with NID operation at the moment are very small, as they are comparable with the ones obtained with ED operation with GEM voltages at 300 V, to be compared with nominal voltage operation of 440V. This is mainly due to difficulties in stripping the primary electron from SF$_6^-$ molecules and from the fact that after the first amplification stage, these electrons undergo again attachment with SF$_6$. The results presented in Sec. \ref{sec:NID} are only the first result and attempt with that gas mixture and amplification system. In the near future, it can be hoped that with a dedicated amplification stage developed for NID, gains similar to those of ED can be achieved. Furthermore, to conduct experiments aimed at detecting rare events at energies typical of Dark Matter, a gain comparable to that achieved with ED is required. For this reason, all the other parameters of the simulation have been left unchanged from ED.
Under this assumption, a set of 10,000 tracks at 10, 20, 30, 40, and 60 keV has been simulated.\\ 
Tracks have been reconstructed with the same algorithm and parameters employed for the ED case. An illustration comparing a track produced with ED diffusion and one with NID diffusion is presented in Figure \ref{fig:EDNIDComparion}.
Starting from these reconstructed tracks, the parameter optimization procedure illustrated in Sec. \ref{sec:paroptimization} for ED has been repeated. Due to the significantly lower diffusion of the tracks, adjustments were made to the range of the $N_{pt}$ parameter, and both the range and step for the $w$ parameter have been modified. With lower diffusion, the expectation is that primary electrons from the interaction point diffuse less, leading to a smaller $N_{pt}$ parameter. Similarly, due to reduced diffusion, a smaller portion of the track is expected to be optimal for accurate direction determination, resulting in a smaller optimal value for the $w$ parameter. The results of the angular resolution as a function of the parameter, obtained as explained in Sec. \ref{sec:paroptimization} are shown in Fig. \ref{fig:AngResVsParNID}.
The optimal parameters found for each energy are reported in Tab. \ref{tab:Parameters_NID}.
\begin{table}[]
    \centering
    \begin{tabular}{|l|lllll|}
    \hline
        E [keV] & 10 & 20 & 30 & 40  & 60 \\ \hline
        $N_{pt}$ & 70 & 70 & 70 & 70 & 70 \\ 
        $w$ & 0.9 & 0.9 & 0.9 & 0.9 & 1.2 \\ \hline
    \end{tabular}
    \caption{In the table, the optimal couples of parameters $N_{pt}$ and $w$ which provide the best angular resolution for each energy for NID diffused tracks are shown.}
    \label{tab:Parameters_NID}
\end{table}
As shown in Fig. \ref{fig:AngResVsParNID} and Tab. \ref{tab:Parameters_NID}, both $N_{pt}$ and $w$ values are smaller compared to those obtained for ED. This is attributed to the reduced diffusion, resulting in tracks comprising fewer pixels. Consequently, the algorithm requires consideration of a smaller number of pixels to identify the track's correct interaction point. Additionally, the parameter $w$ is diminished due to the decreased diffusion, allowing for a smaller segment of the track to be utilized in the direction calculation.

\begin{figure}
    \centering
    \includegraphics[width=0.5\linewidth]{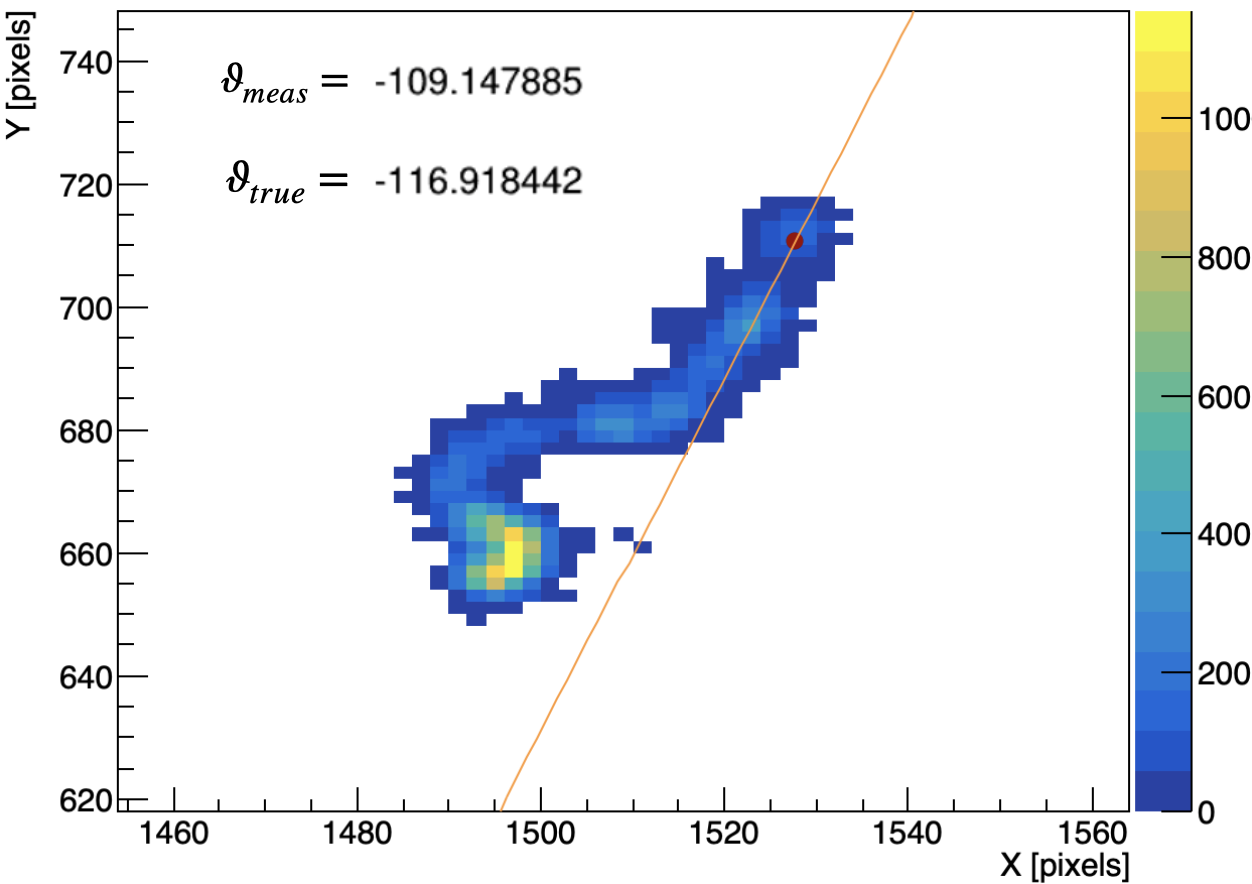}
    \caption{In figure a track simulated with NID diffusion, with the direction measured and the true direction from MC truth is shown.}
    \label{fig:NIDTrack}
\end{figure}

Following the parameter optimization procedure, the angular resolution has been examined using a dataset of tracks generated with a flat energy distribution ranging from 16 to 70 keV. The drift distance was uniformly distributed along the z-axis between 5 and 45 cm, while the angular distribution has been simulated isotropic. An example of a track simulated with NID diffusion featuring the direction measured by the algorithm is shown in Fig. \ref{fig:NIDTrack}.

\begin{figure}
    \centering
    \includegraphics[width=1\linewidth]{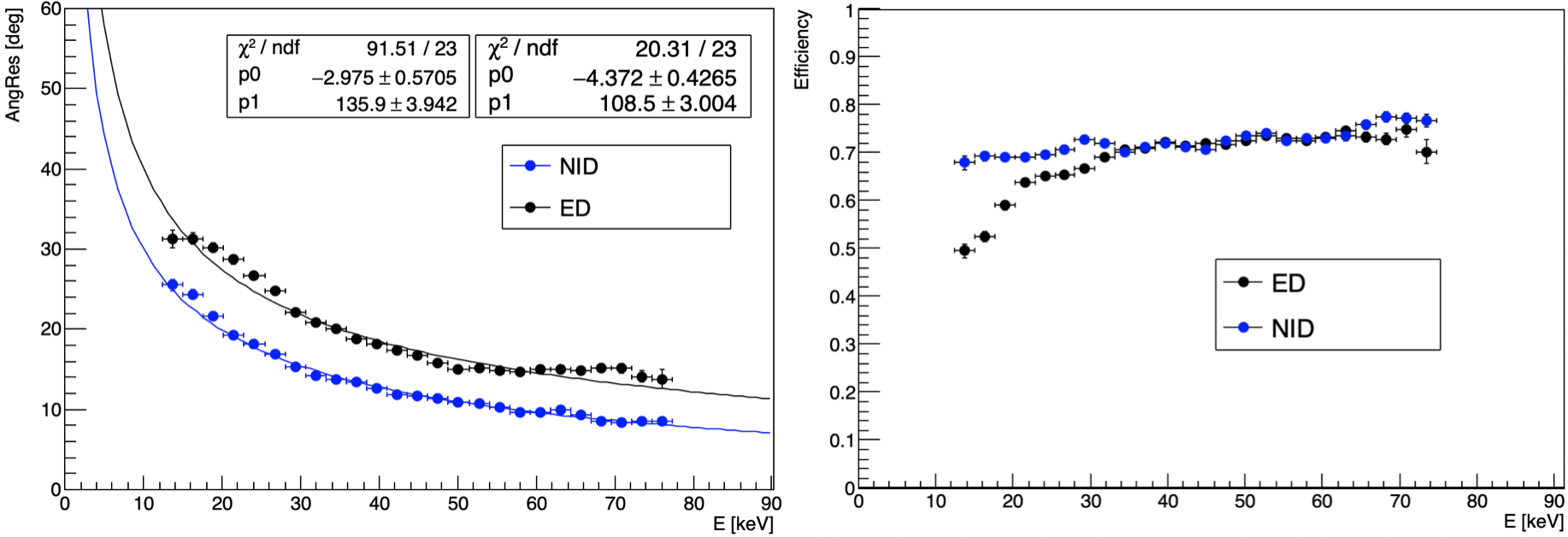}
    \caption{Results of angular resolution (left) and directionality efficiency (right) as a function of the energy under normal ED operation, in black, and with diffusion simulated for NID operations, in blue.}
    \label{fig:NIDAngRes}
\end{figure}

\begin{figure}
    \centering
    \includegraphics[width=1\linewidth]{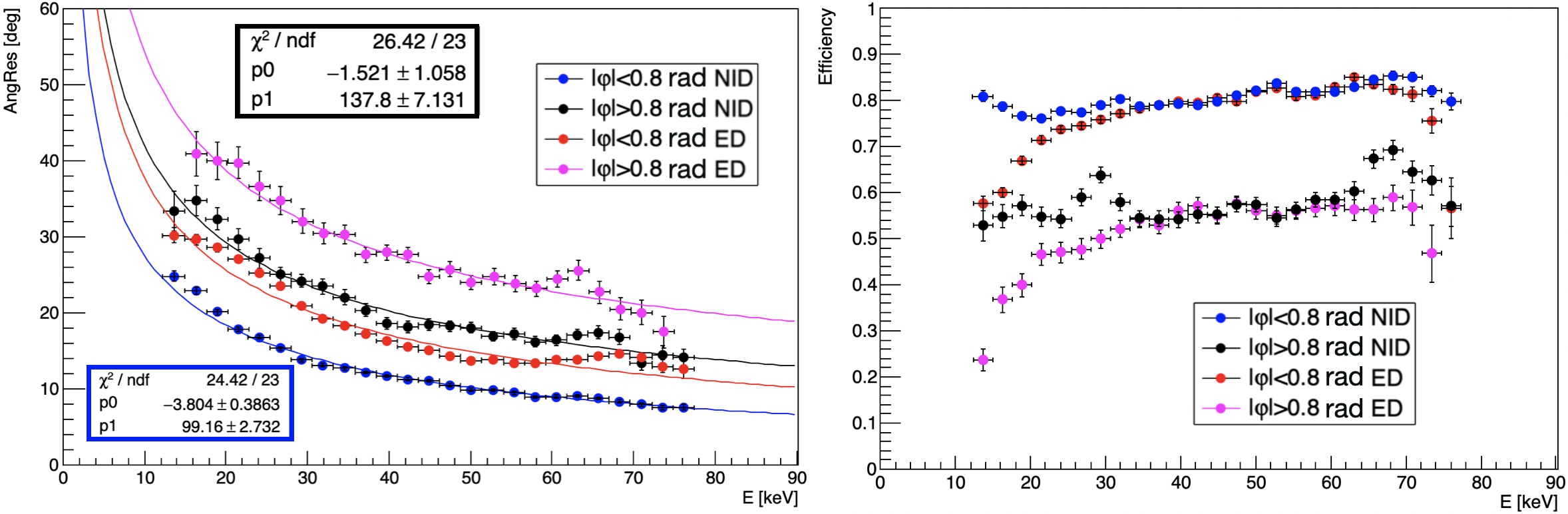}
    \caption{Angular resolution and efficiency for different track inclination with respect to the GEM plane for ED and NID operations. The colors of the contour of the fit results box refer to the lines of the same color. The results of the fit for ED are reported in figure \ref{fig:AngResFurther}.}
    \label{fig:AngVsPhiNID}
\end{figure}

\begin{figure}
    \centering
    \includegraphics[width=1\linewidth]{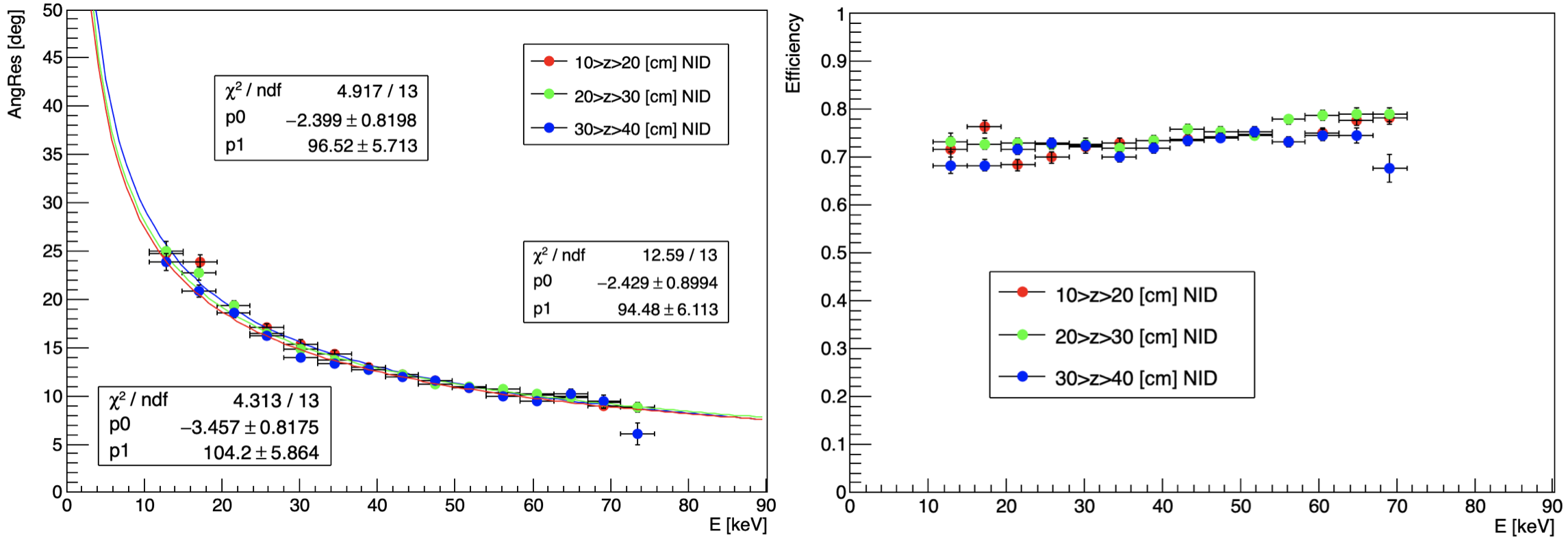}
    \caption{Left: Angular resolution as a function of the energy for tracks at different drift distances. Right: Efficiency as a function of the energy for tracks at different drift distances.}
    \label{fig:angresNIDVsZ}
\end{figure}

The results of angular resolution and efficiency as a function of the energy for NID operations are shown in Fig. \ref{fig:NIDAngRes}, compared to the one obtained with ED.
As expected, a lower diffusion results in improved angular resolution as well as improved efficiency at low energy. At higher energies, both NID and ED operations exhibit similar efficiency levels, since as discussed in Sec. \ref{sec:algoeff}, inefficiency at higher energies is primarily attributed to cases where the algorithm fails to handle track topology correctly and due to the absence of the third coordinate. The IP resolution is found to be constant in energy at $\sim$320 $\mu$m.
Moreover, further investigations of angular resolution have been carried out for different track inclinations relatively to the GEM plane, comparing results with those of ED operations. The plots are shown in Fig. \ref{fig:AngVsPhiNID}.
The plots reveal a degradation in performance for tracks directed towards the GEM plane, even in the case of NID, since even with a reduced diffusion the tail cannot be identified in the absence of third dimension. Additionally, from the efficiency plots (right), the constant behavior of the efficiency at high energy, for both ED and NID, represents further evidence that at high energy there is a component of algorithm inefficiency that does not depend on the diffusion, and a component of tracks produced perpendicularly to the GEM plane, which also does not depend on the diffusion. Finally, as expected, no dependence on angular resolution and efficiency is found for tracks at different drift distances as shown in Fig. \ref{fig:angresNIDVsZ}.

\section{Discussion}
\label{sec:Discussion}
The performances in angular resolution, which quantifies the capability of reconstructing the initial direction of low energy electron recoil, are of paramount importance for a directional detector, particularly for the solar neutrino measurement explored in this thesis.
However, this task is challenging due to multiple scattering experienced by low-energy electron recoils as they traverse the gas volume. These electrons continuously scatter and change direction along their path. Consequently, a simple linear extrapolation from the start to the end of the track is insufficient for determining the recoil direction. Furthermore, this scattering phenomenon results in direction information being concentrated at the very beginning of the track and rapidly lost going forward.\\
To address this, a dedicated algorithm has been implemented on ER tracks from sCMOS images to determine the 2D electron recoil direction. Initially, the algorithm utilizes the track's main axis as a reference and assesses the skewness of the light distribution along the track projected onto this axis to identify the region corresponding to the track's starting point. Once this region is identified, the algorithm calculates the track impact point. From here, since the information on the track direction is contained only at the very beginning of the track, for the subsequent direction calculation each pixel of the track is weighted with a decreasing exponential function which depends on the distance of the pixel from the impact point. Finally, the direction is determined as the main axis of the track with the rescaled pixels passing through the impact point, providing an orientation following the path of light in the pixels.\\
The algorithm operates with two adjustable parameters: the number of points utilized to identify the interaction region and a normalization factor to re-weight the pixels. Since these parameters are not known a priori, and due to the absence of an electron recoil source with a known direction, the model parameters optima values have been determined using simulated tracks generated with the simulation code detailed in Chapter \ref{chap:Simulation}. This simulation, as explained in the chapter, has been validated for accurately replicating ER data. The angular resolution is quantified as the standard deviation of the Gaussian fitting distribution of the difference between the measured angle and the true angle ($\theta_{meas}-\theta_{true}$). In some cases, by construction, the algorithm might fail in determining the impact point. In such instances, the algorithm will reconstruct a random direction, leading to a flat component in the $\theta_{meas}-\theta_{true}$ distribution which is taken into account in the fit. To optimize the parameters, various sets of tracks have been generated at different energies, and for each energy, a scan in angular resolution has been done as a function of the two parameters. Given that within the range of the considered parameters, the flat component of the $\theta_{meas}-\theta_{true}$ distribution remains constant in amplitude, the parameters yielding the smallest standard deviation of the Gaussian were selected as optimal parameters for each energy.\\
To assess the angular resolution performance, a set of tracks uniformly distributed in energy, ranging from 16 to 70 keV was generated. Subsequently, the angular resolution has been estimated in energy bins, employing the optimal parameters determined for each track across the various energy levels.\\
Following parameter optimization, the algorithm has demonstrated efficacy in determining the initial direction of electron recoils, achieving an angular resolution of approximately 28° at 20 keV, which improves to about 13° at 70 keV. Additionally, the resolution performances have been evaluated for tracks generated at different angles with respect to the GEM plane and at different distances from the plane, corresponding to different degrees of track diffusion. As expected, it has been observed that the algorithm is less effective in measuring the track direction when the track is produced more perpendicularly to the GEM plane, due to the absence of a third dimension. Conversely, it performs more effectively when the track is produced parallel to the GEM plane. Additionally, it has been observed that the angular resolution remains unchanged for tracks generated at varying drift distances, due to the fact that the primary contribution to diffusion arises from the GEM and doesn't depend on the drift distance.\\
The efficiency of the algorithm has additionally been investigated at different energies. Efficiency here has been defined as the ratio of events within the Gaussian portion of the $\theta_{meas}-\theta_{true}$ distribution fit to the total number of events. An efficiency different from 1 is determined by cases in which the algorithm doesn't handle correctly the topology of the track, and by the cases in which the track is produced with a direction perpendicular to the GEM plane and the impact point cannot be resolved working on the 2D projection of the track. The efficiency results to be approximately 50\% at 15 keV and increases to 70\% at 30 keV. Beyond this point, the efficiency remains constant at 70\% with further increases in energy. This trend arises because at lower energies, track diffusion has a more pronounced effect, leading to reduced efficiency. However, at higher energies, the contribution from diffusion becomes negligible, and efficiency is primarily limited by algorithm failures and the absence of PMT information.
Moreover, the efficiency has been studied for different production angles of ER with respect to the GEM plane. From this analysis, it resulted in the same behavior of efficiency as a function of the energy, with an improved efficiency from tracks produced parallel to the GEM plane. Specifically, efficiency improves for tracks generated parallel to the GEM plane, reaching up to 80\% above 30 keV. Conversely, efficiency is lower for tracks produced perpendicular to the GEM plane, with values around 50\% above 30 keV.\\ 
A novel algorithm is currently in development, employing a pathlength-based approach to pinpoint the impact point along the track. This new method aims to address instances where the current algorithm struggles to accurately handle the track's topology, with consequent misidentification of the impact point. Additionally, ongoing PMT analysis is expected to enable a three-dimensional reconstruction of ER tracks, allowing for direction measurement also in cases where tracks are generated perpendicular to the GEM plane.\\
Furthermore, as detailed in Chap. \ref{sec:NID}, the collaboration successfully conducted operations with negative ions as charge carriers in the MANGO prototype at atmospheric pressure, utilizing optical readout, with the introduction of a small amount of SF$_6$ into the standard CYGNO gas mixture. However, the achieved gains with this operational mode were considerably lower compared to those attained with ED operations and standard GEM voltages. Nonetheless, this development is only the initial phase of a series of investigations into negative ions operations within the scope of the INITIUM project. In the future, through further testing of new amplification systems and gas mixtures, there is optimism that the same level of gain achievable with ED could be attained with NID while retaining the lower diffusion characteristic of NID operations. Under this assumption, similar studies on angular resolution have been conducted on tracks simulated with the NID measured diffusion and ED gain to examine the impact of reduced diffusion on directional performance.
The study of angular resolution on NID diffused ER reveals an improved angular resolution, ranging from approximately 20° at 20 keV to around 10° at 60 keV. The same features of ED electron recoil tracks have been observed for studies of angular resolution as a function of the angle with respect to the GEM plane and as a function of the diffusion. Regarding efficiency, it is notable that with NID diffused tracks this quantity does not decrease at lower energies as the impact of diffusion at low energy levels is significantly reduced. Consequently, this allows for more accurate track direction determination even at lower energies. Moreover, the efficiency remains constant with energy, comparable to ED tracks at higher energy levels.

%% file: chapters/neutrinodetection.tex
\chapter{CYGNO-30 sensitivity to directional solar neutrino detection}
\label{chap:solarnu}
This chapter illustrates the concluding section of the author's thesis work, where the focus lies on studying the sensitivity of directional detection of solar neutrinos from the pp chain using a 30 $m^3$ detector.
As described in Chap. \ref{chap:CYGNO}, CYGNO/INITIUM is a project aimed at constructing a high-precision 3D TPC with a volume of approximately 30 m$^3$ for the search of rare events. Its unique feature and strength lie in its ability to reconstruct the initial direction of recoils, both from electron recoils (ER) and nuclear recoils (NR). CYGNO/INITIUM is conceived and optimized for the directional search of dark matter (DM). However, directional sensitivity is such a powerful feature that it opens up further avenues for very interesting physics cases, such as solar neutrino spectroscopy. This possibility had already been proposed for high-precision TPCs in the 1990s by Seguinot and Arpesella, as discussed in Sec. \ref{sec:firstproposal}. Therefore, this thesis has been developed to provide an initial assessment of the actual capabilities of the specific approach of CYGNO's 3D TPCs to perform solar neutrino spectroscopy. As shown in Chap \ref{chap:sota}, the performances of a directional TPC for a solar neutrino measurement fundamentally depend on two things: 
\begin{enumerate}
    \item The performance of the detector in terms of energy response, energy resolution, and angular resolution of the relevant electron recoils.
    \item The level of ER background to which the detector is subjected.
\end{enumerate}
To evaluate point number 1) different studies have been done and illustrated in this thesis. 
In Sec. \ref{sec:X-RayAnalisys}, the response of LIME to real ER data ranging from 3.7 to 45 keV has been studied, demonstrating high saturation at low energy and a more linear response at higher energies, and an energy resolution of approximately 13\% for energy greater than 6 keV. Regarding the angular resolution, since its estimation for ER in LIME can not be performed given the lack of a “directional” calibrated source of ER, the directionality performances can only be studied using MC simulation. Thus, in Chap. \ref{chap:Simulation} the simulation of low energy electron recoil track has been illustrated, and an extensive data/MC comparison, not only in terms of energy response but also topological track shape variables, has been conducted as part of this thesis (from which the estimation of directionality depends on).
Furthermore, Chap. \ref{chap:directionality} details the development, optimization, and performance evaluation of an algorithm for direction calculation for ER using MC data. The results of the data-MC comparison straighten the algorithm's validity, given the consistency between data and MC simulations.
As described in Sec. \ref{sec:CYGNOFut}, adopting a modular approach, both CYGNO-04 and CYGNO-30 will be constructed with multiple modules featuring drift lengths and effective granularity similar to those of the largest detector built so far and currently operating underground, LIME, described in Sec. \ref{sec:LIME}. Therefore, to estimate points 1 and 2, the most pragmatic approach is to rely on our current knowledge of LIME.
For the sensitivity study, which represents the last part of the author work carried out for this thesis, a full Bayesian statistical approach has been employed. This chapter will begin by outlining the detector concept and its assumed performances for this application (Sec. \ref{sec:resoandassumption}). 
Following this, the signal model will be detailed together with the calculation of the event rate (\ref{sec:eventrate}). 
Subsequently, the background model, along with an extensive GEANT4 simulation for the internal background studies, will be presented (Sec. \ref{sec:bkgmodel}). Moving forward, the Bayesian statistical approach employed (Sec. \ref{sec:BayesRes}), and the derivation of both the signal template (Sec. \ref{sec:sigtemplate}) and the background template (Sec. \ref{sec:bkgtemplates}) will be discussed.  Finally, the obtained results on the measurement sensitivity, utilizing the Bayesian statistical approach, will be analyzed in the last section (Sec. \ref{sec:resultsBayes}), followed by considerations on optimizing the detector for the measurement (Sec. \ref{sec:furtheroptim}).

\section{Detector concept and physics performances}
\label{sec:resoandassumption}
\begin{figure}
    \centering
    \includegraphics[width=.9\linewidth]{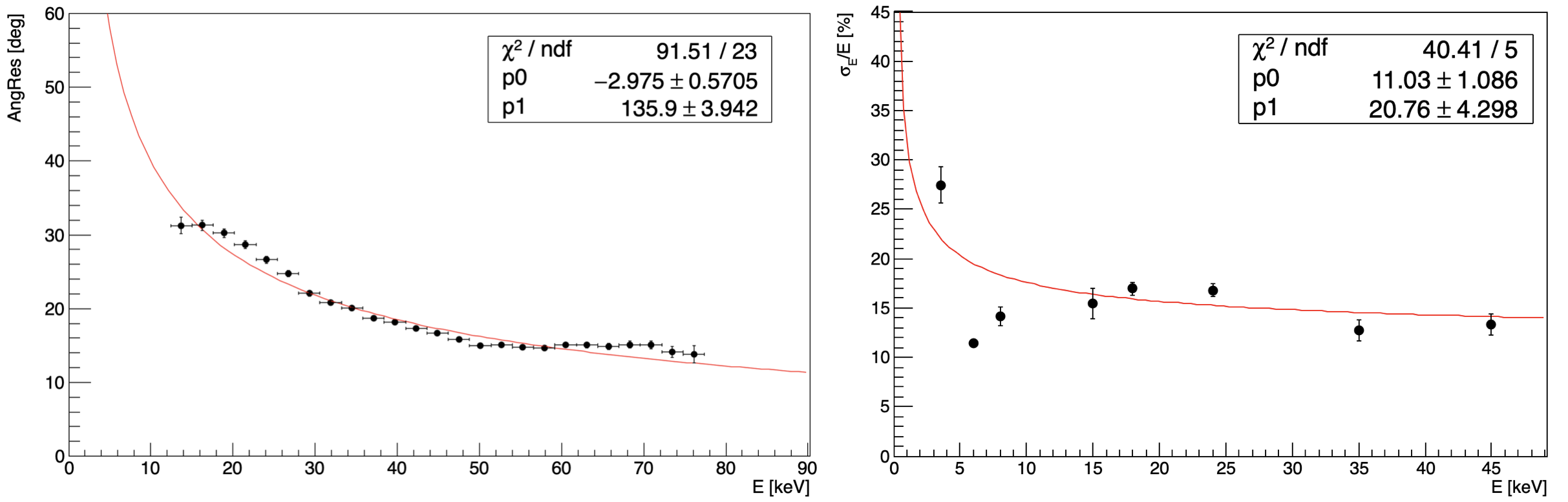}
    \caption{Left: Angular resolution as a function of the energy with LIME. Right: Energy resolution as a function of the energy with LIME.}
    \label{fig:Resos}
\end{figure}

The detector used for this study is the CYGNO-30 detector featuring an active volume of 30 m$^3$ described in detail in Sec. \ref{sec:CYGNOFut}. The current conceptualization of a CYGNO PHASE\_2 30 m$^3$ detector, is believed to be composed of 75 CYGNO-04 detectors with 0.4 m$^3$ volume. These modules are proposed to be arranged in three rows of 25 modules each, stacked on top of one another, and enclosed within the same vessel. Each CYGNO-04 module will be composed of two TPCs positioned back-to-back, featuring a central common cathode measuring 800$\times$500 mm$^2$ and a drift length of 500 mm. Additionally, each TPC will incorporate a triple stack of GEMs measuring 800$\times$500 mm$^2$. Each side of the detector will be readout by 2 sCMOS cameras and 6 PMTs, facing the amplification plane. With this configuration, the detector will be able to maintain a granularity similar to the one of the LIME detector. Since the CYGNO-30 detector will be composed of LIME-like modules, it is reasonable to expect that an individual module can exhibit performance levels at least comparable (if not better) to those obtained with LIME. 
For this reason, it is suitable to assume the performance in energy resolution obtained from the data (Sec. \ref{sec:X-RayAnalisys}) and the one in angular resolution obtained from MC (Sec. \ref{sec:angResResults}), which are reported in Fig. \ref{fig:Resos}.  
Other specific assumptions of the detector performance have been made. As explained in Sec. \ref{sec:overgroundstudies}, the track saturation is more pronounced at low energy $<$10 keV, where the tracks are spot-like and are dominated by very dense charge release. At higher energies, where the tracks are dominated by a low charge density tail, this effect of the saturation reduces. Thus, for this analysis, a linear energy response of the detector is assumed.
Moreover, it is assumed for the angular resolution that this will reach the flat value of 5° for energies exceeding 290 keV, corresponding to the resolution value at that specific energy.
\begin{equation}
    \frac{\sigma_E}{E} = 11.03+\frac{20.76}{\sqrt{E}} \ \ \ \ \ \ \ \ 
    \left\{
    \begin{aligned}
      & \sigma_{\vartheta} = -2.97 + \frac{135.9}{\sqrt{E}} & \ \ \  if \ \ \ E<290 keV \\
      & \sigma_{\vartheta} = 5 \ \ \ & if \ \ \ E\geq 290 keV
    \end{aligned}
  \right.
  \label{eq:resolutions}
\end{equation}
An additional assumption pertains to the resolution in $\varphi$, the angle relative to the GEM plane, where $\varphi=0$ represents a direction on the GEM plane. Considering the velocity of drifting electrons in the electric field, of $\mathcal{O}$(1 cm/$\mu$s), and the system's PMT + readout's sampling capability at $\mathcal{O}$(GS/s), a granularity in $z$ of a similar order of magnitude as the one obtained with the sCMOS is expected. Thus, it is assumed that the angular resolution is the same for both $\vartheta$ and $\varphi$. The measurement of $\varphi$ is related to the ability to perform the tridimensional reconstruction of the track. As shown in Sec. \ref{sec:algoeff}, there are cases in which the track is produced along the line perpendicular to the GEM plane. In this case, the directionality algorithm (Sec. \ref{sec:dirAlgo}) is not able to determine the track direction, leading to an inefficiency in the direction reconstruction. However, if the track can be reconstructed in 3D, these cases are expected to be solved, and the $\varphi$ angle can be measured.  \\
Additionally, it is assumed that electrons not contained in a single module can cross the thin $75\  \mu$m field cage and be reconstructed as a single track with measurable energy and direction. Already in the PHASE\_1 of CYGNO, it is foreseen to systematically study this phenomenon, and perform studies to match and merge tracks present in multiple pictures from different modules belonging to a single particle. For these reasons, it is assumed that the tracks are contained with efficiency 1. \\
The last assumption regards the materials and the geometry of the detector, and in particular its internal radioactivity. Given that the construction of such a detector is expected in approximately $\geq 6 \text{y}$, abundant time is foreseen for the optimization of the material selection based on radioactivity, as well as the dedicated development of low-radioactivity materials/components for CYGNO-30. This step is anyway essential also for dark matter searches to ensure the effectiveness and reliability of the detector.
Thus, for the detector, on one side an optimization of the geometry has been assumed leading to the minimization of the detector materials together with assumptions on the realization of particular detector components. On the other side, with a work of research, the most radiopure material available has been identified for the realization of the detector components. 

\section{Signal model}
\label{sec:eventrate}
\begin{figure}
    \centering
    \includegraphics[width=0.7\linewidth]{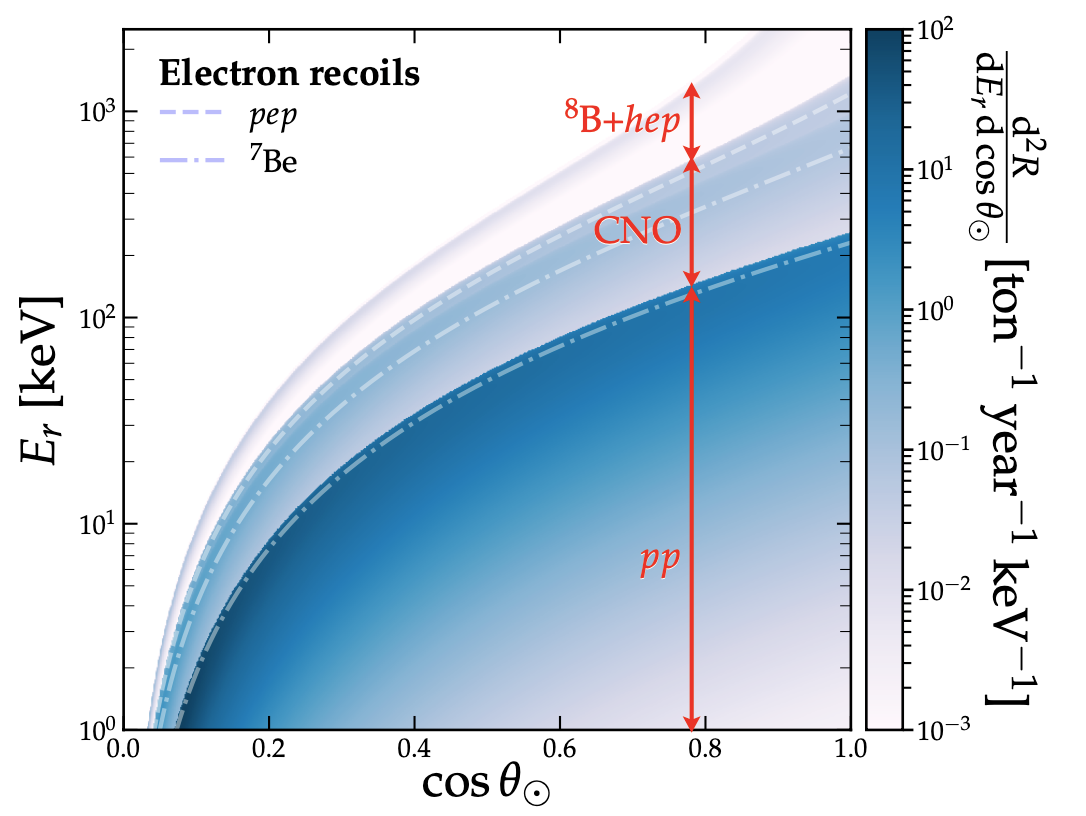}
    \caption{The event rate per ton per year, as a function of recoil energy and angle relative to the direction of the Sun. A cosine value of $\theta_{\odot}=1$ indicates recoils aligned with the Sun. In the plot also interactions for neutrinos from other sources are considered. Plot from \cite{SolrNuDir}.}
    \label{fig:diffRate}
\end{figure}

Solar neutrinos from the pp chain, given the typical energy range, interact with electrons in the detector through the elastic scattering process. In this process, the energy is purely transferred from the incident particle to the target in the form of kinetic energy. This interaction can occur through a charge current process, mediated by a $W^-$ boson, or a neutral current process, mediated by a $Z^0$ boson, as described in Sec. \ref{sec:NueES}.
The differential cross-section $d\sigma/d\cos(\theta)$, which represents the cross-section for which an electron is scattered within an angle $\theta$ and $\theta+d\theta$, with respect to the direction of the incident neutrino is given by:
\begin{align}
\begin{split}
    \frac{d\sigma}{d\cos(\theta)} = &\frac{2G_F^2m_eE_{\nu}^2\cos(\theta)}{\pi}\cdot\frac{(E_{\nu}^2m_e+m_e^3+2m_e^2E_{\nu})}{\left( (E_{\nu}+m_e)^2 - E_{\nu}^2\cos^2(\theta)\right)^2}\cdot\\
    & \left[(2+g_V+g_A)^2+(g_v+g_A)^2\left(1-\frac{2E_{\nu}m_e\cos^2(\theta)}{(E_\nu+m_e)^2-E_\nu^2\cos^2(\theta)}\right)^2\right.\\ 
    &\left.-(g_v-g_a)(g_V+g_A+2)\left(\frac{2m_e^2\cos^2(\theta)}{(m_e+E_\nu)^2-E_\nu^2\cos^2(\theta)}\right) \right]\cdot (\hbar c)^2
    \label{eq:diffcscostheta}
\end{split}
\end{align}
where equation parameters have been previously defined in \ref{eq:DiffCS}, and for dimensional consistency, the term $(\hbar c)^2$ has been included, where $\hbar$ denotes the reduced Planck constant, and $c$ represents the speed of light.
Moreover, the kinematic expression of the angle $\cos(\theta)$ can be written as:
\begin{equation}
    \cos(\theta) = \frac{E_{\nu}+m_e}{E_{\nu}}\sqrt{\frac{T'_e}{T'_e+2m_e}}
\end{equation}
From this relation the differential cross-section $d\sigma_{\nu-e}/dT'_e$ in Eq. \ref{eq:DiffCS} can be rewritten in terms of the measurable electron energy and angle by imposing its conservation through a Dirac $\delta$ function writing:
\begin{equation}
    \frac{d^2\sigma(E_{\nu})}{dT'_e\cdot d\cos{\theta}} = \frac{d\sigma_{\nu-e}(E_{\nu})}{dT'_e} \cdot \delta \left( \cos(\theta)- \frac{E_{\nu}+m_e}{E_{\nu}}\sqrt{\frac{T'_e}{T'_e+2m_e}} \right)
\end{equation}
From here the differential interaction rate with respect to the electron angle and kinetic energy of the electron can be written as: 
\begin{equation}
    \frac{d^2R}{dT'_e d\cos(\theta)} = N_{targ} \int_{E_{\nu,min}}^{E_{\nu,max}} {\frac{d^2\sigma(E_{\nu})}{dT'_e\cdot d\cos{\theta}} \frac{d\phi(E_{\nu})}{dE_{\nu}}}dE_{\nu}
\end{equation}
where $N_{targ}$ is the number target per unit volume, and $E_{\nu,min}$ and $E_{\nu,max}$ are respectively the minimum neutrino energy capable of producing a detectable electron recoil and maximum neutrino energy, with $E_{\nu,min}=\sqrt{m_e^2T'_e/(T'_e+2m_e)}$ \cite{SolrNuDir}.
The differential rate $d^2R/(dE_e\cdot d\cos{\theta_{\odot}})$ [$ton^{-1}\cdot y^{-1}\cdot keV^{-1}$] for the pp chain, form \cite{SolrNuDir}, is shown in Fig. \ref{fig:diffRate} for a gas mixture of He:SF$_6$ 97:3 (with an electron density $\sim$6 times lower than the CYGNO one) and a 1000 m$^3$ volume. In the plot, $\theta_{\odot}$ has the same meaning on theta in the thesis. The x-axis represents the scattering angle of the electron with respect to the Sun direction, while the y-axis represents the recoil energy of the electron. The code of colors in the plot represents the differential cross-section $d^2R/(dE_e\cdot d\cos{\theta_{\odot}})$ is reported.
The total rate of ER produced in the detector from solar neutrino interaction is given by the product between the detector number of targets $N_e$, and the integral of the differential flux of incident particles $\frac{d\phi(E_\nu)}{dE_{\nu}}$ multiplied by the cross-section between the incident particle and the target particle $\sigma_{\nu,e}(E_\nu)$ integrated over the minimal neutrino energy that can cause a detectable recoil and the maximal neutrino energy:
\begin{equation}
    R= N_e \cdot \int_{E_{\nu,min}}^{E_{\nu,max}}{\frac{d\phi(E_{\nu})}{dE_{\nu}} \cdot \sigma_{\nu,e} (E_\nu)} dE_{\nu}
\end{equation}
For the CYGNO gas mixture, composed of a 60:40 mixture of He:CF$_4$ at atmospheric pressure and room temperature, the electron density per m$^3$ and the total number of electrons can be computed. Applying the ideal gas law, the total number of moles contained per m$^3$ is determined as:
\begin{equation}
        n=\frac{PV}{RT}=\frac{1 atm \cdot 1000 l}{0.082 \frac{l \cdot atm}{mol \ K} \cdot 298,15 K }=40.9\  mol
\end{equation}
where P, V, and T are the detector pressure, volume and temperature, and $R=0.0821(\text{l} \cdot \text{atm}) / (\text{mol} \cdot \text{K})]$ is the universal gas constant. 
Within the detector, 60\% of these moles consist of He, with each atom possessing $N_{\text{He}}=2$ electrons, while the remaining 40\% comprises CF$_4$, which contains $N_{\text{CF}_4}=1\cdot 6+4\cdot 9=42$ electrons. Denoting $f_{\text{He}}$ as the fraction of He and $f_{\text{CF}_{4}}$ as the fraction of CF$_4$, the total number of electrons is expressed as:
\begin{equation}
        N_e=n \ f_{He} \ N_A \  N_{He}+ n \ f_{CF_4}\  N_A \ N_{CF_4}=4.43\cdot 10^{26}\  m^{-3} 
        \label{eq:electrontarget}
\end{equation}
where $N_{A}=6.022 \cdot 10^{23}\ mol^{-1}$ is the Avogadro number.\\
As shown in Sec. \ref{sec:nuoscillation}, neutrinos from the sun can oscillate. Thus, an electron neutrino produced in the Sun can arrive at the Earth with a different flavor. Due to the small angle $\theta_{13} \sim 8.7$° compared to the others, and the closeness in mass of two of the mass states compared to the third, it is possible to move in the approximation of two neutrino flavors considering an electron neutrino and a non-electron neutrino (superposition of $\nu_\mu$ and $\nu_\tau$) with $\Delta m_{ij}^2 = \Delta m_{12}^2$ and $\theta_{ij}=\theta_{12}$ \cite{Vissani}. 
The oscillation probability will be given by:
\begin{equation}
    P\left(\nu_e \rightarrow \nu_\mu\right)=\sin ^2\left(2 \theta_{12}\right) \cdot \sin ^2\left(\frac{\Delta m_{12}^2 L}{4 E}\right)
\end{equation}
However, since solar neutrinos are produced in a wide space (solar core), which implies a large variation of L, and with a continuous
energy spectrum, the exact information on L/E is totally lost, and an average on L/E can be considered. 
By averaging these quantities over the distance traveled and the energy, the probability of oscillation and non-oscillation are respectively:
\begin{equation}
    P\left(\nu_e \rightarrow \nu_\mu\right)=P_{e \mu}=\frac{1}{2} \sin ^2\left(2 \theta_{12}\right) \quad P\left(\nu_e \rightarrow \nu_e\right)=P_{e e}=1-\frac{1}{2} \sin ^2\left(2 \theta_{12}\right)
\end{equation}
with $\Delta m_{12}^2=(7.53 \pm 0.18) \times 10^{-5}\ eV^2$ and $\sin^2(2\theta_{12})$ = 0.846 \cite{Vissani}. 
Finally, the interaction rate can be calculated as: 
\begin{equation}
    R=N_e\cdot \int_{E_{min}}^{E_{max}}{\frac{d\varphi_{pp}(E_{\nu})}{dE_{\nu}}(P_{ee}\sigma_{\nu_e}(E_\nu) + P_{e\mu}\sigma_{\nu_\mu}(E_\nu) )dE_\nu}
\end{equation}
where $E_{min}$ and $E_{max}$ are respectively the minimum neutrino energy capable of producing a detectable electron recoil and maximum neutrino energy the maximal energies of the neutrino from $pp$ cycle, $\varphi_{pp}(E)$ represent the flux of solar neutrino from $pp$ cycle as a function of the neutrino energy, $\sigma_{ee}$ and $\sigma_{e\mu}$ are respectively the elastic scattering cross-section of electron neutrino on electron and muon neutrino on electron defined in Eq. \ref{CsIntegrNue} and Eq. \ref{CsIntegrNumu}.
Finally, due to the non-availability of an analytic expression for the fluxes as a function of the neutrino energy, the previous expression is reduced to: 
\begin{equation}
        R=N_e\sum_{i}\varphi(E_i)(P_{ee}\sigma_{\nu_e}(E_{\nu,i}) + P_{e\mu}\sigma_{\nu_\mu}(E_{\nu,i}) )) \Delta E
        \label{rate}
\end{equation}
and the value of $\varphi(E_i)$ and $E_i$ are tabulated in Tab. \ref{TabNu} from \cite{solSpec}.
From the full calculation, it results in a rate of events
\begin{equation}
        R=3.065\cdot 10^{-8} \ \frac{events}{s \cdot m^3}=0.97\ \frac{events}{y\cdot m^3}
\end{equation}
which corresponds to $\sim$30 ev/y in a 30 $m^3$ CYGNO PAHSE\_2 detector. \\ 

\section{Background model}
\label{sec:bkgmodel}
As detailed in Sec. \ref{sec:solarnuchallenges}, the external gamma radiation coming mainly from $^{238}$U, $^{232}$Th, and $^{40}$K in the surrounding rock, interacting in the detector constitute an irreducible background for solar neutrino detection. To overcome this background, an appropriate shielding must be employed for the detector. Additionally to this background, the internal background, coming from the detector components is also present and must be characterized. From the point of view of the model, these backgrounds are assumed to exhibit an isotropic distribution in angle. Also in the circumstance of a localized gamma source in the laboratory reference frame, when the case is translated into solar coordinates, these localized excesses are blurred by the coordinate transformation, considering also the earth rotation.\\
For the external background, it is assumed that it can be reduced up to a negligible level with respect to the internal one by the employment of thick enough copper shielding. For the internal one, a GEANT4 simulation has been developed optimizing different geometry of the detector components and different radiopure material. 
The results obtained at the end of this optimization process are reported in this section.



\subsection{Detector design}
\label{sec:CYGNO30DesignGEANT}
A CYGNO-30 detector (Sec. \ref{sec:CYGNOFut}) is envisioned to be composed of many CYGNO-04 (Sec. \ref{sec:CYGNO04}) modules arranged closely together. In particular, in the simulation, three rows of 25 CYGNO-04 stacked one on top of the other have been considered.\\ 
A CYGNO-04 module comprises two TPCs arranged back-to-back, sharing a common cathode measuring 500$\times$800 mm$^2$. Following the cathode, a field cage, also sized 500$\times$800 mm$^2$ in the x and y dimensions, ensures uniform drift over the entire 50 cm drift length. The field cage is composed of a 75 $\mu$m substrate with 35 $\mu$m thin copper bands printed on top. The copper bands are 50 mm wide in z and spaced by 50 mm. The partition of the tension along the copper of the field cage is guaranteed by a series of small SMD resistors. At the opposite end of the cathode, a stack of three GEMs measuring 50$\times$80 cm$^2$ each forms the amplification stage. The entire detector is enclosed within a gas-tight vessel. Positioned outside the vessel are two sCMOS cameras and six PMTs per side, facing the GEMs. For the construction of CYGNO-04 as outlined in the CYGNO Technical design report \cite{giovanni_mazzitelli_2023_76967}
the development of higher radiopure material for the realization of the detector components has been proposed. A specialized lens is under development in partnership with external firms utilizing radio-pure materials like Suprasil\textsuperscript{TM}, aiming to reduce radioactivity by a factor of 10$^4$ with respect to the current level. Leveraging expertise gained from the DRIFT collaboration, the cathode will consist of a 0.9 $\mu$m aluminized Mylar film \cite{Daw:2011wq}. This thin film minimizes mass, thereby reducing the potential background induced by the gas. It also increases the likelihood of alpha particles from potential radioactive contaminants escaping and depositing their energy in the fiducial volume, enabling their identification and exclusion from event analysis. However, this cathode shows a very high amount of radioactivity due to the presence of $^{235}$U contaminations. The drift field cage will be constructed from a thin Kapton\textsuperscript{TM} foil with copper layers acting as rings connected by SMD resistors, following a technique employed by DRIFT \cite{Daw:2011wq}. This design minimizes copper and resistor mass near the sensitive gas, thus reducing potential background from their radioactive contaminants. Employing a method pioneered by the T-REX Collaboration \cite{Castel_2019}, the GEMs will undergo cleaning with deionized water baths to reduce surface radioactive contamination.\\
The simulation of CYGNO-30 starts from these assumptions, only the primary components of the detector have been considered, as they are likely to contribute the most to the background as resulting from the background studies of LIME. Specifically, in the simulation each module includes:
\begin{itemize}
\item A common 1 mm thick copper cathode measuring 500$\times$800 mm$^2$ shared by both sides of the detector. The copper is used since the chosen one (Sec. \ref{sec:CYGNO30MaterialGEANT}) is more radiopure than the aluminum used for the CYGNO-04 cathode. 
\item On each side of the TPC, a field cage constructed using a 75 $\mu m$ thick acrylic substrate with 5 evenly spaced 35 $\mu m$ thick copper bands printed on the acrylic is present. The field cage has dimensions of 500$\times$800 mm$^2$ and extends 500 mm along the drift direction covering the whole side of the sensitive volume. Along the drift direction, the copper bands have a width of 55 mm. The field cage design draws inspiration from the one employed in the DRIFT collaboration \cite{daw2011drift}, with considerations leading to the use of ultrapure acrylic instead of Kapton. 
\item In the space connecting each copper band, an SMD resistor of 0.32$\times$1.0$\times$0.5 $mm^3$ made by Al$_2$O$_3$ is positioned outside the field cage on the acrylic for the partition of the drift field along the drift direction. The SMD resistor concept is based on the one used in XENON \cite{Aprile_2017}. 
\item At the other extremity of the field cage with respect to the cathode, the amplification stage with the 3 GEMs is present. The GEM constitutes a relatively new technology in low-background physics applications, with its development primarily geared toward high-energy physics. Consequently, these components have never been optimized from the radiopurity point of view. Efforts involving the exploration of alternative materials characterized by high radiopurity for the fabrication of GEMs are in progress. The research discussed in \cite{vahsen2020cygnus} highlights the successful creation of GEMs with an acrylic core in the case of 400 $\mu m$ GEMs. GEMs constructed with a radiopure acrylic core are expected to exhibit lower radioactivity compared to conventional GEMs featuring a Kapton core. The production of 50 $\mu m$ GEMs with an ultrapure acrylic core can be foreseen with the advancement of the production techniques.  Consequently, in the simulation, three GEMs, each measuring 500$\times$800 mm$^2$, consisting of a 40 $\mu m$ acrylic core with armors of 5 $\mu m$ copper are used. 
\item The entire setup is enclosed within a copper vessel of 10 mm thickness. 
\item Positioned outside the vessel, at a distance of 576 mm from the GEMs, are the camera lenses. Each lens is a 10 mm diameter, 1 mm thick disk made of Suprasil \cite{Suprasil}, a highly radiopure material employed in the construction of camera lenses.
\item Positioned at 60 mm behind the lenses, which is approximately the distance of the sensor from the lenses inside the camera, there is a silicon sensor with dimensions 10.6 $\times$ 18.0 mm$^2$ and a thickness of 1 mm. It is anticipated that in the future, only the sensor and lens will be housed within the outer shielding, while the remaining camera electronics will be positioned outside the shielding to mitigate radioactive contributions. This adjustment will likely require customized development.
\end{itemize}
The modules are arranged in close proximity along the rows, with separation between rows dictated by the presence of SMD resistors.
Within each space delimited by the field cage, cathode, and the first GEM, the He:CF$_4$ gas mixture volume is placed and labeled as “sensitive” detector.

\subsection{Detector elements material choice}
\label{sec:CYGNO30MaterialGEANT}
For the realization of the CYGNO-30 detector components, the use of the most radiopure material available is foreseen: 
\begin{itemize}

\item For the realization of all the copper elements, the Electroformed copper (EFCu) described in the Majorana Demonstrator Radioassay Program \cite{EFCopper} is considered. This is an exceptionally high-purity copper produced using a mass spectrometer to eliminate $^{238}$U and $^{232}$Th. Designed for shielding applications for double beta decay searches in $^{76}$Ge detectors \cite{EFCProduction}, EFCu presents a top-tier radiopurity option. In the study conducted in \cite{LaFerriere:2015owa}, the $^{238}$U and $^{232}$Th concentrations in the produced EFCu were measured to be 0.131 $\mu$Bq/kg for $^{238}$U and 0.034 $\mu$Bq/kg for $^{232}$Th. Where the Bq is the Becquerel and represents the number of atomic decays per second. This copper is used in the simulation for the realization of GEM armors, cathodes, field cage copper strips, and the vessel.
\item For insulating material, the selection of acrylic for the GEMs core and support of the copper strip for the field cage has been chosen. Specifically, the most radiopure acrylic, as utilized in the SNO experiment \cite{Leonard_2008}, has been chosen. This acrylic exhibits orders of magnitude greater radiopurity than the most radiopure Kapton available, with measured radioactivity levels of $^{40}$K, $^{232}$Th, and $^{238}$U being $<71.2 \ \mu$Bq/kg, $<56.9\ \mu$Bq/kg, and $<296.0\ \mu$Bq/kg, respectively. In the reference, only an upper limit has been set on the contaminant, but the value reported in the limits will be used in the simulation.
\item The SMD resistors utilized in the simulation are the same as those employed by the Xenon experiment, as documented in \cite{Aprile_2017}. These 7.5 M$\Omega$ resistors, used in XENON1T for the PMTs, are supplied by Vishay Intertechnology and represent the most radiopure option available today.
\item As previously mentioned, the material constituting the lens is Suprasil. The radioactive contamination of this material has been measured by the underground service in LNGS, yielding results for the contaminants $^{40}$K, $^{232}$Th, and $^{238}$U, which are 0.3 mBq/Kg, 40.7 $\mu$Bq/Kg, and 123 $\mu$Bq/Kg, respectively.
\item Ultimately, to simulate the sensor contribution to the background, the radioactive levels of the camera sensor measured at LNGS, have been utilized. This measurement has been done by dismounting a sCMOS camera and measuring the radioactivity of the single camera components. The results for the sensor's $^{40}$K, $^{232}$Th, and $^{238}$U contamination are, respectively, 9 mBq/Kg, 2.8 mBq/Kg, and 2 mBq/Kg.
\end{itemize}
A summary of all detector elements, their materials, and radioactive contamination is provided in Tab. \ref{tab:radioactive}.
\begin{table}
    \centering
    \begin{adjustbox}{max width=\textwidth}
    \begin{tabular}{|l|l|l|l|l|l|l|l|}
    \hline
        Detector element & Material & $^{238}$U & $^{232}$Th & $^{40}$K & $^{235}$U & $^{226}$Ra & $^{228}$Th \\ \hline
        GEM core & Acrylic & $<296.0\  \mu$Bq/Kg & $<56.9\ \mu$Bq/Kg & $<71.2 \ \mu$Bq/Kg & x & eq & eq \\ \hline
        GEM armor & EFCu & 0.131 $\mu$Bq/Kg & 0.034 $\mu$Bq/Kg & x & x & eq & eq \\ \hline
        Field cage support & Acrylic & $<296.0\  \mu$Bq/Kg & $<56.9\ \mu$Bq/Kg & $<71.2 \ \mu$Bq/Kg & x & eq & eq \\ \hline
        Field cage strip & EFCu & 0.131 $\mu$Bq/Kg & 0.034 $\mu$Bq/Kg & x & x & eq & eq \\ \hline
        Cathode & EFCu & 0.131 $\mu$Bq/Kg & 0.034 $\mu$Bq/Kg & x & x & eq & eq \\ \hline
        Vessel & EFCu & 0.131 $\mu$Bq/Kg & 0.034 $\mu$Bq/Kg & x & x & eq & eq \\ \hline
        Camera sensor & Silicon & 2 mBq/Kg & 2.8 mBq/Kg & 9 mBq/Kg & x & eq & eq \\ \hline
        Camera lenses & Suprasil & 123 $\mu$Bq/Kg & 40.7 $\mu$Bq/Kg & 0.3 mBq/Kg & x & eq & eq \\ \hline
        Resistors & $Al_2 O_3$ & 1 $\mu$Bq/pc & 0.14 $\mu$Bq/pc & 1.2 $\mu$ Bq/pc & 0.04 $\mu$Bq/pc & 0.18 $\mu$Bq/pc & 0.13 $\mu$Bq/pc \\ \hline
    \end{tabular}
    \end{adjustbox}
    \caption{In the table, the elements employed for each detector component in the background GEANT4 simulation with the relative activity due to the different contaminants are reported. The activities labeled with “eq.” represent a nuclide in secular equilibrium with the father nucleus of the chain. The one labeled with “x” signals the absence of that nuclide or a concentration too low to be measured. Bq/pc stands for the Becquerel per piece.}
    \label{tab:radioactive}
\end{table}
It must be specified that $^{226}$Ra and $^{228}$Th are present in the $^{238}$U and $^{232}$Th decay chains, respectively, as shown in Fig. \ref{fig:deacay_chain}. Unless explicitly reported with a different radioactive level from the main element of the chain, these contaminants are presumed to be in secular equilibrium (eq.), implying that they are considered to have the same activity. In the case of the resistors, specific activity values are also provided for $^{226}$Ra and $^{228}$Th. In these instances, it is assumed that the equilibrium is broken, and the chains will be simulated with the activity of the primary nucleus up to the indicated point of break. For the rest of the chain, the activity of the nucleus at the point where the chain is broken will be considered in the simulation. If a contaminant is absent or its concentration is too low to be measured it has been labeled with x, and thus it is not included in the simulation.

\subsection{CYGNO-30 internal material radioactivity simulation}
\label{sec:BkgCYGNO30}
\begin{figure}
    \centering
    \includegraphics[scale=.4]{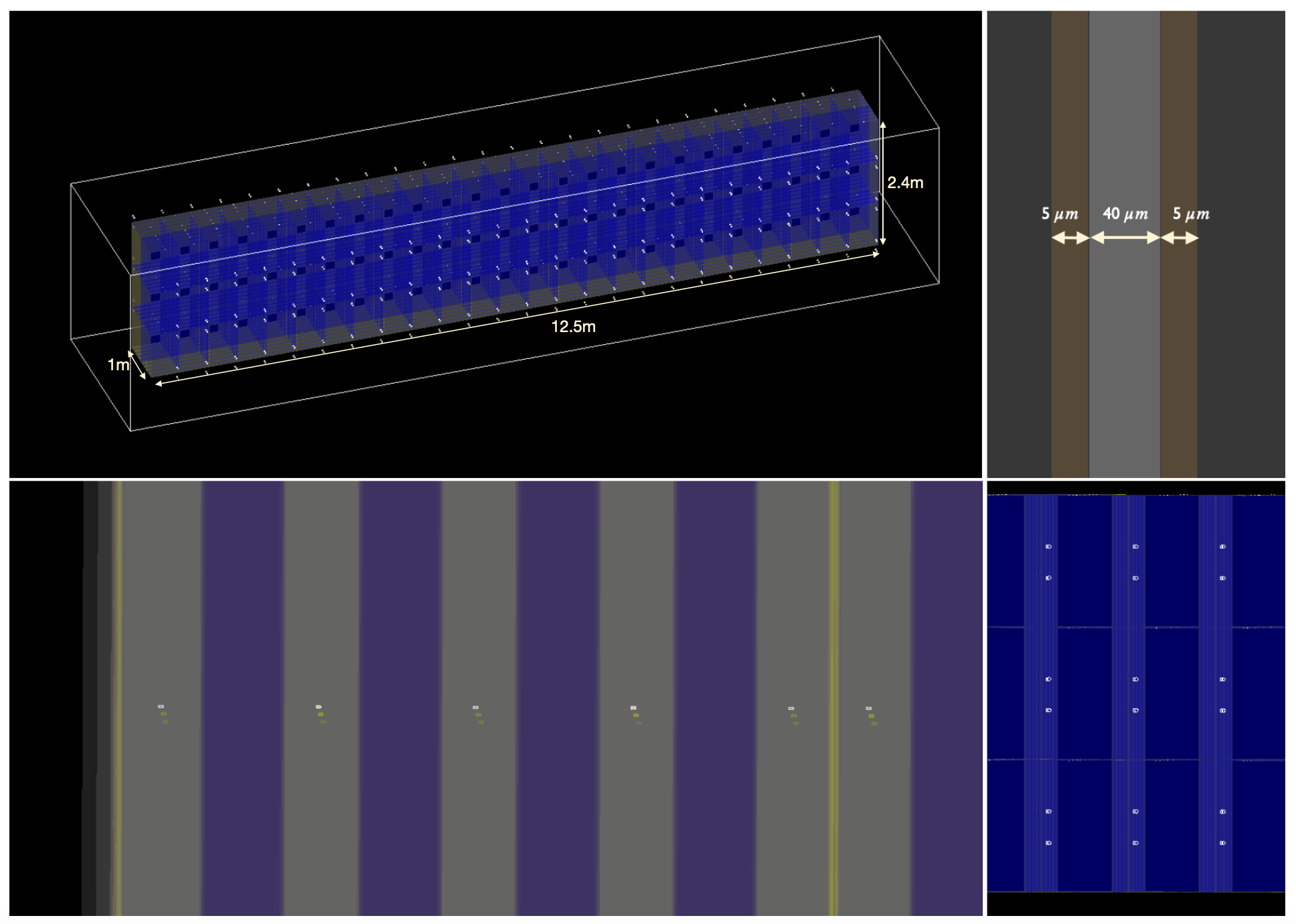}
    \caption{Schematic representation of the detector with details on the detector components. Top Left: Overview of the CYGNO-30 detector. Top Right: Schematic representation of a GEM in the simulation featuring an acrylic core and copper armors. Bottom Left: Schematic representation of the field cage, with white points representing resistors and the purple strips are the copper strips on the support. On the right side of the picture, the orange layer represents the cathode, while on the left, the GEMs are positioned. Bottom Right: View from the longer side of the detector, the white dots are the lenses with sensors situated behind them.}
    \label{fig:GEANT4Det}
\end{figure}

\begin{figure}
    \centering
    \includegraphics[width=0.9\linewidth]{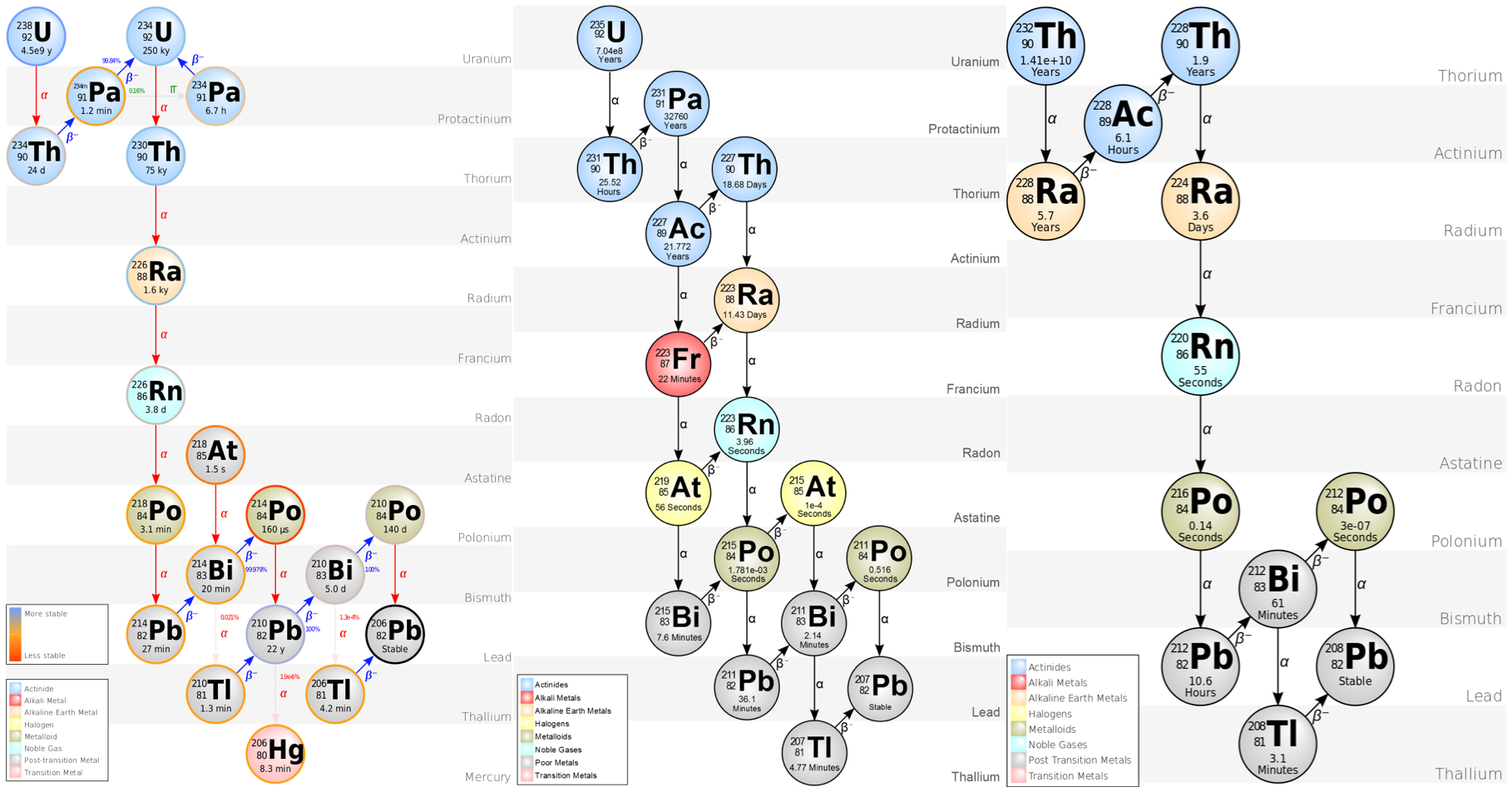}
    \caption{In figure, the radioactive decay chains for $^{238}$U, $^{235}$U, and $^{232}$Th are shown. Plots from \cite{Decay_chain}.}
    \label{fig:deacay_chain}
\end{figure}

\begin{figure}
    \centering
    \includegraphics[width=0.8\linewidth]{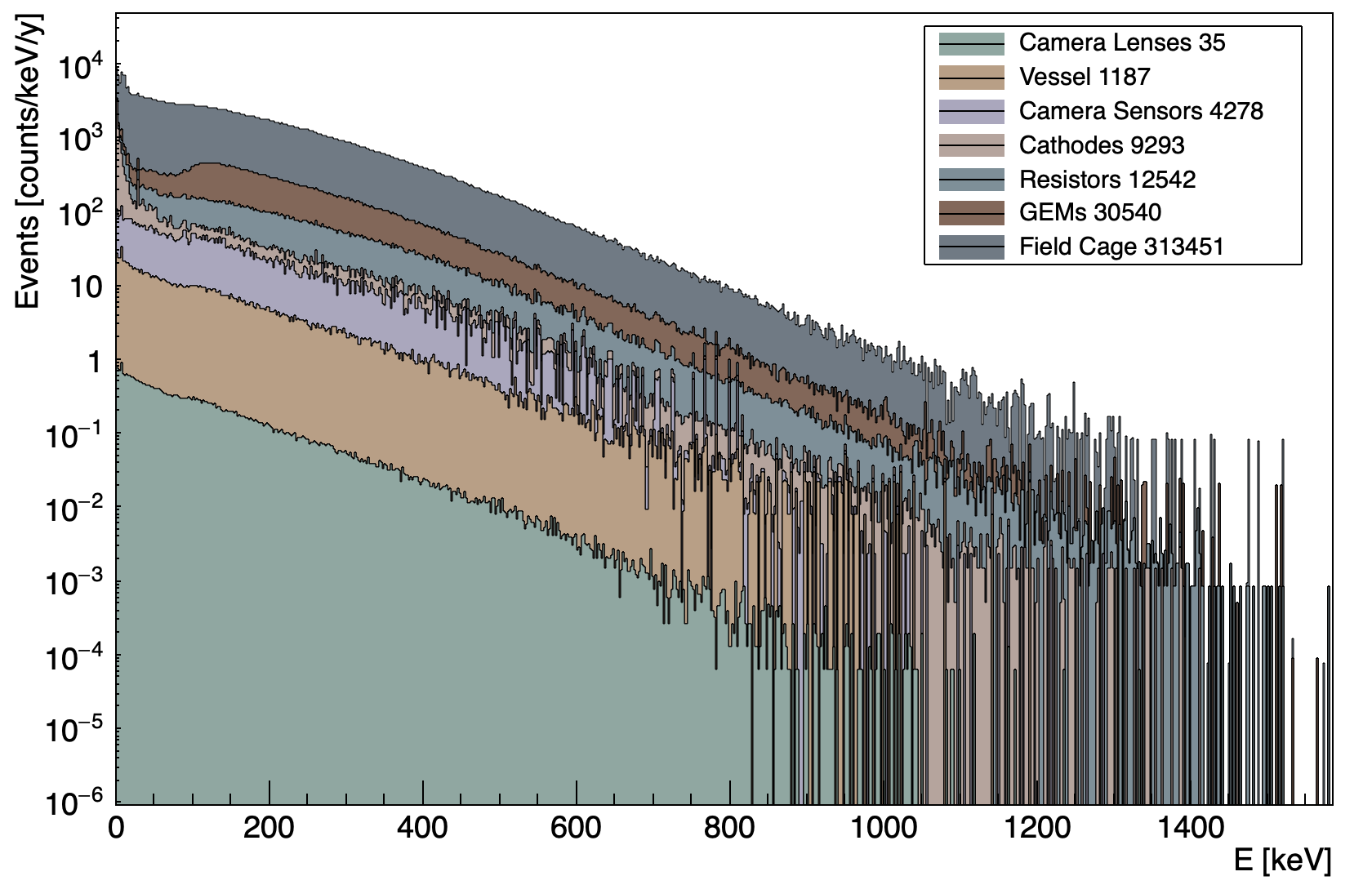}
    \caption{ER recoil simulated spectrum produced by the internal radioactivity of the detector. In the legend, the detector component for each histogram along with the total number of events produced by its radioactive contaminants are shown.}
    \label{fig:IntNoCut}
\end{figure}

\begin{figure}
    \centering
    \includegraphics[width=0.8\linewidth]{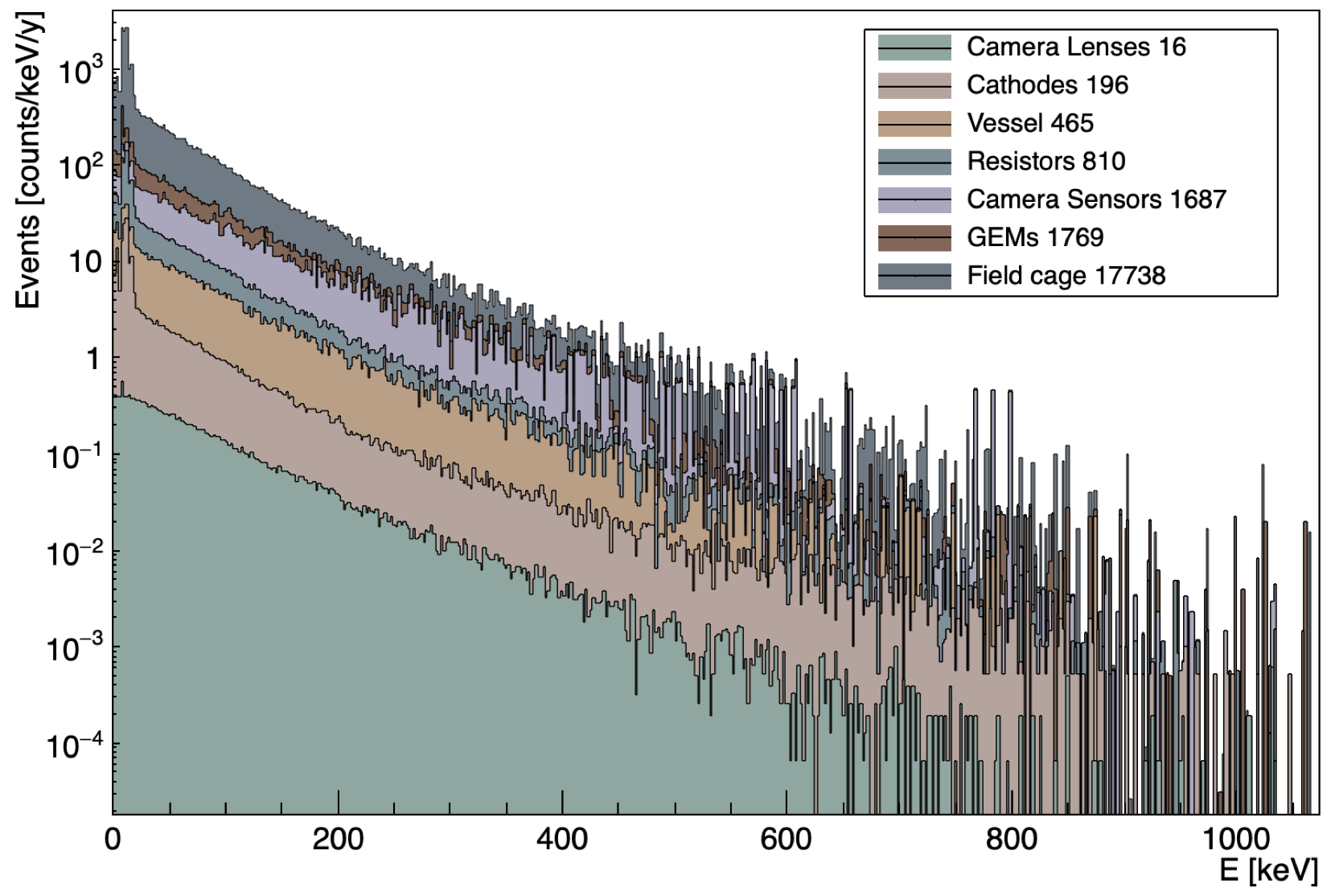}
    \caption{Background spectra after cuts on the particle interaction point. In the legend, the detector component for each histogram along with the total number of events produced by its radioactive contaminants are shown.}
    \label{fig:IntCut}
\end{figure}

To estimate the internal radioactivity of the detector, the geometry described in Sec. \ref{sec:CYGNO30DesignGEANT}, along with the materials outlined in Sec. \ref{sec:CYGNO30MaterialGEANT}, was implemented in GEANT4. Fig. \ref{fig:GEANT4Det} illustrates a detailed view of the detector geometry in GEANT4.
 Subsequently, the radioactivity of each radioactive contaminant present in each detector component was simulated. All charged particles crossing the sensitive volume were recorded, and their energy and position were saved.
In the GEANT4 simulation, the decay of the primary nucleus, as well as the subsequent decays of the daughter nuclei of the chain, including nuclear deexcitations emitting X-Rays, are handled by the physics lists. The physics lists used specifically for radioactive decays are the \lstinline[basicstyle=\large\ttfamily,language=C++]{G4DecayPhysics} which is responsible for handling the decay channels for all unstable nuclei, and the \lstinline[basicstyle=\large\ttfamily,language=C++]{G4RadioactiveDecayPhysics} which is responsible for handling the X-Rays emission from nuclei which left in an excited state after the decay \cite{GEANT4Dev,GEANT4Phys}.\\
Decays have been simulated for each contaminant listed in Tab. \ref{tab:radioactive} while the processes are handled from the physics lists. For each detector component, a random element is selected, and a random point is chosen within the volume of that element. At that specific location, the entire decay chain originating from the parent nucleus is simulated. The simulation includes atomic deexcitations, and their probabilities are managed by the physics list. If the components in the material are in secular equilibrium, the chain is fully simulated until the first stable element. If the secular equilibrium is not present, the radioactive chain is stopped at the daughter nucleus, and the remaining chain is simulated separately, starting from the daughter nucleus measured activity. The schemes of the decay chains for $^{238}$U, $^{235}$U, and $^{232}$Th are reported in Fig. \ref{fig:deacay_chain}.
All the particles crossing the sensitive volumes are recorded. Specifically, the recorded information includes the parent nucleus, the process responsible for generating the interacting particle, the total energy released in the detector (summed if released in different gas volumes), and the interaction point. For each detector component and each contaminant in the list, simulations of $10^7$ primary decays have been performed. Knowing the mass of each component and its radioactive content in Bq/kg, each spectrum has been normalized by scaling it with:
\begin{equation}
    N=\frac{1}{N_{ev}}\cdot A\left[\frac{dec}{s\cdot kg}\right]\cdot M[kg] \cdot 3.15 \cdot  10^7 \left[\frac{s}{y}\right]
    \label{eq:normActivity}
\end{equation}
where $N_{ev}$ is the number of simulated events, $A$ is the activity, $M$ is the element mass, and the last part is the conversion from seconds to years. The electron recoil spectra due to each contaminant present on each detector element are displayed in Fig. \ref{fig:IntNoCut}. Each histogram is shown summed to the previous one, such that the one on the top is the total one. In the legend, the detector components are presented in descending order of their contribution to the total number of events, with the one on top of the legend representing the one that leads the lowest contribution.
For fiducialization purposes, the distribution of interaction points of the particles was analyzed, and 
by applying 2 cm cuts from the border of each gas volume, to suppress the majority of the alpha and beta events, the background spectra reduce to the ones in Fig. \ref{fig:IntCut}. This choice is supported by findings from \cite{Lewis_2022}, demonstrating the feasibility of achieving a 1 cm resolution in z. This suggests that comparable performance levels could be attainable in the future.
In Tab. \ref{tab:totalcontrib} the contributions of the different detector elements to the total background with and without cuts are reported.
Moreover, the total number of events expected in different energy ranges are reported in Tab. \ref{tab:Nevents}.
\begin{table}
    \centering
    \begin{tabular}{|c|c|c|c|c|}
    \hline
        ~ & full range (ev/y) & 1-20 keV (ev/y) & $>$10 keV (ev/y)  & 10-400 keV (ev/y)\\ \hline
        No cut & $3.7\cdot 10^5$ & $5.9\cdot 10^4$ & $3.2\cdot 10^5$ & $3.3\cdot 10^5$\\ \hline
        With cut & $2.26\cdot 10^4$ & $1.28\cdot 10^4$ & $1.77\cdot 10^4$ & $1.76\cdot 10^4$ \\ \hline
    \end{tabular}
    \caption{Number of events in the full range of the histogram, and in other different ranges of energy.}
    \label{tab:Nevents}
\end{table}
\begin{table}[]
    \centering
    \begin{tabular}{|l|l|l|}
    \hline
        Detector element & No cuts [ev/y] & Cuts [ev/y] \\ \hline
        Field cages & 313451 & 17738 \\ \hline
        GEMs & 30540 & 1769 \\ \hline
        Resistors & 12542 & 810 \\ \hline
        Cathodes & 9293 & 196 \\ \hline
        Camera sensor & 4278 & 1687 \\ \hline
        Vessel & 1187 & 465 \\ \hline
        Camera Lenses & 35 & 16 \\ \hline
    \end{tabular}
    \caption{In the table, the total contribution of the different detector elements is reported, with and without fiducial cuts.}
    \label{tab:totalcontrib}
\end{table}
In particular, the 1-20 keV range has been considered since it is interesting for dark matter searches and can be used for a direct comparison of the background needed for solar neutrino and DM searches. 
With a threshold of 10 keV, as considered in the analysis, the expected number of events across the entire range above that threshold value is $1.77 \cdot 10^4\ ev/y$, while 10-400 keV is the range in which the signal is expected.

\section{The Bayesian statistical approach}
\label{sec:BayesRes}
The Bayesian probability approach enables the quantitative determination of probability values associated with statements whose truth or falsity is uncertain prior to repetitions \cite{lista2017statistical}. 
According to Bayes theorem the probability of an event A with the condition that event B has occurred can be expressed as:
\begin{equation}
    P(A|B) = \frac{P(B|A)\cdot P(A)}{P(B)}    
\end{equation}
Where $P(B|A)$ is the probability that the event B happens given the event A, $P(A)$ can be interpreted as the prior probability of the event A to happen, before having any knowledge of B (prior probability), $P(B)$ is the probability of the event $B$ to happen. In this context, $P(A|B)$ can be interpreted as the probability of event A to happen after knowing that event B has occurred. Thus $P(A|B)$ represents an update of knowledge of the probability of A ($P(A)$) to happen, having as additional information that the event B has occurred.\\
For a sample of $n$ observations $\vec{x}=(x_1,...,x_n)$ with a known probability density function (PDF) described by a model $M$ dependent on m parameters $\vec{\theta}=(\theta_1,...,\theta_m)$, the probability density function for the parameters $\vec{\theta}$ given the observation of the data $\vec{x}$, denoted as the posterior Bayesian probability distribution $P(\vec{\theta}|\vec{x})$, can be expressed using the Bayes theorem as follows:
\begin{equation}
P(\vec{\theta}|\vec{x}) = \frac{P(\vec{x}|\vec{\theta})\cdot P(\vec{\theta}) }{P(\vec{x})}
\label{eq:bayescondata}
\end{equation}
Where the $P(\vec{x}|\vec{\theta})$, represents the probability of obtaining a set of observation $\vec{x}$, given the parameters $\vec{\theta}$, which is the likelihood function described in Sec. \ref{sec:likelihood}. The $P(\vec{\theta})$ is the \textit{prior} PDF of the set of parameters $\vec{\theta}$, and represents the $a\  priori$ knowledge on the parameters $\vec{\theta}$. The denominator represents the probability of obtaining a given set of data and can be interpreted as a normalization factor. Finally, the $P(\vec{\theta}|\vec{x})$ represents the updated knowledge on the parameters $\vec{\theta}$ after the observation of the data $\vec{x}$, and it is called \textit{posterior}. On these bases, the Eq. \ref{eq:bayescondata} can be rewritten as:
\begin{equation}
    P(\vec{\theta}|\vec{x}) = \frac{\mathcal{L}(\vec{x}|\vec{\theta}) \cdot \pi(\theta_1)...\pi(\theta_m)}{\int{\mathcal{L}(\vec{x}|\vec{\theta}') \cdot \pi(\theta'_1)...\pi(\theta'_m) \cdot d\theta'_1...d\theta'_m}}
\end{equation}
where $\pi(\theta_i)$ represents the \textit{prior} distribution on the parameter $\theta_i$, which can be interpreted as the \textit{degree of belief} on the value of the parameter before the observation. 
Finally, the posterior on the single parameter $\theta_i$ can be obtained by marginalizing the full posterior distribution.
\begin{equation}
    P(\theta_i|\vec{x}) = \int{P(\vec{\theta'}|\vec{x})d\theta'_1...d\theta'_{i-1}d\theta_{i+1}...\theta'_m }
\end{equation}
In most cases, the exact analytic calculation of the posterior is not possible, instead, there are methods that allow to sample it. This is useful both for obtaining the distribution of the posterior density and for calculating the integral in such a way to obtain a normalized distribution \cite{lambert2018student}. In the Bayesian Analysis Toolkit (BAT), the toolkit used to perform the sensitivity studies in this thesis, an adaptive random walk Metropolis Hasting is implemented (Details in appendix \ref{app:appendixD}).

\subsection{Hypothesis testing}
\label{sec:MOdelCOmp}
The Bayes theorem can also be used to compare the probability of two hypotheses, given the data collected. Defining the two hypotheses $H_0$ and $H_1$, eventually described two sets of parameters $\vec{\theta}_0$ and $\vec{\theta}_1$, the posterior probabilities on the two hypotheses can be written as:
\begin{equation}
    P(H_{0}|\vec{x}) = \frac{\mathcal{L}(\vec{x}|H_{0},\vec{\theta}_0)\cdot \pi(H_{0})\cdot \pi(\vec{\theta}_0)}{P(\vec{x})} \ \ \  P(H_{1}|\vec{x}) = \frac{\mathcal{L}(\vec{x}|H_{1},\vec{\theta}_1)\pi(H_{1})\cdot \pi(\vec{\theta}_1)}{P(\vec{x})}
\end{equation}
by integrating the two likelihoods over the respective parameters of the model $\theta$, and computing the ratio of the posterior the following quantity is obtained:
\begin{equation}
    \frac{P(H_1|\vec{x})}{P(H_0|\vec{x})}= \left(\frac{\int{\mathcal{L}(\vec{x}|H_{1},\vec{\theta}_1) \pi(\vec{\theta}_1)d\vec{\theta}_1}}{\int{\mathcal{L}(\vec{x}|H_{0},\vec{\theta}_0) \pi(\vec{\theta}_0) d\vec{\theta}_0 }}\right) \cdot \frac{\pi(H_1)}{\pi(H_0)} = B_{1/0} \cdot \frac{\pi(H_1)}{\pi(H_0)}
\end{equation}
The ratio of the priors of $H_1$ and $H_0$ represents the relative degree of belief in the models, while the quantity between parentheses is called the Bayes factor \cite{BayesFactor}. The Bayes Factor represents the updated knowledge on the priors induced by the data. Hence, by not assuming any preference on the model, $\pi(H_1)=\pi(H_0)=0.5$, the Bayes Factor represents how much probable is that the data are described by the model of the hypothesis $H_1$ with respect to the one of $H_0$. For example, by assuming the hypothesis $H_1$ as the hypothesis describing the presence of a signal over a background, and $H_0$ the null hypothesis with the absence of the signal, a Bayes factor greater than one would point towards favoring the hypothesis with the presence of the signal. 
According to the scale proposed in \cite{BayesFactor}, the evidence of $H_1$ against the null hypothesis $H_0$ for different values of the Bayes factor can be expressed as in Tab. \ref{tab:BayesFactor}.
\begin{table}
    \centering
    \begin{tabular}{|c|c|}
    \hline
        $B_{1/0}$ & Evidence against $H_0$ \\ \hline
        1-3 & Not worth more than a bare mention \\ \hline
        3-20 & Positive \\ \hline
        20-150 & Strong \\ \hline
        $>$150 & Very Strong \\ \hline
    \end{tabular}
    \caption{Evidence of $H_1$ against the null hypothesis $H_0$ for different values of the Bayes factor.}
    \label{tab:BayesFactor}
\end{table}

\section{Binned Likelihood analysis}
In the sensitivity analysis, both the energy and the angular spectra of the electron recoils are considered. The binned likelihood analysis is conducted in Solar coordinates to enhance the detectability of the anisotropy of electron recoils induced by neutrinos compared to background signals. In this coordinate, the angle $\cos{\theta}$ measured for the electron recoils is thus calculated with respect to the Sun direction. In this reference frame, the background will be flat in $\cos{\theta}$, while the signal will follow the $\cos{\theta}$ distribution described by the differential cross-section in Eq. \ref{eq:diffcscostheta}, and will be peaked towards the sun direction.
As shown in \ref{sec:X-RayAnalisys} to perform an unbinned likelihood fit the analytical form of the distribution is needed. In this case, given the non-trivial analytical form of the background and the angular distributions convoluted with the resolutions, a binned approach has been adopted. 
In conducting the analysis, templates were constructed for both signal and background. These templates were built using theoretical distributions and factoring in detector effects. They provide information on the expected number of events in each energy and angle bin. This template data will serve both as the expected value for the likelihood and for generating toy Monte Carlo simulations to estimate sensitivity.
In this section, the production of the templates will be discussed (Sec.s \ref{sec:sigtemplate}, \ref{sec:bkgtemplates}). The definition of the likelihood for S+B and only B models will follow (Sec. \ref{sec:likelihood_def}) to conclude with the toy MC production, the test of the fitting algorithm, and the final analysis of the sensitivity done with the model comparison (Sec. \ref{sec:MOdelCOmp}).

\subsection{Signal templates}
\label{sec:sigtemplate}
\begin{figure}
    \centering
    \includegraphics[scale=.26]{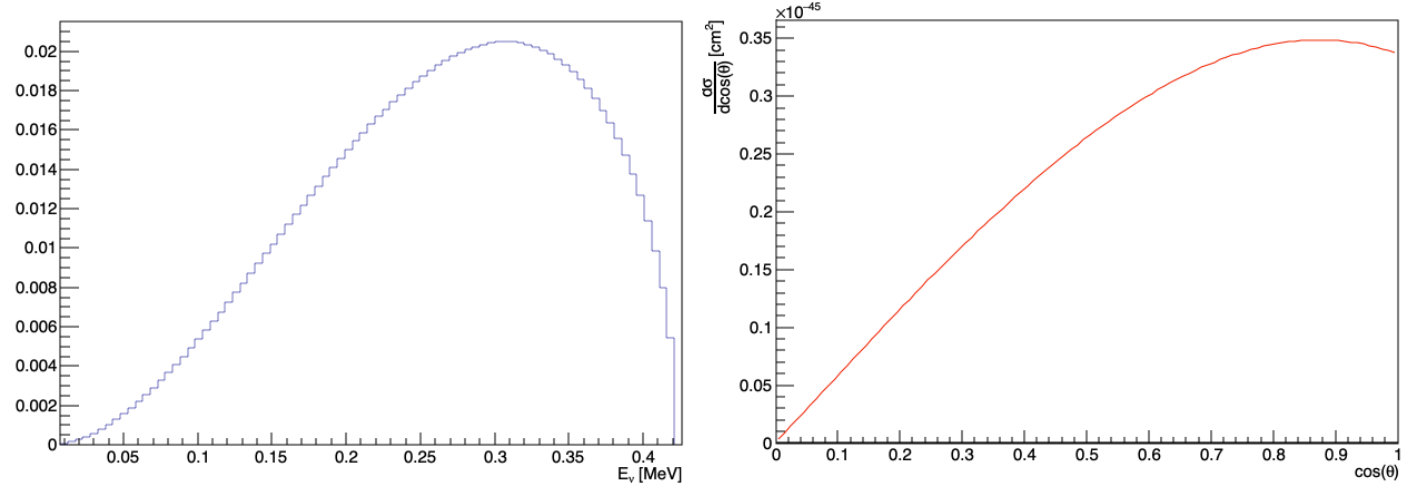}
    \caption{Left: spectral shape of solar neutrino flux from the pp chain normalized. Right: differential cross-section $d\sigma/d\cos(\theta)$ on the electron for a neutrino of 100 keV of energy.}
    \label{fig:NuSpecDiffCs}
\end{figure}

\begin{figure}
    \centering
    \includegraphics[width=0.95\linewidth]{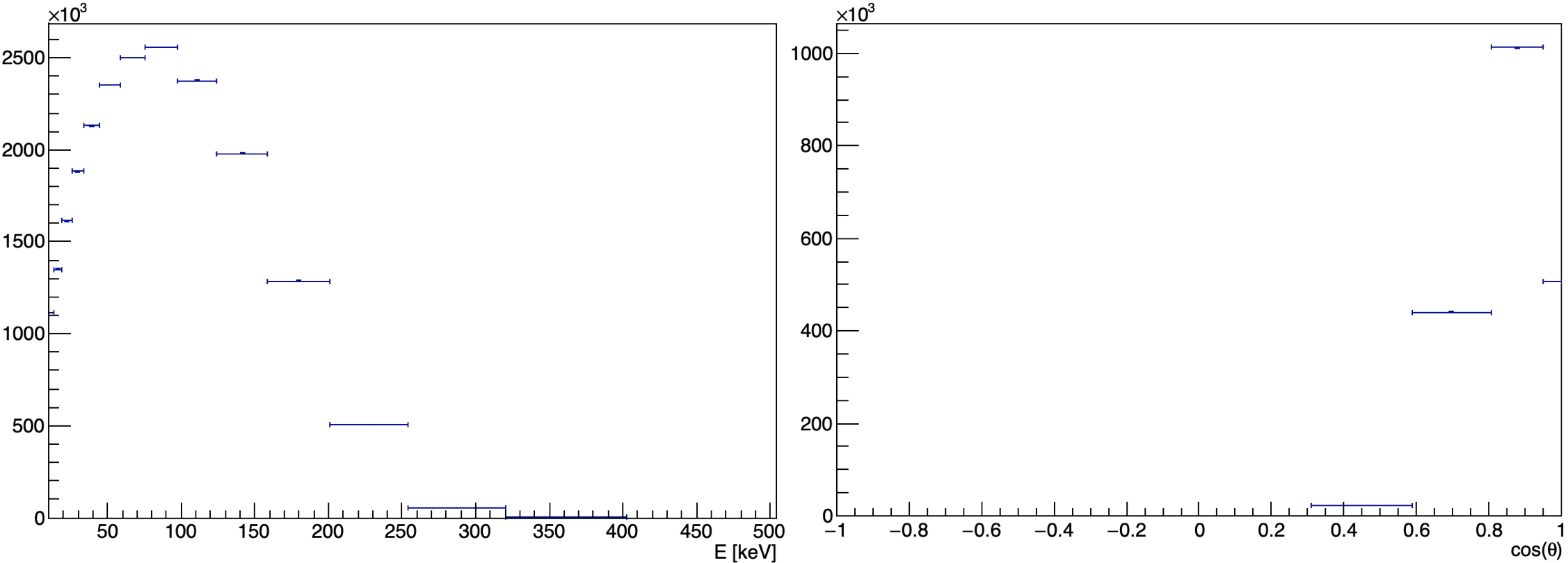}
    \caption{Left: Template of energy distribution for the signal. Right: Template of angular distribution for the signal in the bin of energy centered at 141 keV.}
    \label{fig:templateSign}
\end{figure}

The signal template is derived starting from the solar neutrino spectrum from the pp chain illustrated in the left plot of Fig. \ref{fig:NuSpecDiffCs}. 
A random neutrino energy is sampled from the distribution $E_{\nu}$. Defining $\theta$ as the angle formed by the scattered electron with respect to the original neutrino trajectory, the differential cross-section $d\sigma/d\cos(\theta)$ is computed for a neutrino with energy $E_{\nu}$ using the Eq. \ref{eq:diffcscostheta}.
The right plot in Fig. \ref{fig:NuSpecDiffCs} illustrates an example of the differential cross-section $d\sigma/d\cos(\theta)$ plotted as a function of $\cos(\theta)$ for a 100 keV neutrino. A random $\cos(\theta)$ value is then sampled from the function. \\
Once the values of the electron scattering angle $\cos(\theta)$ and the neutrino energy $E_\nu$ are determined, the kinematics of the process is established, allowing uniquely for the calculation of the electron kinetic energy $T_e$ using the Eq. \ref{eq:kinematic}.
Subsequently, both the energy and the angle undergo smearing based on the resolutions outlined in \ref{eq:resolutions}. Specifically, the smeared energy $T_{e-sm}$ is randomly drawn from a Gaussian distribution with a mean $T_e$ and a standard deviation $\sigma_E(T_e)$ from \ref{fig:Resos}. Simultaneously, the smeared angle, $\theta_{sm}$ (calculated from $\cos(\theta)$), is extracted from a Gaussian distribution with a mean $\theta$ and a standard deviation $\sigma_{\theta}$, still from \ref{fig:Resos}.\\
Subsequently, the signal template histograms were constructed. A histogram for the energy has been generated, starting from 10 keV (the threshold value), and with bins centered at each energy $E$ and a bin width of $2\cdot\sigma_E(E)$ (two times the energy resolution evaluated at the energy $E$). In the same way described above, for each energy bin, an angular distribution histogram, $\cos(\theta)$, was built. The same criteria for bin size were applied in creating the histograms for the recoil angles, ensuring that the range was between -1 and 1.\\ To generate the template, $10^7$ neutrino interaction events have been simulated. For each event, the energy histogram has been populated with the smeared energy $E_{e, sm}$, and the angular histogram corresponding to the energy bin of $E_{e, sm}$ was filled with the corresponding value of $\cos(\theta_{sm})$. The resulting template for the energy distribution, along with the angular distribution template for the bin centered at 141 keV taken as an example, are displayed in Fig. \ref{fig:templateSign}.
It can be observed that as expected the angular distribution of $\cos(\theta)$ is peaked in the opposite direction of the neutrino production point. 

\subsection{Background templates}
\label{sec:bkgtemplates}
\begin{figure}
    \centering
    \includegraphics[width=0.95\linewidth]{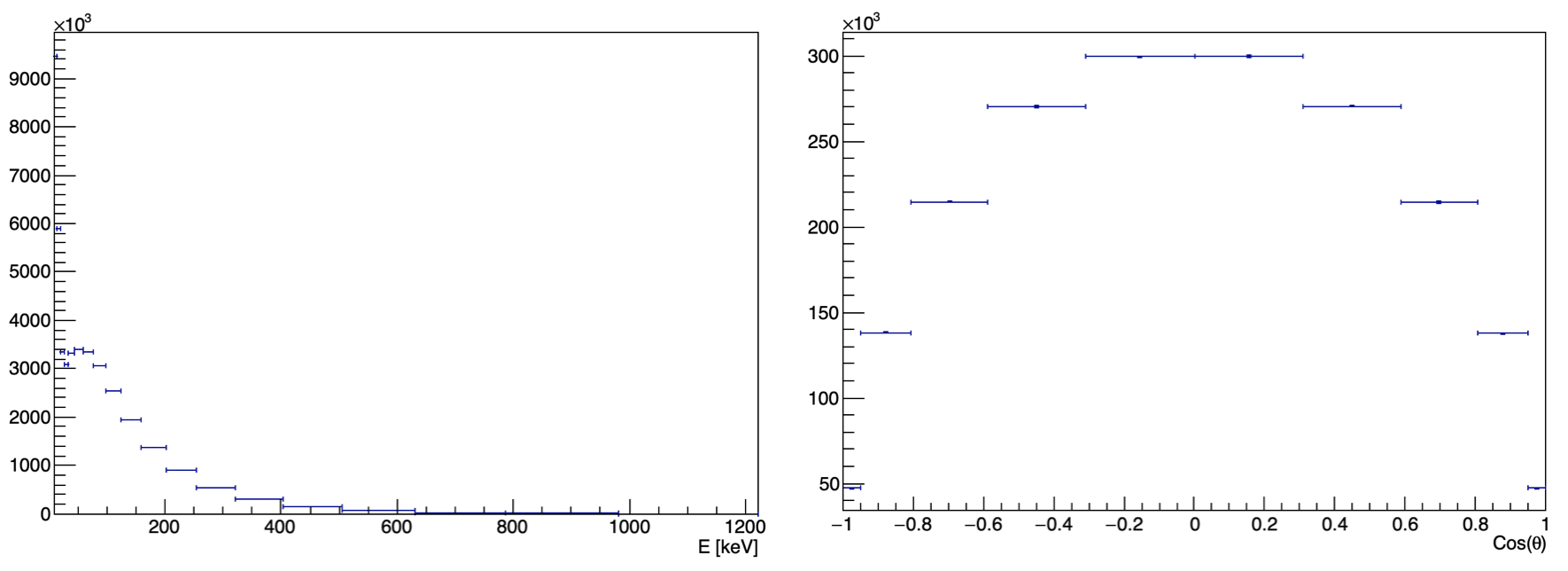}
    \caption{Left: Template of energy distribution for the background. Right: Template of angular distribution for the background in the bin of energy centered at 141 keV. The shape of the angular spectrum in the histogram is given by the choice of a variable bin size.}
    \label{fig:templateBkg}
\end{figure}

The background templates have been generated from the background spectrum obtained from the simulation illustrated in Sec \ref{sec:BkgCYGNO30}. Assuming that the angular distribution of the background is isotropic, templates were created by randomly sampling an energy from the background energy spectrum and an angle from a uniform distribution in $\cos{\theta}$, followed by smearing using the same procedure as before. Similar to the signal, a histogram for energy and a $\cos(\theta)$ histogram for each energy bin have been created with the same binning and have been filled with these values. The resulting energy template and angular template distribution for the bin at $141\ \text{keV}$ are depicted in Fig. \ref{fig:templateBkg}. 

\subsection{Likelihoods definition}
\label{sec:likelihood_def}
As described in Sec. \ref{sec:BayesRes}, the Bayesian statistical approach is based on the concept of likelihood, thus these functions must be defined. The likelihood analysis was conducted on binned histograms.
In this framework, the normalized templates of signal (S) and background (B) offer the expected probability of an event from S or B falling into a specific bin. Given the Poissonian nature of the process and the low expected number of events per bin, the probability of each bin is described by a Poisson distribution:
\begin{equation}
    P(n|\lambda) = \frac{e^{-\lambda}\cdot \lambda^n}{n!}
\end{equation}
where $P(n|\lambda)$ is the probability of observing $n$ events given the expectation value $\lambda$. Hence, the likelihood function can be formulated as the product of the probabilities of observing $n_{i,j}$ events in each bin with expectation value $\lambda_{i,j}$ \cite{Alduino_2017}. Here, $i$ and $j$ represent the indices of the $i^{th}$ bin in of the energy template, and j as the $j^{th}$ bin of the angular template associated with that energy bin. Following this, the likelihood function can be expressed as:
\begin{equation}
    \mathcal{L}(n_{i,j}|\lambda_{i,j}) = \prod_{i,j}{ \frac{e^{-\lambda_{i,j}}\lambda_{i,j}^{n_{i,j}} }{n_{i,j}!} }
\end{equation}
Considering the case in which both signal from neutrino interactions and background are present, the expectation values $\lambda_{i,j}$ can be written as:
\begin{equation}
    \lambda_{i,j}=(\mu_s\cdot P^{s}_{i,j} + \mu_b\cdot P^{b}_{i,j})
\end{equation}
where $\mu_s$ and $\mu_b$ are the number of signal and background events, and $P^{s}_{i,j}$ and $P^{b}_{i,j}$ are the probability that an event of signal and background end up in the bin $i,j$. Defining $H_1$ the hypothesis of signal and background in the data, the likelihood can be rewritten as:
\begin{equation}
    \mathcal{L}(D|\mu_s,\mu_b) = \prod_{i,j}{ \frac{e^{-(\mu_s\cdot P^{s}_{i,j} + \mu_b\cdot P^{b}_{i,j})}(\mu_s\cdot P^{s}_{i,j} + \mu_b\cdot P^{b}_{i,j})^{n_{i,j}} }{n_{i,j}!}  }
    \label{eq:likelihood_sig}
\end{equation}
Finally, the posterior probability, which describes the probability of $\mu_s$ and $\mu_b$, given the data, under the hypothesis $H_1$, can be obtained as:
\begin{equation}
    P(\mu_s,\mu_b|D) =\frac{\mathcal{L}(D|\mu_s,\mu_b)\cdot \pi(\mu_s)\pi(\mu_b)}{P(D)}
\end{equation}
where $\pi(\mu_b)$ and $\pi(\mu_s)$ are the priors on the signal and on the background. For the priors, a Poissonian prior with mean value equal to the number of events expected has been used for the background, while a flat prior has been chosen for the signal to treat every possible $\mu_s$ at the same way:
\begin{equation}
    \pi(\mu_s)= \frac{1}{n_s+n_b}\ \ \ \ \ \ \  \ \ \ \ \ \pi(\mu_{b})=\frac{e^{-n_b} \cdot n_b^{\mu_b}}{\mu_b!}
    \label{eq:priorsdef}
\end{equation}
For the hypothesis $H_0$, under which no signal is present, the likelihood can be written simply as:
\begin{equation}
    \mathcal{L}(D|\mu_s=0,\mu_b) = \prod_{i,j}{ \frac{e^{-(\mu_b\cdot P^{b}_{i,j})}(\mu_b\cdot P^{b}_{i,j})^{n_{i,j}} }{n_{i,j}!}  }
    \label{eq:likelihood_bk}
\end{equation}
Thus the posterior probability reduces to:
\begin{equation}
    P(\mu_b|D) =\frac{\mathcal{L}(D|\mu_s=0,\mu_b)\cdot \pi(\mu_b)}{P(D)}
\end{equation}
The analysis for the sensitivity is then conducted on toy-MC.

\subsection{Toy-MC production and analysis strategy}
The Monte Carlo toys have been generated based on the expected event rates for both the signal and background, considering various exposures $\epsilon$. Additionally, given an eventual further background reduction with respect to what is obtained from the simulation, different hypotheses of a further background reduction $R_{b}$.
Given these values, and considering the Poissonian fluctuations of the number of interactions, for each toy, the actual number of events $n_s$ and $n_b$ have been stochastically sampled from Poisson distributions with respective mean values $\bar{N}_{s}$ and $\bar{N}_{b}$.
Following this, the toy Monte Carlo datasets have been constructed as histograms mirroring the ones used for the templates. These histograms have been populated with $n_s$ events randomly sampled from the signal distribution templates and $n_b$ events from the background distribution templates. For each combination of $\epsilon$ and $R_b$, $10^4$ toys have been generated.
For each of the toys generated, the Bayes factor has been calculated by integrating the likelihood functions under the two different hypotheses and calculating the ratio:
\begin{equation}
    \frac{P(H_1|D)}{P(H_0|D)} = \frac{ \int{\mathcal{L}(D|\mu_s,\mu_b,H_1)\pi(\mu_s)\pi(\mu_b)d\mu_{s}d\mu_{b}}}{\int{\mathcal{L}(D|\mu_b,H_0)\pi(\mu_b)d\mu_{b}}} \cdot \frac{\pi(H_1)}{\pi(H_0)}
\end{equation}
From the operative point of view, the integral of the likelihood function has been computationally calculated with the same likelihoods of Eq. \ref{eq:likelihood_sig} and Eq. \ref{eq:likelihood_bk}, and priors of Eq. \ref{eq:priorsdef}. To not favor any model, the priors on both the hypothesis have been set at 50\%, $\pi(H_1)=\pi(H_0)=0.5$. 

\begin{figure}
    \centering
    \includegraphics[width=0.7\linewidth]{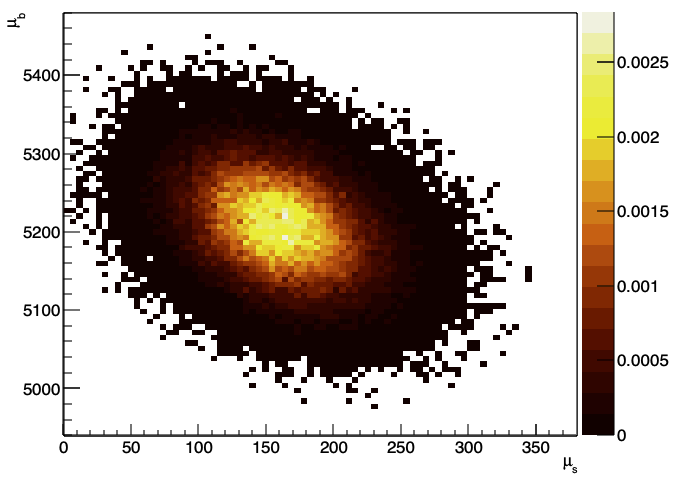}
    \caption{Posterior distribution $P(\mu_{s},\mu_b|D)$ as a function of $\mu_s$ and $\mu_b$ sampled from the MCMC and normalized.}
    \label{fig:PosteriorSamp}
\end{figure}

\begin{figure}
    \centering
    \includegraphics[width=1.\linewidth]{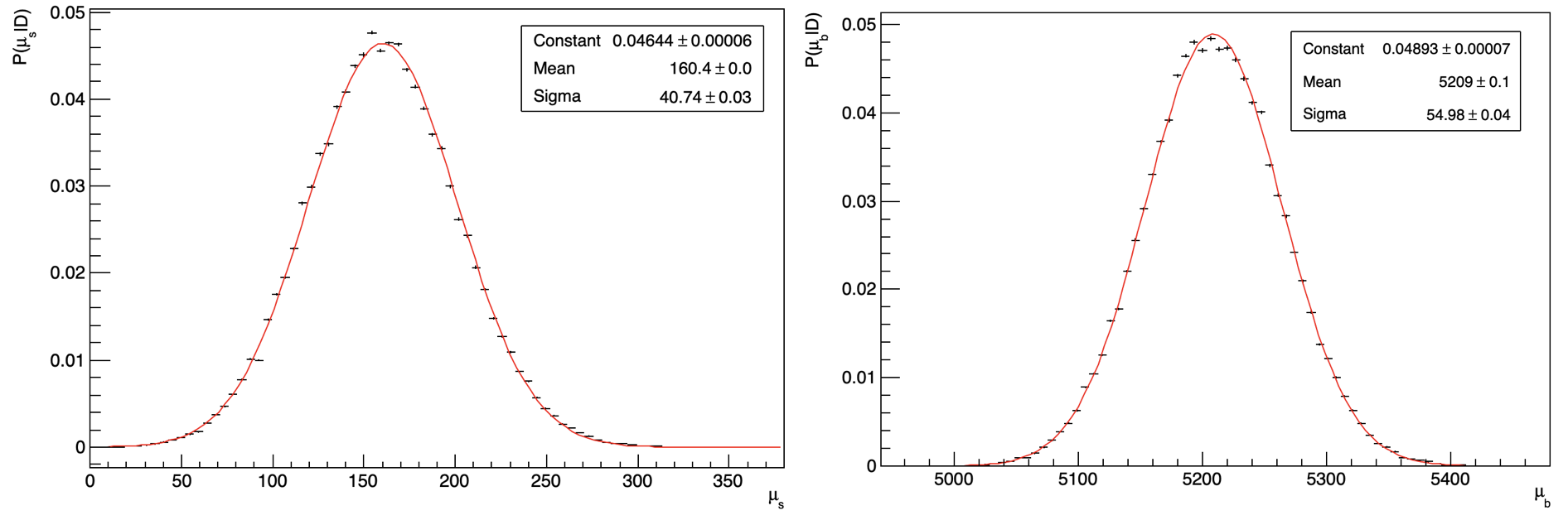}
    \caption{Marginalized posterior distribution $P(\mu_{s}|D)$ (left) and $P(\mu_{b}|D)$ (right) as a function of $\mu_s$ and $\mu_b$, from the marginalization of $P(\mu_{s},\mu_b|D)$. The Gaussian fit is also shown superimposed to the distributions.}
    \label{fig:posteriors}
\end{figure}

To validate the correctness of the likelihood, template, and toy Monte Carlo generation, several tests were conducted by computing the posterior distribution for selected toy MC datasets.
The analysis has been performed with the Bayesian Analysis toolkit (C++) which provides tools for implementing likelihood, conducting integration, marginalization, and an already implemented version of the Metropolis-Hastings Markov Chain Monte Carlo algorithm \cite{caldwell2009bat}. The following results are presented for the scenario with $\epsilon$=4.5 y and $R_b=$15. To calculate the posterior probability, the likelihood function has been built on the toy. A Poissonian prior was employed for the background, assuming it is well-known, with the mean value set to the expected number of events. To give equal probability to every value of $\mu_s$ in the analysis, a flat prior within the range $[0, N=n_s+n_b]$ was adopted for the signal, where $N$ represents the total number of events observed in the toy dataset.
The likelihood was sampled using 8 Metropolis-Hasting Monte Carlo chains. During the pre-run, where the proposal function was adjusted, all chains exhibited good efficiency (approximately 25\%) after 2500 iterations.
Once the optimal proposal function was determined, 100,000 iterations were performed for each chain. All chains converged to the same region. The posterior distribution $P(\mu_{s},\mu_b|D)$, obtained from MCMC sampling, is depicted as a function of $\mu_s$ and $\mu_b$ in Fig. \ref{fig:PosteriorSamp}.
As expected, $\mu_s$ and $\mu_b$ show an anti-correlated behavior since as the number of events is fixed, an increase in $\mu_s$ must lead to a decrease in $\mu_b$. By marginalizing the posterior on the two parameters, the posteriors for $\mu_s$ and $\mu_b$, namely $P(\mu_s|D)$ and $P(\mu_b|D)$, can be obtained. The posterior distributions for $\mu_s$ and $\mu_b$ are shown in the figure \ref{fig:posteriors}.
Given the symmetry of the posterior and the sufficiently high number of events, Gaussian fits were applied to the distributions. For a sample with $R_b=15$ and exposure $\epsilon=4.5$ y, the expected number of events for S and B are $\bar{N}_s=135$ and $\bar{N}_b=5310$, respectively. From the fit, it results a most probable value for $\mu_s = 160.4 \pm 40.7$ and for $\mu_b = 5209\pm55$. Both the number of events obtained are compatible within two $\sigma$ with the expected values, confirming the correctness of the procedure, and $\mu_s$ is found to be at 4$\sigma$ from zero.

\subsection{Sensitivity results}
\label{sec:resultsBayes}
\begin{figure}
    \centering
    \includegraphics[width=1\linewidth]{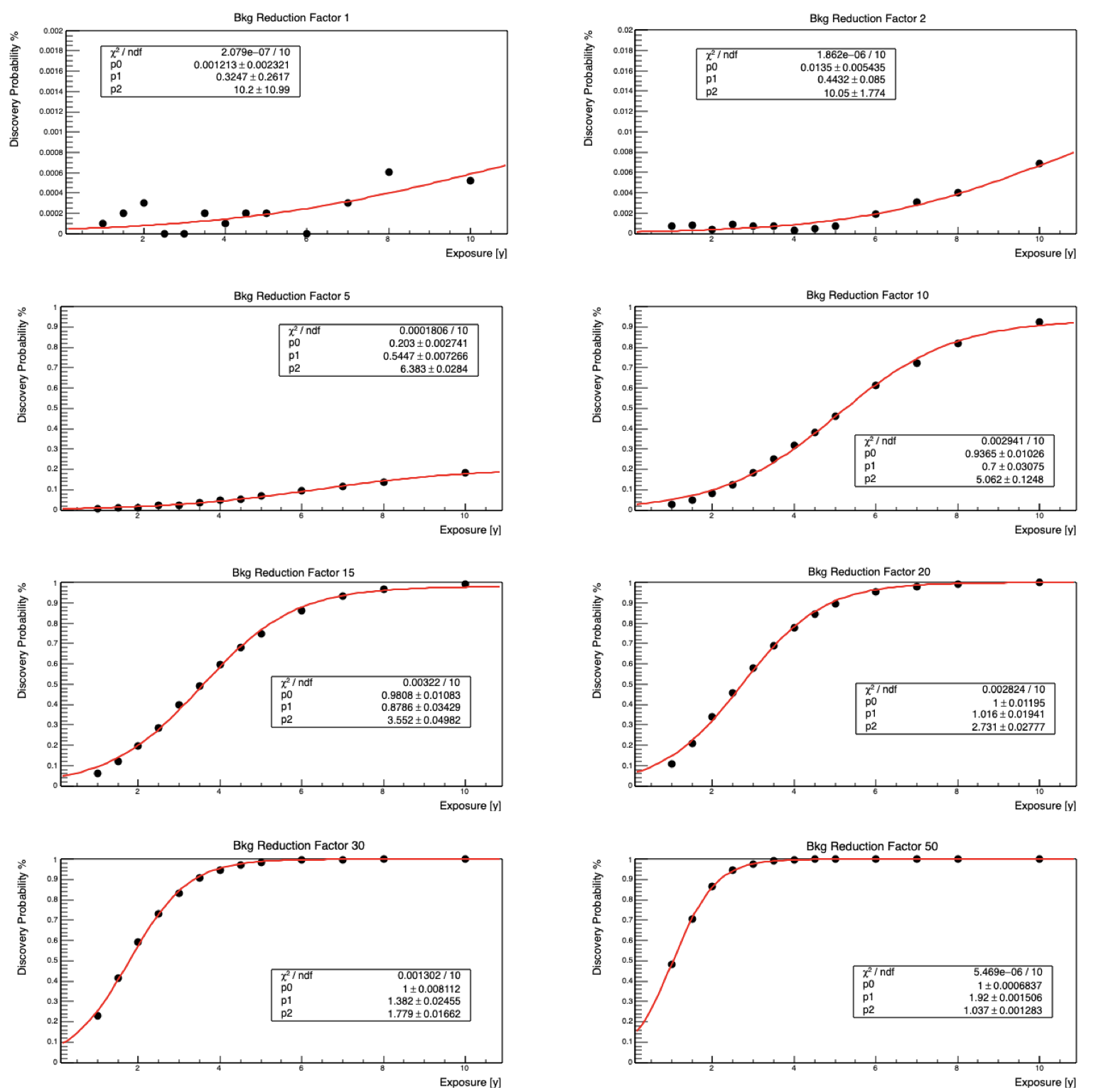}
    \caption{Plots of the discovery probability with $B_f>$20 as a function of the exposure under different hypotheses of further reduction of the background with increasing reduction in order from left to right and from top to bottom. Note the different scales used in the first three plots.}
    \label{fig:DiscoveryProb}
\end{figure}

The sensitivity has been studied for different exposures of the detector $\epsilon$. The exposures utilized are  1, 1.5, 2, 2.5, 3, 3.5, 4, 4.5, 5, 6, 7, 8, and 10 years, and with the standard background rate estimated with GEANT4. Moreover, given the potential improvements that cannot currently be quantified regarding the enhancement of material radiopurity, and to estimate the extent of improvement required, an additional estimation using various background reduction factors has been carried out. The background reduction factors $R_b$ employed are 1, 2, 5, 10, 15, 20, 30 and 50. For each combination of these parameters, the mean expected number of events for signal and background has been calculated as:
\begin{equation}
    \bar{N}_{s} = (30\  ev/y) \cdot \epsilon \ \ \ \ \ \ \ \ \ \ \ \ \bar{N}_b = \frac{(1.77 \cdot 10^4\  ev/y)}{R_b} \cdot \epsilon
\end{equation}
For each set of exposure and background reduction factors, the Bayes factor distribution has been calculated for all the toys. The discovery probability has been defined as the number of toys with a Bayes factor greater than 20 divided by the total number of toys.
\begin{equation}
    \mathcal{D}_{p}(BF>20) = \frac{N_{toys}(B_f>20)}{N_{toys}}
\end{equation}
A Bayes factor $>20$ corresponds to a significance of 3$\sigma$ as reported in \cite{Trotta_2008}.
The discovery probability as a function of the exposure for different $R_f$ is shown in Fig. \ref{fig:DiscoveryProb}.
The points have been fit with a Sigmoid function:
\begin{equation}
    \mathcal{D}_{p}(\epsilon) = \frac{p_0}{1+e^{-p_1\cdot (\epsilon-p_2)}}
\end{equation}
Each point on the plot represents the probability that, at a specific exposure level, the acquired data are at least 20 times more likely to align with the model $H_1$ compared to the model $H_0$. Therefore, the plot indicates that a 50\% probability of this measurement is achievable within $\sim$5.5 years with a Bayes Factor exceeding 20 if the background can be further reduced by a factor of 10. This corresponds to the discrimination of 165 ER from the signal over 9680 ER from the background, with a background-to-signal ratio of $R_{bg}/R_{pp}\simeq60$. This indicates a significantly enhanced background tolerance capability, approximately 26 times higher than that of the Borexino experiment, as detailed in Sec. \ref{sec:bkgRejection}.\\

\section{Possible future studies for the measurement}
\label{sec:furtheroptim}
For this sensitivity measurement, the existing LIME setup was utilized with its detector performance. Consequently, no particular hardware optimization of the detector was undertaken for this specific sensitivity study. Thus, to improve this measurement, different aspects can be considered: 
\begin{itemize}
\item As detailed in Sec. \ref{sec:gemgasmix}, the employment of different amplification systems and voltage configurations can potentially result in reduced track diffusion, offering improved tracking performance and directional measurement enhancement. Different compositions in the gas mixture have the potential to decrease gas density, thereby reducing track diffusion and further enhancing multiple scattering. However, a less dense gas mixture would result in a lower number of events affecting the statistics. 
\item Observations indicate that iron spots, initially perfectly round in a 60:40 He:CF$_4$ gas mixture, exhibit extensions in a gas mixture with 80\% Helium. This phenomenon suggests a potential reduction in the directional threshold for low-energy electron recoils, together with the improvement in directionality due to a decrease in multiple scattering. 
\item Additionally to enhance the radiopurity of materials, a potential strategy for reducing background involves identifying the primary alpha particle that interacts within the detector volume and subsequently rejecting events based on a likelihood analysis as done in \cite{BorexinoCNO}. By tagging the primary alpha, it becomes possible to gather information about the nucleus responsible for the decay and its precise position. With this information, the probability of encountering an electron recoil within a time window $\Delta T$, at a specific position, considering the interaction length, and with known energy based on the decay scheme, can be computed. This detailed information allows for a more precise assessment of the likelihood that the electron recoil is attributable to background. 
The directionality can play a crucial role in this process, since knowing the direction of the ER, it can be inferred if its direction is compatible with the decay of the radioactive contaminant. Moreover, even in scenarios, where Compton interaction is the most probable, especially given the energy range of nuclear deexcitation (on the order of 100s of keV), the precise energy of the emitted gamma can be measured. Indeed, since the direction of the incoming particle can be supposed (decay nucleus previously tagged with the alpha), and the direction and energy of the scattered electron are known, the energy of the primary gamma and the outgoing direction of the gamma (useful for double gamma interaction) can be measured since the kinematic of the process is closed. This would lead to an even more precise background identification, tag, and reduction. A detailed study of the performances of this technique for background reduction must be done.
\end{itemize}
For the purpose of conducting this measurement, all these aspects of optimization may be considered in the future.

\subsection{Sensitivity with negative ions drift diffusion}
\begin{figure}
    \centering
    \includegraphics[width=1\linewidth]{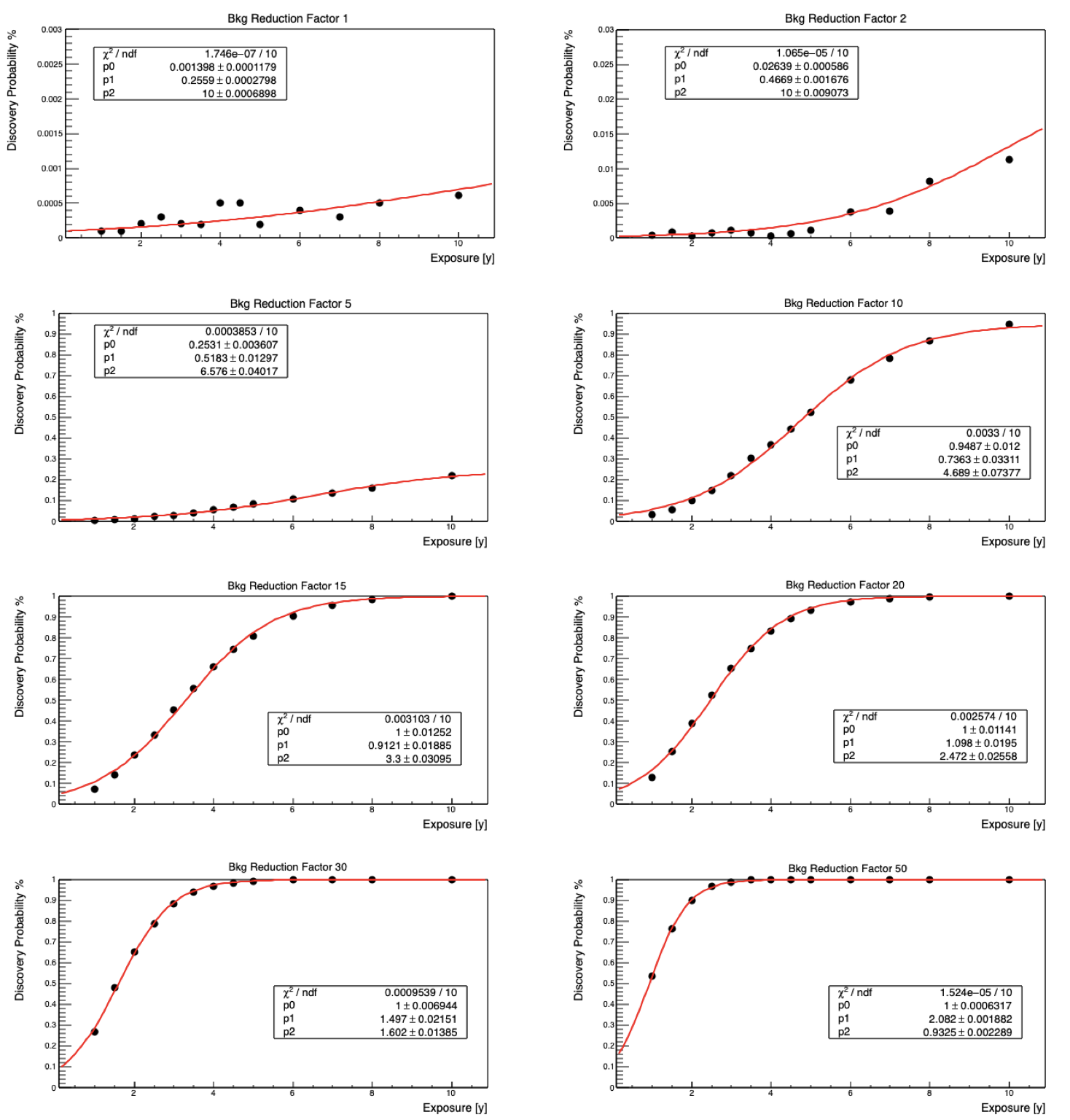}
    \caption{Plots of the discovery probability with $B_f>$20 as a function of the exposure under different hypotheses of further reduction of the background and under the hypothesis of NID diffusion. Note the different scales used in the first three plots.}
    \label{fig:SensitivityNID}
\end{figure}

\begin{figure}
    \centering
    \includegraphics[width=0.75\linewidth]{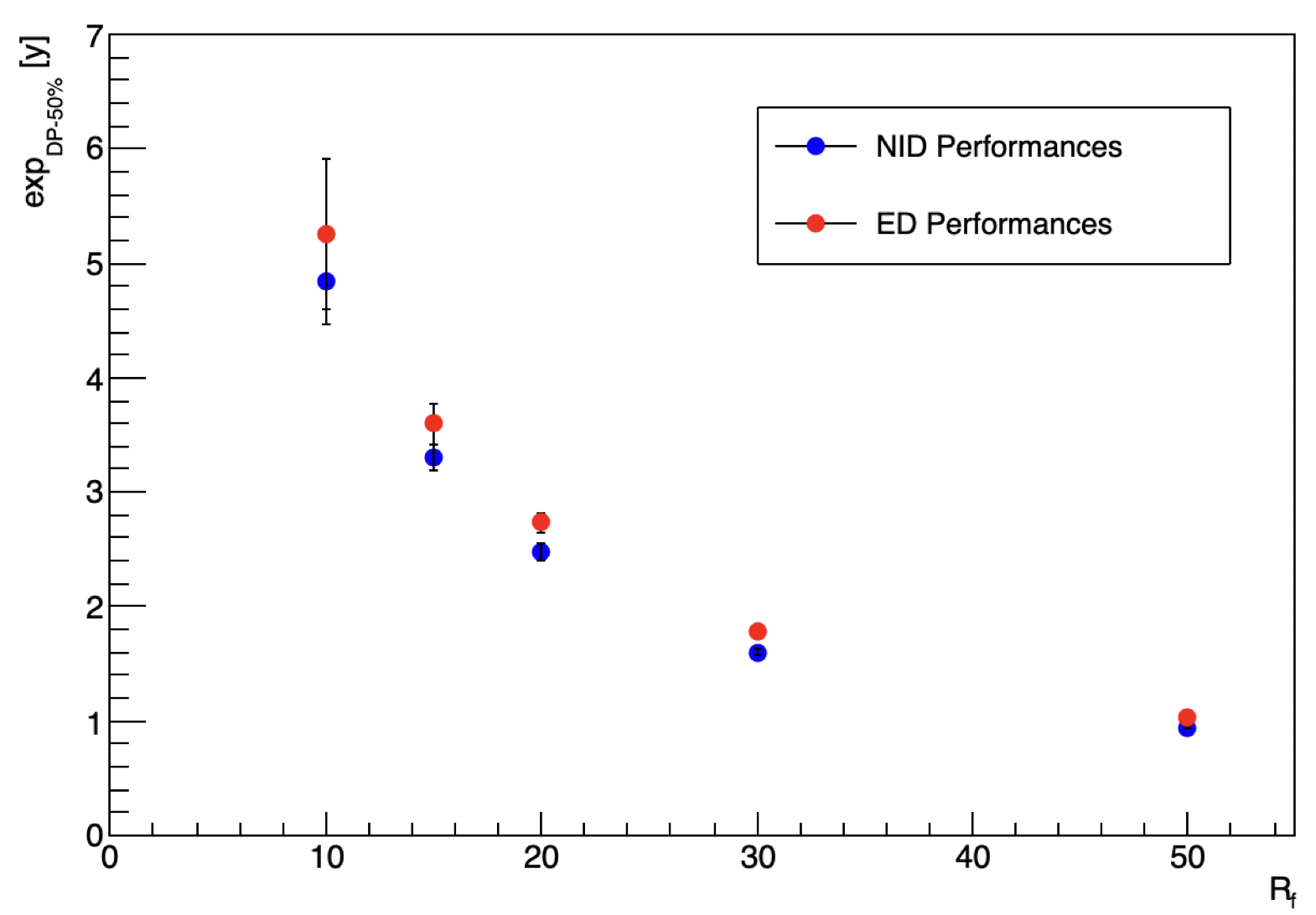}
    \caption{In figure, the exposure for a 30 m$^3$ detector needed to have 50\% probability of observing a signal with BF$>$20 is shown as a function of the background further reduction factor under ED and NID operations.}
    \label{fig:D50EDNID}
\end{figure}

As detailed in Sec. \ref{sec:NID}, the collaboration successfully operated the small prototype utilizing negative ions for charge transport. This method enabled the achievement of one of the lowest diffusions ever observed in a Time Projection Chamber. Furthermore, in Sec. \ref{sec:NIDAngReso} the angular resolution for low-energy electron recoil has been studied on simulated tracks with negative ion diffusion parameters. The data analysis outlined in Sec. \ref{fig:LightVsPar} indicates that the gains achieved through NID operation are notably small, comparable to those attained through ED operation with GEM voltages set at 300 V instead of the nominal voltage operation of 440V. However, it's important to note that the analysis of NID presented represents just the initial stages of exploring its potential. Tests involving variation in the gas mixtures and employing amplification stages more suited for NID operations are already underway. By the time the CYGNO-30 detector is realized, it is expected that this technique will have been extensively investigated. Thus, considering the possibility of reaching similar gains as ED with the use of more appropriate amplification systems for NID, a future 30 m$^3$ detector operating with NID becomes feasible.
In this context, the sensitivity for solar neutrino detection has been evaluated under identical conditions as ED operations, but with the NID angular resolution obtained, utilizing the same assumptions, background estimates, and methodology outlined in this chapter. The results of the sensitivity are shown in Fig. \ref{fig:SensitivityNID}.
For comparison, the exposure needed to have a 50\% probability of observing a signal with BF$>$20 is shown as a function of $R_f$ for ED and NID operation in Fig. \ref{fig:D50EDNID}.
As can be seen from the plot, the reduction of the diffusion and the consecutive improvement of the angular resolution doesn't significantly improve the sensitivity for a discovery. This is because the sensitivity results are still dominated by the low statistics of the signal, from the intrinsic spread of the ER angular distribution, and from the detector energy resolution. 
However, enhancing diffusion with NID could offer significant advantages in this measurement. 
Firstly, it could lead to the realization of a detector with a 1.5 m drift length, while retaining the tracking performance equivalent to a 50 cm drift length detector for tracks generated in the farthest region from the GEMs. This advancement could result in reduced material usage for constructing a 30 m$^3$ detector, thereby lowering background levels and substantially cutting down on construction costs due to 1/3 of the amplification and readout system needed for the detector realization. On the other side, the exposure front, expanding the drift length, while maintaining unchanged the readout and amplification system, offers the potential to enlarge the active volume by a factor of 3 without losing in performance or necessitating an increased complexity in the readout system. 
An improvement in angular resolution together with an improvement in energy resolution would moreover improve significantly the performances on the neutrino energy reconstruction and thus in the neutrino spectroscopy.

\section{Conclusions and outlook}
\begin{figure}
    \centering
    \includegraphics[width=1\linewidth]{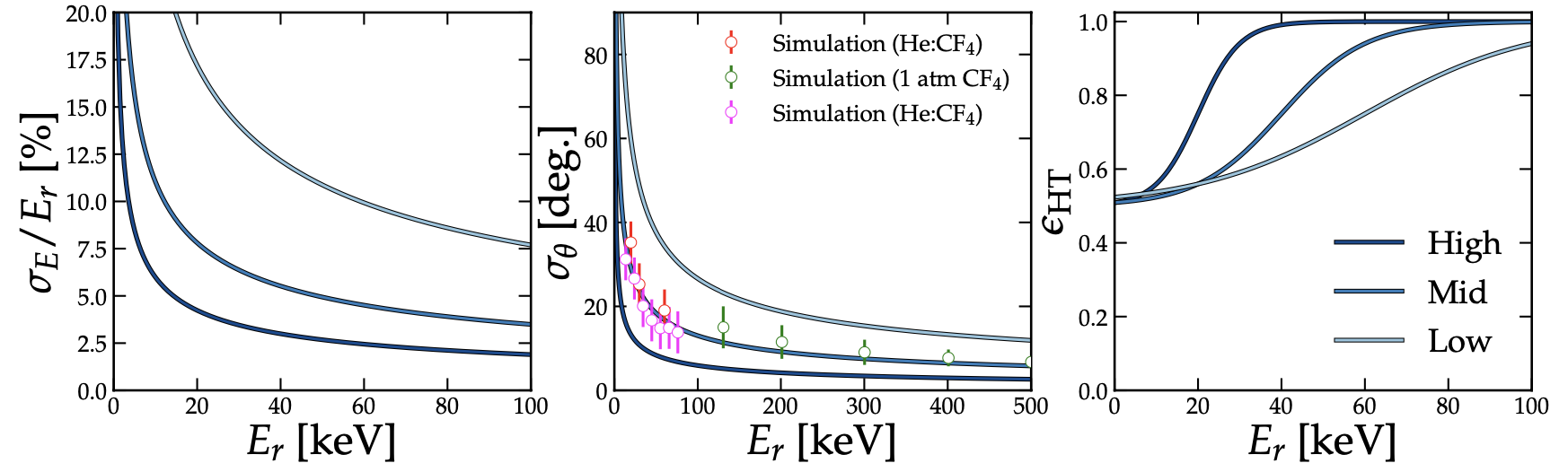}
    \caption{Performance curves dependent on energy for electron recoils showing: energy resolution, angular resolution, and orientation recognition efficiency. In each graph three curves are present representing different benchmark configurations: "low," "mid," and "high," shaded from light to dark blue. The simulated data points in He:CF$_{4}$ (pink) are taken from the results of this thesis. Plots from \cite{SolrNuDir}.}
    \label{fig:performanceBanch}
\end{figure}

\begin{figure}
    \centering
    \includegraphics[width=1.0\linewidth]{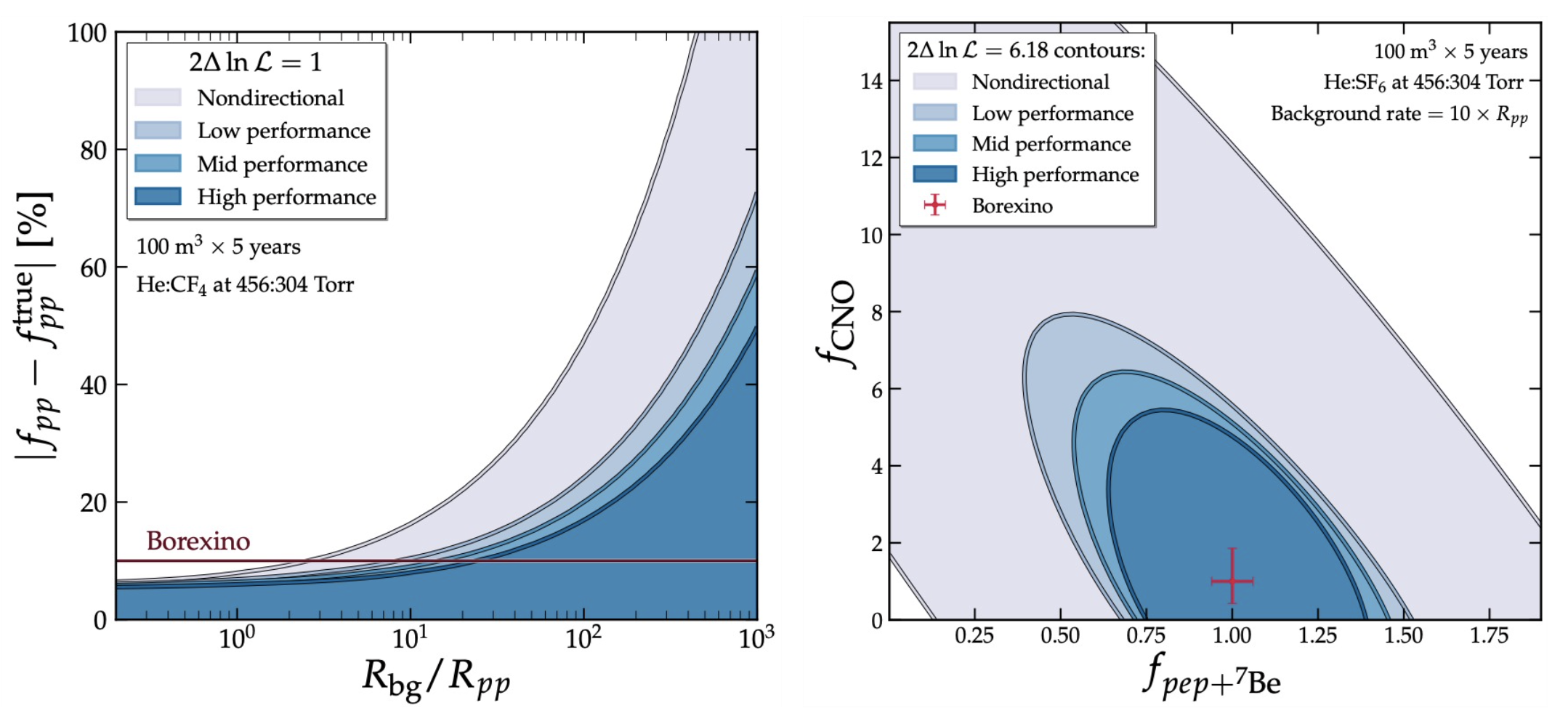}
    \caption{Left: Error on the reconstructed flux of pp neutrinos for a 100 m$^3$ detector, with a 5-year exposure and a He:CF$_4$ 60:40 gas mixture. Right: Predicted 2$\sigma$ sensitivity to a joint measurement of the CNO and pep+7Be fluxes, with the background rate set at 10 times the pp electron recoil rate also for a 100 m$^3$ detector with a 5-year exposure and a He:CF$_4$ 60:40 gas mixture. Different shades represent varying levels of detector performance, as outlined in Fig. \ref{fig:performanceBanch}.  Notably, the inclusion of directionality in the analysis significantly improves upon non-directional approaches, and, superior detector performance results in enhanced precision. Plots granted from C. A. J. O'Hare, C. Lisotti, et al.}
    \label{fig:ppandCNOPrec}
\end{figure}

As outlined in \cite{ohare2022recoil}, prior to the availability of very large-scale gas TPC experiments, there exists an opportunity to conduct intriguing and innovative neutrino physics using solar neutrinos with planned intermediary experiments of approximately $\mathcal{O}$(10) m$^3$. The more favorable kinematics of elastic neutrino-electron scattering result in recoil energies ranging from tens to hundreds of keV. This implies that an experiment with thresholds around O(10 keV), exploiting directionality and the fact that electron recoils are strongly directed back towards the Sun, could detect the strongest, albeit lowest-energy, flux of solar neutrinos from pp reactions. In principle, with well-measured recoil energy and angle relative to the Sun, it would be possible to reconstruct the energy spectrum of each solar neutrino event by event, facilitating an empirical measurement of the solar neutrino energy spectrum. This is crucial when considering scientifically desirable fluxes. While the majority of the event rate would stem from pp neutrinos, whose flux is relatively well-understood, there are several intriguing fluxes, particularly those from the Sun's CNO cycle, which hold significant interest from a solar physics perspective but are obscured by other fluxes and could be probed with ton-scale TPC detectors.\\
Several critical considerations must be addressed when evaluating the feasibility of using gas detectors for solar neutrino physics with electron recoils.
\begin{figure}
    \centering
    \includegraphics[width=0.5\linewidth]{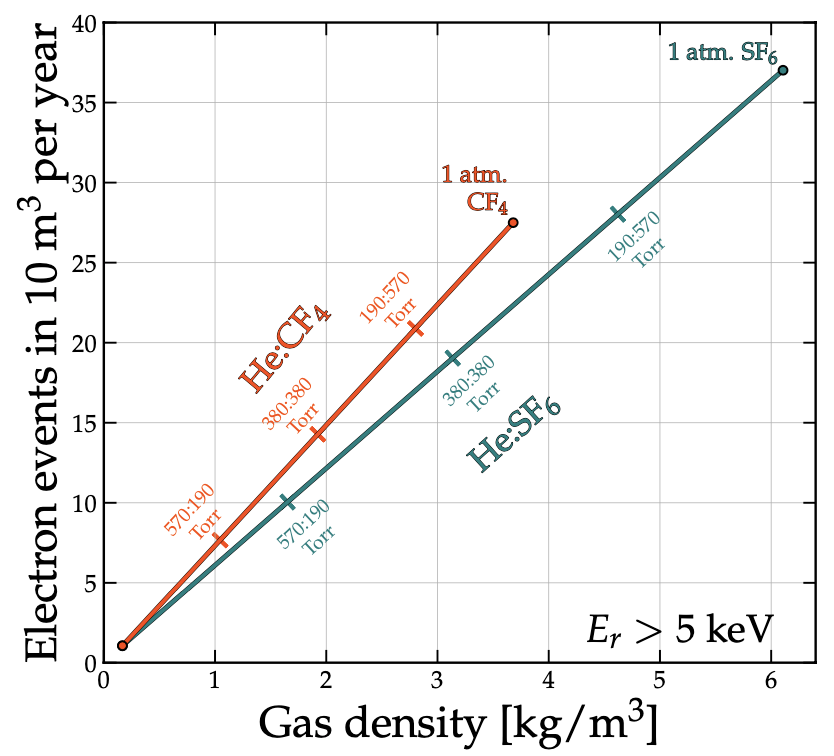}
    \caption{In figure, the number of neutrino induced electron recoil events as a function of the gas density above 5 keV in a 10 m$^3$ detector over one year are shown. In the plot, He:SF$_6$ and He:CF$_4$ gas mixtures with different He percentages are considered. Plot from \cite{SolrNuDir}.}
    \label{fig:NERVsDensity}
\end{figure}
One such consideration is the optimization of gas density. The ideal balance is to maximize the event rate while maintaining adequate angular and energy resolution to accurately reconstruct electron recoils. As depicted in Fig. \ref{fig:NERVsDensity}, the quantity of neutrino interactions correlates directly with the gas density and the number of electron target density within the gas mixture. Increasing both the electron density and the gas density lead to an improvement in statistics for neutrino interaction. Nevertheless, higher gas densities lead to increased multiple scattering for electron recoil and a shorter track length, resulting in lower performances in angular resolution. Another issue that could significantly impact the potential sensitivity of such a search is the amount of electron background. 
However, when the directional capability is strong, it becomes feasible to tolerate a background rate significantly higher than that of neutrinos, while still achieving a competitive measurement. Furthermore, validating the angular resolution performance of such a detector is not straightforward and requires simulations capable of accurately reproducing the data \cite{ohare2022recoil}.
In this thesis, these quantities, crucial for a solar neutrino measurement, have been studied in detail for a CYGNO detector. It includes a detailed examination of the energy resolution for electron recoils across different energy levels using real data (Chap. \ref{chap:LIMEDetector}). Additionally, a simulation capable of accurately replicating tracks as they would appear in a CYGNO-30 detector has been developed and validated through extensive comparison with Monte Carlo data (Chap. \ref{chap:Simulation}). Building upon this simulated data, an algorithm for determining the initial direction of ER has been devised, and the angular resolution performance has been thoroughly investigated (Chap. \ref{chap:directionality}). Furthermore, an assessment of the internal background within a plausible CYGNO-30 detector has been conducted across the entire energy spectrum to enhance the reliability of sensitivity measurements (Sec. \ref{sec:bkgmodel}). The measurements utilized for this specific physics case in a CYGNO-30 detector would also hold significant importance for the entire CYGNUS collaboration, as they would serve as a benchmark, providing valuable insights to enhance the study of directional physics cases and improve performance and background assessments using real detector information.\\
The study conducted in this thesis on the feasibility of a directional solar neutrino measurement of neutrino originating from the pp-chain with a CYGNO-30 experiment has demonstrated that such a measurement is achievable in $\sim$5.5y with a 50\% probability and Bayes factor greater than 20, provided that the background can be constrained below 1770 events per year. Additionally, this analysis underscores the significant advantage of directionality in terms of background tolerance, allowing for the endurance of a background event count approximately 60 times higher than that of the signal. This number of background tolerance, as shown in Chap. \ref{chap:sota}, is approximately 26 times greater than the background-to-signal ratio obtained in the Borexino detector. Furthermore, the angular resolution achievable under lower diffusion conditions through NID operations has been evaluated, and the sensitivity to solar neutrinos from the pp chain has been investigated with enhanced angular resolution capabilities. Simply increasing the angular resolution did not result in a significant enhancement in performance, as the energy resolution and intrinsic angular spread of the tracks still dominate in this context. Nevertheless, the improvement in angular resolution along with an improvement also in the energy resolution could prove crucial in enhancing the precision of the measurements.\\ 
In the reference \cite{SolrNuDir}, three distinct tiers of performance concerning energy resolution, angular resolution, and track orientation recognition have been delineated as a benchmark. The performances are presented in Fig. \ref{fig:performanceBanch}.
Based on the studies conducted in this thesis, the placement for a CYGNO detector is suggested to be:
\begin{itemize}
    \item Energy resolution: between mid and low performances below 40 keV, and overtaking low performances at higher energies.
    \item Angular resolution: between mid and high performances under ED operations, and at high performances under NID operations.
    \item Orientation recognition: Challenging to quantify due to the absence of PMT and instances where the directionality algorithm fails (refer to Sec. \ref{sec:algoeff}), but tends toward high performances in both NID and ED operations.
\end{itemize}
Given these assumptions regarding resolutions, the foreseen uncertainty in the reconstructed flux of pp neutrinos and the expected 2$\sigma$ sensitivity for a joint measurement of the CNO and pep+$^{7}$Be fluxes, with a background rate set at 10 times the pp electron recoil rate, are depicted in Fig. \ref{fig:ppandCNOPrec} for a CYGNO-like detector of 100 $m^3$ volume, together with the Borexino performances.
As depicted in the left plot of Fig. \ref{fig:ppandCNOPrec}, a CYGNO-like detector with a volume of 100 m$^3$, assuming a background level corresponding to the one used for the sensitivity study, corresponding to a ratio of $R_{bg}/R_{pp}=60$, and featuring medium to high detector performance over a period of 5 years, could achieve a precision in the measurement of pp flux similar to the one of Borexino. However, CYGNO can perform the measurement in an energy range significantly more extended towards lower energies, down to 50 keV in the neutrino energy, with 300 times less exposure. This underscores the considerable efficacy of the directional approach. Furthermore, if the background can be reduced to $R_{bg}/R_{pp}<20$, such a detector has the potential to surpass the precision of Borexino in the measurement of the pp flux. Moreover, it implies the possibility of substantially improving our knowledge of the solar neutrino flux beyond current and future experimental endeavors (see below) with a detector about 10000 times smaller than typical solar neutrino experiments. The right plot of Figure \ref{fig:ppandCNOPrec} illustrates how directionality significantly enhances the experiment's ability to measure fluxes. This improvement arises primarily from enhanced background rejection and the additional capability to distinguish between different neutrino fluxes.
On the other side, these results imply that a 1000 m$^3$ CYGNO-like detector would yield improved performance in both measurements, likely surpassing the Borexino precision on the pp flux measurement, and enhancing the precision of the CNO flux alongside the Borexino measurement and operating in a complementary manner, eventually solving the solar metallicity problem.\\
From a purely technological perspective, directional TPCs, as demonstrated extensively, offer a significantly improved signal-to-noise ratio compared to “traditional” neutrino detectors. With equivalent exposure, these detectors notably enhance precision in measuring solar neutrinos.\\  
Currently, there are no active experiments capable of measuring neutrinos from both the pp chain and the CNO cycle following the decommissioning of Borexino. Additionally, existing dark matter experiments, such as XENONnT, lack sensitivity to solar neutrinos due to substantial background noise \cite{slides_on_ABC}, and no other dark matter experiment has confirmed a neutrino detection.\\ 
In the future, larger experiments have been proposed for both dark matter and solar neutrino studies. The DARWIN experiment, featuring 50 tons of ultrapure xenon, aims to measure solar neutrinos from the pp chain within a range of [2-30] keV on the electron recoil energy with exceptional precision, leveraging on an extremely low background in that energy range \cite{DARWIN:2020bnc}. Moreover, on the DM side, DARWIN is expected to reach the neutrino floor \cite{DARWINDM} and this should enable the observation $^{8}$B neutrino through CE$\nu$NS. However, any definitive identification of dark matter at that point would likely require a directional detector capable of distinguishing DM from a dark matter signal, as outlined in Sec. \ref{sec:diradvantagesDM}. On the higher energy side, DARWIN would be unable to observe CNO neutrinos due to the radioactivity stemming from the detector materials and the natural presence of $^{136}$Xe within the 50-ton target.
Another experiment proposed primarily for reactor neutrino measurements, which could also conduct solar neutrino measurements, is the JUNO experiment \cite{Guo:2024wzi}. However, this experiment, comprising 20 kilotons of organic liquid scintillator, would encounter similar challenges regarding thresholds and contamination as the Borexino experiment. The performance in threshold and precision for CNO and pp measurements will heavily rely on the purification level achievable of the organic scintillator \cite{JUNOSOLAR}. Achieving a higher level of radiopurity than that achieved by Borexino may pose a significant challenge, potentially requiring several years to address.\\
On the contrary, given the sensitivity on the differential energy release, with gaseous TPC detectors, it is possible to reach a very low threshold and be sensitive at the same time to DM and solar neutrino. Indeed, these detectors can identify low-energy nuclear recoil with directional capabilities and electron recoil ranging from tens to hundreds of keV, thereby allowing the detection of both neutrinos from the pp chain and the CNO cycle.\\
Hence, directional TPCs stand as groundbreaking technological advances with the potential to unlock multiple frontiers in physics. In the future, they could enable a level of solar neutrino characterization far surpassing current capabilities.

%% file: chapters/Conclusions.tex
\chapter*{Conclusions}
\addcontentsline{toc}{chapter}{Conclusions}
\label{chap:Conclusions}
Over the last five decades, solar neutrinos have played a crucial role in fostering notable scientific advancements, enhancing our comprehension of neutrino characteristics and the Sun's dynamics. While several experiments have conducted measurements of solar neutrinos, a segment of the lower pp spectrum remains uninvestigated, and the limited precision with which this flux has been measured allows the possibility of exotic energy emission mechanisms in the Sun's core. Additionally, the accuracy of measurements from the CNO cycle is inadequate to resolve the solar metallicity issue.
In the 1990s, a proposal emerged for a directional measurement of solar neutrinos through neutrino-electron elastic scattering with a gaseous TPC approach. This method offers several advantages for such measurements. Firstly, it enables background reduction by correlating the direction of the scattered electron with the emission direction of the neutrino. Secondly, it allows for the measurement of neutrino energy on an event-by-event basis. By combining the measurement of electron energy with the electron scattering angle relative to the neutrino direction, the full kinematics can be determined, enabling the measurement of neutrino energy. However, this proposal has not been carried out anymore since the TPC technology was not mature enough, in favor of more solid technologies like liquid scintillators and water Cherenkov experiments.\\
With the latest advancements in the TPC technology, specifically in Micro Pattern Gas Detectors and high-segmented readouts, as well as research on the properties of different gas mixtures, gas TPC is now a mature technology for rare events searches and precision measurement. In this context, the CYGNUS collaboration aims at developing a nuclear recoil observatory with directional sensitivity to Dark Matter and neutrinos at the ton scale. The CYGNUS primary objective would be the detection of nuclear recoils resulting from Dark Matter and neutrinos interacting via CE$\nu$NS. However, recent studies have shown that such a detector could offer a significant contribution also by studying interactions of solar neutrinos within the detector via elastic scattering with electrons in the gas medium. In particular, it is suggested that an  $\mathcal{O}$(10) m$^3$ detector could observe neutrinos from the pp chain at very low energies, while a $\mathcal{O}$(1000) m$^3$ experiment could significantly contribute to measuring the neutrino flux from the CNO cycle, potentially resolving the solar metallicity problem. At larger scales, such a detector could also observe geoneutrinos and neutrinos from supernovae, with pointing capabilities. An essential aspect of this experiment is its unique sensitivity to both pp and CNO neutrinos through elastic scattering on electrons, as well as to Dark Matter and neutrinos interacting via CE$\nu$NS from $^8$B.\\ 
Within the framework of the CYGNUS collaboration, the CYGNO/INITIUM project is dedicated to developing a TPC for rare event searches, utilizing a Fluorine-Helium based gas mixture at atmospheric pressure with optical readout. CYGNO aims at constructing a 30 m$^3$ detector for directional Dark Matter searches. Concurrently, in synergy with CYGNO, INITIUM aims at implementing negative ions operations at atmospheric pressure within the CYGNO optical readout approach. Such a CYGNO detector would have the ability to explore an uncharted region of the cross-section-mass parameter space for spin-dependent DM interaction. Furthermore, CYGNO-30 could pioneer the first directional measurement of pp neutrinos from the Sun down to $\mathcal{O}$(10) keV on the electron recoil energy, corresponding to $\sim$ 50 keV on the neutrino energy. This breakthrough could pave the way for the development of ton-scale TPCs for high-precision neutrino measurements.\\
In this thesis, the feasibility of a directional measurement of solar pp neutrino utilizing a CYGNO-30 detector has been explored in deep.\\
The first step in assessing the sensitivity of a CYGNO-30 detector to solar neutrinos involves characterizing the detector's performance.
As outlined in Chap. \ref{chap:CYGNO}, the CYGNO-30 detector (Section \ref{sec:CYGNOFut}), based on the current project status, will comprise multiple CYGNO-04 modules (Section \ref{sec:CYGNO04}). These modules are expected to share the same drift length and granularity as the LIME prototype (Section \ref{sec:LIME}). It is expected that each CYGNO-04 module within the CYGNO-30 setup will perform comparably, if not better, than LIME. Consequently, a thorough investigation into the performance of LIME for low-energy electron recoils has been undertaken.
In Chap. \ref{chap:LIMEDetector} the characterization of the LIME response to low-energy electron recoils, conducted by the thesis author, has been described. This investigation involved utilizing a $^{55}Fe$ and a multi-energy X-ray source to generate electron recoils within the detector volume at fixed energy levels within 3.7 and 44.48 keV. The resulting electron recoil tracks have been then reconstructed and analyzed using a Maximum Likelihood fit approach on the track integral distribution, allowing for the extrapolation of the desired signal from the background of electron recoils originating from natural radioactivity. Moreover, the $_{s}\mathcal{P}\textrtaill{lot}$ tool has been employed within the framework of the Maximum Likelihood fit to unfold a set of topological track shape variables associated with the signal component of the spectrum. Through this analysis, the energy response of the detector has been characterized, revealing a non-linear response at very low energies due to saturation effects, while a more linear response is observed at higher energy levels due to reduced saturation of the track tail. Additionally, an energy resolution of approximately 13\%, constant across various energy levels above 6 keV, is achieved and assessed. This energy resolution results to be compatible with the one obtained typically with gas detectors.\\
In the context of the INITIUM project, with a contribution from the thesis author on the hardware side and analysis side, an extensive research has been conducted on the MANGO prototype to perform operations utilizing negative ions as charge carriers at atmospheric pressure, with optical readout. By introducing a small quantity of SF$_6$ into the conventional CYGNO gas mixture, the collaboration achieved successful operation under these conditions. The outcomes of these endeavors are detailed in Chap. \ref{sec:NID}. Through a study of ion mobility conducted using PMT, it was determined that the ions' mobility in the gas aligns with the anticipated mobilities for negative ion operations. This finding is further supported by previous charge-based analyses of the same gas mixture, confirming the viability of negative ion operation with optical readout at atmospheric pressure. Furthermore, subsequent systematic investigations into diffusion under negative ions operations have revealed the achievement of an exceptionally low diffusion of 45 $\mu$m/$\sqrt{cm}$ with a drift field of 600 V/cm, marking one of the lowest diffusions ever recorded in a TPC. Within the CYGNO experiment framework, this achievement carries considerable significance. With the prospect of larger gains or the availability of sCMOS sensors boasting heightened sensitivity, there exists the potential to extend the drift region while preserving comparable or superior tracking capabilities.\\
For an in-depth analysis of the detector response and the assessment of angular resolution performance with low-energy electron recoils, a simulation capable of generating electron/nuclear recoil tracks resembling those in actual sCMOS images has been developed by the collaboration. Detailed in Chap. \ref{chap:Simulation}, this simulation process begins with a simulated GEANT4/SRIM track and incorporates various detector effects, including the stochastic nature of processes. Optimization of the simulation parameters has been carried out through a continuous comparison of the simulated tracks and data acquired with a $^{55}$Fe source at different drift distances with the LIME detector. Through extensive comparisons between simulated tracks and real data at different energies carried out by the thesis author, the simulation has been validated. This comparison, conducted between simulated tracks and multi-energy electron recoils from the data, demonstrates consistency in light response, energy resolution, as well as in the reproduction of nine topological track shape variables.\\
Studying the angular resolution performances on low-energy electron recoils in a directional solar neutrino measurement is of paramount importance. However, due to the lack of a low-energy electron recoil source with known direction, directionality studies have relied on simulated electron recoils generated by a simulation able to produce tracks resembling the data. Chap. \ref{chap:directionality} details the algorithm devised from the thesis author, based on an algorithm developed for X-Ray polarimetry, for determining the directionality of electron recoils and the results in angular resolution. Following an algorithm parameter optimization using a set of tracks with known energy, the angular resolution was investigated across energy bins using a second sample of tracks generated with uniformly distributed random energies. These studies yielded an angular resolution of approximately 28° at 20 keV, improving to about 13° at 70 keV. Additionally, to assess the impact of reduced diffusion on angular resolution, a similar analysis of parameter optimization and angular resolution was conducted on tracks simulated with NID diffusion. This examination revealed an enhanced angular resolution, ranging from approximately 20° at 20 keV to around 10° at 60 keV. Further investigation into angular resolution carried on in this thesis work included a detailed analysis of simulated tracks at varying drift distances and angles relative to the GEM plane. Additionally, the efficiency of the algorithm has been studied across different track production conditions, revealing certain limitations. An enhanced algorithm is currently under development, expected to significantly improve directionality efficiency. Moreover, in the future, with the addition of the PMT information, also cases in which the track interaction point cannot be resolved due to the lack of 3D information can be resolved.\\
Finally, as the core result of this thesis work, a sensitivity analysis has been conducted for a solar neutrino measurement using a CYGNO-30 experiment within a Bayesian framework, as outlined in Chap. \ref{chap:solarnu}. The analysis took into account the energy and angular signatures of both the signal and background. The interaction rate for neutrinos from the pp chain was calculated to be $\sim$1 event/m$^3$/year, equivalent to approximately 30 events/year in a CYGNO-30 detector. To characterize the internal background level, a detailed GEANT4 simulation of the detector was developed by the thesis author. This involved optimizing the detector components and materials to minimize background levels. Following the resolution determination, background and signal templates were created accordingly. A Bayesian analysis was then conducted using toy Monte Carlo (MC) data generated under various exposure scenarios and further reductions of the background. Finally, the discovery probability was computed based on the Bayes factor, considering values $>$20 in the toy analysis.
The analysis revealed that such a measurement is feasible at more than 3$\sigma$ confidence level with a 50\% probability in 5.5 years if the background can be constrained to below 1760 ev/y. Repeating the study with NID angular resolution yielded only marginal improvements, as the low statistics and intrinsic angular spread of the electron recoil largely influenced the sensitivity. Nonetheless, employing NID in the measurement would offer several advantages. It would enable the construction of a more compact detector with a 1.5 m drift length, resulting in enhanced tracking capabilities. This, in turn, would significantly reduce background levels and overall detector costs, requiring only 1/3 of the cathodes, resistors, GEMs, and cameras, along with reduced vessel material. Since, as demonstrated by this study, the most limiting factor to pp observation probability and precision of the pp flux measurement is represented by the internal background level, NID operation could substantially help to achieve the needed reduction (which would directly translate into a reduction of the background levels also for DM searches).
This analysis underscores the remarkable efficacy of the directional approach, revealing that a 30 m$^3$ CYGNO detector could feasibly measure solar neutrinos while tolerating background levels approximately 60 times higher than the signal. This translates to a noise-to-signal tolerance capability roughly 23 times greater than that of Borexino of $\sim$ 2.3. Additionally, at the moment, there are no operative experiments that conduct solar neutrino measurements from the pp chain and the CNO cycle. Future experiments like Darwin, designed for Dark Matter searches, are not designed to detect neutrinos from the CNO cycle. Moreover, the success of the future experiment JUNO in measuring CNO neutrinos will heavily hinge on the background levels achievable by the detector in terms of scintillator target radiopurity. Anyway, this detector will rely on a threshold at least comparable to the one of the Borexino experiment, preventing it from measuring pp neutrino in the lower part of the spectrum. The arguments discussed in this thesis, and the versatility of a CYGNO detector, offering sensitivity to both Dark Matter and solar neutrinos from the pp chain and the CNO cycle, present a strong rationale for developing ton-scale gas TPCs as prospective detectors for future research in both DM and neutrino studies. \\
Hence, this thesis comprehensively analyzed the detector's energy resolution and background levels for a CYGNO-30 m$^3$ detector, assessing its feasibility for a directional solar neutrino measurement. The findings underscore the capability of such a detector to conduct this measurement alongside significant Dark Matter research. The performances achieved with the detector, obtained for the first time in the collaboration, represent also a significative result for the whole CYGNUS collaboration, potentially serving as a benchmark for future investigations.

%% file: appendices/appendixA.tex
\chapter{The $_{s}\mathcal{P}\textrtaill{lot}$ weight determination}
\label{app:appendixA}

Let's consider a log-likelihood expressed as:
\begin{equation}
    log(\mathcal{L})=\sum_{e=1}^{N}{log\left(\sum_{i=1}^{N_s}N_if_i(y_e)\right)}-\sum_{i=1}^{N_s}N_i
\end{equation}
where:
\begin{itemize}
    \item $N$ is the total number of events in the data sample.
    \item $N_s$ is the total number of species (e.g. background and signal $N_s=2$).
    \item $N_i$ is the total number of events expected for each species. 
    \item y is the set of discriminating variables
    \item $f_i$ is the p.d.f. for the discriminating variable for the $i^{th}$ species.
    \item $f_i(y_e)$ is the value of the p.d.f. $f_i$ for the event e, it can be associated with a set of $y_e$ in the case of a set of discriminating variables.
    \item x is the control variable with an unknown p.d.f.
\end{itemize}
The log-likelihood is in this way a function of ($N_1,N_2,...,N_s$) yields of events per species, and some eventual parameter to build the p.d.f. on the data.
The aim of the $_{s}\mathcal{P}\textrtaill{lot}$ is to provide a method to unfold the true distribution of a control variable x for the $n^{th}$ species denoted $M_{n}(x)$. The distribution of $M_{n}(x)$ can be derived with the only knowledge of the p.d.f.s of the discriminating variables $f_i$.
The initial step to achieve this is to perform a fit, through which the yields $N_i$ are determined for each species. After that, some weights are assigned to each event, and these weights are subsequently used to reconstruct the $M_{n}(x)$.  

If there is a correlation between the control variable x and the discriminating variable y, so x belongs to the set of y, the following weights can be assigned for each event of the species $n$:
\begin{equation}
    \mathcal{P}_n(y_e) = \frac{N_nf_n(y_e)}{\sum_{k=1}^{N_s}{N_kf_k(y_e)}}
    \label{eq:equationP}
\end{equation}

For the case in which only $s$ and $b$ are present as species in the dataset, the previous equation for the signal class reduces to
\begin{equation}
    \mathcal{P}_s(y_e) = \frac{N_sf_s(y_e)}{N_sf_s(y_e)+N_bf_b(ye)} = \frac{1}{1+\frac{N_bf_b(y_e)}{N_sf_s(y_e)}}
\end{equation}
From this, it becomes evident that the weight assigned for the event depends on the ratio of the number of $b$ events and the number of $s$ events and on the ratio of the p.d.f. of $b$ and $s$ calculated for that specific event. 
If the quantity x is correlated with y the $\mathcal{P}_n(y_e)$ can be used to construct the distribution of x denoted $\tilde{M}_n$ defined as:
\begin{equation}
    N_n\tilde{M}_n(\bar{x})\delta x = \sum_{e\in \delta x}{\mathcal{P}_n(y_e)}
\label{eq:MDist}
\end{equation}

where the the sum is over all the $N_{\delta x}$ events for which the value of the variable $x$ for each event $e$ belong to the bin centered at $\bar{x}$ with width $\delta x$. The resulting quantity $N_n\tilde{M}_n(\bar{x})\delta x$ is essentially the histogram where each entry is weighted by $\mathcal{P}_n(y_e)$.
It can be demonstrated that, on average, this procedure reproduces the real distribution $M_n(x)$.
This can be accomplished by replacing the sum with the integral, and the role of the binning is realized with a Dirac delta function $\delta(x(y)-\bar{x})$. Moreover, the summation over different species is to account for all the events with their p.d.f.s .
\begin{equation}
    \left< \sum_{e\in \delta x} \right> \rightarrow \int{dy \sum_{j=1}^{N_s} N_jf_j(y)\delta(x(y)-\bar{x}) \delta x }
\label{eq:average}
\end{equation}
In this way, by taking the average of both sides of \ref{eq:MDist} and simplifying $\delta x$ the equation becomes
\begin{equation}
   \left< N_n\tilde{M}_n(\bar{x}) \right> =  \int{dy \sum_{j=1}^{N_s} N_jf_j(y)\delta(x(y)-\bar{x})} \mathcal{P}_{n}(y)
\label{eq:EqMAverage}
\end{equation}
by substituting $\mathcal{P}$ with its value in \ref{eq:equationP}, replacing $y_e$ with $y$
\begin{equation}
    \left< N_n\tilde{M}_n(\bar{x}) \right> =  \int{dy \sum_{j=1}^{N_s} N_jf_j(y)\delta(x(y)-\bar{x})\frac{N_nf_n(y)}{\sum_{k=1}^{N_s}{N_kf_k(y)}} }
\end{equation}
and finally simplifying the summation, and bringing $N_n$ out of the integral it is obtained
\begin{equation}
    \left< N_n\tilde{M}_n(\bar{x}) \right> = N_{n} \int{dy \delta(x(y)-\bar{x})f_n(y)} 
\end{equation}
Here the integral is identically equal to the theoretical distribution $M_n(x)$ for the variable x. From this we get
\begin{equation}
    \left< N_n\tilde{M}_n(\bar{x}) \right> = N_n M_n(\bar{x})
    \label{eq:final}
\end{equation}
This demonstrates that the $\mathcal{P}_n$ can be used to reproduce the distribution of the variable $x$ for the $n^{th}$ species of the dataset.

The most interesting scenario is when the control variable $x$ and the discriminating variable $y$ are uncorrelated. In this situation the overall p.d.f. can be factorized as $f_i(x,y)=M_i(x)f_i(y)$.
By rewriting the equation \ref{eq:EqMAverage} with the weight defined in equation \ref{eq:equationP}, it can be seen that \ref{eq:final} doesn't hold anymore due to the factorization of $f_i(x,y)$:
\begin{equation}
    \left< N_n\tilde{M}_n(\bar{x}) \right> = \iint{dydx \sum_{j=1}^{N_s}{N_jM_j(x)f_j(y)\delta(x-\bar{x})}\mathcal{P}_n}
    \label{eq:uncorrave}
\end{equation}
Integrating over $dx$, substituting $\mathcal{P}_n$ it is obtained
\begin{equation}
    \left< N_n\tilde{M}_n(\bar{x}) \right> = \int{ dy \sum_{j=1}^{N_s}{N_jM_j(\bar{x})f_j(y)} \frac{N_nf_n(y)}{\sum_{k=1}^{N_s}{N_kf_k(y)}} }
\end{equation}
Finally by rearranging the terms
\begin{equation}
\label{eq:newcase}
    \left< N_n \tilde{M}_n(\bar{x}) \right>  = 
    N_n \sum_{j=1}^{Ns}{ M_j(\bar{x})\left(N_j \int{ dy \frac{f_n(y)f_j(y)}{\sum_{k=1}^{Ns}{N_kf_k(y)}} }\right)}
\end{equation}
which is $\neq N_n M_n(\bar{x})$ due to the presence of a correction term between parenthesis.
However, it can be noticed, that this correction term looks similar to the inverse of the covariance matrix $V_{nj}^{-1}$  of $-\mathcal{L}$, indeed
\begin{equation}
    V_{nj}^{-1} = \frac{\partial^2(-\mathcal{L})}{\partial N_n \partial N_j} = \sum_{e=1}^{N}{ \frac{f_n(y_e)f_j(y_e)}{ (\sum_{k=1}^{N_s}{N_kf_k(y_e)})^2 } }
\end{equation}
By averaging this quantity, replacing the sum with the integral as done in \ref{eq:average} it is obtained:

\begin{equation}
     \left< V_{nj}^{-1} \right> = \iint{ dxdy \sum_{l=1}^{N_s}{N_lM_l(x)f_l(y) \frac{f_n(y)f_j(y)}{ ( \sum_{k=1}^{N_s}{N_kf_k(y)} )^2 } } } = \int{dy  \frac{f_n(y)f_j(y)}{\sum_{k=1}^{N_s}{N_kf_k(y)} } }
\end{equation}
where the last integral has been obtained factorizing the double integral in $dxdy$, using the property $\int{M_l(x)dx}=1$, and simplifying the summation over $k$ and $l$.
In light of this, eq. \ref{eq:newcase} can be rewritten as
\begin{equation}
\label{eq:int}
    \left< \tilde{M}_n(\bar{x}) \right> = \sum_{j=1}^{N_s}{M_j(\bar{x})N_j \left< V_{nj}^{-1} \right>}
 \end{equation}

by inverting the matrix, multiplying by $V_{jn}$ both members of eq. \ref{eq:int} the equation of interest in obtained

\begin{equation}
    N_n M_n(\bar{x}) = \sum_{j=1}^{N_s}{ \left< V_{nj} \right> \left< \tilde{M}_j(\bar{x}) \right>}
\end{equation}

Hence, if the control variable $x$ is uncorrelated with the discriminating variable $y$, the true distribution of $x$ can be reconstructed using the sWeights, which are a covariance-weighted weights
\begin{equation}
    \label{eq:splots}
    _s\mathcal{P}_n(y_e) = \frac{ \sum_{j=1}^{N_s}{V_{nj}f_j(ye)} }{ \sum_{k=1}^{N_s}{ N_k f_k(ye) } }
\end{equation}

By substituting this sWeight in \ref{eq:uncorrave} it is obtained that, on average, the true distribution for the uncorrelated variable is correctly reproduced. Finally, the distribution of the control variable x for the species n can be obtained from the $_{s}\mathcal{P}\textrtaill{lot}$ histogram:
\begin{equation}
    N_n \ _s\tilde{M}_n(\bar{x})\delta x = \sum_{e\in \delta x}{ _s\mathcal{P}_n(y_e)}
\end{equation}
As a final consideration, it is worth noting that due to the inclusion of the covariance matrix elements in the calculation of the sWeight (eq. \ref{eq:splots}), it is possible to obtain negative weights associated to negative elements in the covariance matrix. These negative weights play a fundamental role in the $_{s}\mathcal{P}\textrtaill{lot}$ production of a given variable, contributing to canceling the components associated to other species \cite{Pivk_2005}\cite{PIVK_2006}.

%% file: appendices/appendixB.tex

\chapter{Multi energy X-Ray production}
\label{app:appendixB}

In this section, a detailed description of the radioactive calibration sources utilized will be provided. Specifically, the $^{55}Fe$ source and the Amersham $^{241}$Am based source will be described, together with the methodologies employed to generate monochromatic X-rays at various energies starting from these two sources.

\section{The Amersham multi-energy X-Ray source}
\begin{figure}
    \centering
    \includegraphics[scale=.3]{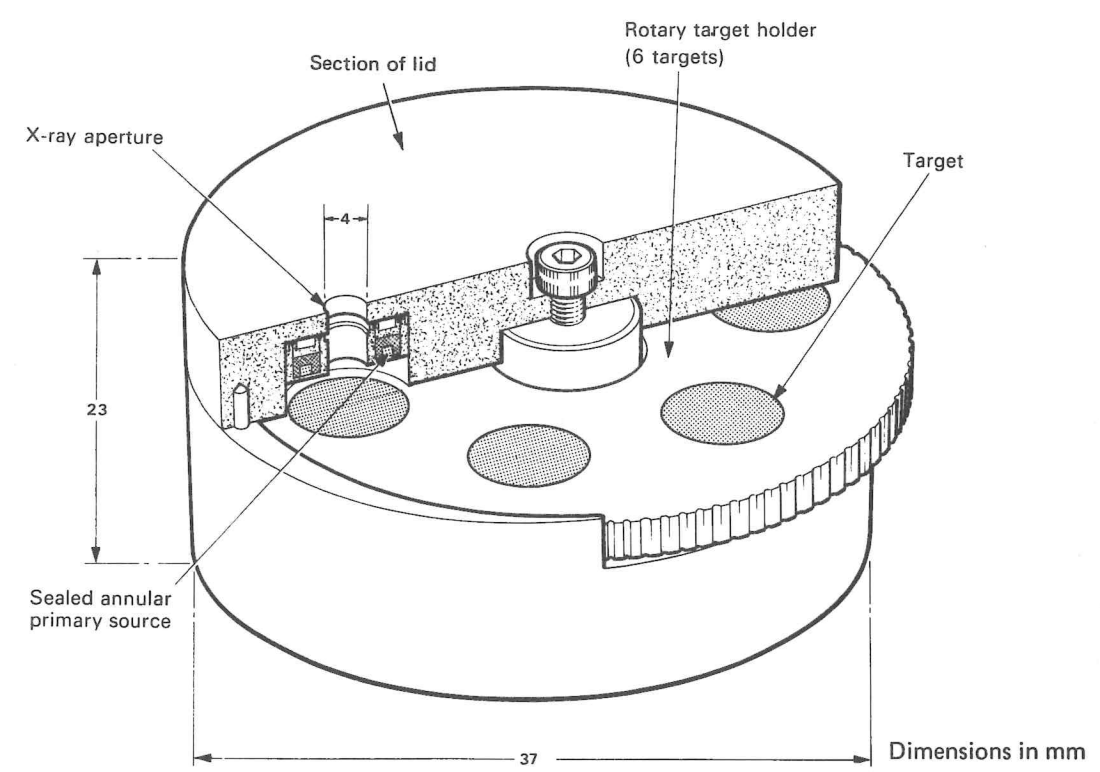}
    \caption{Scheme of the Amersham AMC.2084 variable energy X-ray source}
    \label{fig:source1}
\end{figure}

One of the multi-energy X-ray sources employed is the Amersham AMC.2084. This consists of a 370 MBq $^{241}$Am source producing $\sim$5 MeV alpha particles, directed onto disks composed of different materials placed on a rotating support. The materials used are copper (Cu), rubidium (Rb), molybdenum (Mo), silver (Ag), barium (Ba), and terbium (Tb), and they can be selected through the rotation of a wheel. The wheel rotation causes the disk to be in correspondence with an aperture for the passage on the X-Ray. A scheme of the source is reported in figure \ref{fig:source1}.

\begin{figure}
    \centering
    \includegraphics[scale=.14]{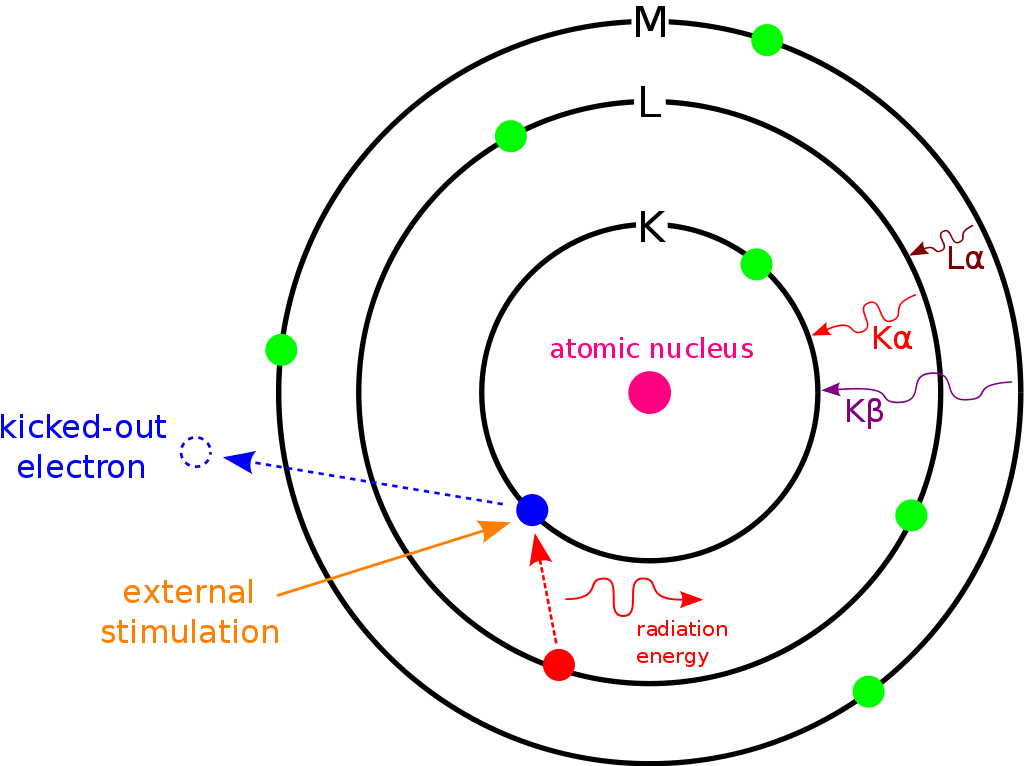}  
    \caption{In figure the scheme of X-ray production for different electronic transitions after an external stimulation is shown.}
    \label{fig:CharRad}
\end{figure}

Upon interaction with the materials, the alpha particles, through ionization, cause the ejection of an electron from the atom's inner shell, K-shell. An outer-shell electron, from the M-shell or L-shell, filling the vacancy causes the production of a monochromatic X-ray, resulting in what is respectively named $k_{\alpha}$ and $k_{\beta}$ emissions. A schematic view of the process is shown in figure \ref{fig:CharRad}. 

The energy of the emitted gamma will be equal to $ E_{\gamma}=E_{K-shell}-E_{M/L-shell}$, where $E_{X-shell}$ represents the bounding energy of the electron in the X-shell. Given that $\Delta E_{M-L}<<\Delta E_{K-ML}$, the resulting emission spectrum will consist of two closely spaced lines, each centered at the respective $E_{\gamma}$. 
The spectrum resulting from the emission of the materials described above is shown in figure \ref{fig:gamma_emission_E}, and the energies of the photons for both $k_{\alpha}$ and $k_{\beta}$ emissions are reported in table \ref{tab:gamma_emission_E}. As can be seen from the plot the yield for the $k_{\beta}$ emission is $\sim$20\% of the $k_{\alpha}$. This happens because the selection rules for electronic transition favor the $k_{\alpha}$ emission.

  \begin{minipage}{\textwidth}
  \begin{minipage}[!t]{0.45\textwidth}
    \centering
\includegraphics[scale=.5]{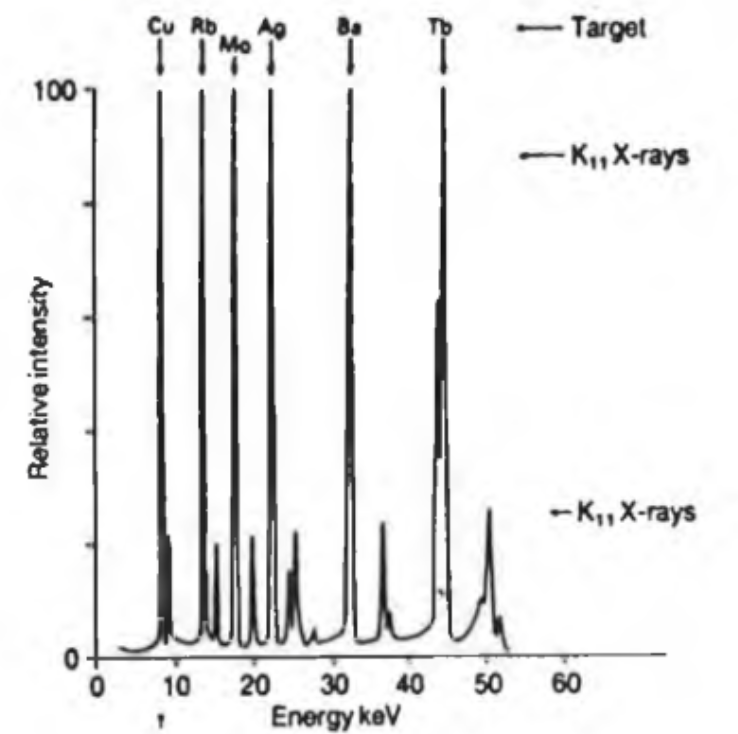}  
    \captionof{figure}{In the figure the spectrum of the source is shown for all the target materials}
    \label{fig:gamma_emission_E}
  \end{minipage}
  \hfill
  \begin{minipage}[!t]{0.45\textwidth}
    \centering
    \begin{tabular}[]{|c|c|c|}                          
    \hline                                                     
        Element & $k_{\alpha}$ \ [keV] & $k_{\beta}$ \ [keV] \\ \hline   Cu & 8.04 & 8.91 \\ \hline                                       
        Rb & 13.37 & 14.97 \\ \hline                                     
        Mo & 17.44 & 19.63 \\ \hline                                     
        Ag & 22.10 & 24.99 \\ \hline                                     
        Ba & 32.06 & 36.55 \\ \hline                                     
        Tb & 44.23 & 50.65 \\ \hline                                     
    \end{tabular}
      \captionof{table}{In the table the energy of the photons for the different $k_{\alpha}$ and $k_{\beta}$ emissions are \
reported}
    \label{tab:gamma_emission_E}
    \end{minipage}
  \end{minipage}

\begin{figure}
    \centering
    \includegraphics[scale=.45]{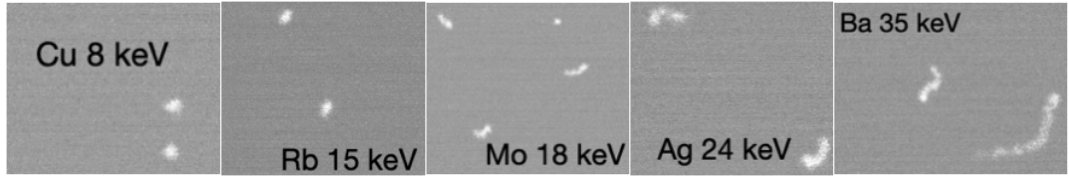}
    \caption{In figure the electron recoils produced by gammas at different energy }
    \label{fig:multiEtracks}
\end{figure}

The gammas interact in the detector through the photoelectric effect and will produce electron recoils. Pictures of the ER resulting from X-ray interaction at different energies are shown in figure \ref{fig:multiEtracks}.

 It can be observed that the tracks appear like a spot at the energy of 8 keV. This occurs because at that energy the distance covered by the electron traveling in the gas is considerably smaller than the size of the diffused tracks. The tracks start to exhibit a significant extension for energies above 15 keV. The source was positioned at the upper section of the detector in correspondence of the thin EFTE window, at the middle point of the sensitive volume along the drift direction.

\section{Ti and Ca X-ray stimulated emission}

\begin{figure}
    \centering
    \includegraphics[scale=.3]{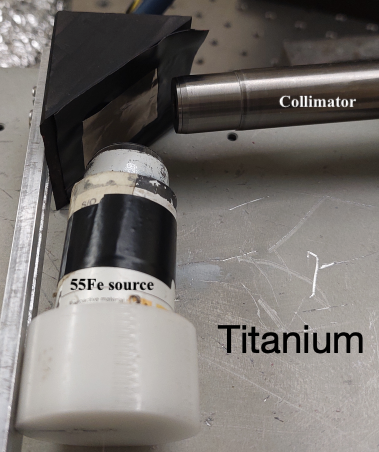}
    \includegraphics[scale=.5]{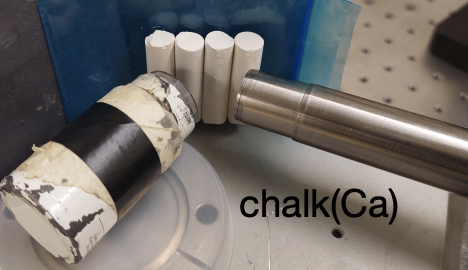}
    \caption{The pictures show the experimental configurations used for the production of 4.5 e 3.7 keV X-ray from titanium and Calcium.}
    \label{fig:CaTiSource}
\end{figure}

To investigate the detector's response at energies below 8 keV the 740 MBq source of $^{55}$Fe has been employed.
For the production of further lower energy X-ray, the same $^{55}$Fe source has been used to irradiate with gammas a titanium (Ti) sheet and chalk pieces containing Calcium (Ca). These elements have an energy for the $k_{\alpha}$ emission of 4.5 keV and 3.7 keV, respectively. The setup in these configurations is shown in figure \ref{fig:CaTiSource} where the collimator for the X-ray from the target material to the detector is also shown.

%% file: appendices/appendixC.tex

\chapter{$\nu_e e$ES cross-section: a detailed calculation}
\label{app:appendixC}

\begin{figure}
    \centering
    \includegraphics[scale=.3]{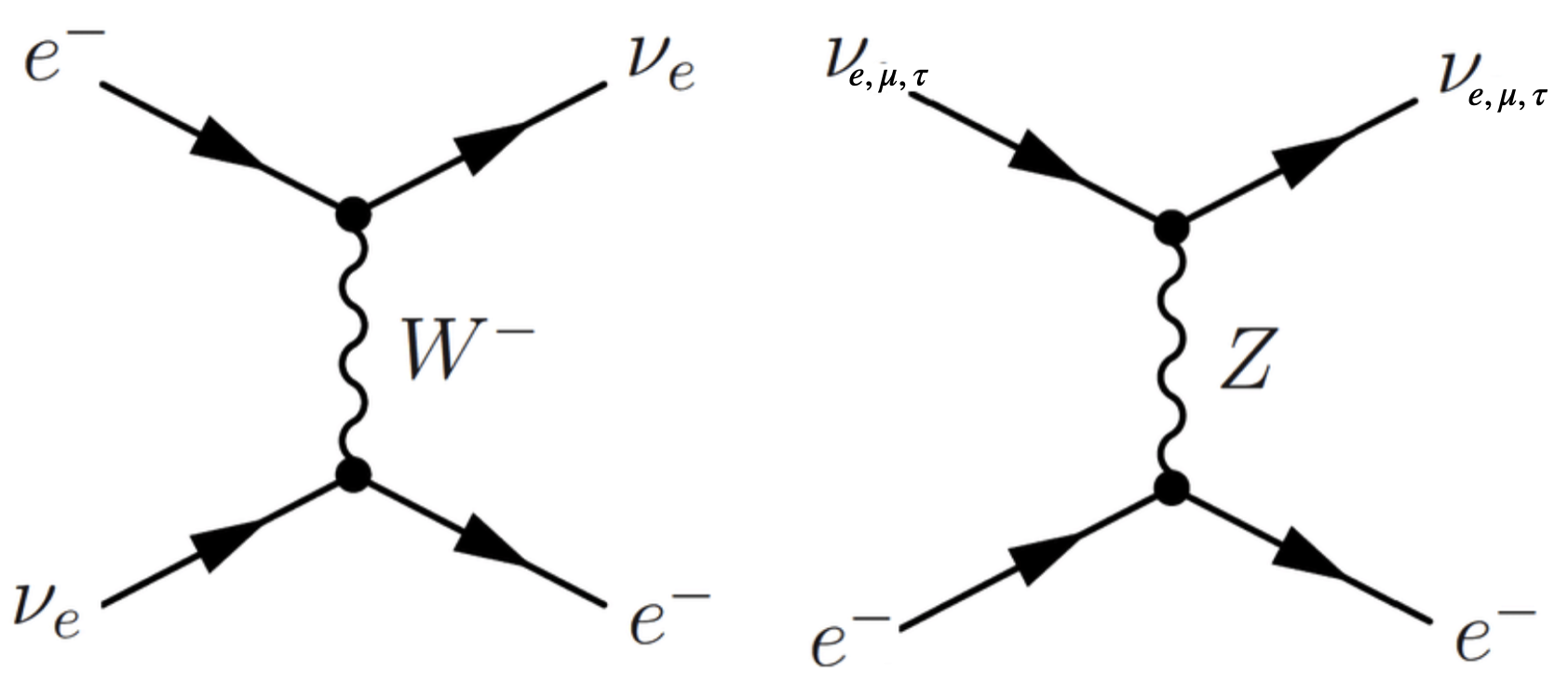}
    \caption{Left: Feynman diagram of electron neutrino interacting on an electron through a charge current interaction. Right: neutrino interacting on an electron through a neutral current interaction.}
    \label{fig:nuIntOneApp}
\end{figure}

 In this appendix, we provide a detailed calculation of electron neutrino scattering on an electron, which can occur through elastic scattering processes mediated by charged current (CC) or neutral currents (NC) (figure \ref{fig:nuIntOneApp}). 
 For a non-electron neutrino, the interaction occurs solely through a neutral current interaction, and the calculation can be straightforwardly derived by excluding the CC interaction from the matrix element.

\section*{General cross section}
 For a generic interaction process where two particles (A and B) interact, resulting in N particles in the final state, the cross section can be formulated as:
\begin{equation}
    d\sigma=\frac{1}{s_0}\frac{1}{s_1}...\frac{1}{s_N}\frac{1}{\varphi}|M_{i\rightarrow f}|^2(2\pi)^4\delta^4\left(P_A+P_B-\sum_{i=0}^{i=N}P_i\right)\cdot \prod_{i=0}^{i=N}\left(\frac{d^3P_i}{2E_i(2\pi)^3}\right)
    \label{CrossSection}
\end{equation}
    Where $S_i$ represents the number of possible spin values of the outgoing particles, $\varphi$ denotes the flux, $M_{i\rightarrow f}$ is the amplitude of the scattering process or matrix element, $\delta^4()$ ensures conservation of the 4-momenta, and the final part in the product denotes the phase space element. 

    \section*{The matrix element}
    From the Lagrangian of the Electroweak interaction (Sec. \ref{sec:stdmodelNu}), the matrix elements of the charge current process and the neutral current process can be derived. Defining $k$ and $k'$ the 4-momenta of the neutrino in the initial and final state respectively, and $p$ and $p'$ the 4-momenta of the electron in the initial and final state respectively, the matrix element of the CC process can be expressed, following the rules of Feynman diagrams, as:   
    \begin{equation}
        M_{CC}=\frac{g^2_{w}}{8}[\bar{e}(p')\gamma^\mu(1-\gamma^5)\nu_e(k)]\frac{-g_{\mu\nu}+\frac{q^\mu q^\nu}{M_W^2}}{q^2-M_W^2}[\bar{\nu_e}(k')\gamma^\nu(1-\gamma^5)e(p)]
    \end{equation}
    In the equation, $g_{w}$ represents the coupling constant of the weak interaction, $q=k-p'=k'-p$ denotes the transferred 4-momentum, and $M_w$ stands for the mass of the W boson. In this interaction, both the neutrino and the electron are left-handed, resulting in only one possible value for the spin of each particle.
    Similarly, the matrix element of the NC process can be written as:
    \begin{align}
    \begin{split}   
        M_{NC}=&\frac{g^2_{\omega}}{8 \cos^2(\theta_W)}\bar{\nu_e}(k')\gamma^\mu(1-\gamma^5)\nu_e(k)\frac{-g_{\mu\nu}+\frac{q^\mu q^\nu}{M_Z^2}}{q^2-M_Z^2} \cdot \\ &[g_L\bar{e}(p')\gamma^\nu(1-\gamma^5)e(p)+g_R\bar{e}(p')\gamma^\nu(1+\gamma^5)e(p)] 
    \end{split}
    \end{align}    
    Given that solar neutrinos from the $ppI$ cycle possess energy $E_\nu << M_W$, it's evident that the transferred 4-momentum to the electron, $q^2$, is also very small compared to the W boson mass ($q^2<<M_W$). Consequently, the quantities in the propagators simplify to:
    \begin{equation}
         \frac{q^\mu q^\nu}{M_{W/Z}^2}\sim 0\ \ \ \ \ \ \ \  q^2- M_{W/Z}^2\sim -M_{W/Z}^2
    \end{equation}
    Moreover defining: 
    \begin{equation}
    \frac{G_F}{\sqrt{2}}=\frac{g_\omega^2}{8 M_W^2}=\frac{g_\omega^2}{8 M_Z^2\ cos^2(\theta_W)} 
    \end{equation}
    the amplitudes of the two processes can be rewritten as:
    \begin{equation}
        M_{CC}=\frac{G_F}{\sqrt{2}}[\bar{e}(p')\gamma^\mu(1-\gamma^5)\nu_e(k)]g_{\mu \nu}[\bar{\nu_e}(k')\gamma^\nu(1-\gamma^5)e(p)]
    \end{equation}
    
    \begin{align}
    \begin{split}
        M_{NC}=&\frac{G_F}{\sqrt{2}}[\bar{\nu_e}(k')\gamma^\mu(1-\gamma^5)\nu_e(k)g_{\mu\nu}]\\ & \cdot [g_L\bar{e}(p')\gamma^\nu(1-\gamma^5)e(p)+g_R\bar{e}(p')\gamma^\nu(1+\gamma^5)e(p)]
     \end{split}   
    \end{align}
    where $g_L=-1/2+sin^2(\theta_W)$ and $g_R=sin^2(\theta_W)$
    The final states of the NC and CC processes are indistinguishable, so they enter in the cross section calculation as a superposition of the two processes. Thus to calculate the total cross section, the amplitude of the entire process is taken as the modulus squared of the sum of the amplitudes of the two processes, as $|M_{i\rightarrow f}|^2=|M_{CC}+M_{NC}|^2$.
    To simplify the calculation, it is useful to consider the following Fierz identities \cite{Nieves_2004}:
    \begin{equation}
        \bar{u_1}\gamma^\mu(1-\gamma^5)u_2\cdot \bar{u_3}\gamma_\mu(1-\gamma^5)u_4=-\bar{u_1}\gamma^\mu(1-\gamma^5)u_4\cdot \bar{u_3}\gamma_\mu(1-\gamma^5)u_2
    \label{FID}    
    \end{equation}
    Utilizing the Fierz identity in Eq. \ref{FID} to rewrite the term $M_{CC}$, and defining $g_V=g_L+g_R$, $g_A=g_L-g_R$, $V=1+g_V$, and $A=1+g_A$, the total amplitude for the process can be expressed as:
    \begin{equation}
        M_{i\rightarrow f}=M_{CC}+M_{NC}=\frac{G_F}{\sqrt{2}}\cdot \bar{\nu_e}(k')\gamma^\mu(1-\gamma^5)\nu_e(k) \cdot \bar{e}(p')\gamma_\mu(V-A\gamma^5)e(p)
    \end{equation}    
    By taking the square modulus of $M_{i\rightarrow f}$ and noting that $u_\alpha(p') \bar{u}{\beta}(p')=(\slashed{p}'+m){\alpha\beta}$, where $\slashed{p}=\gamma^a p_a$, it can be obtained, expliciting all the spinorial indices:
    \begin{align}
\begin{split} 
        |M_{i\rightarrow f}|^2 =& \frac{G_F^2}{2}
        \bar{\nu_e}(k')\gamma^\mu(1-\gamma^5)\nu_e(k) \cdot \bar{e}(p')\gamma_\mu(V-A\gamma^5)e(p) \\ &\bar{e}(p)\gamma_\lambda(V-A\gamma^5)e(p')\bar{\nu_e}(k)\gamma^\lambda(1-\gamma^5)\nu_e(k')= \\
        =& \frac{G_F^2}{2}
        (\bar{\nu_e}(k'))_\alpha(\gamma^\mu(1-\gamma^5))_{\alpha\beta}(\nu_e(k))_{\beta} \cdot (\bar{e}(p'))_\eta(\gamma_\mu(V-A\gamma^5))_{\eta\theta}(e(p))_\theta \\ &\cdot (\bar{e}(p))_{\delta}(\gamma_\lambda(V-A\gamma^5))_{\delta\epsilon}(e(p'))_{\epsilon}\cdot(\bar{\nu_e}(k))_{\rho}(\gamma^\lambda(1-\gamma^5))_{\rho\sigma}(\nu_e(k'))_{\sigma}
    \end{split}
    \end{align}
    By rearranging the terms to couple all spinorial fields in the form  $u_\alpha(p') \bar{u}_{\beta}(p')$, and considering that, given A,B,C and D matrices, $A_{\alpha\beta}B_{\beta\delta}C_{\delta\epsilon}D_{\epsilon\alpha}=(ABCD)_{\alpha\alpha}=Tr(ABCD)$. The $|M_{i\rightarrow f}|^2$ become:
    \begin{align}
    \begin{split}
        |M_{i\rightarrow f}|^2 =\frac{G_F^2}{2}\{Tr(\slashed{k}'\gamma^{\mu}(1-\gamma^5)\slashed{k}\gamma^\lambda(1-\gamma^5))\}\cdot \\  \{Tr((\slashed{p}'+m_e)\gamma_{\mu}(V-A\gamma^5)(\slashed{p}+m_e)\gamma^\lambda(V-A\gamma^5))\}
    \end{split}
    \end{align}
    
    Recalling that the trace of an odd number of gamma matrices is 0 and utilizing the standard properties of the gamma matrices, specifically:
    \begin{itemize}
        \item $\gamma^5$ anticommute with any $\gamma$ matrix
        \item $Tr(\gamma^a\gamma^b)=4g^{ab}$
        \item $Tr(\gamma^a\gamma^b\gamma^c\gamma^d)=4(g^{ab}g^{cd}-g^{ac}g^{bd}+g^{ad}g^{bc})$
        \item $Tr(\gamma^a\gamma^b\gamma^c\gamma^d\gamma^5)=4i\epsilon^{abcd}$
        \item $\epsilon^{abcd}\epsilon_{aecf}=-2(\delta^b_e\delta^d_f-\delta^b_f\delta^d_e)$
    \end{itemize}
    The matrix element reduces to
    \begin{align}
\begin{split}
        |M_{i\rightarrow f}|^2=64\cdot\frac{G_F^2}{2}\{&(V^2+A^2)\cdot[(k'p')(kp)+(k'p)(kp')]- \\
        &(V^2-A^2)\cdot[k'k]+2VA\cdot[(k'p')(kp)-(k'p)(kp')] \}
    \end{split}
    \end{align}
   In the electron's reference frame, the following equations, derived from the conservation of 4-momentum $p+k=p'+k'$ and subsequently squared, hold true:
    \begin{itemize}
        \item $k'p'=kp=m_e E_\nu$ where $m_e$ is the mass of the electron and $E_\nu$ is the energy of the incoming neutrino
        \item $k'p=kp'=m_eE'_{\nu}=me(E_\nu-T'_e)$, where it has been used the relation $T'_e=E_\nu-E'_\nu=E'_e-m_e$, and $T'_e$, and $E'_e$ are respectively the kinetic energy and the energy of the electron in the final state.
        \item $k'k=-m_e^2+p'p=-m_e^2+E'_em_e=m_eT'_e$
    \end{itemize}
    
    With some calculation the modulus square of the amplitude of the process reduces to 
    \begin{equation}
        |M_{i\rightarrow f}|^2=64\frac{G_F^2 E_\nu^2 m_e^2}{2}\left\{(V+A)^2+(V-A)^2\left(1-\frac{T'_e}{E_\nu}\right)^2-(V^2-A^2)\left(\frac{m_eT'_2}{E_\nu^2}\right)\right\}
    \end{equation}
    
    \section*{The Flux}
    The flux $\varphi$ can be evaluated in the following way
    \begin{equation}
        \varphi=4E_\nu E_e|\vec{v}_\nu-\vec{v}_e|
    \end{equation}
    Due to the fact that the electron is at rest, and the neutrino travels at the speed of light, the flux factor reduces to
    \begin{equation}
        \varphi=4m_eE_\nu
    \end{equation}
    
    \section*{The phase space factor}
    The phase space factor in case of $e-\nu$ scattering consists of the following pieces
    \begin{equation}
        d\lambda =\frac{1}{(2\pi)^2} \delta^4(p+k-p'-k')\frac{d^3p'}{2E_e'}\frac{d^3k'}{2E_\nu'}
    \end{equation}
    where $d^3p'$ and $d^3k'$ represent the infinitesimal elements of 3-momentum for the electron and neutrino, while $E'_\nu$ and $E'_e$ denote the energies of the neutrino and electron in the final state, respectively.
    Since the modulus square of the matrix element depends solely on the initial neutrino energy and the kinetic energy of the final electron, integration over all parameters except the final electron energy becomes straightforward.
    Utilizing the properties of the Dirac delta function, the phase space factor can be expressed as follows:  
    \begin{equation}
        d\lambda =\frac{1}{(2\pi)^2} \delta^4(k'-(p+k-p'))\frac{d^3p'}{2E_e'}\delta(k'^2)d^4k'
    \end{equation}
    The integral over $d^4k'$ can be readily computed, and upon expanding the square in the second Dirac delta, the above equation simplifies to:
    \begin{equation}
        d\lambda= \frac{1}{2(2\pi)^2}\delta(E_\nu P'_e cos(\theta)-T'_e(m_e+E_\nu))\frac{d^3p'}{2E'_e}
    \end{equation}
    Expressing the integration element in polar coordinates as $d^3p' = p'^2 \ dp \ d\cos(\theta)\ d\eta$, and integrating over $d\cos(\theta)$ and $d\eta$, while also considering the equality $p' \ dp' = E_e' \ dE_e'$ and $dE_e' = dT_e'$, the final expression for the phase space element is:
    \begin{equation}
        d\lambda=\frac{1}{8\pi}\frac{dT'_e}{E_\nu}
    \end{equation}
    \section*{Total Cross Section}
    Combining all components and simplifying the terms, the total differential cross-section, with $V$ and $A$ rewritten as $V=1+g_V$ and $A=1+g_A$, can be expressed as:
    \begin{align}
\begin{split}
        \frac{d\sigma_{\nu_e-e}(E_\nu,T'_e)}{dT'_e}=\frac{G_F^2m_e}{2\pi}&\left\{(2+g_V+g_A)^2 +(g_V-g_A)^2\left(1-\frac{T'_e}{E_\nu}\right)^2\right.-\\
        &-\left.(g_V-g_A)(g_V+g_A+2)\frac{m_eT'_e}{E_\nu^2}\right\}
\end{split}
\end{align} 
    The cross section for muon and tau neutrinos can be obtained in the same manner by considering $|M_{i\rightarrow f}|^2=|M_{NC}|^2$. This yields:
    \begin{equation}
\small
        \frac{d\sigma_{\nu_\mu-e}(E_\nu,T'_e)}{dT'_e}=\frac{G_F^2m_e}{2\pi}\left\{(g_V+g_A)^2+(g_V-g_A)^2\left(1-\frac{T'_e}{E_\nu}\right)^2-(g_V^2-g_A^2)\frac{m_eT'_e}{E_\nu^2}\right\}
\end{equation}

It can be noticed that in case of no threshold and $E_\nu>>m_e$, the previous expression reduces to 
    
    \begin{equation}
        \sigma(E_\nu)\simeq \frac{G_F^2 m_e}{2\pi}\left\{(g_V+g_A+2)^2+\frac{(g_V-g_A)^2}{3}\right\}E_\nu
    \end{equation}
    
    By substituting the numerical values into the equation and multiplying by $(\hbar c)^2=3.8\cdot 10^{-28}, \text{GeV}^2, \text{cm}^2$, the previous expression becomes:
    
    \begin{equation}
        \sigma(E_\nu)\simeq 9.7\cdot 10^{-42} E_\nu \frac{cm^2}{GeV} = 0.97 \cdot 10^{-44} E_\nu \frac{cm^2}{MeV}
    \end{equation}
    giving a compatible result with \cite{Vissani}.

%% file: appendices/appendixD.tex
\chapter{Markov Chain Monte Carlo: Random Walk Metropolis-Hasting}
\label{app:appendixD}
\begin{figure}
    \centering
    \includegraphics[scale=.5]{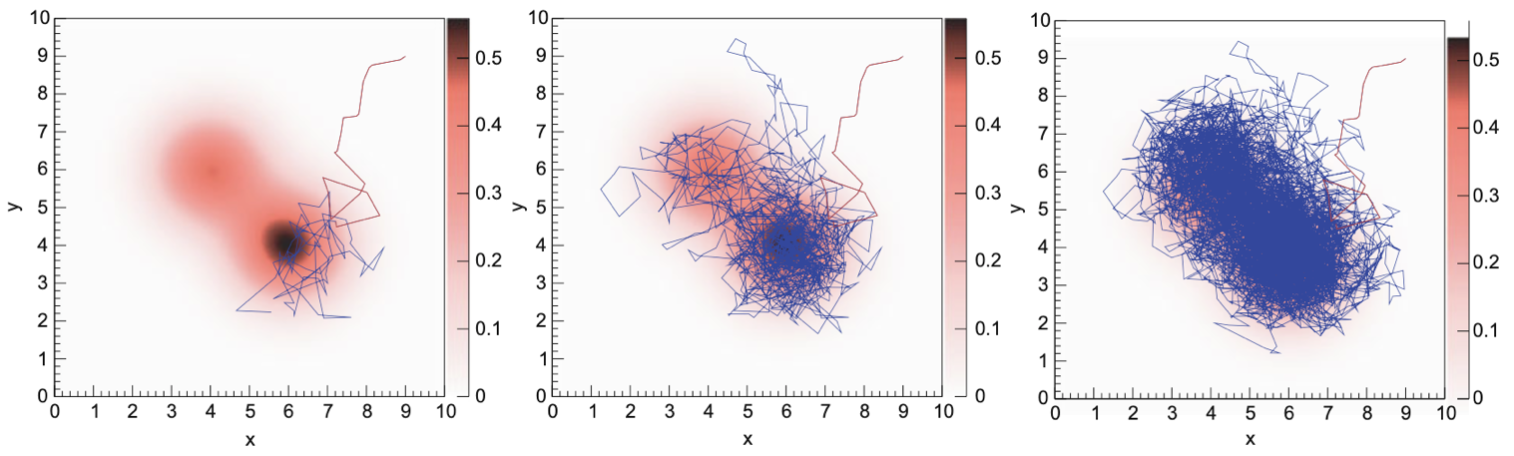}
    \caption{Monte Carlo Markov Chain random walk Metropolis-Hasting applied to the sampling of a double Gaussian distribution. The plots show in order a chain of 100, 2000, and 5000 points.}
    \label{fig:MCMC}
\end{figure}

Markov Chain Monte Carlo is a class of sampling algorithms that can be used to sample distributions with unknown analytical forms. In particular, a given a sequence of random points in the space ($\vec{x}_0$,...,$\vec{x}_n$), this sequence is called a Markov-Chain if the probability distribution for $\vec{x}_{n+1}$, $f_n$, is such that:
\begin{equation}
    f_n(\vec{x}_{n+1}|\vec{x}_0,...,\vec{x}_n )=f_n(\vec{x}_{n+1}| \vec{x}_n)
\end{equation}
which means that the probability density function of the point $\vec{x}$ at the step $n+1$ depends only on the point at the step $n$. The Markov Chain is said to be homogeneous if the function $f$ is the same at every step.
\begin{equation}
f_n(\vec{x}_{n+1}|\vec{x}_n)=f(\vec{x}_{n+1}|\vec{x}_n)
\end{equation}
Given a PDF to sample $h(\vec{x})$, the method starts from a point $\vec{x}_0$ in the domain of the PDF. A second point $\vec{x}$ is generated according to a PDF $g(\vec{x}|\vec{x}_0)$. The PDF $g$ is called the proposal function, which is essentially the function that is used to generate the new proposed points to sample $h$. For the new points $\vec{x}$ proposed the Hasting test ratio $r$ is computed:
\begin{equation}
    r=min\left(1\ , \ \frac{h(\vec{x}) \cdot g(\vec{x}_0|\vec{x})}{ h(\vec{x}_0) \cdot g(\vec{x}|\vec{x}_0)}  \right)
\end{equation}
Subsequently, a random value $u$ between 0 and 1 is randomly extracted. The new point $\vec{x}$ is accepted if $u\leq r$, otherwise it is rejected, and a new extraction of $\vec{x}$ is performed. If the new point is accepted it is labeled as $\vec{x}_1$, and the algorithm is iterated performing a new extraction from $\vec{x}_1$. The process is repeated indefinitely until the desired number of points is sampled. The proposal function and the number of points to be sampled are two important model parameters. A proposal function too spread will not sample the distribution correctly and will result in being inefficient. Contrarily, a proposal function excessively thin might result in a computation time too long.  
Except for particular cases the proposal function is chosen to be symmetric, such that $g(\vec{x}_0|\vec{x}) = g(\vec{x}|\vec{x}_0)$ and the Hasting test ratio simplifies into the Metropolis-Hasting ratio which does not depend on $g$ \cite{lista2017statistical,Robert2004}.
It can be further demonstrated that if only a finite set of values is discarded, the sequence of $\vec{x}_i$ follows the PDF, and covers it completely after a sufficiently high number of repetitions due to the ergodicity of the process \cite{lista2017statistical,MCConvErg}. An example of Monte Carlo Markov Chain random walk Metropolis-Hasting applied to a double Gaussian distribution is shown in figure \ref{fig:MCMC}.
In the Bayesian Analysis Toolkit (BAT), the toolkit used to perform the sensitivity studies in this thesis, an adaptive random walk Metropolis Hasting is implemented. In this implementation a symmetric proposal function with an additional parameter $\xi$, $g(\vec{x}_{i+1},\xi |\vec{x}_{i}) $ is used. Then in a first prerun phase, the chain is run and periodically the value of $\xi$ is changed based on the past iterations. Subsequently, once the optimal $\xi$ is found a proper Markov Chain is used in the main run with fixed $\xi$. To declare the convergence of the Markov Chain, multiple chains are run from different starting points and the convergence is declared if the chains explore the same region of the PDF domain. In BAT, used for the sensitivity analysis in this thesis, for the posterior sampling a symmetric proposal function with an additional parameter $\xi$, $g(\vec{x}_{i+1},\xi |\vec{x}_{i}) $ is used. Then in a first prerun phase, the chain is run and periodically the value of $\xi$ is changed based on the past iterations. Subsequently, once the optimal $\xi$ is found a proper Markov Chain is used in the main run with fixed $\xi$. To declare the convergence of the Markov Chain, multiple chains are run from different starting points and the convergence is declared if the chains explore the same region of the PDF domain. 

%% file: appendices/AppendixE.tex
\chapter{Table of pp solar neutrino flux}

\begin{table}[]
\centering
\begin{tabular}{|cc|cc|cc|cc|}
\hline
q [MeV] & $\phi(q)$ $[cm^{-2}s^{-1}]$ & q & $\phi(q)$  & q  & $\phi(q)$ & q  & $\phi(q)$ \\ \hline
0.00504 & 0.0035 & 0.11089 & 1.2477 & 0.21675 & 3.2300 & 0.32260 & 4.0356 \\
0.01008 & 0.0138 & 0.11593 & 1.3417 & 0.22179 & 3.3094 & 0.32764 & 4.0114 \\
0.01512 & 0.0307 & 0.12097 & 1.4370 & 0.22683 & 3.3859 & 0.33268 & 3.9794 \\
0.02016 & 0.0538 & 0.12601 & 1.5335 & 0.23187 & 3.4594 & 0.33772 & 3.9391 \\
0.02520 & 0.0830 & 0.13106 & 1.6310 & 0.23691 & 3.5298 & 0.34276 & 3.8900 \\
0.03024 & 0.1179 & 0.13610 & 1.7291 & 0.24195 & 3.5966 & 0.34780 & 3.8316 \\
0.03528 & 0.1582 & 0.14114 & 1.8278 & 0.24699 & 3.6599 & 0.35284 & 3.7632 \\
0.04032 & 0.2038 & 0.14618 & 1.9267 & 0.25203 & 3.7194 & 0.35788 & 3.6842 \\
0.04537 & 0.2543 & 0.15122 & 2.0258 & 0.25707 & 3.7749 & 0.36292 & 3.5937 \\
0.05041 & 0.3094 & 0.15626 & 2.1247 & 0.26211 & 3.8262 & 0.36796 & 3.4907 \\
0.05545 & 0.3691 & 0.16130 & 2.2233 & 0.26715 & 3.8731 & 0.37300 & 3.3740 \\
0.06049 & 0.4329 & 0.16634 & 2.3214 & 0.27219 & 3.9154 & 0.37804 & 3.2422 \\
0.06553 & 0.5006 & 0.17138 & 2.4187 & 0.27723 & 3.9529 & 0.38309 & 3.0932 \\
0.07057 & 0.5721 & 0.17642 & 2.5151 & 0.28227 & 3.9854 & 0.38813 & 2.9246 \\
0.07561 & 0.6469 & 0.18146 & 2.6105 & 0.28731 & 4.0127 & 0.39317 & 2.7330 \\
0.08065 & 0.7250 & 0.18650 & 2.7044 & 0.29235 & 4.0344 & 0.39821 & 2.5136 \\
0.08569 & 0.8061 & 0.19154 & 2.7969 & 0.29740 & 4.0505 & 0.40325 & 2.2589 \\
0.09073 & 0.8899 & 0.19658 & 2.8877 & 0.30244 & 4.0605 & 0.40829 & 1.9567 \\
0.09577 & 0.9761 & 0.20162 & 2.9766 & 0.30748 & 4.0644 & 0.41333 & 1.5832 \\
0.10081 & 1.0647 & 0.20666 & 3.0634 & 0.31252 & 4.0617 & 0.41837 & 1.0783 \\
0.10585 & 1.1553 & 0.21171 & 3.1479 & 0.31756 & 4.0522 & 0.42341 & 0.0000 \\ \hline
\end{tabular}
\caption{In table the neutrino solar spectrum from the $pp$ cycle is reported from \cite{bahcall1989neutrino}. The values q and $\phi(q)$ are respectively the energy of the neutrino in $MeV$ and the flux scaled by a factor $10^{11}$.}
\label{TabNu}
\end{table}